\pdfoutput=1

\documentclass[final,3p,times]{elsarticle}

\usepackage{epsfig}
\usepackage{amssymb}
\usepackage{amsmath,amsfonts}
\usepackage{bm}
\usepackage{color,soul} 
\usepackage{booktabs}

\usepackage{nomencl}
\usepackage{subcaption}
\usepackage[percent]{overpic}
\usepackage{dcolumn}
\usepackage[export]{adjustbox} 
\usepackage{caption}
\usepackage[T1]{fontenc}
\usepackage{lipsum}
\usepackage{dcolumn}

\makenomenclature

\usepackage[colorlinks,
            linkcolor=blue,
            anchorcolor=green,   citecolor=blue]{hyperref}
\usepackage{float}
\usepackage{placeins}

\journal{Journal of Computational Physics}

\begin{document}

\begin{frontmatter}



\title{Uncertainty quantification and stability of neural operators for prediction of three-dimensional turbulence}

\author[1,2]{Xintong Zou}
\author[3]{Zhijie Li}
\author[1,2]{Yunpeng Wang}
\author[1,2]{Huiyu Yang}
\author[1,2]{Jianchun Wang\corref{cor1}}
\cortext[cor1]{Corresponding author.}
\ead{wangjc@sustech.edu.cn}

\address[1]{Department of Mechanics and Aerospace Engineering, Southern University of Science and Technology, Shenzhen 518055, Guangdong, China}
\address[2]{Guangdong Provincial Key Laboratory of Turbulence Research and Applications, Southern University of Science and Technology, Shenzhen, 518055, China}
\address[3]{Department of Biomedical Engineering Applications, National University of Singapore, Singapore, 117583, Singapore}
\begin{abstract}
The uncertain and chaotic nature of turbulence poses significant challenges for numerical simulation, particularly in capturing multiscale dynamics and managing high computational costs. Conventional numerical methods with turbulence modeling face limitations in accuracy, long-term stability, and efficiency, especially in complex physical scenarios.
Recent advances in scientific machine learning (SciML), including surrogate models such as the Fourier neural operator (FNO), have shown promise in solving partial differential equations (PDEs). FNO-based models typically adopt a one-step-ahead prediction framework, where the model predicts the system state at the next time step based solely on the current state. However, many of these models primarily focus on short-term pointwise accuracy and often struggle to maintain stability over long temporal horizons, particularly in three-dimensional turbulent flows.
To evaluate the reliability of neural operator approaches in general turbulent flow problems, this paper presents a theoretical framework for assessing the trustworthiness and robustness of neural operator models. Three-dimensional forced homogeneous isotropic turbulence (HIT) is employed as a representative example to demonstrate and validate the proposed methodology.
Uncertainty quantification (UQ) is conducted to assess predictive variability across different time resolutions and spatial Fourier modes, where the statistical distributions of prediction errors are utilized to compare the performance and reliability of different models.
Stability is further assessed from a statistical perspective, focusing on error propagation through time-marching and sensitivity to initial perturbations. Notably, certain FNO-based models demonstrate superior statistical stability compared to conventional large eddy simulation (LES) methods.
The autocorrelation function (ACF) is utilized to explore the connection between temporal coherence in the flow field and the reliability of model predictions. 
The results demonstrate that the proposed factorized-implicit FNO (F-IFNO) offers a trade-off between accuracy, long-term stability, and computational efficiency, outperforming conventional numerical solvers and other FNO-based models. In particular, modifying the one-step-ahead paradigm through implicit factorization helps mitigate error accumulation and enhances long-term prediction capability.
The findings underscore that incorporating prediction constraints and selecting appropriate time intervals are crucial for enhancing the robustness and reliability of FNO-based models.
The study highlights the importance of UQ, stability, and temporal correlation in the development of robust operator learning frameworks for turbulent flows and other multi-scale nonlinear dynamic systems.
\end{abstract}


\begin{highlights}
\item A unified framework is proposed to assess the trustworthiness and robustness of FNO-based models in three-dimensional turbulence, incorporating both accuracy and long-term robustness under complex flow dynamics.
\item A comprehensive uncertainty quantification and stability analysis is performed across different temporal intervals and spatial Fourier scales, considering error accumulation in time-marching and sensitivity to initial perturbations. The role of temporal autocorrelation in characterizing flow coherence is systematically explored, revealing its strong correlation with model reliability.
\item The findings highlight that incorporating prediction constraints and optimizing time interval choices are critical to improve the robustness and reliability of FNO-based models. In particular, the proposed F-IFNO achieves a favorable trade-off among accuracy, long-term stability, and computational efficiency.
\end{highlights}


\end{frontmatter}



\section{Introductuion}
\label{sec1}

Due to the inherently chaotic nature, turbulent flows present significant challenges in numerical simulation. Nevertheless, accurately capturing their behavior is essential for enhancing predictive models in a broad spectrum of scientific and engineering applications \cite{pope2000}. To effectively predict turbulence, three major challenges must be addressed. First, the intrinsic multi-scale nature of turbulence demands high-resolution methods to ensure accuracy. Second, long-term iterative simulations suffer from numerical instability. Third, accurately resolving all relevant scales leads to prohibitive computational costs.
Direct numerical simulation (DNS) can fully resolve all turbulent scales, providing highly accurate solutions. However, its computational cost becomes impractical for complex geometries or high Reynolds number flows \cite{fanNeuralDifferentiableModeling2025}. To mitigate this, traditional approaches including Reynolds-averaged Navier-Stokes (RANS) simulation \cite{ROELOFS2019213} and large eddy simulation (LES) \cite{Piomelli1996} have been developed. RANS models employ averaging to simplify turbulence modeling, while LES resolves large-scale energy-carrying structures and models the effects of small-scale turbulence.
Although RANS and LES reduce computational demands by modeling the effects of unresolved turbulence, they still face limitations in capturing the multi-scale turbulence structures accurately and remain computationally expensive for many practical engineering problems.

In recent years, scientific machine learning (SciML) has made significant strides in addressing complex problems in numerical modeling and scientific computation. In particular, deep learning techniques have been increasingly adopted to solve partial differential equations (PDEs) in computational fluid dynamics (CFD), offering promising alternatives to conventional numerical methods including DNS, RANS, and LES \cite{vinuesaEnhancingComputationalFluid2022,JIN2021109951,heMultilevelPhysicsInformed2024}. The key advantage of deep learning lies in its efficiency, enabling faster predictive simulations of turbulence.
One approach in this domain involves using neural networks to directly approximate the solution of PDEs. A representative example is the physics-informed neural network (PINN) \cite{raissiPhysicsInformedDeep2017h,raissiPhysicsInformedDeep2017g}, a class of universal function approximators capable of embedding the governing physical laws expressed as PDEs into the learning process \cite{RAISSI2019686}.
Beyond function approximation, operator learning frameworks have been proposed to model mappings between infinite-dimensional function spaces. Notable examples include the deep operator network (DeepONet) \cite{Lu_2021deeponet} and the Fourier neural operator (FNO) \cite{li2021fourierneuraloperatorparametric,LI2022100389FNO}. These neural operators take functions (such as initial physical fields or parameter distributions) as input and output the corresponding solution fields \cite{KovachkiNeuraloperator}, making them well-suited for problems with variable conditions (e.g., initial or boundary conditions) and enhancing the reusability of pretrained models \cite{yeLocalityLocalNeural2024b}. Importantly, pretrained neural operators can solve the Navier-Stokes equations hundreds of times faster than conventional solvers \cite{li2021fourierneuraloperatorparametric}. Benefiting from the promising performance and computational efficiency of FNO, numerous advanced variants have been developed to further enhance its capabilities.
The implicit Fourier neural operator (IFNO) formulates the solution operator as an implicitly defined mapping by employing a shallow-to-deep training strategy \cite{YOU2022115296}.
The hybrid U-Net enhanced Fourier neural operator (U-FNO) integrates a U-Net pathway within each Fourier layer to perform local convolutions, thereby enhancing the model’s ability to represent high-frequency components \cite{WEN2022104180}.
The implicit U-Net enhanced Fourier neural operator (IUFNO) incorporates a U-Net architecture as a residual component to effectively capture small-scale flow features, while adopting an implicit strategy that enables a shallow-to-deep training approach \cite{liLongtermPredictionsTurbulence2023a}.
The factorized Fourier neural operator (F-FNO) employs a Fourier factorization strategy, learning features independently along each dimension in Fourier space, which significantly reduces model complexity \cite{tranFactorizedFourierNeural2023d}.

Despite their computational advantages, purely data-driven deep learning surrogate models encounter significant challenges when applied to complex turbulent flows. The inherently chaotic and nonlinear dynamics of turbulence lead to the sensitivity to perturbations and rapid error accumulation due to time stepping, which severely compromises long-term predictive stability \cite{liLearningDissipativeDynamics2022a,fanNeuralDifferentiableModeling2025,leiLongtimeIntegrationNonlinear2025}. Most existing models are optimized for short-term trajectory accuracy, yet often fail to maintain reliability over extended time horizons.
Stability, therefore, becomes a key criterion for evaluating the trustworthiness of surrogate models in CFD. Both traditional numerical methods and pretrained neural operators suffer from instability during time-marching procedures, as small prediction errors accumulate over time \cite{sivashinskyNonlinearAnalysisHydrodynamic1977}. As a result, ensuring consistent long-term statistical accuracy and robustness is more critical than achieving short-term pointwise precision \cite{chattopadhyayChallengesLearningMultiscale2024c, matsumotoStableReproducibilityTurbulence2024}. 
These limitations highlight the urgent need for developing turbulence surrogate models that can simultaneously offer computational efficiency, long-term stability, and reliable statistical performance across chaotic regimes.

As turbulent flows are inherently uncertain, the SciML paradigm also faces significant challenges related to uncertainty \cite{ZOU2025117479,zouLeveragingViscousHamilton2024,zouMultiheadPhysicsinformedNeural2025,toscanoInferringVivoMurine,mengFastMultifidelityMethod2021,yinOnedimensionalModelingFractional2019}. The chaotic dynamics of turbulence introduce substantial variability in surrogate-model predictions, making uncertainty quantification (UQ) essential for establishing model reliability and trustworthiness. 
For pretrained neural operators, overall predictive uncertainty is commonly decomposed into two main components: aleatoric and epistemic uncertainty \cite{PSAROS2023111902,zouNeuralUQComprehensiveLibrary2024}. Aleatoric uncertainty stems from the intrinsic randomness or noise present in the data, and is fundamentally irreducible \cite{PSAROS2023111902,zouNeuralUQComprehensiveLibrary2024}. In contrast, epistemic uncertainty arises from model limitations, including insufficient or noisy training data, architectural choices, or overparameterization of the neural network \cite{PSAROS2023111902,zouNeuralUQComprehensiveLibrary2024}. Furthermore, physical model misspecification when the neural model fails to adequately capture the governing physics can further contribute to predictive uncertainty \cite{PSAROS2023111902}.
Given these challenges, rigorous UQ becomes a cornerstone for ensuring robust and interpretable predictions in SciML-based modeling of turbulent flows.

\begin{figure}[ht!]\centering
	\includegraphics[width=1\textwidth]{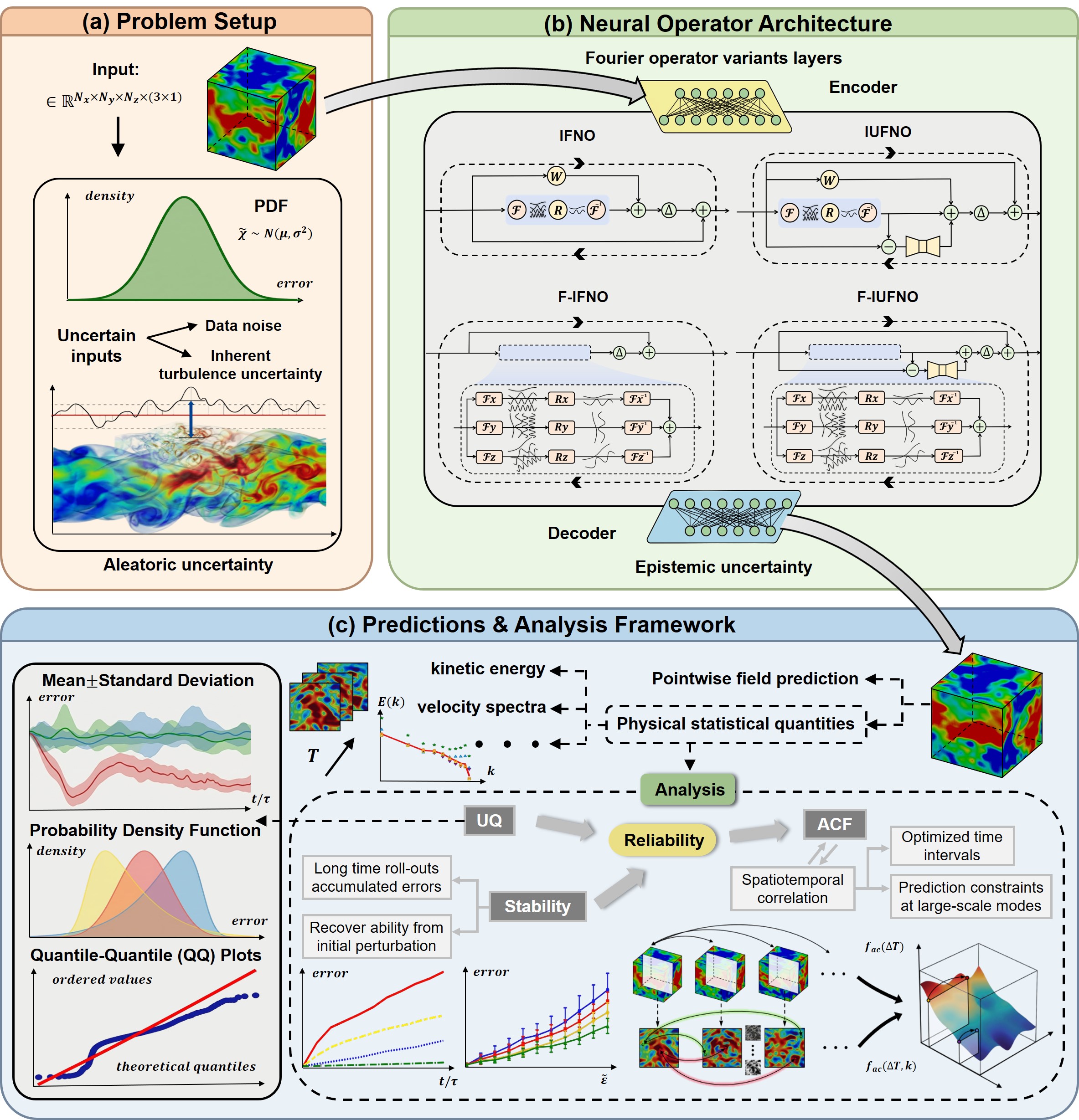}
	\caption{A schematic overview of the problem setup, neural operator architecture, and analysis framework for prediction results. In \textbf{(a)}, we present the problem setup, where a three-dimensional flow field exhibiting aleatoric uncertainty (due to inherent turbulence and data noise) is used as input for surrogate models to make predictions. In \textbf{(b)}, we present four FNO-based model architectures used as surrogate models in this work. In \textbf{(c)}, we present the analysis framework for the posterior prediction results, including uncertainty quantification (UQ) for physical statistical quantities (such as the mean $\pm$ standard deviation of prediction errors, probability density function of prediction errors, and quantile-quantile (QQ) plots), stability analysis for long-term rollouts and accumulated errors, recovery from initial perturbations, and autocorrelation function (ACF) analysis to explore the temporal coherence of turbulent dynamics and error distribution across spatial scales.}\label{fig:1}
\end{figure}

To address the aforementioned challenges, this study presents a systematic analysis of the trustworthiness and reliability of FNO-based models by investigating their uncertainty quantification and long-term predictive stability in the context of three-dimensional forced homogeneous isotropic turbulence (HIT). To further interpret the varying performance of different FNO-based models under different temporal resolutions, we examine the relationship between predictive accuracy and the temporal correlation structure of the underlying flow. In particular, the autocorrelation function (ACF) is adopted as a meaningful proxy to characterize the temporal coherence of turbulent dynamics and to assess model stability in operator learning frameworks \cite{sardarConcerningUseTurbulent2024b, wangFurtherInvestigationData2023}.
Moreover, this study analyzes the error distribution across spatial scales, distinguishing between low-frequency (large-scale) and high-frequency (small-scale) modes. A schematic overview of the work is shown in Fig.~\ref{fig:1}, which integrates the problem setup, the neural-operator architecture, and the framework for posterior analysis. Fig.~\ref{fig:1} illustrates how the systematic analysis framework integrates and organizes the methodology, outlining the entire process of this study. This includes inputting a flow field with aleatoric uncertainty into the surrogate models (introducing epistemic uncertainty) to make predictions, and applying stability analysis and UQ to assess reliability; ACF is also used to analyze the temporal coherence of turbulent dynamics and the distribution of errors across spatial scales. The results of the study emphasize that incorporating prediction constraints and optimizing time intervals are essential for enhancing the robustness and reliability of FNO-based models. Notably, the proposed factorized-implicit Fourier neural operator (F-IFNO) strikes a favorable balance between predictive robustness and computational efficiency, outperforming both traditional numerical solvers and other FNO-based models. 

The remainder of this paper is organized as follows. Section~\ref{sec2} formulates the problem in detail, including the governing equations used to generate data, the methods for UQ and stability analysis, the application of ACF analysis, and quantile-quantile (QQ) plot technique for UQ analysis. Section~\ref{sec3} describes the numerical approaches adopted for both conventional numerical solver and the training of neural operator models. Section~\ref{sec4} presents and discusses the results of numerical experiments. Section~\ref{sec5} discusses broader implications, key findings, limitations, and future work. Section~\ref{sec6} concludes the paper.

\section{Problem statement}
\label{sec2}
This section provides an introduction to governing equations, the methods for uncertainty quantification and stability analysis, and the analysis of ACF.

\subsection{Governing equations}
\label{subsec2.1}
The Navier-Stokes equations in conservative form for incompressible three-dimensional turbulence are given by \cite{pope2000,sagaut2006,annurev:/content/journals/10.1146/annurev.fluid.010908.165203}
\begin{equation}
\frac{\partial u_{i}}{\partial x_{i}} =0,
\label{eq:1}
\end{equation}
\begin{equation}
\frac{\partial u_{i}}{\partial t} + \frac{\partial (u_{i}u_{j})}{\partial x_{j}}=-\frac{\partial p}{\partial x_{i}}+\nu \frac{\partial^2 u_{i}}{\partial x_{j}\partial x_{j}}+F_{i}.
\label{eq:2}
\end{equation}
Here $u_{i}$ denotes the $i$-th component of the velocity field, $p$ represents the pressure, $\nu$ is the kinematic viscosity, and $F_{i}$ corresponds to a large-scale external forcing term acting on the momentum equation in the $i$-th coordinate direction. Throughout this work, the Einstein summation convention is adopted.

Additionally, the kinetic energy $E_{k}$ is defined as \cite{pope2000}
\begin{equation}
E_{k}=\int_{0}^{\infty } E(k)\mathrm{d}k =\frac{1}{2} (u^{rms} )^{2}, 
\label{eq:3}
\end{equation}
\begin{equation}
u^{rms}=\sqrt{\left \langle u_{i}  u_{i}\right \rangle },
\label{eq:4}
\end{equation}
where $E(k)$ denotes the energy spectrum, and $u^{rms}$ represents the root-mean-square velocity, and $\left \langle \cdot \right \rangle $ indicates a spatial average taken along the homogeneous directions. Besides, the Taylor length scale $\lambda$, the Taylor-scale Reynolds number $Re_{\lambda}$ and the Kolmogorov length scale $\eta$ are defined, respectively, as \cite{pope2000,wangConstantcoefficientSpatialGradient2022}
\begin{equation}
\lambda =\sqrt{\frac{5\nu }{\varepsilon } } u^{rms},
\label{eq:5}
\end{equation}
\begin{equation}
Re_{\lambda }=\frac{u^{rms}\lambda }{\sqrt{3} \nu }, 
\label{eq:6}
\end{equation}
\begin{equation}
\eta =\left (\frac{\nu^{3} }{\varepsilon }\right )^{1/4},
\label{eq:7}
\end{equation}
where $\varepsilon=2\nu\left \langle S_{ij} S_{ij} \right \rangle $ stands for the average dissipation rate of kinetic energy and $S_{ij}=\frac{1}{2} (\frac{\partial u_{i}}{\partial x_{j}}+\frac{\partial u_{j}}{\partial x_{i}} )$ denotes the strain rate tensor. Furthermore, the integral length scale $L_{I}$ and the large-eddy turnover time $\tau$ are given by \cite{pope2000}
\begin{equation}
L_{I}=\frac{3\pi }{2(u^{rms})^{2}} \int_{0}^{\infty } \frac{E(k)}{k}\mathrm{d}k,  
\label{eq:8}
\end{equation}
\begin{equation}
\tau=\frac{L_{I}}{u^{rms}}. 
\label{eq:9}
\end{equation}

The physical fields in turbulent flows can be partitioned into resolved large-scale components and sub-filter-scale quantities by employing a spatial filtering technique \cite{lesieur1996,meneveau2000,leonard1975}. The filtering operation is mathematically expressed as
\begin{equation}
\bar{f}(\mathbf{x}) = \int_{\Omega} f(\mathbf{x} - \mathbf{r}) G(\mathbf{r}, \mathbf{x}; \Delta)\, \mathrm{d}\mathbf{r},
\label{eq:10}
\end{equation}
where \( f \) denotes a physical-space variable and \( \Omega \) represents the spatial domain of interest. The function \( G \) is the filter kernel, and \( \Delta \) characterizes the filter width. In Fourier space, the filtered counterpart of a physical variable \( f \) can be written as
\begin{equation}
\bar{\hat{f}}(\textit{\textbf{k}}) = \hat{G}(\textbf{\textit{k}}) \hat{f}(\textbf{\textit{k}}),
\label{eq:11}
\end{equation}
where \( \hat{G} \) denotes the Fourier transform of the filter kernel. In this study, a sharp spectral cutoff filter defined by
\begin{equation}
\hat{G}(\textbf{\textit{k}}) = H(k_c - |\textbf{\textit{k}}|)
\label{eq:12}
\end{equation}
is adopted for homogeneous isotropic turbulence \cite{pope2000}, with the cutoff wavenumber given by \( k_c = \pi / \Delta \). Here \( H(x) \) is the Heaviside step function, satisfying \( H(x) = 1 \) for \( x \geq 0 \) and \( H(x) = 0 \) otherwise \cite{pope2000}.

Therefore, the filtered incompressible Navier-Stokes equations for the resolved variables can be expressed as \cite{changEffectSubfilterScale2022}
\begin{equation}
\frac{\partial \bar{u} _{i}}{\partial \bar{x} _{i}} =0,
\label{eq:13}
\end{equation}
\begin{equation}
\frac{\partial \bar{u} _{i}}{\partial t} + \frac{\partial (\bar{u} _{i}\bar{u} _{j})}{\partial x_{j}}=-\frac{\partial \bar{p} }{\partial x_{i}}-\frac{\partial \tau_{ij} }{\partial x_{j}} +\nu  \frac{\partial^2 \bar{u} _i}{\partial x_j\partial x_j} +\bar{F} _i.  
\label{eq:14}
\end{equation}
Here, the sub-grid scale (SGS) stress tensor $\tau_{ij}$ is defined by 
\begin{equation}
\tau_{ij}=\overline{u_i u_j} -\bar{u} _i \bar{u} _j.
\label{eq:15}
\end{equation}

In this paper, a pseudo-spectral method is employed to simulate incompressible homogeneous isotropic turbulence (HIT) within a cubic domain of size \((2\pi)^3\), uniformly discretized with periodic boundary conditions. This method is used to generate filtered direct numerical simulation (fDNS) data, which serves as the basis for training and evaluating various neural operator models. Further details on the generation of fDNS data are provided in Section~\ref{sec4}. Moreover, Subsection~\ref{subsec3.1} presents a detailed introduction to the dynamic Smagorinsky model (DSM), a classical eddy viscosity model widely used in LES of turbulent flows.

\subsection{Methods for uncertainty quantification}
\label{subsec2.2}

Uncertainty quantification (UQ) is critical in assessing the trustworthiness and reliability of machine learning models, particularly in complex scientific applications such as turbulence prediction \cite{ZOU2025117479,ABDAR2021243,stahlEvaluationUncertaintyQuantification2020,YANG2021109913,LINKA2022115346,zouLearningDiscoveringMultiple2025,deflorioQuantificationTotalUncertainty2025}. Due to the inherently chaotic nature, turbulent flows introduce substantial uncertainty into the predictions made by surrogate models \cite{pope2000}. The total uncertainty can be attributed to data (noisy, gappy), physical models (misspecification, stochasticity), neural networks (architecture, hyperparameters, overparameterization), and posterior inference \cite{PSAROS2023111902,zouNeuralUQComprehensiveLibrary2024}. For pretrained neural operators, this total uncertainty is commonly decomposed into aleatoric and epistemic components \cite{hullermeierAleatoricEpistemicUncertainty2021,PSAROS2023111902}. The aleatoric uncertainty arises from inherent noise in the data, which is irreducible \cite{PSAROS2023111902,zouNeuralUQComprehensiveLibrary2024}. In contrast, epistemic uncertainty stems from model limitations, including insufficient or noisy training data and the overparameterization of the neural network \cite{PSAROS2023111902,zouNeuralUQComprehensiveLibrary2024}. Additionally, physical model misspecification can further contribute to predictive uncertainty in neural network models \cite{PSAROS2023111902,zouCorrectingModelMisspecification2024}. In this subsection, methods to quantify uncertainty in predictions for pretrained neural operators will be introduced.

For a noisy input-output model: 
\begin{equation}
\tilde{y} =\mathcal{H} (\chi )+\epsilon _{o},
\label{eq:16}
\end{equation}
\begin{equation}
\tilde{\chi} =\chi +\epsilon _{in}.
\label{eq:17}
\end{equation}
Here $\mathcal{H}$ denotes a pretrained neural operator as a surrogate model whose parameters are known and fixed. $\chi$ is the input, and $\tilde{y}$ along with $\tilde{\chi}$ are the observed value of the output and input of $\mathcal{H}$, respectively. $\epsilon _{o}$ and $\epsilon _{in}$ stand for output and input measurement error or noise. Then the Bayesian framework can be introduced to address the noisy inputs-outputs in pretrained neural operators to solve hybrid problems \cite{vandeschootBayesianStatisticsModelling2021,Wright19991261}. Then, following Eqs.~\eqref{eq:16} and~\eqref{eq:17}, the input $\chi$, the model parameters $\theta$ are modeled and likelihood functions for measurements of both the inputs $\chi$ and the outputs $\tilde{y}$ of $\mathcal{H}$ are established. According to Bayes' theorem, we have \cite{dellaportasBayesianAnalysisErrorsinvariables1995, Gleser198124}:
\begin{equation}
p(\theta ,\chi \mid \mathcal{D} )\propto p(\mathcal{D}\mid \theta ,\chi)p(\theta ,\chi),
\label{eq:18}
\end{equation}
where $p(\mathcal{D}\mid \theta ,\chi)$ is the likelihood, and $p(\theta ,\chi)$ is the prior. By the assumption of independence between the parameter $\theta$ and actual input $\tilde{\chi}$, the prior can be expressed as $p(\theta ,\chi)=p(\theta)p(\chi)$. And $\mathcal{D}$ denotes the observed data $\mathcal{D}=\left \{ \tilde{\chi} _{i},\tilde{y} _{i} \right \}_{i=1}^{N}$, where $N$ is the number of data \cite{ZOU2025117479,dellaportasBayesianAnalysisErrorsinvariables1995}. 
In this paper, the noise in inputs and outputs is considered to facilitate a clear comparison of the predictive accuracy among different surrogate models. This approach allows for the evaluation of UQ that arises from model structures and parameters $\theta$. The quantified uncertainties are subsequently analyzed to assess the predictive robustness and reliability of each surrogate model, providing critical insights into their generalization capabilities in turbulent flow predictions. 

Although Eq.~\eqref{eq:18} offers a rigorous Bayesian framework by modeling the joint posterior of model parameters $\theta$ and inputs $\chi$, full Bayesian inference is computationally expensive for high-dimensional surrogate models in turbulence. To address this challenge, we employ an empirical UQ strategy grounded in prediction error statistics (Eqs.~\eqref{eq:19}-\eqref{eq:20}), serving as a frequentist approximation to posterior variability.
This empirical framework aligns with the Bayesian rationale while remaining tractable. Importantly, in three-dimensional turbulence, the key quantities of interest are statistical measures, including kinetic energy $E_k$ and the velocity spectra $E(k)$, rather than pointwise flow fields. While Eq.~\eqref{eq:18} models uncertainty at the field level, our method focuses on the uncertainty in such statistics, which are more physically meaningful and robust.
Hence, the proposed UQ approach can be interpreted as a data-driven, computationally feasible approximation of the Bayesian uncertainty in Eq.~\eqref{eq:18}, suited for turbulence modeling.

In this study, uncertainty quantification (UQ) focuses on statistical measures, including $E_k$ and $E(k)$, rather than pointwise flow fields, reflecting the physical relevance of global quantities in turbulence modeling. Accordingly, UQ is defined based on the prediction errors across $N$ samples as follows \cite{man2025uncertaintyerrorquantificationdatadriven}:
\begin{equation}
\text{UQ} = \frac{1}{N} \sum_{i=1}^{N} \epsilon_i \, \pm \, \sqrt{ \frac{1}{N} \sum_{i=1}^{N} \left( \epsilon_i - \overline{\epsilon} \right)^2 },
\label{eq:19}
\end{equation}
where $\epsilon_i$ denotes the prediction error of the $i$-th sample, and $\overline{\epsilon}$ is the mean prediction error, defined by
\begin{equation}
\overline{\epsilon} = \frac{1}{N} \sum_{i=1}^{N} \epsilon_i.
\label{eq:20}
\end{equation}
The second term, $\sqrt{ \frac{1}{N} \sum_{i=1}^{N} \left( \epsilon_i - \overline{\epsilon} \right)^2 }$, represents the standard deviation of the prediction errors across all the samples, which quantifies the variability or dispersion of the errors.

In this work, UQ of the prediction error $\epsilon$ is evaluated for both the input data and the surrogate model outputs across different physical quantities. Specifically, under the autoregressive prediction framework used in this study, the effective input at each step combines the intrinsic fluctuation of the flow field (fDNS) and the accumulated prediction error from the previous steps. This composite uncertainty corresponds to the input noise term $\epsilon_{\text{in}}$ in Eq.~\eqref{eq:17}. Meanwhile, the output noise $\epsilon_{\text{o}}$ in Eq.~\eqref{eq:16} represents the prediction error at the current step.

This interpretation bridges the empirical error-based UQ with the Bayesian framework introduced in Eq.~\eqref{eq:18}, where both input and output uncertainties are explicitly modeled. While the full Bayesian treatment is computationally expensive, the empirical strategy adopted here offers a practical approximation, grounded in prediction error statistics. In Section~\ref{sec4}, the UQ results are visualized using the mean $\pm$ standard deviation (Mean $\pm$ Std), errorbars, and the probability density function (PDF) of the prediction errors. These consistent representations provide a comprehensive and interpretable view of predictive uncertainty.

\subsection{Methods for stability analysis}
\label{subsec2.3}

Stability is a crucial criterion for evaluating the trustworthiness of a surrogate model in computational fluid dynamics. For conventional numerical methods, including finite difference methods (FDM), finite volume methods (FVM), and spectral methods, numerical stability is a concept that is strictly applicable to time-marching problems. A numerical scheme is considered stable if errors introduced by round-off, truncation, or implementation inaccuracies remain bounded as the computation proceeds from one time step to the next \cite{ComputationalFluidMechanics2020}. To assess the stability of numerical methods for solving nonlinear partial differential equations, such as the Navier-Stokes equations, the von Neumann stability analysis is one of the most widely utilized techniques \cite{doi:10.1137/1011025,crankPracticalMethodNumerical1996,Charney01011950}. This method systematically introduces a means to introduce an error into the solution structure, and study its amplification in time \cite{KONANGI2018643}. In a stable scheme, any initial perturbation remains bounded as the simulation advances, whereas in an unstable scheme, perturbations may grow exponentially and eventually corrupt the solution. Moreover, the von Neumann analysis leads to the derivation of constraints on the Courant-Friedrichs-Lewy (CFL) number \cite{zbMATH02577222,5391985CourantPDE,alma995860290001771}, which provides critical guidance on the selection of mesh resolution and timestep size required to maintain numerical stability \cite{KONANGI2018643,ZANDSALIMY2024113195,alma995860290001771}.

Similar to the concept of stability in conventional numerical methods, long-term predictive stability is also a critical property for machine learning approaches, which perform time-marching predictions in a manner analogous to traditional numerical schemes \cite{wuTurbL1AchievingLongterm2025}. Due to the inherently chaotic nature of turbulence, it is unrealistic to expect machine learning models to produce accurate trajectories at long time scales. Instead, the primary objective is to ensure that the models can capture desirable long-term statistical properties and obtain short-term trajectory accuracy \cite{chattopadhyayChallengesLearningMultiscale2024c}. For neural operator-based models to achieve long-term stable predictions, several critical factors play a fundamental role. First, the architecture and parameters of the neural operator models are essential, which includes the expressiveness of the model and the richness of the training datasets, as well as the adoption of appropriate training strategies, such as the incorporation of historical information \cite{majid2024mixture}. Second, the choice of time discretization is equally important. In most cases, using a smaller time interval $\Delta T$ allows the model to capture sufficient information between adjacent flow fields, thereby reducing prediction errors. However, if the time interval used for training is excessively small, it can lead to the accumulation of errors during the time-marching process, ultimately degrading model performance \cite{liLearningDissipativeDynamics2022a,yeLocalityLocalNeural2024b}.

In this paper, the analysis of model stability consists of two parts. First, the model is evaluated by making predictions over a time horizon significantly longer than the training time, and observing whether it can reach a statistically steady state after sufficient time rollouts. Second, an initial perturbation is introduced to the input data, and the model's ability to recover and return to a stable predictive condition is assessed. Here, the initial perturbation is defined as follows
\begin{equation}
\hat{\textbf{\textit{u}}} (k,t_{0})=(1+\tilde{\varepsilon} )\hat{\textbf{\textit{u}}}^{ref} (k,t_{0}),
\label{eq:21}
\end{equation}
where $\tilde{\varepsilon}$ is a small real number (0.1, 0.5, 1, 2, 5, and 10), used to evaluate model stability under different perturbation levels, with the same value applied uniformly across all wavenumbers $k$. $\hat{\textbf{\textit{u}}}$ stands for the velocity field in frequency domain, $k$ is Fourier mode and $t_{0}$ denotes the initial time. This represents a perturbation in the magnitude of the Fourier modes \cite{wangTemporallySparseData2022}. Based on our tests, perturbations in magnitude, phase, and both magnitude and phase of the Fourier modes exhibit similar effects on the stability of model predictions. Therefore, perturbations in magnitude are used in this study as a representative form of Fourier mode perturbations.

\subsection{Analysis of ACF}
\label{subsec2.4}

As discussed in Subsection~\ref{subsec2.3}, stability is closely related to time discretization. A large time interval can result in increased testing loss and prediction error due to insufficient correlation between adjacent flow fields, whereas a small time interval may cause error accumulation during the time-marching process, ultimately leading to instability in long-term predictions. To further illustrate and logically explain this phenomenon, we introduce the concept of the autocorrelation function (ACF) \cite{sardarConcerningUseTurbulent2024b,wangFurtherInvestigationData2023}. With ACF as a diagnostic tool, we can better understand why machine learning-based models exhibit different performances with varying time intervals \(\Delta T\).

Machine learning approaches typically assume that samples within a dataset are independent and identically distributed (IID). However, when applying machine learning-based models to turbulence, this soft IID assumption is often violated. This is because turbulence inherently does not satisfy the independence condition: turbulent structures exhibit spatial correlations, which can be quantified by two-point correlation functions, and temporal correlations may also exist between individual snapshots obtained from DNS of turbulent flows \cite{pope2000}. We define temporal correlations in the velocity field using the autocorrelation function (ACF), which is :
\begin{equation}
f_{ac}(\Delta T)=\frac{\left \langle \textbf{\textit{u}}(t) \cdot  \textbf{\textit{u}}(t+\Delta T) \right \rangle }{\left \langle \textbf{\textit{u}}(t)\cdot \textbf{\textit{u}}(t) \right \rangle},
\label{eq:22}
\end{equation}
where $f_{ac}$ is the autocorrelation function, $\textbf{\textit{u}}(t)$ is the velocity field in physical space at time $t$, $\Delta T$ is the time lag, and $\left \langle \cdot  \right \rangle $ denotes a spatial averaging operator over the entire three-dimensional domain \cite{sardarConcerningUseTurbulent2024b}.

To evaluate temporal coherence at different spatial scales, the velocity field is first decomposed in Fourier space and filtered to retain only the components within a specific wavenumber shell $k$. The filtered field is then transformed back into physical space and denoted as $\textbf{\textit{u}}^{(k)}(t)$. The temporal ACF corresponding to the Fourier mode $k$ is defined as
\begin{equation}
f_{ac}(\Delta T,k)=\frac{\left \langle \textbf{\textit{u}}^{(k)} (t) \cdot  \textbf{\textit{u}}^{(k)}(t+\Delta T) \right \rangle }{\left \langle \textbf{\textit{u}}^{(k)}(t) \cdot \textbf{\textit{u}}^{(k)}(t) \right \rangle},
\label{eq:23}
\end{equation}
where $\left \langle \cdot  \right \rangle $ denotes spatial averaging. This formulation characterizes the temporal coherence of flow structures associated with the wavenumber $k$ over a time lag $\Delta T$ \cite{wangFurtherInvestigationData2023}.

By analyzing $f_{ac}(\Delta T)$ and $f_{ac}(\Delta T, k)$ for the training data (fDNS), we can interpret the differences in prediction stability under varying $\Delta T$, identify the Fourier modes that contribute most significantly to the stability of surrogate models, and determine the optimal range of time intervals for achieving stable predictive performance.

\subsection{Quantile-quantile (QQ) plot technique}
\label{subsec2.5}

We employ quantile-quantile (QQ) plots to assess how well the chosen distributions fit the errors of the physical statistical quantities, which is crucial for the UQ analysis and will be discussed in detail in Subsection~\ref{subsec4.2} \cite{ProbabilityPlottingMethodsQQ,gnanadesikan1997methods,thode2002testing}.
A QQ plot is a graphical technique used to compare the quantiles of a sample distribution against those of a specified theoretical distribution. The theoretical distribution can be symmetric (e.g., Gaussian) or skewed (e.g., Gamma, Log-normal), making the QQ plot a versatile tool for assessing various types of data distributions. Given a sorted sample $\{x_{(1)}, x_{(2)}, \dots, x_{(n)}\}$ of size $n$, the $i$-th theoretical quantile is computed as:
\begin{equation}
z_i = F^{-1}\left( \frac{i - 0.5}{n} \right), \quad i = 1, 2, \dots, n,
\label{eq:24}
\end{equation}
where $F^{-1}$ is the inverse cumulative distribution function (quantile function) of the chosen reference distribution. In constructing the QQ plot, each point corresponds to a pair $(z_i, x_{(i)})$, where $z_i$ is the theoretical quantile and $x_{(i)}$ is the sample quantile. If the sample distribution aligns well with the theoretical distribution, the points should closely follow a straight line, indicating a good fit. Deviations from this straight line signal discrepancies between the sample and the theoretical distribution, providing insights into the nature of the data’s distribution \cite{gnanadesikan1997methods,thode2002testing}. This visual inspection is particularly useful for identifying skewness, kurtosis, or outliers in the data.

\section{Numerical methods}
\label{sec3}

This section presents the detailed numerical methodologies used to solve the incompressible filtered Navier-Stokes equations, as defined in Eqs.~\eqref{eq:13} and~\eqref{eq:14}, including both the dynamic Smagorinsky model and neural operator-based surrogate models.

\subsection{Dynamic Smagorinsky model}
\label{subsec3.1}

In large-eddy simulation (LES), the flow field is decomposed into resolved large scales and unresolved small scales. Subgrid-scale (SGS) models are introduced to approximate the effects of unresolved small-scale motions on the resolved flow, thereby achieving the closure of the filtered incompressible Navier-Stokes equations \cite{annurev:/content/journals/10.1146/annurev-fluid-060420-023735,johnsonPhysicsinspiredAlternativeSpatial2022}. 
One of the most widely used LES models is the Smagorinsky model, which is defined as follows \cite{GENERALCIRCULATIONEXPERIMENTSWITHTHEPRIMITIVEEQUATIONS,germanoTurbulenceFilteringApproach1992,lilly1967representation}:
\begin{equation}
\tau _{ij} - \frac{\delta _{ij}}{3} \tau_{kk} = -2C_{s}^{2}\Delta ^{2} \left | \bar{S} \right | \bar{S}_{ij},
\label{eq:25}
\end{equation}
where $\bar{S}_{ij} = \frac{1}{2} \left( \frac{\partial \bar{u}_{i}}{\partial x_{j}} + \frac{\partial \bar{u}_{j}}{\partial x_{i}} \right)$ denotes the strain-rate tensor of the filtered velocity field, and $\left | \bar{S} \right | = (2\bar{S}_{ij}\bar{S}_{ij})^{1/2}$ represents the magnitude of the strain rate. Here, $\delta_{ij}$ is the Kronecker delta, and $\Delta$ is the filter width.

The Smagorinsky coefficient $C_{s}$ can be determined either through theoretical analysis or empirical calibration \cite{pope2000,lilly1967representation}. A widely adopted approach is the dynamic procedure, which computes $C_s$ adaptively based on the flow field by employing the Germano identity \cite{germanoDynamicSubgridscaleEddy1991a,lillyProposedModificationGermano1992a}, resulting in the so-called dynamic Smagorinsky model (DSM). In DSM, $C_{s}^{2}$ is evaluated by the following least-squares formulation:
\begin{equation}
C_{s}^{2} = \frac{ \left \langle \mathcal{L}_{ij} \mathcal{M}_{ij} \right \rangle }{ \left \langle \mathcal{M}_{kl} \mathcal{M}_{kl} \right \rangle },
\label{eq:26}
\end{equation}
where the Leonard stress is defined as $\mathcal{L}_{ij} = \widetilde{\bar{u}_{i} \bar{u}_{j}} - \tilde{\bar{u}}_{i} \tilde{\bar{u}}_{j}$, and the tensor $\mathcal{M}_{ij}$ is given by $\mathcal{M}_{ij} = \tilde{\alpha}_{ij} - \beta_{ij}$. Here, $\alpha_{ij} = 2\Delta^{2} \left | \bar{S} \right | \bar{S}_{ij}$ and $\beta_{ij} = 2\tilde{\Delta}^{2} \left | \tilde{\bar{S}} \right | \tilde{\bar{S}}_{ij}$, with an overbar denoting filtering at scale $\Delta$, and a tilde indicating the test filtering operation at the larger scale $\tilde{\Delta} = 2\Delta$.

\subsection{Neural operators}
\label{subsec3.2}

The Fourier neural operator (FNO) is one of the most widely used neural operator architectures for solving partial differential equations and has demonstrated significant potential in modeling turbulent flows \cite{KovachkiNeuraloperator,li2021fourierneuraloperatorparametric,liFourierNeuralOperator2024a}. Different from other neural networks that operate on discretized pointwise data, FNO leverages the Fourier transform to project high-dimensional input functions into the frequency domain, where convolution operations can be performed more efficiently and globally. By learning complex-valued weight matrices in the Fourier domain, FNO effectively captures both local and global dependencies, allowing it to generalize across different resolutions and domains \cite{li2021fourierneuraloperatorparametric}. However, when applied to three-dimensional turbulence, the vanilla FNO does not perform as well as in other scenarios. To address this limitation, several enhanced variants of FNO have been developed, demonstrating significantly improved performance. In this subsection, we introduce four FNO-based models: implicit Fourier neural operator (IFNO), implicit U-Net enhanced Fourier neural operator (IUFNO), factorized-implicit Fourier neural operator (F-IFNO), and factorized-implicit U-Net enhanced Fourier neural operator (F-IUFNO).

\subsubsection{Implicit Fourier neural operator}
\label{subsubsec3.2.1}

The Fourier neural operator (FNO) aims to learn mappings between infinite-dimensional function spaces using a finite set of input-output data pairs \cite{li2021fourierneuraloperatorparametric, LI2022100389FNO}. By leveraging the Fourier transform in its Fourier layers, FNO effectively captures global features. It has been shown that increasing the depth of the Fourier layers enables FNO to act as a universal approximator \cite{JMLR:v22:21-0806Nikola}. However, deeper Fourier layers may lead to gradient-related issues \cite{YOU2022111536}. To address this, the idea of a shared hidden layer has been proposed as a solution \cite{ImplicitDL20M1358517}. Based on this idea, You \emph{et al.} \cite{YOU2022115296} proposed the implicit Fourier neural operator (IFNO), which formulates the solution operator as an implicitly defined mapping by employing a shallow-to-deep training strategy. This approach has been shown to be less prone to overfitting and achieves better performance on noisy experimental datasets \cite{YOU2022115296}. The detailed IFNO architecture is defined in ~\ref{subappendix1.1}.

The key distinction between FNO and IFNO lies in their kernel integration schemes: FNO adopts an explicit update mechanism, whereas IFNO employs an implicit iterative integration. By reusing parameters across the \( L \) iterative Fourier layers, the IFNO framework enhances prediction accuracy while significantly reducing memory overhead. This iterative design enables deeper modeling capacity without incurring prohibitive computational costs \cite{YOU2022115296}.

\subsubsection{Implicit U-Net enhanced Fourier neural operator}
\label{subsubsec3.2.2}

The implicit U-Net enhanced Fourier neural operator (IUFNO) was proposed by Li \emph{et al.} \cite{liLongtermPredictionsTurbulence2023a}, in which a U-Net architecture is employed as a residual component to effectively capture small-scale flow features \cite{WEN2022104180}. The only distinction between IFNO and IUFNO lies in the design of the implicit Fourier layers: IUFNO integrates the U-Net as a residual module within each Fourier layer to enhance multiscale feature extraction. The detailed IUFNO architecture is defined in ~\ref{subappendix1.2}.

\subsubsection{Factorized-implicit Fourier neural operator}
\label{subsubsec3.2.3}

Tran \emph{et al.} \cite{tranFactorizedFourierNeural2023d} proposed the factorized Fourier neural operator (F-FNO) for solving partial differential equations (PDEs), demonstrating improved accuracy and efficiency compared to the vanilla FNO. The main innovation of F-FNO lies in the factorization of Fourier transforms across problem dimensions, which significantly reduces the number of trainable parameters. Inspired by the Fourier transform factorization strategy and the design of the Euler-form implicit iterative structure in IFNO, we configure a factorized version of IFNO, namely the factorized-implicit Fourier neural operator (F-IFNO). Unlike the F-FNO architecture, which directly updates high-dimensional features between consecutive Fourier layers, the F-IFNO framework models the feature increment in an integrator-like manner. Similar to the F-FNO model, the weights in F-IFNO are shared between different layers to reduce the trainable parameters. The detailed F-IFNO architecture is defined in ~\ref{subappendix1.3}.

\subsubsection{Factorized-implicit U-Net enhanced Fourier neural operator}
\label{subsubsec3.2.4}

We also apply the idea of factorization to modify the existing IUFNO architecture, which is termed as factorized-implicit U-Net enhanced Fourier neural operator (F-IUFNO). The detailed F-IUFNO architecture is defined in ~\ref{subappendix1.4}.

To ensure a fair comparison among the four neural operator architectures, we adopt consistent experimental settings across all models. For a given neural operator, these settings remain unchanged regardless of the different choice of training time intervals. Specifically, we construct the training dataset using the velocity field at the previous time step as input and the velocity field at the current time step as the label, i.e., the projection pair is defined as \( U_t \rightarrow U_{t+1}^{\text{pre}} \).
The Adam optimizer is employed for model training \cite{kingma2017adammethodstochasticoptimization}, and the activation function used throughout is the hyperbolic tangent function, \(\tanh(x)\), defined as \cite{inproceedingsEfficientHardware}:
\begin{equation}
\tanh(x) = \frac{\exp(x) - \exp(-x)}{\exp(x) + \exp(-x)}.
\label{eq:27}
\end{equation}
The loss function adopted in this study is the normalized \(L_{2}\) loss, which represents the relative error between the predicted (\(u\)) and ground truth (\(\hat{u}\)) velocity fields:
\begin{equation}
L_{2}(u, \hat{u}) = \frac{\left\| u - \hat{u} \right\|_2}{\left\| \hat{u} \right\|_2} = \frac{\sqrt{\sum_{i=1}^{N} (u_i - \hat{u}_i)^2}}{\sqrt{\sum_{i=1}^{N} \hat{u}_i^2}},
\label{eq:28}
\end{equation}
where \(N\) denotes the total number of spatial grid points.

All hyperparameters are carefully tuned for each model to minimize both training and testing loss, ensuring a consistent and fair evaluation. Table~\ref{tab:1} lists the detailed hyperparameter settings. For a given model, the same hyperparameter settings are applied across different training time intervals $\Delta T$. Among them, the factorization ratio $\gamma$ controls the channel reduction rate in the feedforward network, where the hidden dimension is reduced from $C$ to $C/\gamma$ before being projected back \cite{tranFactorizedFourierNeural2023d}. This design improves efficiency without significantly compromising model capacity.

\begin{table}[ht!]
	\begin{center}
		\caption{Tuned hyperparameter settings for FNO-based models.}\label{tab:1}
		\begin{tabular*}{1\textwidth}{@{\extracolsep{\fill}} lcccccc }
			\hline\hline
			\small    
			Model & Modes & Width & Hidden layers $L$  & Weight decay & Learning rate & Factorization ratio $\gamma$\\ \hline    
			IFNO &  12 &  90 & 40  & 1e-8 & 5e-4 & N/A \\ 
			IUFNO &  12 &  90 & 40  & 1e-8 & 5e-4 & N/A \\
			F-IFNO &  12 &  90 & 40  & 1e-8 & 5e-4 & 4\\
			F-IUFNO &  12 &  90 & 40  & 1e-8 & 5e-4 & 4\\ \hline\hline
		\end{tabular*}%
	\end{center}
\end{table}

\subsection{Method for prediction constraints}
\label{subsec3.3}

For both fDNS and DSM methods, the solenoidal large-scale forcing is implemented by fixing the total kinetic energy within the two lowest wavenumber shells \cite{wang2012a}. As a result, while pointwise discrepancies may exist, the energy contained in the low-wavenumber components of both fDNS and DSM shows minimal deviation from the reference, indicating accurate large-scale behavior.
To ensure a fair comparison between conventional numerical methods and neural operator approaches, we adopt the same constraint strategy during the prediction phase of the pretrained neural operators. Specifically, we enforce that the total kinetic energy in the first two wavenumber shells ($k=1,2$) remains equal to the corresponding values from the reference fDNS, thereby mimicking the external forcing mechanism.

This constraint is derived from the implementation in the original fDNS forcing routine, where the velocity field is rescaled in Fourier space to maintain prescribed energy levels at selected wavenumbers. In particular, given a predicted velocity field $\textbf{\textit{u}}(x,y,z)$ at each time step, we first compute its discrete Fourier transform $\hat{\textbf{\textit{u}}}_{x,y,z}(\mathbf{k})$ using orthonormal normalization. The total kinetic energy $E_k$ in shell $k$ is calculated as:
\begin{equation}
E_k = \sum_{\mathbf{k} \in \mathcal{S}_k} \left( |\hat{u}_x(\mathbf{k})|^2 + |\hat{u}_y(\mathbf{k})|^2 + |\hat{u}_z(\mathbf{k})|^2 \right),
\label{eq:29}
\end{equation}
where $\mathcal{S}_k = \left\{ \mathbf{k} = (k_x, k_y, k_z) \, \middle| \, \sqrt{k_x^2 + k_y^2 + k_z^2} \in (k - 0.5, k + 0.5] \right\}$ denotes the set of wavevectors belonging to the $k$-th spherical shell in Fourier space. A rescaling factor $f_k$ is then defined for each shell $k=1,2$:
\begin{equation}
f_k = \sqrt{\frac{E_k^{\text{target}}}{E_k}}, 
\label{eq:30}
\end{equation}
where $E_k^{\text{target}}$ is the reference energy level obtained from fDNS, with $E_1^{\text{target}}=1.242477$ and $E_2^{\text{target}}=0.391356$, corresponding to the energy content in shells $k=1$ and $k=2$, respectively. This factor is applied uniformly to all Fourier modes in the shell:
\begin{equation}
\hat{\textbf{\textit{u}}}_{x,y,z}(\mathbf{k}) \leftarrow f_k \cdot \hat{\textbf{\textit{u}}}_{x,y,z}(\mathbf{k}), \quad \forall \mathbf{k} \in \mathcal{S}_k.
\label{eq:31}
\end{equation}

This procedure is repeated at each prediction step, ensuring that the low-wavenumber energy content remains consistent with that of the fDNS simulation. 
By enforcing this spectral constraint, we effectively eliminate long-term drift in the large-scale modes and ensure that the neural operator preserves the correct energy injection rate.
In the next section, we present and compare the results obtained from both constrained and unconstrained neural operator predictions to evaluate the impact of the constraints on model performance.

\section{Numerical experiments}
\label{sec4}

In this section, all FNO-based models are trained using the filtered direct numerical simulation (fDNS) datasets of forced homogeneous isotropic turbulence (HIT), and compared against the conventional dynamic Smagorinsky model (DSM). For evaluation, fDNS, DSM, and all FNO-based models are tested using thirty different random initializations of HIT, which is sufficient to obtain statistically convergent results due to the abundance of samples. This section is organized into four subsections to provide a comprehensive analysis of model reliability in terms of uncertainty quantification (UQ), stability, and their relationship with the autocorrelation function (ACF), followed by an additional subsection dedicated to the study of computational efficiency. Before that, we first present a comprehensive introduction to the HIT training and prediction setups.

The direct numerical simulation (DNS) of forced homogeneous isotropic turbulence (HIT) is performed on a uniform grid with a resolution of $256^3$ within a cubic domain of size $(2\pi)^3$, subject to periodic boundary conditions \cite{PhysRevFluids.5.054606}. To eliminate aliasing errors, the two-thirds truncation rule is employed \cite{patterson1971}. Temporal integration is performed using an explicit two-step Adams-Bashforth method, offering second-order accuracy in time \cite{wang2012}. The solenoidal large-scale forcing is implemented by fixing the total kinetic energy in the two lowest wavenumber shells of the velocity spectrum, in order to maintain the turbulence in a statistically steady state \cite{wang2012a}. The kinematic viscosity is set to $\nu = 0.00625$, resulting in a Taylor-scale Reynolds number of $Re_{\lambda} \approx 100$. To ensure that the flow reaches a statistically steady state, data are recorded after a sufficiently long simulation time (exceeding $10\tau$, where $\tau = L_{I}/u^{\mathrm{rms}} \approx 1.0$ represents the large-eddy turnover time). 

To obtain the fDNS data, the DNS results are filtered into a large-scale flow field with a grid resolution of $32^{3}$ using a sharp spectral filter (described in Subsection~\ref{subsec2.1}) with a cutoff wavenumber of $k_{c} = 10$. To train the models, 50 independent random velocity fields are employed as initial conditions. For each realization, 600 temporal snapshots are recorded, resulting in an fDNS dataset of size $[50\times 600\times32\times32\times32\times3]$. This dataset comprises 50 groups, each containing 600 time steps, where each time step corresponds to a filtered velocity field with a spatial resolution of $32^3$ and three velocity components. By adopting a one-step-ahead prediction framework, 599 input-output pairs can be extracted from each group, yielding a total of 29,950 samples. Among these, $80\%$ are allocated for training and the remaining $20\%$ for testing. 
To investigate the influence of temporal resolution on the performance of neural operators and to explore its relationship with the autocorrelation function (ACF), we employ different time intervals $\Delta T$. The base time step in the fDNS simulations is set to $ 0.001\tau$ (DSM uses the same time step). Snapshots of the numerical solution are extracted every [20, 40, 100, 200, 300, 400] steps, corresponding to time intervals of $\Delta T = 0.02\tau, 0.04\tau, 0.1\tau, 0.2\tau, 0.3\tau, 0.4\tau$, respectively.

A comparison between FNO-based methods on the training loss and testing loss through various time intervals is given in Table~\ref{tab:2}. As mentioned in Subsection~\ref{subsec3.2}, we use the same number of Fourier modes, the same number of hidden layer loop iterations $L$, the same initial learning rate, and the same optimizer for each model across different time intervals to ensure a fair comparison. For each model trained with a specific time interval, hyperparameters such as the learning rate and decay rate are individually tuned to ensure optimal performance under the corresponding conditions. The reported loss corresponds to that of the final epoch, with all models trained for a total of 40 epochs. It is observed that shorter time intervals result in lower training and testing losses. This phenomenon can be attributed to the hypothesis that smaller time intervals preserve stronger temporal correlations and consistent patterns, which the model can exploit more effectively. For a fixed time interval, IUFNO achieves the lowest training loss, while F-IUFNO yields the lowest testing loss. This is because the incorporation of the U-Net as a residual component enhances the model's ability to capture small-scale features, and the factorization of Fourier transforms reduces the number of parameters, thereby improving generalization capability. Nevertheless, a lower loss does not always translate to improved predictive accuracy, as will be further analyzed in the subsequent discussion.
\begin{table}[ht!]
	\begin{center}
		\caption{Comparison of (Traning loss$\times \%$, Testing loss$\times \%$) through various time intervals for FNO-based models in forced homogeneous isotropic turbulence.}\label{tab:2}
		\begin{tabular*}{1\textwidth}{@{\extracolsep{\fill}} lcccccc }
			\hline\hline
			\small    
			Model & $\Delta T=0.02\tau$ & $\Delta T=0.04\tau$ & $\Delta T=0.1\tau$ & $\Delta T=0.2\tau$ & $\Delta T=0.3\tau$ & $\Delta T=0.4\tau$ \\ \hline
			IFNO & (1.340,1.392)    & (2.639,2.752)   & (5.906,6.134)  & (10.173,10.586)   & (14.207,14.793)  & (19.028,19.820)\\
			IUFNO & (\textbf{0.434},\textbf{0.541})    & (\textbf{1.150},\textbf{1.586})   & (\textbf{3.372},4.863)  & (\textbf{6.206},8.839)   & (\textbf{8.519},11.809)  & (\textbf{10.390},14.850)\\
			F-IFNO & (1.419,1.422)    & (2.494,2.499)   & (5.234,5.245)  & (9.017,9.014)   & (11.912,11.932)  & (14.664,14.709)\\           
			F-IUFNO & (0.908,0.990)    & (1.768,1.980)   & (4.008,\textbf{4.438})  & (6.872,\textbf{7.554})   & (10.257,\textbf{10.875})  & (11.489,\textbf{12.668})\\ \hline\hline
		\end{tabular*}%
	\end{center}
\end{table}

In the following three subsections, the fDNS data is used as the baseline for evaluation. The DSM model is implemented on a uniform grid with a resolution of $32^3$ within a cubic domain of size $(2\pi)^3$, using the same numerical method as in the DNS. The DSM simulation is initialized with the instantaneous velocity field obtained from the fDNS data. The base time step for DSM is $ 0.001\tau$, same as fDNS. Moreover, we present and compare the results obtained from both constrained and unconstrained neural operator predictions to assess the impact of the imposed constraints on model performance.

\subsection{A posteriori study}
\label{subsec4.1}

In the \textit{a posteriori} study, thirty independent data groups generated from different initial conditions are used for evaluation. All models trained with varying time intervals require a certain amount of prediction time to reach a statistically steady state. At this state, the mean values (averaged over thirty cases) of key statistical quantities remain nearly constant, exhibiting only negligible fluctuations due to the inherent properties of forced HIT. These steady-state values serve as indicators of the long-term stability of each method. In the following, we present the results of five representative physical statistics at the statistically steady state, along with the long-term evolution of vorticity fields.

\begin{figure}[ht!]
    \centering
    \begin{subfigure}[b]{0.32\textwidth}
        \begin{overpic}[width=1\linewidth]{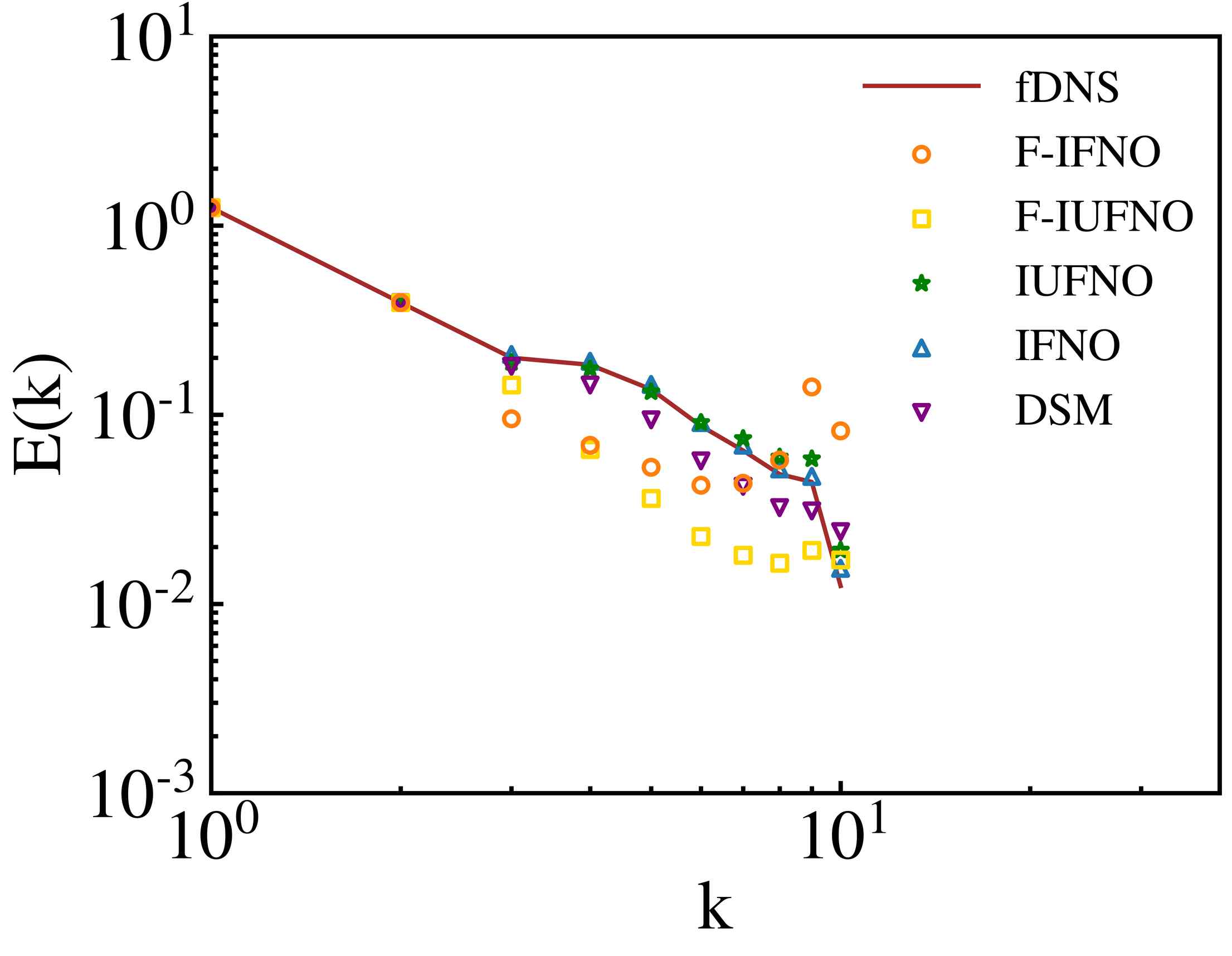}
            \put(-3,68){\small (a)}  
        \end{overpic}
    \end{subfigure}
    \hfill
    \begin{subfigure}[b]{0.32\textwidth}
        \begin{overpic}[width=1\linewidth]{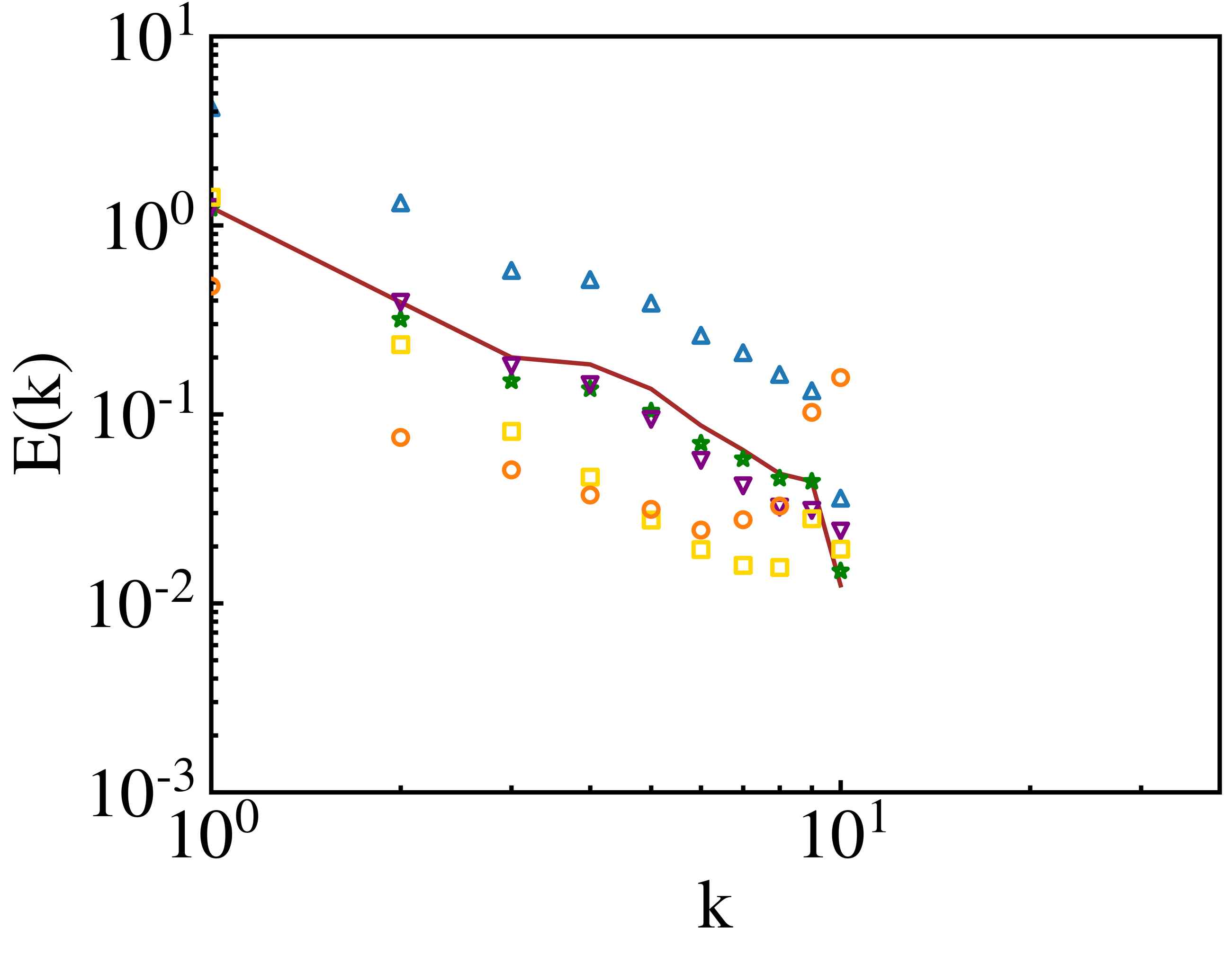}
            \put(-3,68){\small (b)} 
        \end{overpic} 
    \end{subfigure}
    \hfill
    \begin{subfigure}[b]{0.32\textwidth}
        \begin{overpic}[width=1\linewidth]{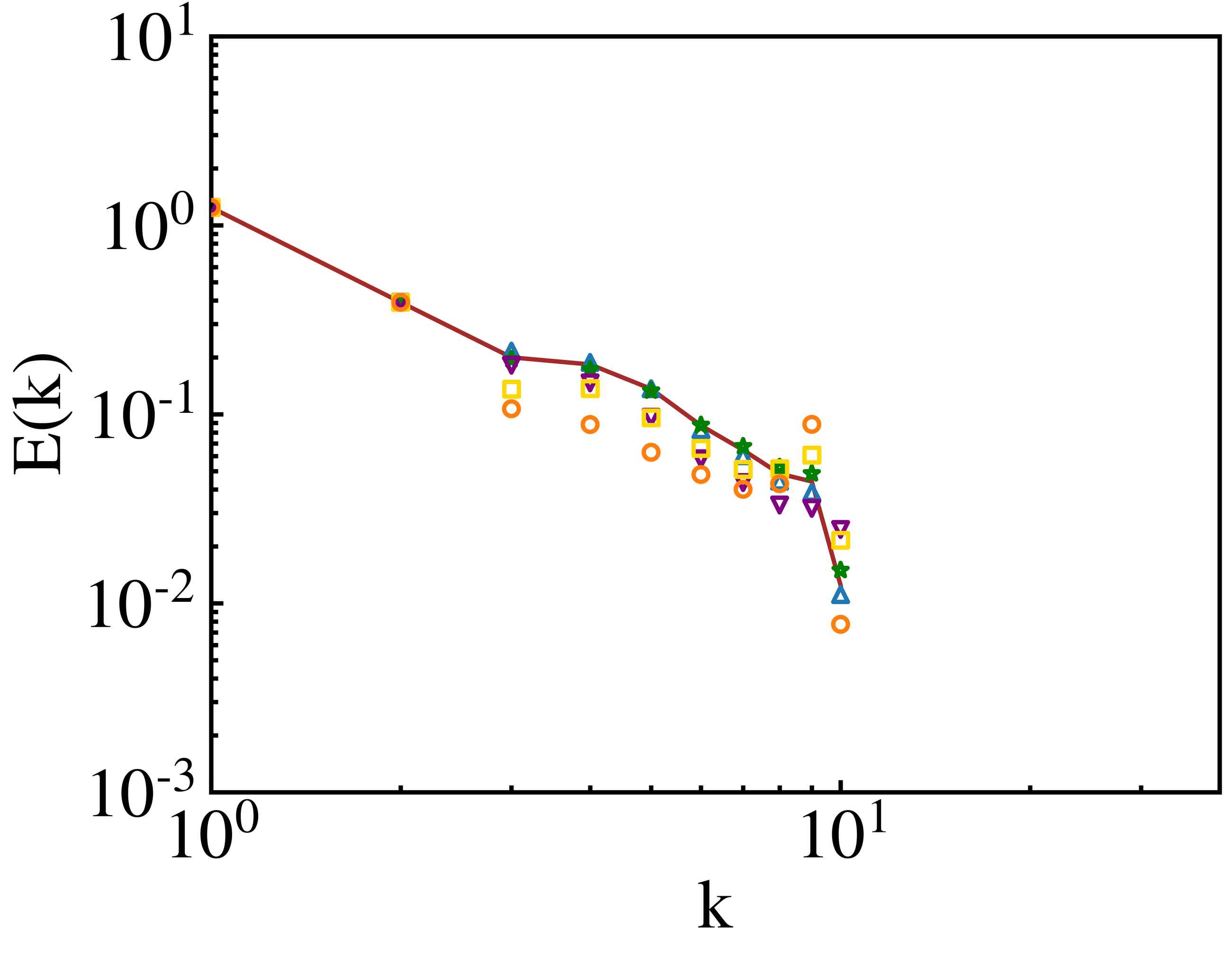}
            \put(-3,68){\small (c)} 
        \end{overpic}
    \end{subfigure}
    \vspace{0.1cm}
    \begin{subfigure}[b]{0.32\textwidth}
        \begin{overpic}[width=1\linewidth]{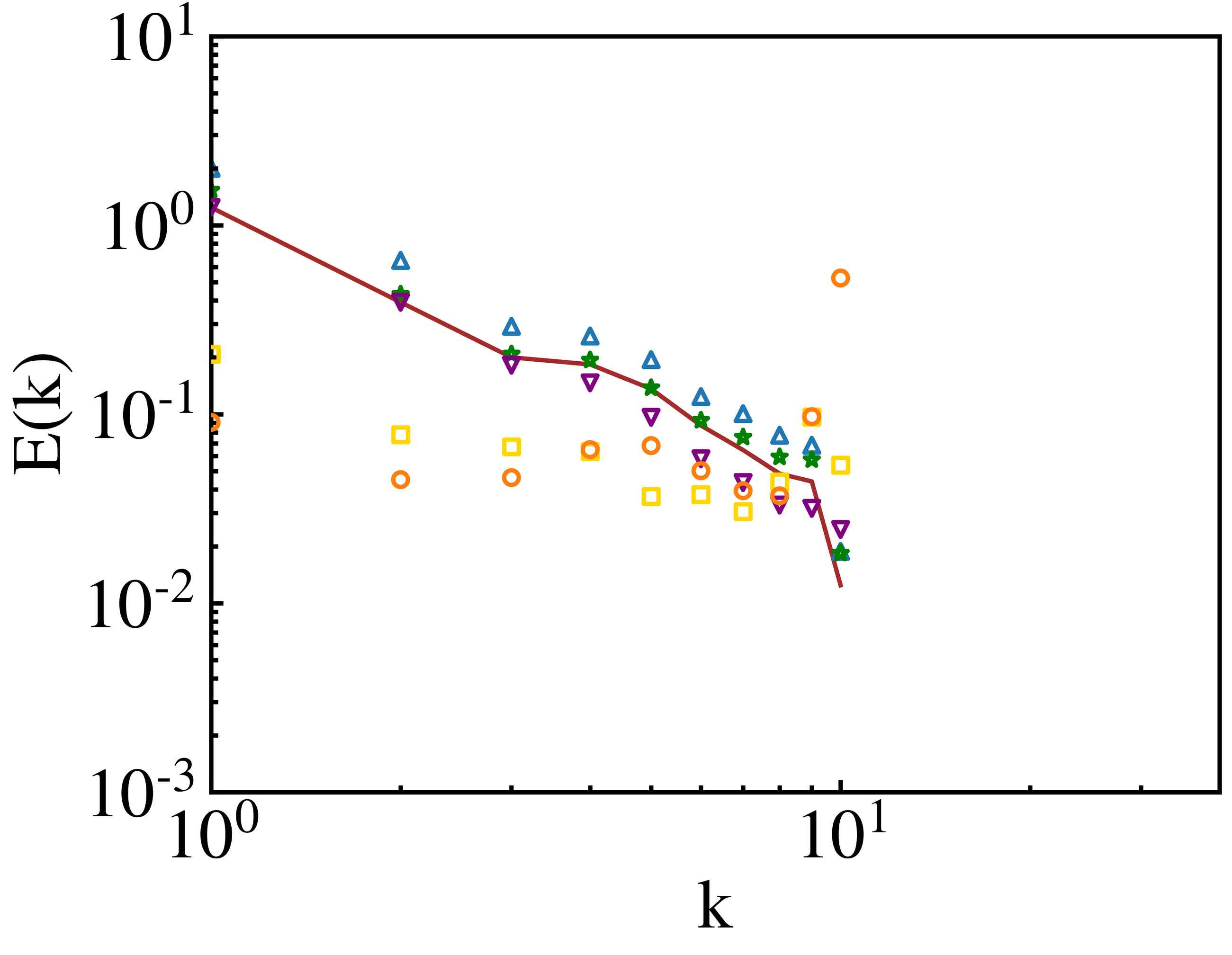}
            \put(-3,68){\small (d)} 
        \end{overpic}
    \end{subfigure}
    \hfill
    \begin{subfigure}[b]{0.32\textwidth}
        \begin{overpic}[width=1\linewidth]{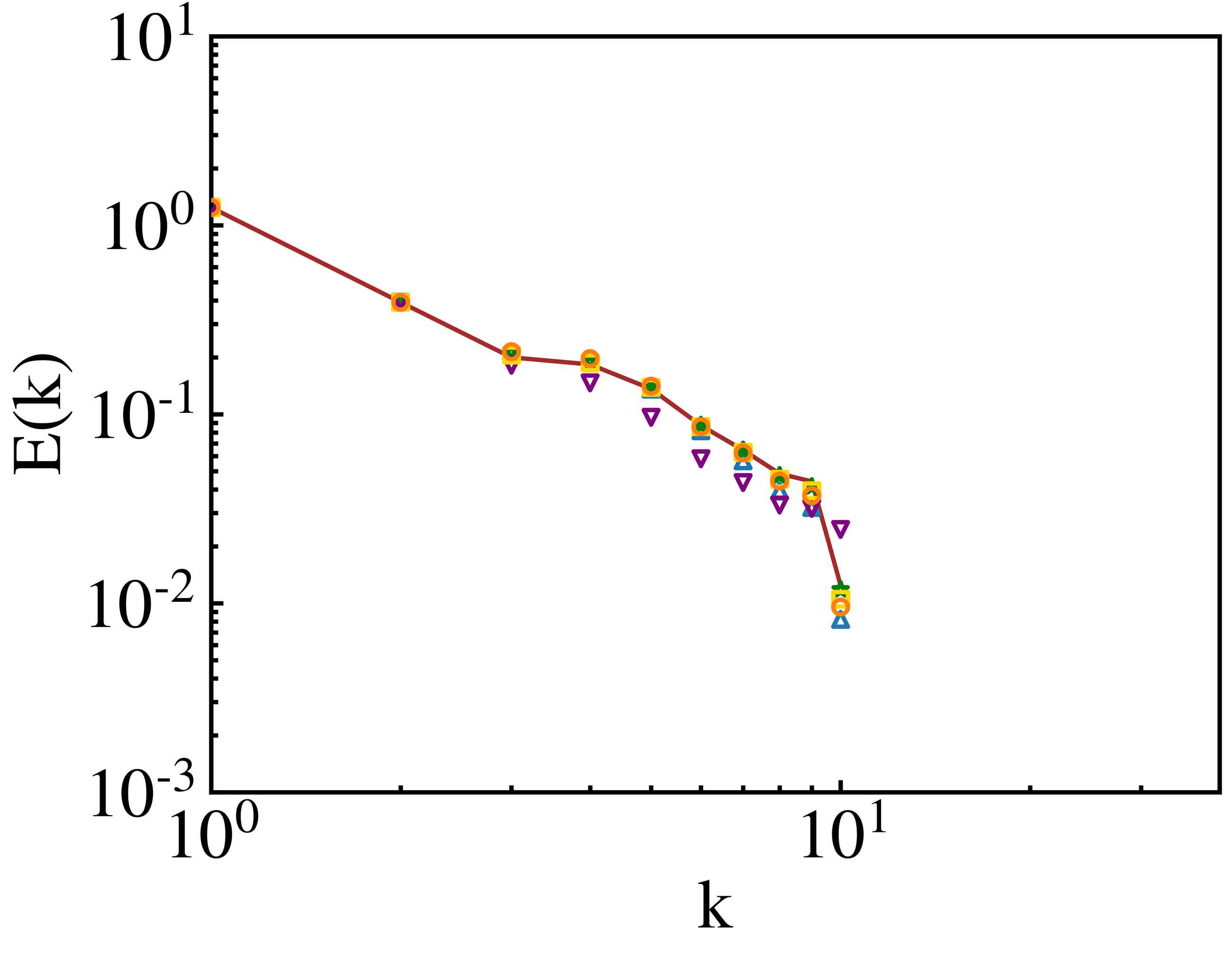}
            \put(-3,68){\small (e)} 
        \end{overpic}
    \end{subfigure}
    \hfill
    \begin{subfigure}[b]{0.32\textwidth}
        \begin{overpic}[width=1\linewidth]{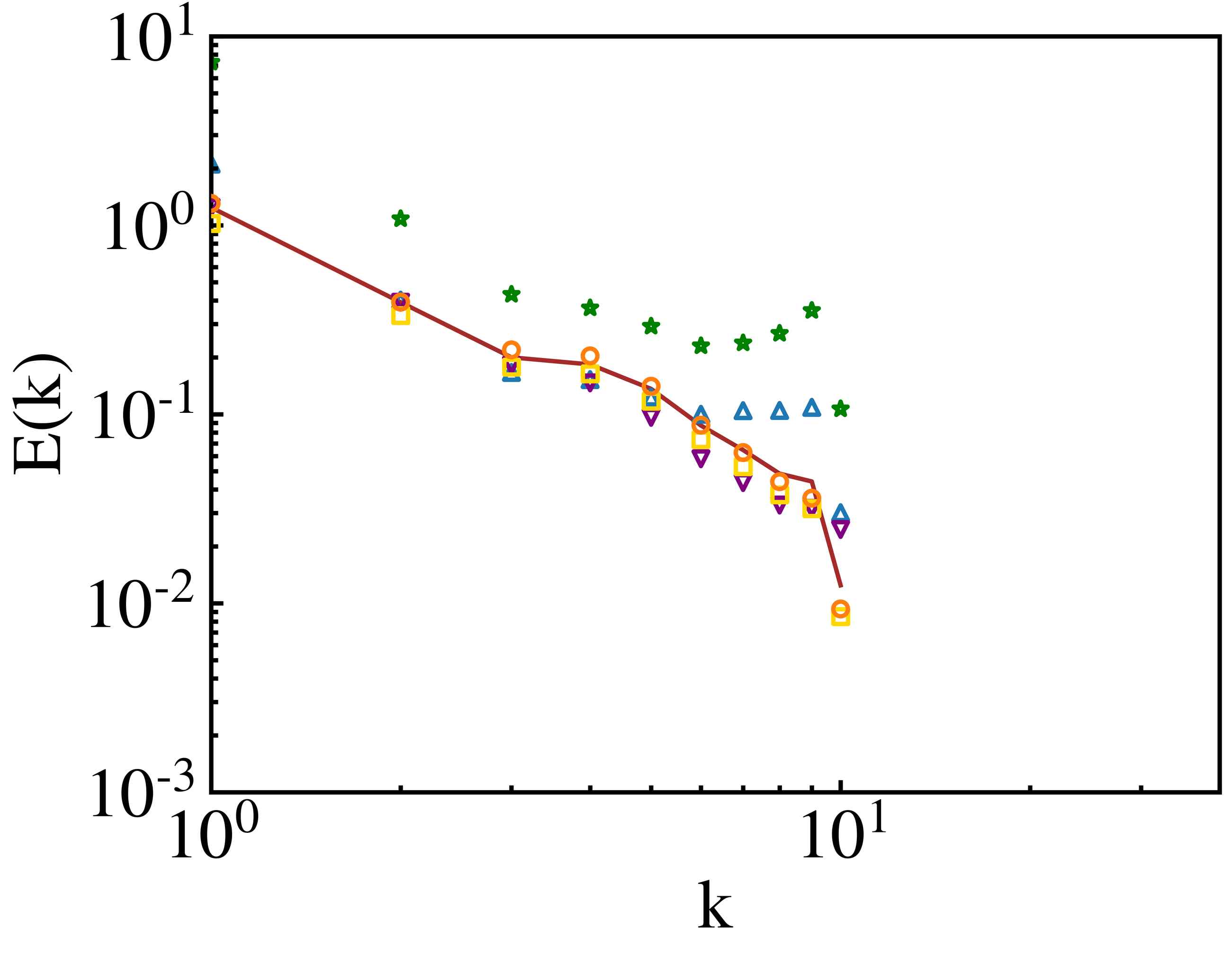}
            \put(-3,68){\small (f)} 
        \end{overpic}
    \end{subfigure}
    \vspace{0.1cm}
    \begin{subfigure}[b]{0.32\textwidth}
        \begin{overpic}[width=1\linewidth]{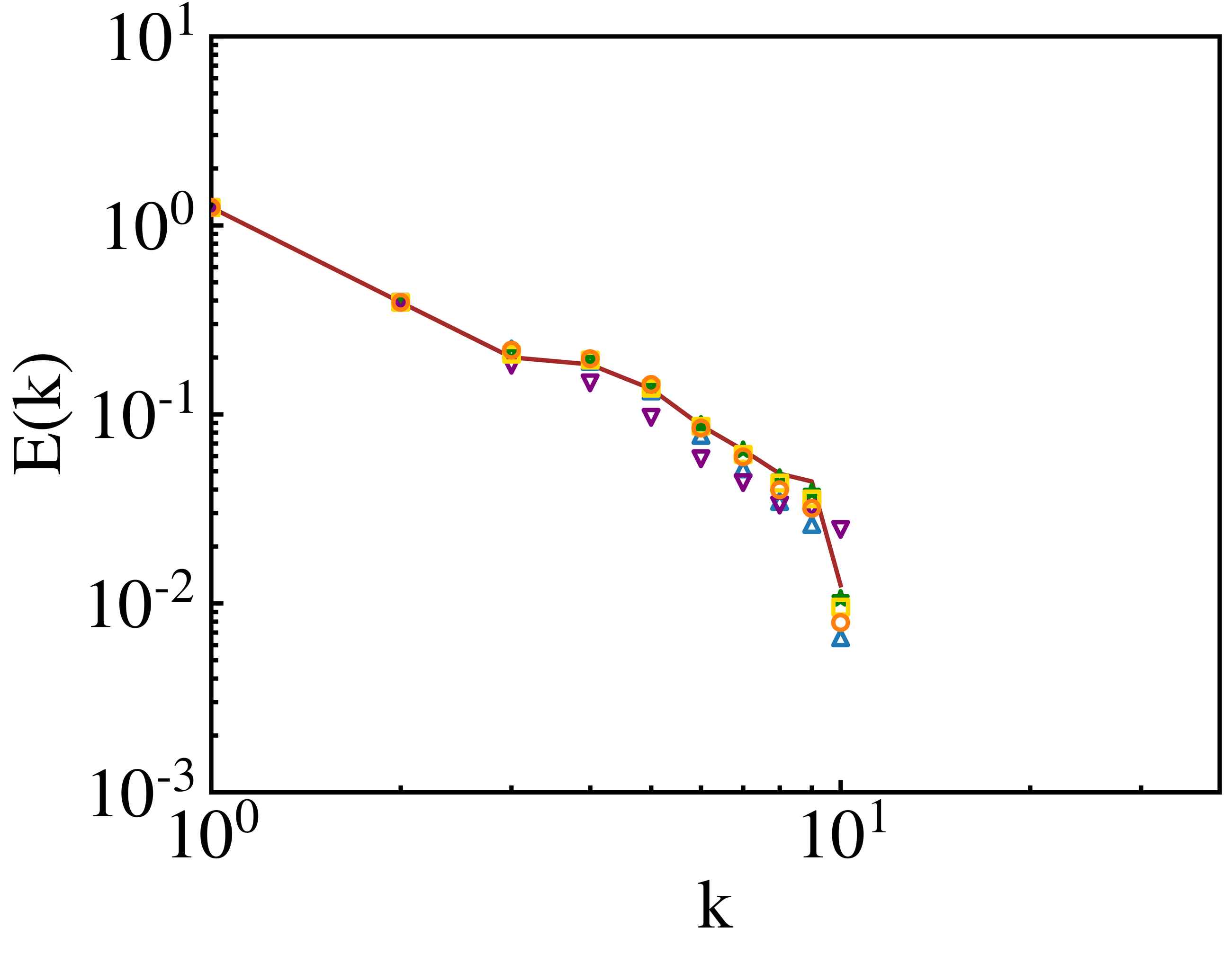}
            \put(-3,68){\small (g)} 
        \end{overpic}
    \end{subfigure}
    \hfill
    \begin{subfigure}[b]{0.32\textwidth}
        \begin{overpic}[width=1\linewidth]{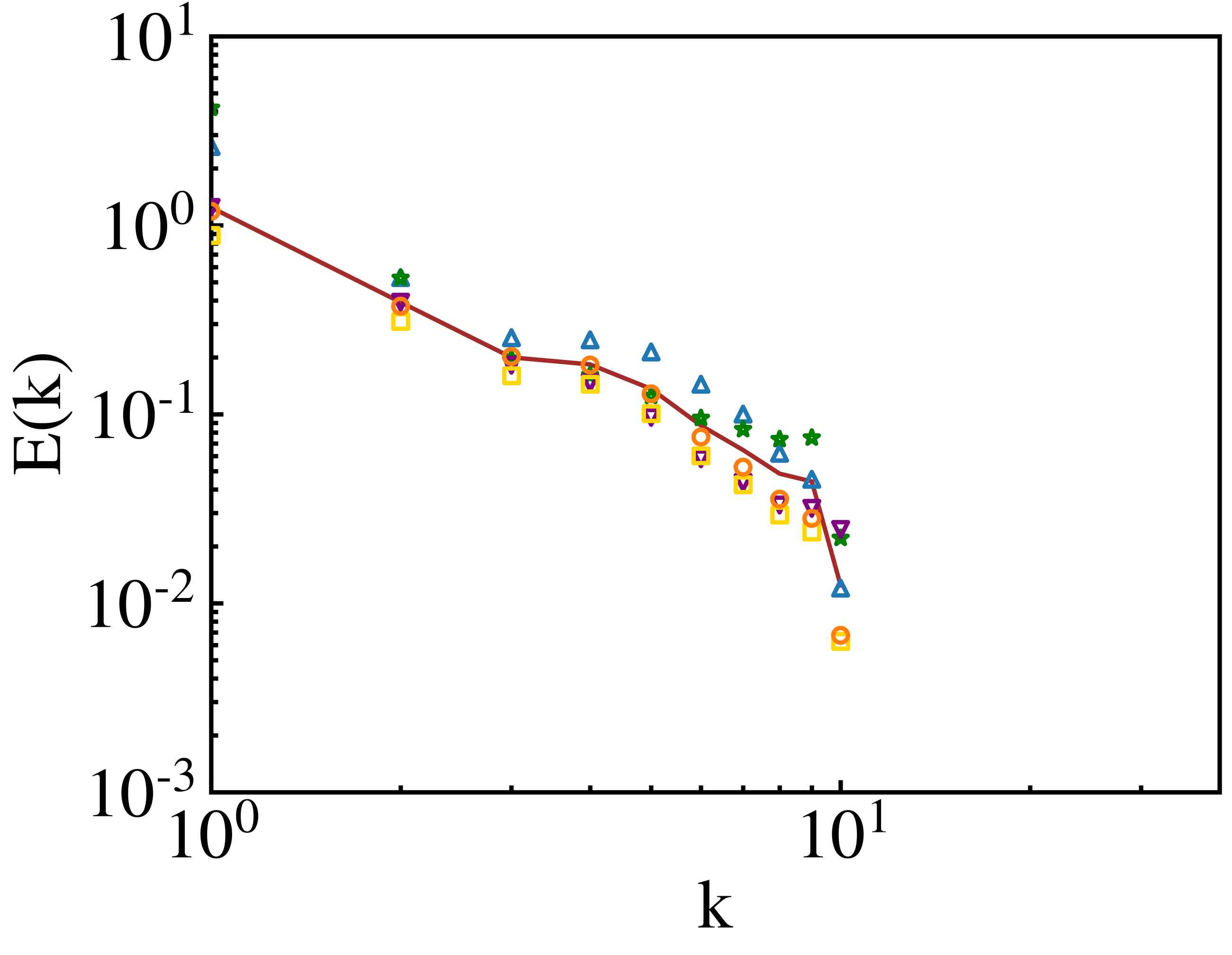}
            \put(-3,68){\small (h)} 
        \end{overpic}
    \end{subfigure}
    \hfill
    \begin{subfigure}[b]{0.32\textwidth}
        \begin{overpic}[width=1\linewidth]{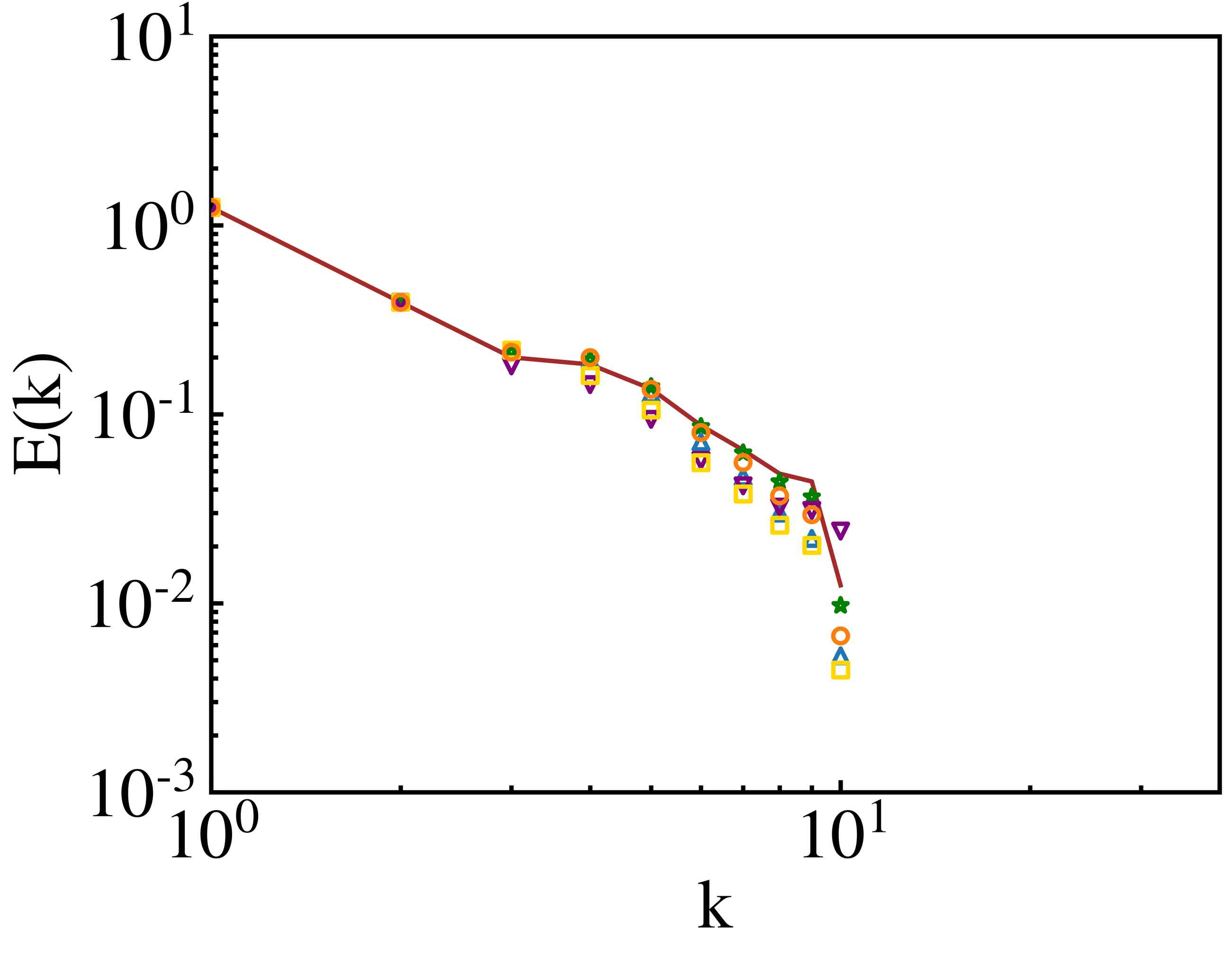}
            \put(-3,68){\small (i)} 
        \end{overpic}
    \end{subfigure}
    \vspace{0.1cm}
    \begin{subfigure}[b]{0.32\textwidth}
        \begin{overpic}[width=1\linewidth]{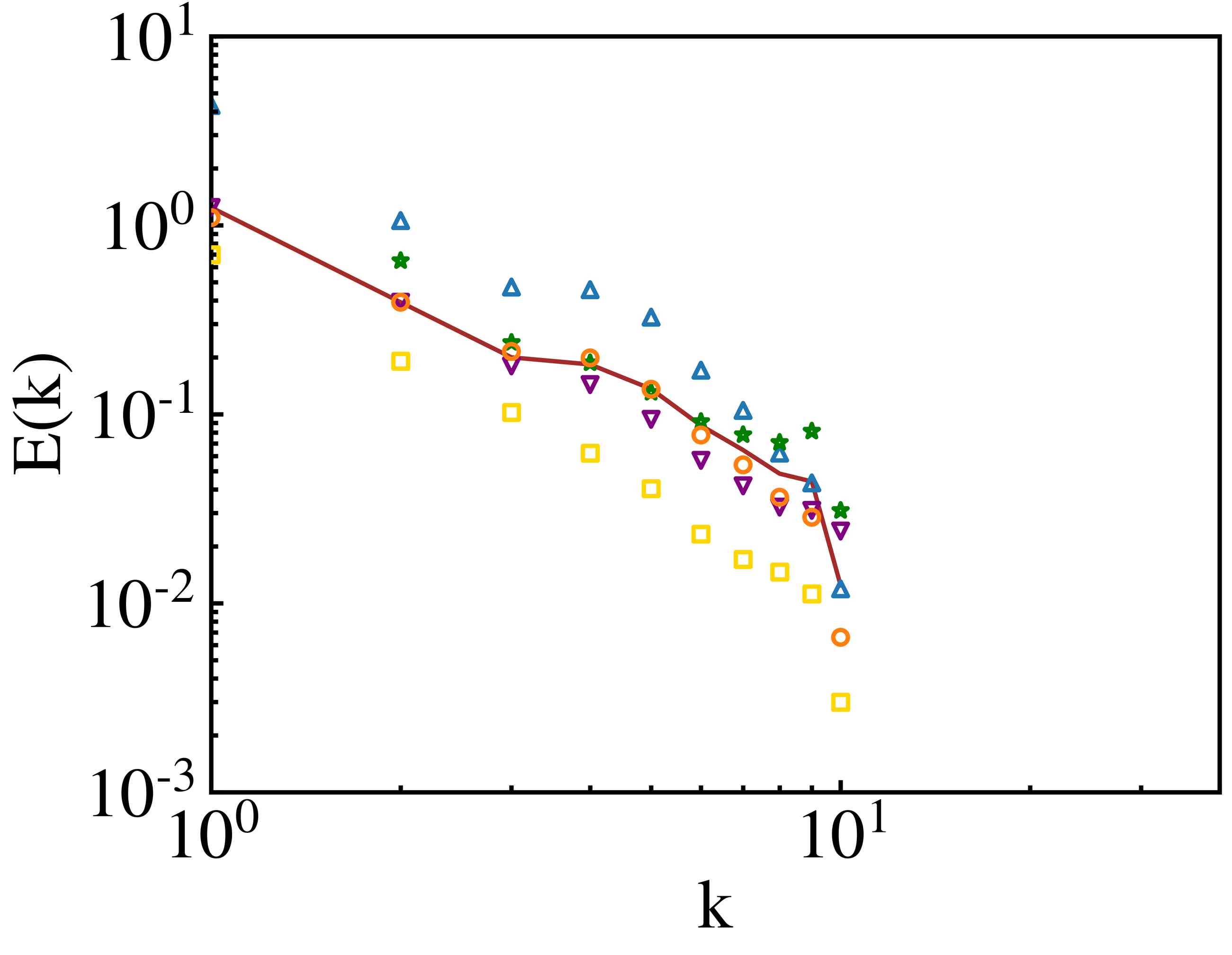}
            \put(-3,68){\small (j)} 
        \end{overpic}
    \end{subfigure}
    \hfill
    \begin{subfigure}[b]{0.32\textwidth}
        \begin{overpic}[width=1\linewidth]{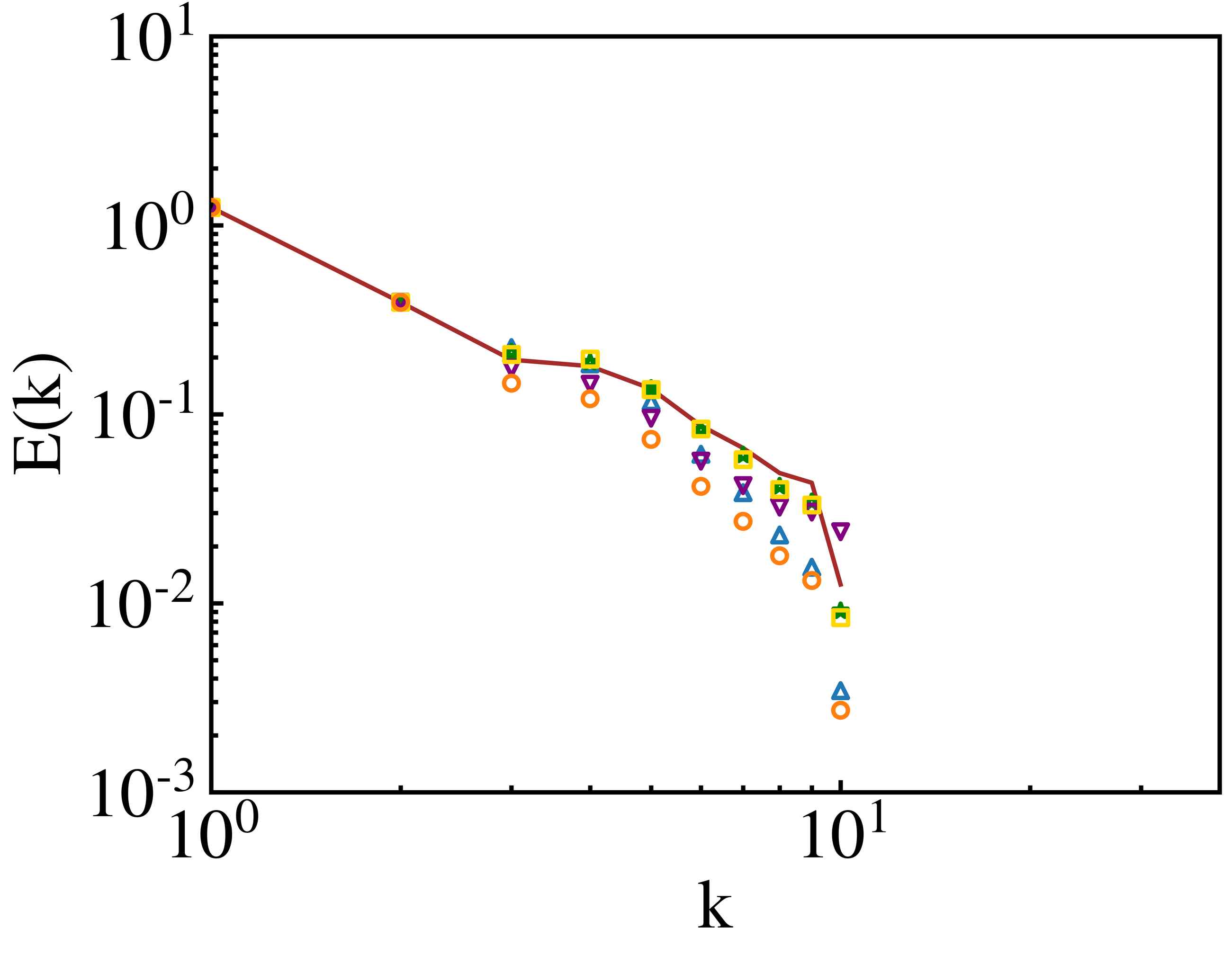}
            \put(-3,68){\small (k)} 
        \end{overpic}
    \end{subfigure}
    \hfill
    \begin{subfigure}[b]{0.32\textwidth}
        \begin{overpic}[width=1\linewidth]{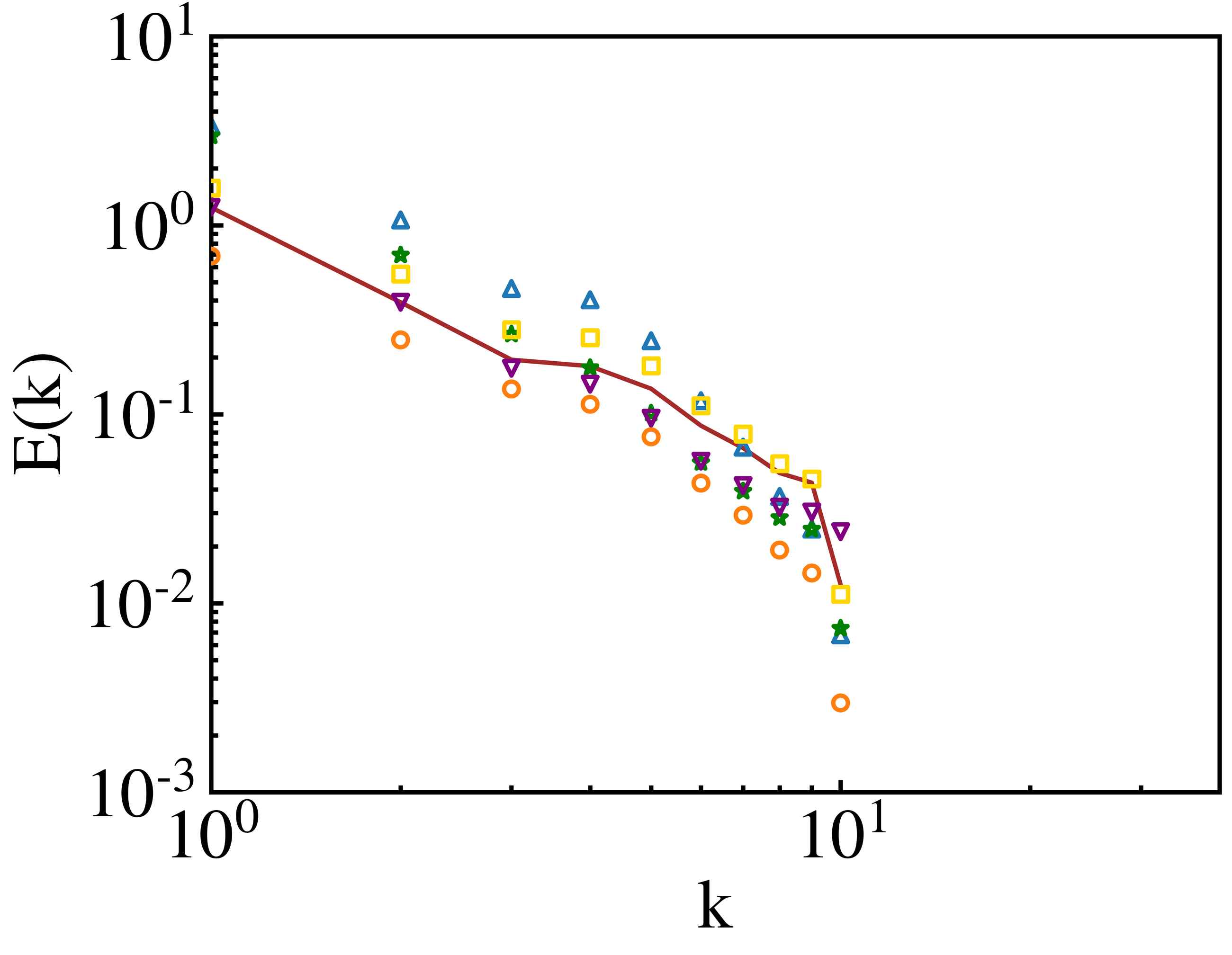}
            \put(-3,68){\small (l)}  
        \end{overpic}
    \end{subfigure}
	\caption{The velocity spectra of LES obtained using different models in forced HIT under various training and prediction time intervals at the statistically steady state: (a) $\Delta T=0.02\tau$ constrained; (b) $\Delta T=0.02\tau$ unconstrained; (c) $\Delta T=0.04\tau$ constrained; (d) $\Delta T=0.04\tau$ unconstrained; (e) $\Delta T=0.1\tau$ constrained; (f) $\Delta T=0.1\tau$ unconstrained; (g) $\Delta T=0.2\tau$ constrained; (h) $\Delta T=0.2\tau$ unconstrained; (i) $\Delta T=0.3\tau$ constrained; (j) $\Delta T=0.3\tau$ unconstrained; (k) $\Delta T=0.4\tau$ constrained; (l) $\Delta T=0.4\tau$ unconstrained. Here, the time instance shown for (a)-(l) is $t/\tau=120$.}\label{fig:2}
\end{figure}

The velocity spectra predicted by different methods at the statistically steady state, corresponding to different training and prediction time intervals $\Delta T$, are shown in Fig.~\ref{fig:2}. For $\Delta T = 0.02\tau, 0.04\tau, 0.1\tau, 0.2\tau, 0.3\tau, 0.4\tau$, the spectra are evaluated at $t/\tau=120$, where all models give the statistically steady results.
As illustrated in Fig.~\ref{fig:2}, all methods with prediction constraints exhibit optimal performance when $\Delta T = 0.1\tau, 0.2\tau$, suggesting that the time interval range $\Delta T \in [0.1\tau, 0.2\tau]$ is optimal for FNO-based models. Within this range, FNO-based models with prediction constraints achieve significantly higher accuracy at wavenumbers $3 \leq k \leq 10$ than the DSM model. Additionally, constrained FNO-based models maintain results close to fDNS even after long-term prediction.
Furthermore, within this optimal time interval range, constrained F-IFNO and F-IUFNO demonstrate a better accuracy than constrained IFNO and IUFNO. Among unconstrained models, our proposed F-IFNO and F-IUFNO still demonstrate competitive accuracy at $3 \leq k \leq 8$ after long-term prediction compared to DSM, whereas IFNO and IUFNO diverge significantly from fDNS.
In contrast, at $\Delta T = 0.02\tau, 0.04\tau, 0.3\tau, 0.4\tau$, both constrained and unconstrained versions of F-IFNO and F-IUFNO exhibit poor performance after long-term prediction when compared to DSM. Similarly, IFNO and IUFNO also perform poorly at these time intervals, with the only exception being the constrained versions at $\Delta T = 0.02\tau, 0.04\tau$, which show relatively improved performance. Notably, even the best performance of IFNO and IUFNO within the time interval range $\Delta T \in [0.02\tau, 0.4\tau]$ still falls short of the optimal results achieved by F-IFNO and F-IUFNO in the same range. It is evident that FNO-based models achieve better accuracy when prediction constraints are applied compared to their unconstrained counterparts.

\begin{figure}[ht!]
    \centering
    \begin{subfigure}[b]{0.32\textwidth}
        \begin{overpic}[width=1\linewidth]{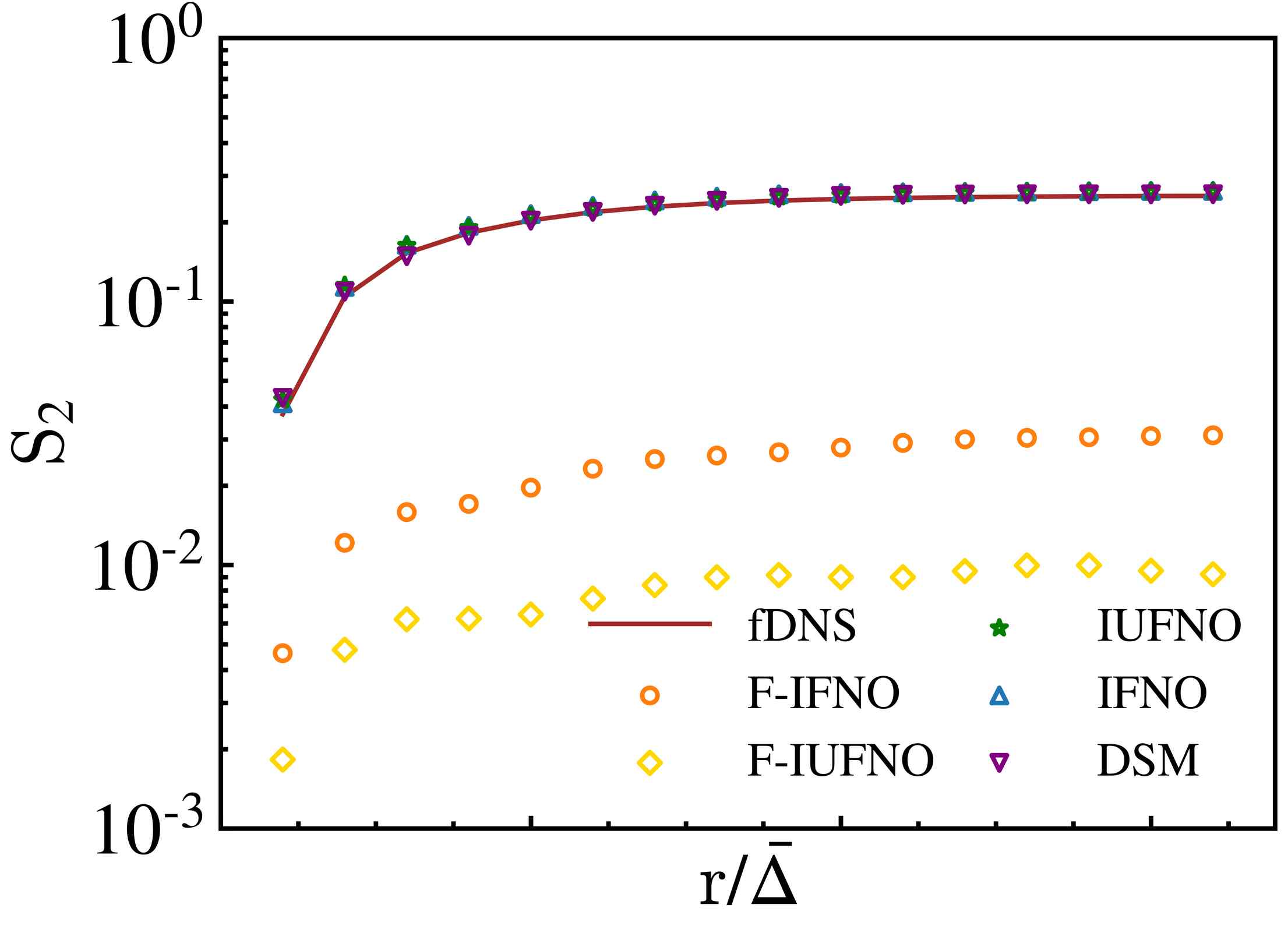}
            \put(-3,64){\small (a)}  
        \end{overpic}
    \end{subfigure}
    \hfill
    \begin{subfigure}[b]{0.32\textwidth}
        \begin{overpic}[width=1\linewidth]{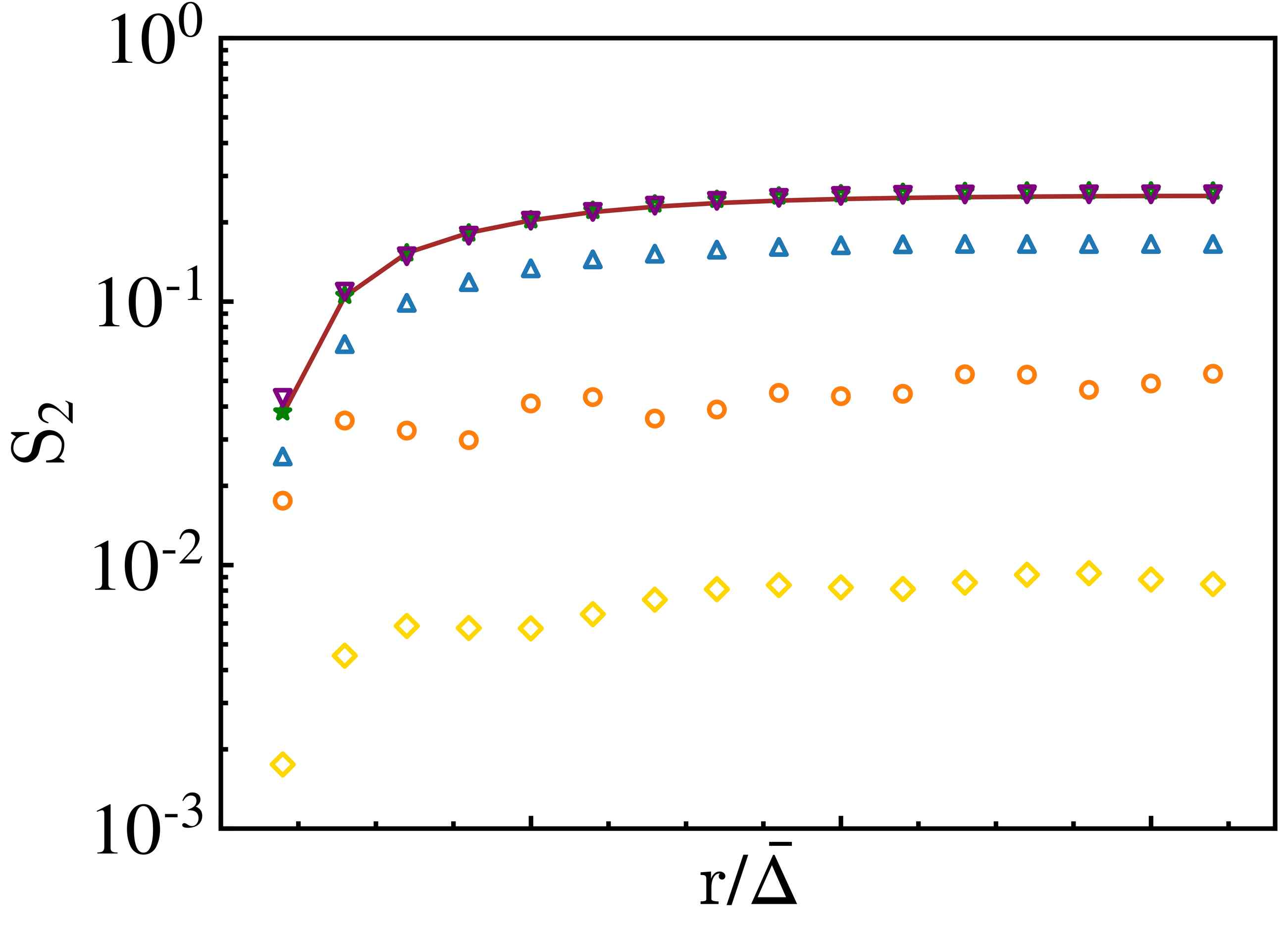}
            \put(-3,64){\small (b)} 
        \end{overpic} 
    \end{subfigure}
    \hfill
    \begin{subfigure}[b]{0.32\textwidth}
        \begin{overpic}[width=1\linewidth]{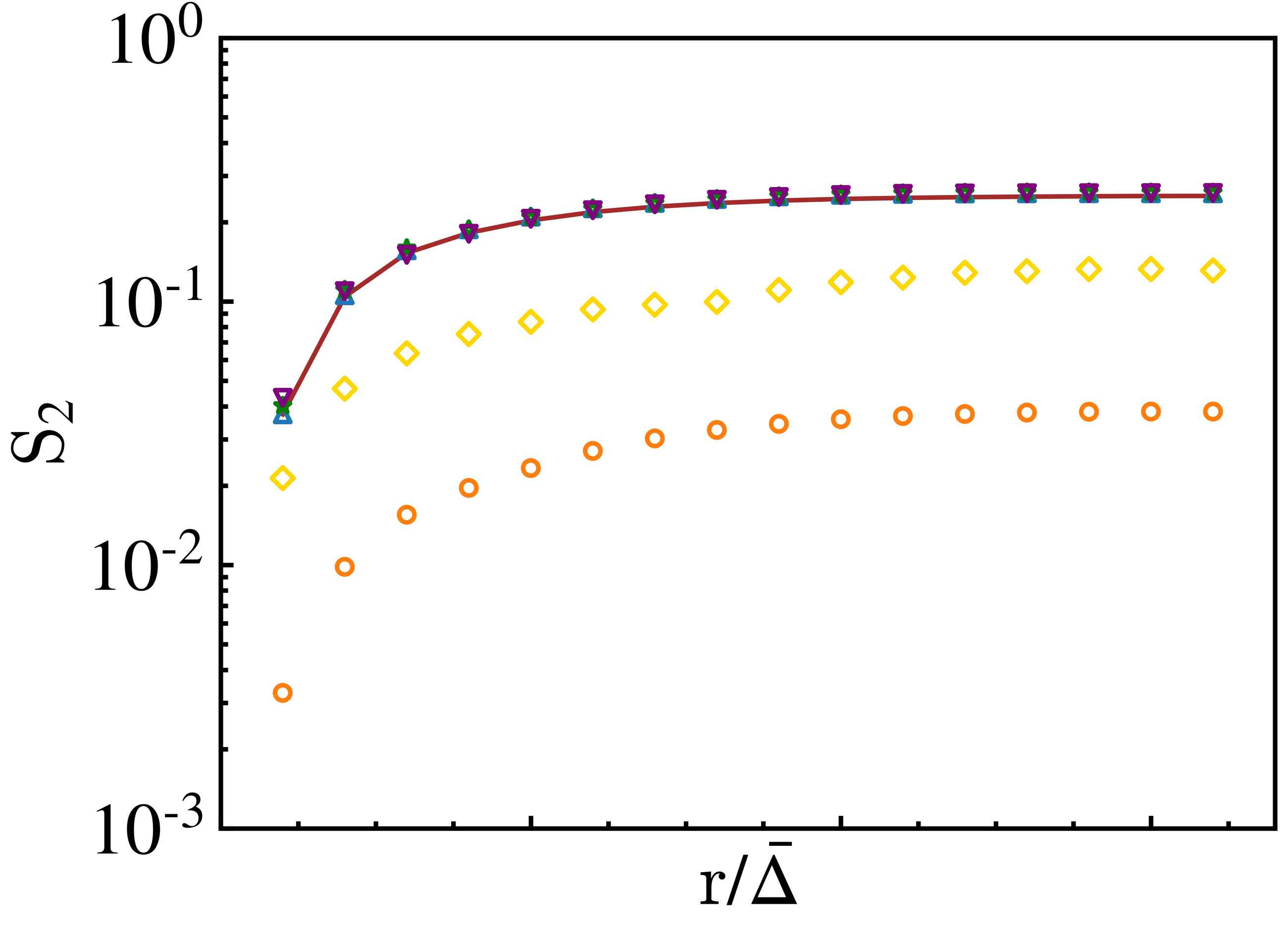}
            \put(-3,64){\small (c)} 
        \end{overpic}
    \end{subfigure}
    \vspace{0.1cm}
    \begin{subfigure}[b]{0.32\textwidth}
        \begin{overpic}[width=1\linewidth]{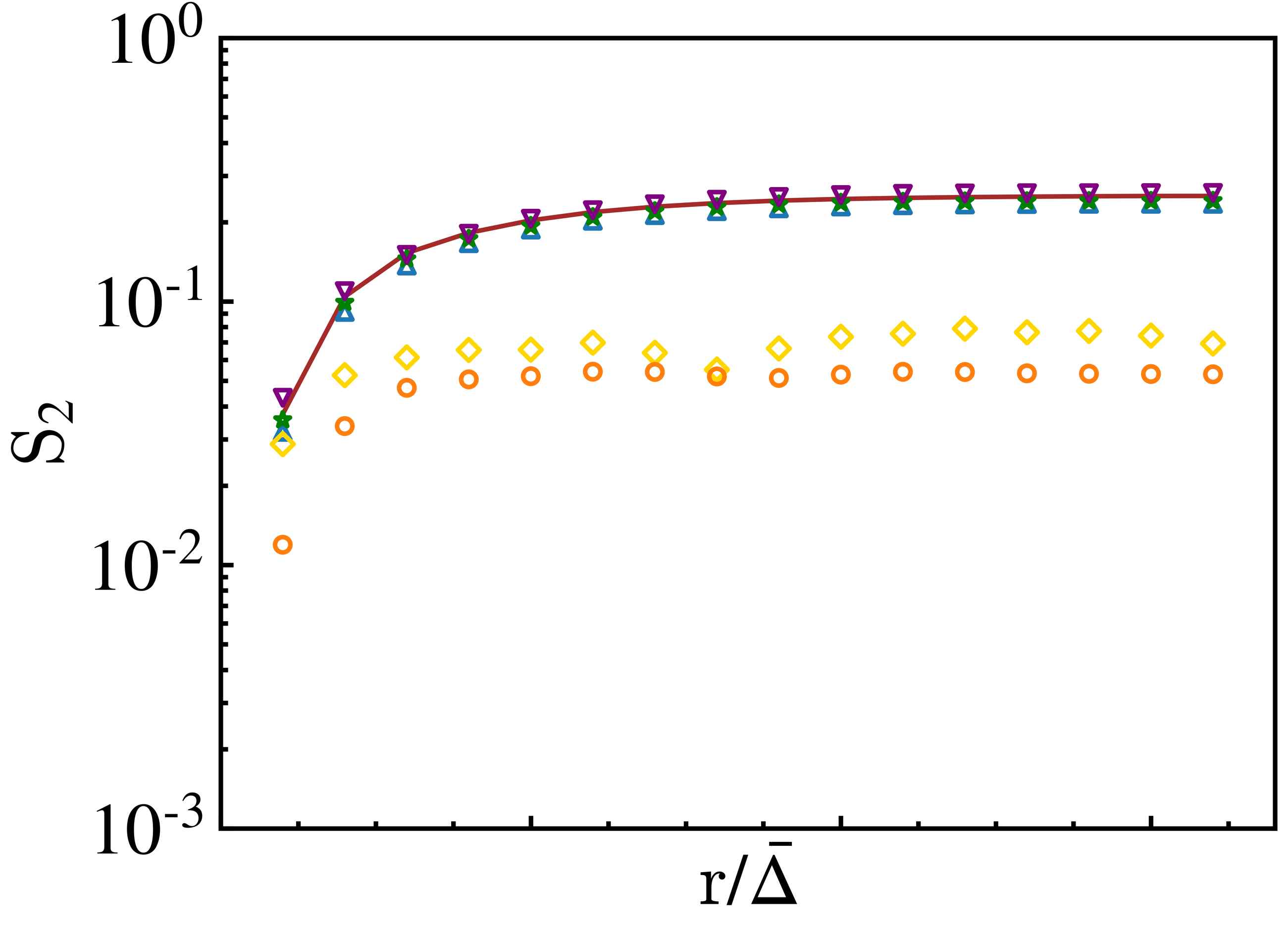}
            \put(-3,64){\small (d)} 
        \end{overpic}
    \end{subfigure}
    \hfill
    \begin{subfigure}[b]{0.32\textwidth}
        \begin{overpic}[width=1\linewidth]{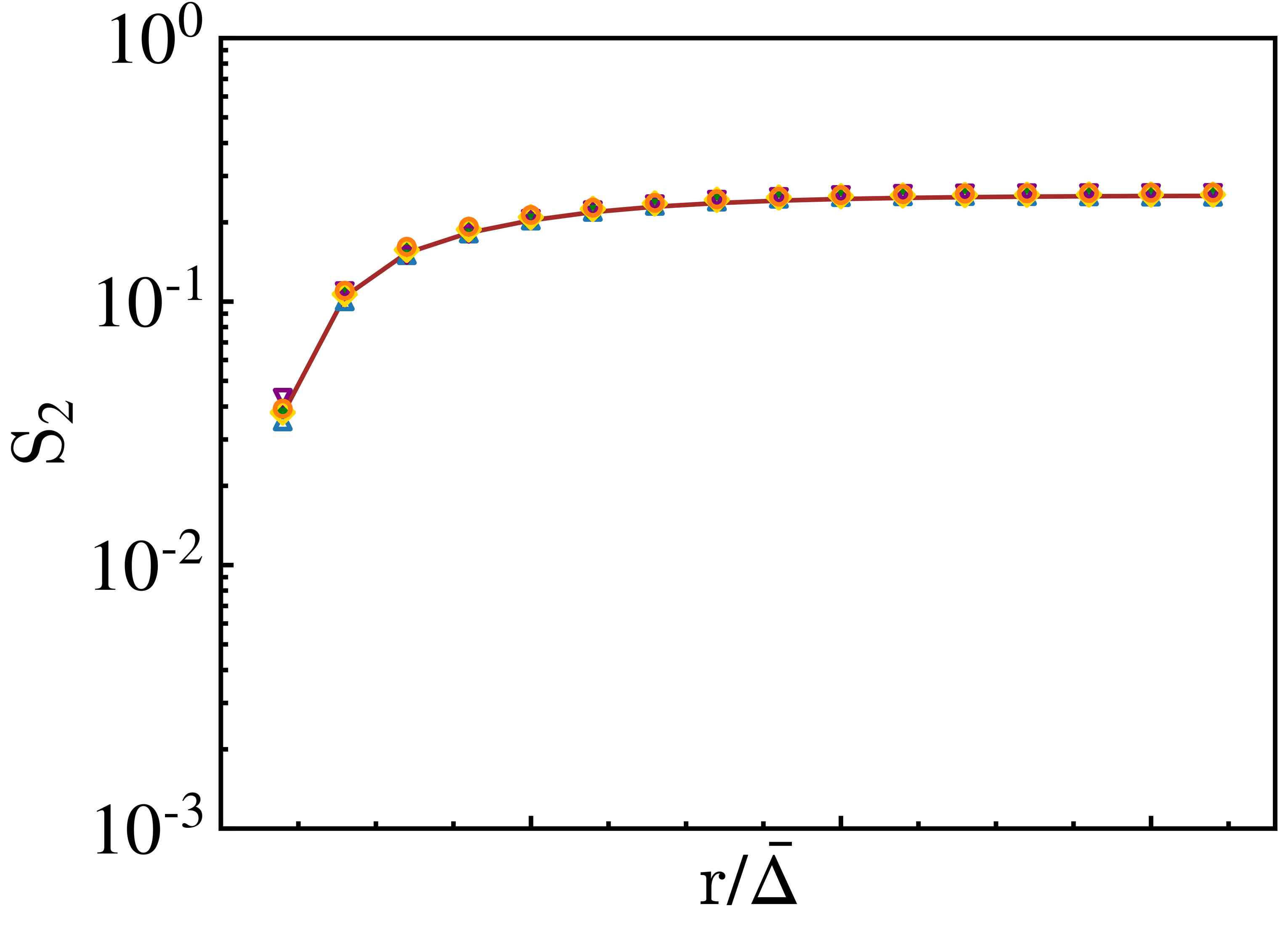}
            \put(-3,64){\small (e)} 
        \end{overpic}
    \end{subfigure}
    \hfill
    \begin{subfigure}[b]{0.32\textwidth}
        \begin{overpic}[width=1\linewidth]{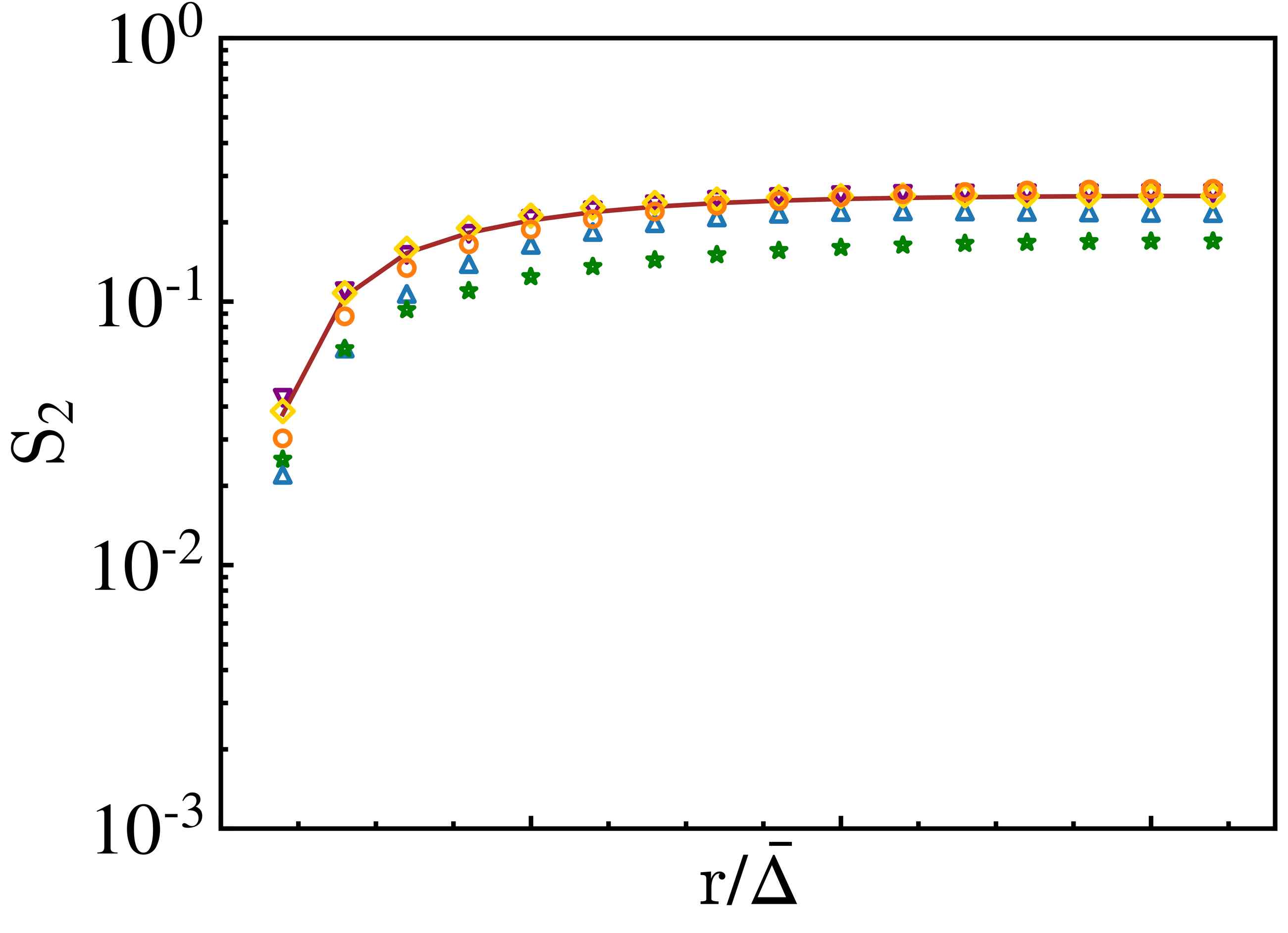}
            \put(-3,64){\small (f)} 
        \end{overpic}
    \end{subfigure}
    \vspace{0.1cm}
    \begin{subfigure}[b]{0.32\textwidth}
        \begin{overpic}[width=1\linewidth]{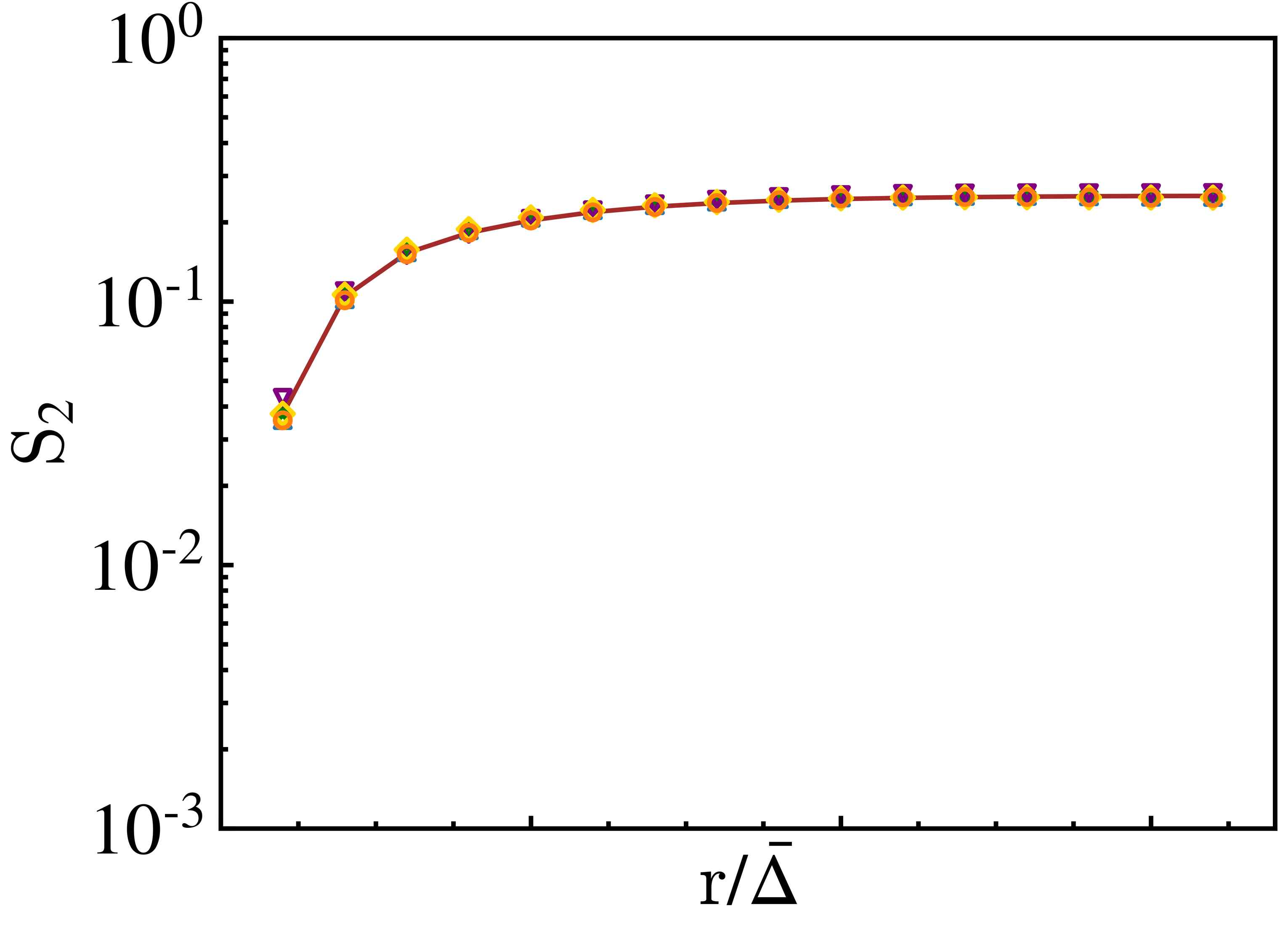}
            \put(-3,64){\small (g)} 
        \end{overpic}
    \end{subfigure}
    \hfill
    \begin{subfigure}[b]{0.32\textwidth}
        \begin{overpic}[width=1\linewidth]{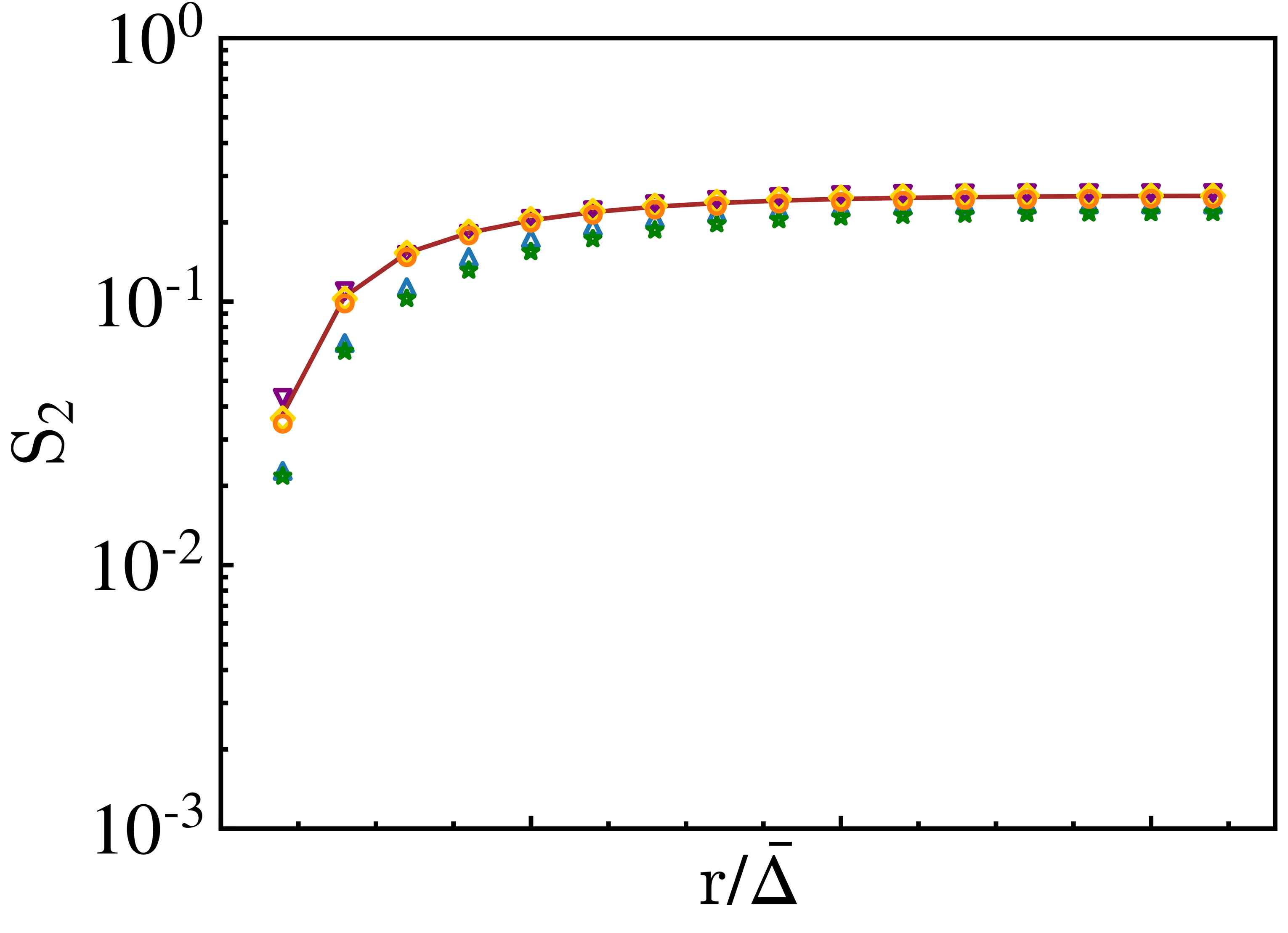}
            \put(-3,64){\small (h)} 
        \end{overpic}
    \end{subfigure}
    \hfill
    \begin{subfigure}[b]{0.32\textwidth}
        \begin{overpic}[width=1\linewidth]{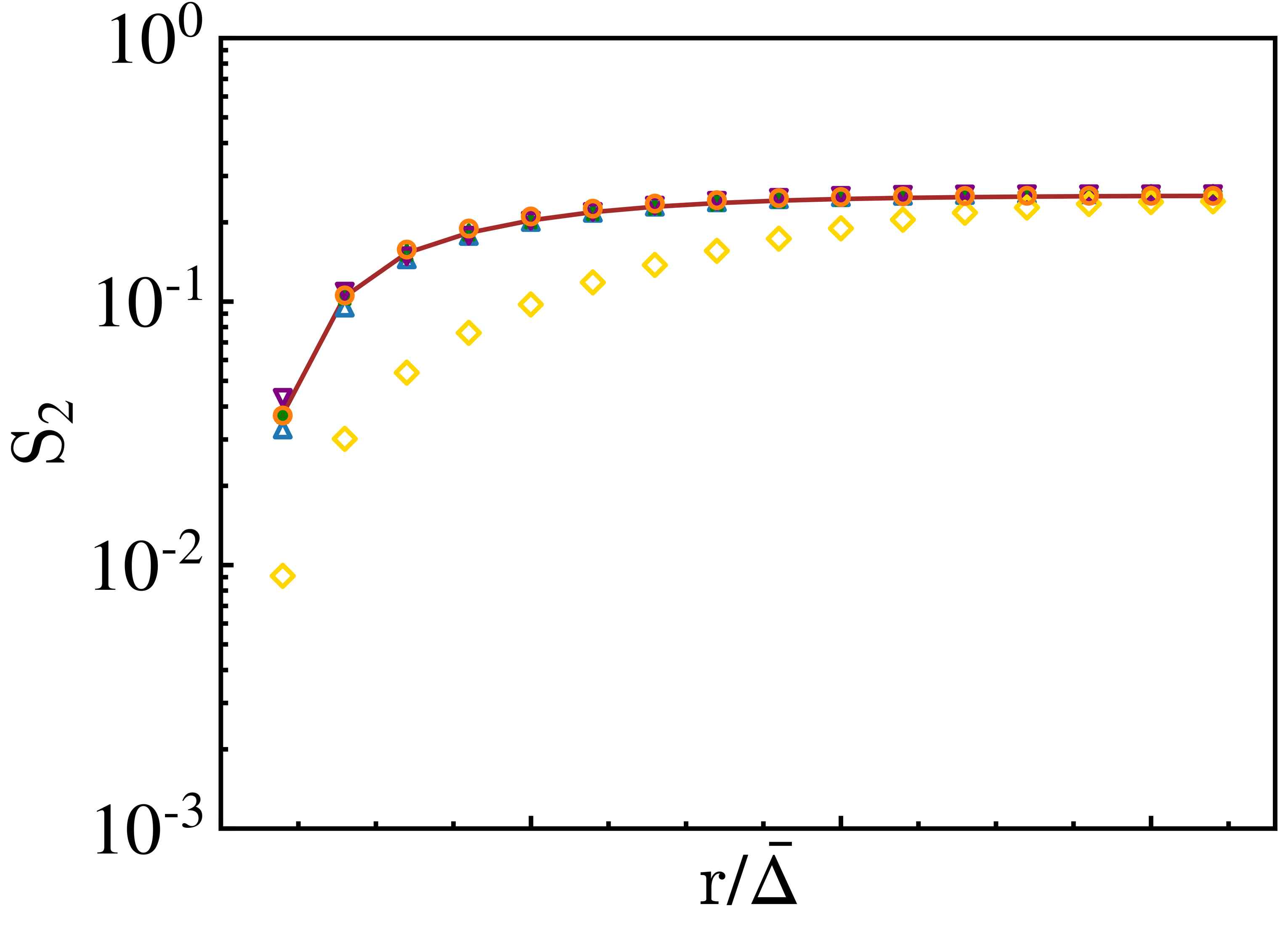}
            \put(-3,64){\small (i)} 
        \end{overpic}
    \end{subfigure}
    \vspace{0.1cm}
    \begin{subfigure}[b]{0.32\textwidth}
        \begin{overpic}[width=1\linewidth]{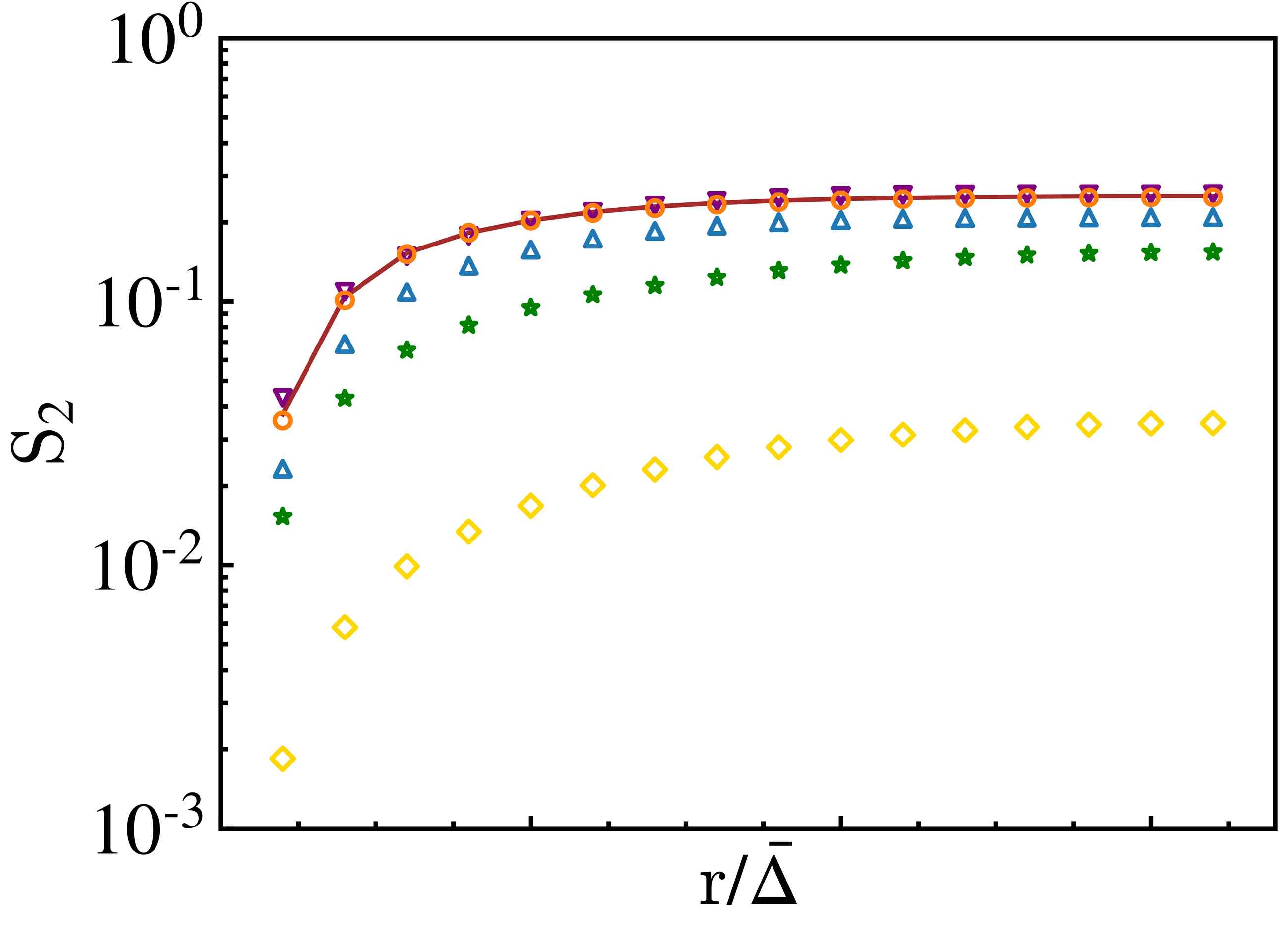}
            \put(-3,64){\small (j)} 
        \end{overpic}
    \end{subfigure}
    \hfill
    \begin{subfigure}[b]{0.32\textwidth}
        \begin{overpic}[width=1\linewidth]{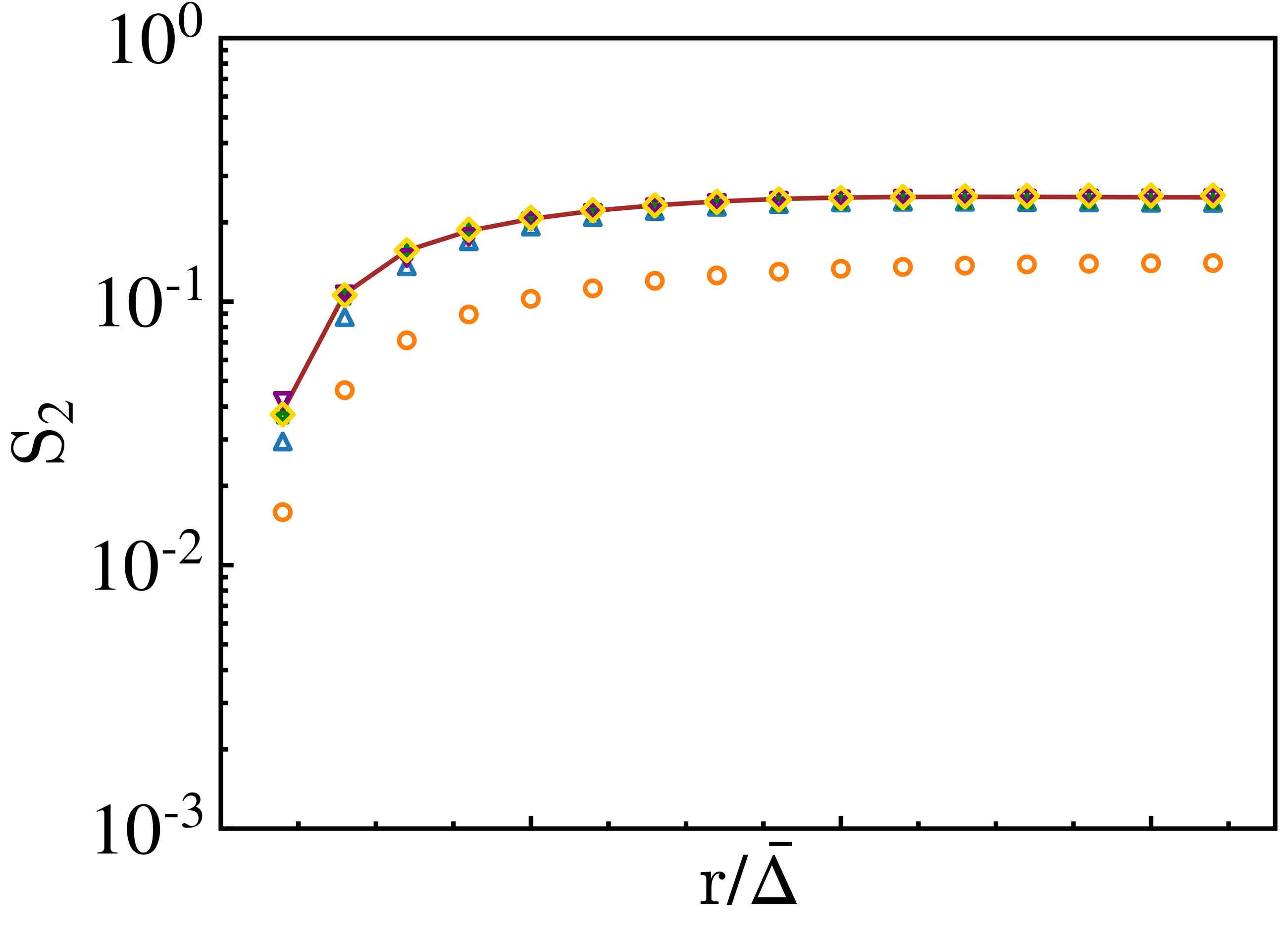}
            \put(-3,64){\small (k)} 
        \end{overpic}
    \end{subfigure}
    \hfill
    \begin{subfigure}[b]{0.32\textwidth}
        \begin{overpic}[width=1\linewidth]{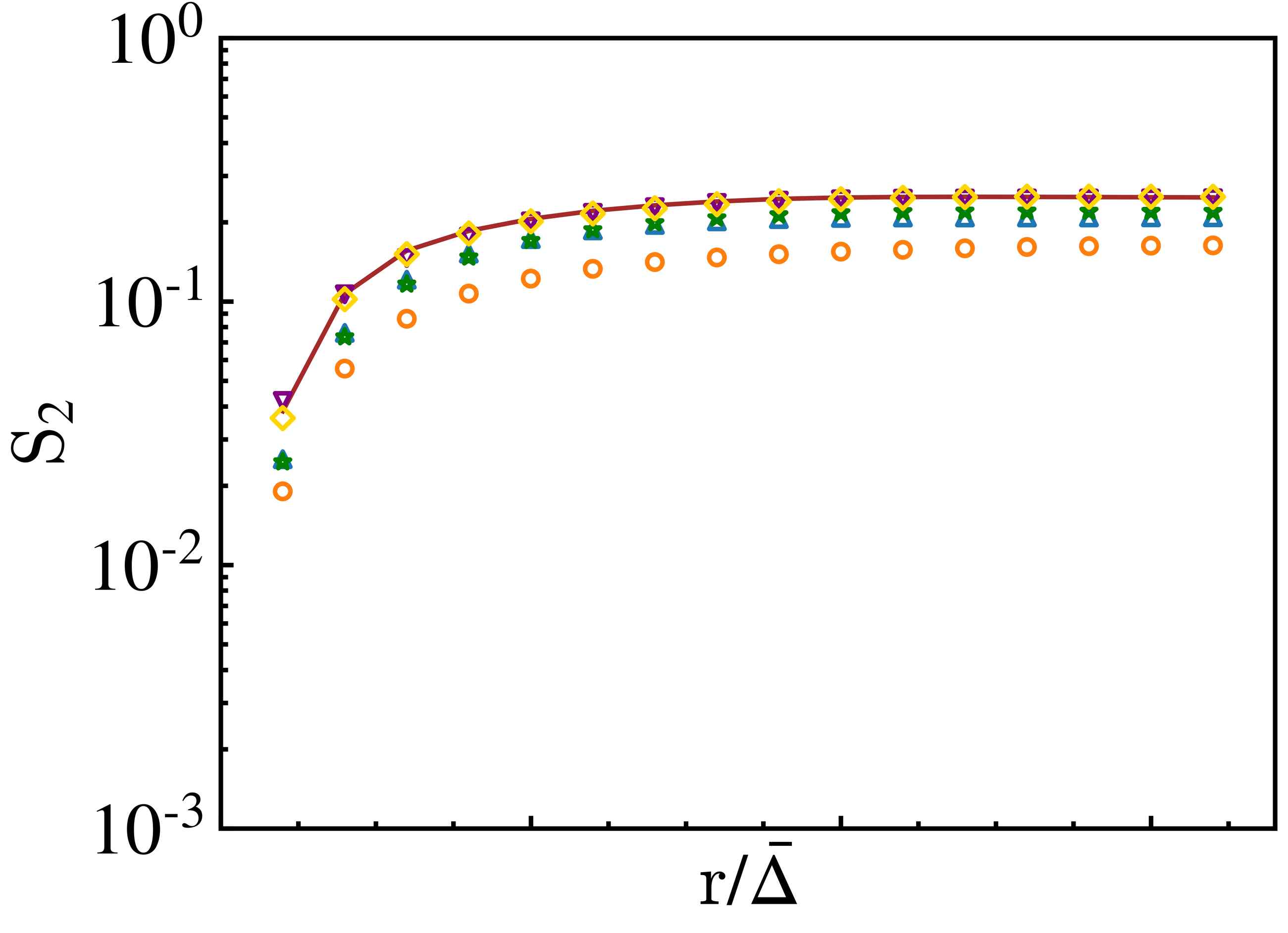}
            \put(-3,64){\small (l)}  
        \end{overpic}
    \end{subfigure}
	\caption{Second-order structure functions of LES obtained using different models in forced HIT under various training and prediction time intervals at the statistically steady state: (a) $\Delta T=0.02\tau$ constrained; (b) $\Delta T=0.02\tau$ unconstrained; (c) $\Delta T=0.04\tau$ constrained; (d) $\Delta T=0.04\tau$ unconstrained; (e) $\Delta T=0.1\tau$ constrained; (f) $\Delta T=0.1\tau$ unconstrained; (g) $\Delta T=0.2\tau$ constrained; (h) $\Delta T=0.2\tau$ unconstrained; (i) $\Delta T=0.3\tau$ constrained; (j) $\Delta T=0.3\tau$ unconstrained; (k) $\Delta T=0.4\tau$ constrained; (l) $\Delta T=0.4\tau$ unconstrained. Here, the time instance shown for (a)-(l) is $t/\tau=120$.}\label{fig:3}
\end{figure}

To further investigate the ability of different methods to predict multi-scale properties of turbulence, we compute the longitudinal structure functions of the filtered velocity, defined as~\cite{xieModifiedOptimalModel2018b,articlexiecicp10.4208LESCIT,Wang_Wan_Chen_Chen_2018KETinCIT}:
\begin{equation}
\bar{S}_n(r) = \left\langle \left| \frac{\delta_r \bar{u}}{\bar{u}^{\mathrm{rms}}} \right|^n \right\rangle,
\label{eq:32}
\end{equation}
where $n$ denotes the order of the structure function, and $\delta_r \bar{u} = [\bar{\mathbf{u}}(\mathbf{x}+\mathbf{r}) - \bar{\mathbf{u}}(\mathbf{x})] \cdot \hat{\mathbf{r}}$ represents the longitudinal velocity increment at spatial separation $\mathbf{r}$. Here, $\hat{\mathbf{r}} = \mathbf{r} / |\mathbf{r}|$ is the unit vector in the direction of $\mathbf{r}$.
Fig.~\ref{fig:3} compares the second-order structure functions of the filtered velocity predicted by different models with those from fDNS at the statistically steady state. It can be observed that, within the time interval range $\Delta T \in [0.1\tau, 0.2\tau]$, FNO-based models with prediction constraints achieve accurate long-term predictions. In particular, unconstrained F-IFNO and F-IUFNO demonstrate good accuracy in this range, whereas unconstrained IFNO and IUFNO perform poorly.
For $\Delta T = 0.02\tau, 0.04\tau, 0.3\tau, 0.4\tau$, the performance of F-IFNO and F-IUFNO declines compared to their results within $\Delta T \in [0.1\tau, 0.2\tau]$. However, IFNO and IUFNO show improved accuracy at these time intervals $\Delta T = 0.02\tau, 0.04\tau, 0.3\tau, 0.4\tau$, except for the unconstrained IFNO at $\Delta T = 0.02\tau$, which performs poorly.
Additionally, the DSM model overestimates the structure functions at small separations and underestimates them at large separations, in comparison with fDNS data. These observations further validate that the optimal time interval for F-IFNO and F-IUFNO lies within $\Delta T \in [0.1\tau, 0.2\tau]$, while the optimal range for IFNO and IUFNO is slightly smaller.

\begin{figure}[ht!]
    \centering
    \begin{subfigure}[b]{0.32\textwidth}
        \begin{overpic}[width=1\linewidth]{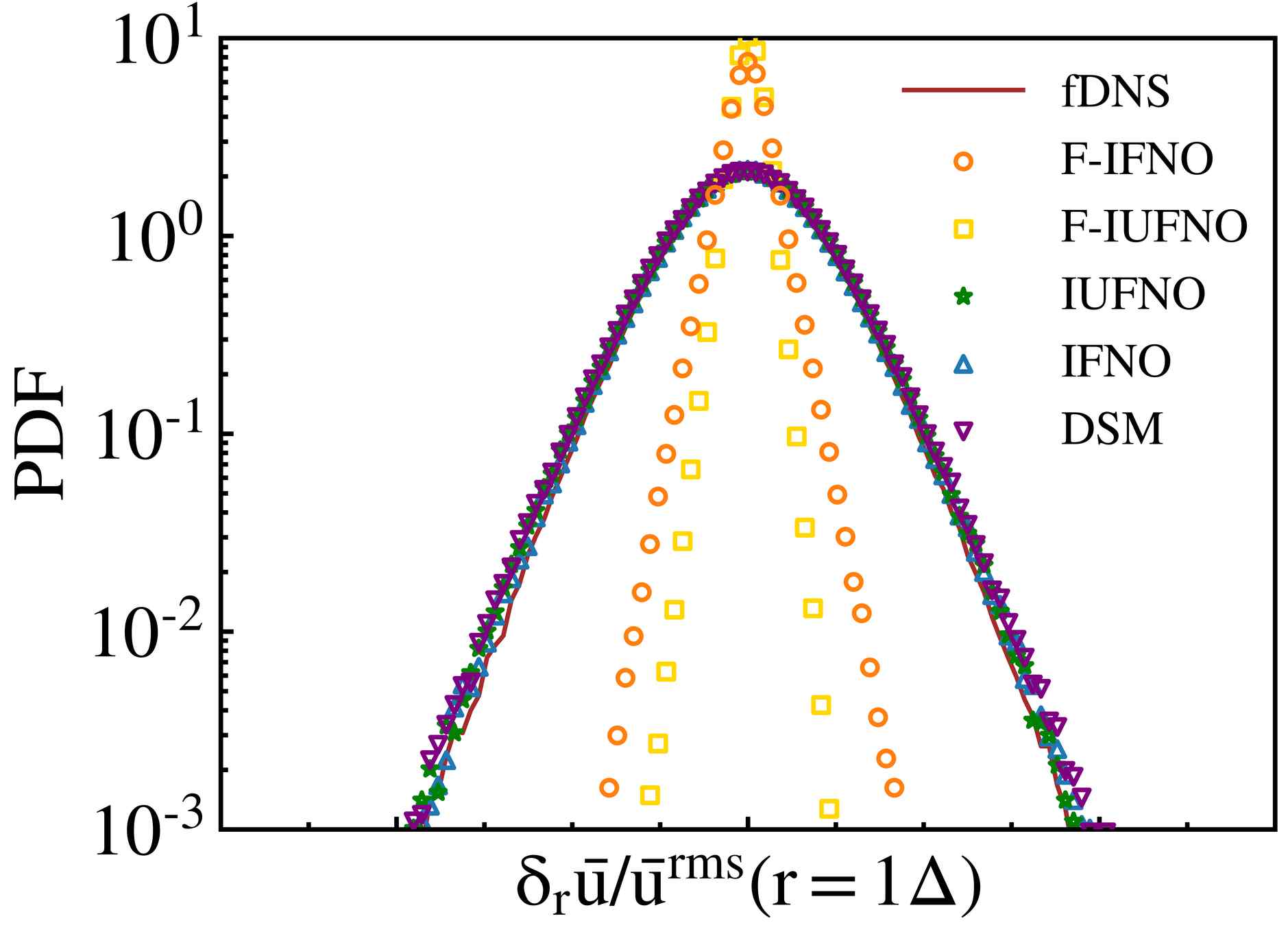}
            \put(-3,65){\small (a)}  
        \end{overpic}
    \end{subfigure}
    \hfill
    \begin{subfigure}[b]{0.32\textwidth}
        \begin{overpic}[width=1\linewidth]{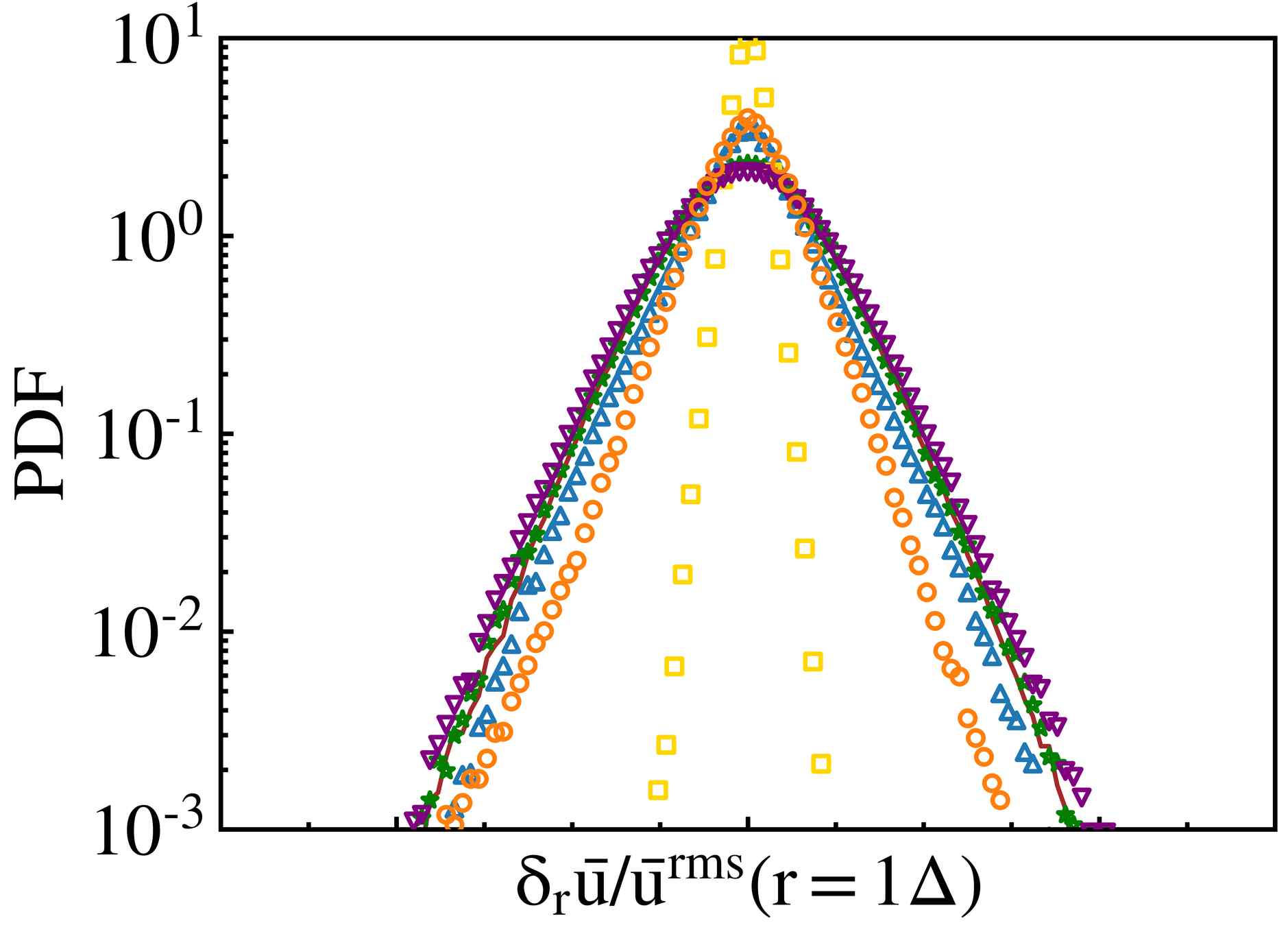}
            \put(-3,65){\small (b)} 
        \end{overpic} 
    \end{subfigure}
    \hfill
    \begin{subfigure}[b]{0.32\textwidth}
        \begin{overpic}[width=1\linewidth]{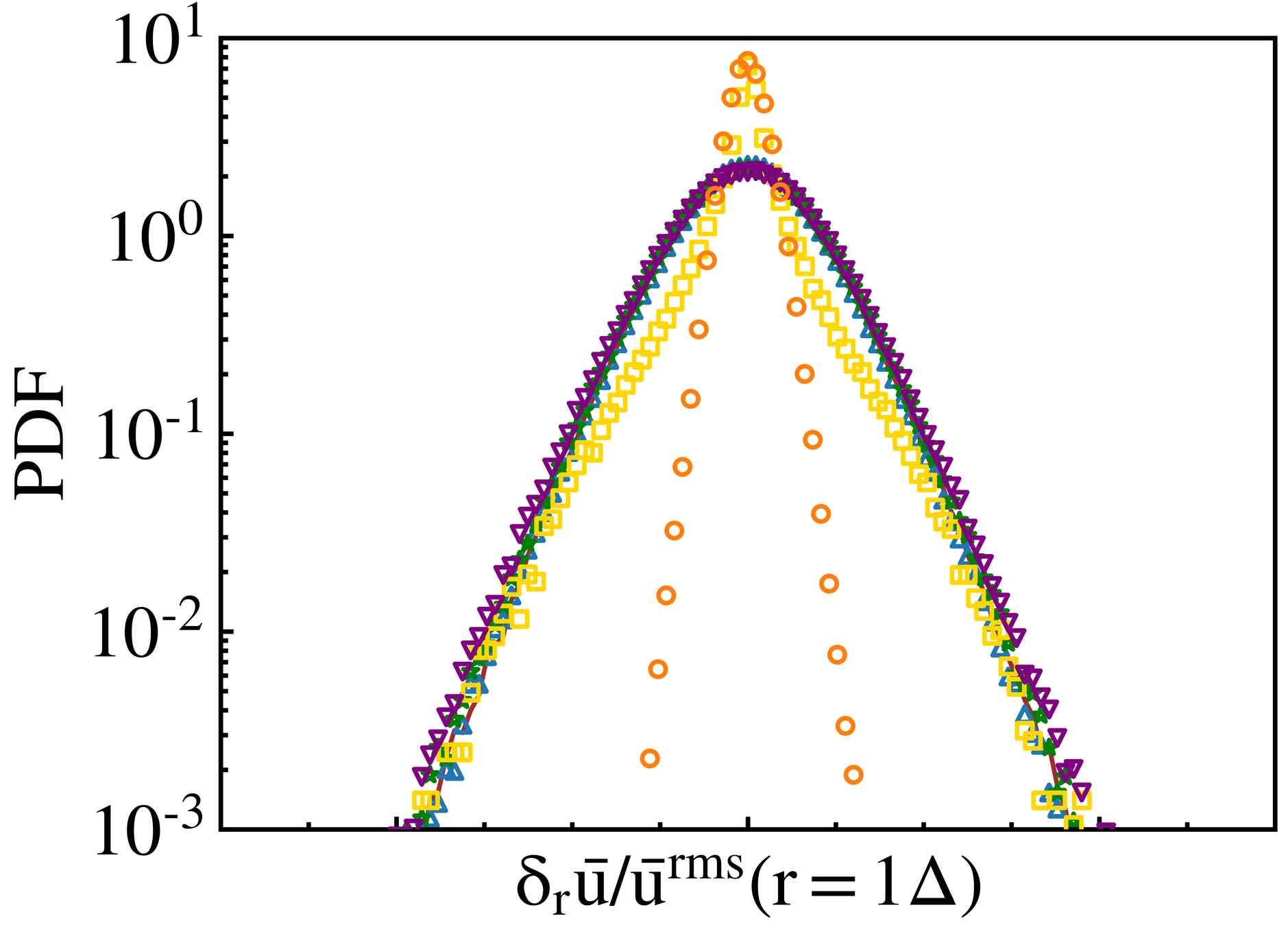}
            \put(-3,65){\small (c)} 
        \end{overpic}
    \end{subfigure}
    \vspace{0.1cm}
    \begin{subfigure}[b]{0.32\textwidth}
        \begin{overpic}[width=1\linewidth]{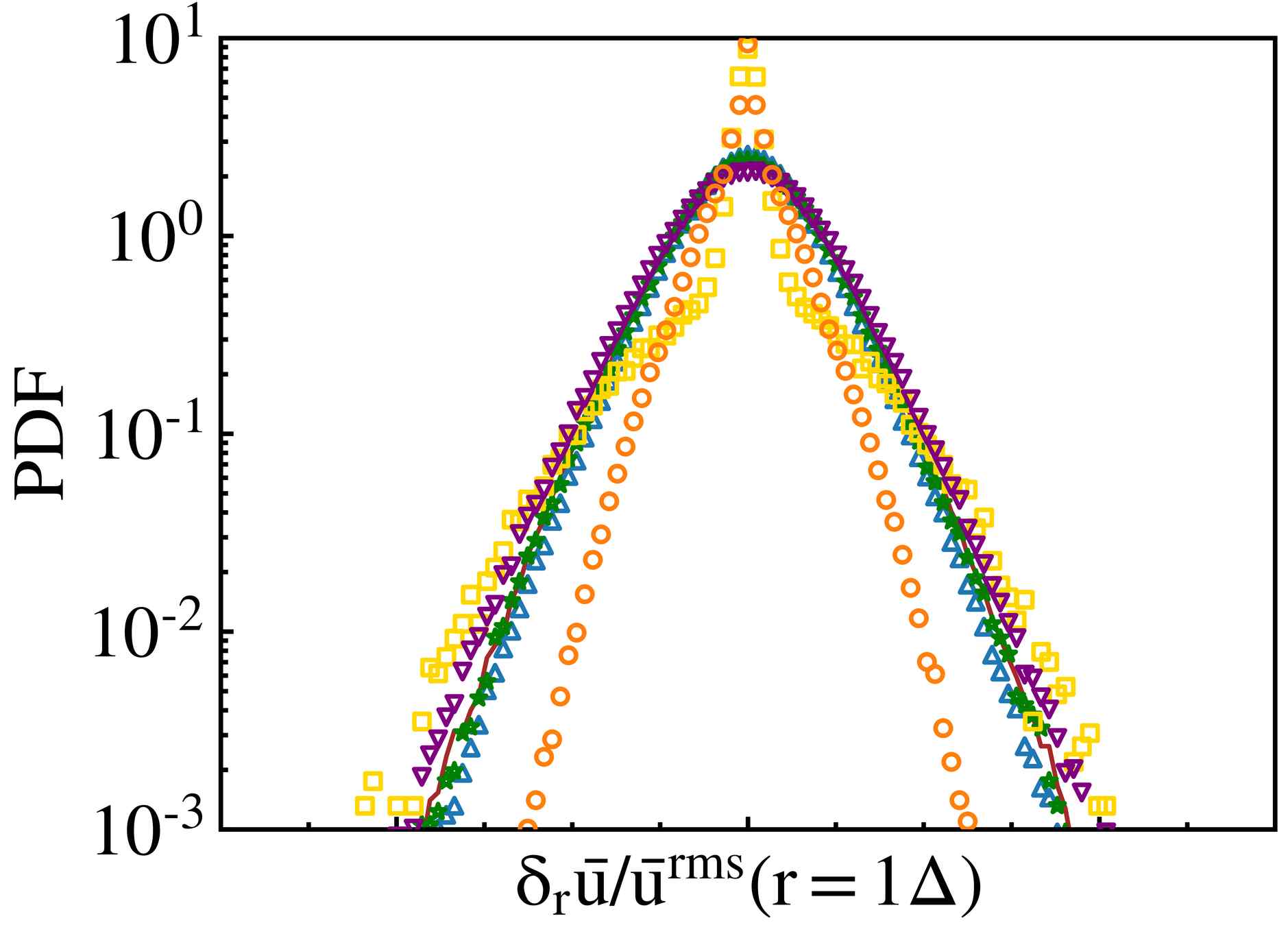}
            \put(-3,65){\small (d)} 
        \end{overpic}
    \end{subfigure}
    \hfill
    \begin{subfigure}[b]{0.32\textwidth}
        \begin{overpic}[width=1\linewidth]{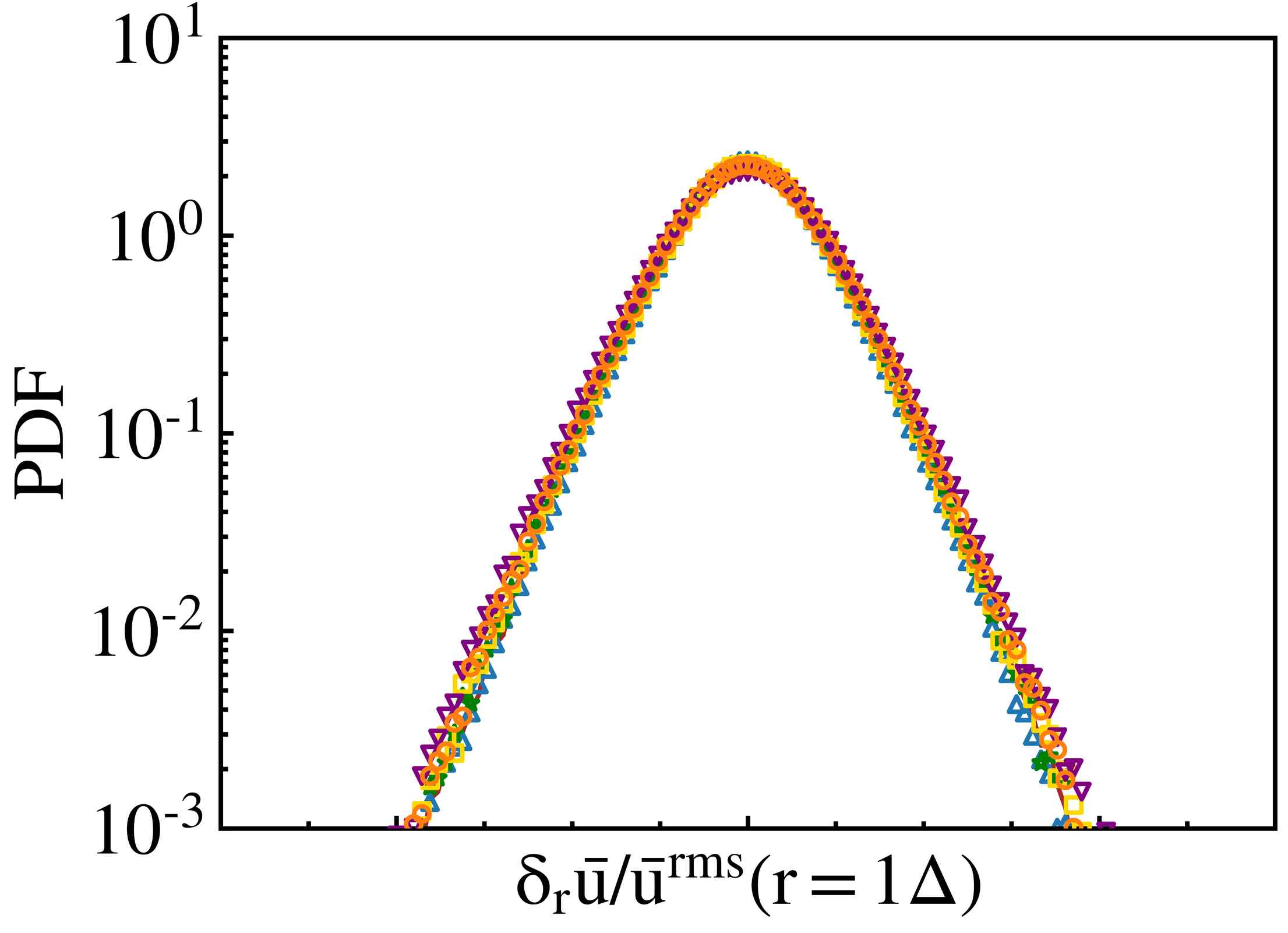}
            \put(-3,65){\small (e)} 
        \end{overpic}
    \end{subfigure}
    \hfill
    \begin{subfigure}[b]{0.32\textwidth}
        \begin{overpic}[width=1\linewidth]{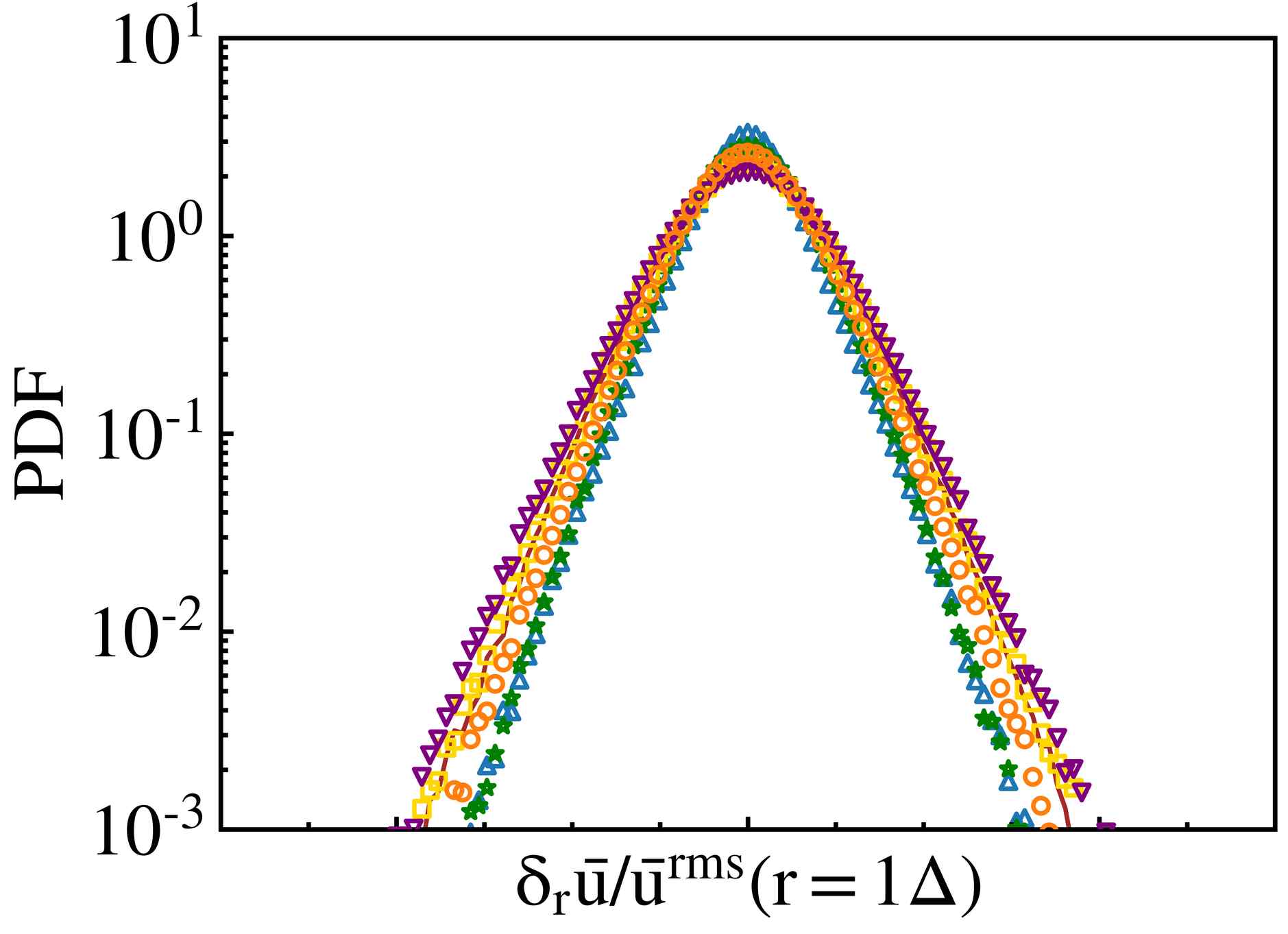}
            \put(-3,65){\small (f)} 
        \end{overpic}
    \end{subfigure}
    \vspace{0.1cm}
    \begin{subfigure}[b]{0.32\textwidth}
        \begin{overpic}[width=1\linewidth]{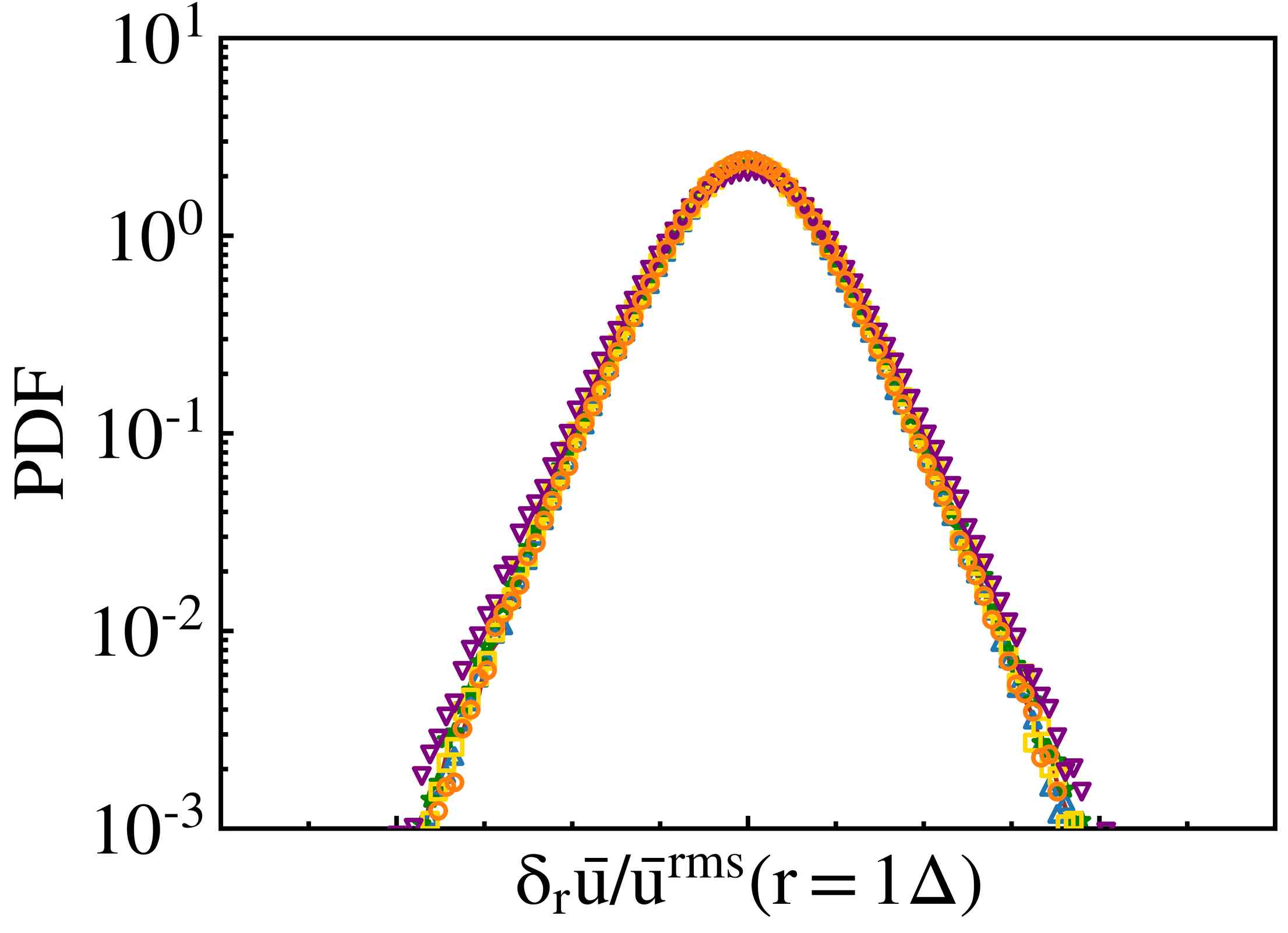}
            \put(-3,65){\small (g)} 
        \end{overpic}
    \end{subfigure}
    \hfill
    \begin{subfigure}[b]{0.32\textwidth}
        \begin{overpic}[width=1\linewidth]{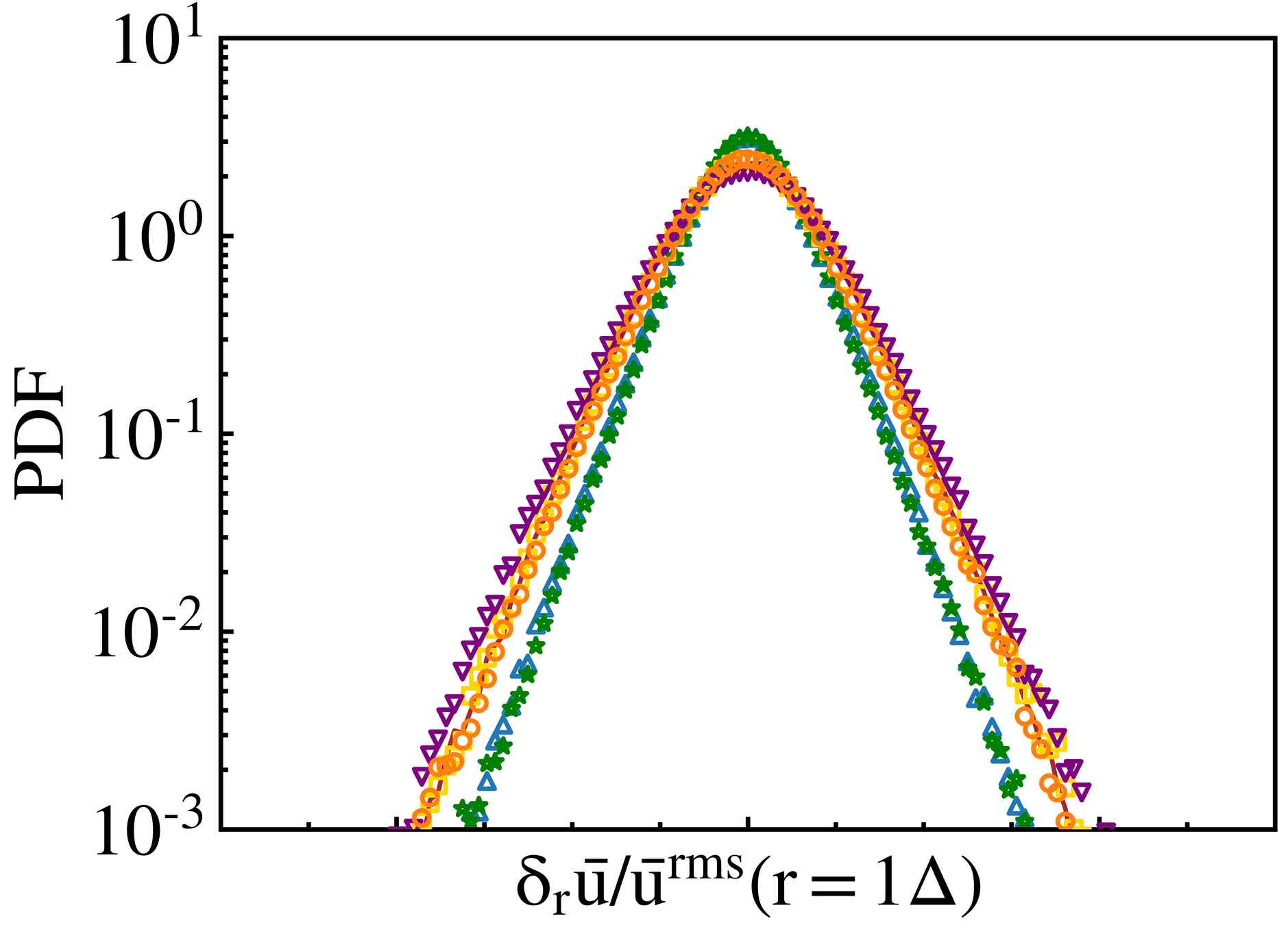}
            \put(-3,65){\small (h)} 
        \end{overpic}
    \end{subfigure}
    \hfill
    \begin{subfigure}[b]{0.32\textwidth}
        \begin{overpic}[width=1\linewidth]{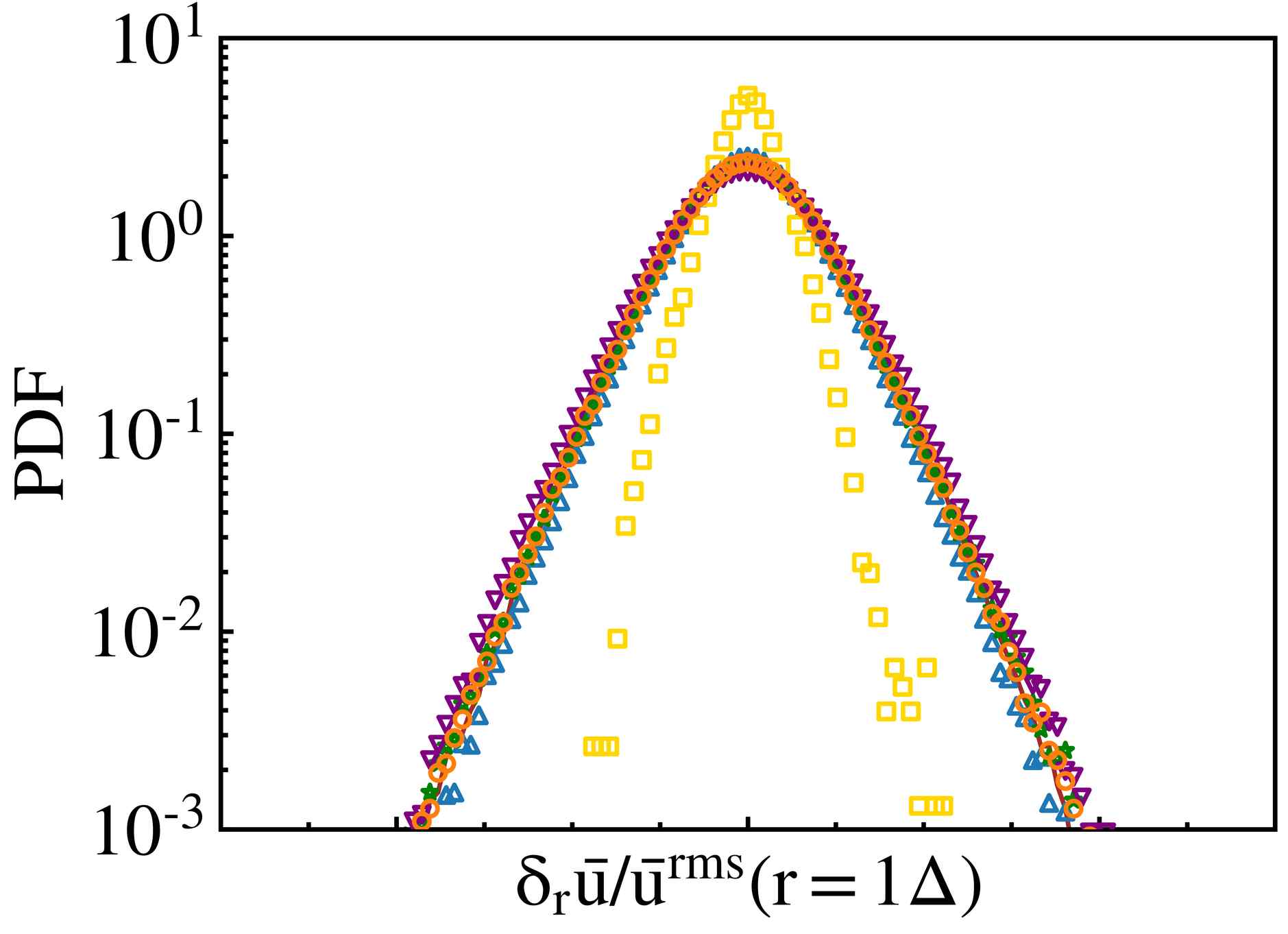}
            \put(-3,65){\small (i)} 
        \end{overpic}
    \end{subfigure}
    \vspace{0.1cm}
    \begin{subfigure}[b]{0.32\textwidth}
        \begin{overpic}[width=1\linewidth]{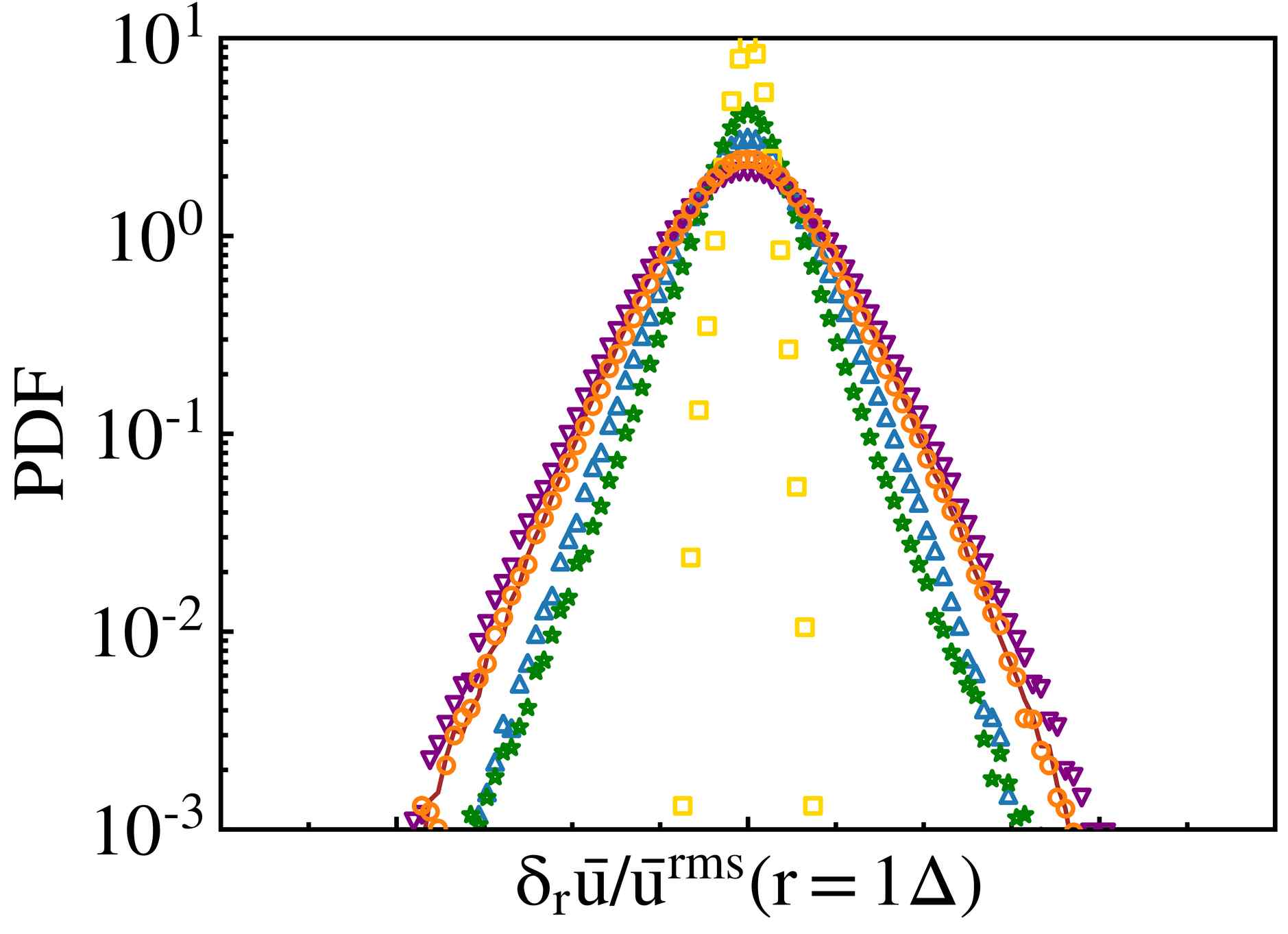}
            \put(-3,65){\small (j)} 
        \end{overpic}
    \end{subfigure}
    \hfill
    \begin{subfigure}[b]{0.32\textwidth}
        \begin{overpic}[width=1\linewidth]{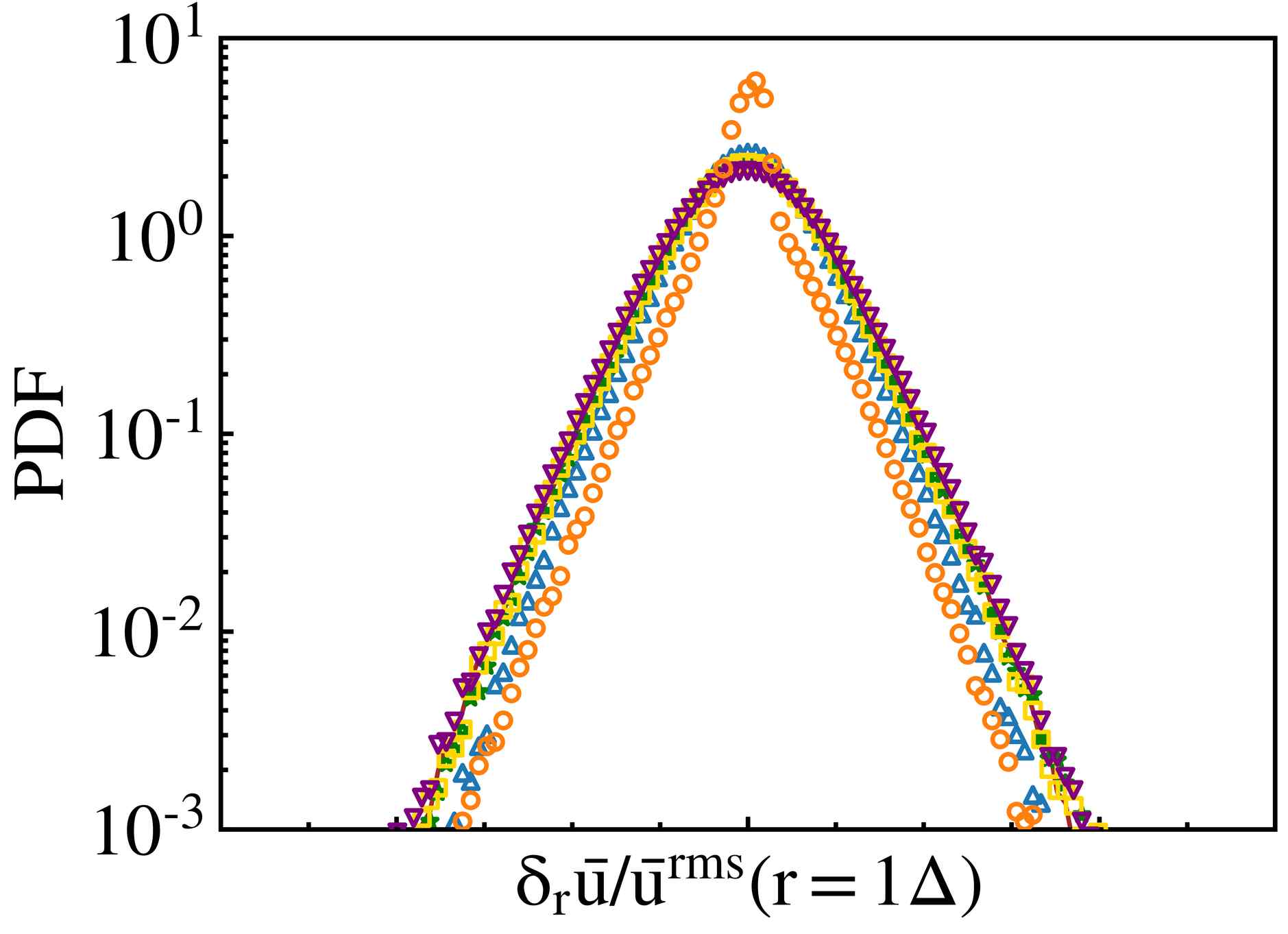}
            \put(-3,65){\small (k)} 
        \end{overpic}
    \end{subfigure}
    \hfill
    \begin{subfigure}[b]{0.32\textwidth}
        \begin{overpic}[width=1\linewidth]{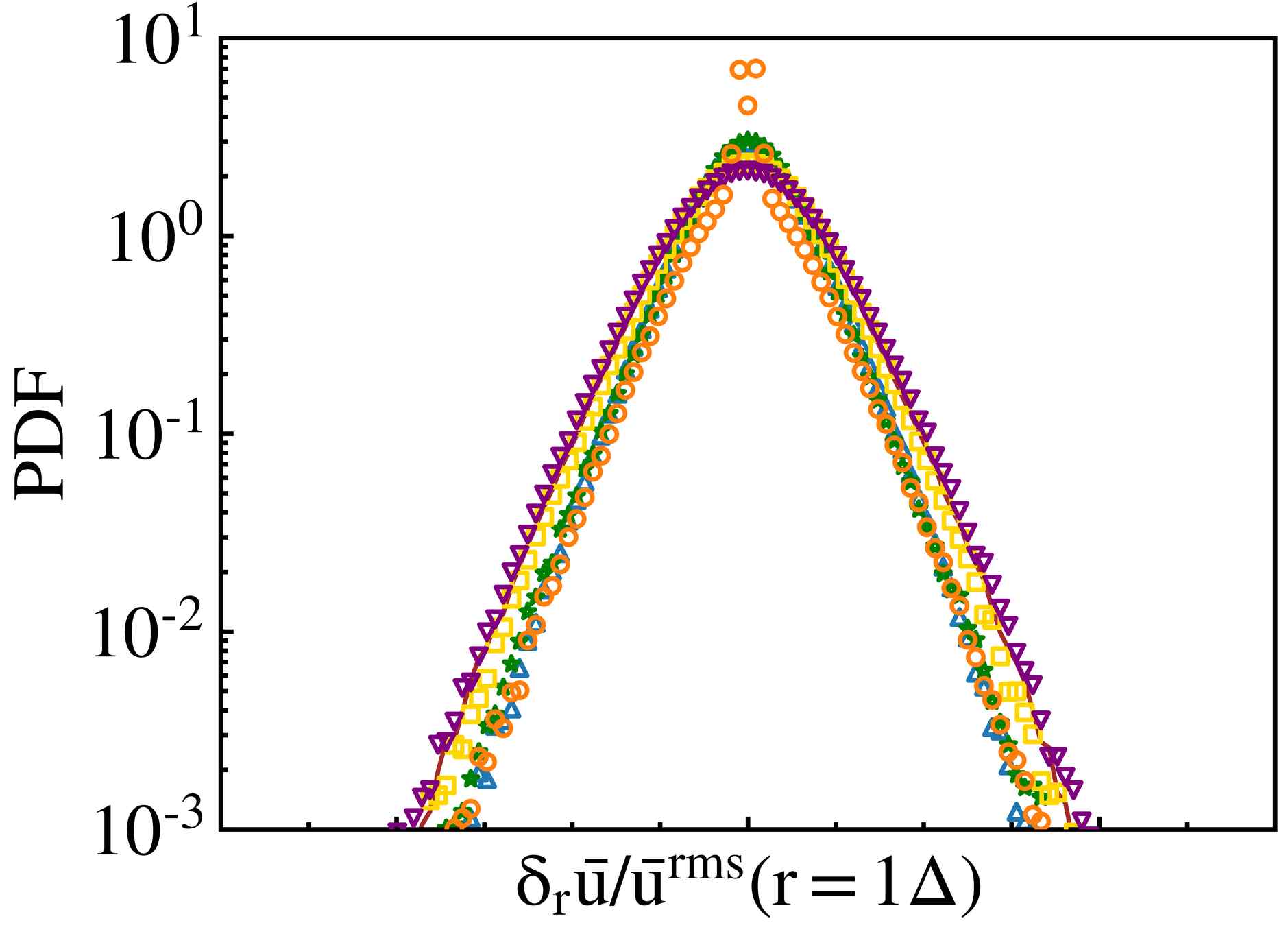}
            \put(-3,65){\small (l)}  
        \end{overpic}
    \end{subfigure}
	\caption{The PDFs of the normalized velocity increments $\delta_r \bar{u}/\bar{u}^{rms}$ of LES obtained using different models in forced HIT under various training and prediction time intervals at the statistically steady state: (a) $\Delta T=0.02\tau$ constrained; (b) $\Delta T=0.02\tau$ unconstrained; (c) $\Delta T=0.04\tau$ constrained; (d) $\Delta T=0.04\tau$ unconstrained; (e) $\Delta T=0.1\tau$ constrained; (f) $\Delta T=0.1\tau$ unconstrained; (g) $\Delta T=0.2\tau$ constrained; (h) $\Delta T=0.2\tau$ unconstrained; (i) $\Delta T=0.3\tau$ constrained; (j) $\Delta T=0.3\tau$ unconstrained; (k) $\Delta T=0.4\tau$ constrained; (l) $\Delta T=0.4\tau$ unconstrained. Here, the time instance shown for (a)-(l) is $t/\tau=120$.}\label{fig:4}
\end{figure}

Moreover, we compare the probability density functions (PDFs) of the normalized velocity increments $\delta_r \bar{u}/\bar{u}^{\mathrm{rms}}$ at a spatial separation of $r = \Delta$ under statistically steady conditions, as shown in Fig.~\ref{fig:4}. Both constrained and unconstrained versions of F-IFNO and F-IUFNO show a good agreement with the fDNS data when the training and prediction time intervals lie within the range $\Delta T \in [0.1\tau, 0.2\tau]$. In contrast, for IFNO and IUFNO, only the constrained models align well with the fDNS results in this interval, while their unconstrained counterparts perform poorly.
Outside this optimal interval, specifically at $\Delta T = 0.02\tau, 0.04\tau, 0.3\tau, 0.4\tau$, F-IFNO and F-IUFNO exhibit significantly degraded performance. At theses time intervals, for IFNO, both constrained and unconstrained versions perform relatively well at $\Delta T = 0.02\tau, 0.04\tau$ but show a substantial decline in performance at $\Delta T = 0.3\tau, 0.4\tau$. Similarly, IUFNO exhibits comparable behavior at $\Delta T = 0.02\tau, 0.04\tau, 0.3\tau, 0.4\tau$; however, with constraints, it manages to maintain reasonable performance even at $\Delta T = 0.3\tau, 0.4\tau$. Nevertheless, the best performances of IFNO and IUFNO across $\Delta T \in [0.02\tau, 0.4\tau]$ still do not match the superior performance of F-IFNO and F-IUFNO within the optimal range.
Additionally, compared to the FNO-based models and fDNS, the DSM model displays a slightly broader PDF of normalized velocity increments. These observations further reinforce the conclusion that the time interval range $\Delta T \in [0.1\tau, 0.2\tau]$ is optimal for the training and deployment of F-IFNO and F-IUFNO models.

\begin{figure}[ht!]
    \centering
    \begin{subfigure}[b]{0.32\textwidth}
        \begin{overpic}[width=1\linewidth]{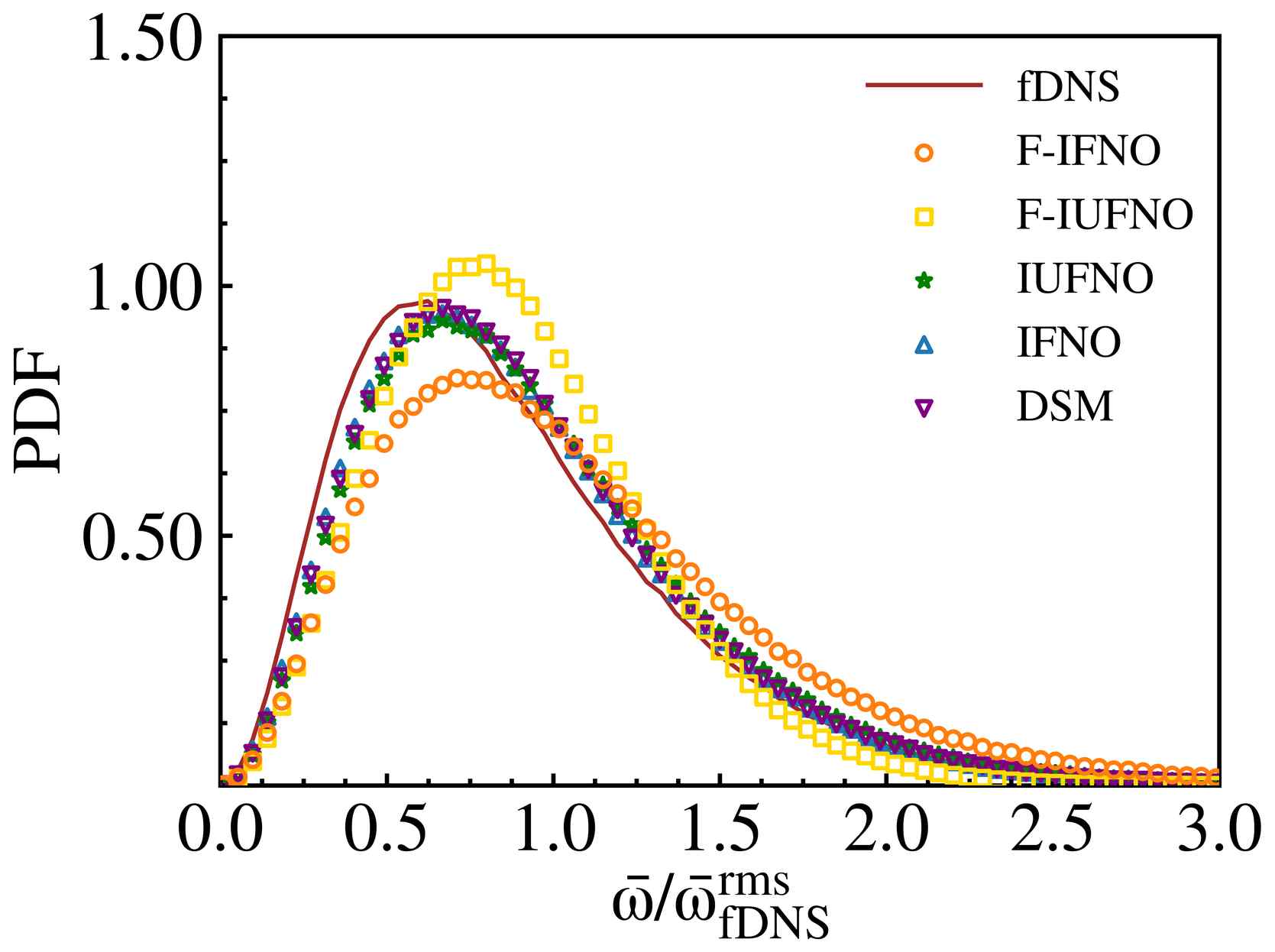}
            \put(-3,65){\small (a)}  
        \end{overpic}
    \end{subfigure}
    \hfill
    \begin{subfigure}[b]{0.32\textwidth}
        \begin{overpic}[width=1\linewidth]{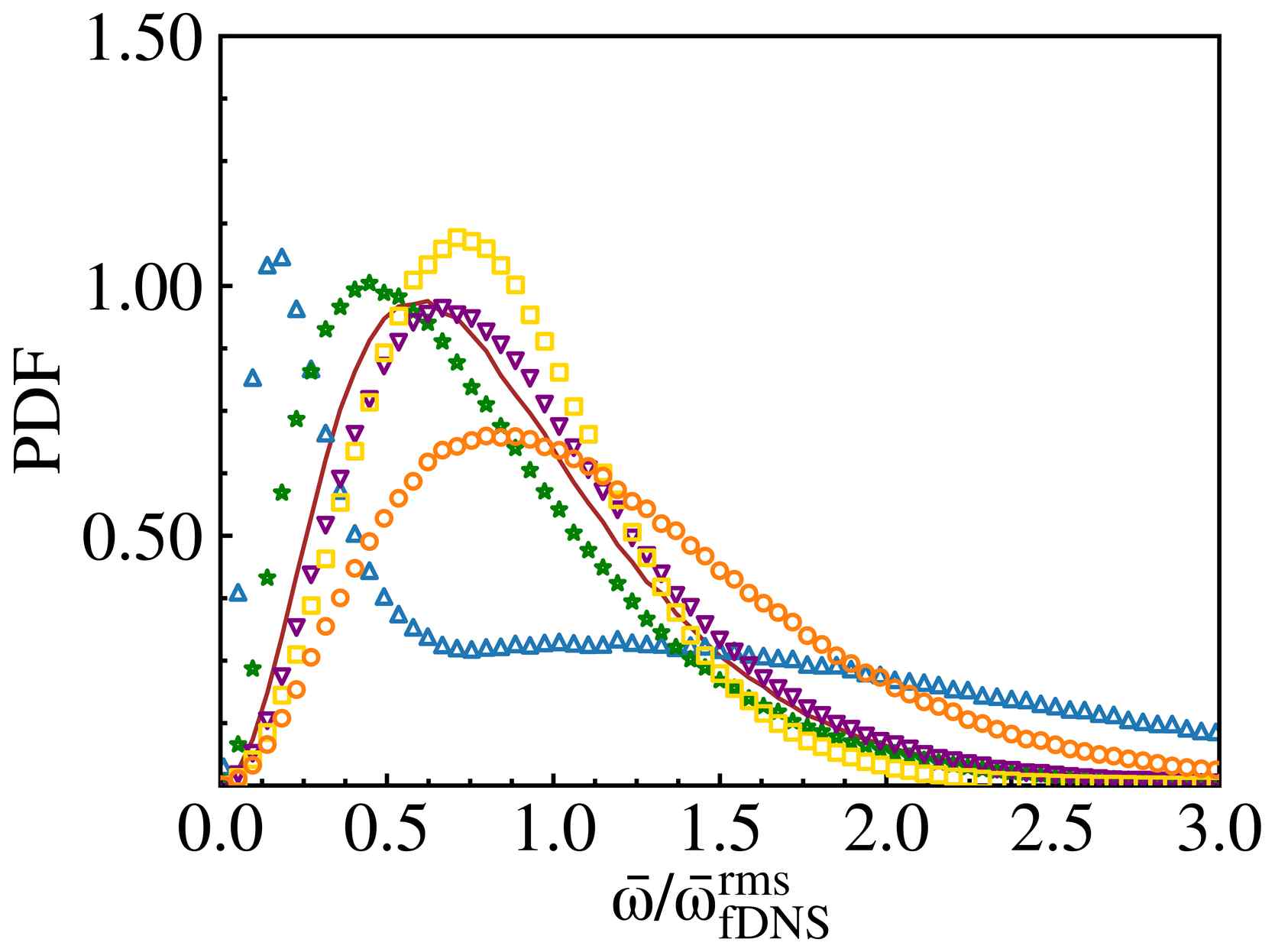}
            \put(-3,65){\small (b)} 
        \end{overpic} 
    \end{subfigure}
    \hfill
    \begin{subfigure}[b]{0.32\textwidth}
        \begin{overpic}[width=1\linewidth]{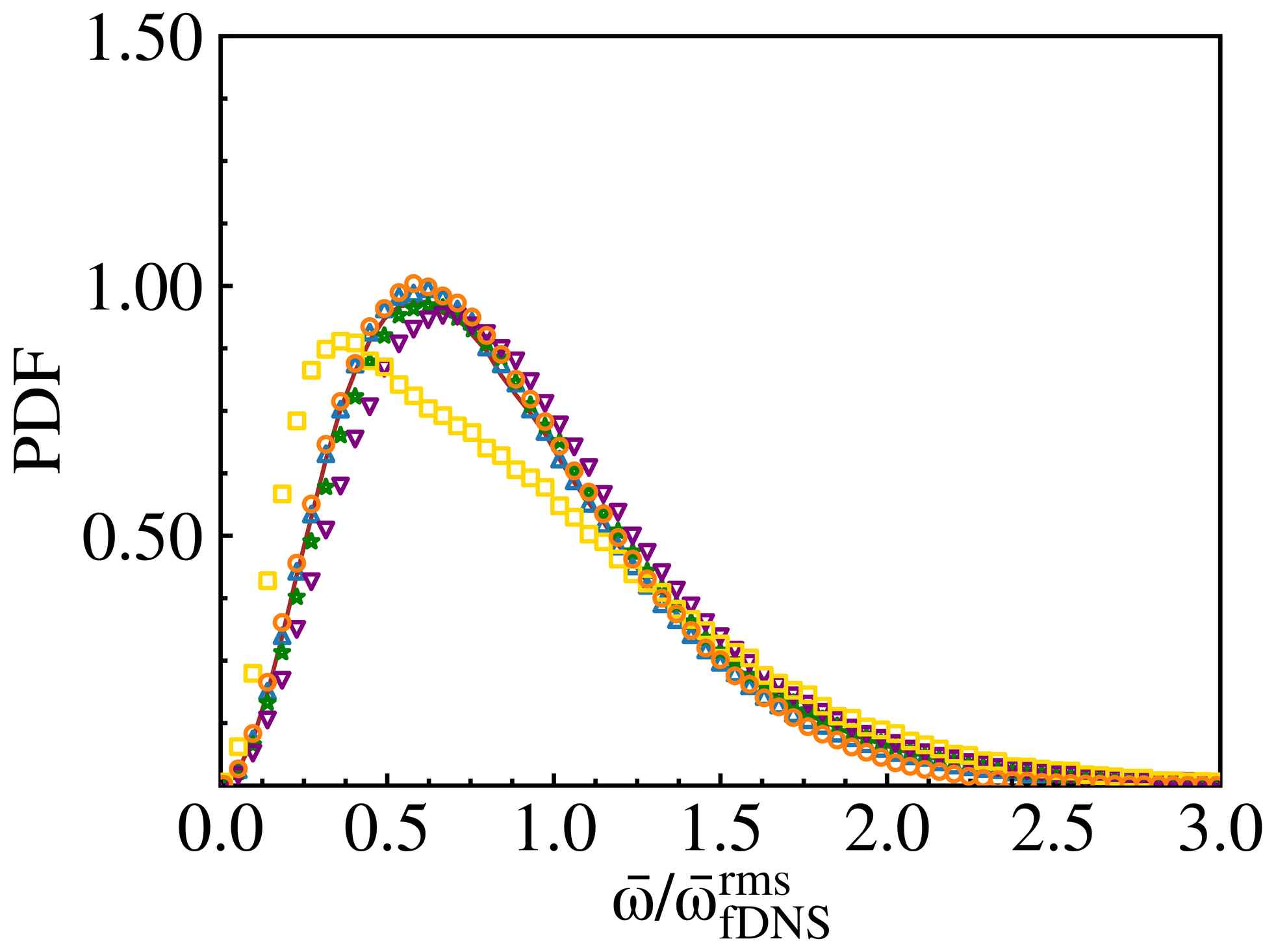}
            \put(-3,65){\small (c)} 
        \end{overpic}
    \end{subfigure}
    \vspace{0.1cm}
    \begin{subfigure}[b]{0.32\textwidth}
        \begin{overpic}[width=1\linewidth]{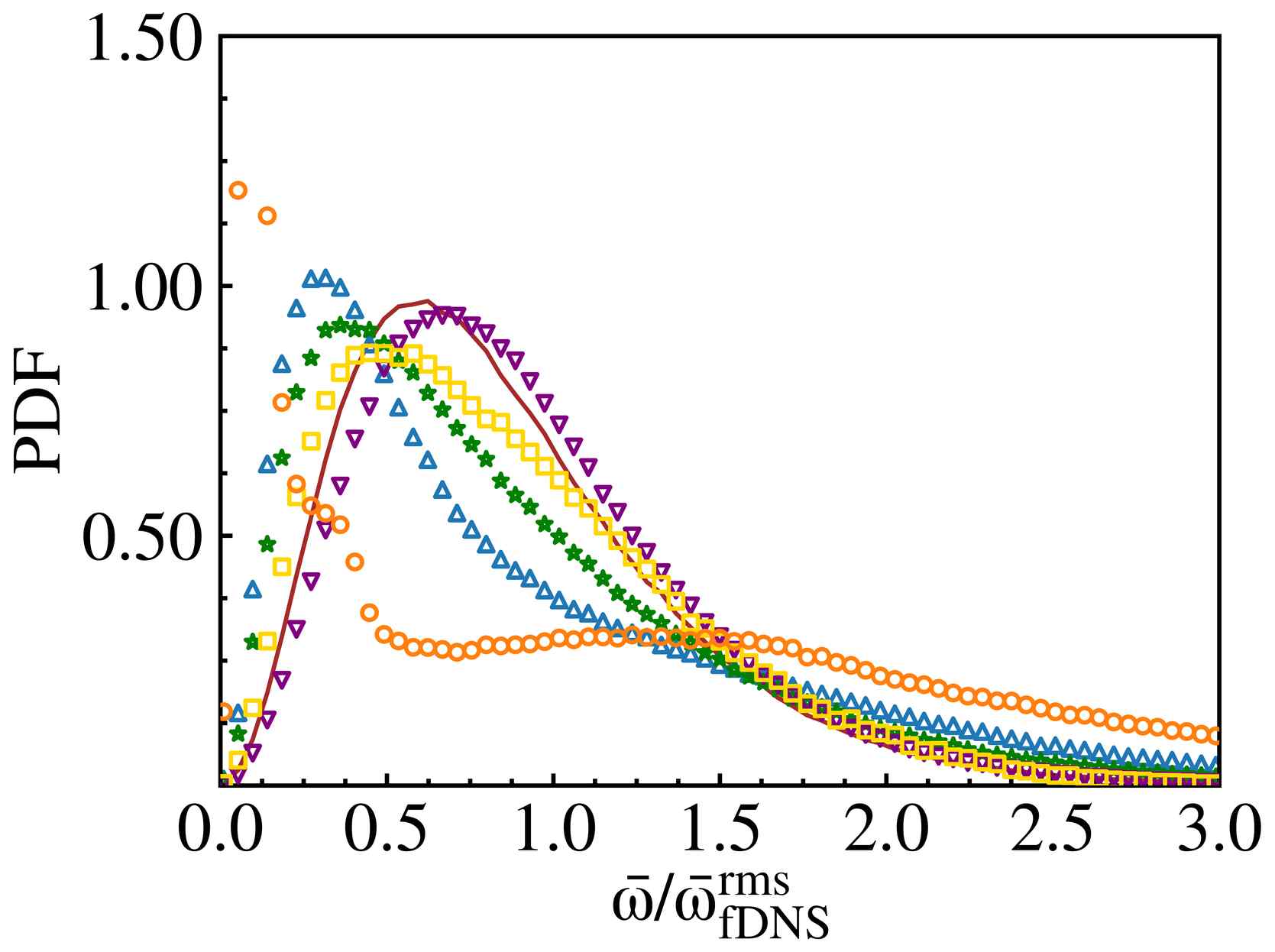}
            \put(-3,65){\small (d)} 
        \end{overpic}
    \end{subfigure}
    \hfill
    \begin{subfigure}[b]{0.32\textwidth}
        \begin{overpic}[width=1\linewidth]{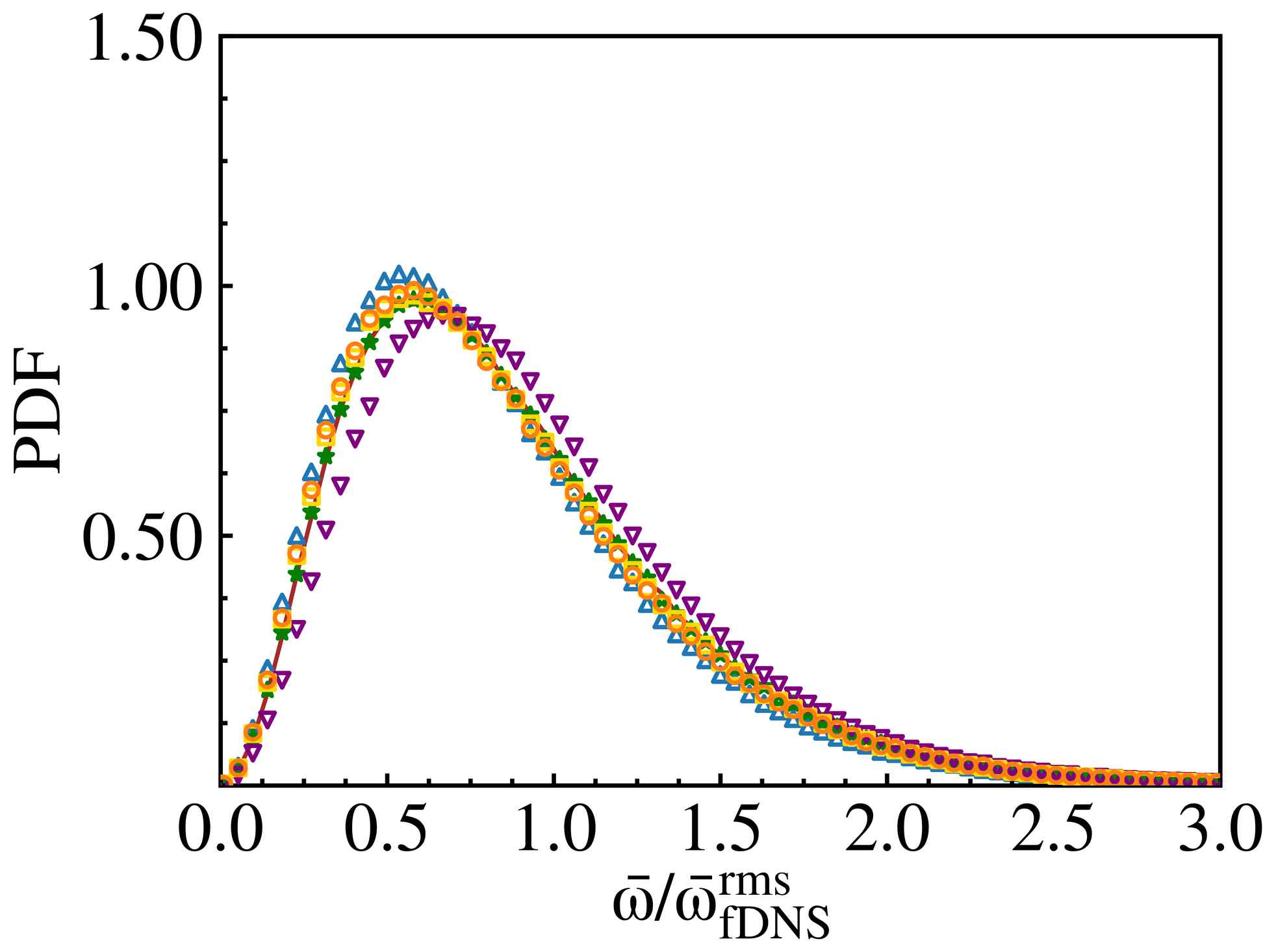}
            \put(-3,65){\small (e)} 
        \end{overpic}
    \end{subfigure}
    \hfill
    \begin{subfigure}[b]{0.32\textwidth}
        \begin{overpic}[width=1\linewidth]{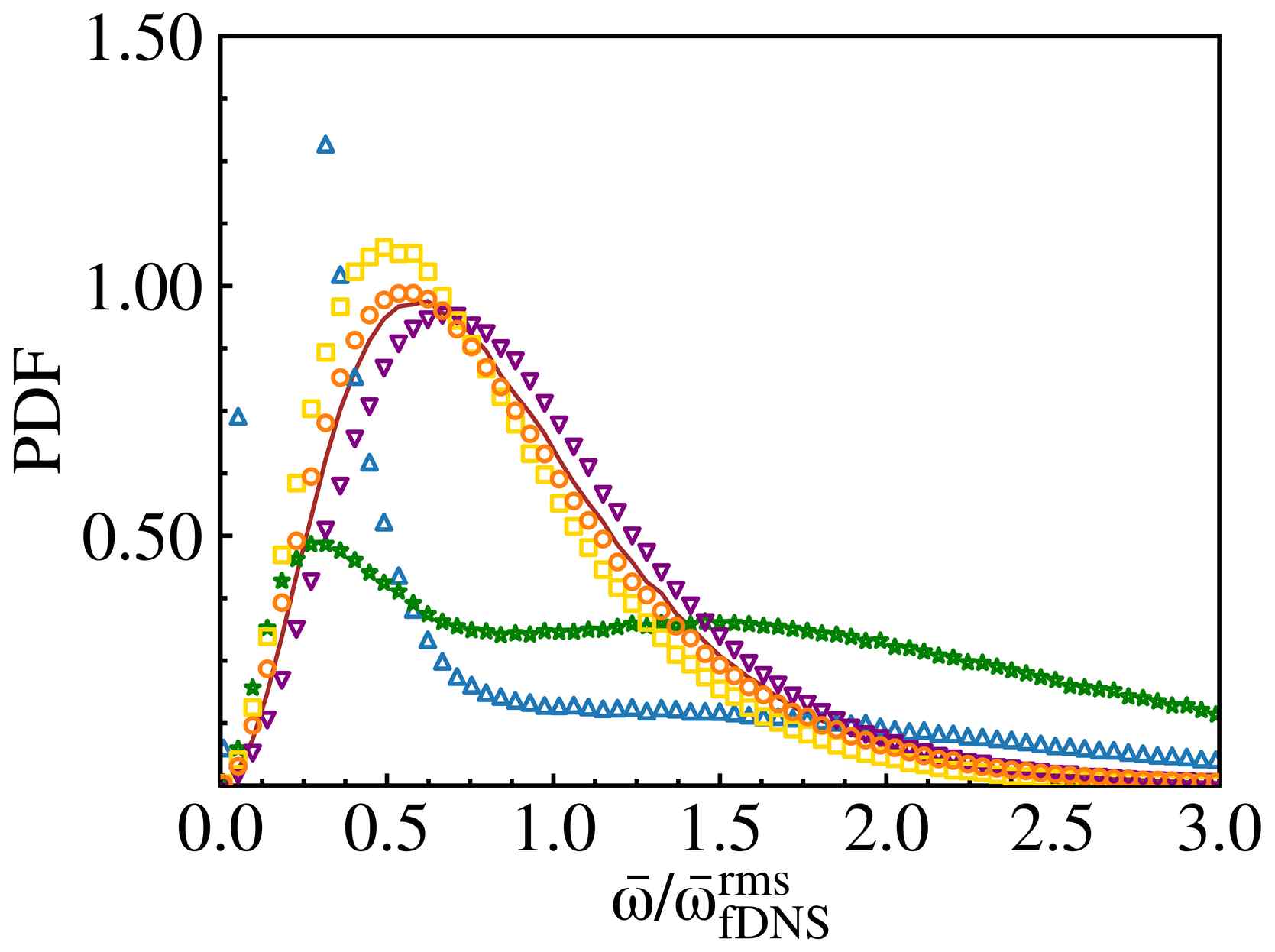}
            \put(-3,65){\small (f)} 
        \end{overpic}
    \end{subfigure}
    \vspace{0.1cm}
    \begin{subfigure}[b]{0.32\textwidth}
        \begin{overpic}[width=1\linewidth]{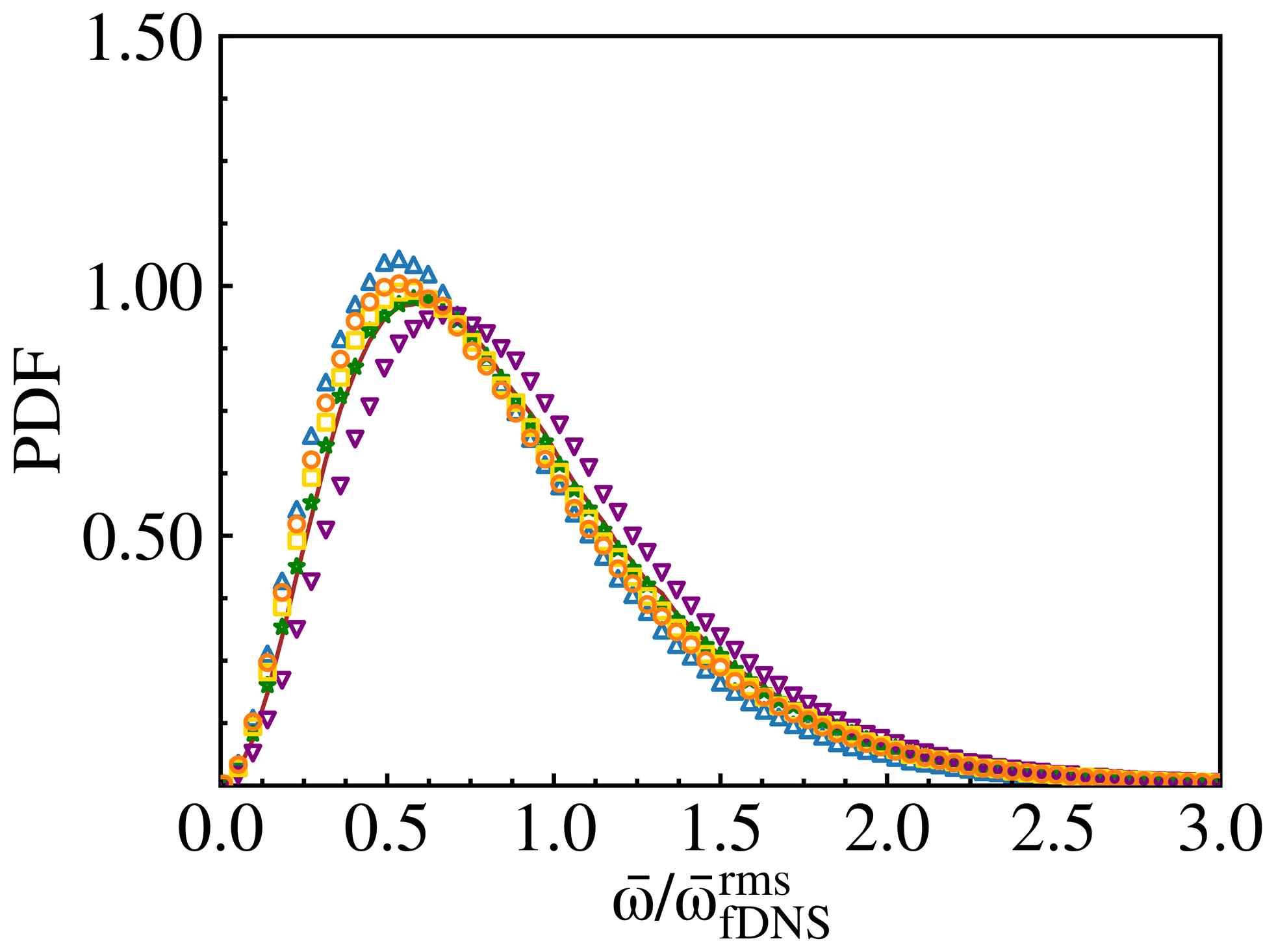}
            \put(-3,65){\small (g)} 
        \end{overpic}
    \end{subfigure}
    \hfill
    \begin{subfigure}[b]{0.32\textwidth}
        \begin{overpic}[width=1\linewidth]{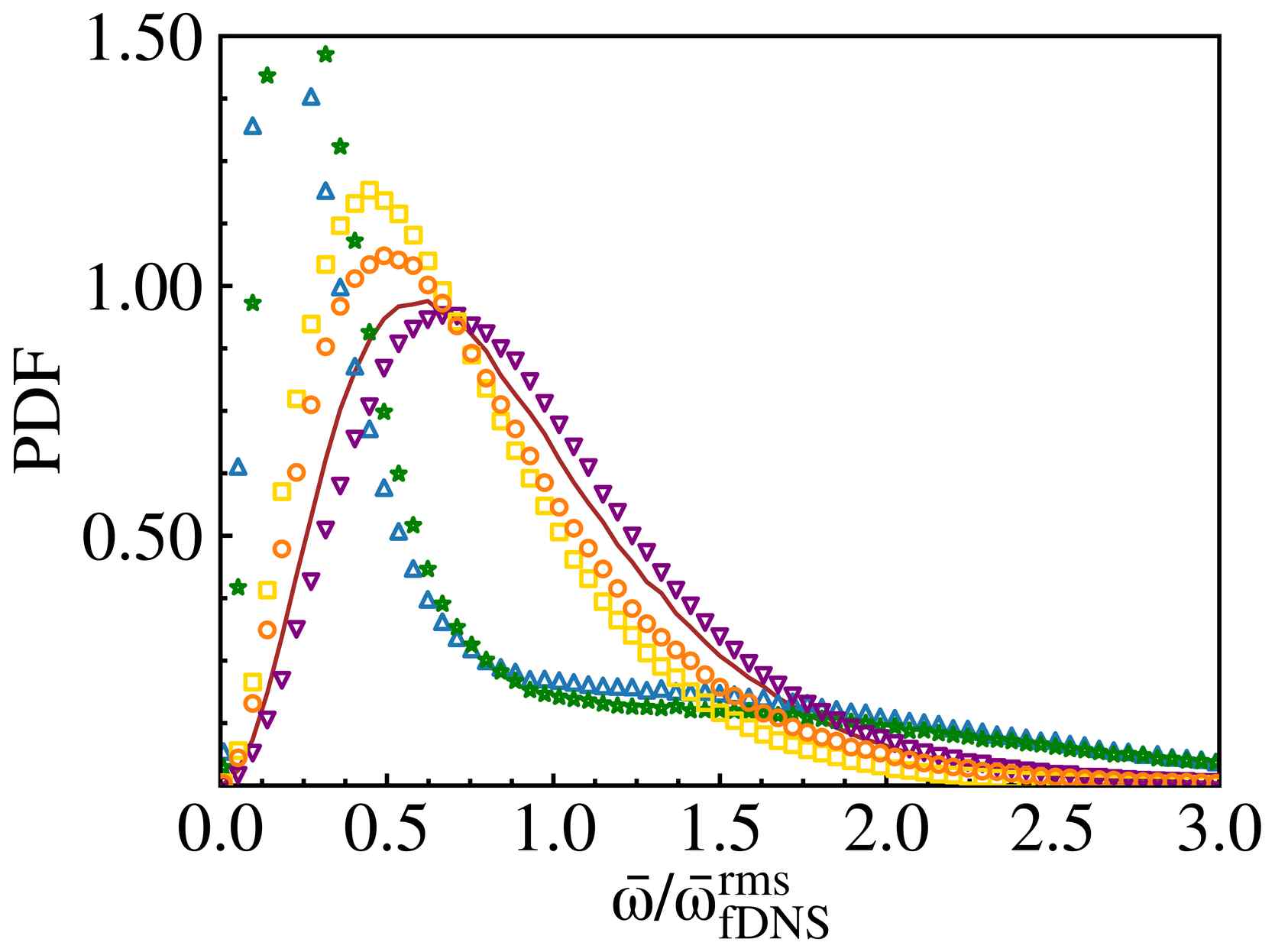}
            \put(-3,65){\small (h)} 
        \end{overpic}
    \end{subfigure}
    \hfill
    \begin{subfigure}[b]{0.32\textwidth}
        \begin{overpic}[width=1\linewidth]{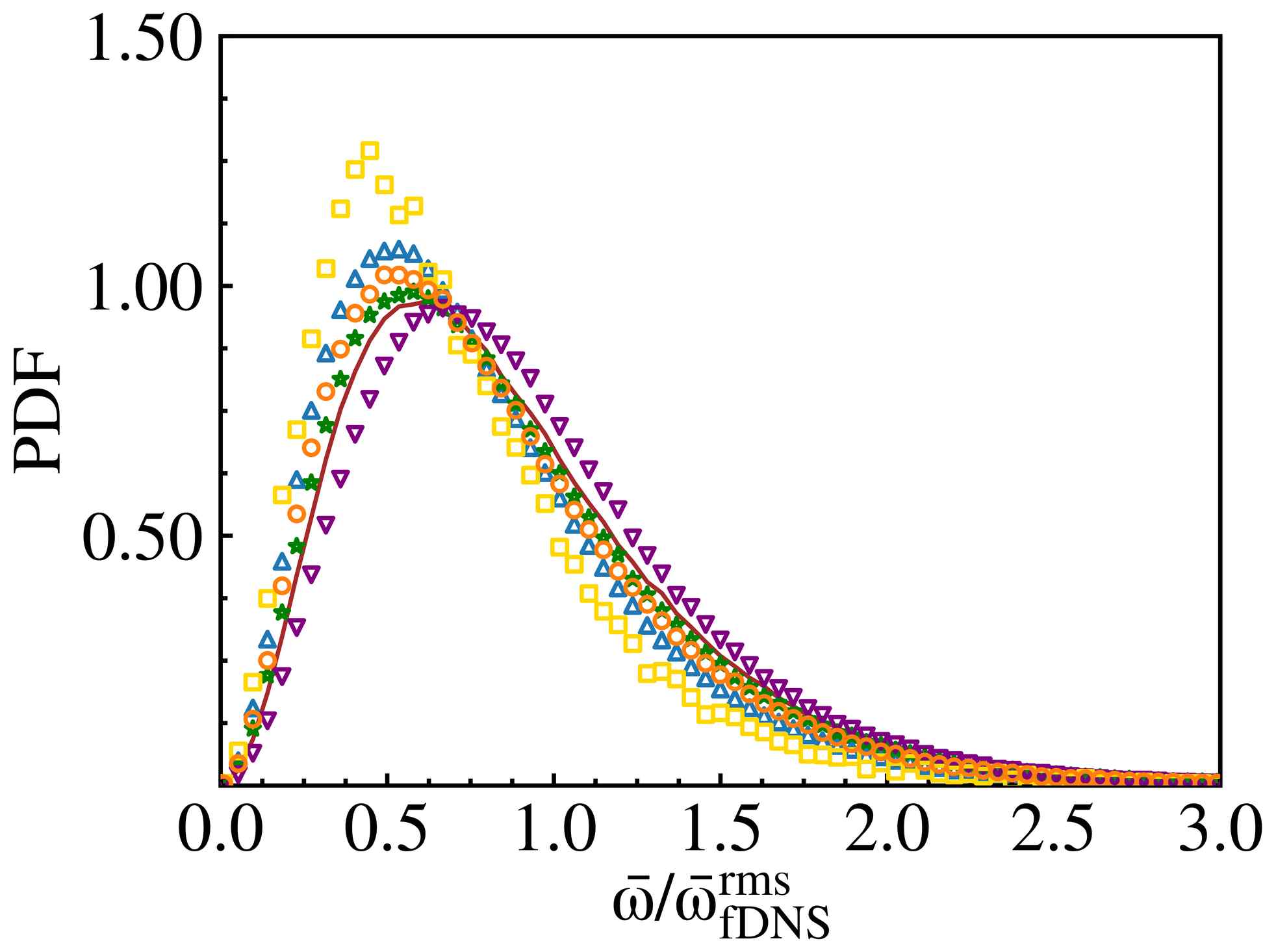}
            \put(-3,65){\small (i)} 
        \end{overpic}
    \end{subfigure}
    \vspace{0.1cm}
    \begin{subfigure}[b]{0.32\textwidth}
        \begin{overpic}[width=1\linewidth]{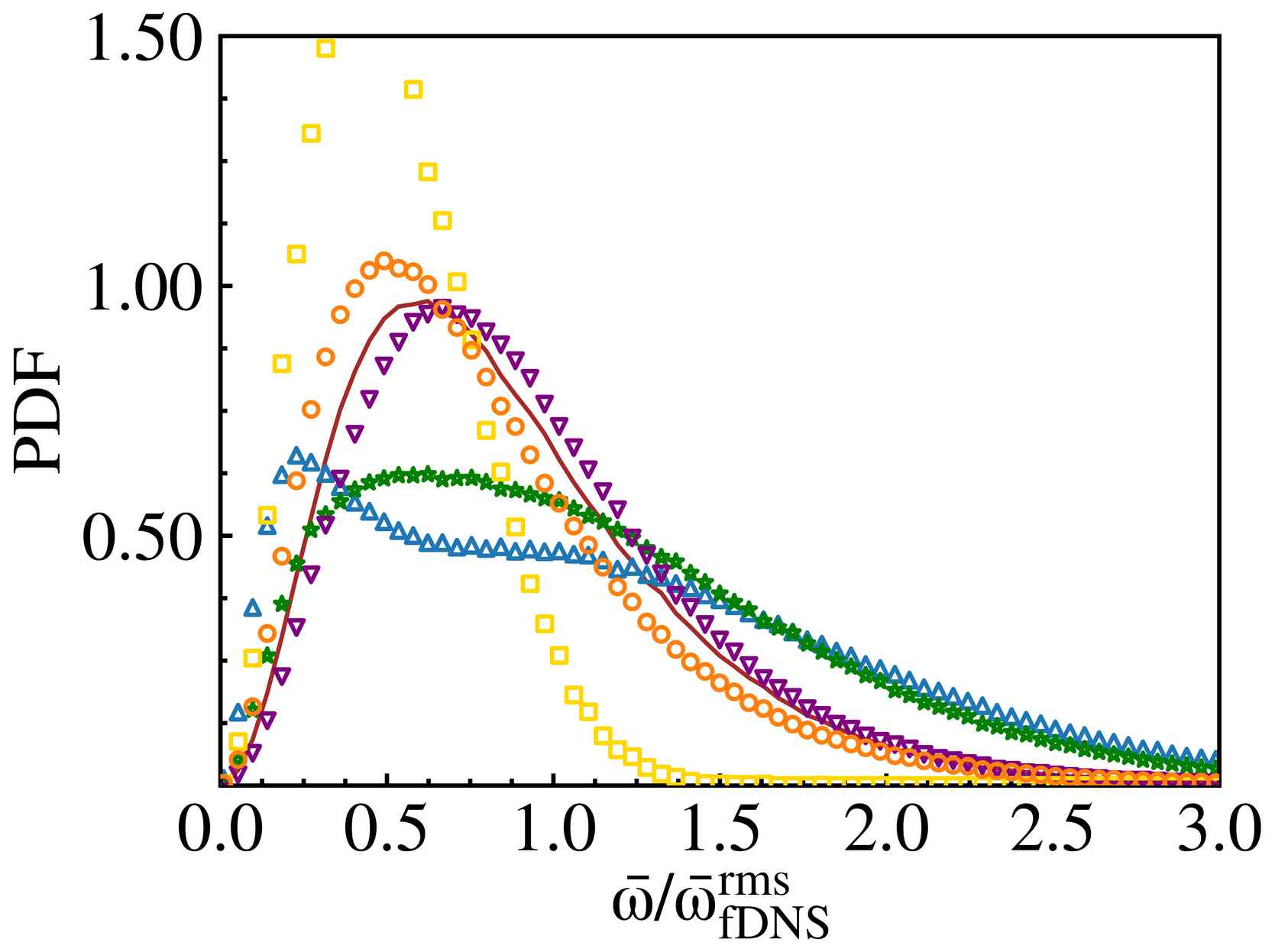}
            \put(-3,65){\small (j)} 
        \end{overpic}
    \end{subfigure}
    \hfill
    \begin{subfigure}[b]{0.32\textwidth}
        \begin{overpic}[width=1\linewidth]{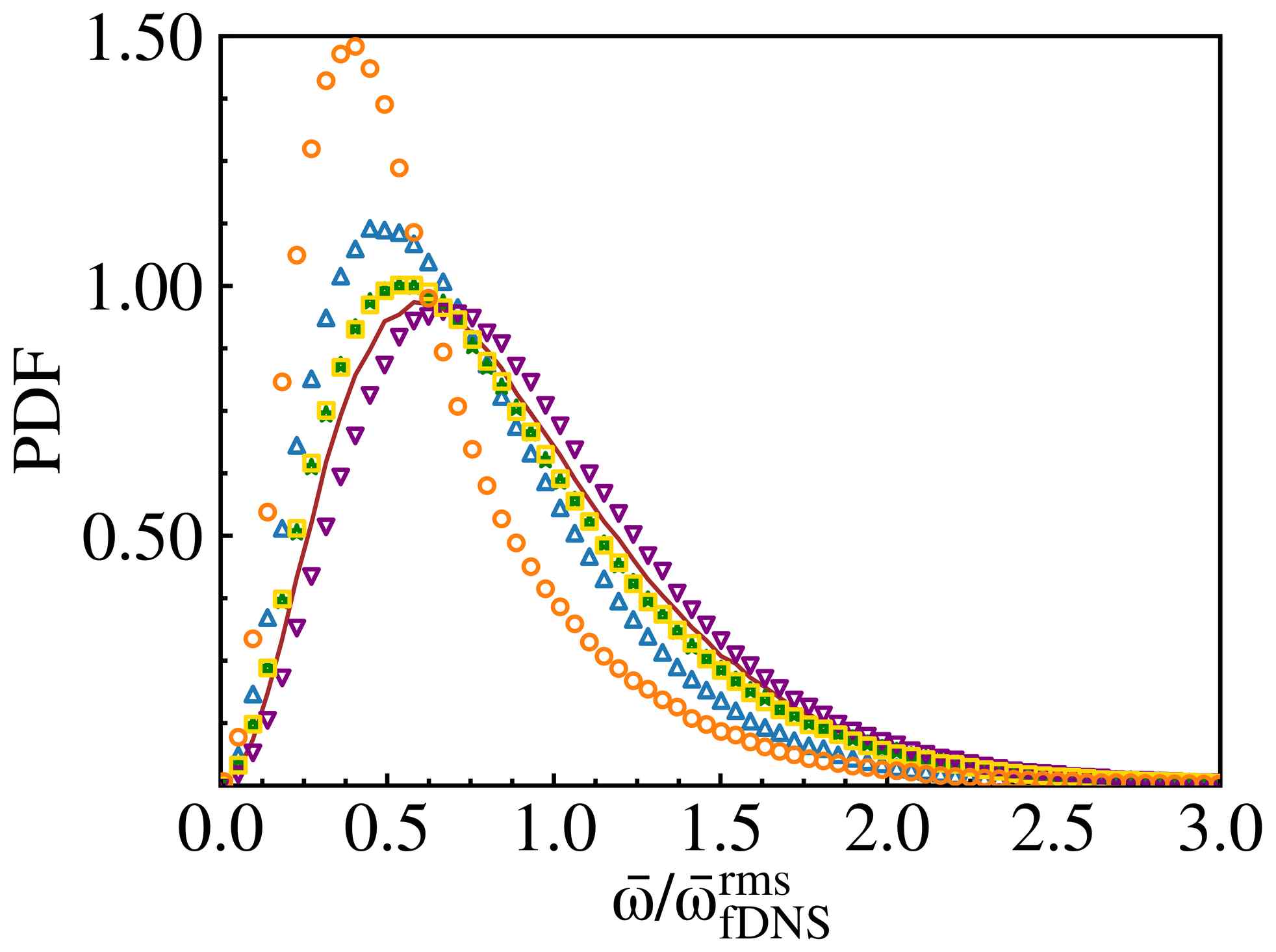}
            \put(-3,65){\small (k)} 
        \end{overpic}
    \end{subfigure}
    \hfill
    \begin{subfigure}[b]{0.32\textwidth}
        \begin{overpic}[width=1\linewidth]{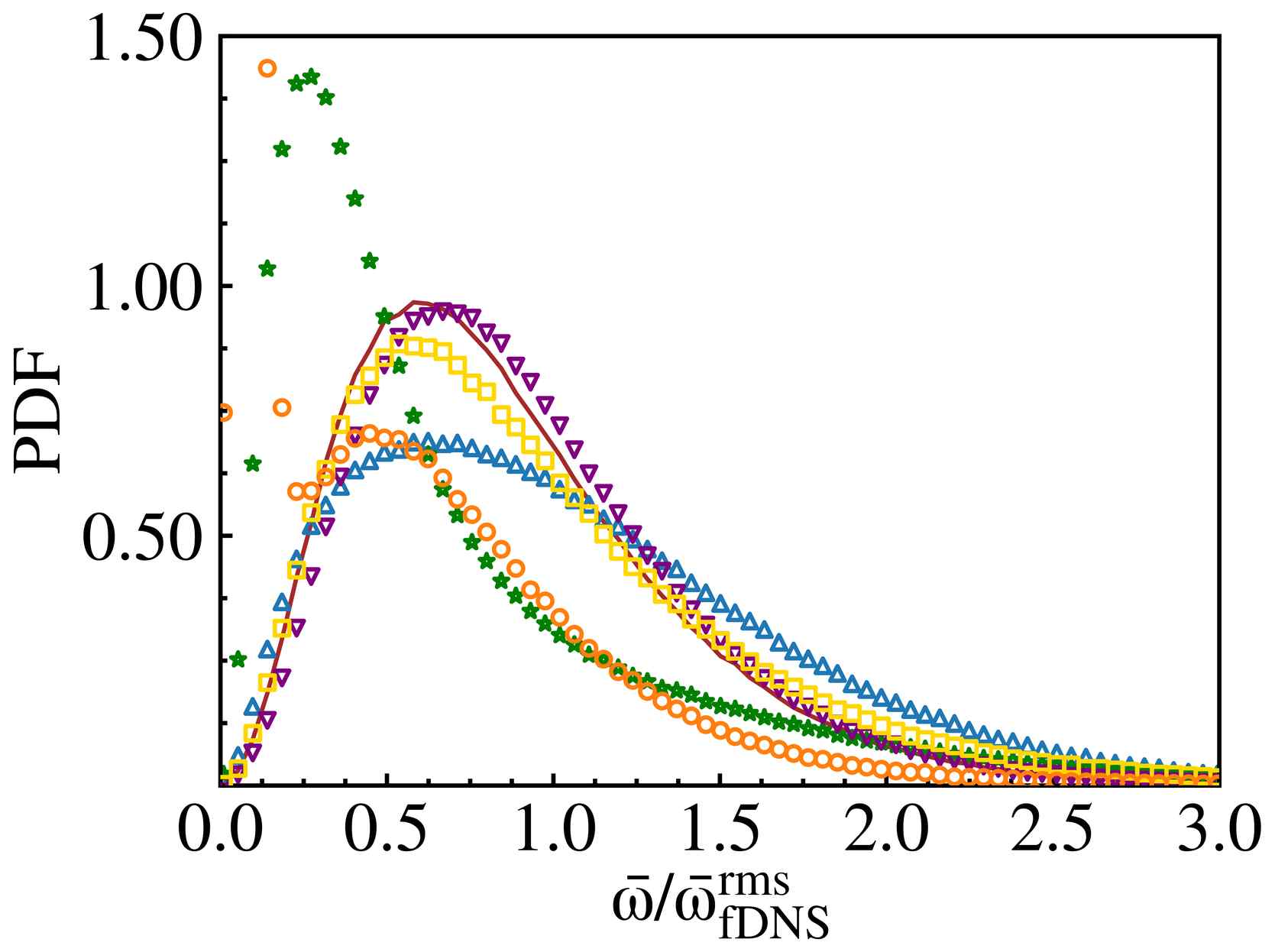}
            \put(-3,65){\small (l)}  
        \end{overpic}
    \end{subfigure}
	\caption{The PDFs of the normalized vorticity $\bar{\omega}/\bar{\omega}^{rms}_{fDNS}$ of LES obtained using different models in forced HIT under various training and prediction time intervals at the statistically steady state: (a) $\Delta T=0.02\tau$ constrained; (b) $\Delta T=0.02\tau$ unconstrained; (c) $\Delta T=0.04\tau$ constrained; (d) $\Delta T=0.04\tau$ unconstrained; (e) $\Delta T=0.1\tau$ constrained; (f) $\Delta T=0.1\tau$ unconstrained; (g) $\Delta T=0.2\tau$ constrained; (h) $\Delta T=0.2\tau$ unconstrained; (i) $\Delta T=0.3\tau$ constrained; (j) $\Delta T=0.3\tau$ unconstrained; (k) $\Delta T=0.4\tau$ constrained; (l) $\Delta T=0.4\tau$ unconstrained. Here, the time instance shown for (a)-(l) is $t/\tau=120$.}\label{fig:5}
\end{figure}

The PDFs of the normalized vorticity magnitude at the statistically steady state are shown in Fig.~\ref{fig:5}, where the vorticity is normalized by the root mean square (rms) value obtained from the fDNS data. Consistent with previous observations, the time interval range $\Delta T \in [0.1\tau, 0.2\tau]$ is optimal for the F-IFNO and F-IUFNO models. Within this range, the constrained versions of F-IFNO and F-IUFNO yield PDFs that closely match those of the fDNS data, while the unconstrained versions exhibit a slight leftward deviation.
When $\Delta T = 0.02\tau, 0.04\tau, 0.3\tau, 0.4\tau$, the performance of F-IFNO and F-IUFNO deteriorates, leading to significant divergence from the expected PDF distributions. In contrast, the IFNO and IUFNO models fail to produce reasonable PDFs across almost all the time intervals, except for the constrained IFNO within $\Delta T = 0.04\tau, 0.1\tau, 0.2\tau$ and the constrained IUFNO within $\Delta T = 0.04\tau, 0.1\tau, 0.2\tau, 0.3\tau$.
These results indicate that the range $\Delta T \in [0.1\tau, 0.2\tau]$ is optimal for FNO-based models. The DSM model, on the other hand, consistently exhibits a slight rightward deviation from the fDNS PDFs, regardless of the value of $\Delta T$.
In conclusion, FNO-based models with prediction constraints within the range $\Delta T \in [0.1\tau, 0.2\tau]$ can achieve high accuracy in reproducing the PDFs of normalized vorticity magnitude over long time rollouts, outperforming the DSM model.

\begin{figure}[ht!]
    \centering
    \begin{subfigure}[b]{0.32\textwidth}
        \begin{overpic}[width=1\linewidth]{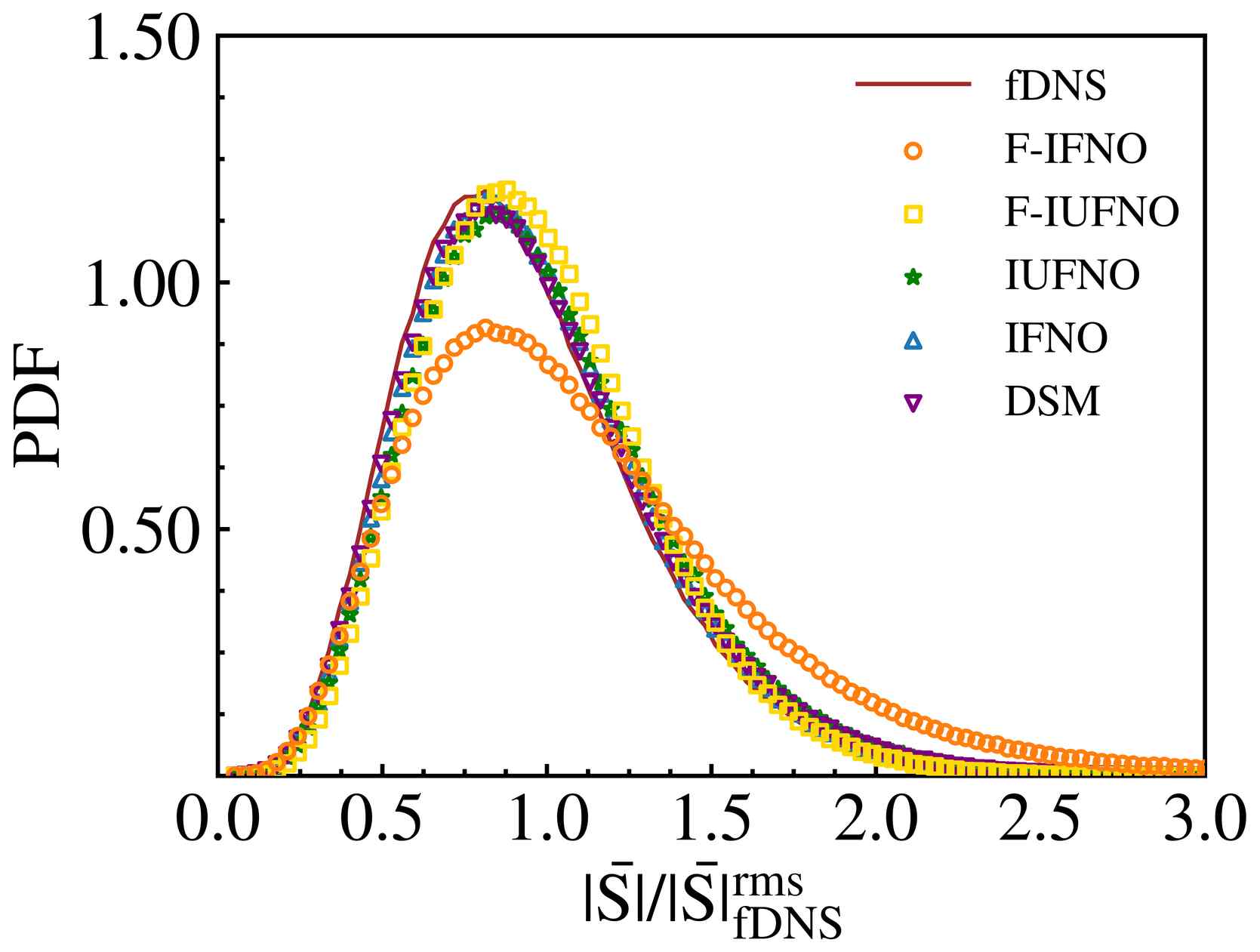}
            \put(-3,65){\small (a)}  
        \end{overpic}
    \end{subfigure}
    \hfill
    \begin{subfigure}[b]{0.32\textwidth}
        \begin{overpic}[width=1\linewidth]{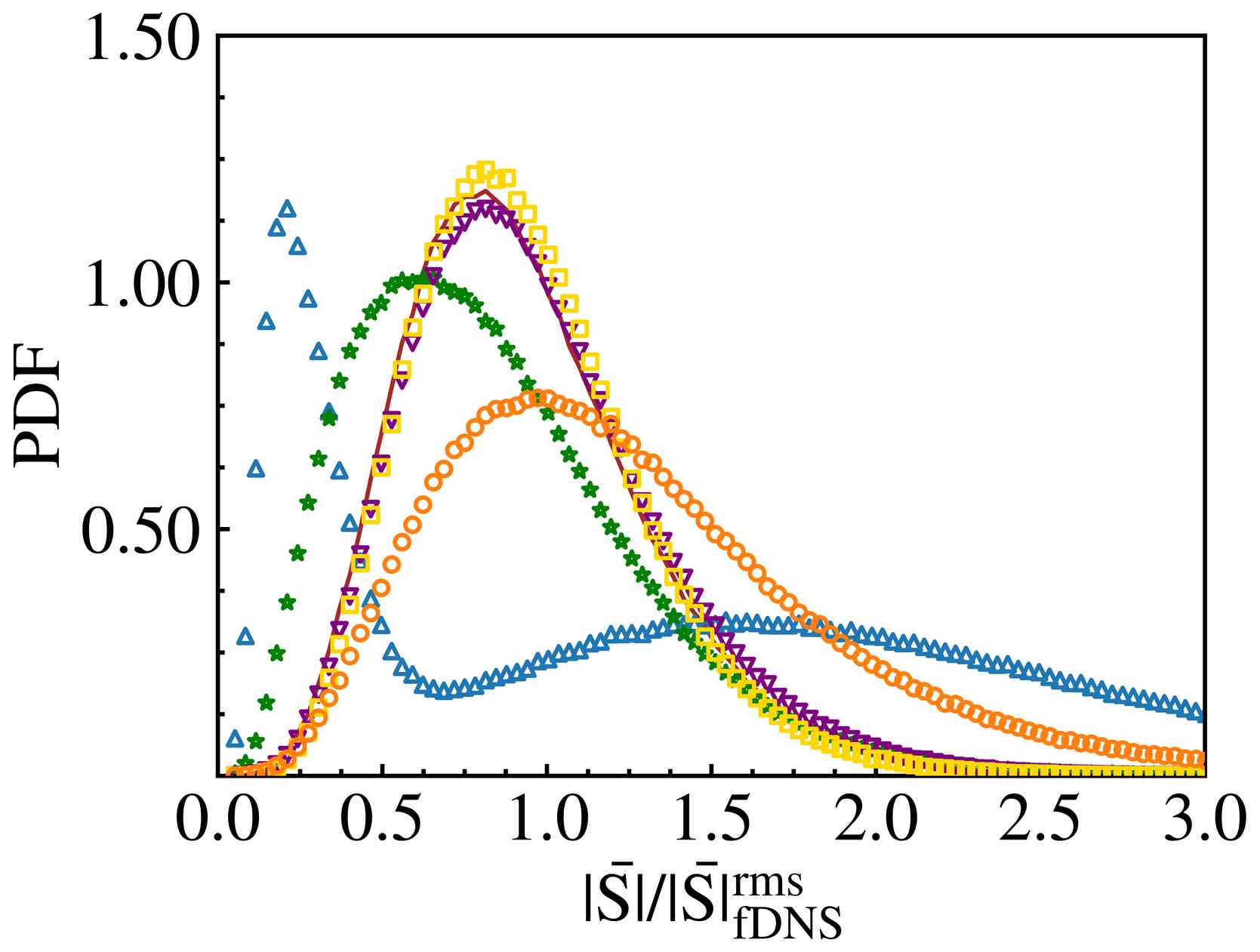}
            \put(-3,65){\small (b)} 
        \end{overpic} 
    \end{subfigure}
    \hfill
    \begin{subfigure}[b]{0.32\textwidth}
        \begin{overpic}[width=1\linewidth]{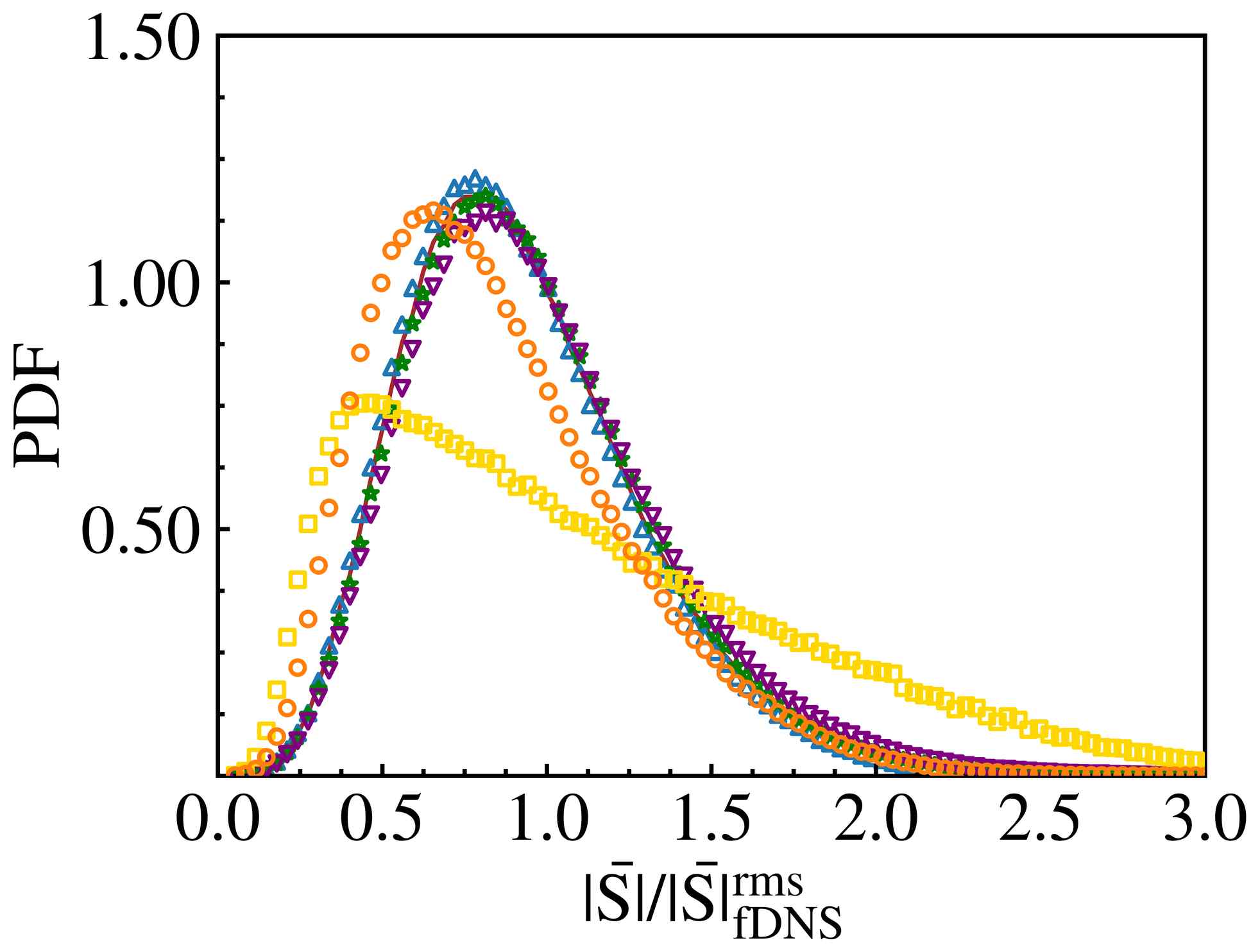}
            \put(-3,65){\small (c)} 
        \end{overpic}
    \end{subfigure}
    \vspace{0.1cm}
    \begin{subfigure}[b]{0.32\textwidth}
        \begin{overpic}[width=1\linewidth]{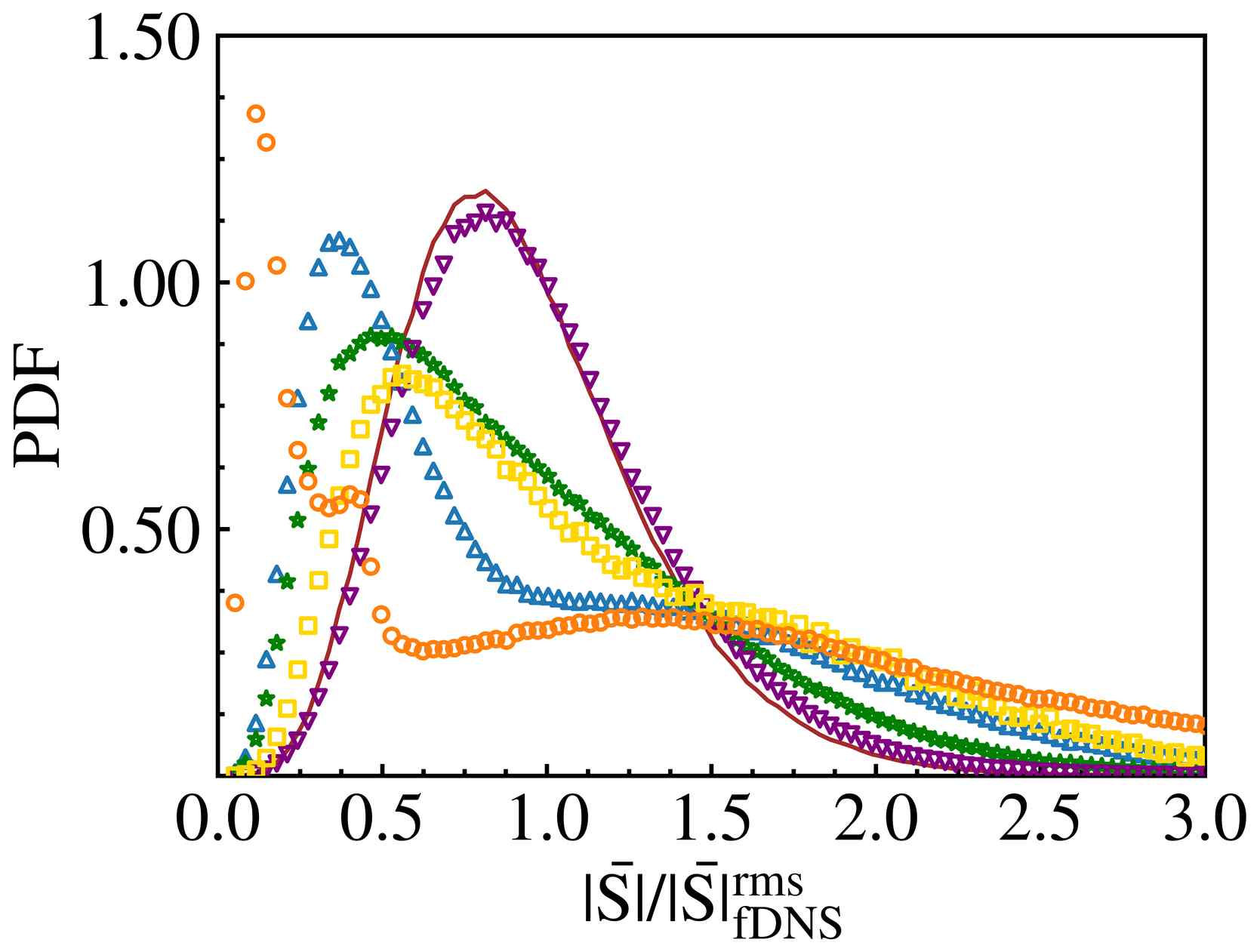}
            \put(-3,65){\small (d)} 
        \end{overpic}
    \end{subfigure}
    \hfill
    \begin{subfigure}[b]{0.32\textwidth}
        \begin{overpic}[width=1\linewidth]{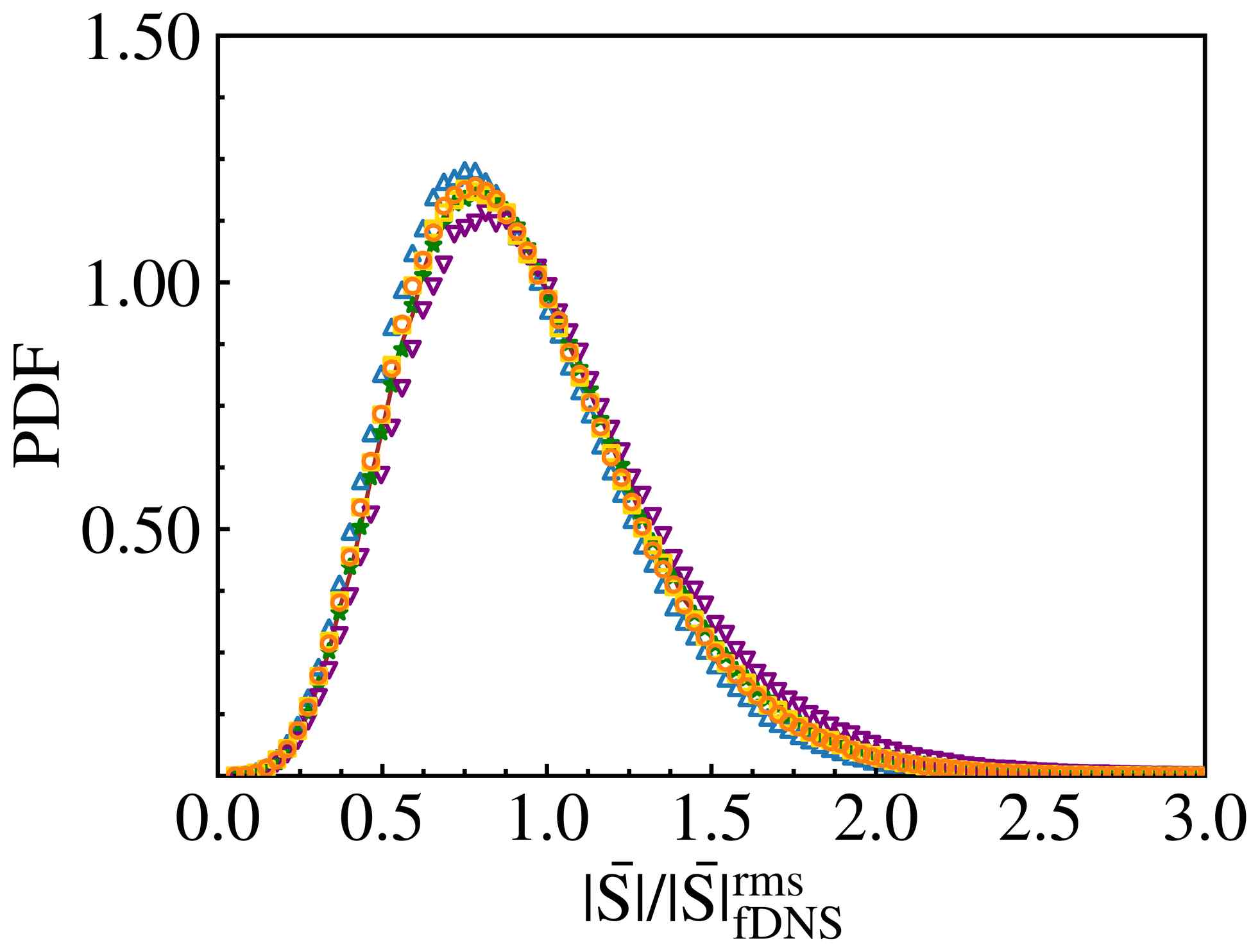}
            \put(-3,65){\small (e)} 
        \end{overpic}
    \end{subfigure}
    \hfill
    \begin{subfigure}[b]{0.32\textwidth}
        \begin{overpic}[width=1\linewidth]{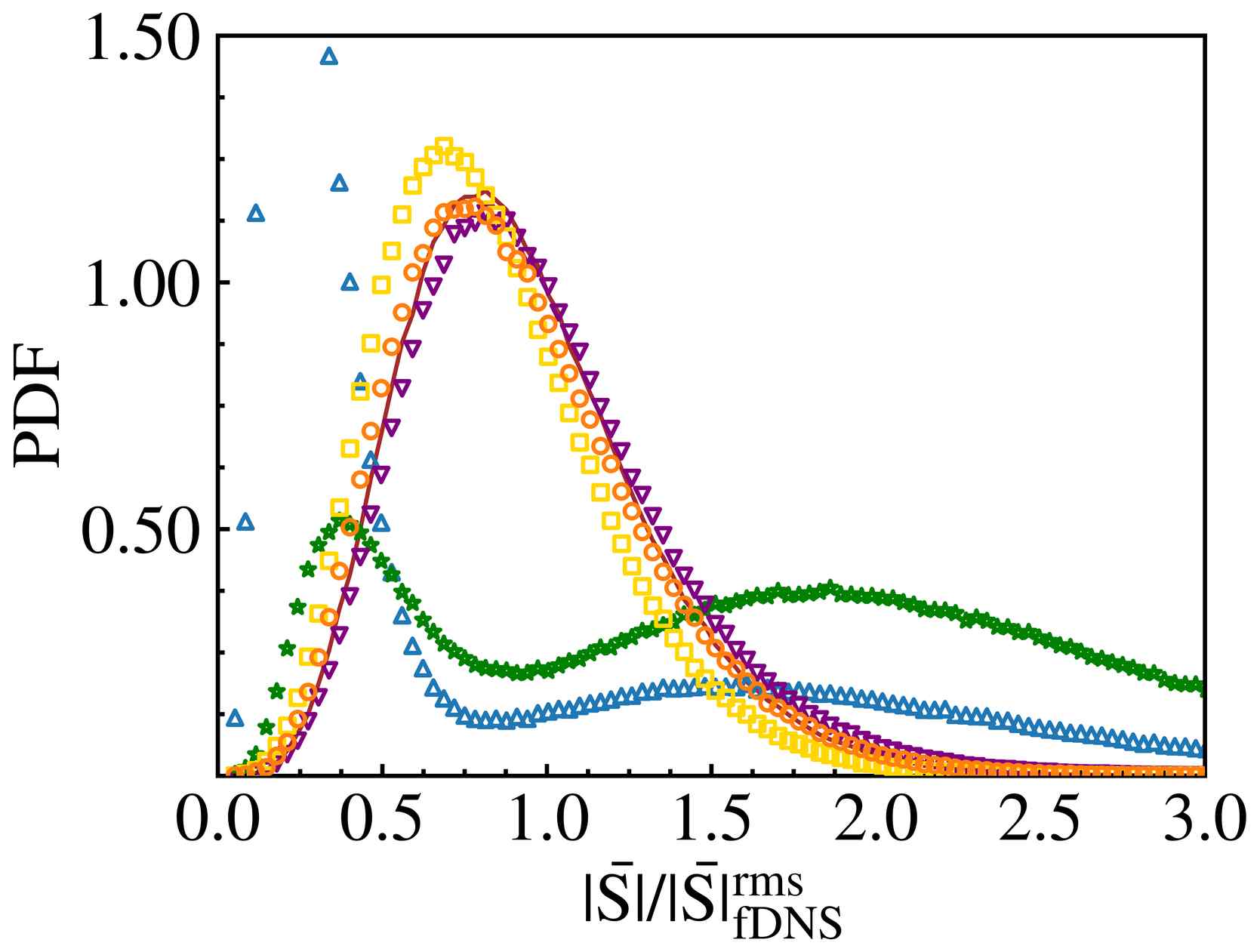}
            \put(-3,65){\small (f)} 
        \end{overpic}
    \end{subfigure}
    \vspace{0.1cm}
    \begin{subfigure}[b]{0.32\textwidth}
        \begin{overpic}[width=1\linewidth]{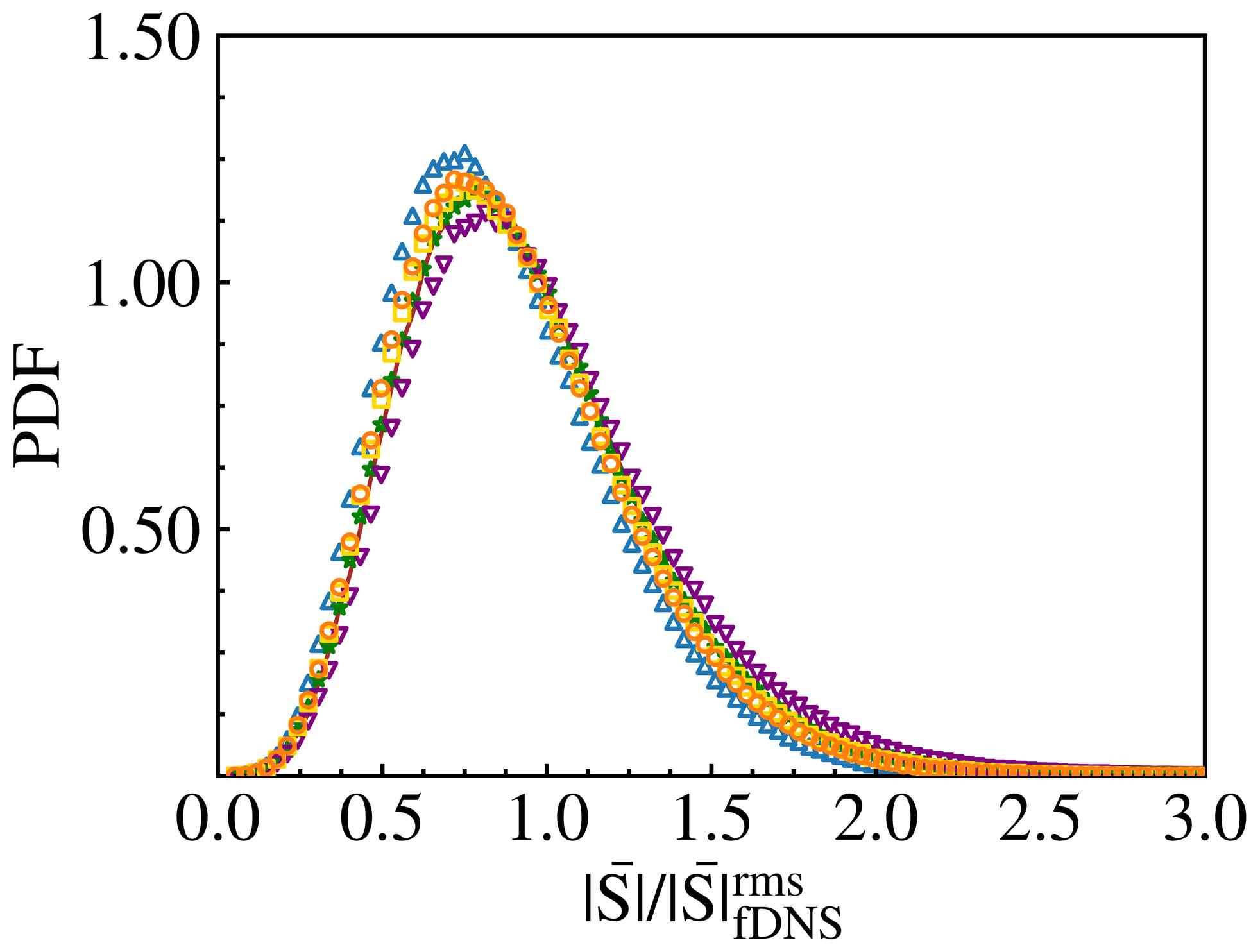}
            \put(-3,65){\small (g)} 
        \end{overpic}
    \end{subfigure}
    \hfill
    \begin{subfigure}[b]{0.32\textwidth}
        \begin{overpic}[width=1\linewidth]{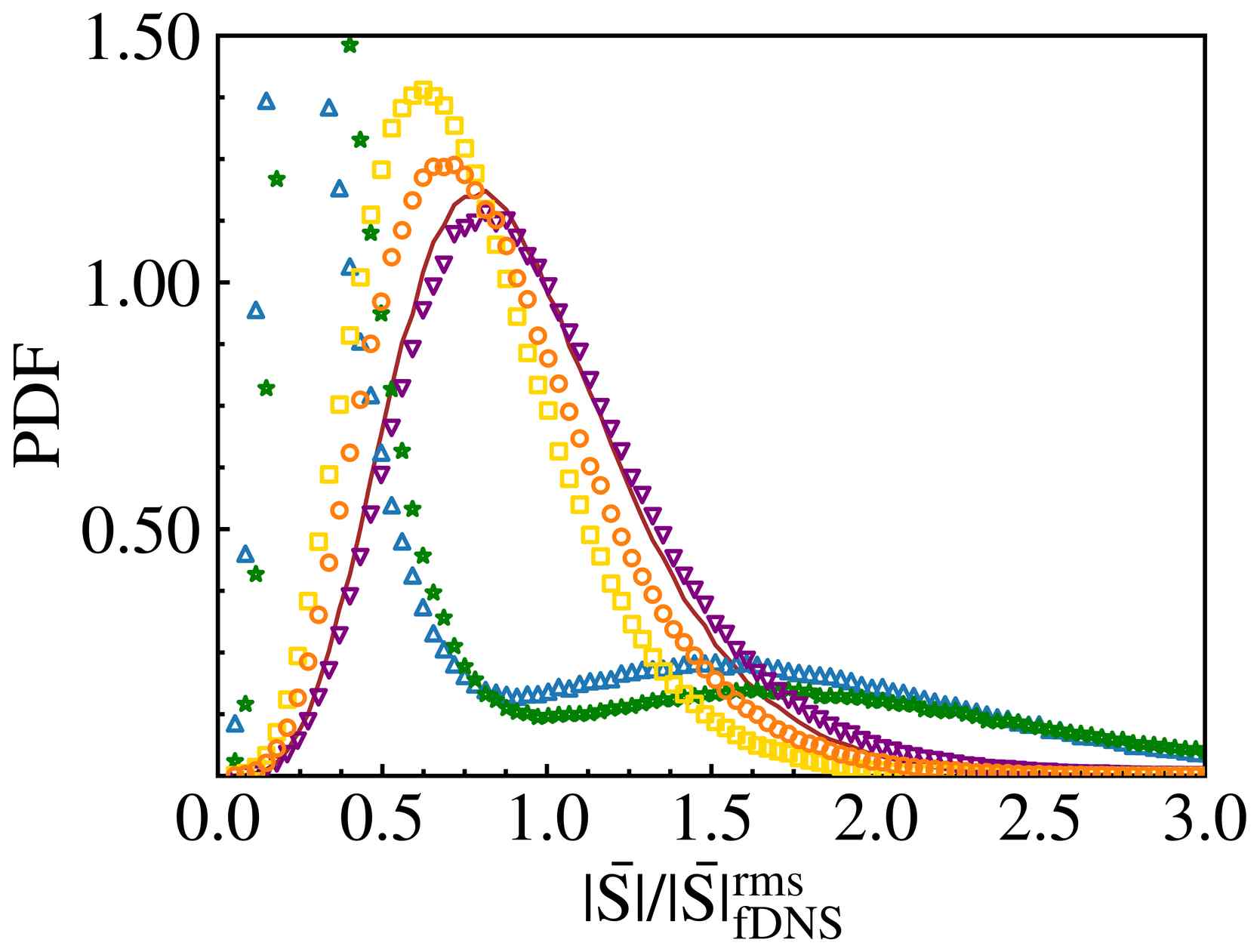}
            \put(-3,65){\small (h)} 
        \end{overpic}
    \end{subfigure}
    \hfill
    \begin{subfigure}[b]{0.32\textwidth}
        \begin{overpic}[width=1\linewidth]{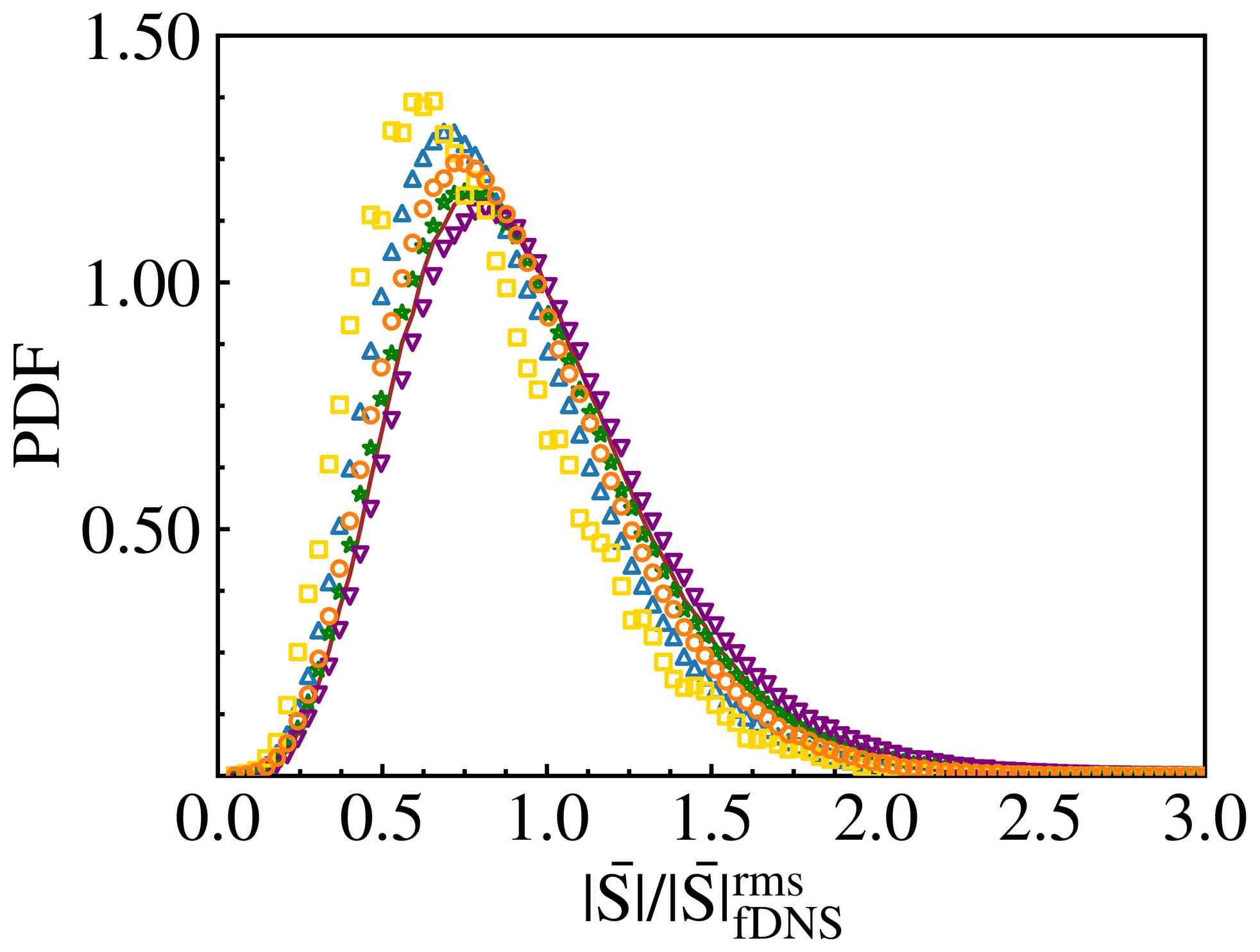}
            \put(-3,65){\small (i)} 
        \end{overpic}
    \end{subfigure}
    \vspace{0.1cm}
    \begin{subfigure}[b]{0.32\textwidth}
        \begin{overpic}[width=1\linewidth]{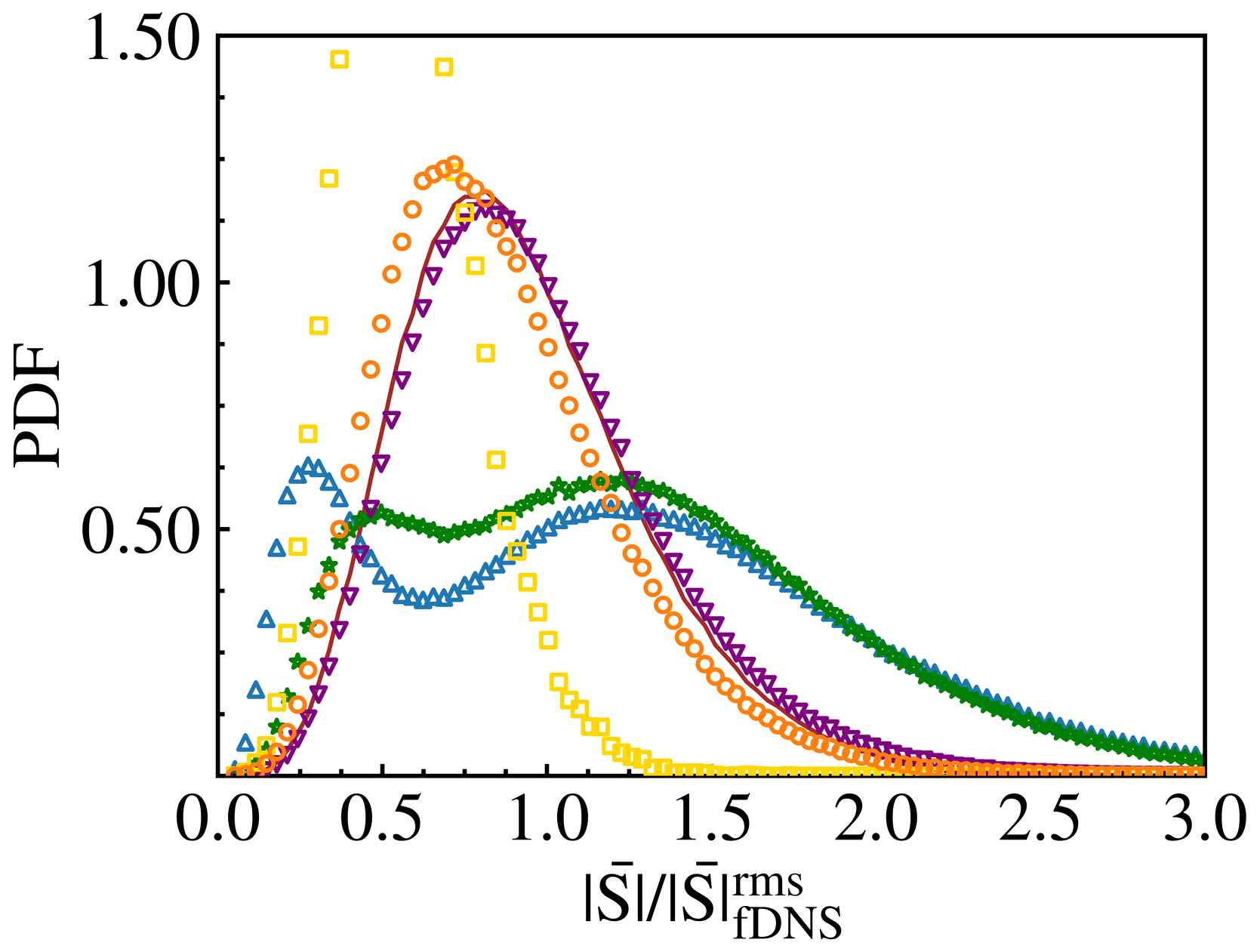}
            \put(-3,65){\small (j)} 
        \end{overpic}
    \end{subfigure}
    \hfill
    \begin{subfigure}[b]{0.32\textwidth}
        \begin{overpic}[width=1\linewidth]{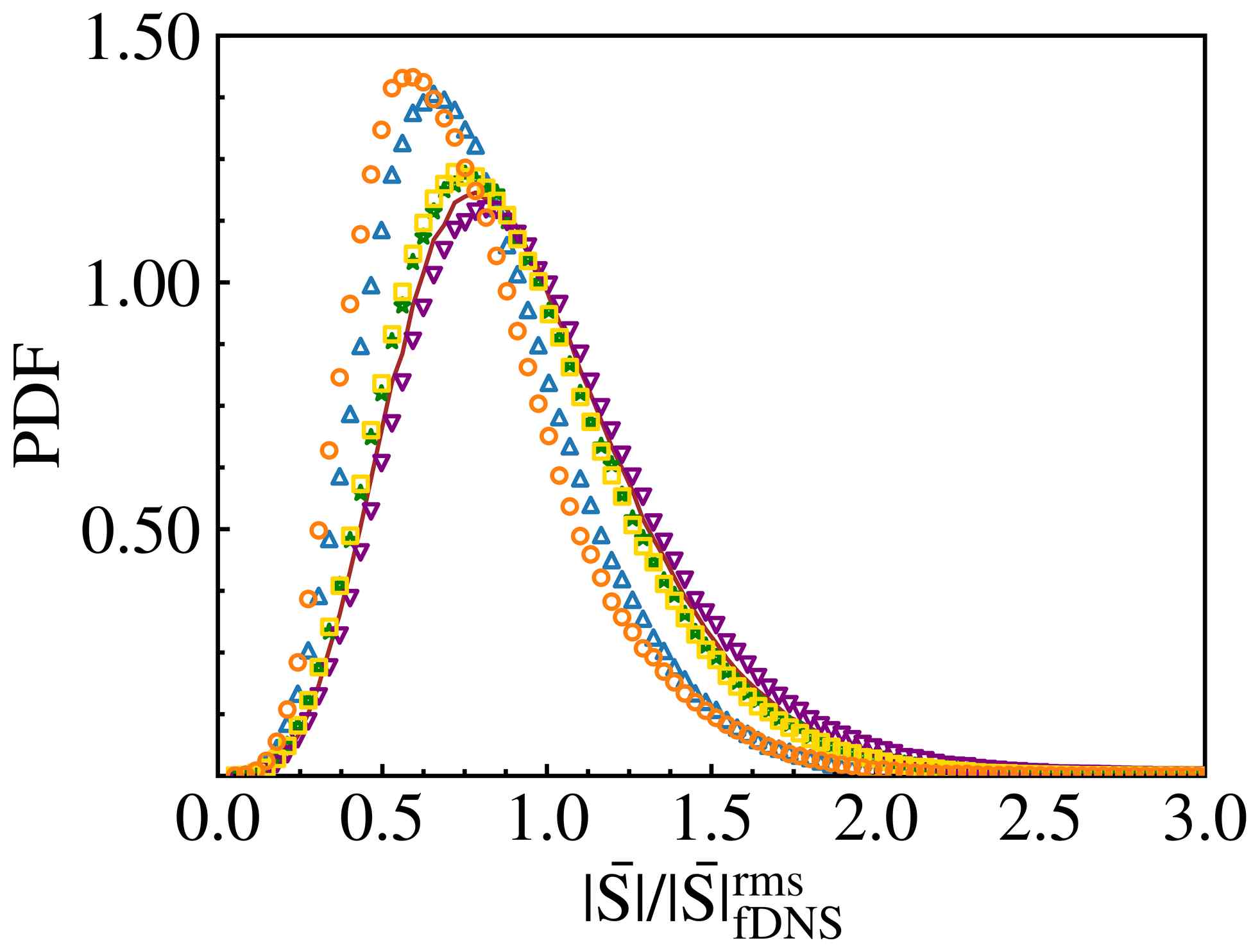}
            \put(-3,65){\small (k)} 
        \end{overpic}
    \end{subfigure}
    \hfill
    \begin{subfigure}[b]{0.32\textwidth}
        \begin{overpic}[width=1\linewidth]{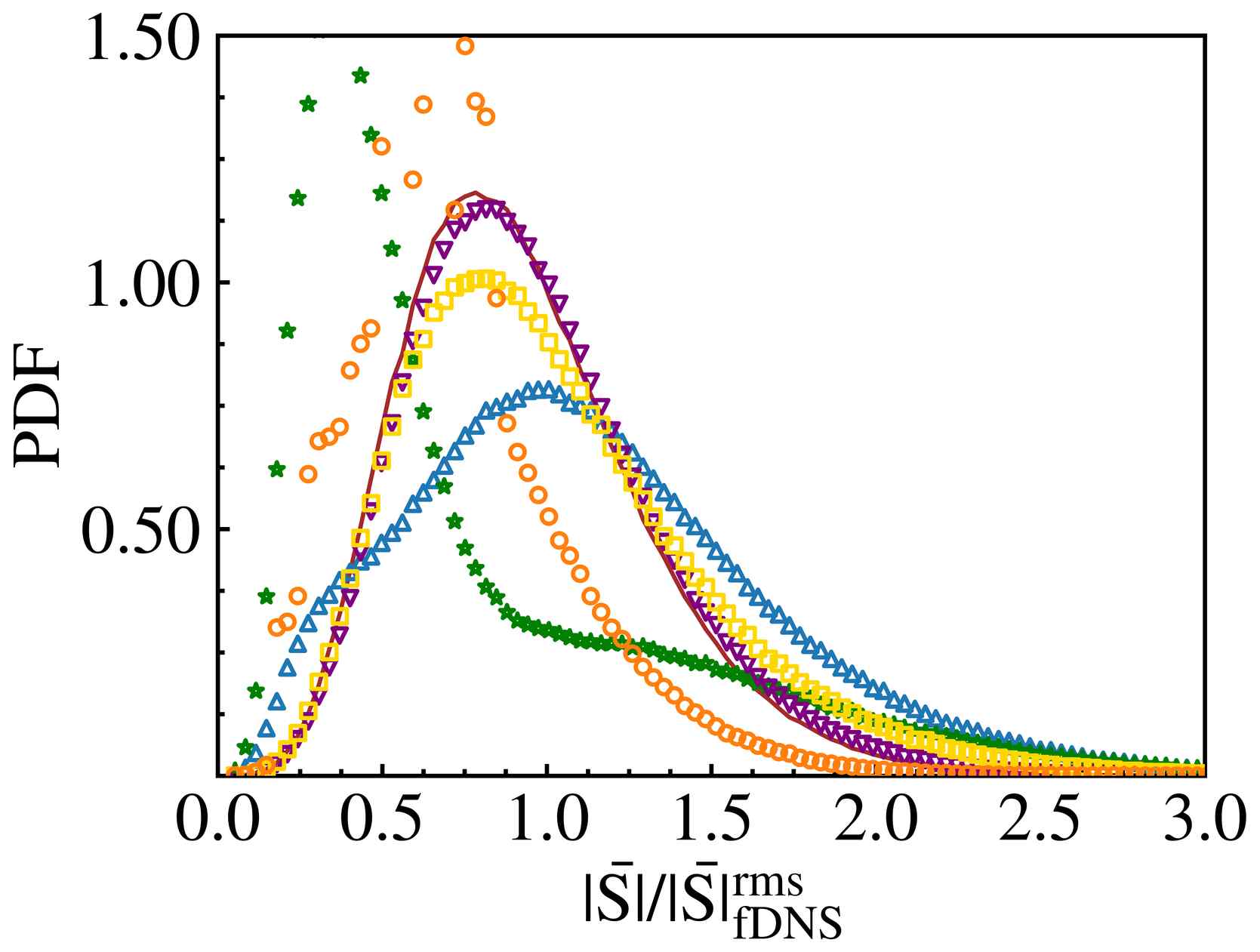}
            \put(-3,65){\small (l)}  
        \end{overpic}
    \end{subfigure}
	\caption{The PDFs of the normalized characteristic strain rate $\left | \bar{S}  \right | / \left | \bar{S}  \right |^{rms}_{fDNS}$ of LES obtained using different models in forced HIT under various training and prediction time intervals at the statistically steady state: (a) $\Delta T=0.02\tau$ constrained; (b) $\Delta T=0.02\tau$ unconstrained; (c) $\Delta T=0.04\tau$ constrained; (d) $\Delta T=0.04\tau$ unconstrained; (e) $\Delta T=0.1\tau$ constrained; (f) $\Delta T=0.1\tau$ unconstrained; (g) $\Delta T=0.2\tau$ constrained; (h) $\Delta T=0.2\tau$ unconstrained; (i) $\Delta T=0.3\tau$ constrained; (j) $\Delta T=0.3\tau$ unconstrained; (k) $\Delta T=0.4\tau$ constrained; (l) $\Delta T=0.4\tau$ unconstrained. Here, the time instance shown for (a)-(l) is $t/\tau=120$.}\label{fig:6}
\end{figure}

Furthermore, the PDFs of the normalized characteristic strain rate in forced HIT at the statistically steady state are shown in Fig.~\ref{fig:6}. The characteristic strain rate is defined as $\left| \bar{S} \right| = \sqrt{2\bar{S}_{ij} \bar{S}_{ij}}$ and is normalized by the rms values computed from the corresponding fDNS data. Similar to previous observations, within the time interval range $\Delta T \in [0.1\tau, 0.2\tau]$, both F-IFNO and F-IUFNO models exhibit a good agreement with the fDNS data. In particular, the constrained variants achieve significantly better accuracy than the DSM model (which exhibits a slight rightward deviation), whereas the unconstrained variants show a slight leftward deviation. 
For $\Delta T = 0.02\tau, 0.04\tau, 0.3\tau, 0.4\tau$, the performance of F-IFNO and F-IUFNO deteriorates compared to that observed at $\Delta T \in [0.1\tau, 0.2\tau]$. Regarding the IFNO and IUFNO models, the constrained IFNO within $\Delta T \in [0.02\tau, 0.2\tau]$ and the constrained IUFNO within $\Delta T \in [0.02\tau, 0.4\tau]$ deliver acceptable accuracy. These results indicate that the optimal time interval range for F-IFNO and F-IUFNO is $\Delta T \in [0.1\tau, 0.2\tau]$, while for IFNO and IUFNO, a slightly smaller interval around $\Delta T = 0.04\tau$ is preferable.

In conclusion, based on the results of five representative physical statistics evaluated at the statistically steady state, we can determine that the optimal time interval for the F-IFNO and F-IUFNO models is $\Delta T \in [0.1\tau, 0.2\tau]$. Within this range, the constrained versions of F-IFNO and F-IUFNO achieve superior long-term predictive accuracy compared to the DSM model, while the unconstrained versions yield a comparable performance to DSM.
In contrast, the IFNO and IUFNO models only deliver satisfactory accuracy when prediction constraints are applied. Specifically, the optimal time interval for constrained IFNO is approximately $\Delta T = 0.04\tau$, and for constrained IUFNO, it lies within $\Delta T \in [0.04\tau, 0.2\tau]$. Within their respective optimal intervals, constrained IFNO and IUFNO show a good agreement with the fDNS data, but still fall short of the best performance achieved by F-IFNO and F-IUFNO.
Unconstrained IFNO and IUFNO, on the other hand, perform poorly across all tested time intervals. Overall, our proposed F-IFNO and F-IUFNO models, when operated within their optimal time interval range, outperform the conventional DSM approach and other machine learning-based models (IFNO and IUFNO) in terms of long-term prediction accuracy and stability in reproducing physical statistics.

To further evaluate the long-term vorticity prediction performance of FNO-based models, we adopt $\Delta T = 0.2\tau$ (which is the optimal time interval $\Delta T$ for F-IFNO and F-IUFNO) as the representative time interval, and present the corresponding results below.

\begin{figure}[ht!]\centering
	\includegraphics[width=1\textwidth]{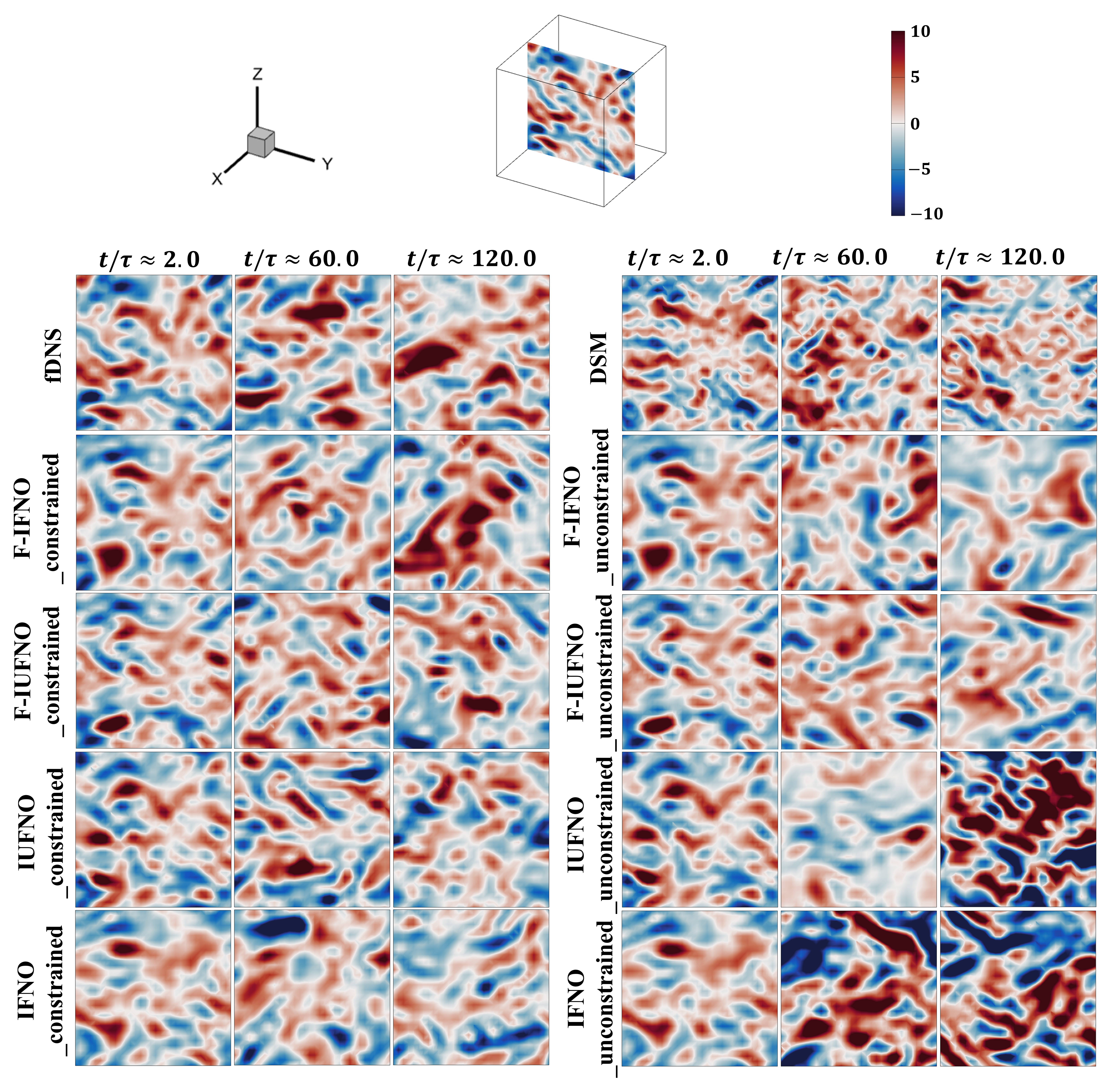}
	\caption{Evolution of the predicted vorticity fields in the $z$-direction at the mid-plane along the $x$-direction of the cubic domain as a function of time for forced HIT at time $t/\tau \approx 2.0, 60.0, 120.0$. Here, the prediction time interval $\Delta T=0.2\tau$.}\label{fig:7}
\end{figure}

Fig.~\ref{fig:7} illustrates the vorticity field contours predicted by different models. The instantaneous snapshots are taken on the central $y$-$z$ plane at three representative time instances: $t/\tau \approx 2.0$, $60.0$, and $120.0$. The slice is taken at the midpoint along the $x$-direction of the cubic domain, and the contours are colored by the $z$-component of the vorticity.
At the short-term time instance $t/\tau \approx 2.0$, constrained and unconstrained versions of each FNO-based model exhibit similar performance. Among them, F-IUFNO closely matches the fDNS vorticity field, while F-IFNO and IUFNO capture the overall distribution but lack fine-scale detail. In contrast, IFNO barely reconstructs the fDNS vorticity field.
As time progresses to the long-term instances $t/\tau \approx 60.0$ and $120.0$, the effect of constraints becomes more pronounced. Although all FNO-based models deviate from the fDNS results at these later times, the constrained F-IFNO and F-IUFNO still provide reasonable magnitude predictions. In comparison, unconstrained IFNO and IUFNO fail to produce reasonable results. 
The DSM model, however, consistently produces spurious small-scale structures throughout the entire prediction horizon, which significantly deviate from the corresponding fDNS results. Overall, compared to DSM and other FNO-based models, our proposed F-IFNO and F-IUFNO demonstrate not only accurate short-term vorticity field prediction but also relatively stable long-term prediction performance.

\begin{figure}[ht!]\centering
	\includegraphics[width=1\textwidth]{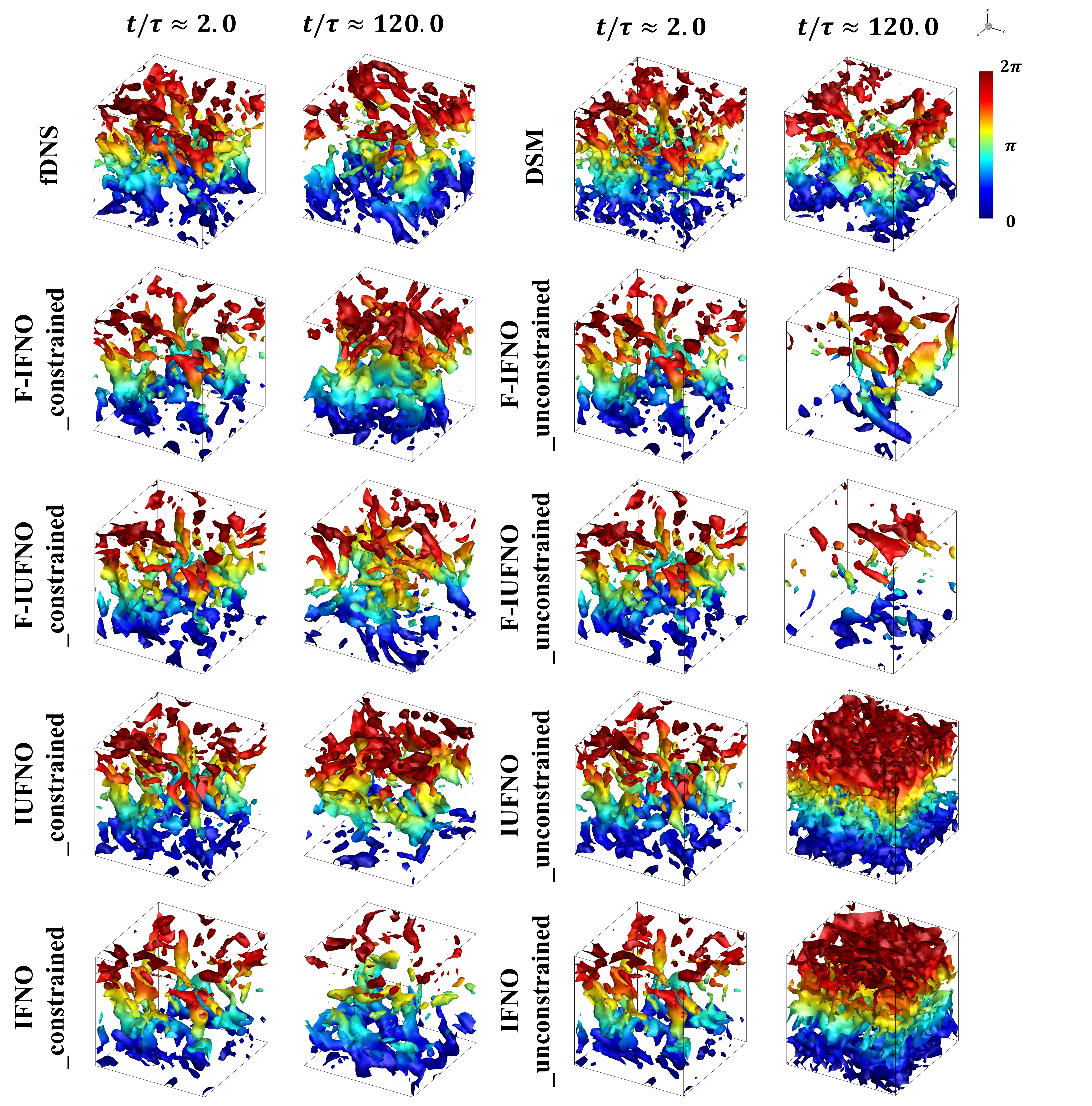}
	\caption{Isosurface of the normalized vorticity $\bar{\omega}/\bar{\omega}^{rms}_{fDNS}=1.0$ (colored by altitude of $z$-direction) at time $t/\tau \approx 2.0, 120.0$ for forced HIT. Here, the prediction time interval $\Delta T=0.2\tau$.}\label{fig:8}
\end{figure}

We further examine the isosurfaces of the normalized vorticity magnitude $\bar{\omega}/\bar{\omega}^{\mathrm{rms}}_{\mathrm{fDNS}} = 1.0$, colored by the $z$-direction coordinate, as shown in Fig.~\ref{fig:8}. The time interval $\Delta T = 0.2\tau$ is used for the spatial structure predictions, with two representative time instances selected: $t/\tau \approx 2.0$ and $t/\tau \approx 120.0$. At the early time $t/\tau \approx 2.0$, both constrained and unconstrained FNO-based models accurately capture the overall flow structures of the vorticity field, outperforming the DSM model. However, at the later time $t/\tau \approx 120.0$, only the constrained FNO-based models continue to predict physically reasonable flow structures, while the unconstrained versions fail to do so. 
In summary, compared to the DSM model, FNO-based models demonstrate superior accuracy in predicting flow structures over short-term horizons, but only the constrained variants maintain reasonable predictions over long-term rollout.

From Fig.~\ref{fig:7} and Fig.~\ref{fig:8}, it can be observed that even under the optimal time interval, FNO-based models achieve accurate short-term predictions of vorticity fields but struggle to maintain stability over long-term rollouts. This behavior differs from their performance in predicting physical statistics. Nevertheless, in the context of turbulence modeling, the long-term stability of physical statistical quantities is generally of greater importance than the instantaneous accuracy of vorticity or velocity field predictions. 
In this regard, the proposed F-IFNO and F-IUFNO models effectively fulfill the core requirements of turbulence prediction: they deliver stable and accurate long-term predictions of physical statistics, while also preserving short-term accuracy in the reconstruction of vorticity (velocity) fields. 
Moreover, incorporating prediction constraints significantly enhances the ability of FNO-based models to reconstruct accurate and physically plausible turbulent structures, both for long-term statistics and for short-term field evolution.

\subsection{Statistical analysis of posterior results}
\label{subsec4.2}

In this subsection, we present a statistical analysis of the posterior results. We use the same prediction data as in Subsection~\ref{subsec4.1}. For each model and its corresponding fDNS data under a given prediction time interval $\Delta T$, thirty independent data groups are collected, each containing six hundred time steps, resulting in a total of 18,000 samples per method for each time interval $\Delta T$. We focus on two key physical quantities for turbulent flows: the kinetic energy $E_k$ and the velocity spectra $E(k)$, where k= 1, 2 ,... 10. 

In the statistical analysis, we evaluate the errors of kinetic energy $E_k$ and velocity spectra $E(k)$ for each method with different prediction time interval $\Delta T$. Since the data is obtained from forced HIT, which is a statistically steady turbulent flow, the fDNS results naturally exhibit small fluctuations at each time step and for each independent data group.
To establish a consistent statistical reference, we define the ground truth as the average values of $E_k$ and $E(k)$ computed over all 18,000 fDNS samples. Although the flow is statistically stationary, the finite size of the system inevitably introduces temporal fluctuations in these quantities. Therefore, the variation observed in fDNS reflects the intrinsic statistical variability rather than prediction error. Based on this reference, we compute the prediction errors for all surrogate models, while the fDNS results are included only to illustrate the magnitude of natural statistical fluctuations over time.

\begin{figure}[ht!]
    \centering
    \begin{subfigure}[b]{0.32\textwidth}
        \begin{overpic}[width=1\linewidth]{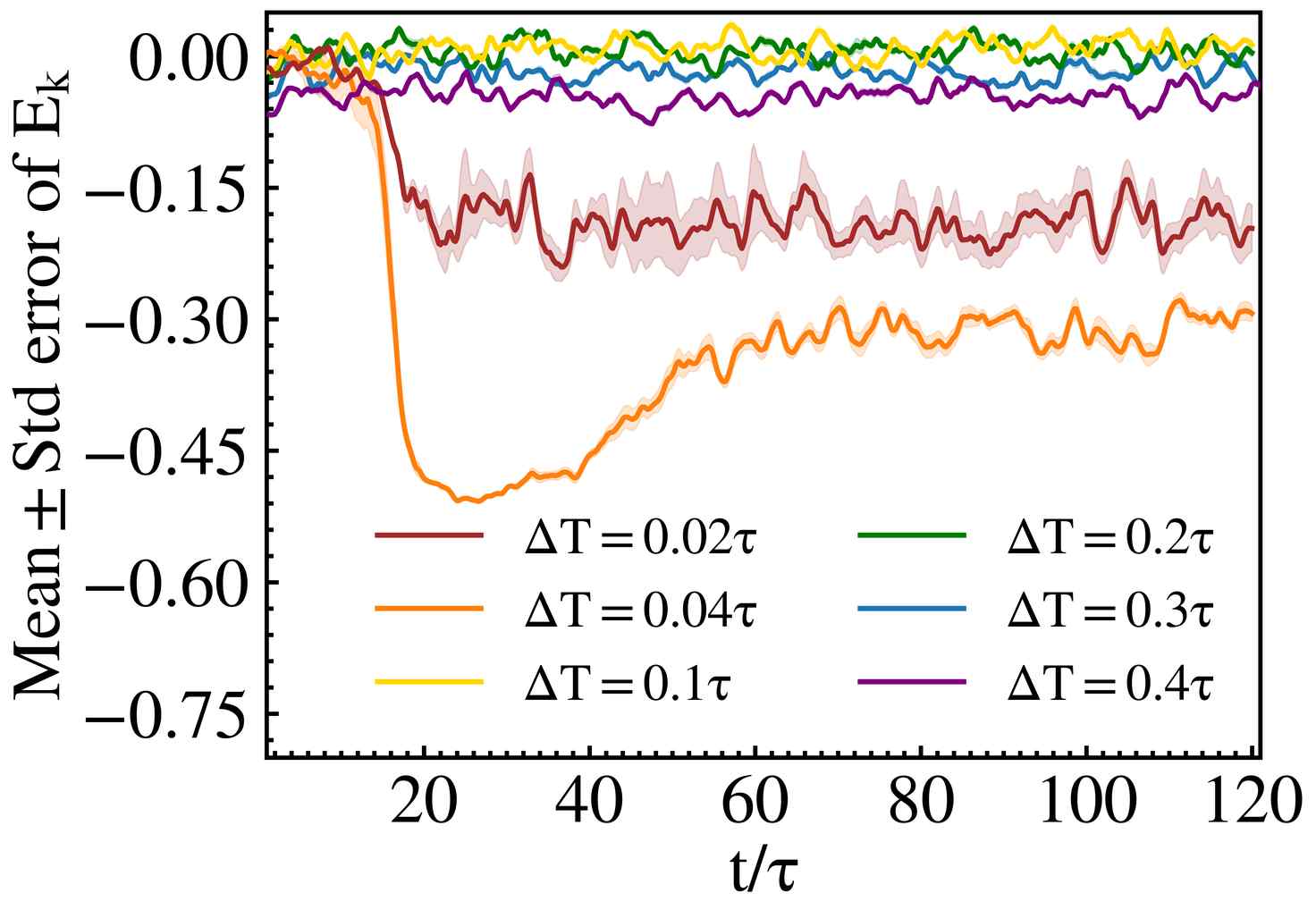}
            \put(-6,68){\small (a)}  
        \end{overpic}
    \end{subfigure}
    \hfill
    \begin{subfigure}[b]{0.32\textwidth}
        \begin{overpic}[width=1\linewidth]{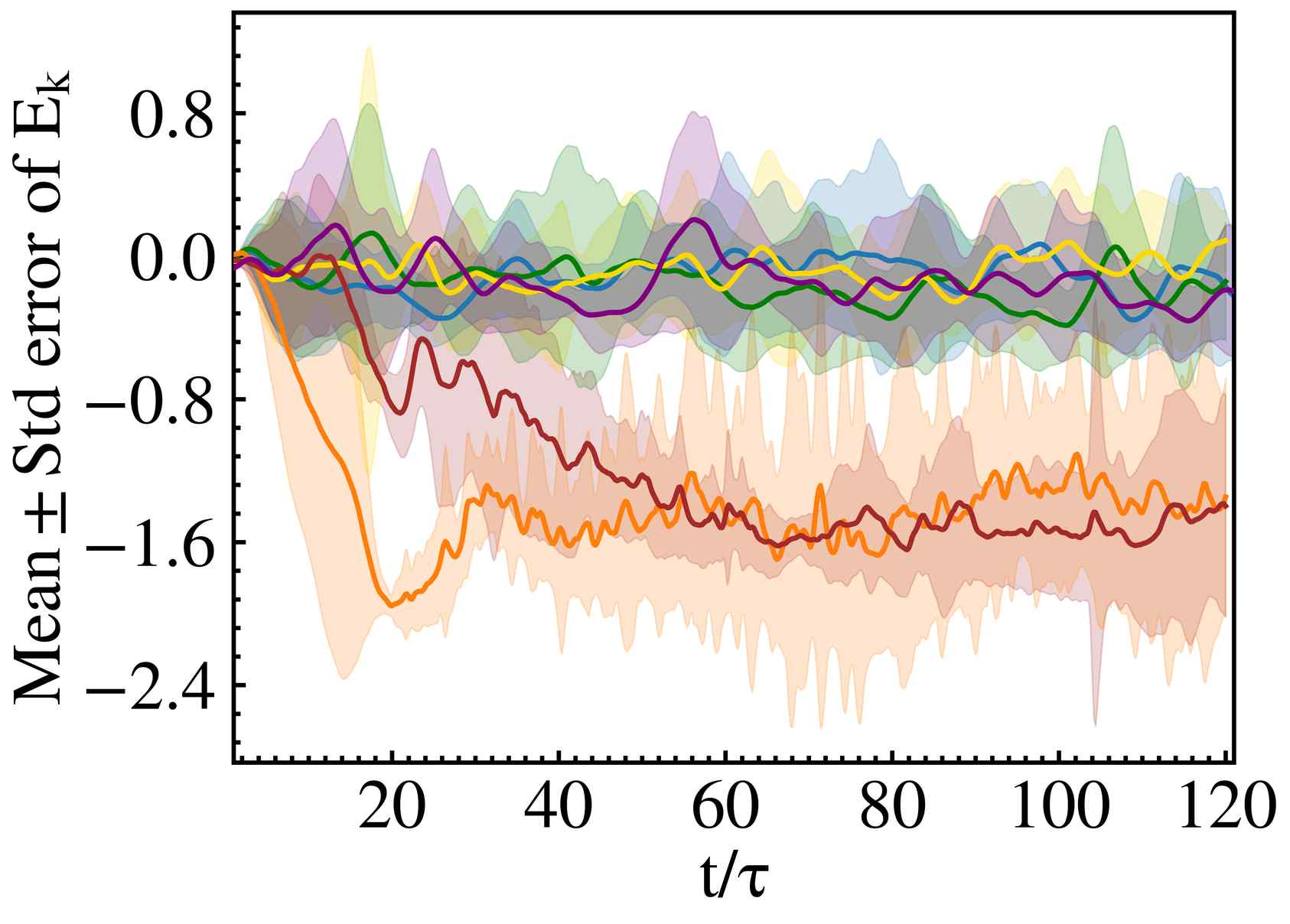}
            \put(-6,68){\small (b)} 
        \end{overpic} 
    \end{subfigure}
    \hfill
    \begin{subfigure}[b]{0.32\textwidth}
        \begin{overpic}[width=1\linewidth]{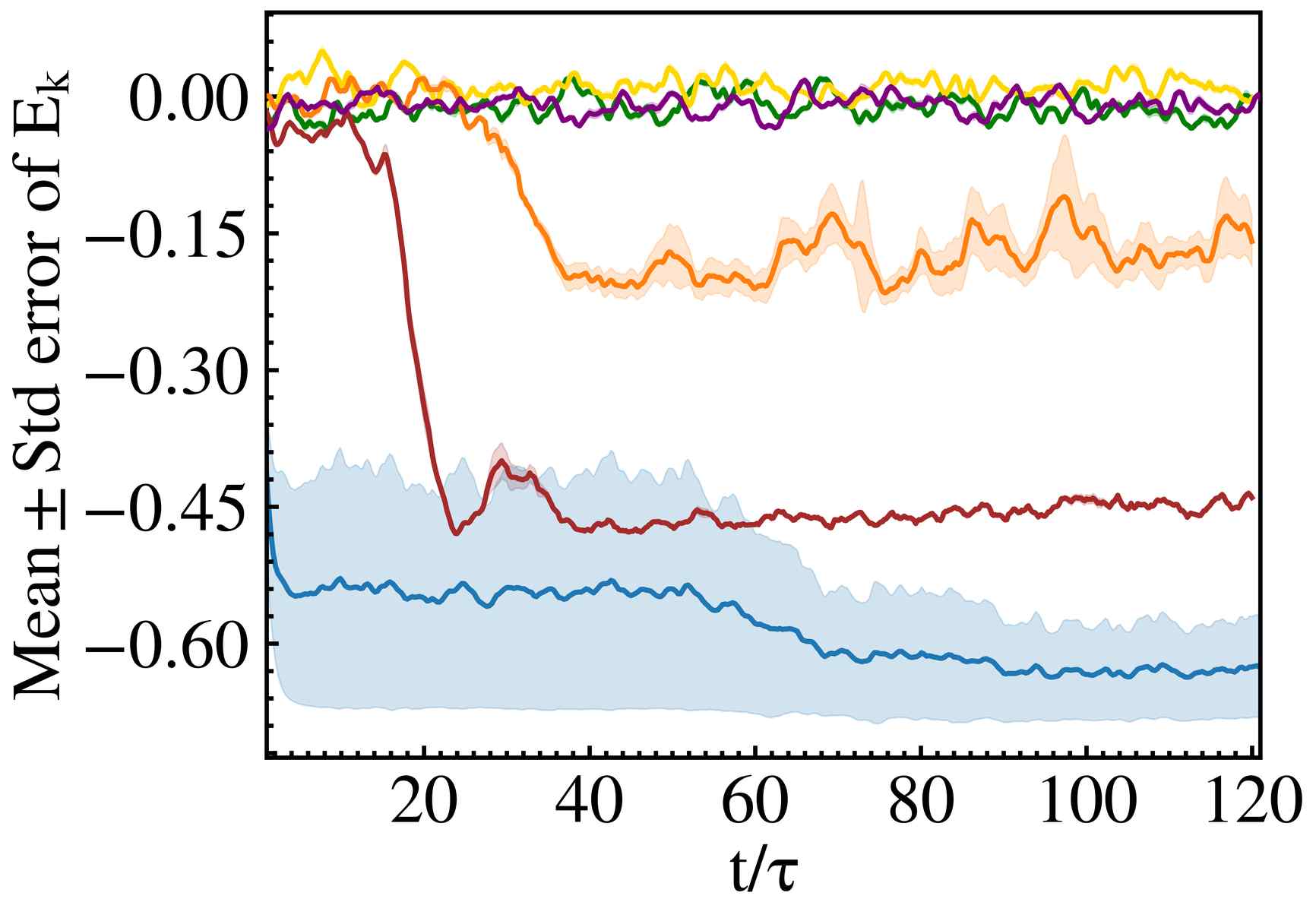}
            \put(-6,68){\small (c)} 
        \end{overpic}
    \end{subfigure}
    \vspace{0.1cm}

    \begin{subfigure}[b]{0.32\textwidth}
        \begin{overpic}[width=1\linewidth]{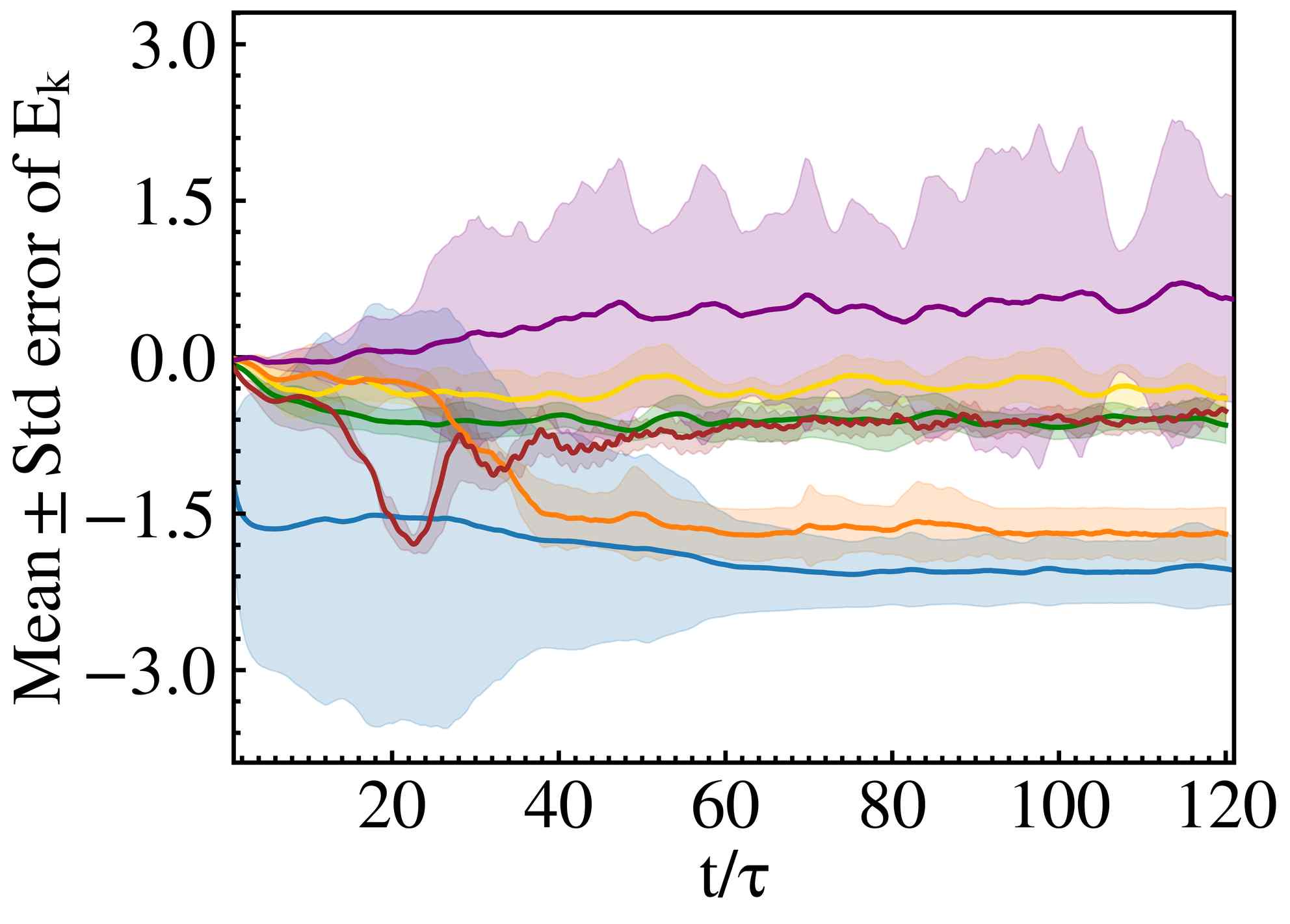}
            \put(-6,68){\small (d)}  
        \end{overpic}
    \end{subfigure}
    \hfill
    \begin{subfigure}[b]{0.32\textwidth}
        \begin{overpic}[width=1\linewidth]{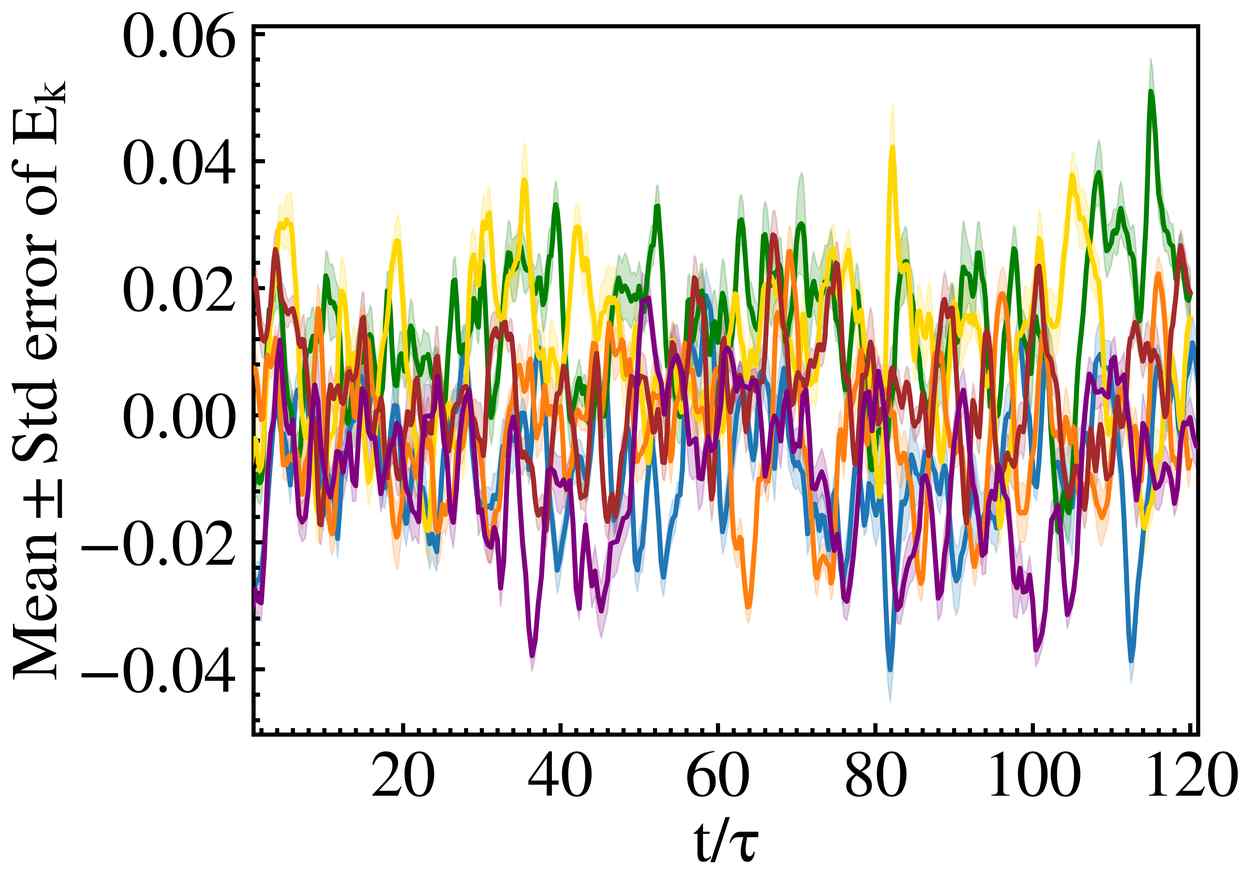}
            \put(-6,68){\small (e)} 
        \end{overpic} 
    \end{subfigure}
    \hfill
    \begin{subfigure}[b]{0.32\textwidth}
        \begin{overpic}[width=1\linewidth]{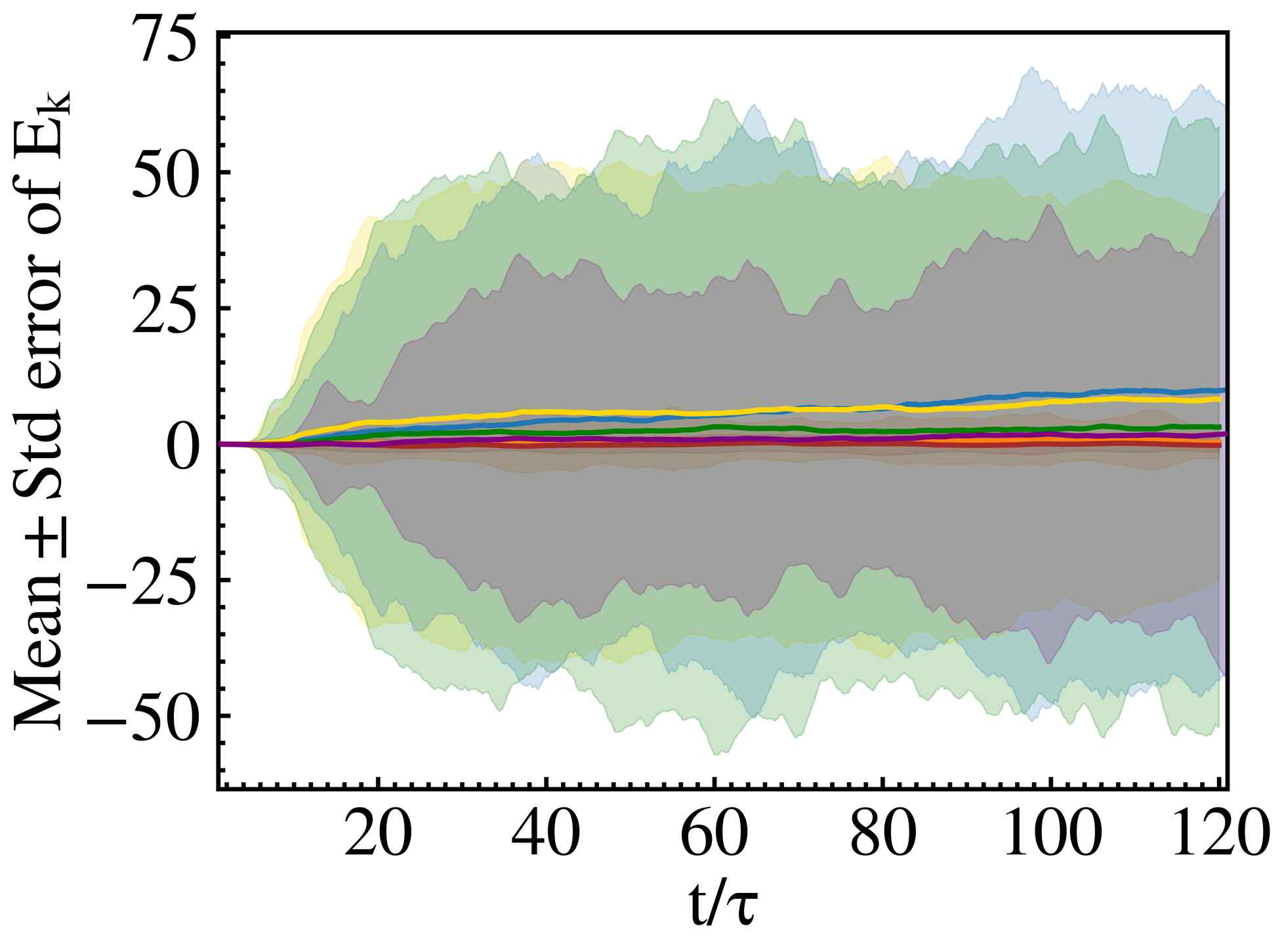}
            \put(-6,68){\small (f)} 
        \end{overpic}
    \end{subfigure}
    \vspace{0.1cm}

    \begin{subfigure}[b]{0.32\textwidth}
        \begin{overpic}[width=1\linewidth]{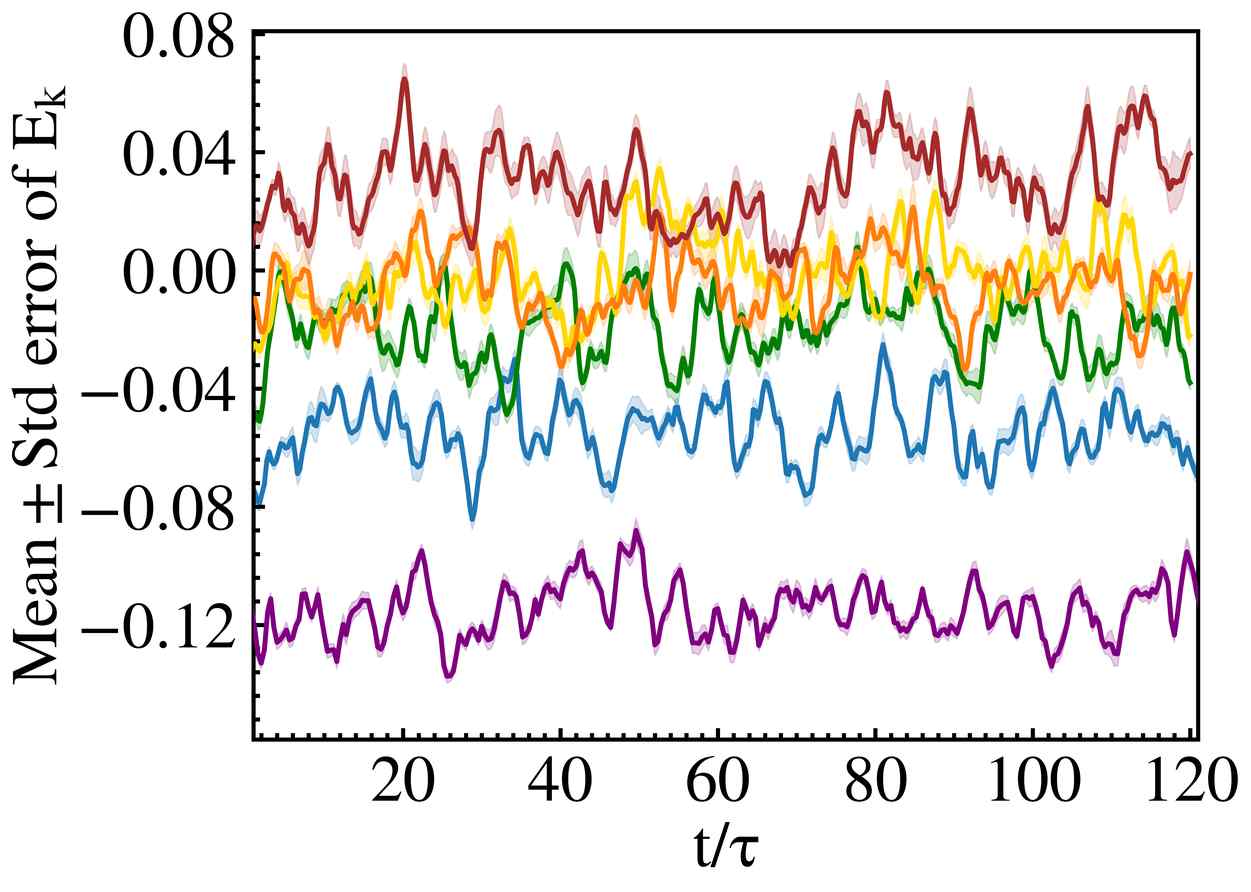}
            \put(-6,68){\small (g)}  
        \end{overpic}
    \end{subfigure}
    \hfill
    \begin{subfigure}[b]{0.32\textwidth}
        \begin{overpic}[width=1\linewidth]{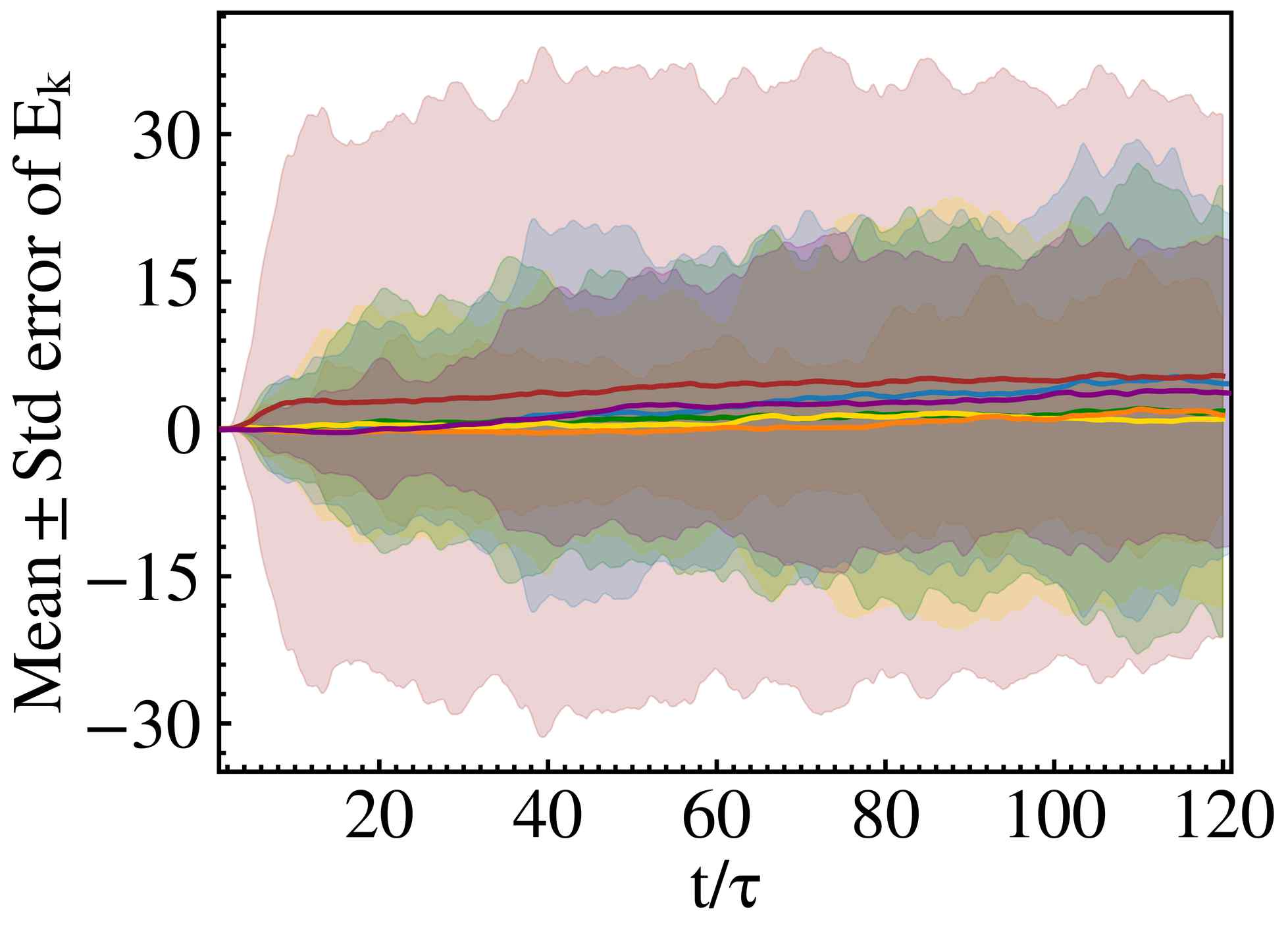}
            \put(-6,68){\small (h)} 
        \end{overpic} 
    \end{subfigure}
    \hfill
    \begin{subfigure}[b]{0.32\textwidth}
        \begin{overpic}[width=1\linewidth]{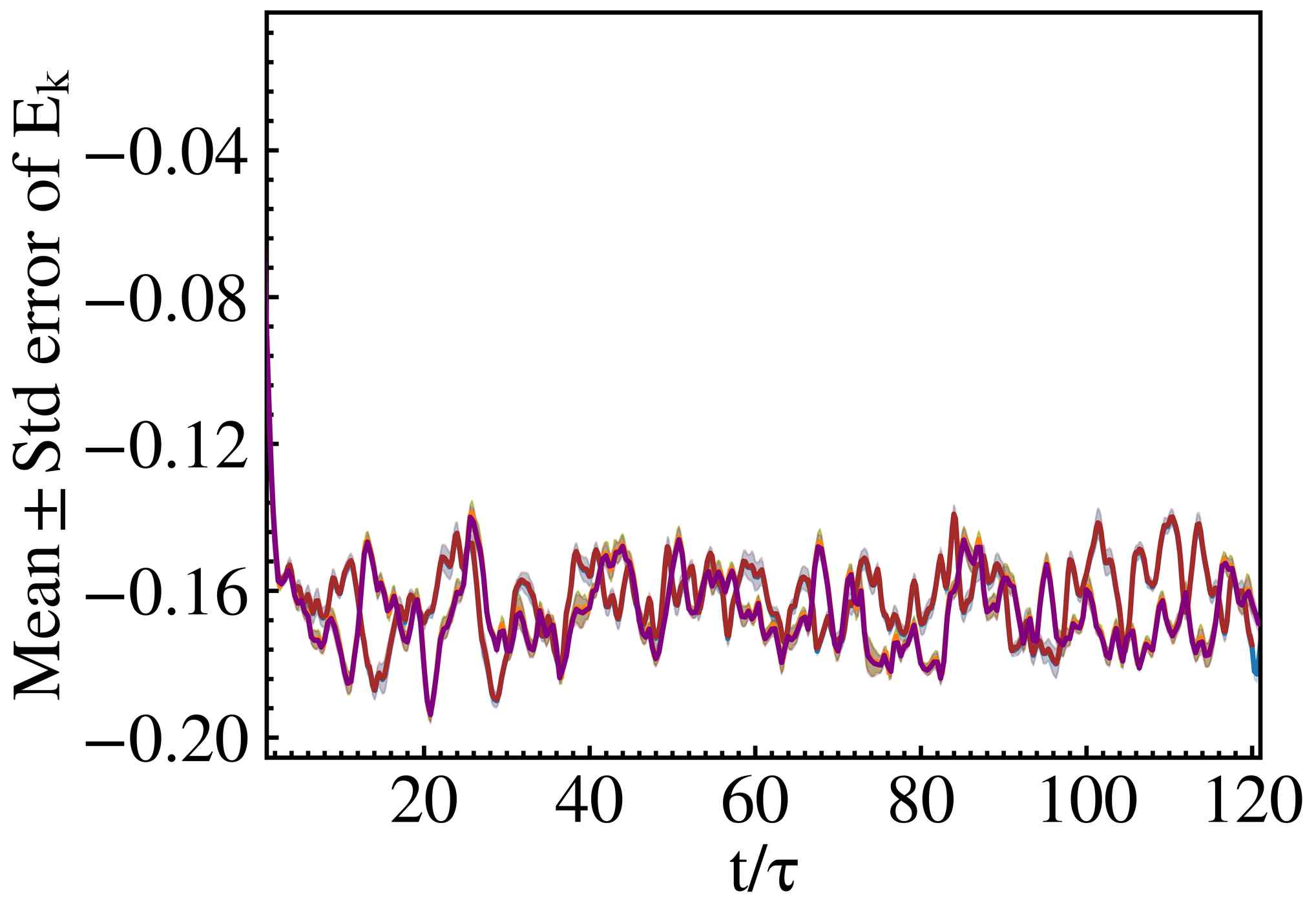}
            \put(-6,68){\small (i)} 
        \end{overpic}
    \end{subfigure}
    \vspace{0.1cm}

    \begin{subfigure}[b]{1\textwidth}
        \centering
        \begin{overpic}[width=0.32\linewidth]{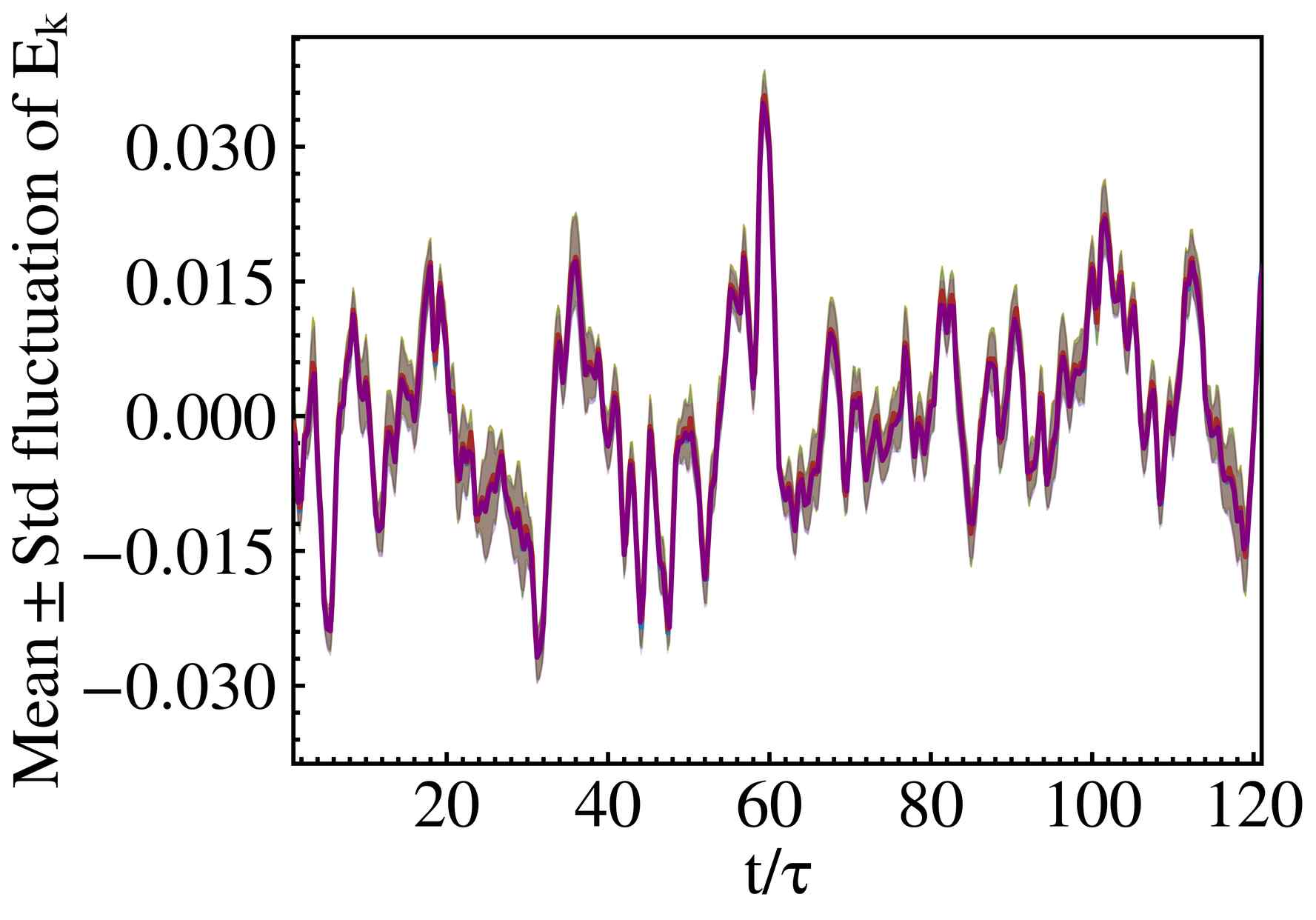}
            \put(-6,68){\small (j)}  
        \end{overpic}
    \end{subfigure}

	\caption{Temporal evolutions of $E_k$ error value consisting of mean and standard deviation (Mean $\pm$ Std) using different models under various time intervals $\Delta T$: (a) F-IFNO constrained; (b) F-IFNO unconstrained; (c) F-IUFNO constrained; (d) F-IUFNO unconstrained; (e) IUFNO constrained; (f) IUFNO unconstrained; (g) IFNO constrained; (h) IFNO unconstrained; (i) DSM; (j) fDNS. Note that for fDNS, the values represent natural statistical fluctuations over time, not prediction errors.}\label{fig:9}
\end{figure}

Fig.~\ref{fig:9} shows the temporal evolution of the $E_k$ error (mean $\pm$ standard deviation) across thirty independent data groups. Specifically, Figs.~\ref{fig:9}(a) and \ref{fig:9}(b) correspond to the constrained and unconstrained F-IFNO models, respectively; Figs.~\ref{fig:9}(c) and \ref{fig:9}(d) to F-IUFNO; Figs.~\ref{fig:9}(e) and \ref{fig:9}(f) to IUFNO; Figs.~\ref{fig:9}(g) and \ref{fig:9}(h) to IFNO; and Figs.~\ref{fig:9}(i) and \ref{fig:9}(j) present the DSM and fDNS results.
It can be observed that the fluctuations of fDNS and errors of DSM remain nearly constant over time, whereas the errors of FNO-based models evolve and eventually stabilize. When comparing the constrained and unconstrained versions of the same model, the constrained versions consistently show improved accuracy and stability. Notably, the fluctuations of fDNS and errors of DSM are both independent of the prediction interval $\Delta T$, but for FNO-based models, the error clearly varies with $\Delta T$.
Among the FNO-based models, F-IFNO performs best at $\Delta T = 0.1\tau$ and $0.2\tau$, suggesting that the optimal time interval range for both F-IFNO and F-IUFNO lies within $[0.1\tau, 0.2\tau]$. Although unconstrained IFNO and IUFNO exhibit larger error fluctuations, applying constraints significantly reduces both the mean error and its variability.
Moreover, consistent with the observations in Subsection~\ref{subsec4.1}, models with lower mean errors also tend to have smaller standard deviations. For example, all models in Fig.~\ref{fig:9}(e) demonstrate low mean and standard deviation, whereas those in Fig.~\ref{fig:9}(f) show much higher values. This aligns with the performance observed previously, where unconstrained IUFNO underperforms, while its constrained counterpart exhibits better reliability. Compared to other FNO-based models, in both constrained and unconstrained situations, our proposed F-IFNO and F-IUFNO exhibit the highest accuracy and lowest error when the optimal time interval $\Delta T$ is used.

\begin{figure}[ht!]
    \centering
    \begin{subfigure}[b]{0.49\textwidth}
        \begin{overpic}[width=1\linewidth]{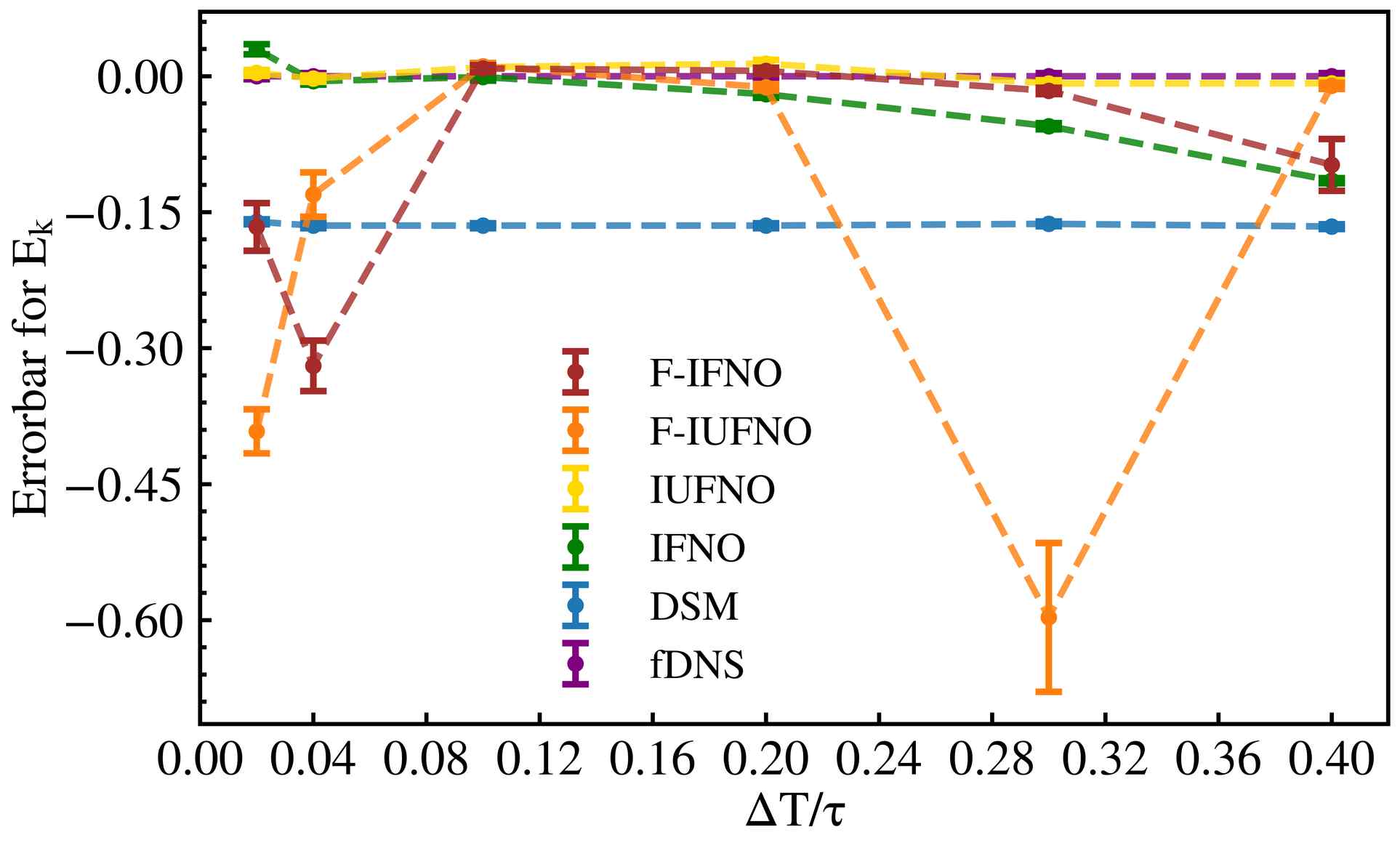}
            \put(0,55){\small (a)}  
        \end{overpic}
    \end{subfigure}
    \hfill
    \begin{subfigure}[b]{0.49\textwidth}
        \begin{overpic}[width=1\linewidth]{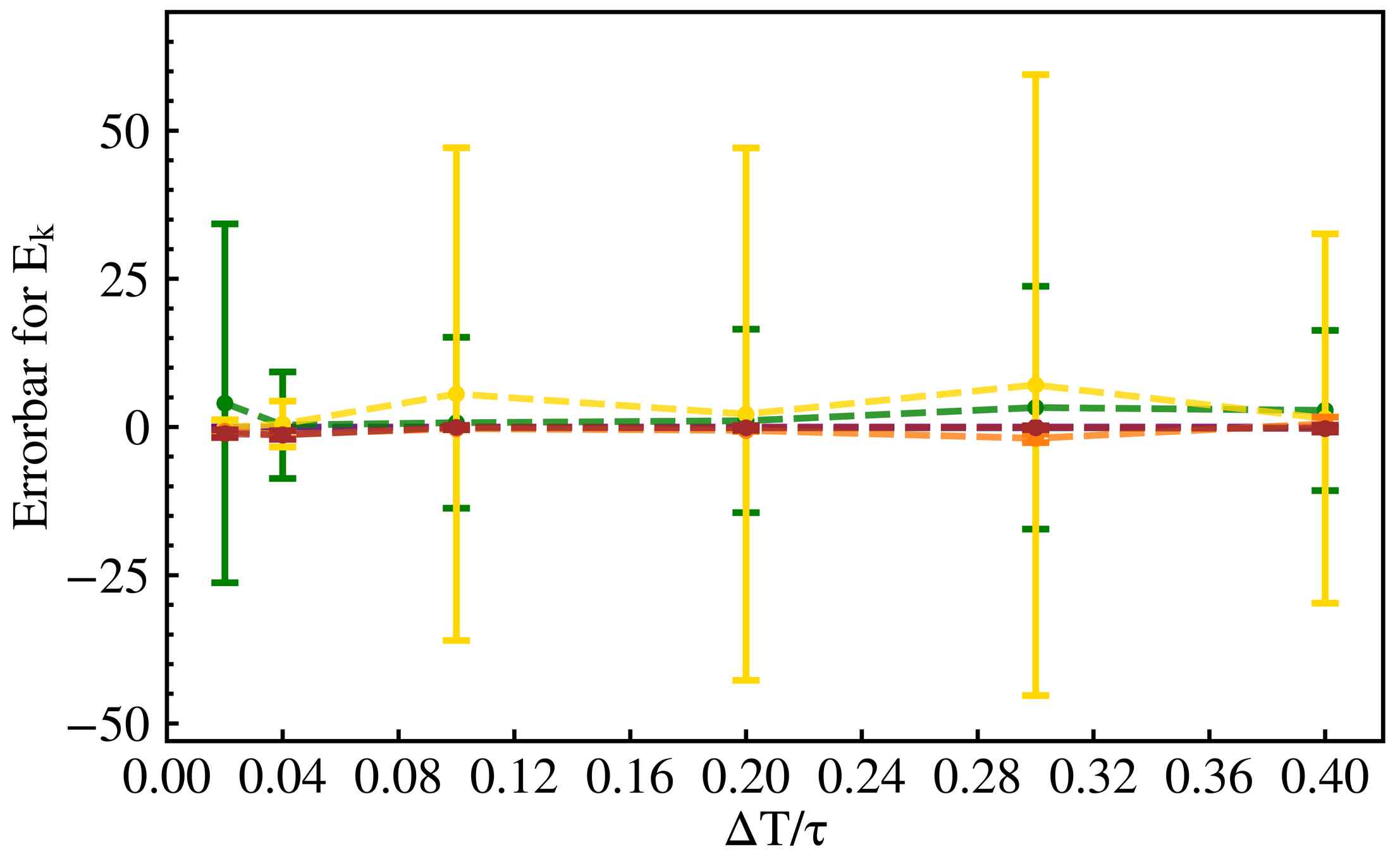}
            \put(0,55){\small (b)} 
        \end{overpic} 
    \end{subfigure}

	\caption{Errorbars of kinetic energy $E_k$ for various methods as a function of the time interval $\Delta T$: (a) constrained FNO-based models; (b) unconstrained FNO-based models. Note that for fDNS, the values represent natural statistical fluctuations over time, not prediction errors.}\label{fig:10}
\end{figure}

Since the turbulent flow is a statistically steady forced HIT, each time snapshot from fDNS can be regarded as ground truth with only minor fluctuations. Therefore, the errors from all cases across all time steps (a total of 18,000 samples per method) are used for statistical analysis via errorbars.
Fig.~\ref{fig:10} presents the errorbars of the kinetic energy $E_k$ for various methods as a function of the time interval $\Delta T$. Fig.~\ref{fig:10}(a) illustrates results for constrained FNO-based models, while Fig.~\ref{fig:10}(b) shows those for unconstrained counterparts.
From Fig.~\ref{fig:10}(a), we observe that the constrained F-IFNO and F-IUFNO models yield low error and standard deviation when $\Delta T=0.1\tau$ and $0.2\tau$, indicating accurate and stable predictions. In addition, the constrained IUFNO performs consistently well across all tested $\Delta T$ values, and the constrained IFNO achieves good accuracy when $\Delta T=0.04\tau$, $0.1\tau$, and $0.2\tau$. Notably, with their respective optimal $\Delta T$ values, FNO-based models significantly outperform DSM, exhibiting both lower mean errors and reduced variance.
In contrast, Fig.~\ref{fig:10}(b) shows that the unconstrained IFNO and IUFNO models suffer from significantly larger errorbars across all $\Delta T$, whereas F-IFNO and F-IUFNO still maintain competitive performance with error levels comparable to or better than DSM.
In summary, both versions of F-IFNO and F-IUFNO, when used with their optimal time intervals, not only achieve superior prediction accuracy but also demonstrate greater stability and lower variability compared to DSM and other FNO-based models.

\begin{figure}[ht!]
    \centering
    \begin{subfigure}[b]{0.49\textwidth}
        \begin{overpic}[width=1\linewidth]{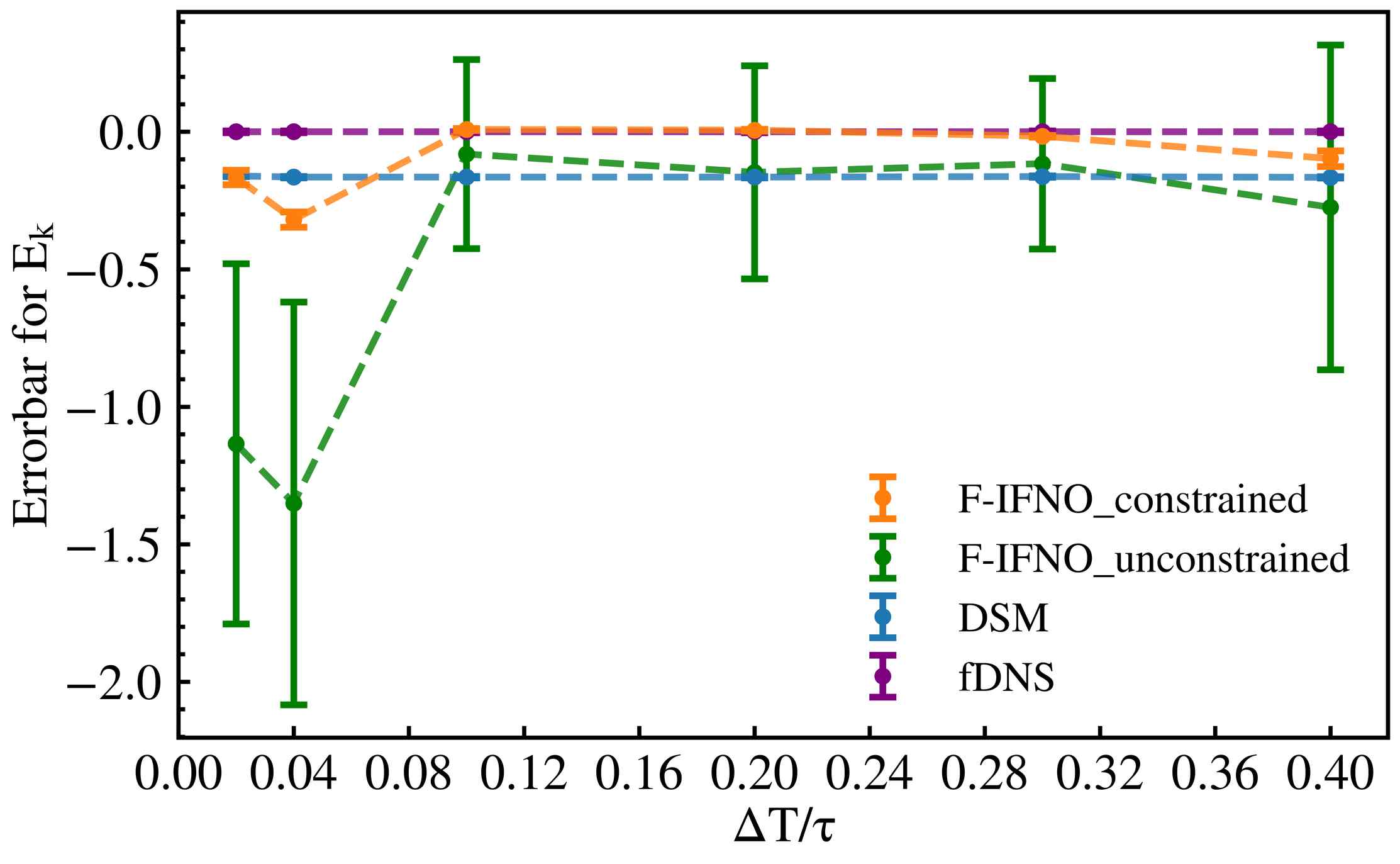}
            \put(0,55){\small (a)}  
        \end{overpic}
    \end{subfigure}
    \hfill
    \begin{subfigure}[b]{0.49\textwidth}
        \begin{overpic}[width=1\linewidth]{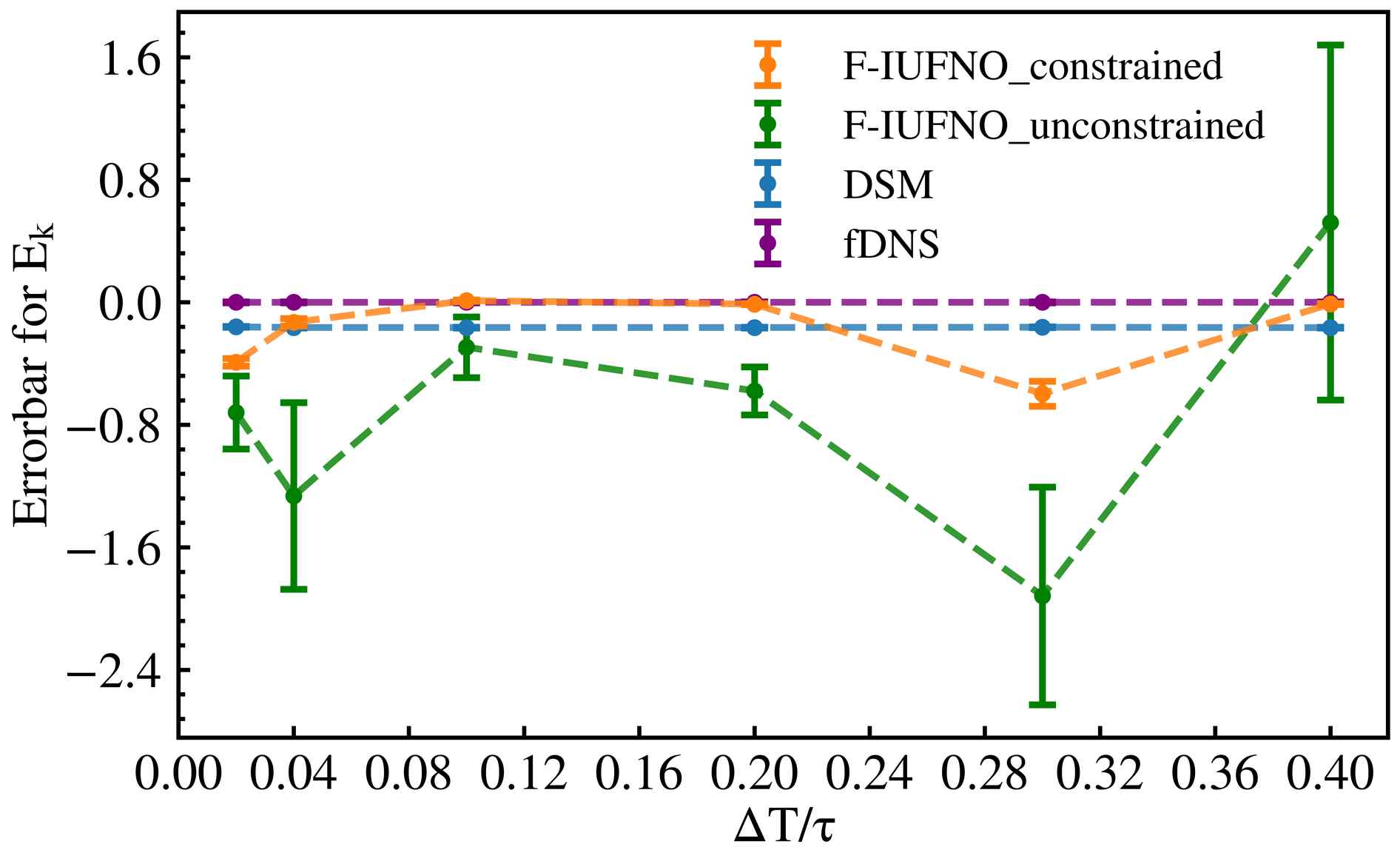}
            \put(0,55){\small (b)} 
        \end{overpic} 
    \end{subfigure}
    \vspace{0.1cm}

    \begin{subfigure}[b]{0.49\textwidth}
        \begin{overpic}[width=1\linewidth]{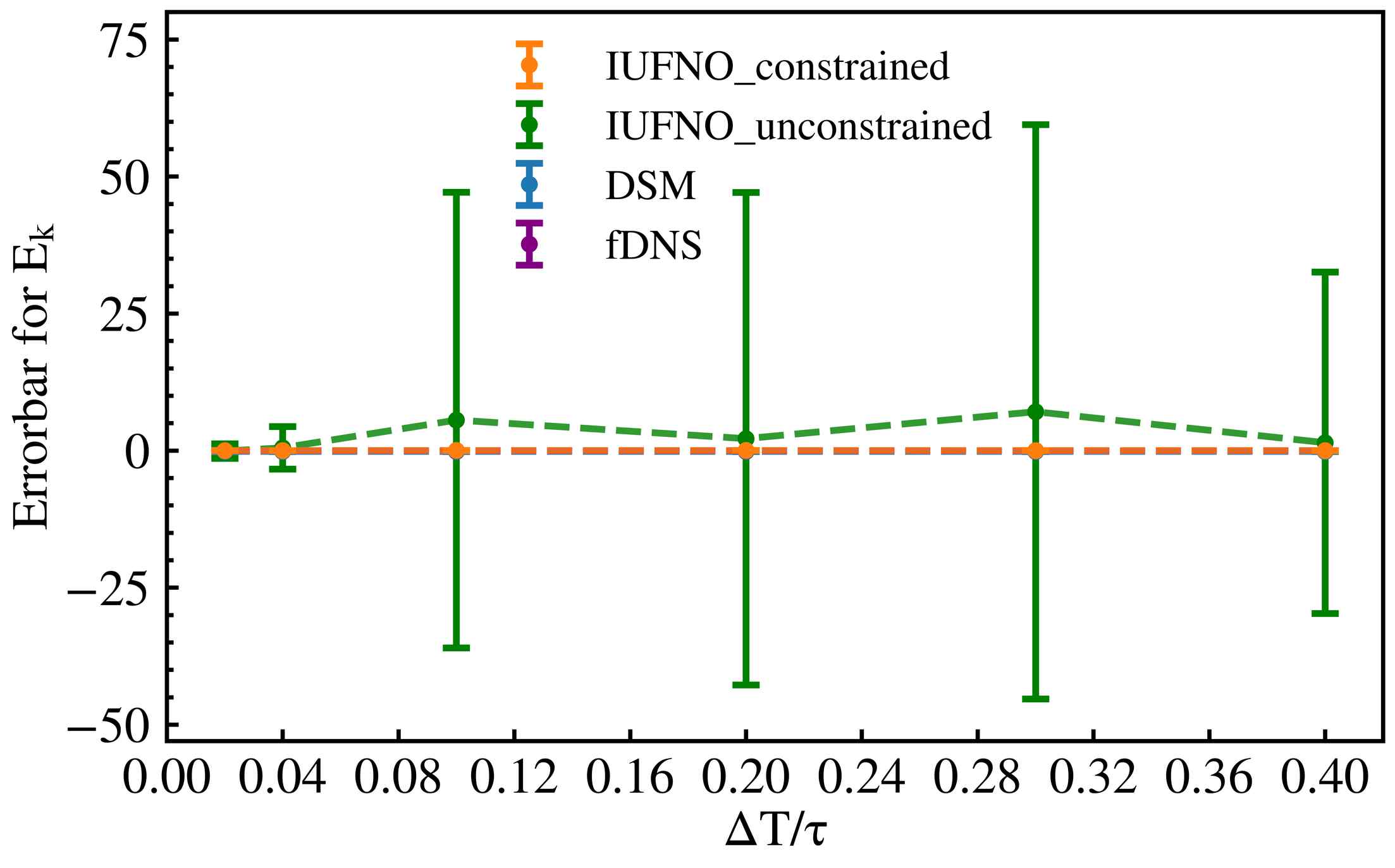}
            \put(0,55){\small (c)}  
        \end{overpic}
    \end{subfigure}
    \hfill
    \begin{subfigure}[b]{0.49\textwidth}
        \begin{overpic}[width=1\linewidth]{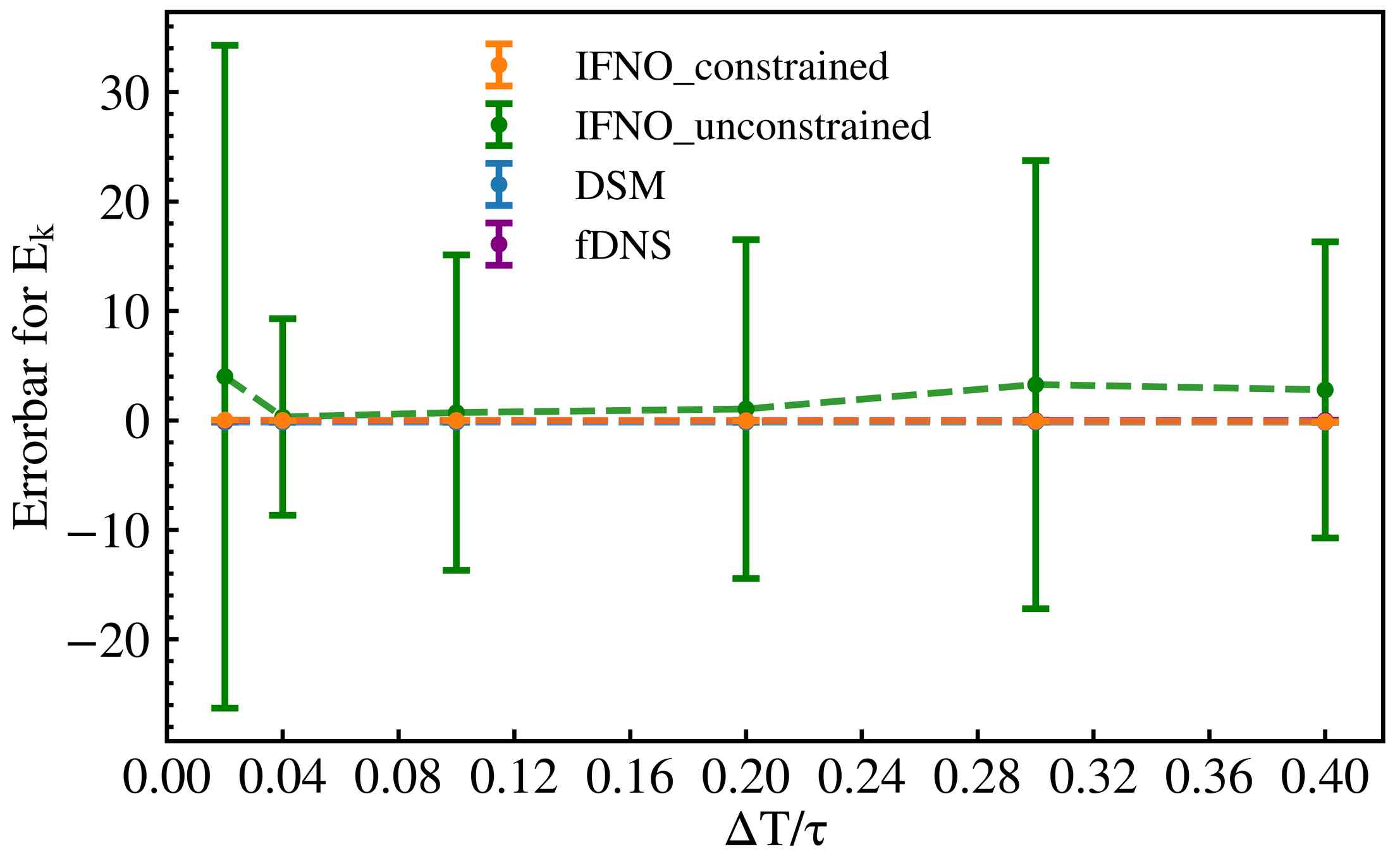}
            \put(0,55){\small (d)} 
        \end{overpic} 
    \end{subfigure}
   
	\caption{Errorbars of kinetic energy $E_k$ for constrained and unconstrained FNO-based models: (a) F-IFNO; (b) F-IUFNO; (c) IUFNO; (d) IFNO. Note that for fDNS, the values represent natural statistical fluctuations over time, not prediction errors.}\label{fig:11}
\end{figure}

Fig.~\ref{fig:11} further illustrates the errorbars of the kinetic energy $E_k$ for different FNO-based models, comparing constrained and unconstrained versions. Specifically, Fig.~\ref{fig:11}(a) shows the results for F-IFNO, Fig.~\ref{fig:11}(b) for F-IUFNO, Fig.~\ref{fig:11}(c) for IUFNO, and Fig.~\ref{fig:11}(d) for IFNO.
In Fig.~\ref{fig:11}(a), it is evident that the constrained F-IFNO significantly outperforms the unconstrained version. The constrained variant achieves both a lower mean error and a smaller standard deviation across all time intervals, whereas the unconstrained F-IFNO only exhibits a slightly lower mean error than DSM within the optimal range of $\Delta T=0.1\tau$ to $0.2\tau$.
Similar trends are observed in Figs.~\ref{fig:11}(b), (c) and (d), where the constrained versions consistently yield better performance than their unconstrained counterparts. In particular, the unconstrained IFNO and IUFNO demonstrate notably poor accuracy across all $\Delta T$ values, while the constrained ones remain stable and accurate.
These results clearly highlight that applying constraints during prediction can substantially enhance the accuracy and robustness of FNO-based models.

To further investigate the error characteristics of different models, we select a representative time interval of $\Delta T = 0.2\tau$ to compare model performance. 
The probability density function (PDF) of the normal distribution is defined by \cite{TSOKOS2016231}
\begin{equation}
f(x;\mu ,\sigma ) = \frac{1}{\sqrt{2\pi \sigma^2}} \exp\left(-\frac{(x-\mu)^2}{2\sigma^2}\right),
\label{eq:33}
\end{equation}
where $\mu$ denotes the mean (or average), which determines the center of the distribution, and $\sigma > 0$ is the standard deviation, which characterizes the spread (or width) of the distribution.

The probability density function of the skew normal distribution is given by \cite{10.1093/biomet/63.1.201,ASHOUR2010341,MUDHOLKAR2000291} 
\begin{equation}
f(x;\alpha ,\mu ,\sigma ) = \frac{2}{\sigma} \, \phi\left(\frac{x - \mu}{\sigma}\right) \, \Phi\left(\alpha \, \frac{x - \mu}{\sigma}\right),
\label{eq:34}
\end{equation}
where $\mu$ is the location parameter (controlling the center of the distribution), $\sigma > 0$ is the scale parameter (controlling the width of the distribution), and $\alpha$ is the shape parameter (controlling the skewness of the distribution). Here, $\phi(z) = \frac{1}{\sqrt{2\pi}} \exp\left(-\frac{z^2}{2}\right)$ is the standard normal PDF, and $\Phi(z) = \int_{-\infty}^{z} \phi(t)\,\mathrm{d}t$ is its cumulative distribution function (CDF).
When $\alpha = 0$, the skew normal distribution reduces to the standard normal distribution. A positive $\alpha$ implies right skewness (longer tail on the right), while a negative $\alpha$ implies left skewness (longer tail on the left).

\begin{figure}[ht!]
    \centering
    \begin{subfigure}[b]{0.32\textwidth}
        \begin{overpic}[width=1\linewidth]{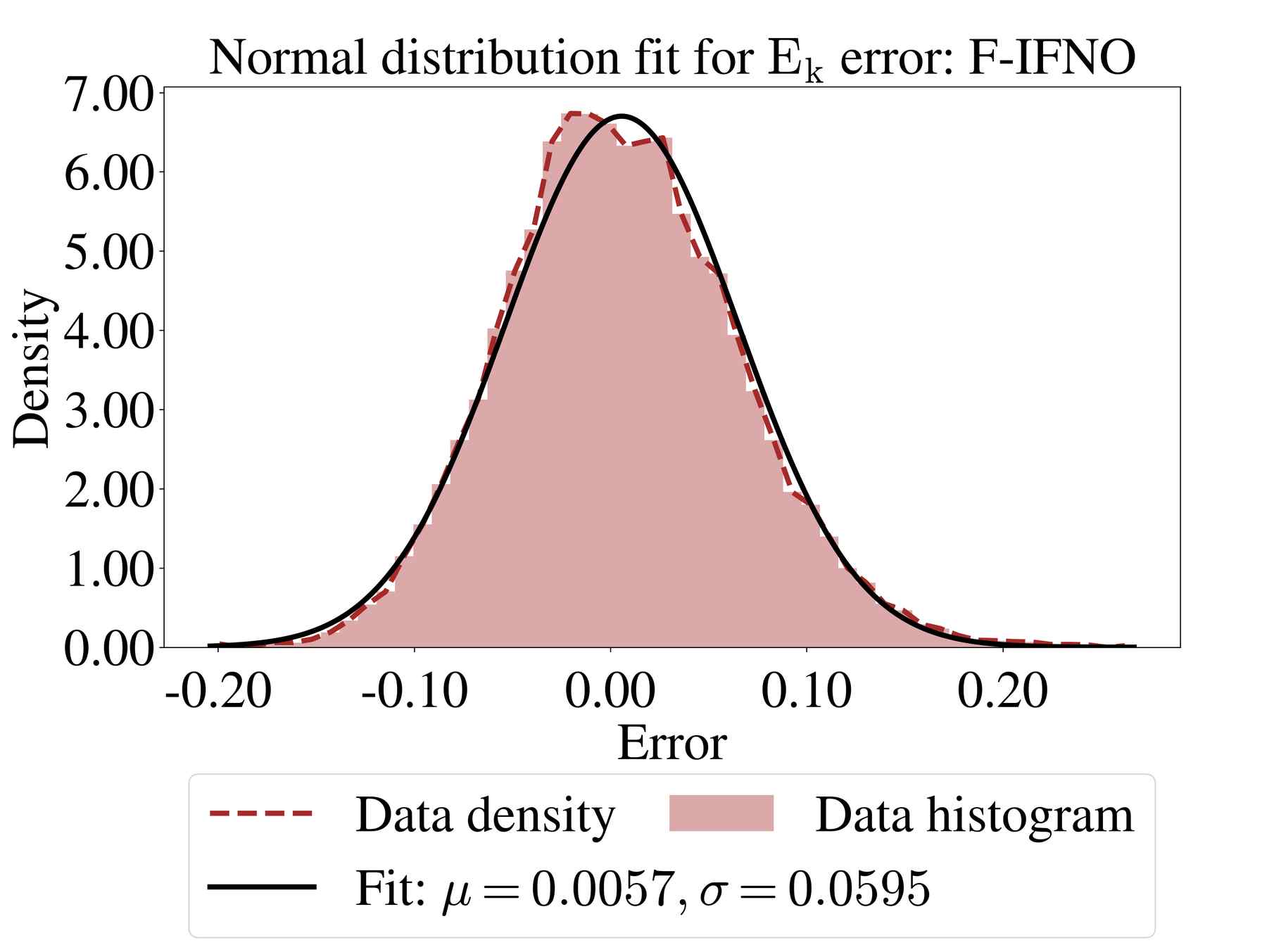}
            \put(-5,60){\small (a)}  
        \end{overpic}
    \end{subfigure}
    \hfill
    \begin{subfigure}[b]{0.32\textwidth}
        \begin{overpic}[width=1\linewidth]{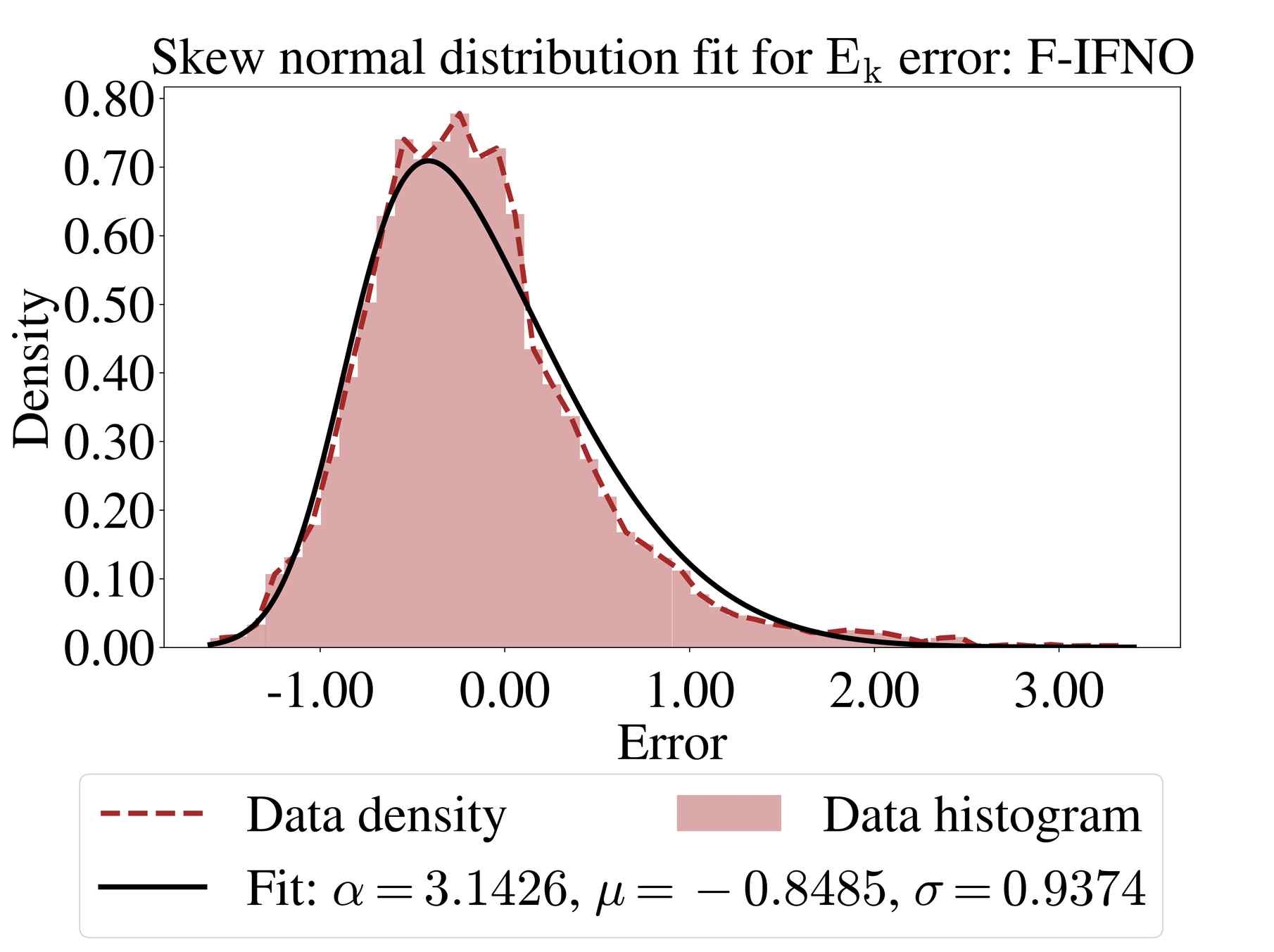}
            \put(-5,60){\small (b)} 
        \end{overpic} 
    \end{subfigure}
    \hfill
    \begin{subfigure}[b]{0.32\textwidth}
        \begin{overpic}[width=1\linewidth]{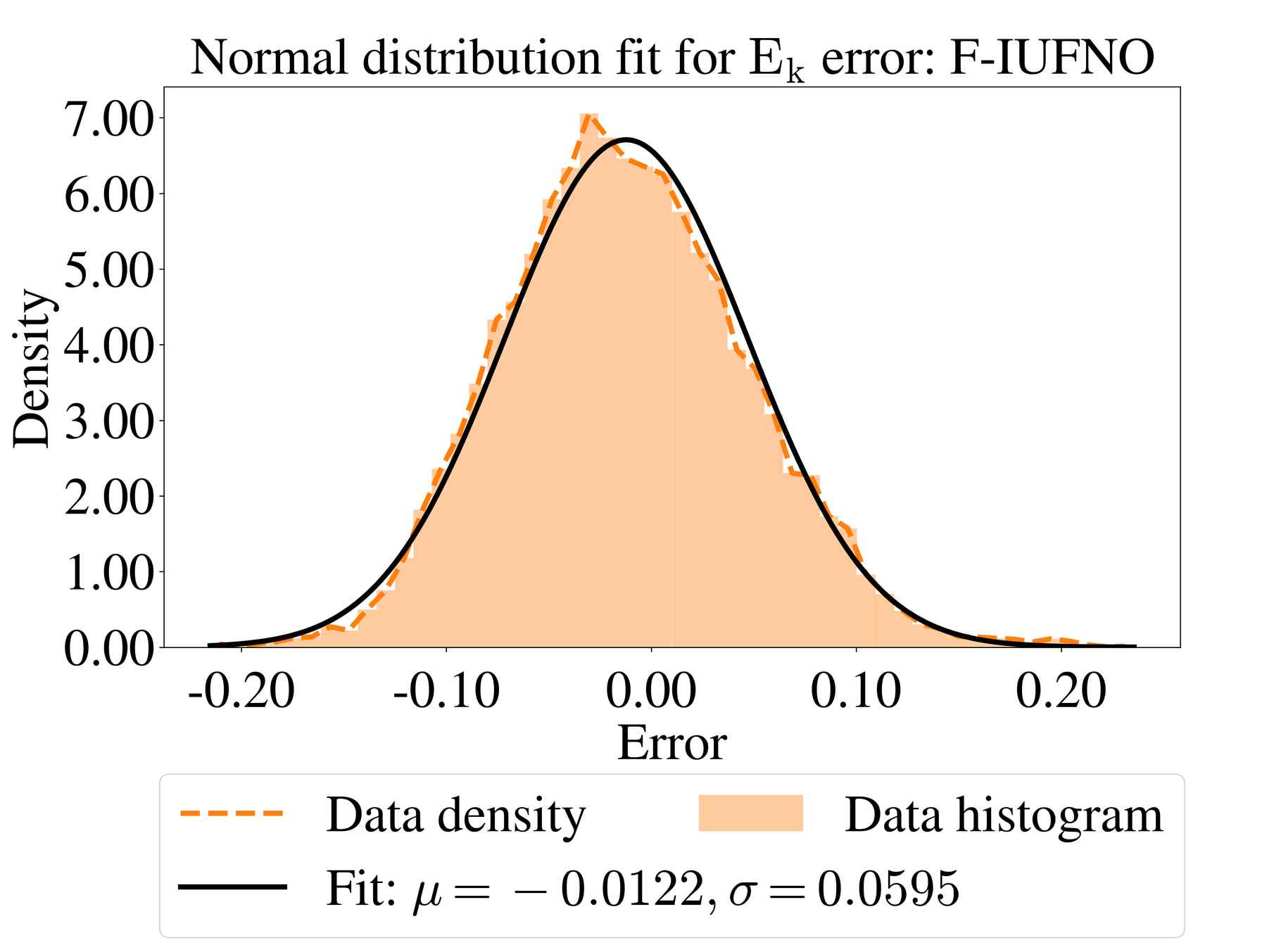}
            \put(-5,60){\small (c)} 
        \end{overpic}
    \end{subfigure}
    \vspace{0.1cm}

    \begin{subfigure}[b]{0.32\textwidth}
        \begin{overpic}[width=1\linewidth]{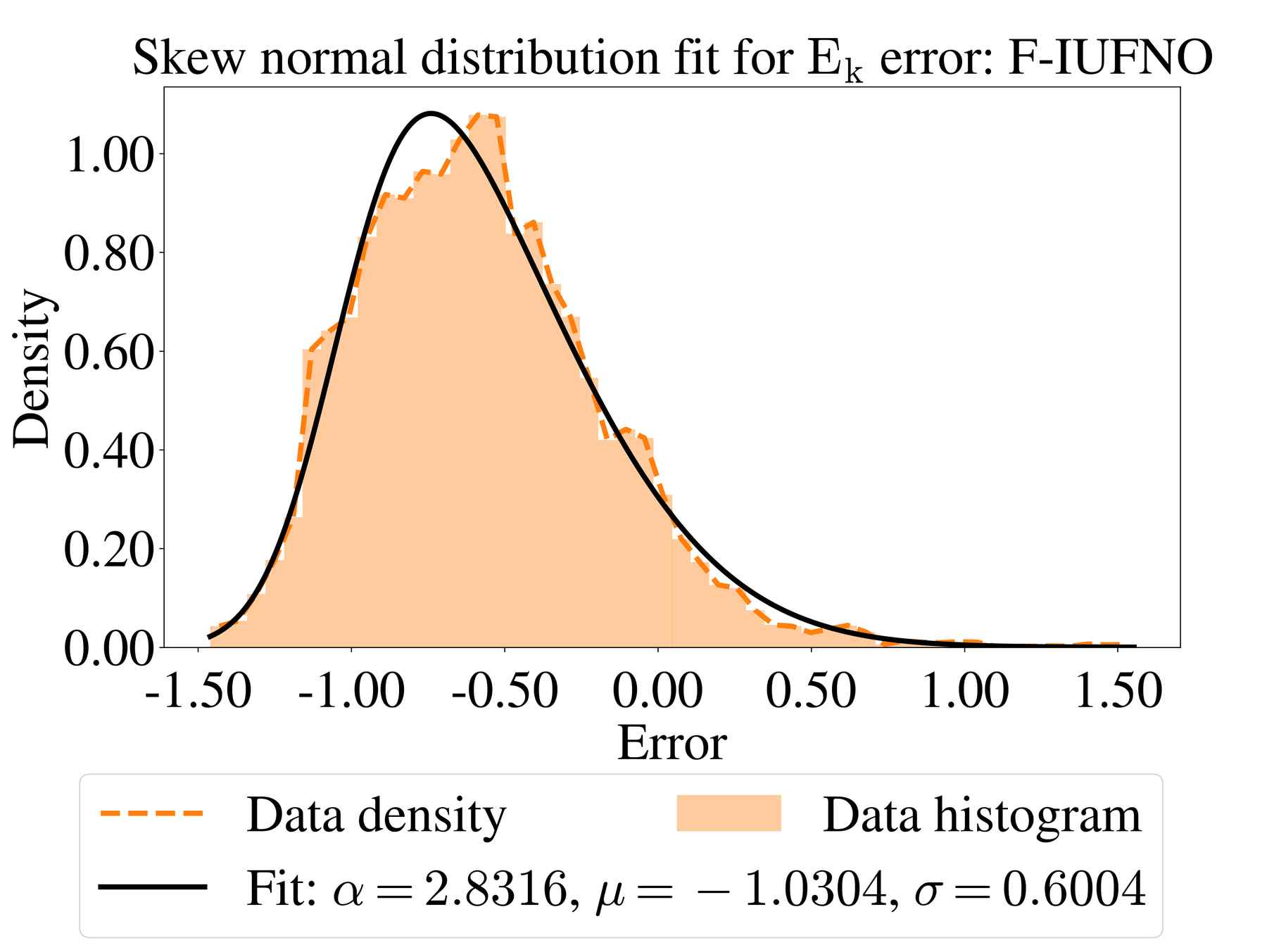}
            \put(-5,60){\small (d)}  
        \end{overpic}
    \end{subfigure}
    \hfill
    \begin{subfigure}[b]{0.32\textwidth}
        \begin{overpic}[width=1\linewidth]{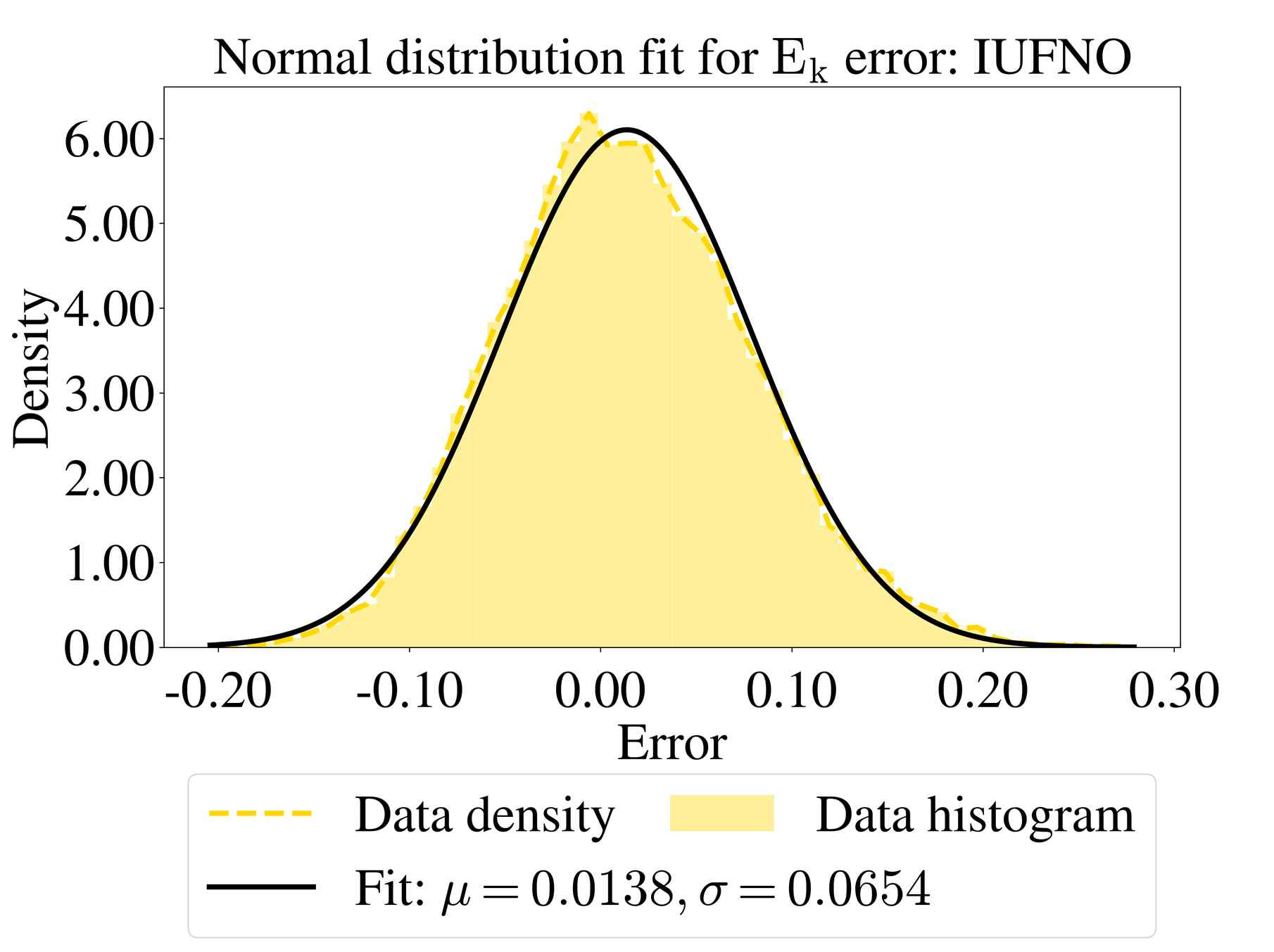}
            \put(-5,60){\small (e)} 
        \end{overpic} 
    \end{subfigure}
    \hfill
    \begin{subfigure}[b]{0.32\textwidth}
        \begin{overpic}[width=1\linewidth]{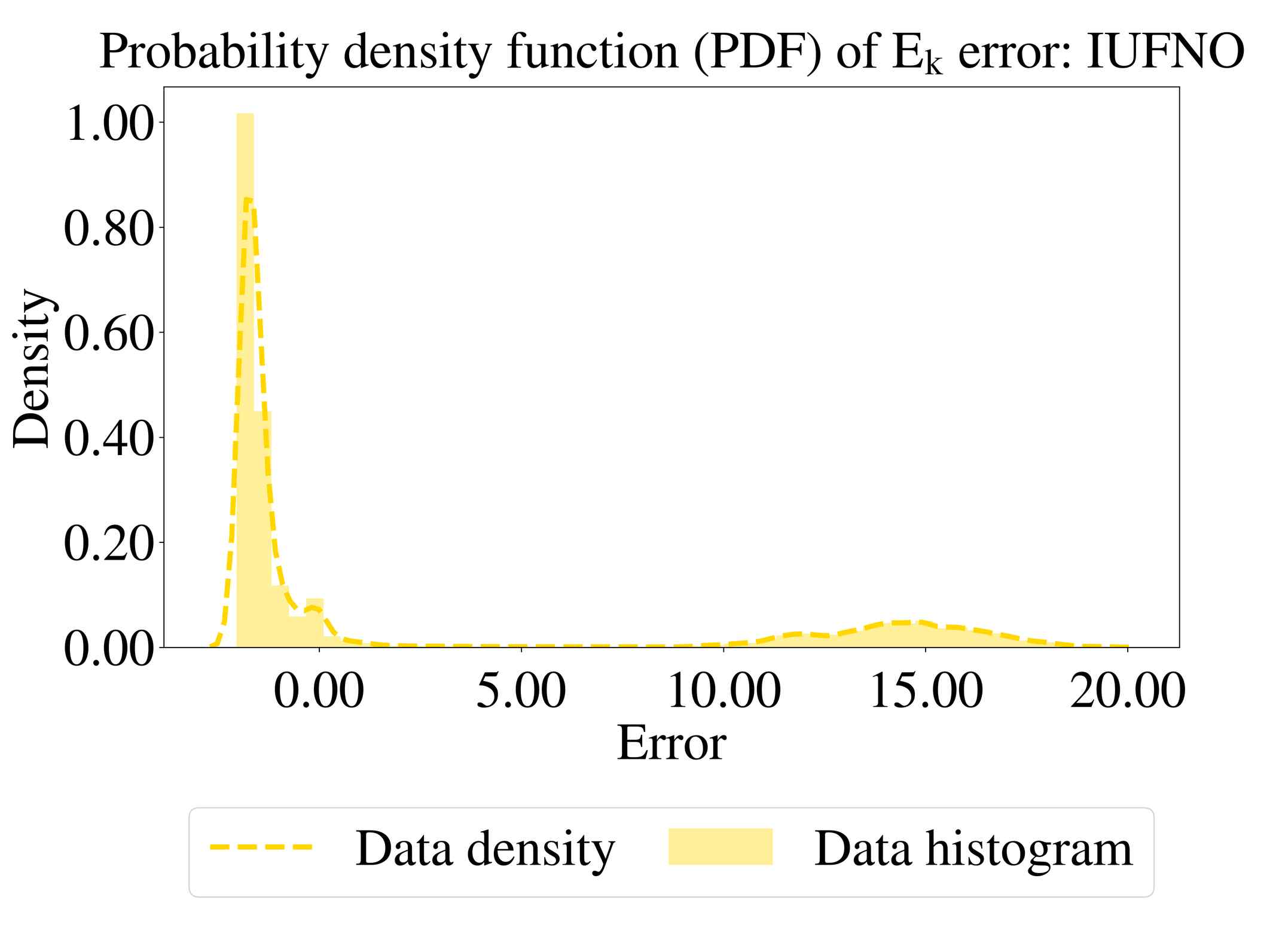}
            \put(-5,60){\small (f)} 
        \end{overpic}
    \end{subfigure}
    \vspace{0.1cm}

    \begin{subfigure}[b]{0.32\textwidth}
        \begin{overpic}[width=1\linewidth]{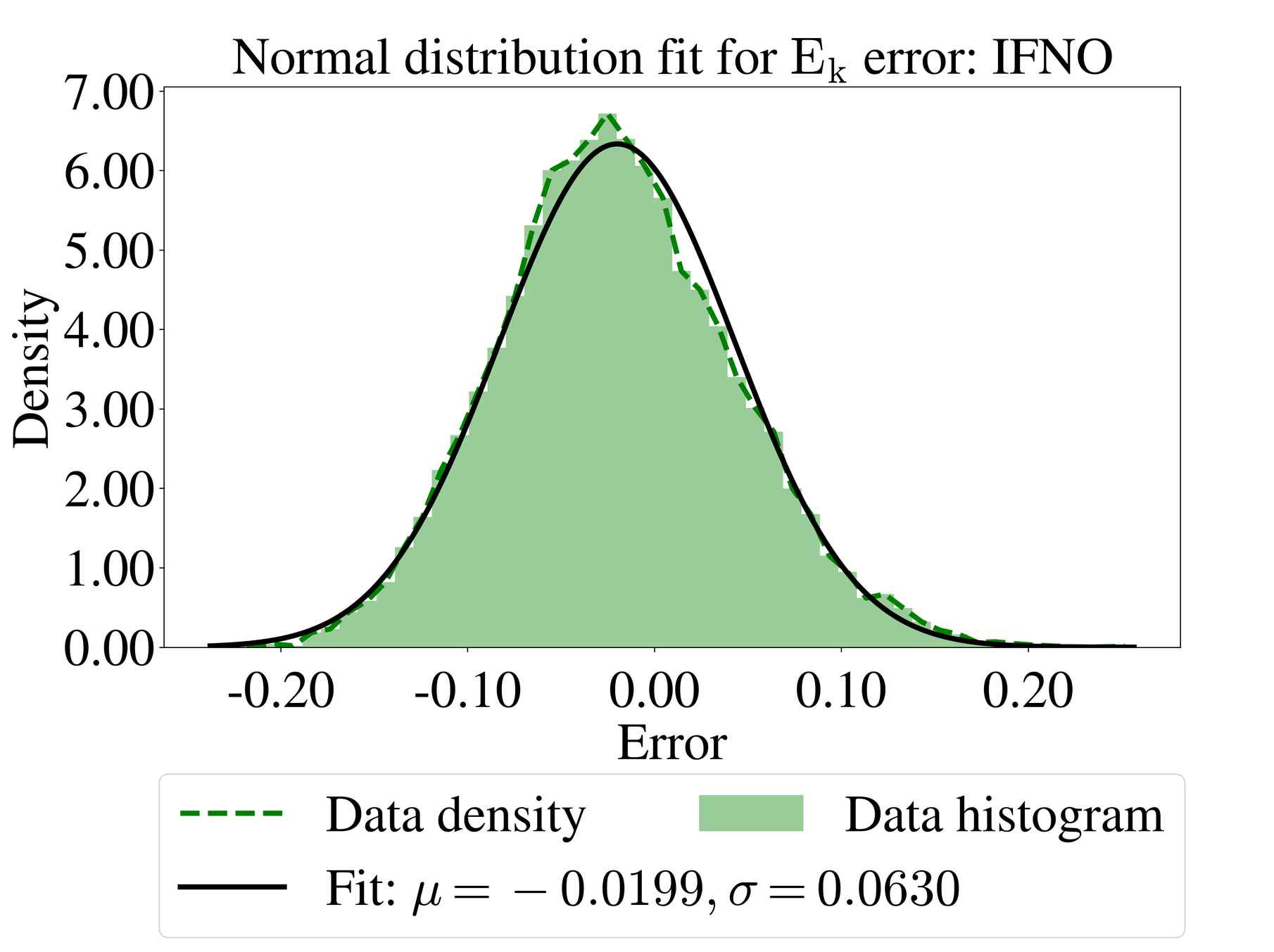}
            \put(-5,60){\small (g)}  
        \end{overpic}
    \end{subfigure}
    \hfill
    \begin{subfigure}[b]{0.32\textwidth}
        \begin{overpic}[width=1\linewidth]{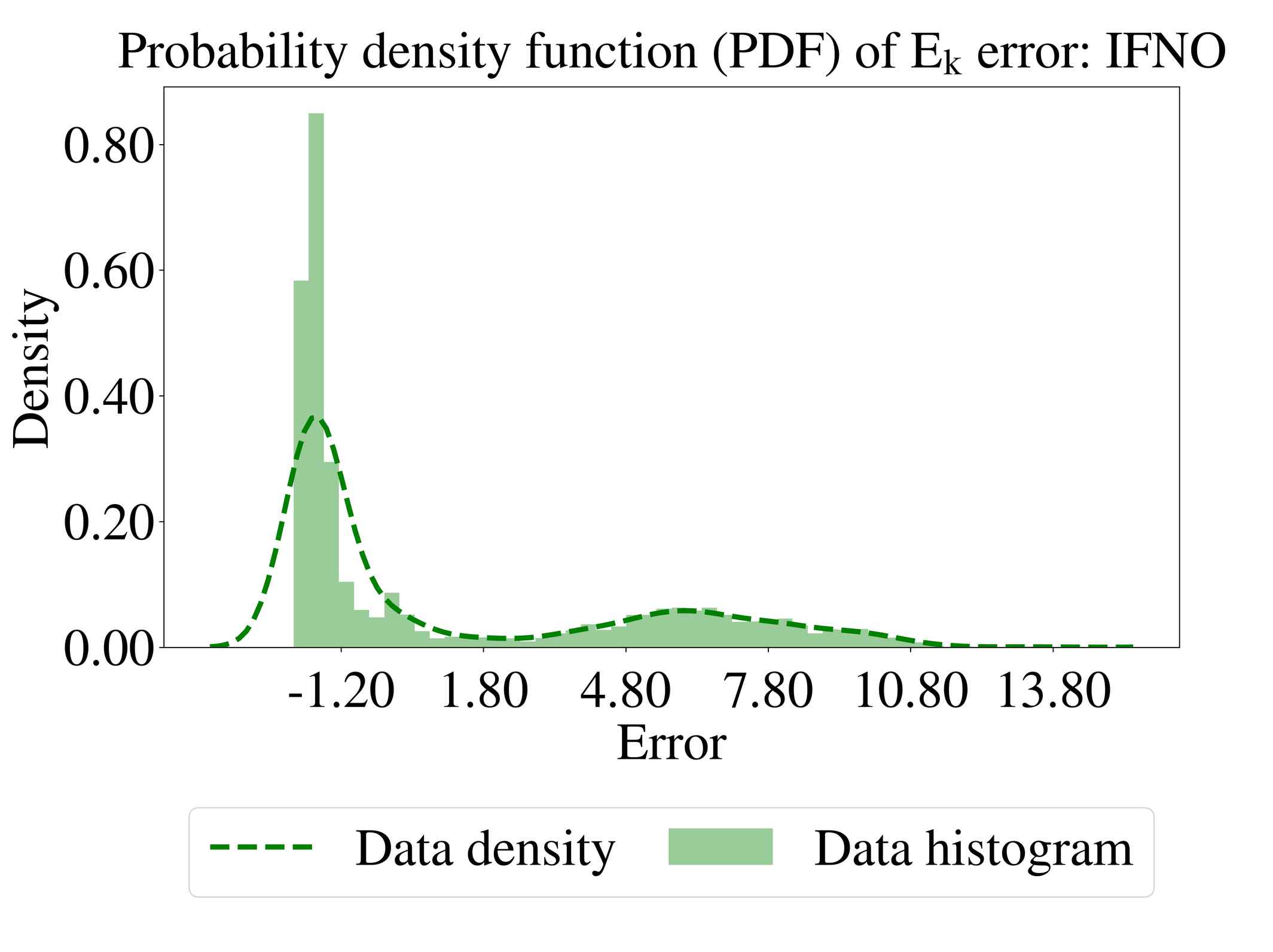}
            \put(-5,60){\small (h)} 
        \end{overpic} 
    \end{subfigure}
    \hfill
    \begin{subfigure}[b]{0.32\textwidth}
        \begin{overpic}[width=1\linewidth]{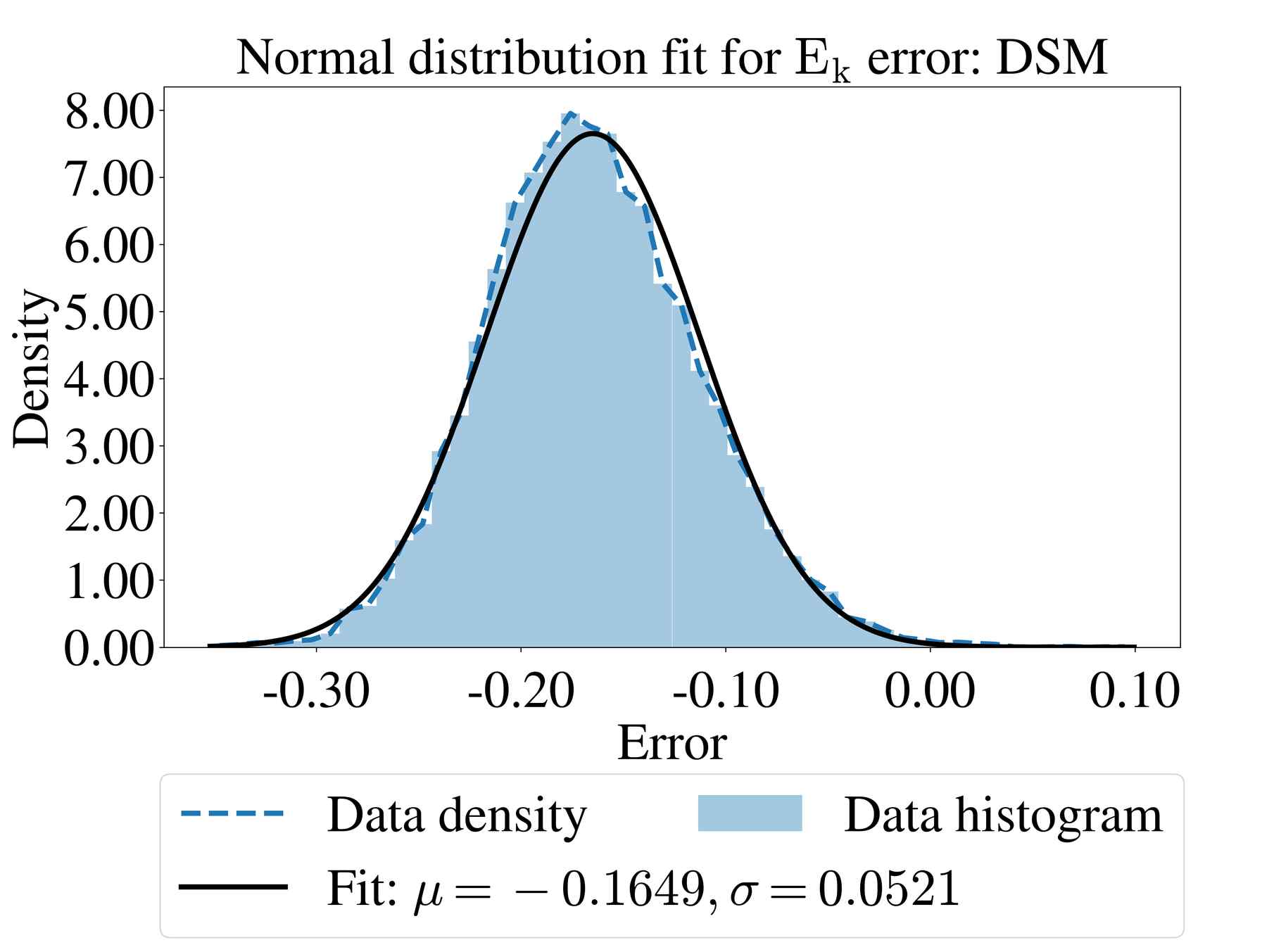}
            \put(-5,60){\small (i)} 
        \end{overpic}
    \end{subfigure}
    \vspace{0.1cm}

    \begin{subfigure}[b]{1\textwidth}
        \centering
        \begin{overpic}[width=0.32\linewidth]{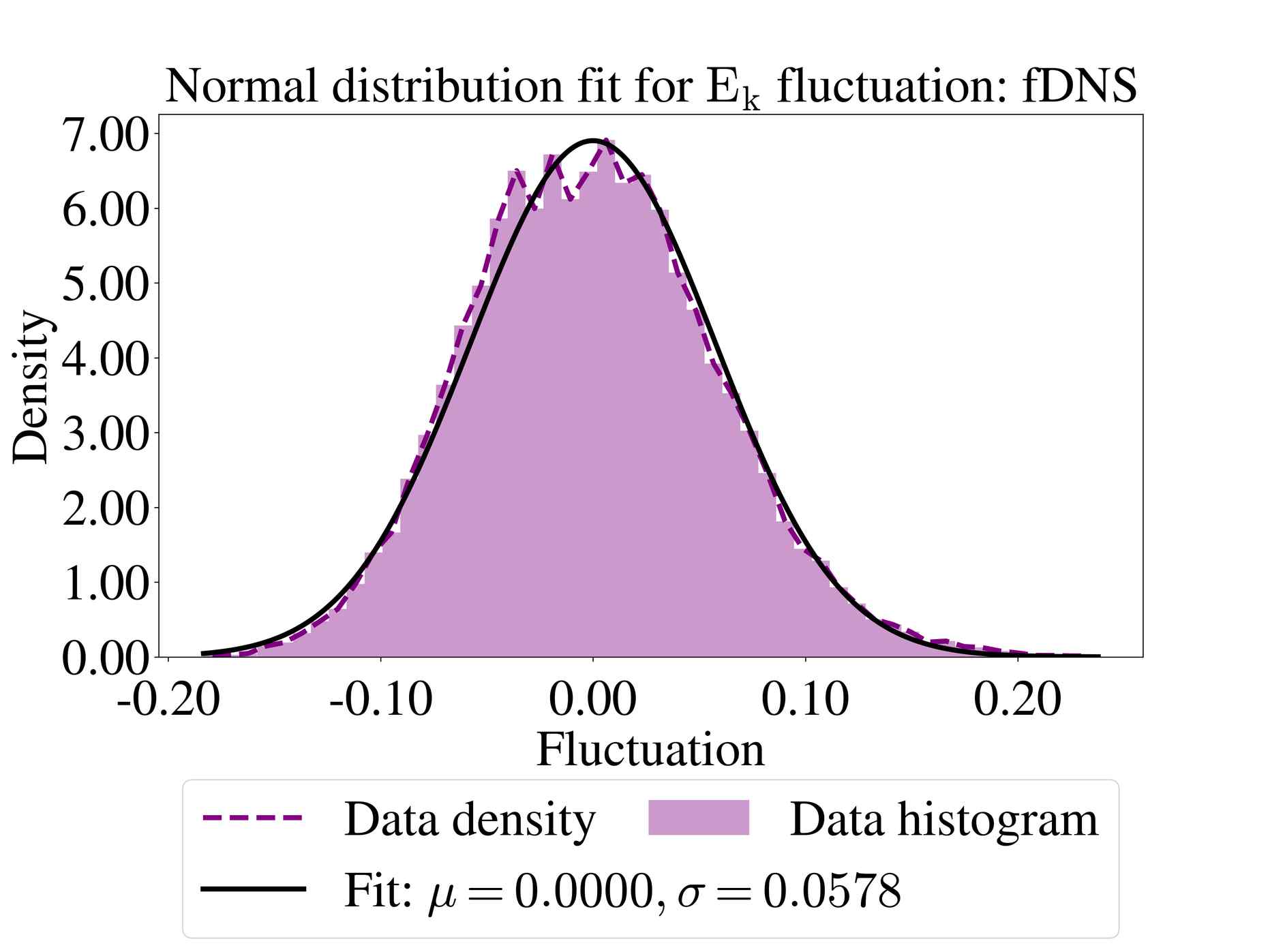}
            \put(-5,60){\small (j)}  
        \end{overpic}
    \end{subfigure}

	\caption{The PDFs of the kinetic energy $E_k$ errors for each method at time interval $\Delta T=0.2\tau$: (a) F-IFNO constrained; (b) F-IFNO unconstrained; (c) F-IUFNO constrained; (d) F-IUFNO unconstrained; (e) IUFNO constrained; (f) IUFNO unconstrained; (g) IFNO constrained; (h) IFNO unconstrained; (i) DSM; (j) fDNS. Note that for fDNS, the values represent natural statistical fluctuations over time, not prediction errors.}\label{fig:12}
\end{figure}

Fig.~\ref{fig:12} presents the probability density functions (PDFs) of the kinetic energy $E_k$ errors for each method at this specific time interval. Additionally, we fit the distributions using either normal or skew normal distributions where appropriate, as also shown in Fig.~\ref{fig:12}.
The results indicate that the error distributions of the constrained FNO-based models, DSM, and fluctuation distribution of fDNS follow a normal distribution, while the unconstrained F-IFNO and F-IUFNO follow a skew normal distribution. In contrast, the error distributions of the unconstrained IFNO and IUFNO cannot be reasonably fitted by either normal or skew normal distributions.
Based on these observations, we infer that models whose error distributions conform to either a normal or skew normal form tend to produce more reliable predictions, exhibiting better accuracy and stability. Moreover, the constrained FNO-based models, whose error distributions closely resemble the natural fluctuation distribution of fDNS and are narrower than the error distributions of DSM, demonstrate superior performance. Compared to the unconstrained F-IFNO and F-IUFNO, which follow skew normal distributions, the constrained variants exhibit narrower error ranges.
On the other hand, the unconstrained IFNO and IUFNO models show the widest error distribution ranges, reflecting their inferior predictive capability. These observations reinforce the conclusion that the proposed F-IFNO and F-IUFNO, when used with optimal time intervals, achieve the best overall performance in terms of both accuracy and consistency.

\begin{figure}[ht!]
    \centering
    \begin{subfigure}[b]{0.32\textwidth}
        \begin{overpic}[width=1\linewidth]{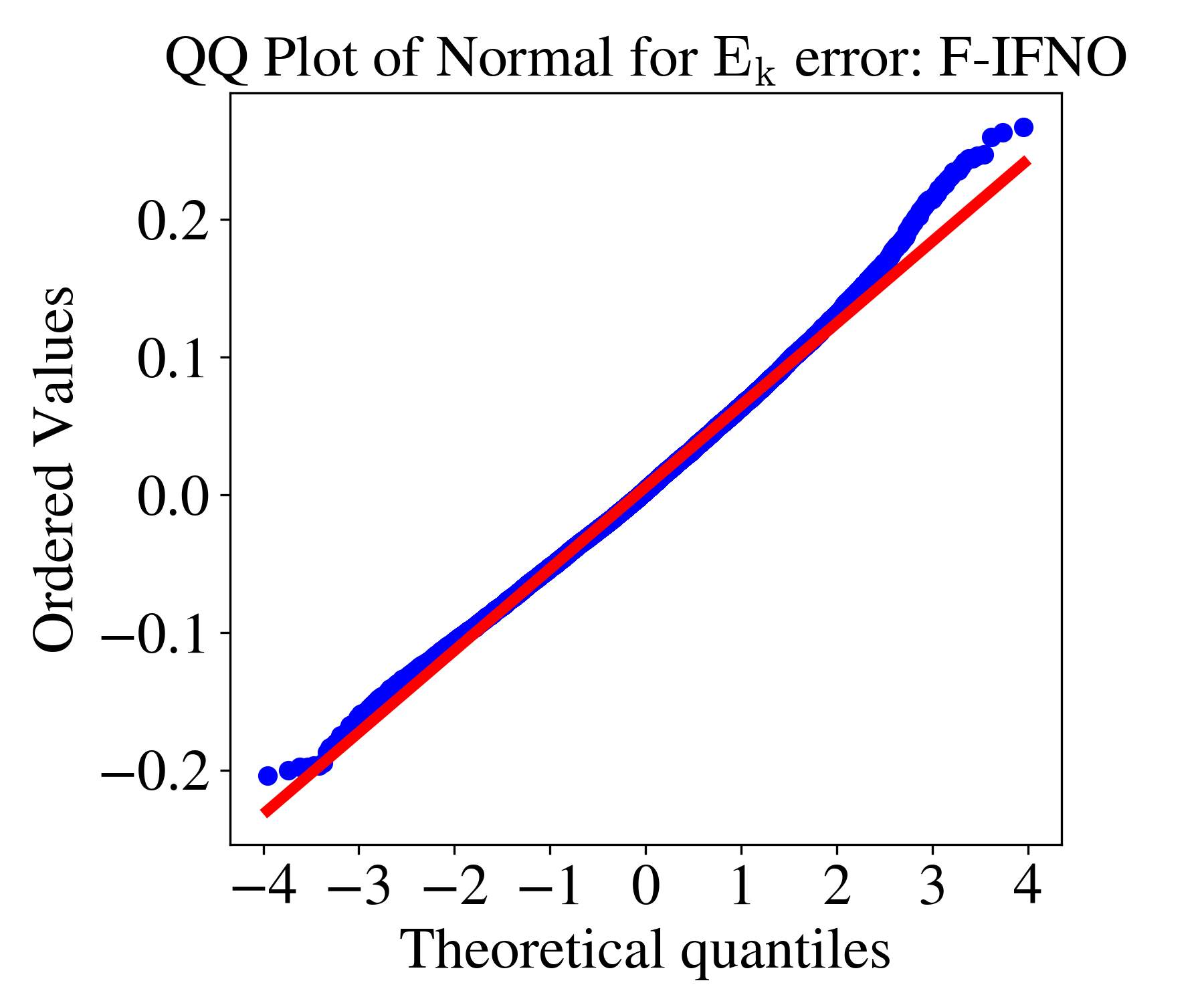}
            \put(-3,65){\small (a)}  
        \end{overpic}
    \end{subfigure}
    \hfill
    \begin{subfigure}[b]{0.32\textwidth}
        \begin{overpic}[width=1\linewidth]{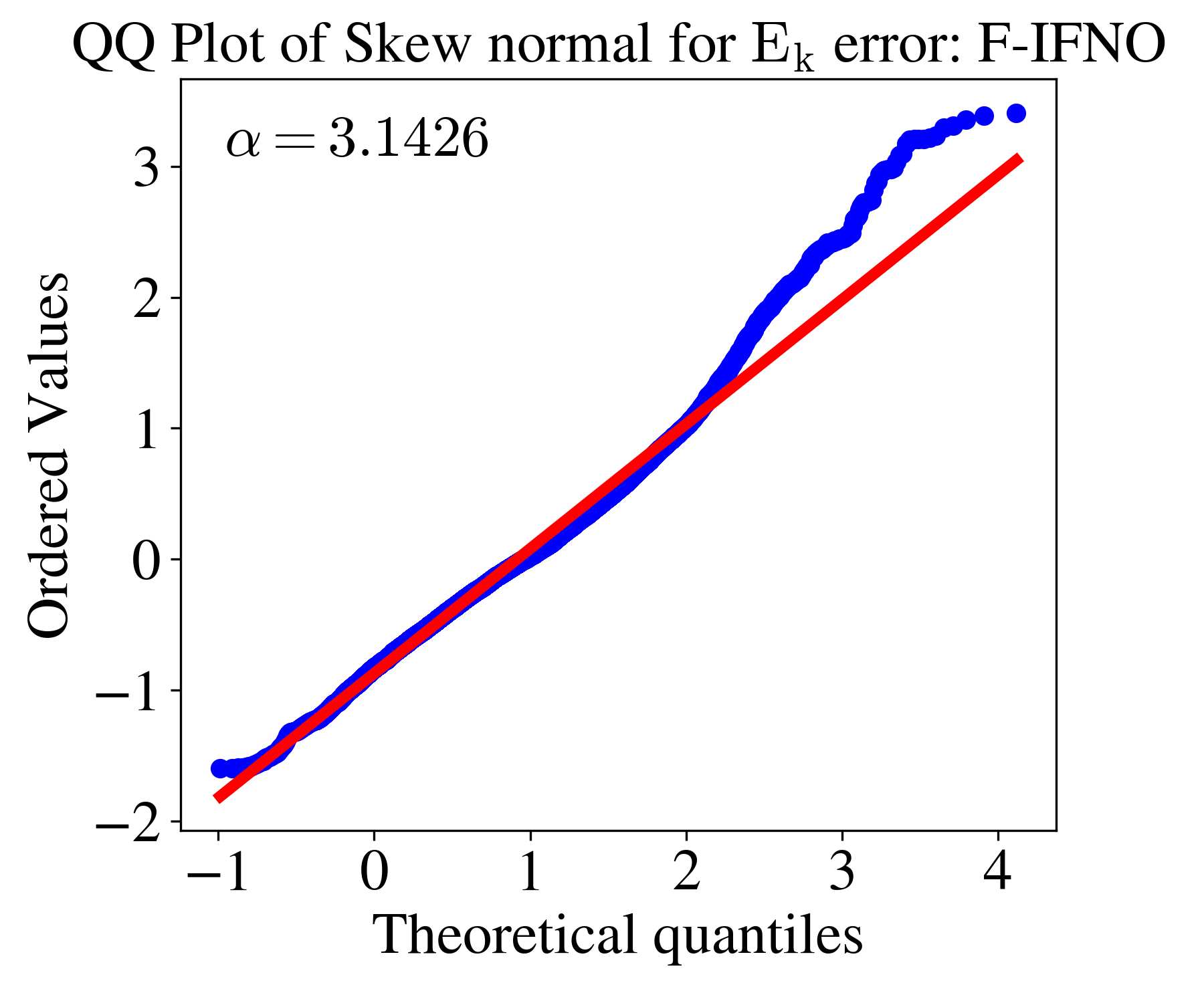}
            \put(-3,65){\small (b)} 
        \end{overpic} 
    \end{subfigure}
    \hfill
    \begin{subfigure}[b]{0.32\textwidth}
        \begin{overpic}[width=1\linewidth]{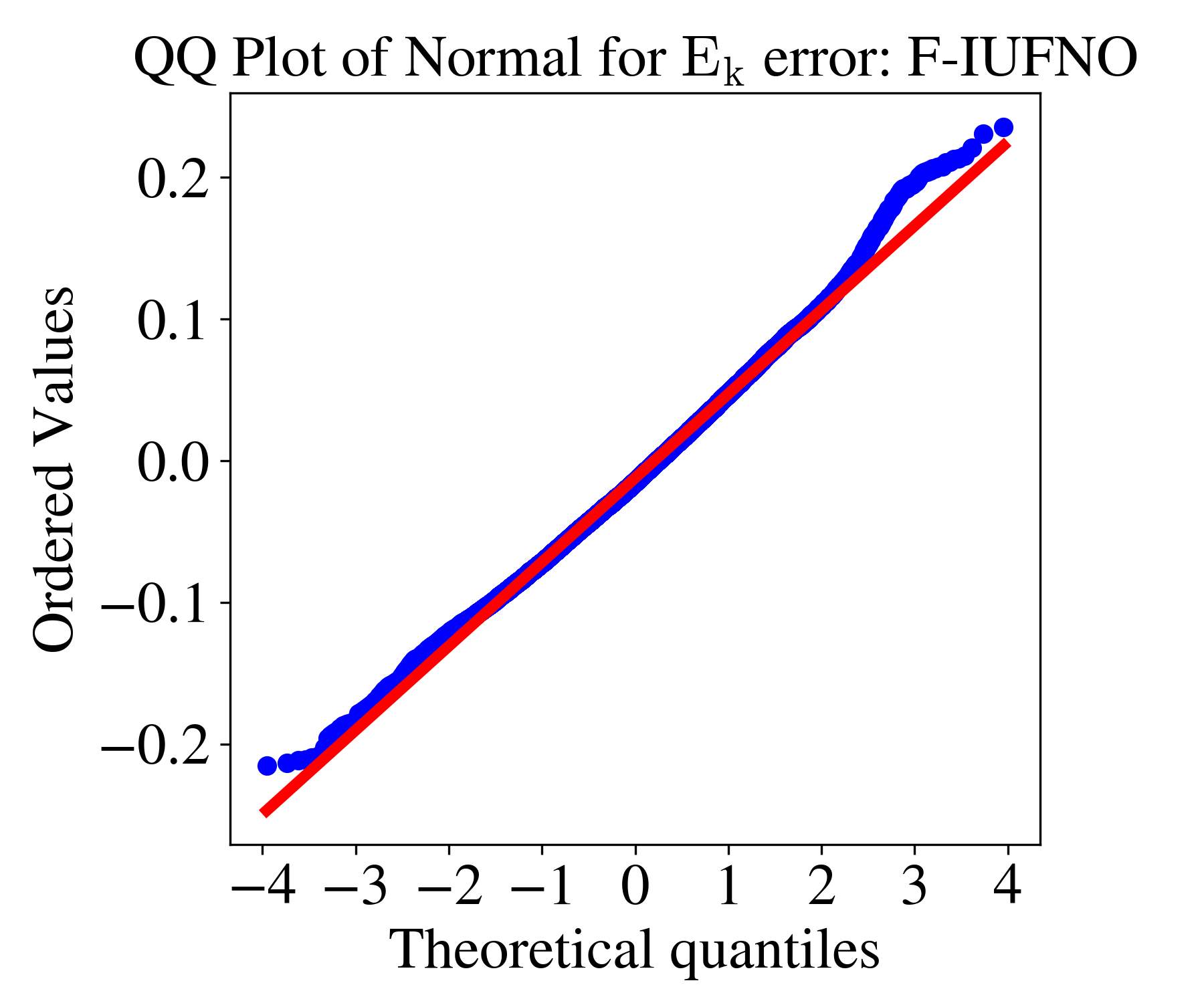}
            \put(-3,65){\small (c)} 
        \end{overpic}
    \end{subfigure}
    \vspace{0.1cm}

    \begin{subfigure}[b]{0.32\textwidth}
        \begin{overpic}[width=1\linewidth]{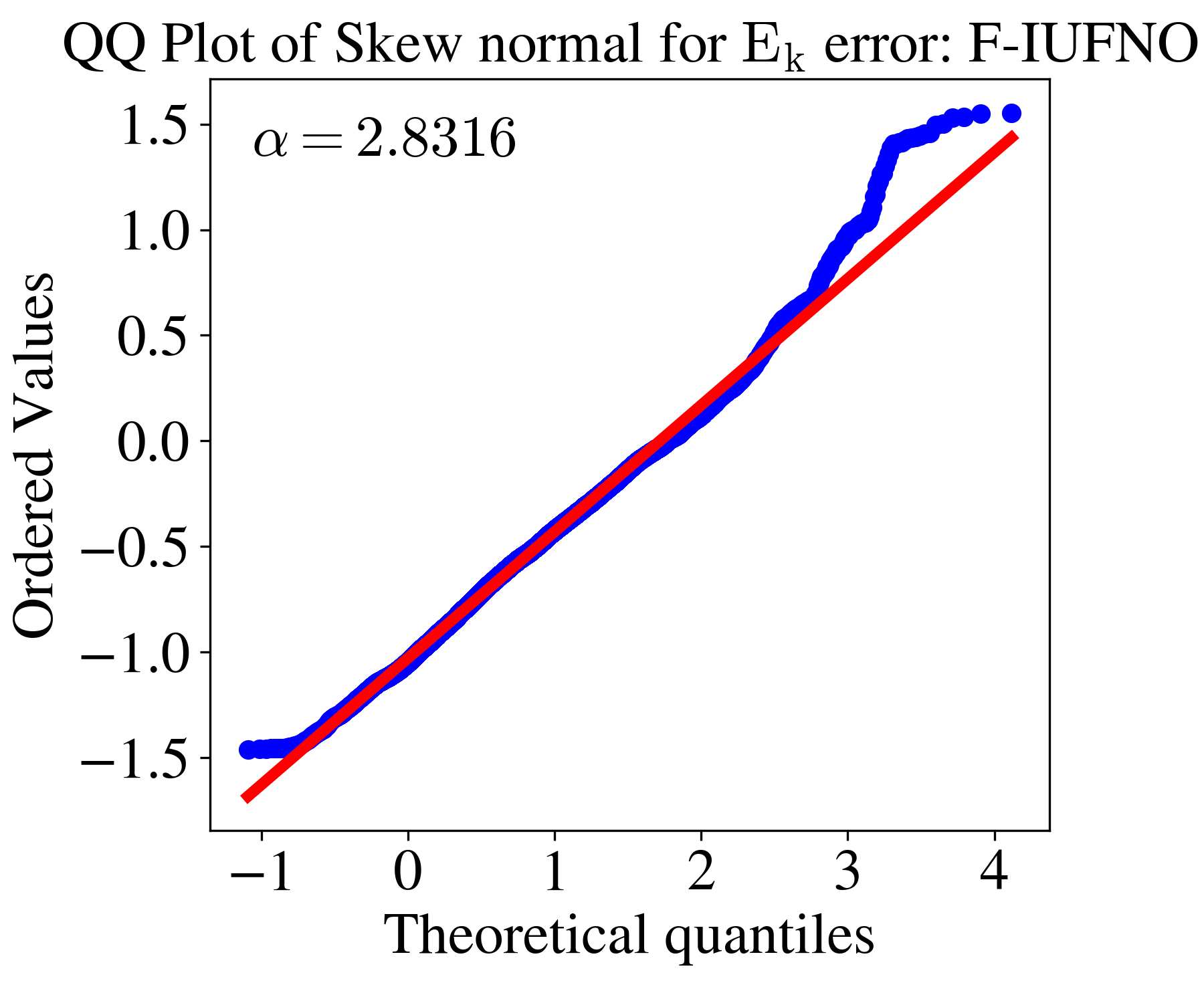}
            \put(-3,65){\small (d)}  
        \end{overpic}
    \end{subfigure}
    \hfill
    \begin{subfigure}[b]{0.32\textwidth}
        \begin{overpic}[width=1\linewidth]{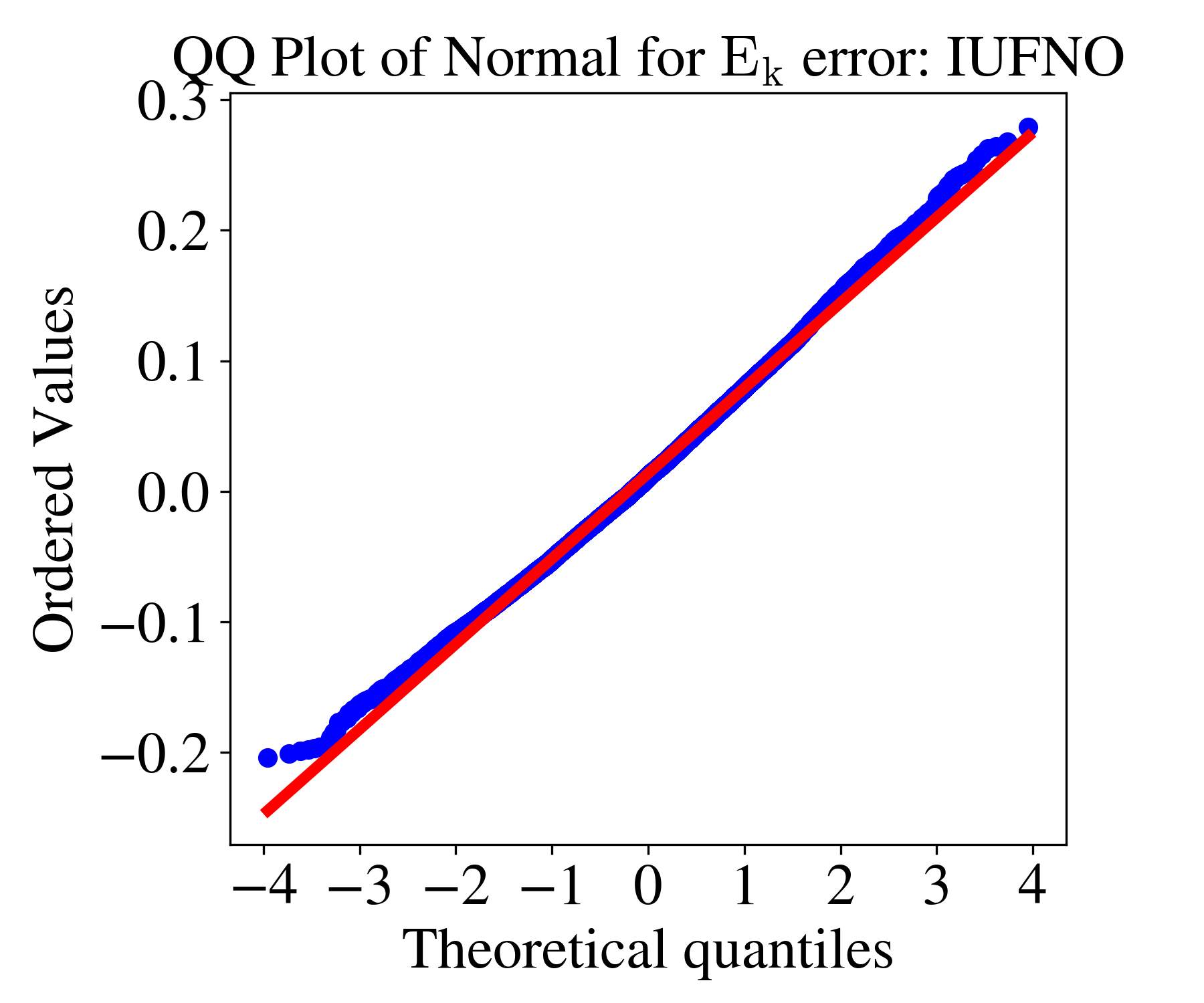}
            \put(-3,65){\small (e)} 
        \end{overpic} 
    \end{subfigure}
    \hfill
    \begin{subfigure}[b]{0.32\textwidth}
        \begin{overpic}[width=1\linewidth]{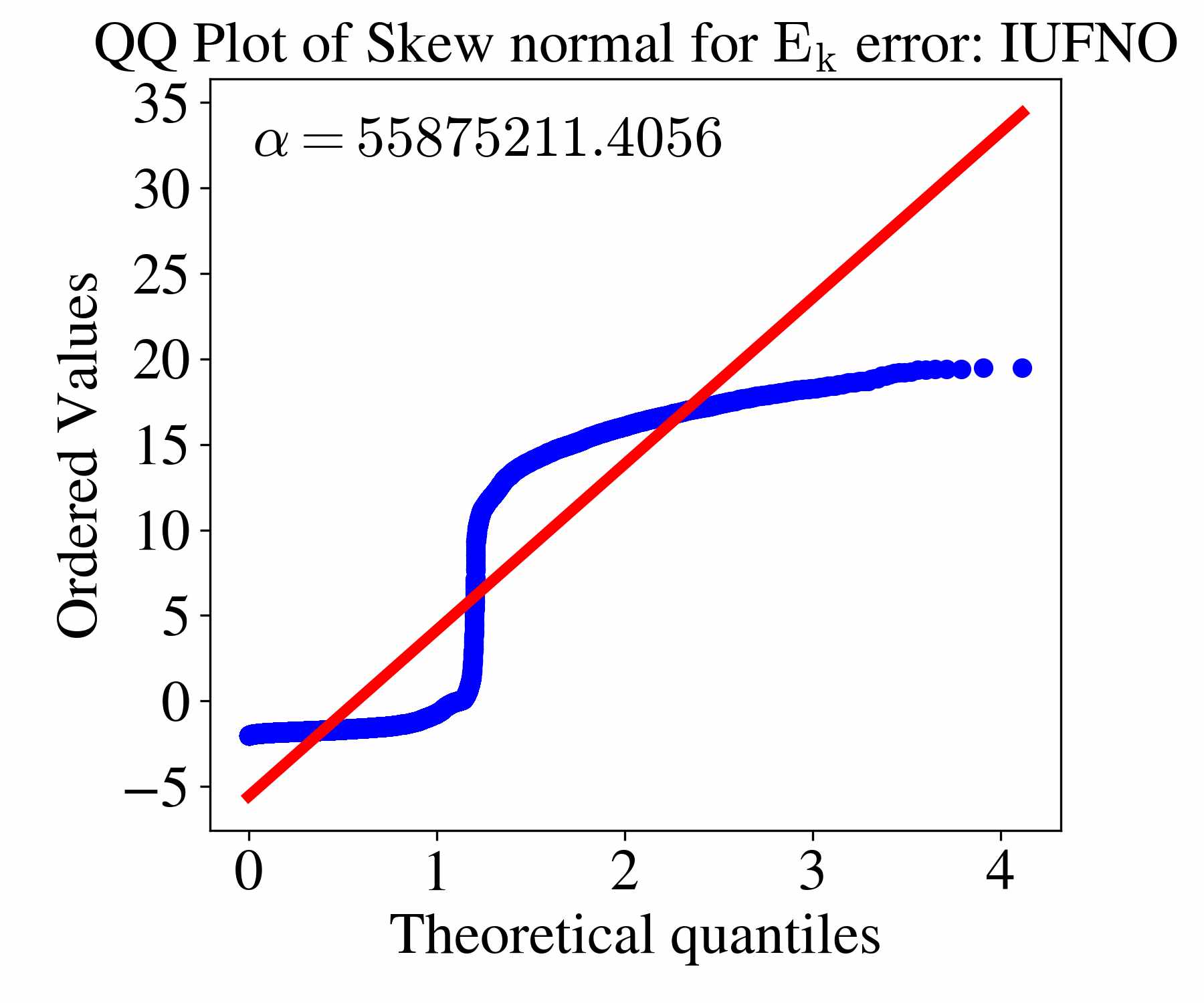}
            \put(-3,65){\small (f)} 
        \end{overpic}
    \end{subfigure}
    \vspace{0.1cm}

    \begin{subfigure}[b]{0.32\textwidth}
        \begin{overpic}[width=1\linewidth]{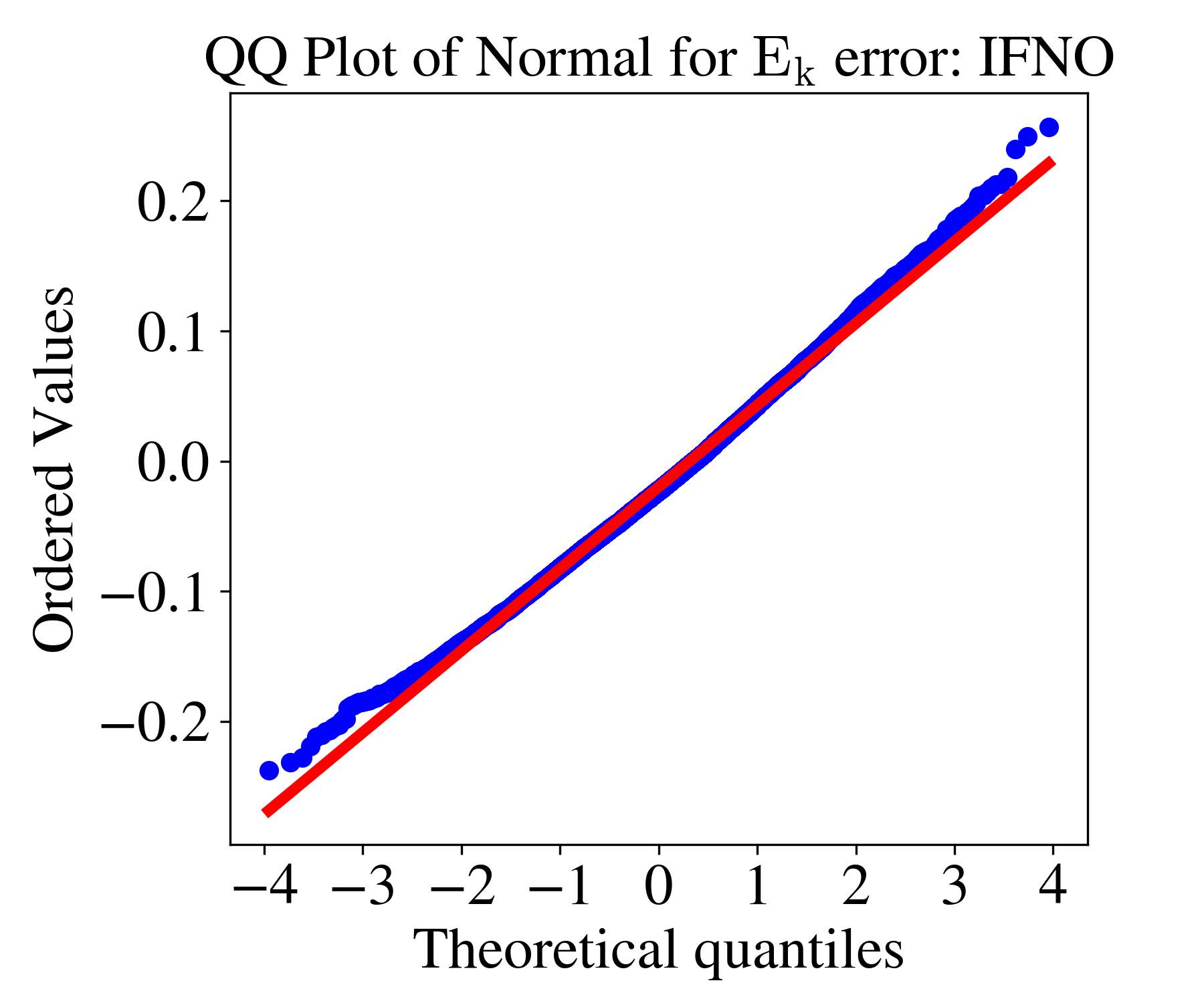}
            \put(-3,65){\small (g)}  
        \end{overpic}
    \end{subfigure}
    \hfill
    \begin{subfigure}[b]{0.32\textwidth}
        \begin{overpic}[width=1\linewidth]{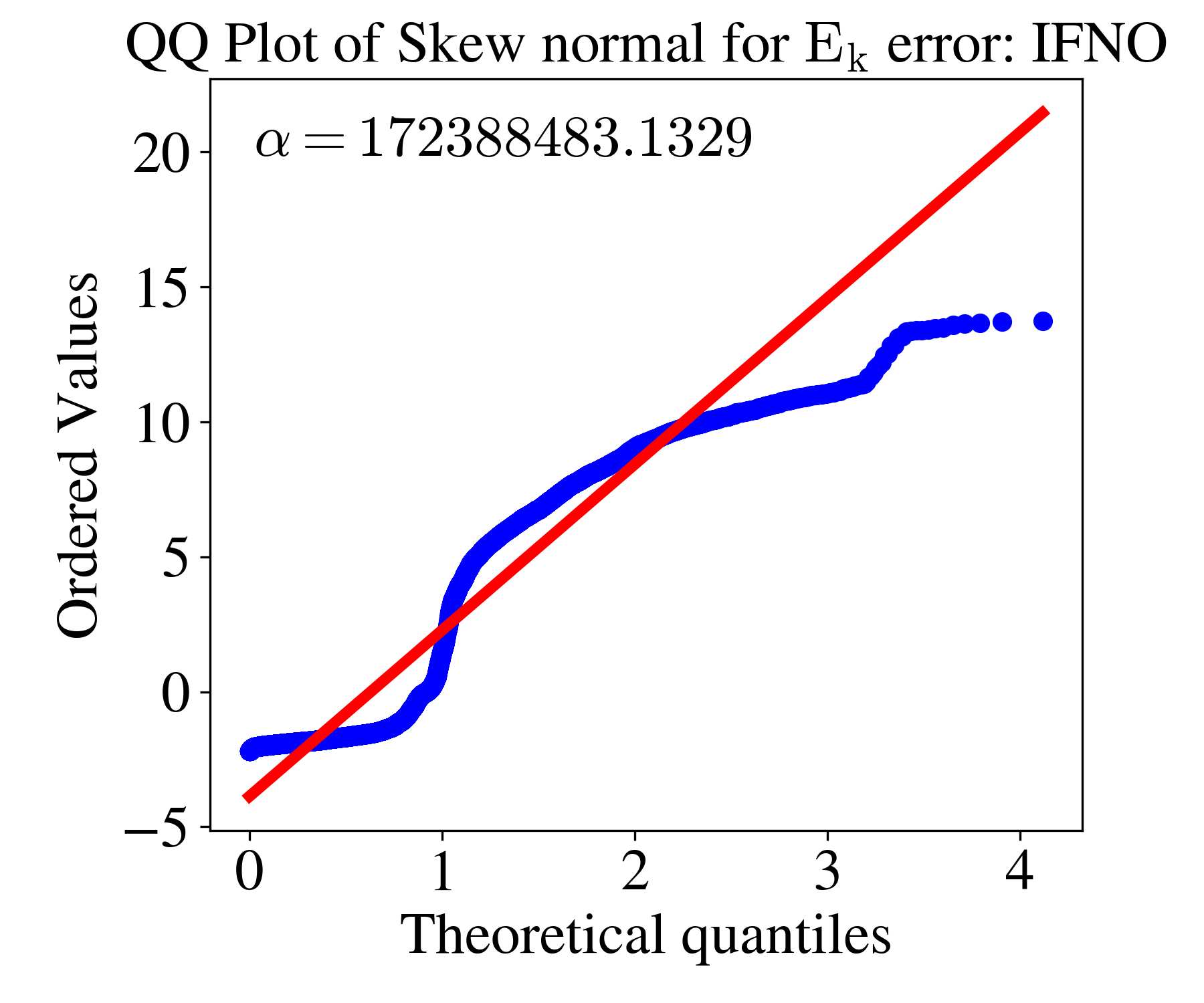}
            \put(-3,65){\small (h)} 
        \end{overpic} 
    \end{subfigure}
    \hfill
    \begin{subfigure}[b]{0.32\textwidth}
        \begin{overpic}[width=1\linewidth]{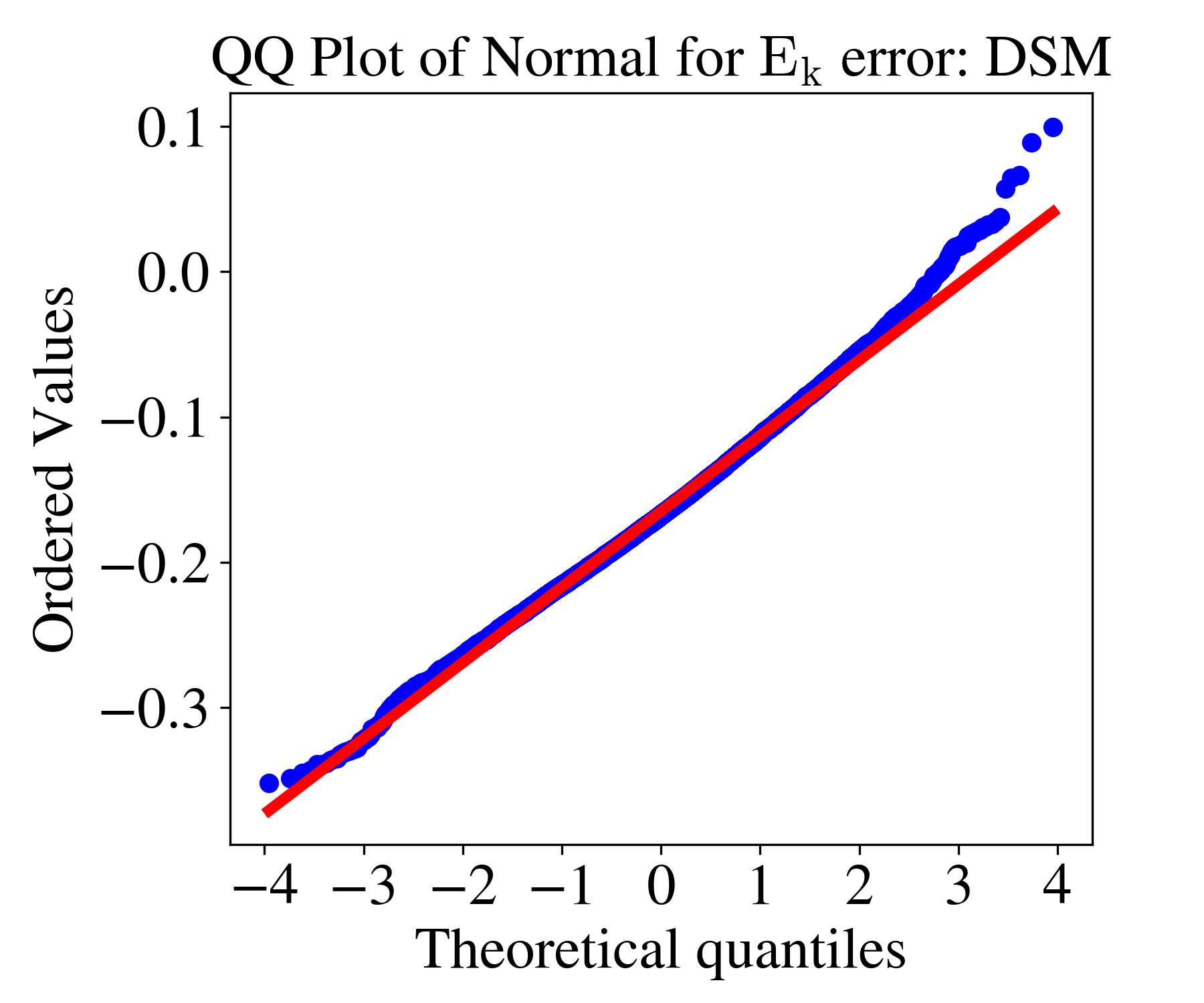}
            \put(-3,65){\small (i)} 
        \end{overpic}
    \end{subfigure}
    \vspace{0.1cm}

    \begin{subfigure}[b]{1\textwidth}
        \centering
        \begin{overpic}[width=0.32\linewidth]{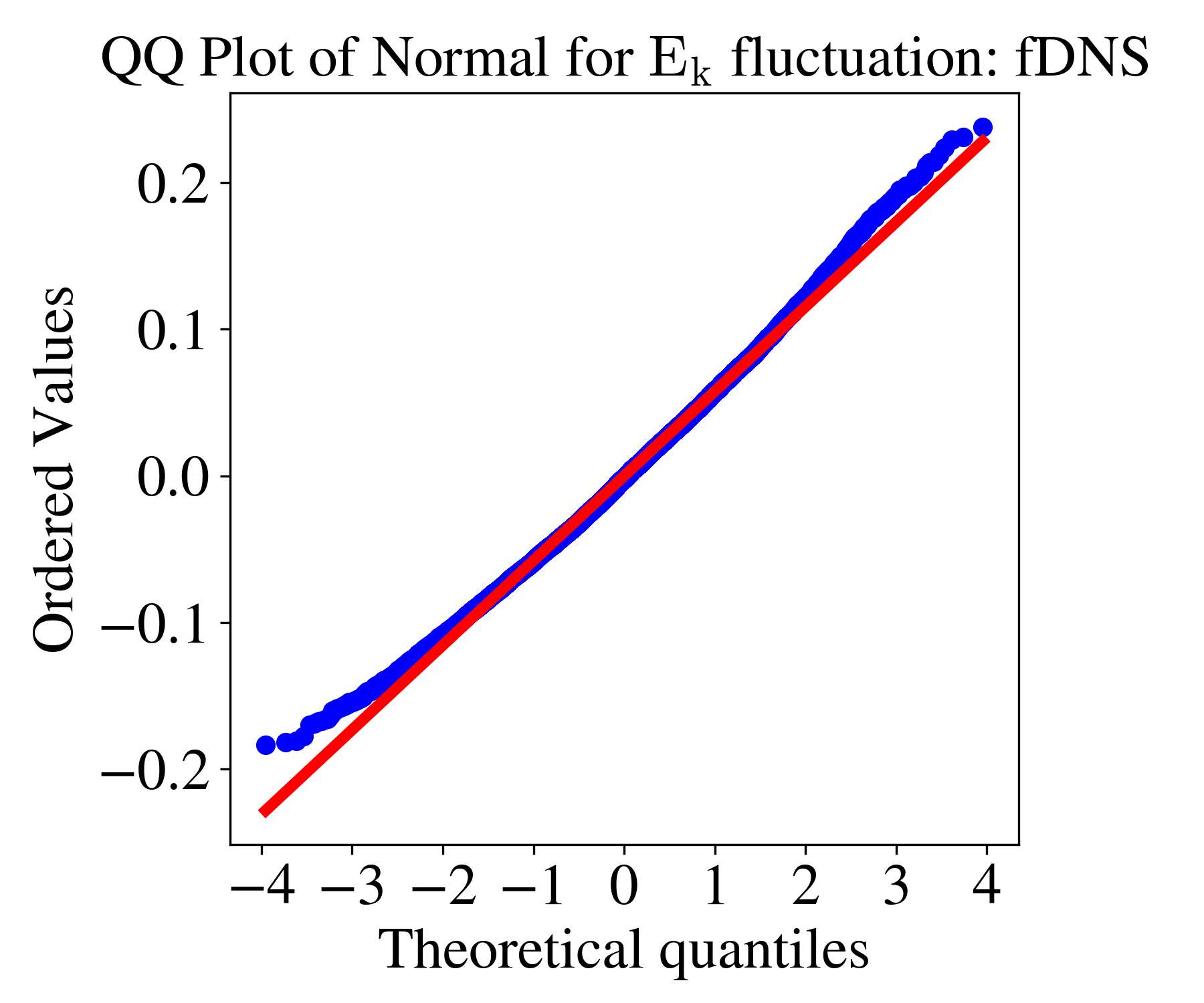}
            \put(-3,65){\small (j)}  
        \end{overpic}
    \end{subfigure}

	\caption{The QQ plots of kinetic energy $E_k$ errors for each method at the representative time interval $\Delta T = 0.2\tau$: (a) F-IFNO constrained; (b) F-IFNO unconstrained; (c) F-IUFNO constrained; (d) F-IUFNO unconstrained; (e) IUFNO constrained; (f) IUFNO unconstrained; (g) IFNO constrained; (h) IFNO unconstrained; (i) DSM; (j) fDNS. Note that for fDNS, the values represent natural statistical fluctuations over time, not prediction errors.}\label{fig:13}
\end{figure}

To further verify the correctness of the distribution fitting in Fig.~\ref{fig:12}, we also employ quantile-quantile (QQ) plots to evaluate the goodness of fit \cite{ProbabilityPlottingMethodsQQ,gnanadesikan1997methods,thode2002testing}. 
Fig.~\ref{fig:13} presents the QQ plots of kinetic energy $E_k$ errors for each method at the representative time interval $\Delta T = 0.2\tau$, corresponding to the error distributions shown in Fig.~\ref{fig:12}. 
In the QQ plots of Fig.~\ref{fig:13}, the blue dots represent the empirical quantiles of the error values, while the red line denotes the theoretical quantiles from the assumed distribution. If the blue dots align closely with the red line, it indicates that the assumed distribution provides a good fit to the data.
From Figs.~\ref{fig:13}(a), (c), (e), (g), (i), and (j), it can be observed that the constrained FNO-based models, as well as DSM and fDNS, have error or fluctuation distributions that conform well to the normal distribution. In contrast, for the unconstrained F-IFNO and F-IUFNO models, shown in Figs.~\ref{fig:13}(b) and (d), the error distributions exhibit skewness and match well with the skew normal distribution. However, for the unconstrained IUFNO and IFNO models, shown in Figs.~\ref{fig:13}(f) and (h), the QQ plots indicate that their error distributions cannot be satisfactorily described even by a skew normal distribution.
These observations provide strong support that the distribution assumptions made in Fig.~\ref{fig:12} are valid.

\begin{figure}[ht!]
    \centering
    \begin{subfigure}[b]{0.32\textwidth}
        \begin{overpic}[width=1\linewidth]{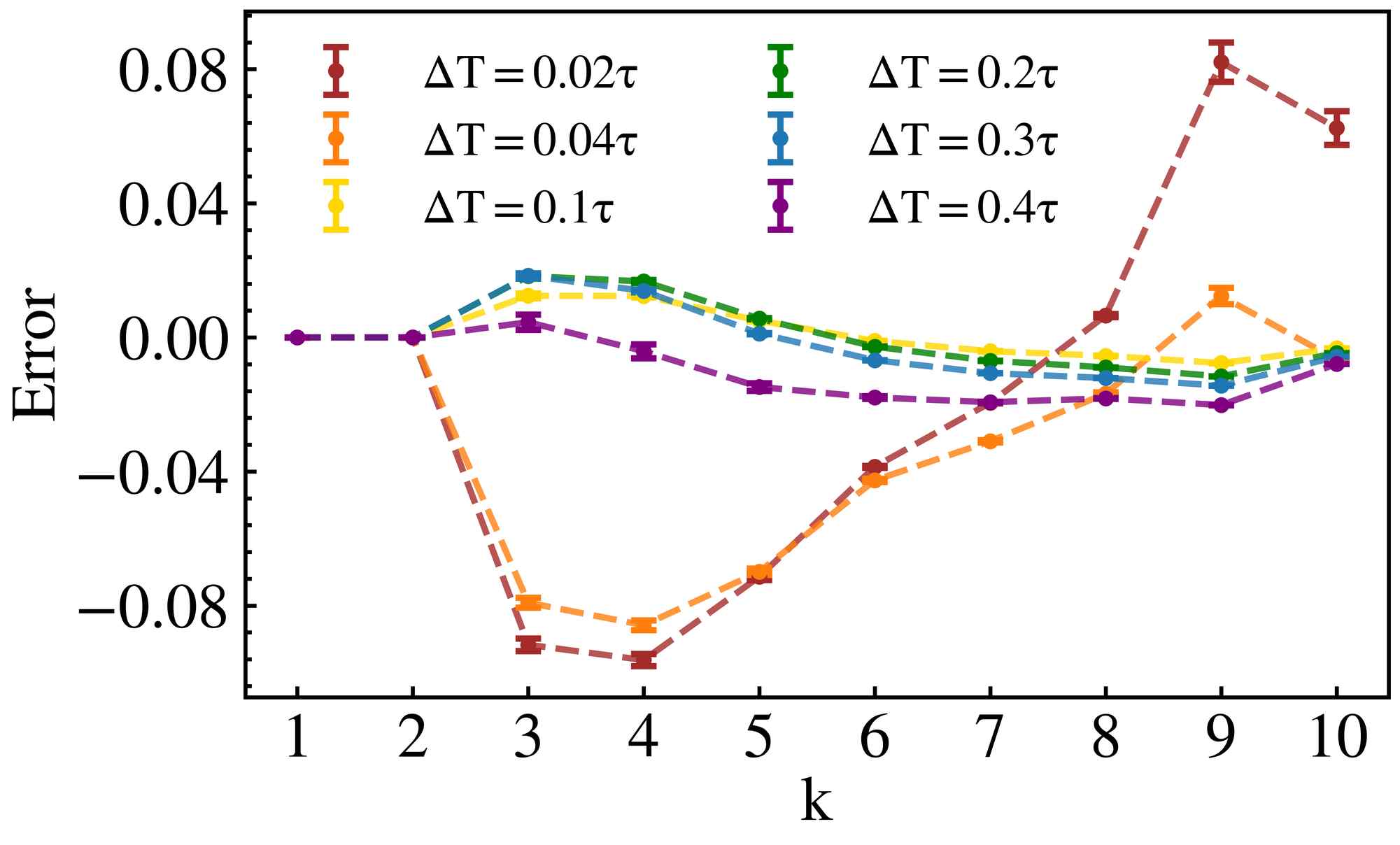}
            \put(-2,55){\small (a)}  
        \end{overpic}
    \end{subfigure}
    \hfill
    \begin{subfigure}[b]{0.32\textwidth}
        \begin{overpic}[width=1\linewidth]{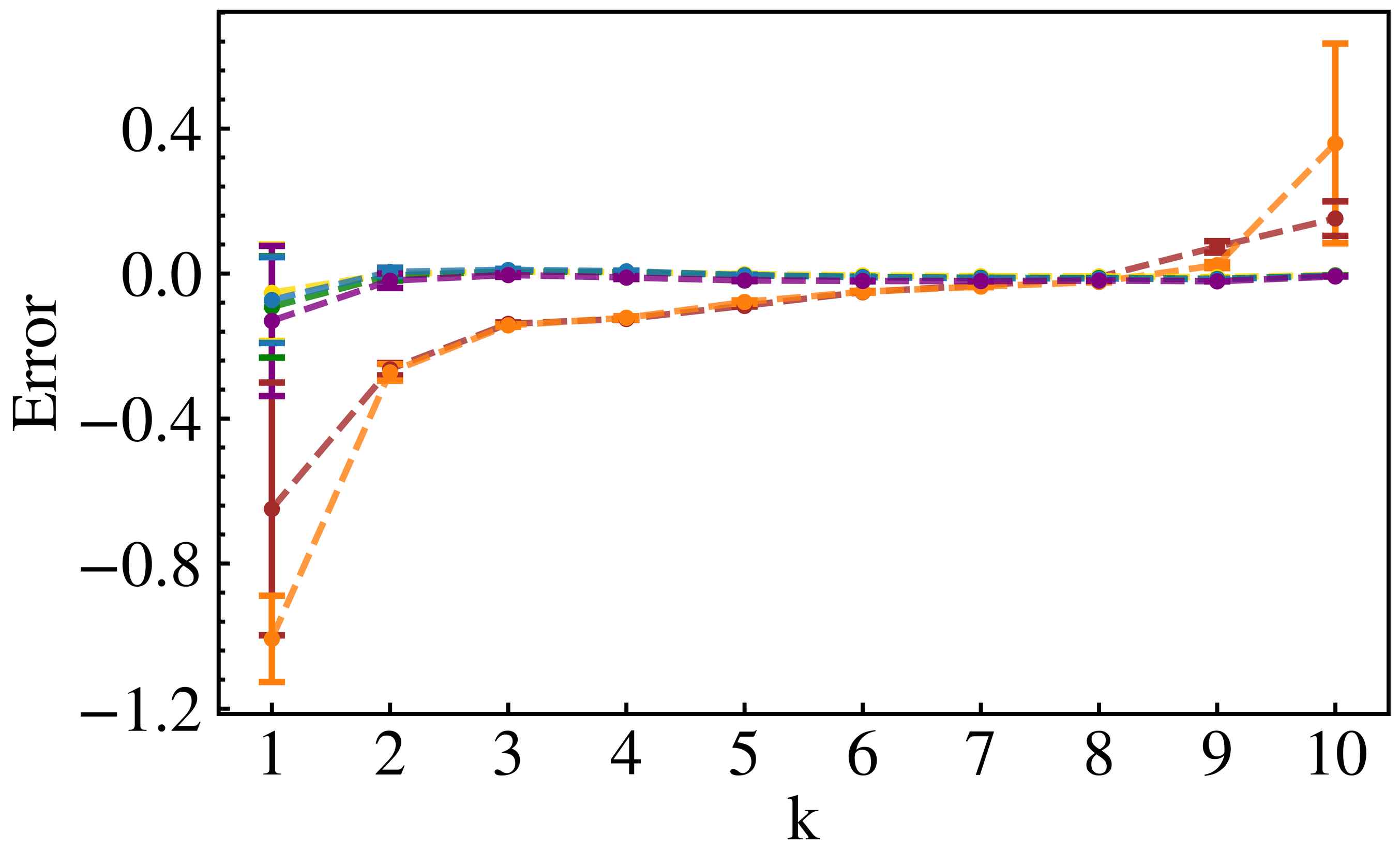}
            \put(-2,55){\small (b)} 
        \end{overpic} 
    \end{subfigure}
    \hfill
    \begin{subfigure}[b]{0.32\textwidth}
        \begin{overpic}[width=1\linewidth]{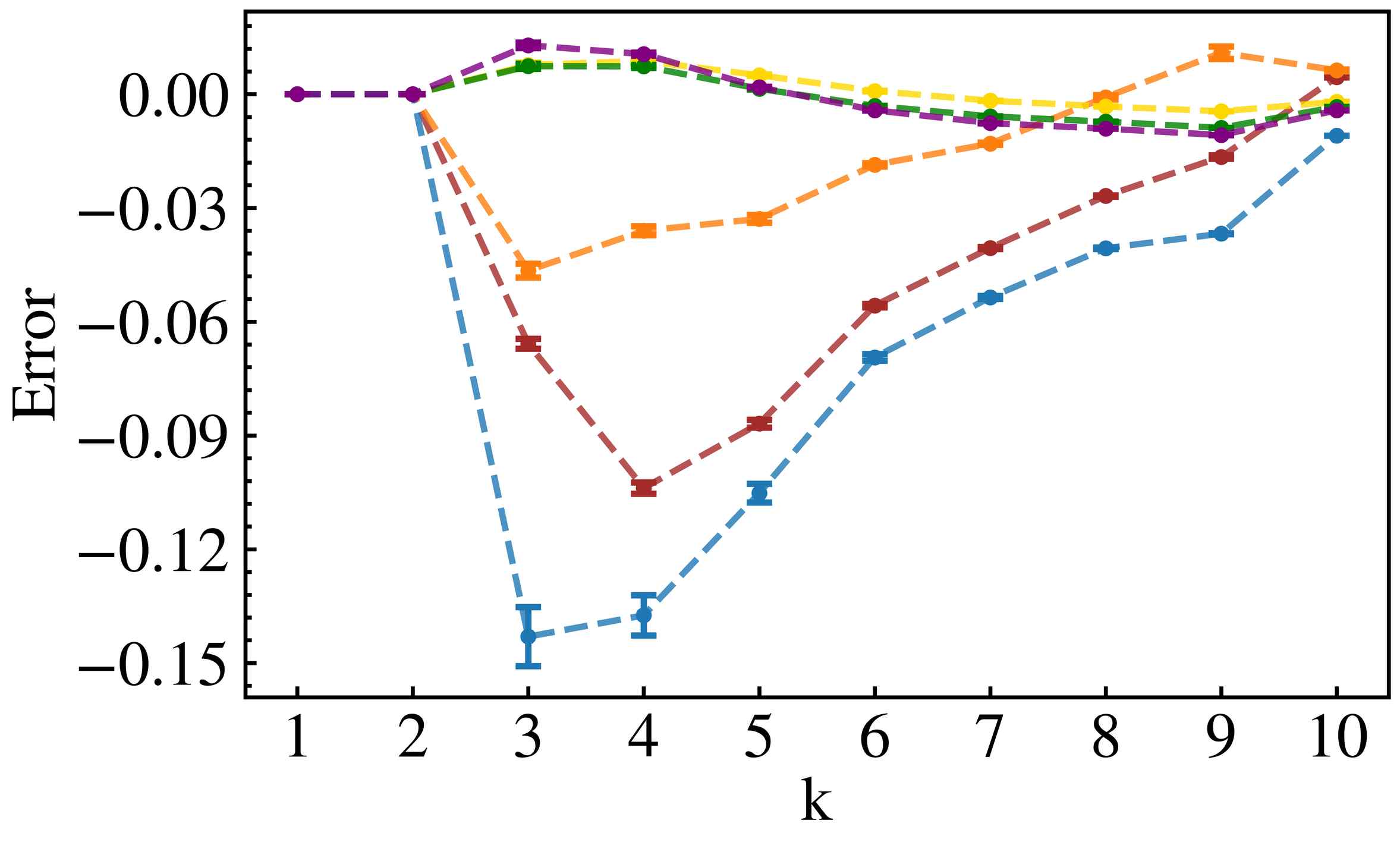}
            \put(-2,55){\small (c)} 
        \end{overpic}
    \end{subfigure}
    \vspace{0.1cm}

    \begin{subfigure}[b]{0.32\textwidth}
        \begin{overpic}[width=1\linewidth]{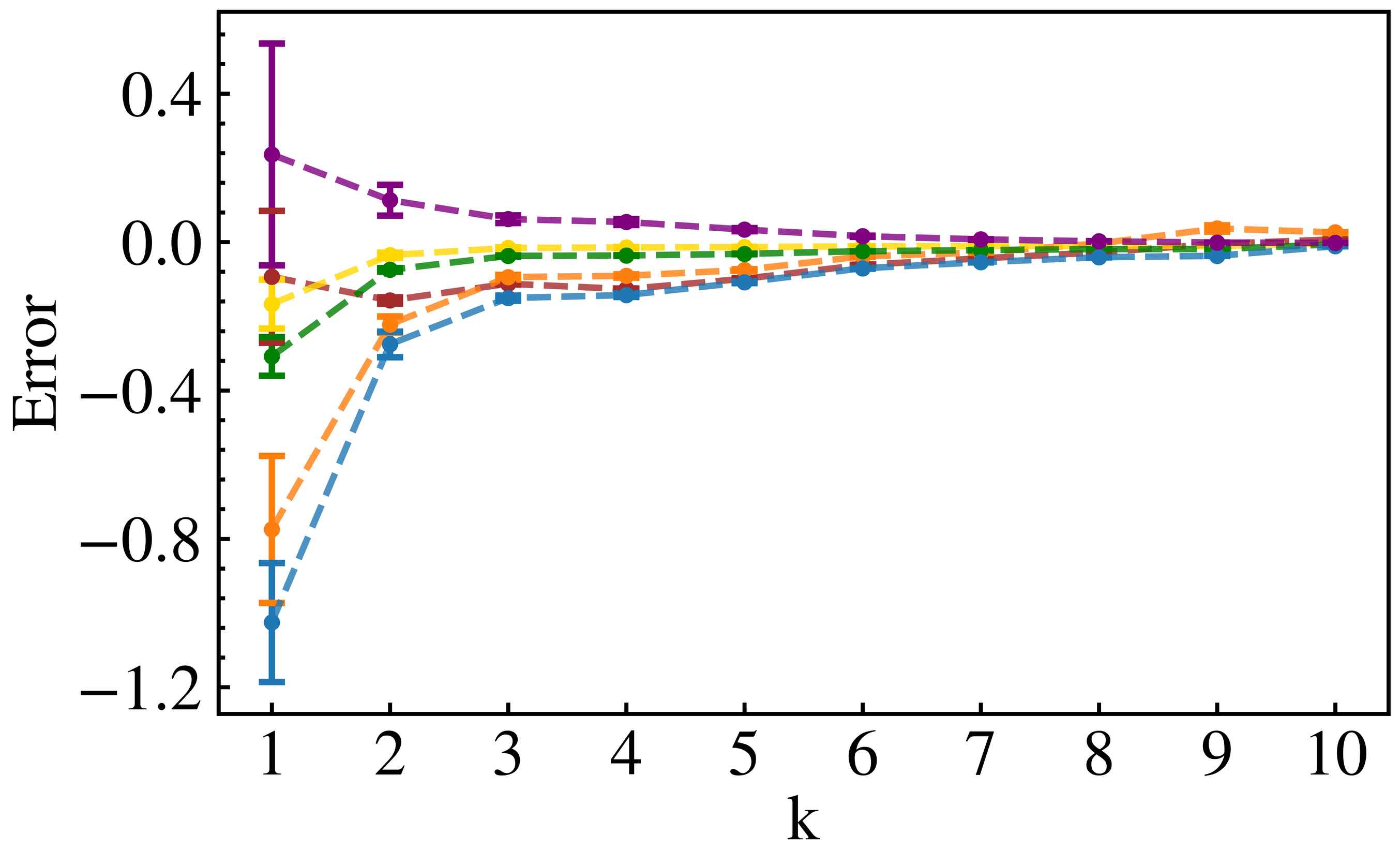}
            \put(-2,55){\small (d)}  
        \end{overpic}
    \end{subfigure}
    \hfill
    \begin{subfigure}[b]{0.32\textwidth}
        \begin{overpic}[width=1\linewidth]{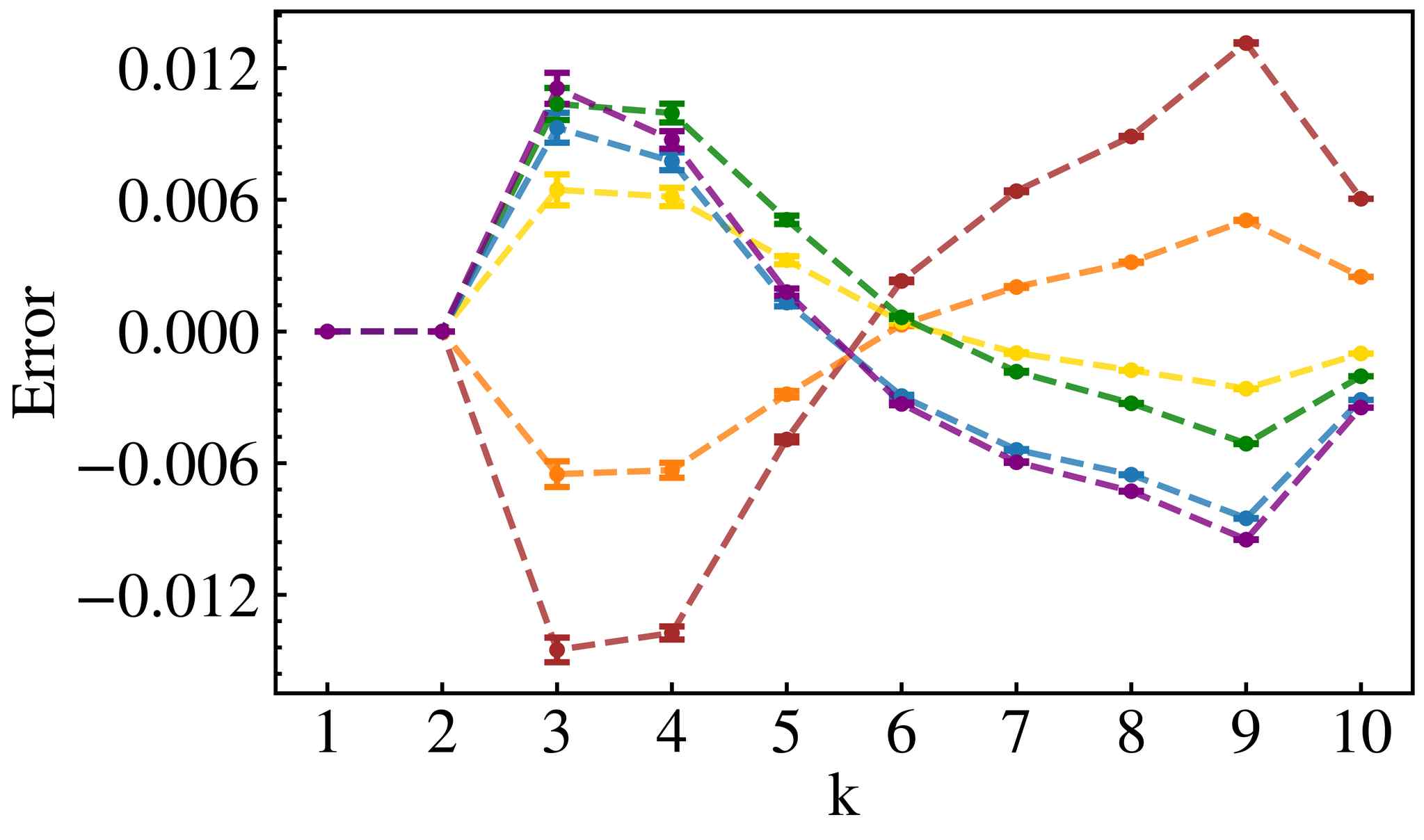}
            \put(-2,55){\small (e)} 
        \end{overpic} 
    \end{subfigure}
    \hfill
    \begin{subfigure}[b]{0.32\textwidth}
        \begin{overpic}[width=1\linewidth]{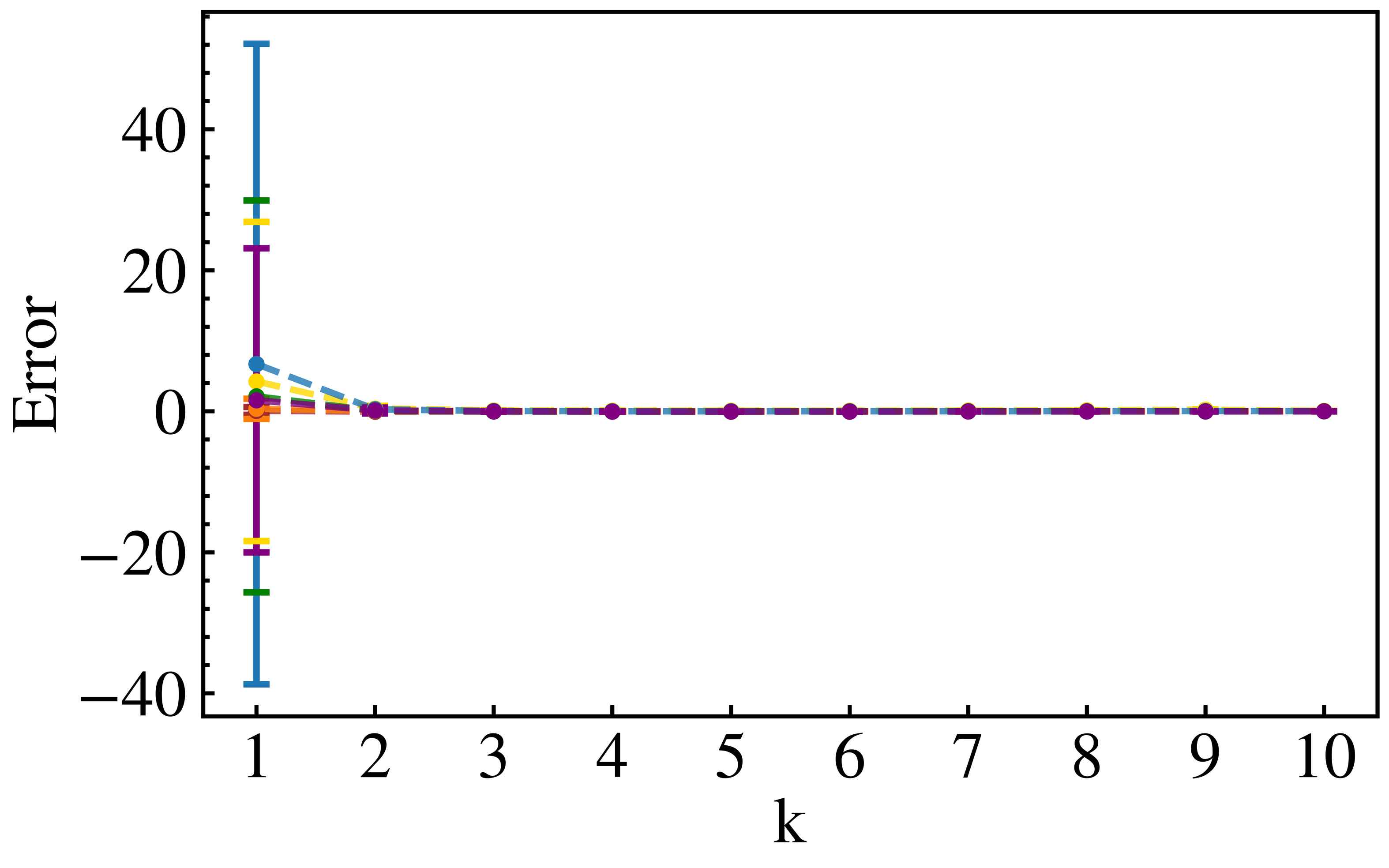}
            \put(-2,55){\small (f)} 
        \end{overpic}
    \end{subfigure}
    \vspace{0.1cm}

    \begin{subfigure}[b]{0.32\textwidth}
        \begin{overpic}[width=1\linewidth]{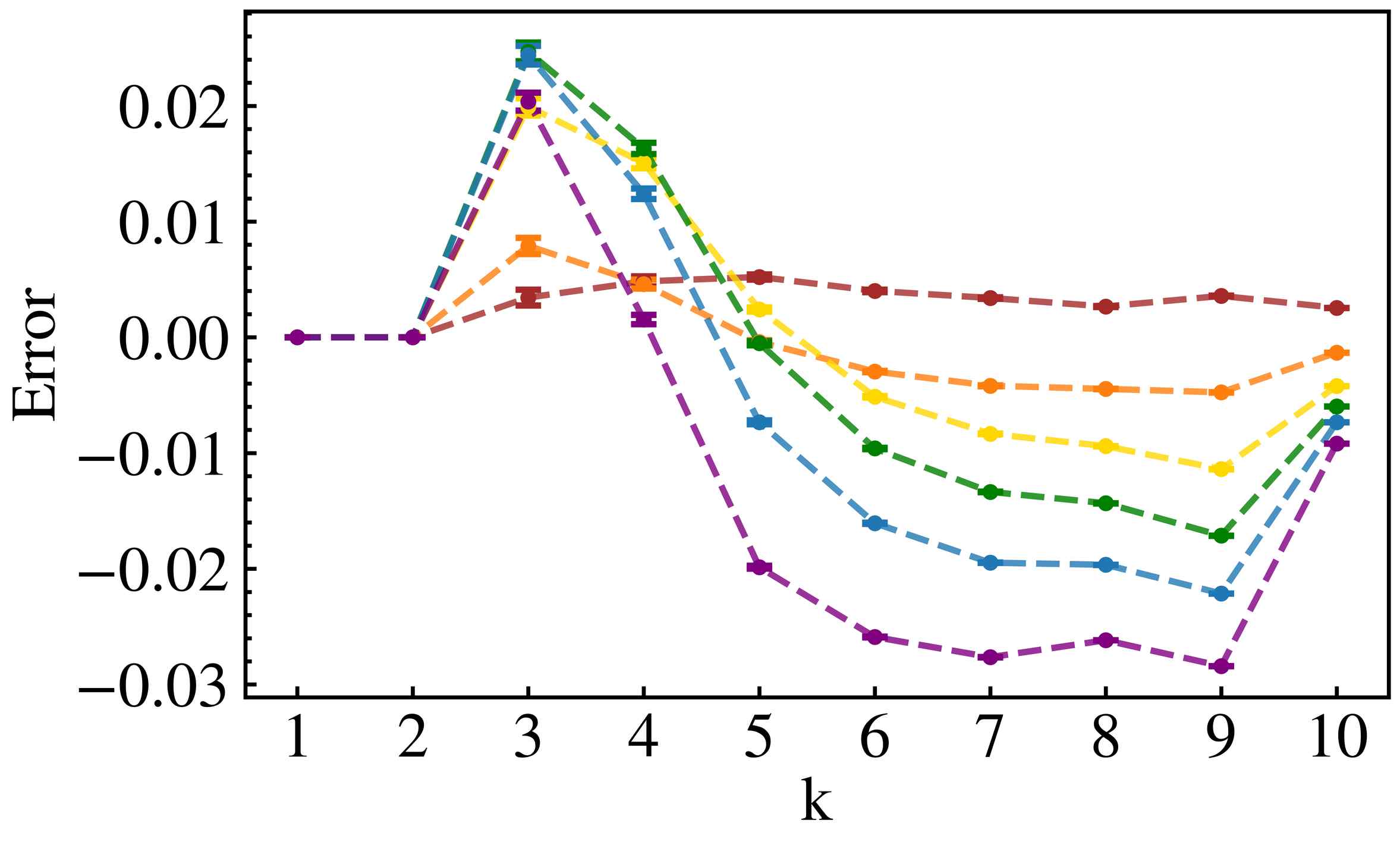}
            \put(-2,55){\small (g)}  
        \end{overpic}
    \end{subfigure}
    \hfill
    \begin{subfigure}[b]{0.32\textwidth}
        \begin{overpic}[width=1\linewidth]{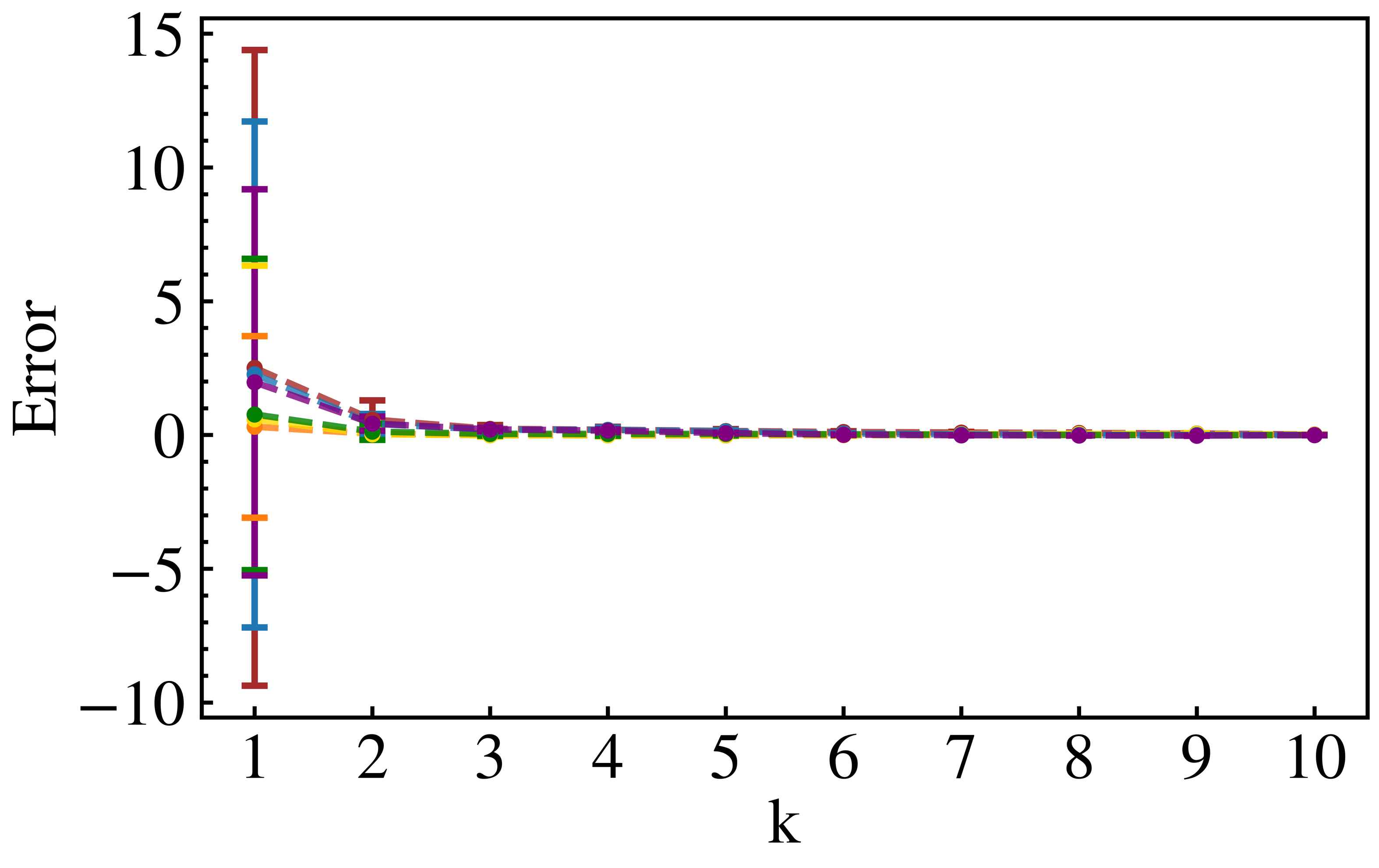}
            \put(-2,55){\small (h)} 
        \end{overpic} 
    \end{subfigure}
    \hfill
    \begin{subfigure}[b]{0.32\textwidth}
        \begin{overpic}[width=1\linewidth]{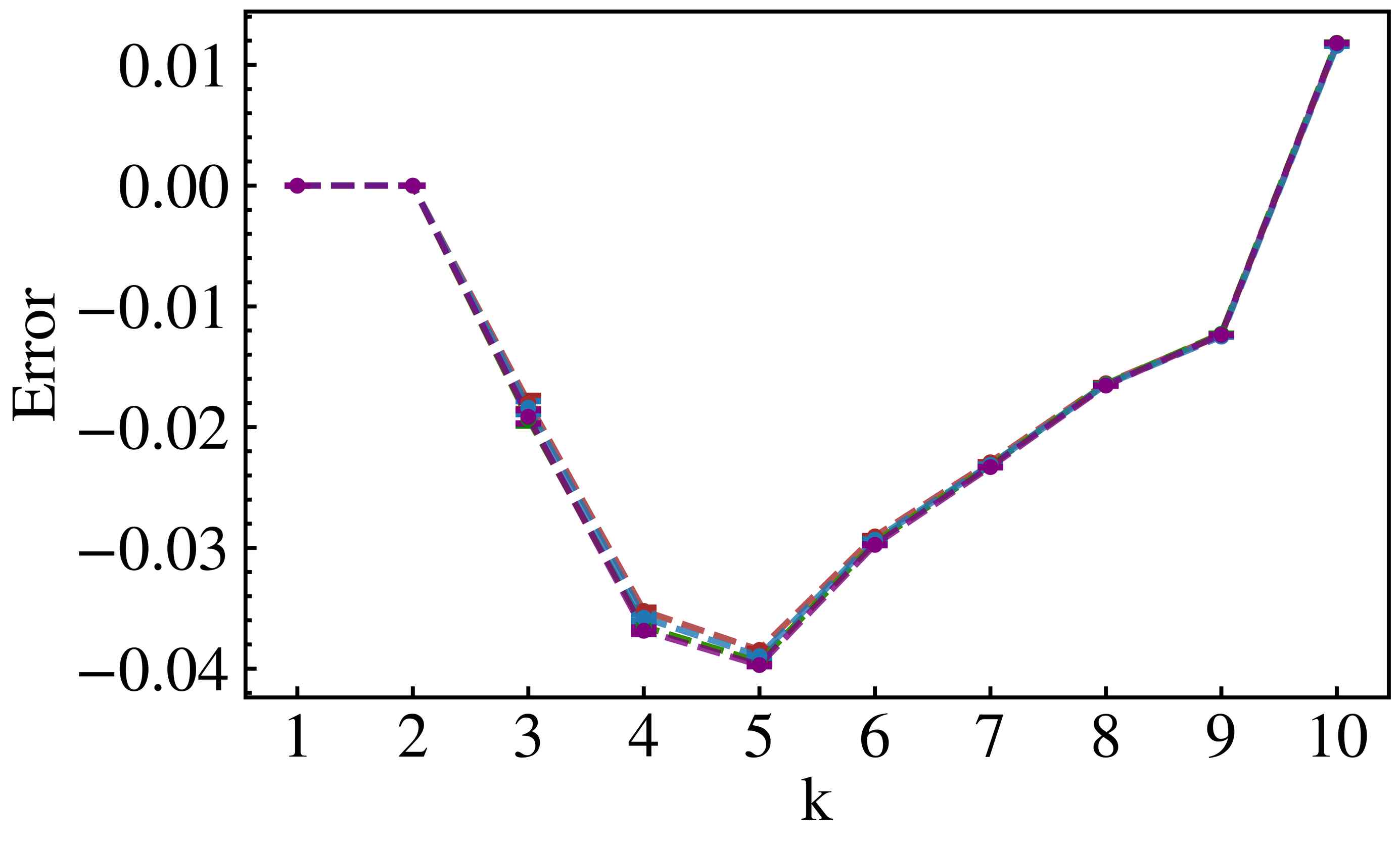}
            \put(-2,55){\small (i)} 
        \end{overpic}
    \end{subfigure}
    \vspace{0.1cm}

    \begin{subfigure}[b]{1\textwidth}
        \centering
        \begin{overpic}[width=0.32\linewidth]{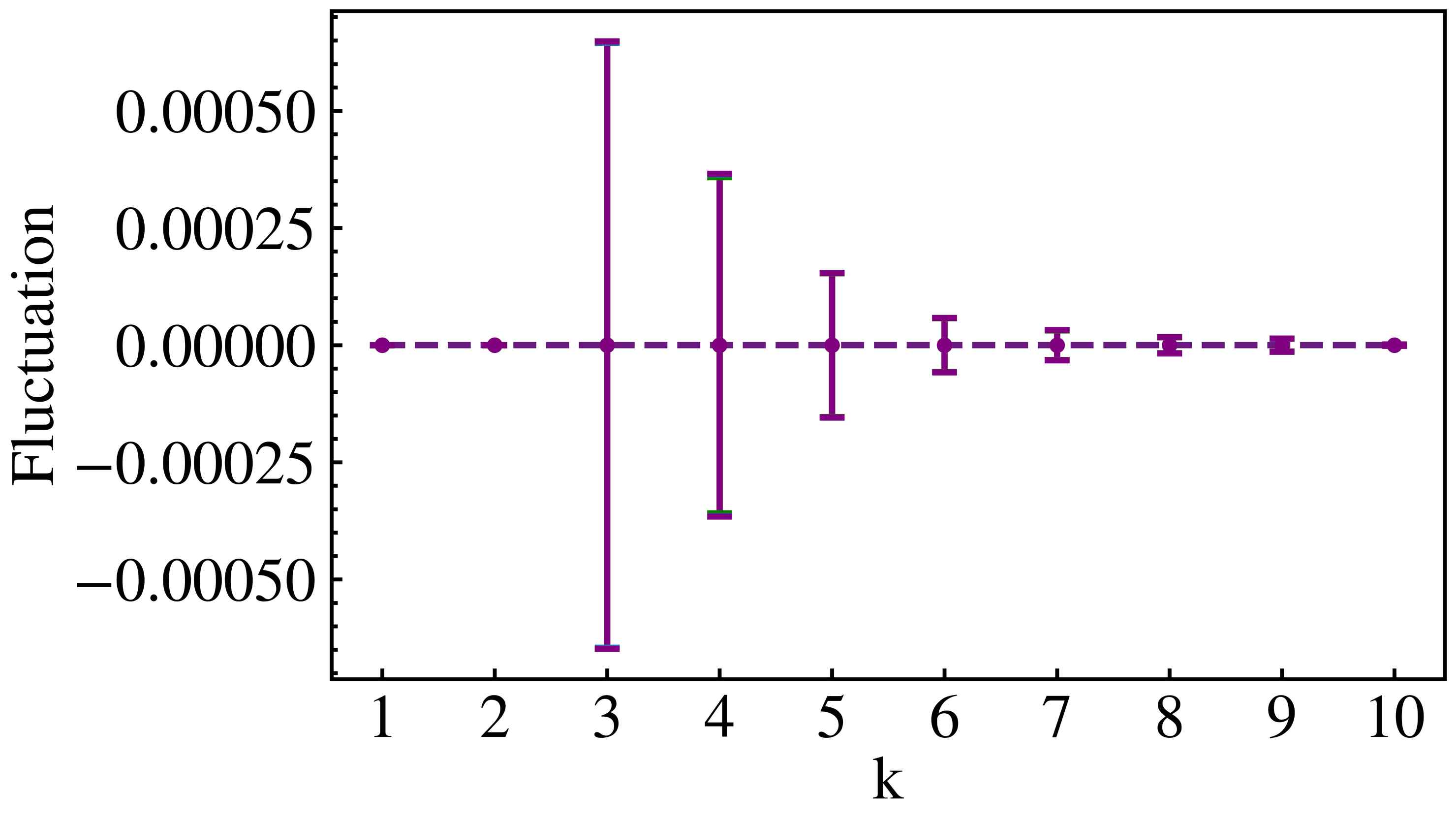}
            \put(-2,55){\small (j)}  
        \end{overpic}
    \end{subfigure}

	\caption{Errorbars of velocity spectra $E(k)$ for various methods with different time interval $\Delta T$: (a) F-IFNO constrained; (b) F-IFNO unconstrained; (c) F-IUFNO constrained; (d) F-IUFNO unconstrained; (e) IUFNO constrained; (f) IUFNO unconstrained; (g) IFNO constrained; (h) IFNO unconstrained; (i) DSM; (j) fDNS. Note that for fDNS, the values represent natural statistical fluctuations over time, not prediction errors.}\label{fig:14}
\end{figure}

Fig.~\ref{fig:14} illustrates the errorbars of velocity spectra $E(k)$ for various methods under different time intervals $\Delta T$. As expected, the errorbars for DSM and fDNS remain unchanged across different $\Delta T$ values, while those for FNO-based models vary significantly with $\Delta T$. For fDNS, the mean value remains centered around zero, and the standard deviation decreases as the wavenumber $k$ increases from 3 to 10, indicating higher stability in smaller-scale components. In the case of DSM, the mean error exhibits a concave profile with respect to $k$, reaching a minimum at $k=5$.
For unconstrained FNO-based models, both the mean and standard deviation of errors tend to decrease as $k$ increases, suggesting that the dominant error contributions originate from the large scales (i.e., low-$k$ modes). When constraints are applied specifically to the large-scale components ($k=1,2$), the error is significantly reduced compared to the unconstrained versions.
Figs.~\ref{fig:14}(a) and (c) show that, for constrained F-IFNO and F-IUFNO at $\Delta T=0.1\tau$ and $0.2\tau$ where both models achieve reliable predictions (the errors are smallest among all the $\Delta T$), the mean error initially increases from zero at $k=3$, then drops below zero, and finally rises again to a value near zero as $k$ increases. A similar trend is observed in constrained IUFNO within $\Delta T \in [0.04\tau, 0.2\tau]$, which is also within its reliable prediction range. In contrast, for constrained IFNO at $\Delta T = 0.02\tau$ and $0.04\tau$ (where the model achieves its most accurate predictions with the smallest errors), the mean error initially decreases from zero at $k=3$, then rises to a positive value, and finally declines again toward zero as $k$ increases.
These observations demonstrate that the imposed constraints are particularly effective at reducing errors in large-scale components and that different FNO-based models exhibit distinct error distribution behaviors across spatial scales depending on the time interval $\Delta T$.

\begin{figure}[ht!]
    \centering
    \begin{subfigure}[b]{0.49\textwidth}
        \begin{overpic}[width=1\linewidth]{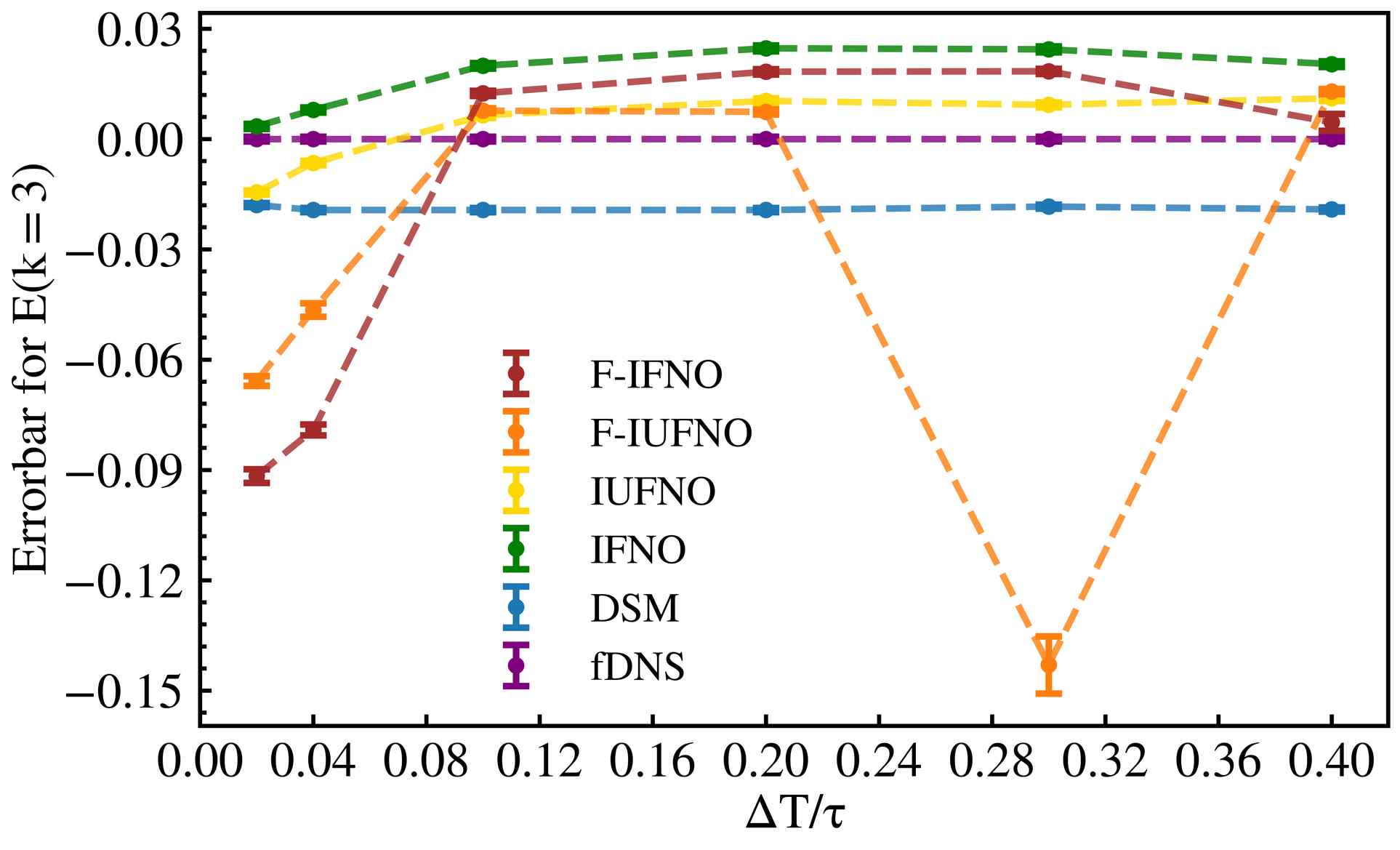}
            \put(0,55){\small (a)}  
        \end{overpic}
    \end{subfigure}
    \hfill
    \begin{subfigure}[b]{0.49\textwidth}
        \begin{overpic}[width=1\linewidth]{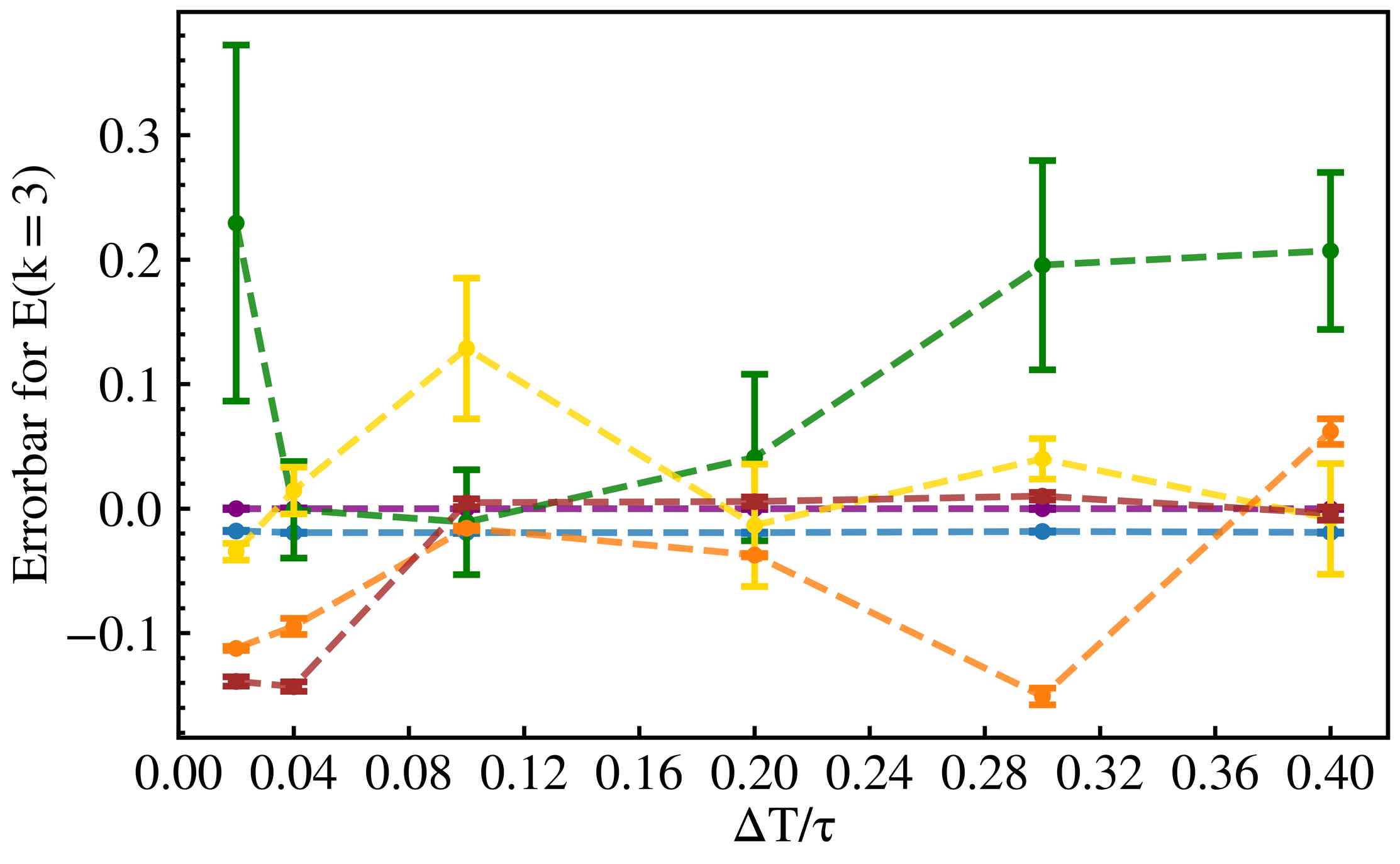}
            \put(0,55){\small (b)} 
        \end{overpic} 
    \end{subfigure}
    \vspace{0.1cm}

    \begin{subfigure}[b]{0.49\textwidth}
        \begin{overpic}[width=1\linewidth]{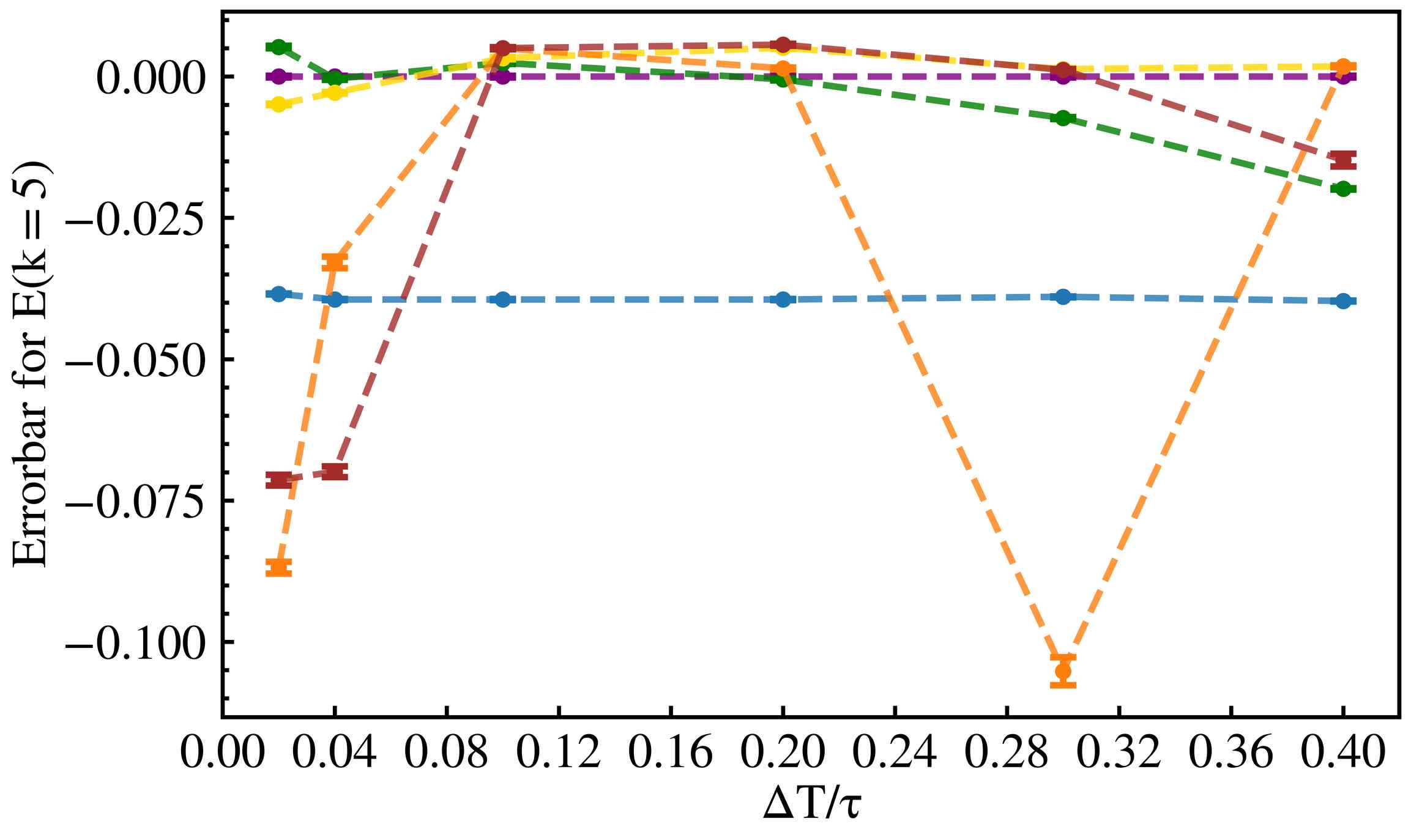}
            \put(0,55){\small (c)}  
        \end{overpic}
    \end{subfigure}
    \hfill
    \begin{subfigure}[b]{0.49\textwidth}
        \begin{overpic}[width=1\linewidth]{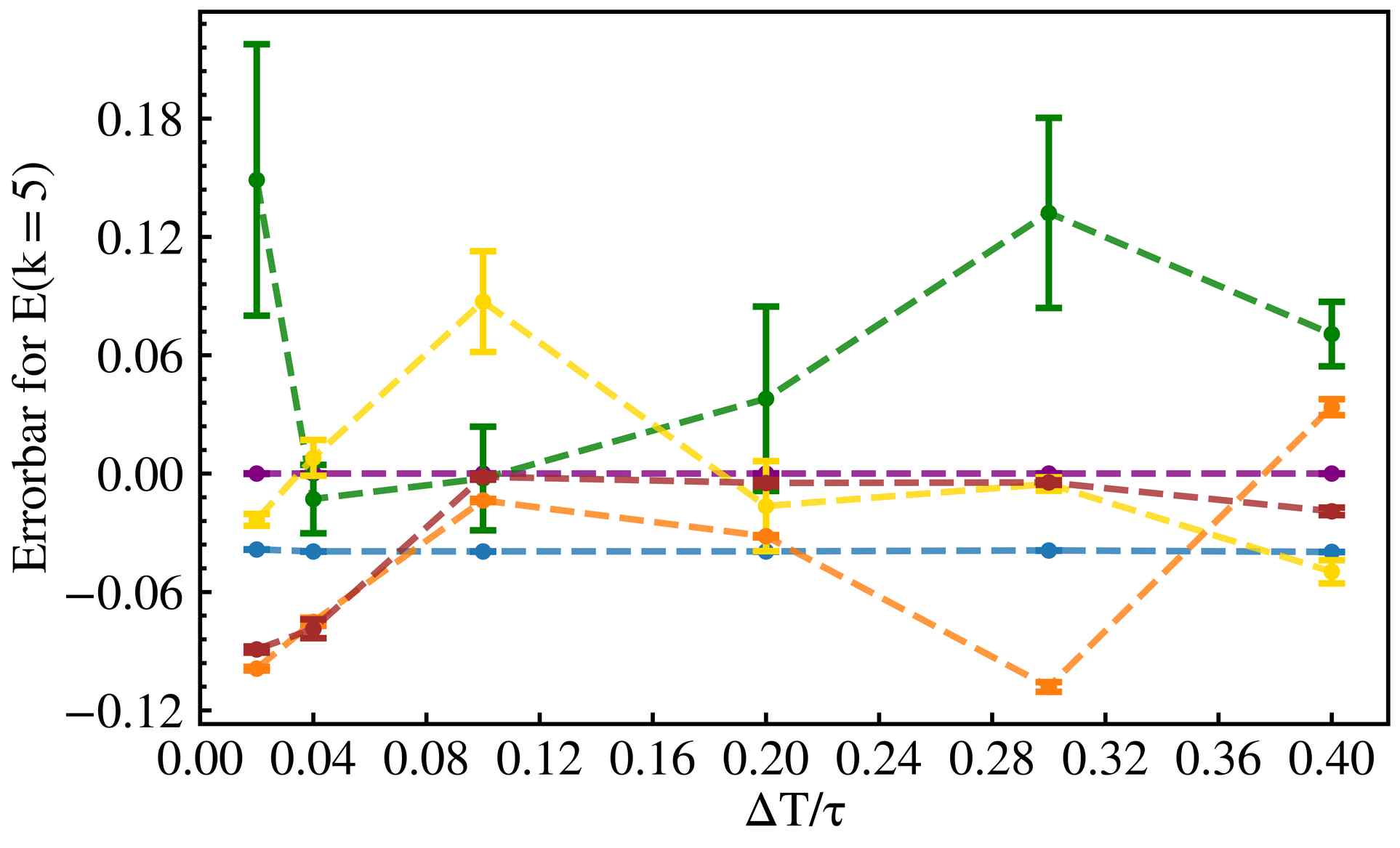}
            \put(0,55){\small (d)} 
        \end{overpic} 
    \end{subfigure}
    \vspace{0.1cm}

    \begin{subfigure}[b]{0.49\textwidth}
        \begin{overpic}[width=1\linewidth]{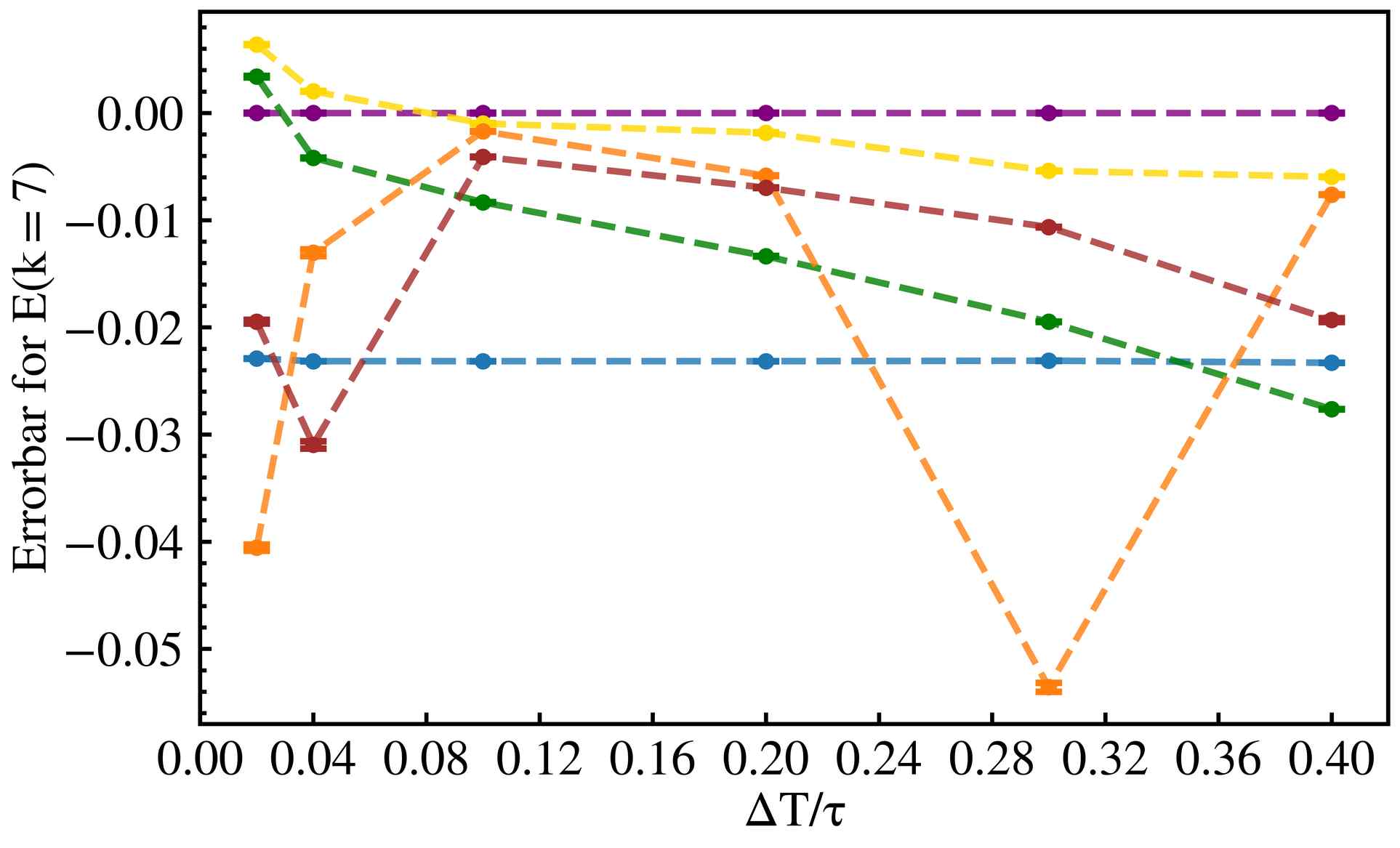}
            \put(0,55){\small (e)}  
        \end{overpic}
    \end{subfigure}
    \hfill
    \begin{subfigure}[b]{0.49\textwidth}
        \begin{overpic}[width=1\linewidth]{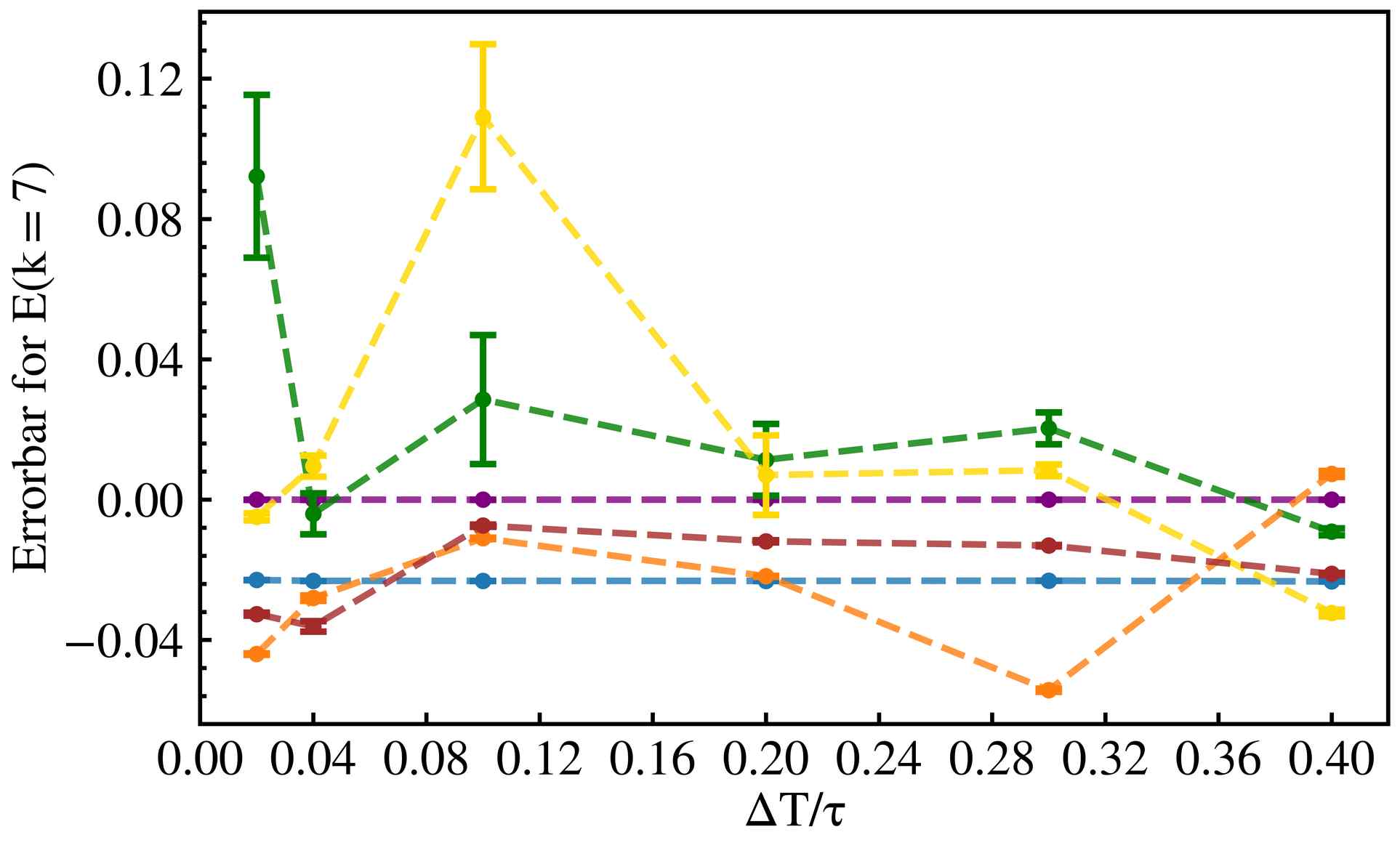}
            \put(0,55){\small (f)} 
        \end{overpic} 
    \end{subfigure}
    
	\caption{Errorbars of $E(k=3, 5, 7)$ for various methods as a function of the time interval $\Delta T$: (a) $E(k=3)$ for FNO-based models constrained; (b) $E(k=3)$ for FNO-based models unconstrained; (c) $E(k=5)$ for FNO-based models constrained; (d) $E(k=5)$ for FNO-based models unconstrained; (e) $E(k=7)$ for FNO-based models constrained; (f) $E(k=7)$ for FNO-based models unconstrained. Note that for fDNS, the values represent natural statistical fluctuations over time, not prediction errors.}\label{fig:15}
\end{figure}

Specifically, we select $k = 3$, $5$, and $7$ to further examine the performance of different methods. Fig.~\ref{fig:15} presents the errorbars of $E(k = 3, 5, 7)$ for various methods as a function of the time interval $\Delta T$, where Figs.~\ref{fig:15}(a)-(b) correspond to $E(k = 3)$, (c)-(d) to $E(k = 5)$, and (e)-(f) to $E(k = 7)$. Similar trends can be observed across $k = 3$, $5$, and $7$: for a given method, the variation pattern of the error remains consistent across different $k$ values.
For fDNS and DSM, the fluctuations or errors remain constant regardless of $\Delta T$, while for FNO-based models, the errors clearly vary with $\Delta T$. Among the constrained FNO-based models, F-IFNO and F-IUFNO with optimal time intervals ($\Delta T \in [0.1\tau, 0.2\tau]$), IUFNO over the entire $\Delta T$ range, and IFNO with $\Delta T = 0.04\tau$ all demonstrate significantly smaller errors compared to DSM. In contrast, the unconstrained versions of these models generally exhibit larger errors than DSM, with the exception of F-IFNO and F-IUFNO under optimal $\Delta T$, which still outperform DSM.
In conclusion, the proposed F-IFNO and F-IUFNO models consistently achieve superior performance compared to DSM, particularly under optimal time intervals.

\begin{figure}[ht!]
    \centering
    \begin{subfigure}[b]{0.32\textwidth}
        \begin{overpic}[width=1\linewidth]{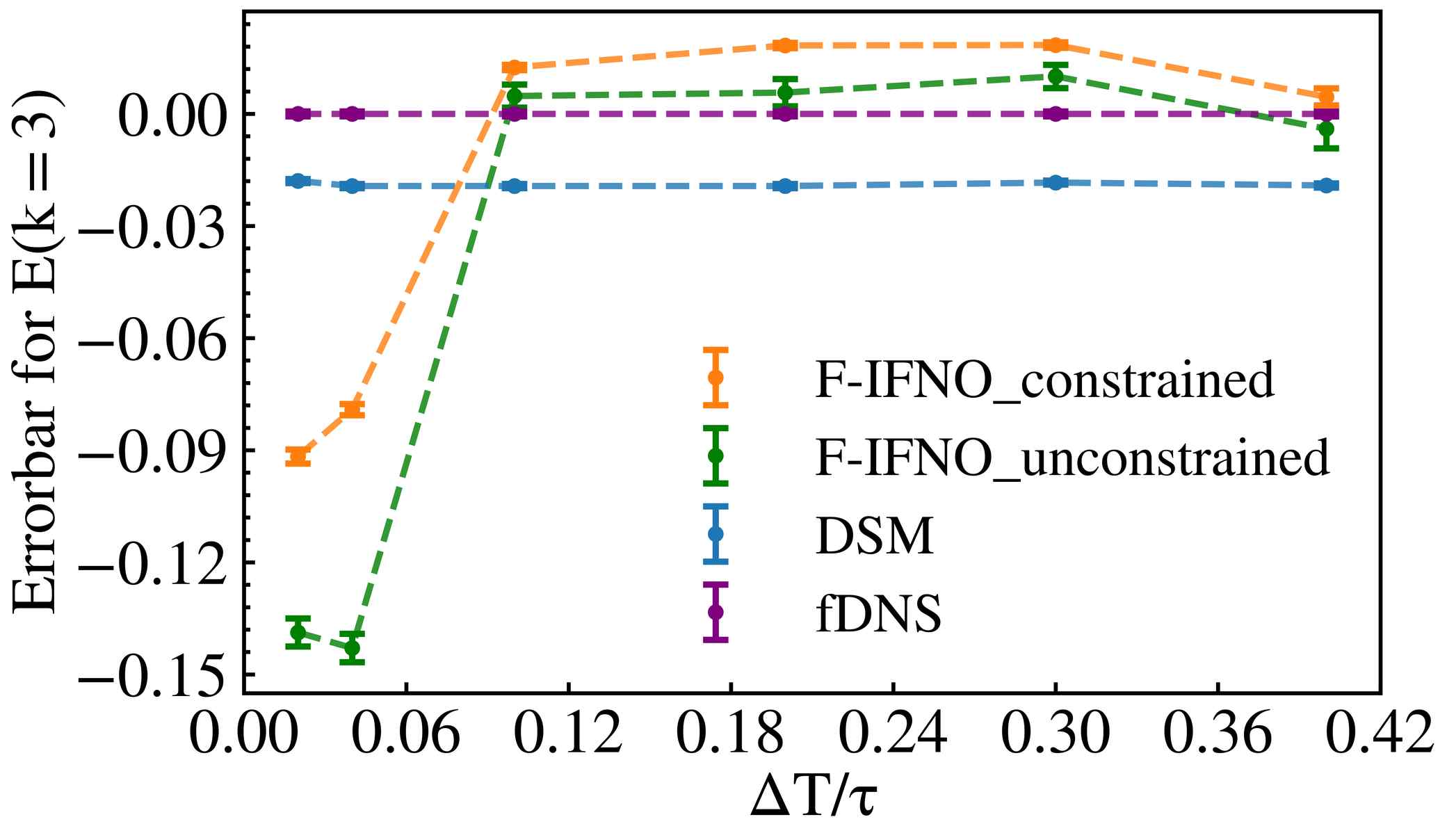}
            \put(-2,55){\small (a)}  
        \end{overpic}
    \end{subfigure}
    \hfill
    \begin{subfigure}[b]{0.32\textwidth}
        \begin{overpic}[width=1\linewidth]{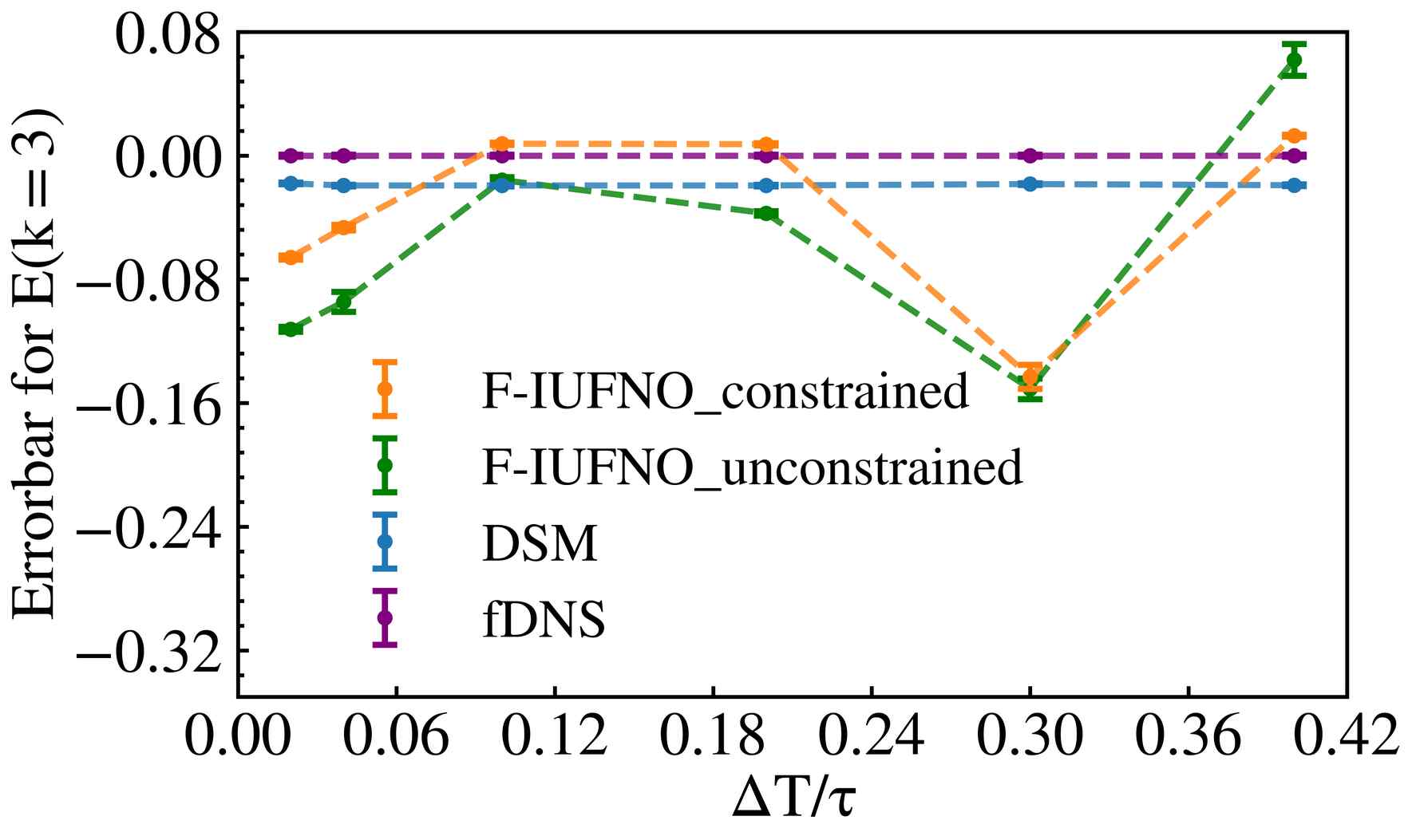}
            \put(-2,55){\small (b)} 
        \end{overpic} 
    \end{subfigure}
    \hfill
    \begin{subfigure}[b]{0.32\textwidth}
        \begin{overpic}[width=1\linewidth]{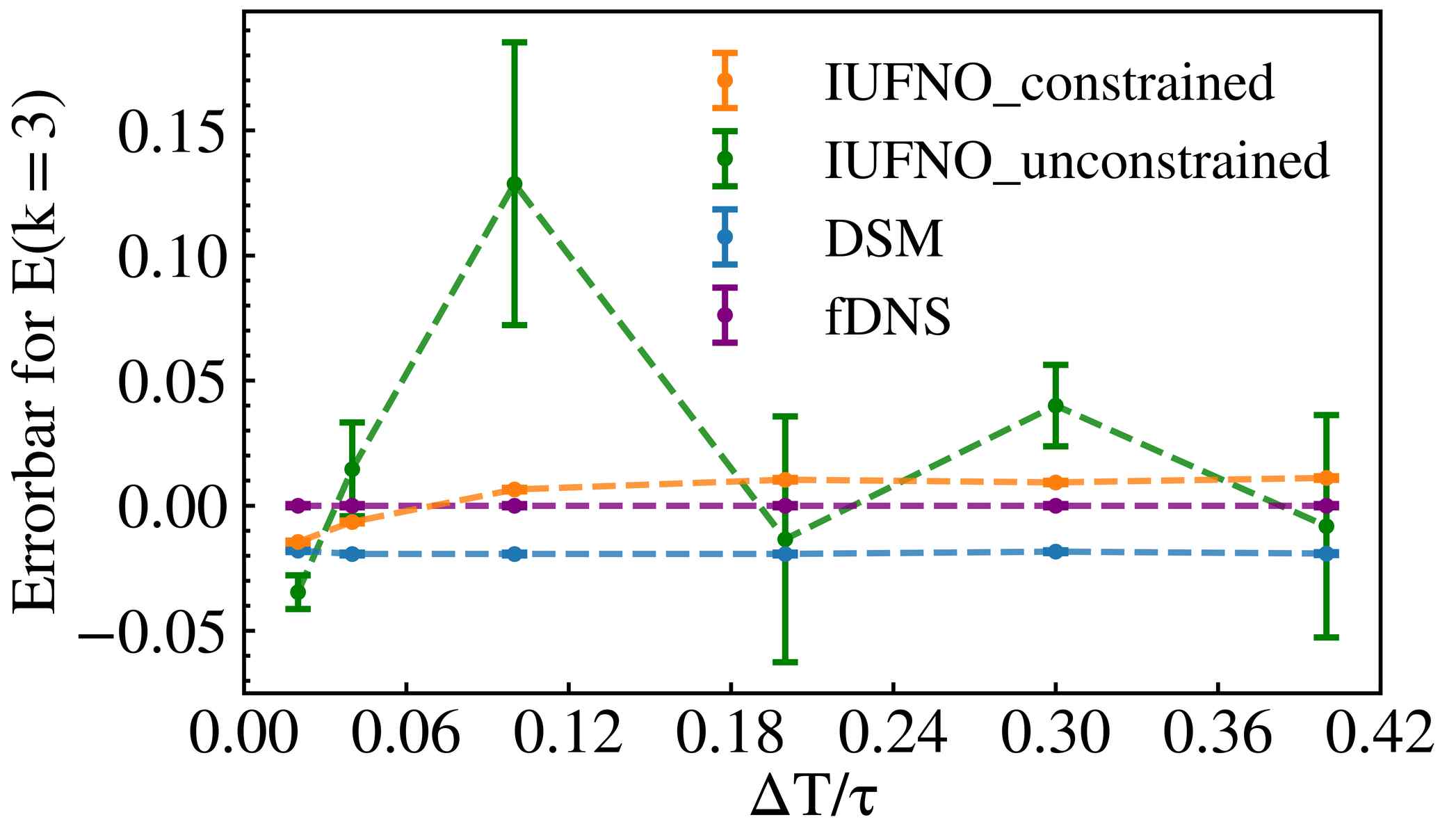}
            \put(-2,55){\small (c)} 
        \end{overpic}
    \end{subfigure}
    \vspace{0.1cm}
    \begin{subfigure}[b]{0.32\textwidth}
        \begin{overpic}[width=1\linewidth]{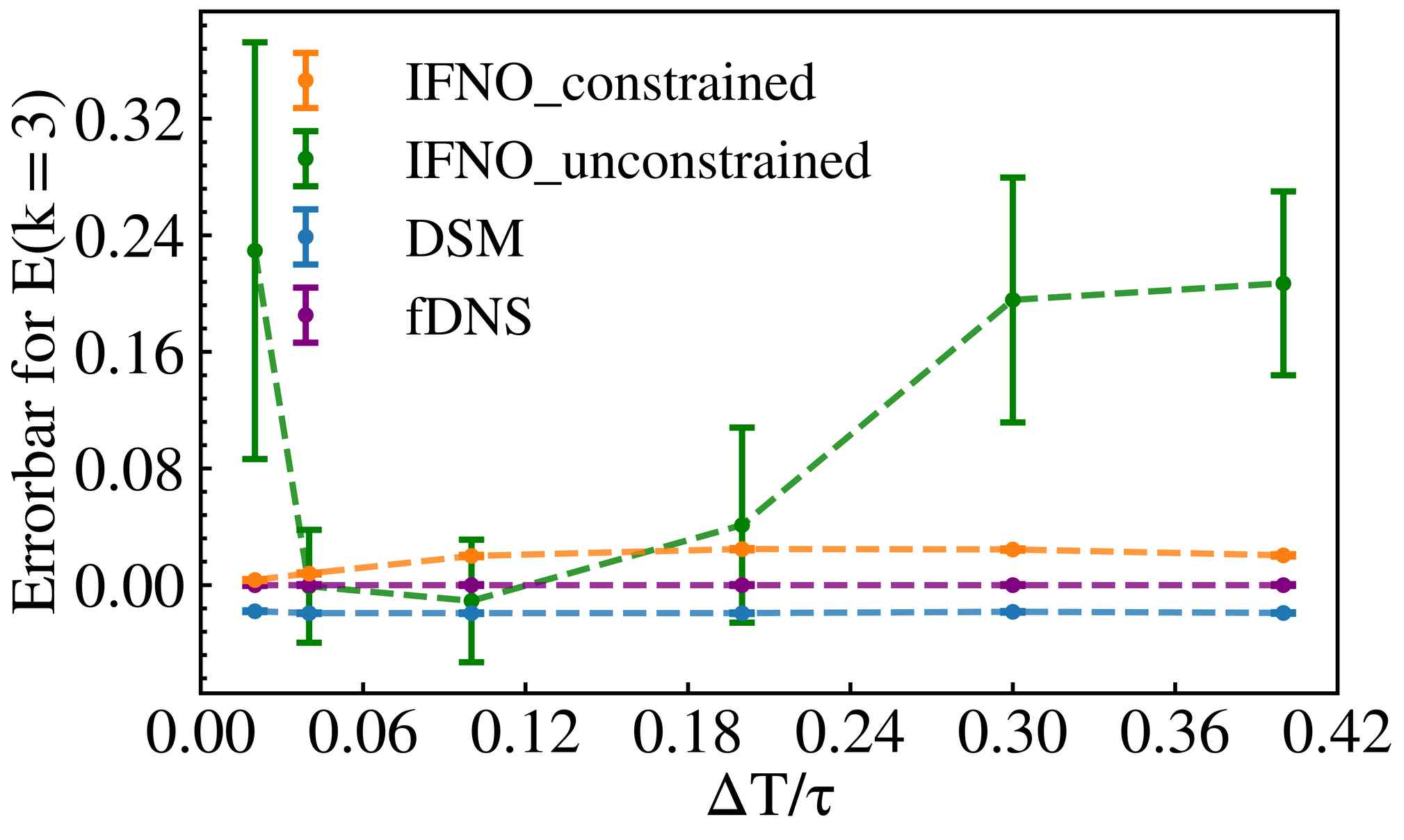}
            \put(-2,55){\small (d)} 
        \end{overpic}
    \end{subfigure}
    \hfill
    \begin{subfigure}[b]{0.32\textwidth}
        \begin{overpic}[width=1\linewidth]{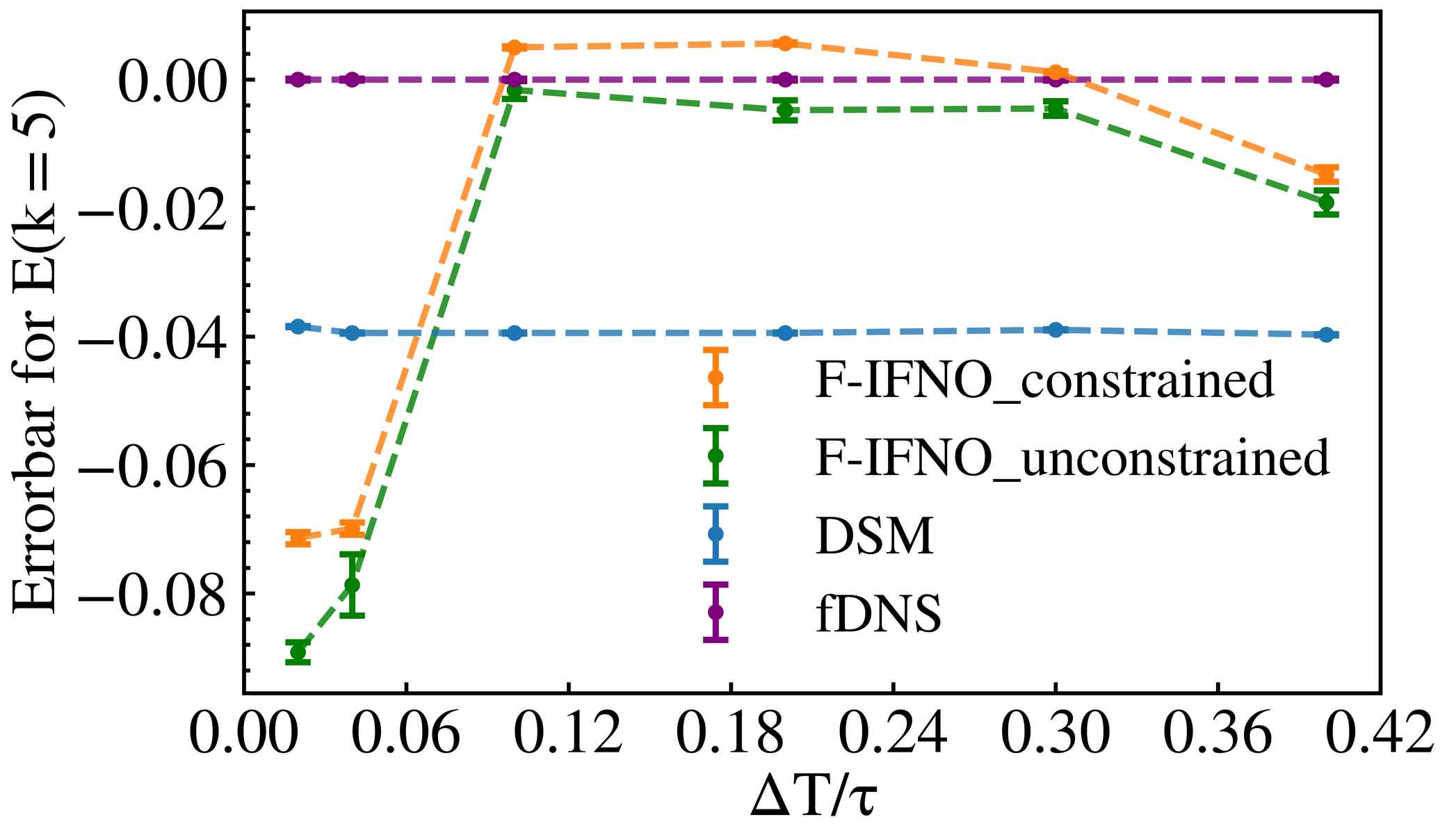}
            \put(-2,55){\small (e)} 
        \end{overpic}
    \end{subfigure}
    \hfill
    \begin{subfigure}[b]{0.32\textwidth}
        \begin{overpic}[width=1\linewidth]{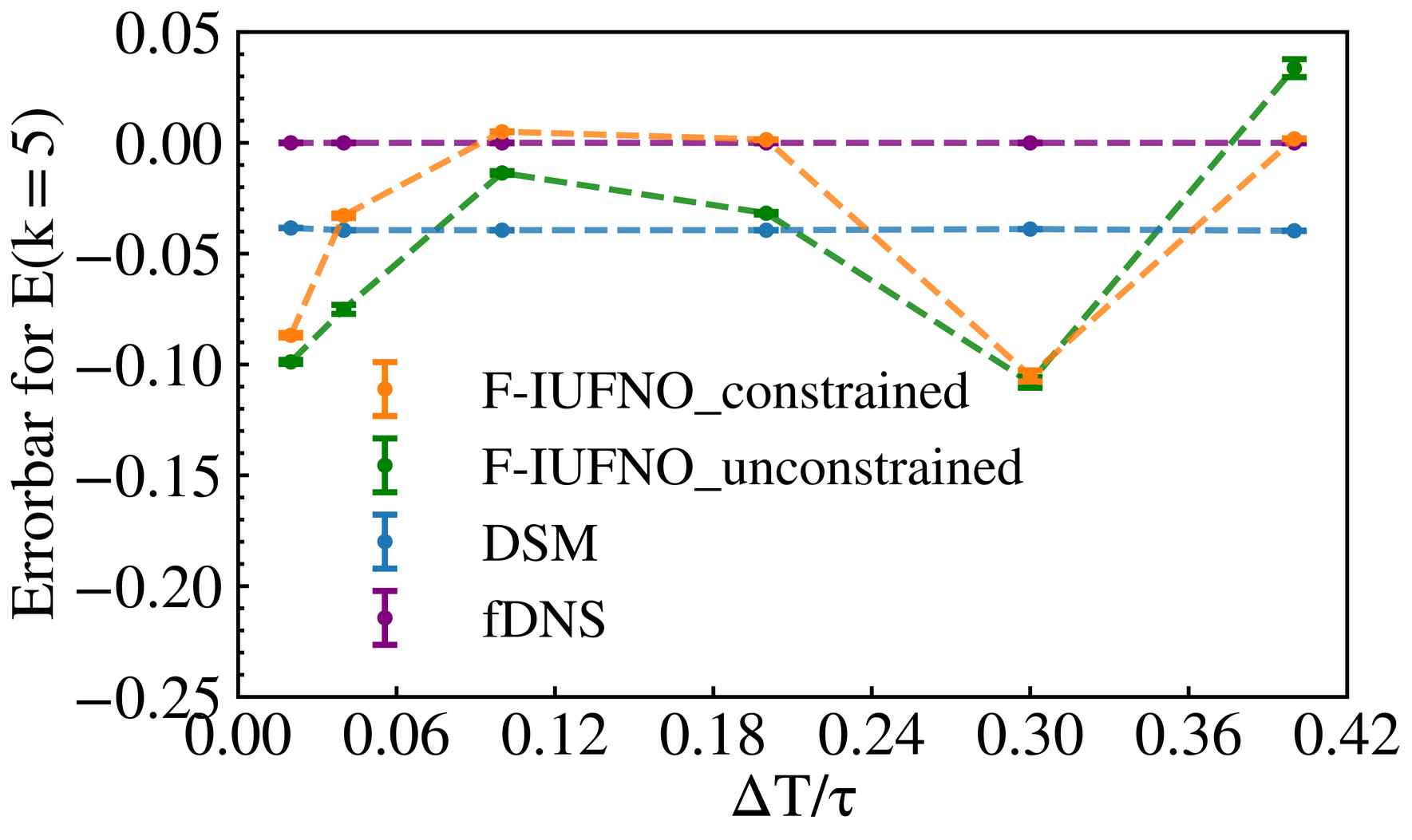}
            \put(-2,55){\small (f)} 
        \end{overpic}
    \end{subfigure}
    \vspace{0.1cm}
    \begin{subfigure}[b]{0.32\textwidth}
        \begin{overpic}[width=1\linewidth]{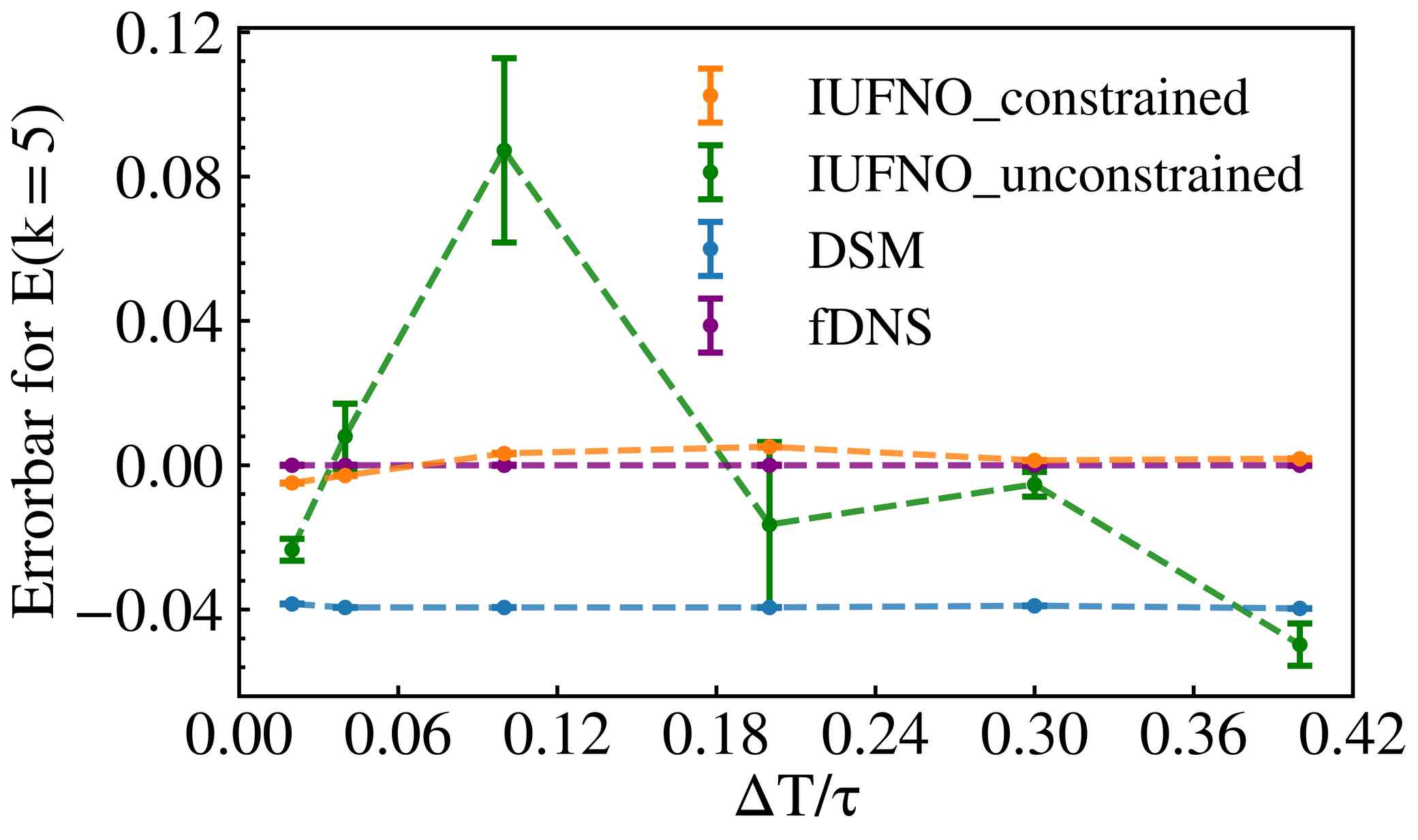}
            \put(-2,55){\small (g)} 
        \end{overpic}
    \end{subfigure}
    \hfill
    \begin{subfigure}[b]{0.32\textwidth}
        \begin{overpic}[width=1\linewidth]{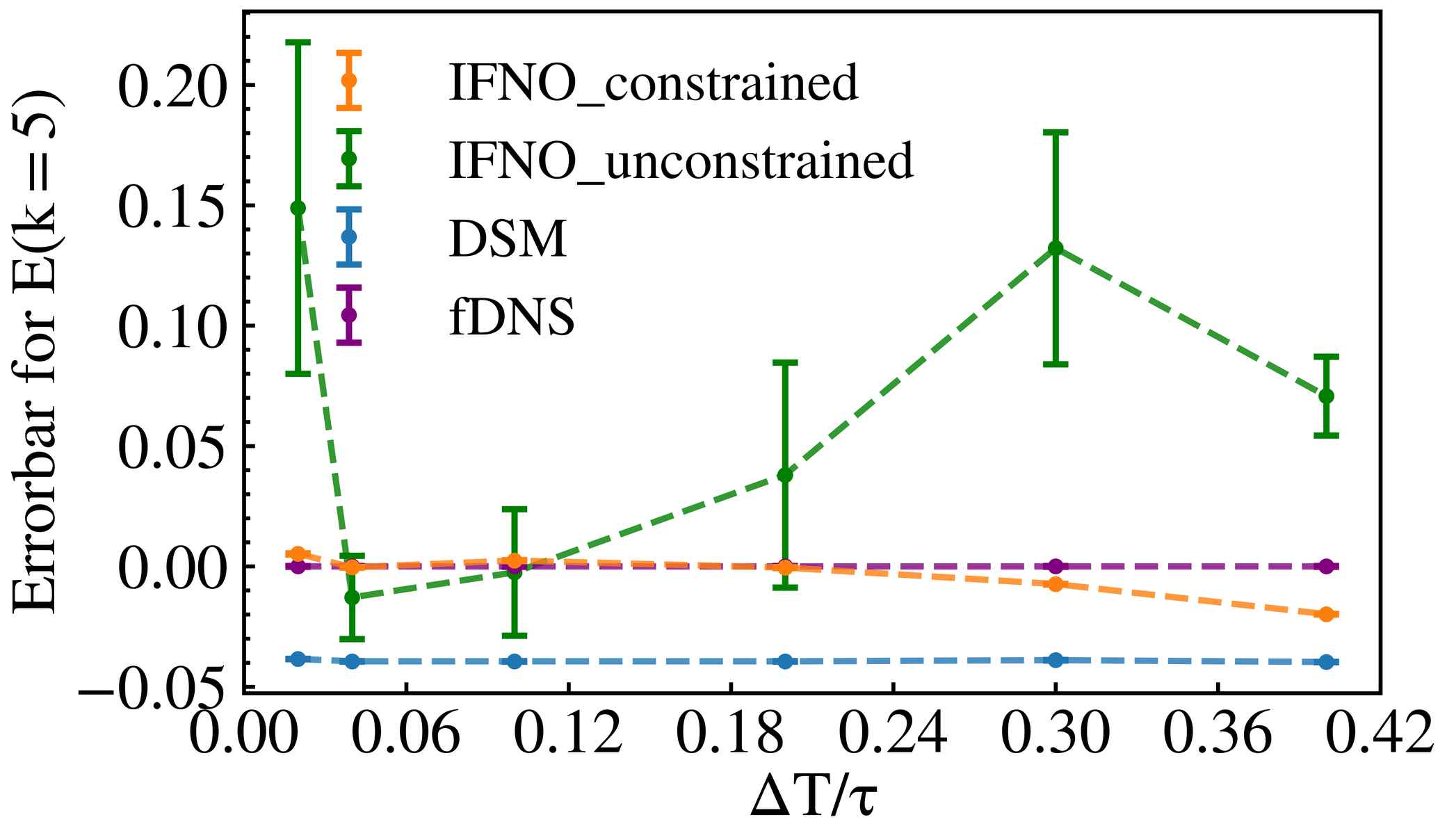}
            \put(-2,55){\small (h)} 
        \end{overpic}
    \end{subfigure}
    \hfill
    \begin{subfigure}[b]{0.32\textwidth}
        \begin{overpic}[width=1\linewidth]{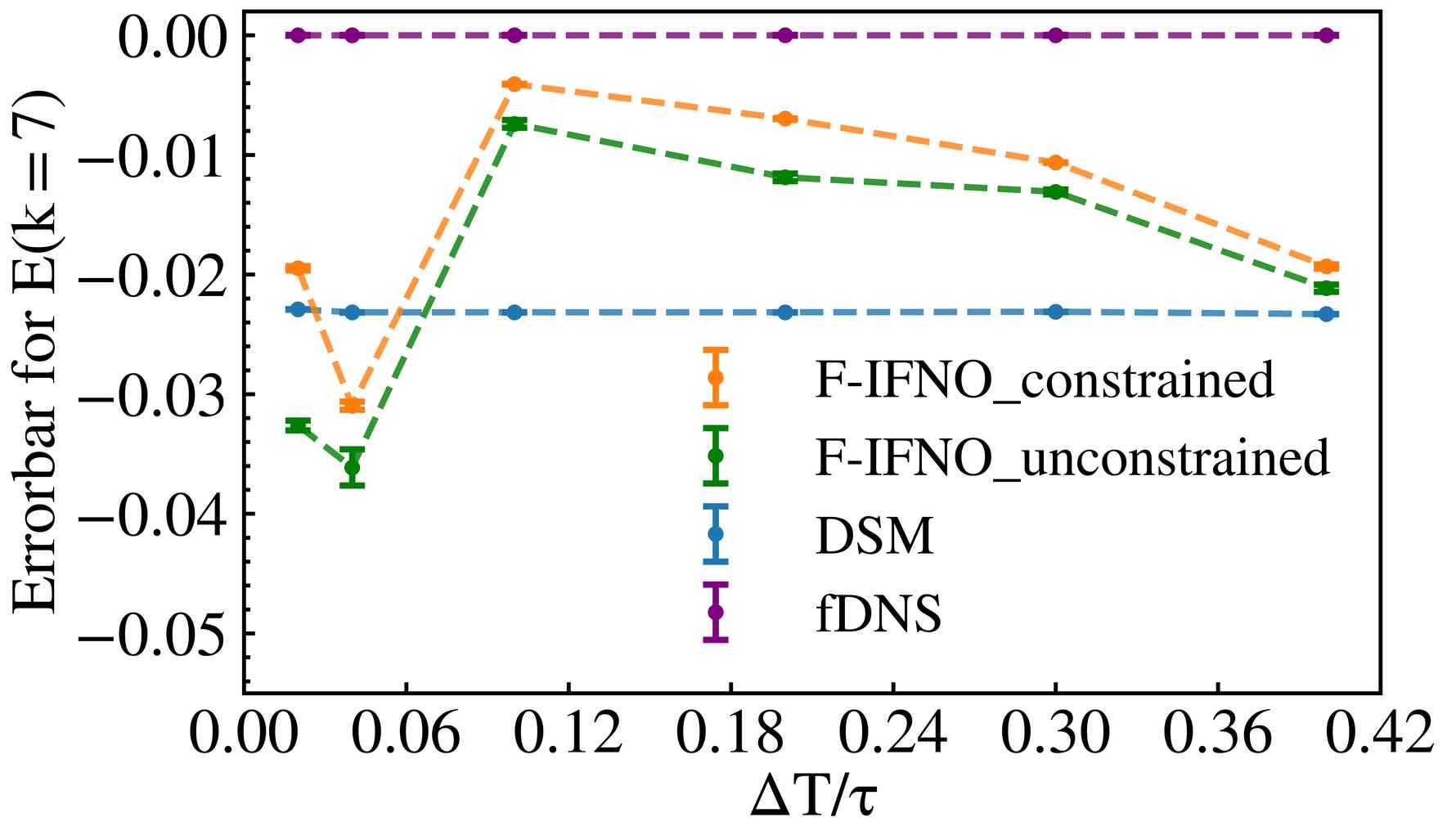}
            \put(-2,55){\small (i)} 
        \end{overpic}
    \end{subfigure}
    \vspace{0.1cm}
    \begin{subfigure}[b]{0.32\textwidth}
        \begin{overpic}[width=1\linewidth]{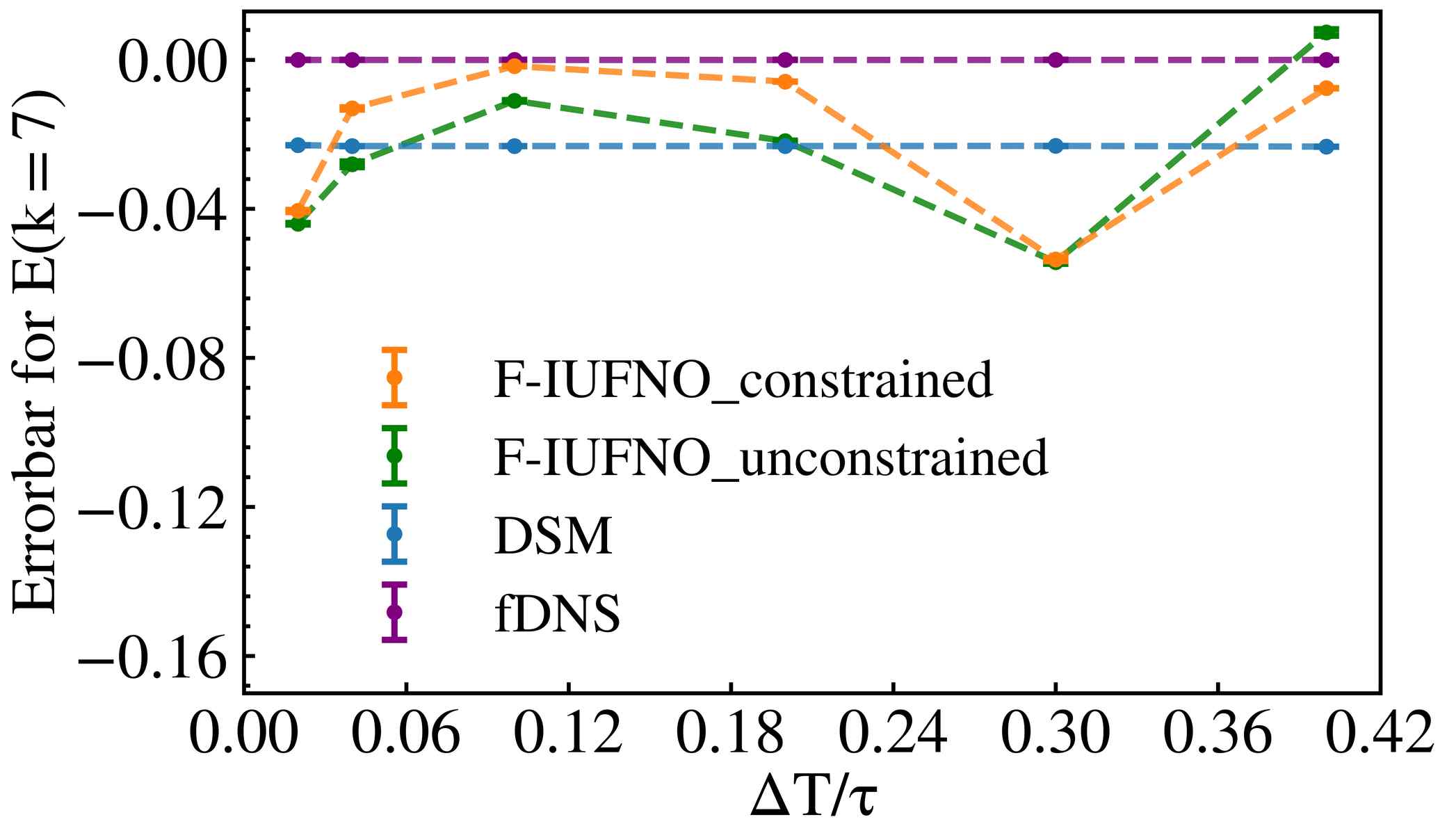}
            \put(-2,55){\small (j)} 
        \end{overpic}
    \end{subfigure}
    \hfill
    \begin{subfigure}[b]{0.32\textwidth}
        \begin{overpic}[width=1\linewidth]{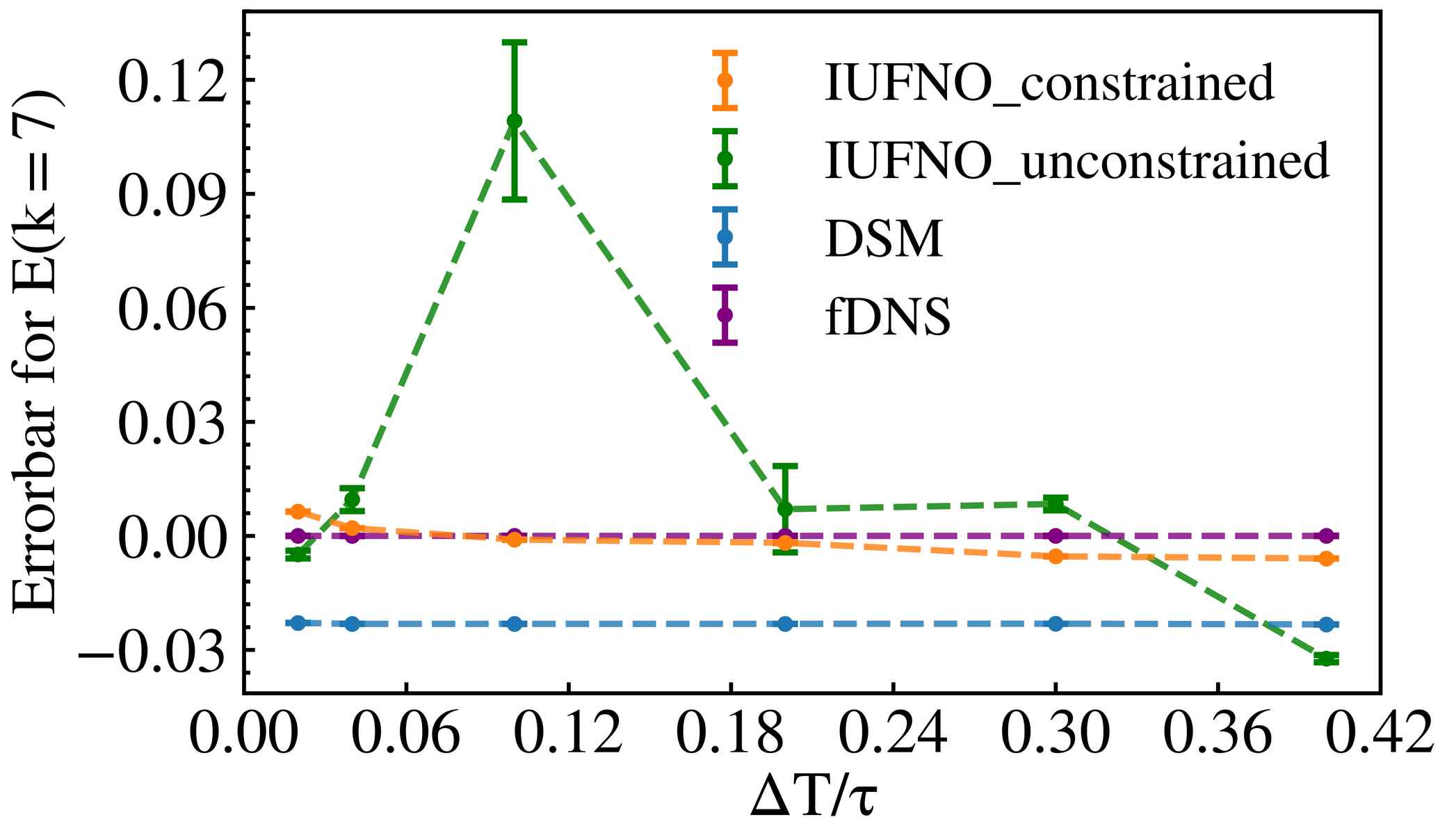}
            \put(-2,55){\small (k)} 
        \end{overpic}
    \end{subfigure}
    \hfill
    \begin{subfigure}[b]{0.32\textwidth}
        \begin{overpic}[width=1\linewidth]{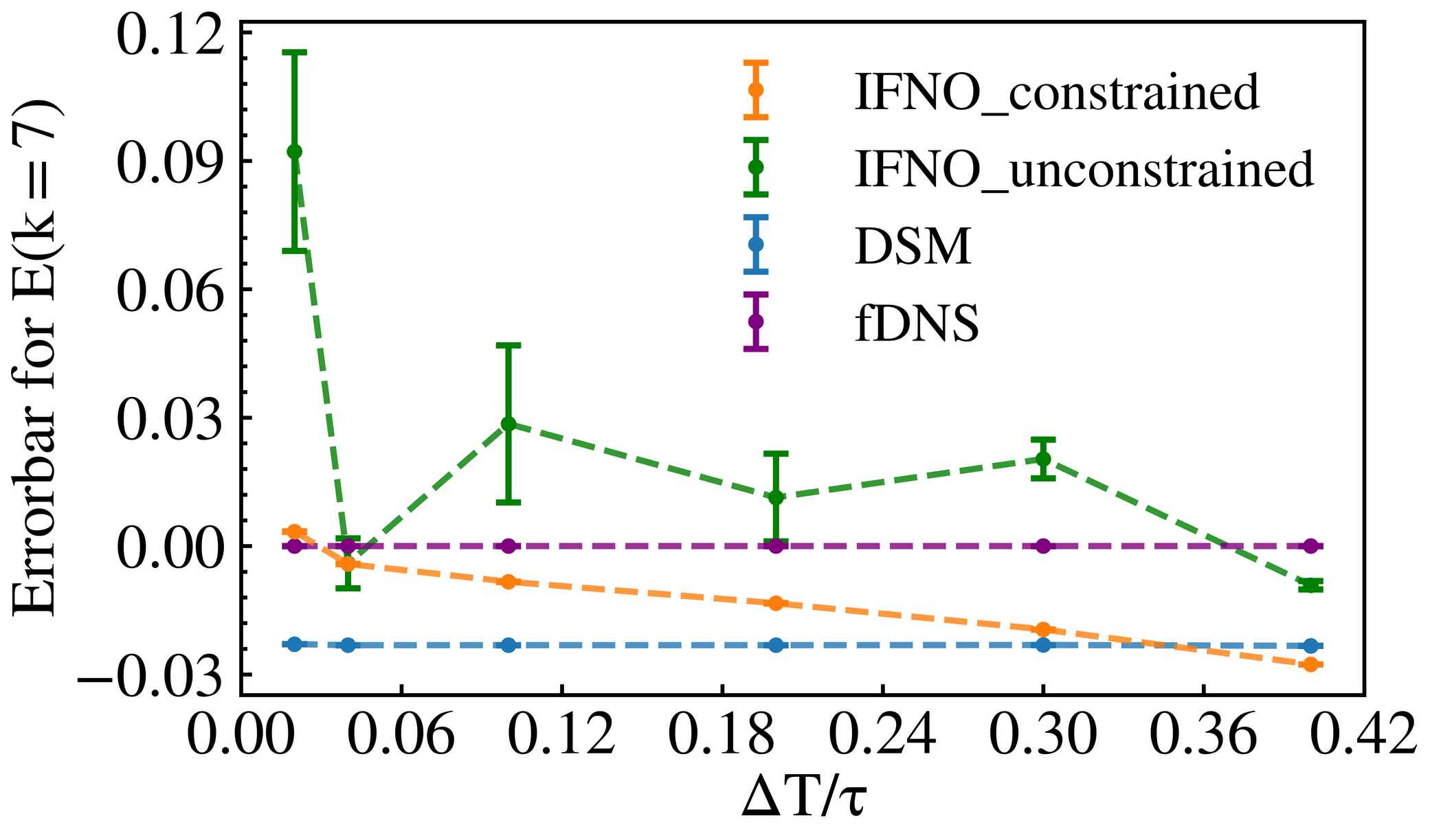}
            \put(-2,55){\small (l)}  
        \end{overpic}
    \end{subfigure}
	\caption{Errorbars of $E(k=3, 5, 7)$ for constrained and unconstrained FNO-based models: (a) $E(k=3)$ for F-IFNO; (b) $E(k=3)$ for F-IUFNO; (c) $E(k=3)$ for IUFNO; (d) $E(k=3)$ for IFNO; (e) $E(k=5)$ for F-IFNO; (f) $E(k=5)$ for F-IUFNO; (g) $E(k=5)$ for IUFNO; (h) $E(k=5)$ for IFNO; (i) $E(k=7)$ for F-IFNO; (j) $E(k=7)$ for F-IUFNO; (k) $E(k=7)$ for IUFNO; (l) $E(k=7)$ for IFNO. Note that for fDNS, the values represent natural statistical fluctuations over time, not prediction errors.}\label{fig:16}
\end{figure}

Fig.~\ref{fig:16} illustrates the errorbars of $E(k = 3, 5, 7)$ for both constrained and unconstrained FNO-based models. Specifically, Figs.~\ref{fig:16}(a)-(d) correspond to $E(k = 3)$, Figs.~\ref{fig:16}(e)-(h) to $E(k = 5)$, and Figs.~\ref{fig:16}(i)-(k) to $E(k = 7)$. Consistent with the observations in Fig.~\ref{fig:15}, it is evident that, for each method, the constrained version significantly outperforms its unconstrained counterpart. 
These results highlight the necessity of applying prediction constraints to improve the accuracy and robustness of FNO-based models.

\begin{figure}[ht!]
    \centering
    \begin{subfigure}[b]{0.32\textwidth}
        \begin{overpic}[width=1\linewidth]{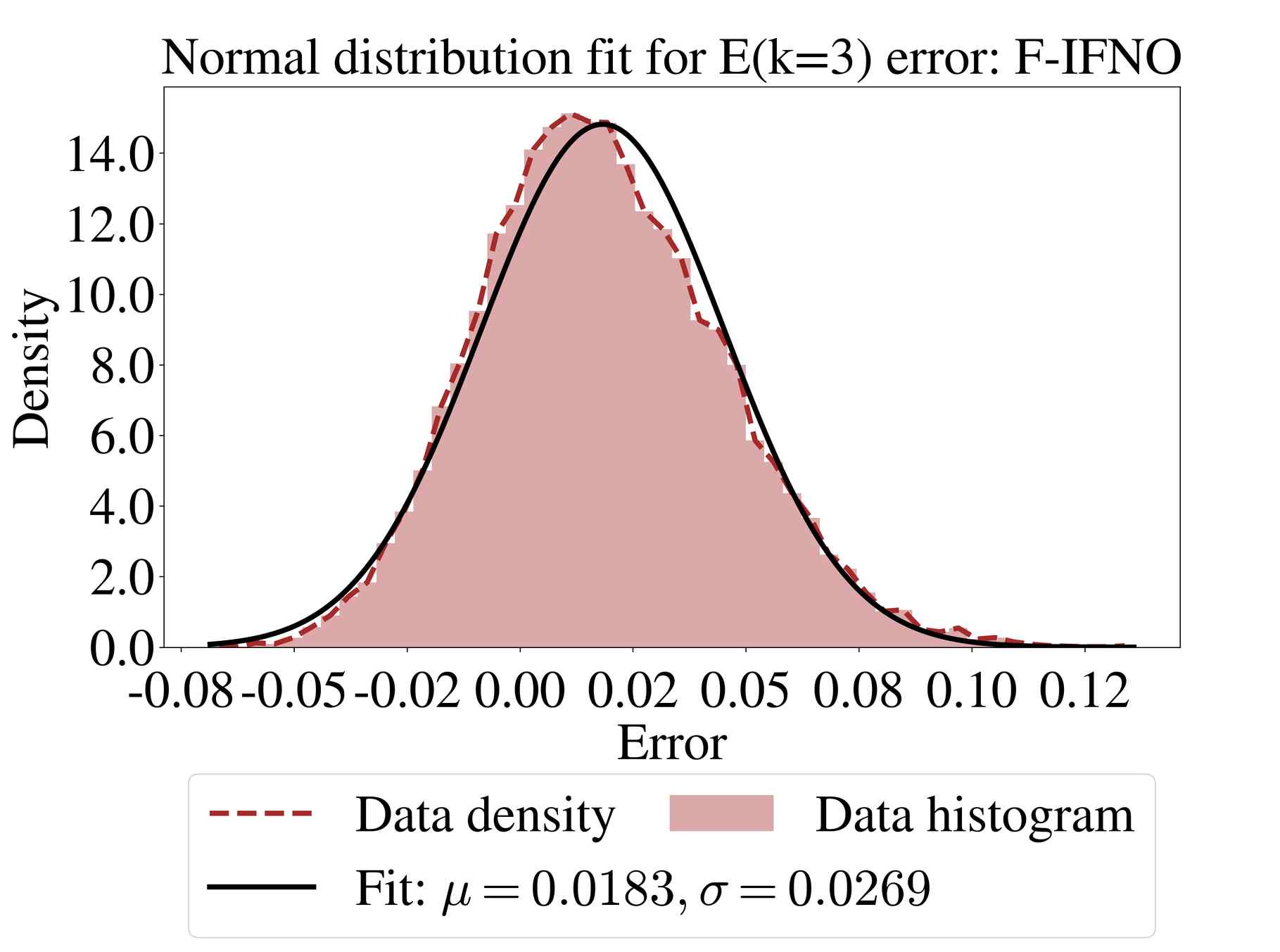}
            \put(-3,60){\small (a)}  
        \end{overpic}
    \end{subfigure}
    \hfill
    \begin{subfigure}[b]{0.32\textwidth}
        \begin{overpic}[width=1\linewidth]{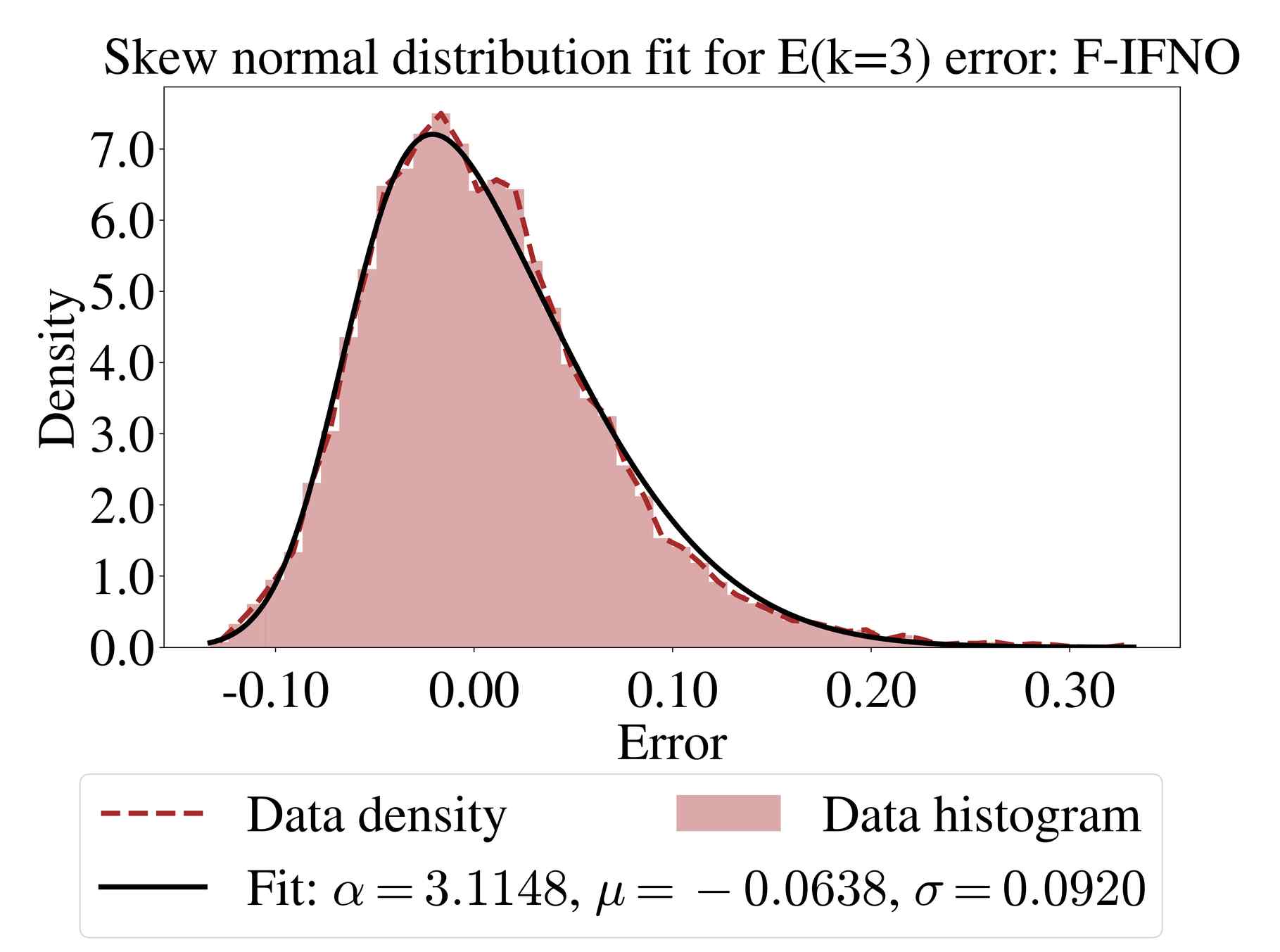}
            \put(-3,60){\small (b)} 
        \end{overpic} 
    \end{subfigure}
    \hfill
    \begin{subfigure}[b]{0.32\textwidth}
        \begin{overpic}[width=1\linewidth]{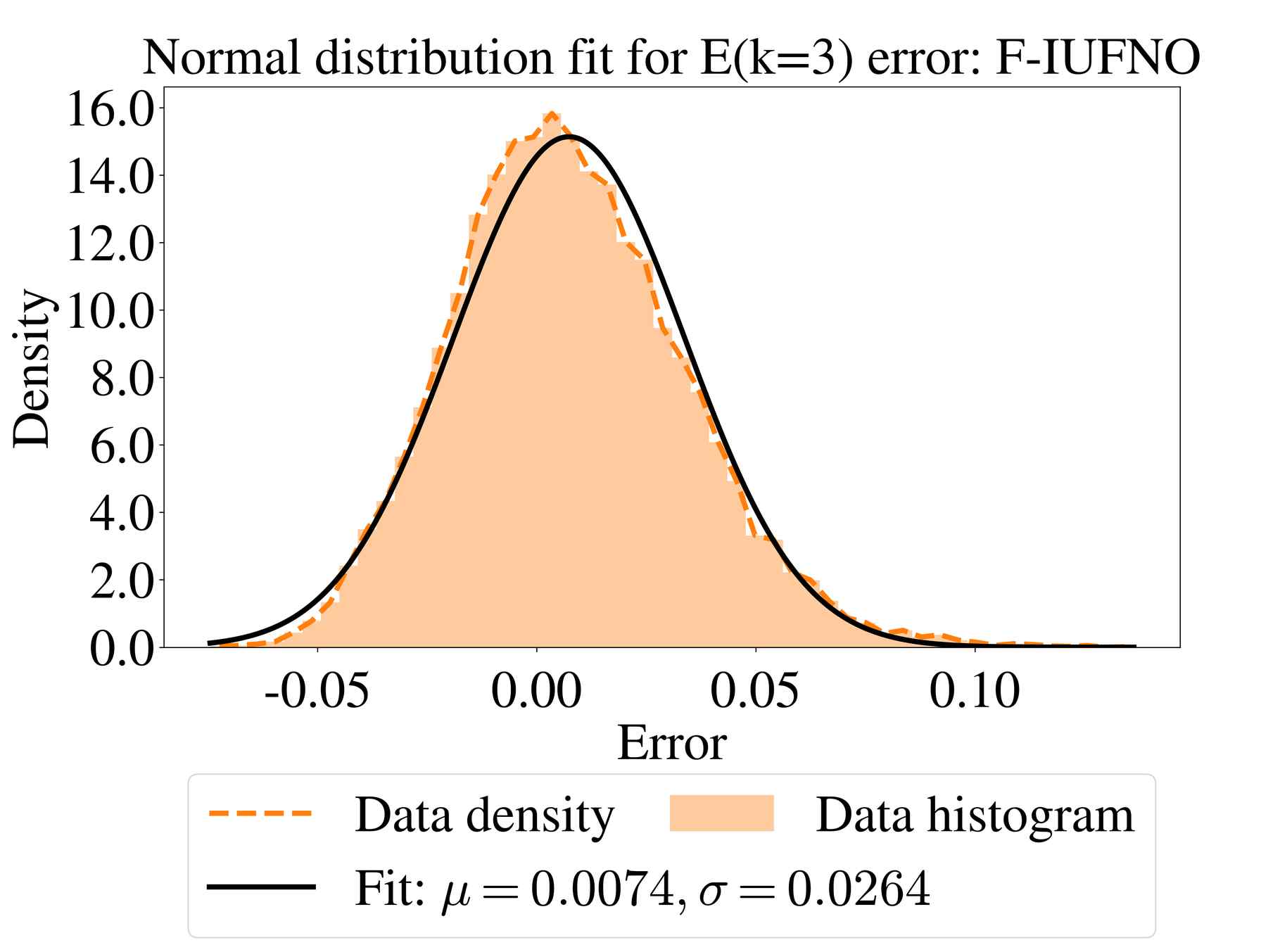}
            \put(-3,60){\small (c)} 
        \end{overpic}
    \end{subfigure}
    \vspace{0.1cm}

    \begin{subfigure}[b]{0.32\textwidth}
        \begin{overpic}[width=1\linewidth]{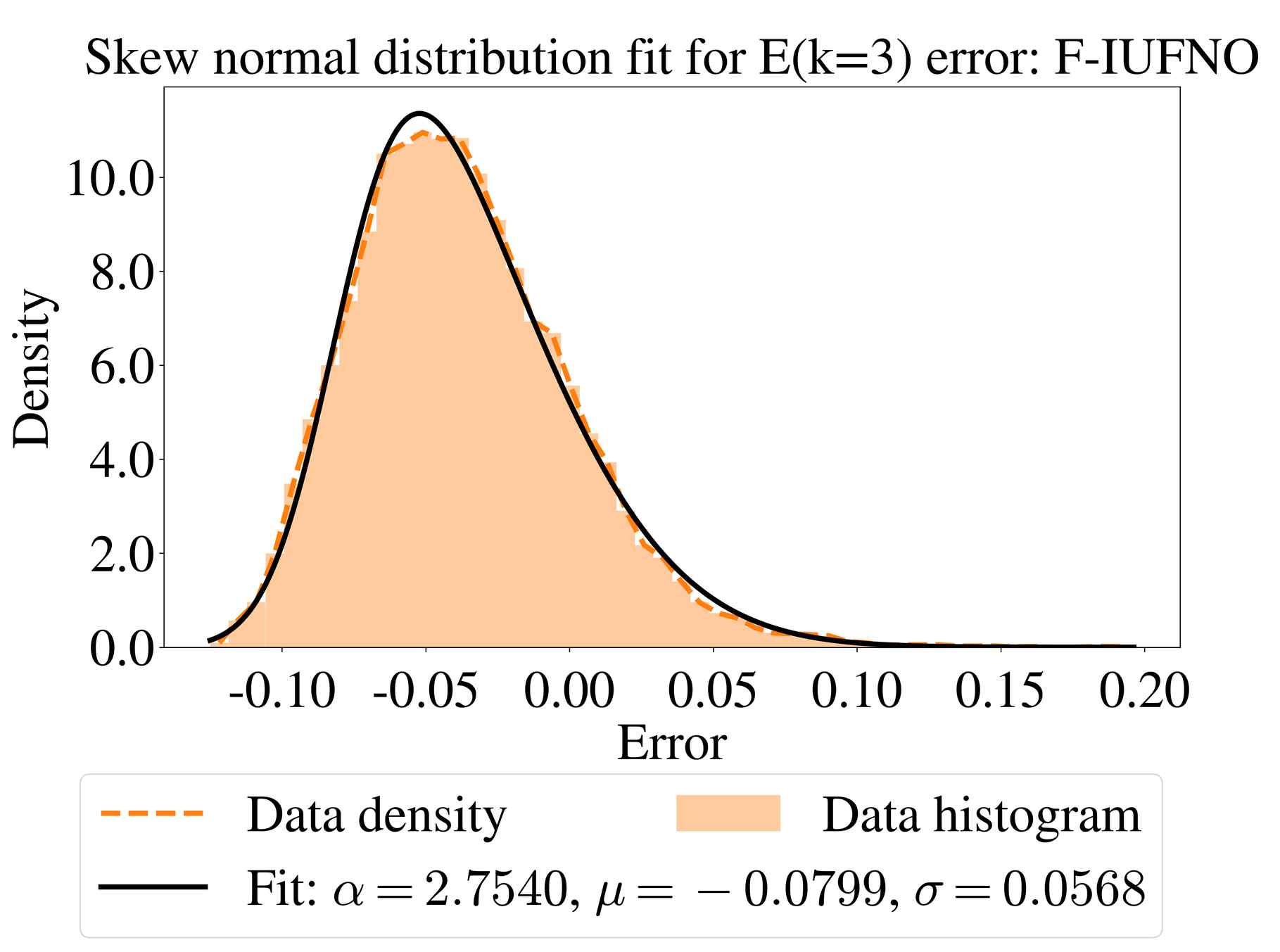}
            \put(-3,60){\small (d)}  
        \end{overpic}
    \end{subfigure}
    \hfill
    \begin{subfigure}[b]{0.32\textwidth}
        \begin{overpic}[width=1\linewidth]{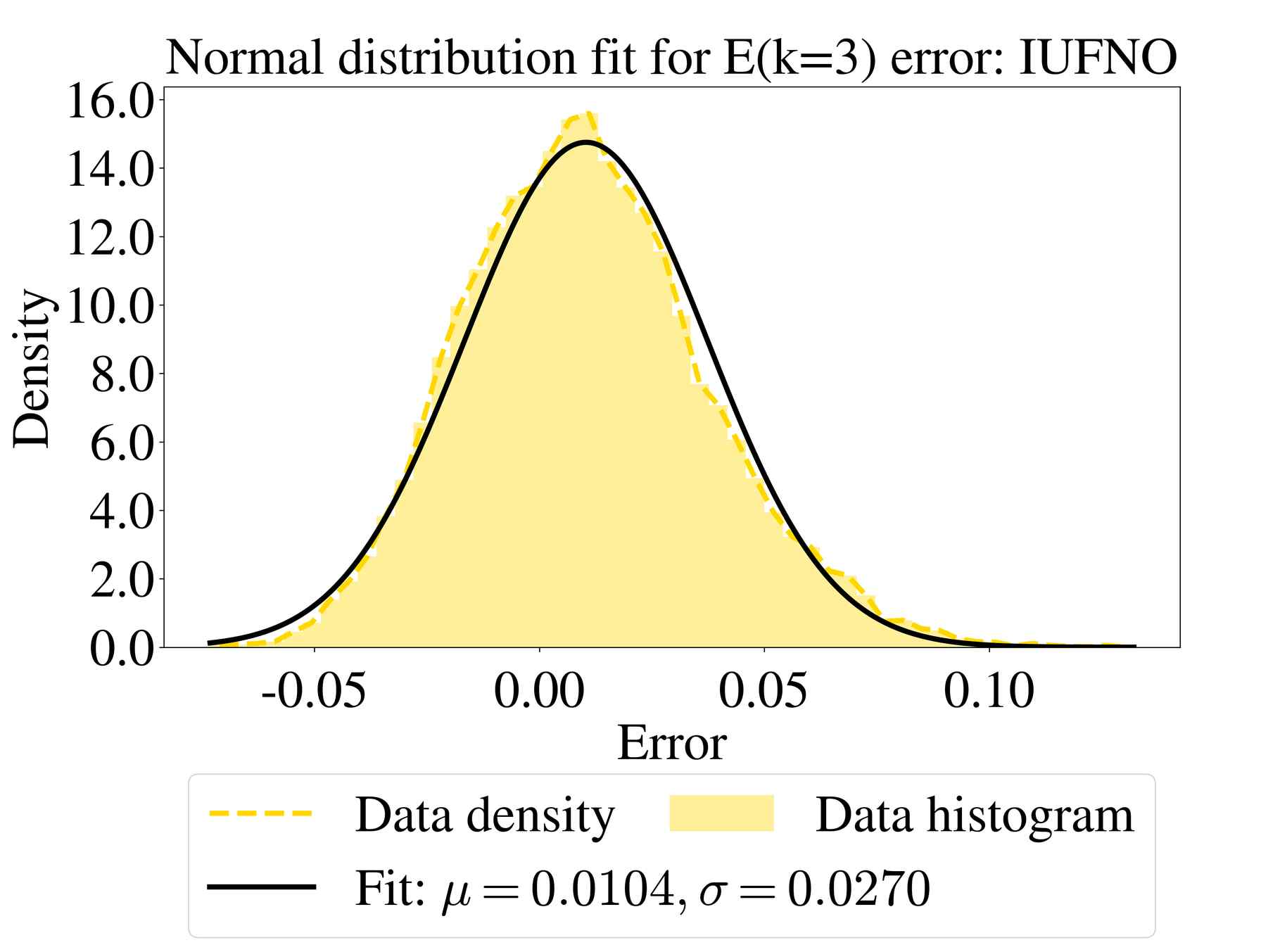}
            \put(-3,60){\small (e)} 
        \end{overpic} 
    \end{subfigure}
    \hfill
    \begin{subfigure}[b]{0.32\textwidth}
        \begin{overpic}[width=1\linewidth]{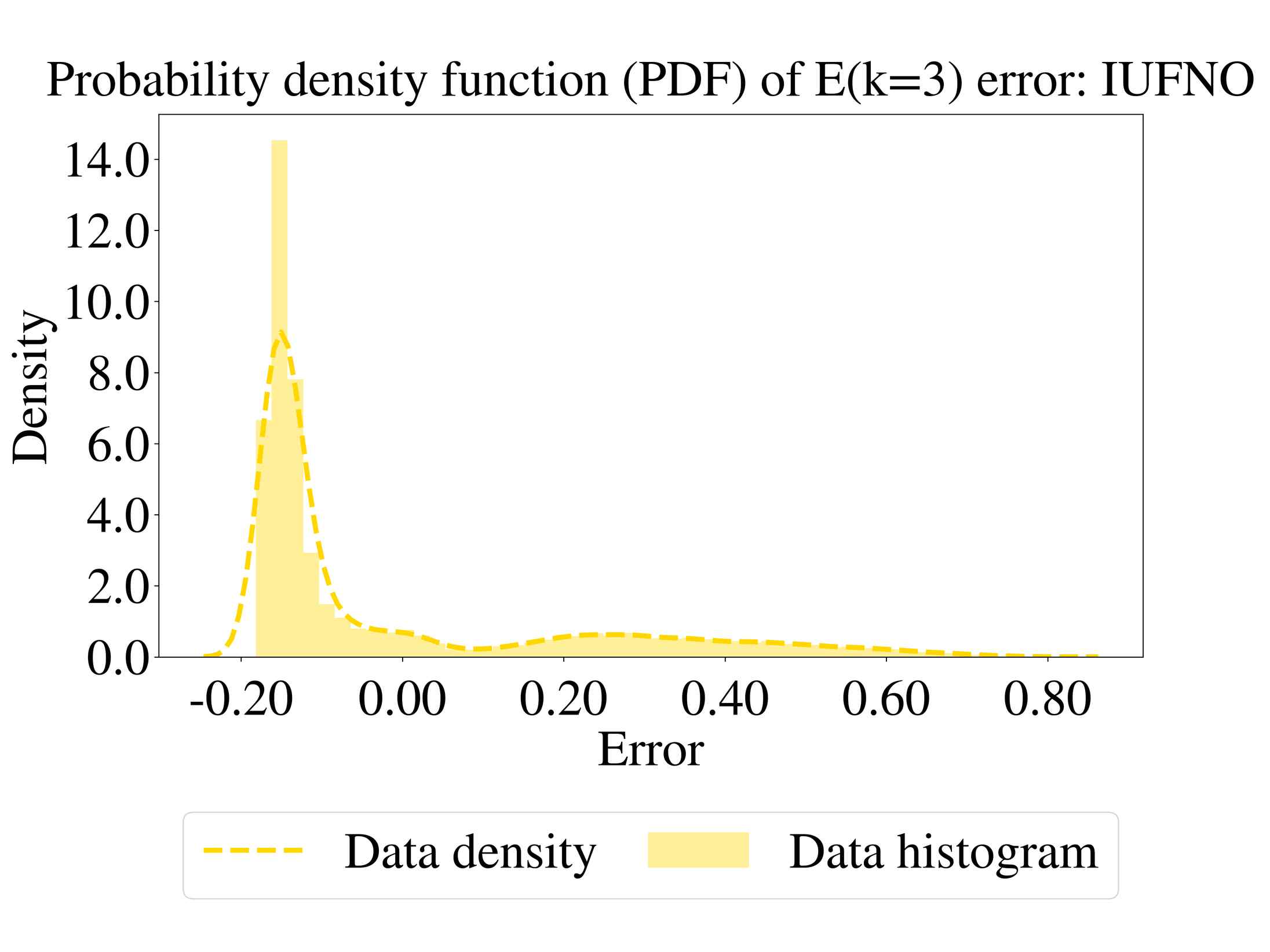}
            \put(-3,60){\small (f)} 
        \end{overpic}
    \end{subfigure}
    \vspace{0.1cm}

    \begin{subfigure}[b]{0.32\textwidth}
        \begin{overpic}[width=1\linewidth]{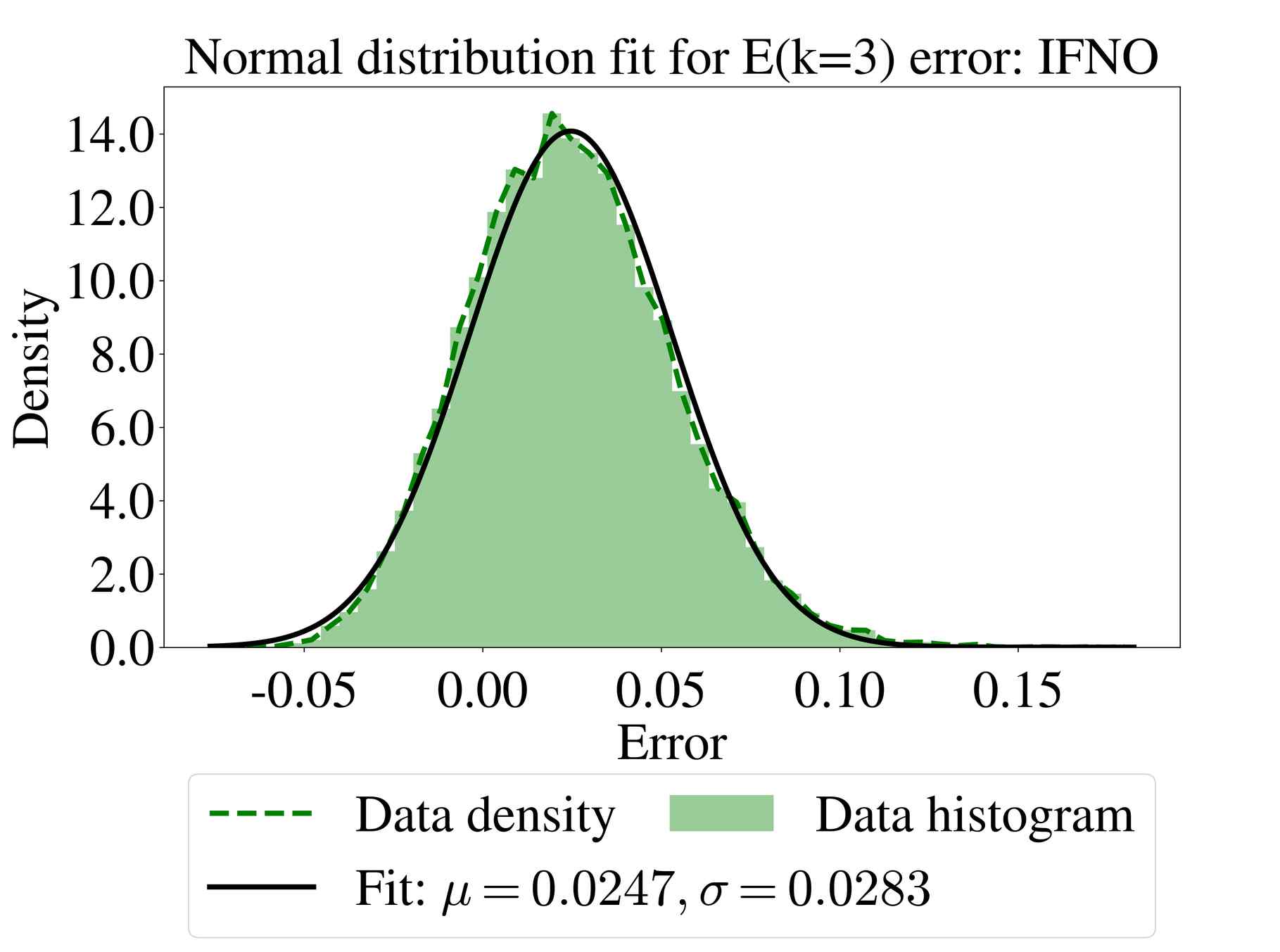}
            \put(-3,60){\small (g)}  
        \end{overpic}
    \end{subfigure}
    \hfill
    \begin{subfigure}[b]{0.32\textwidth}
        \begin{overpic}[width=1\linewidth]{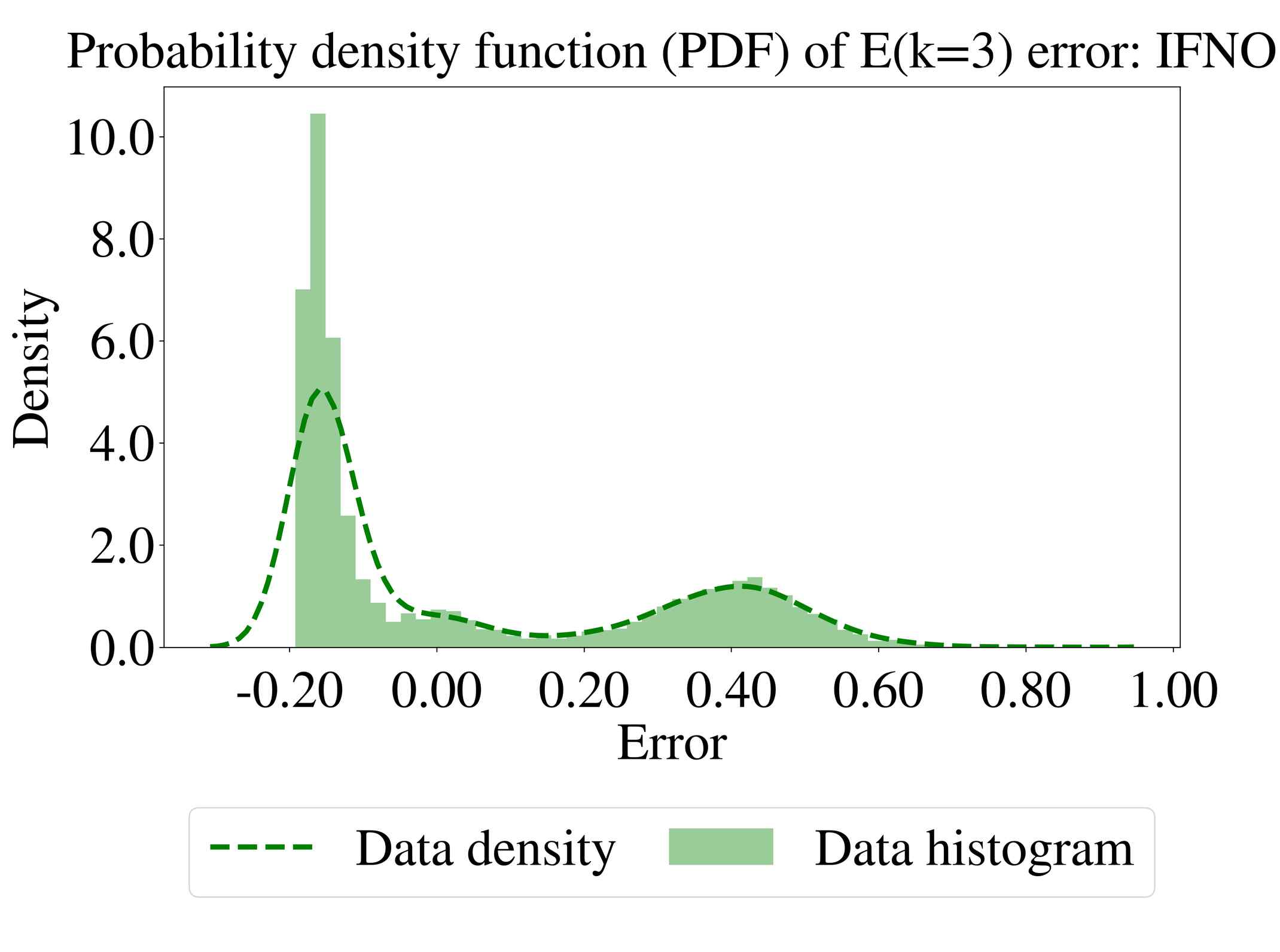}
            \put(-3,60){\small (h)} 
        \end{overpic} 
    \end{subfigure}
    \hfill
    \begin{subfigure}[b]{0.32\textwidth}
        \begin{overpic}[width=1\linewidth]{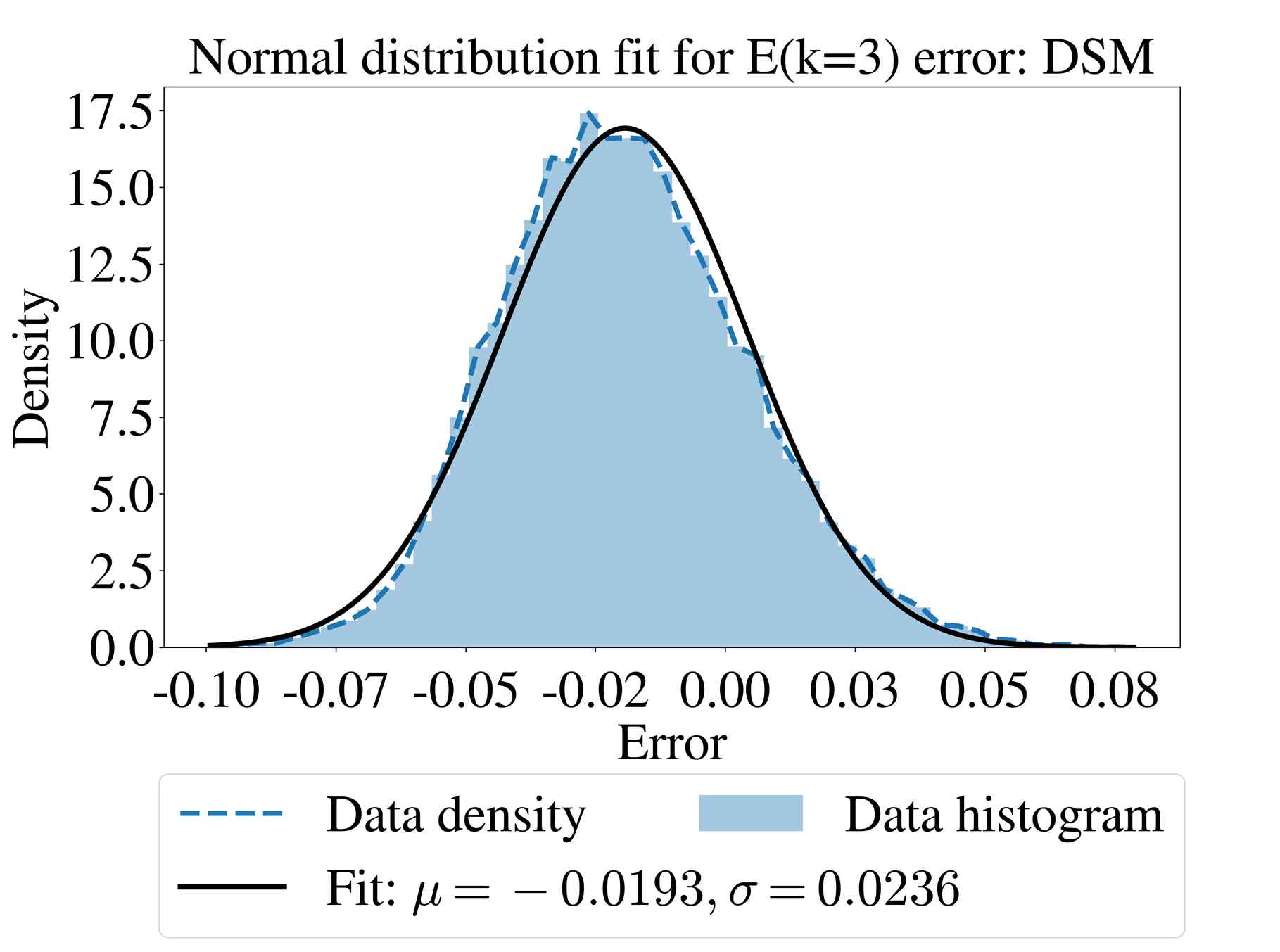}
            \put(-3,60){\small (i)} 
        \end{overpic}
    \end{subfigure}
    \vspace{0.1cm}

    \begin{subfigure}[b]{1\textwidth}
        \centering
        \begin{overpic}[width=0.32\linewidth]{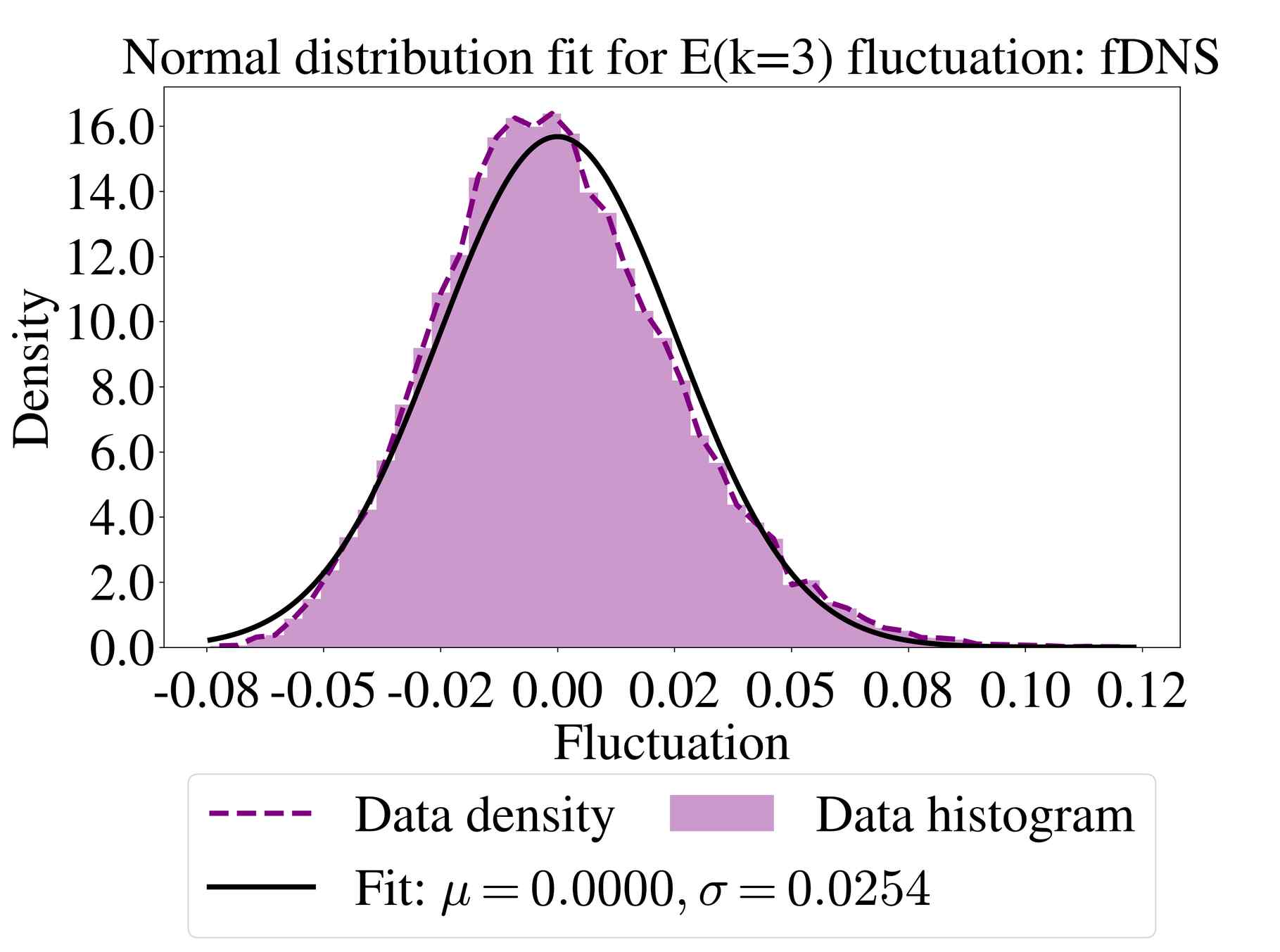}
            \put(-3,60){\small (j)}  
        \end{overpic}
    \end{subfigure}

	\caption{The PDFs of the $E(k=3)$ errors for each method at time interval $\Delta T=0.2\tau$: (a) F-IFNO constrained; (b) F-IFNO unconstrained; (c) F-IUFNO constrained; (d) F-IUFNO unconstrained; (e) IUFNO constrained; (f) IUFNO unconstrained; (g) IFNO constrained; (h) IFNO unconstrained; (i) DSM; (j) fDNS. Note that for fDNS, the values represent natural statistical fluctuations over time, not prediction errors.}\label{fig:17}
\end{figure}

\begin{figure}[ht!]
    \centering
    \begin{subfigure}[b]{0.32\textwidth}
        \begin{overpic}[width=1\linewidth]{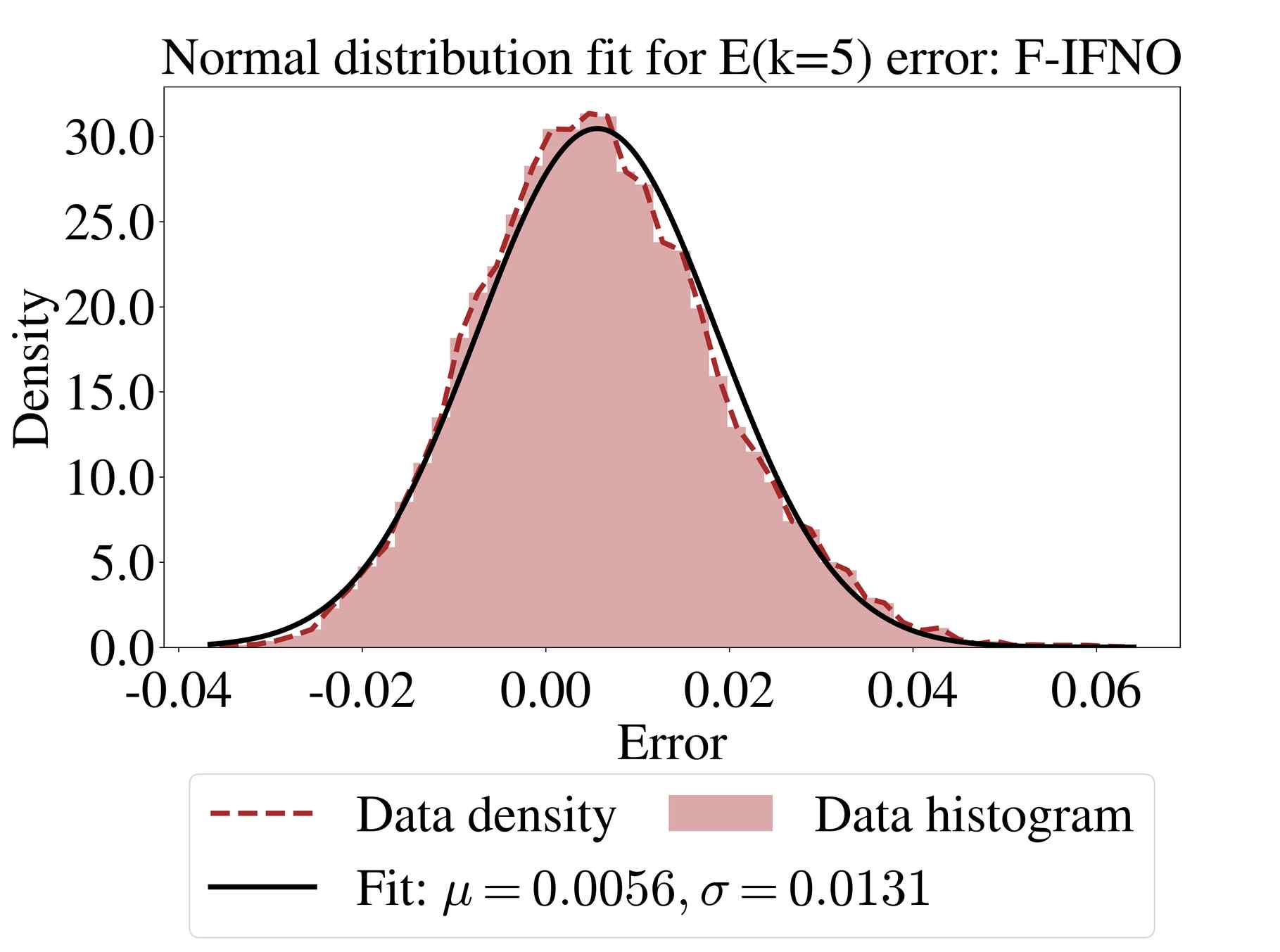}
            \put(-4,60){\small (a)}  
        \end{overpic}
    \end{subfigure}
    \hfill
    \begin{subfigure}[b]{0.32\textwidth}
        \begin{overpic}[width=1\linewidth]{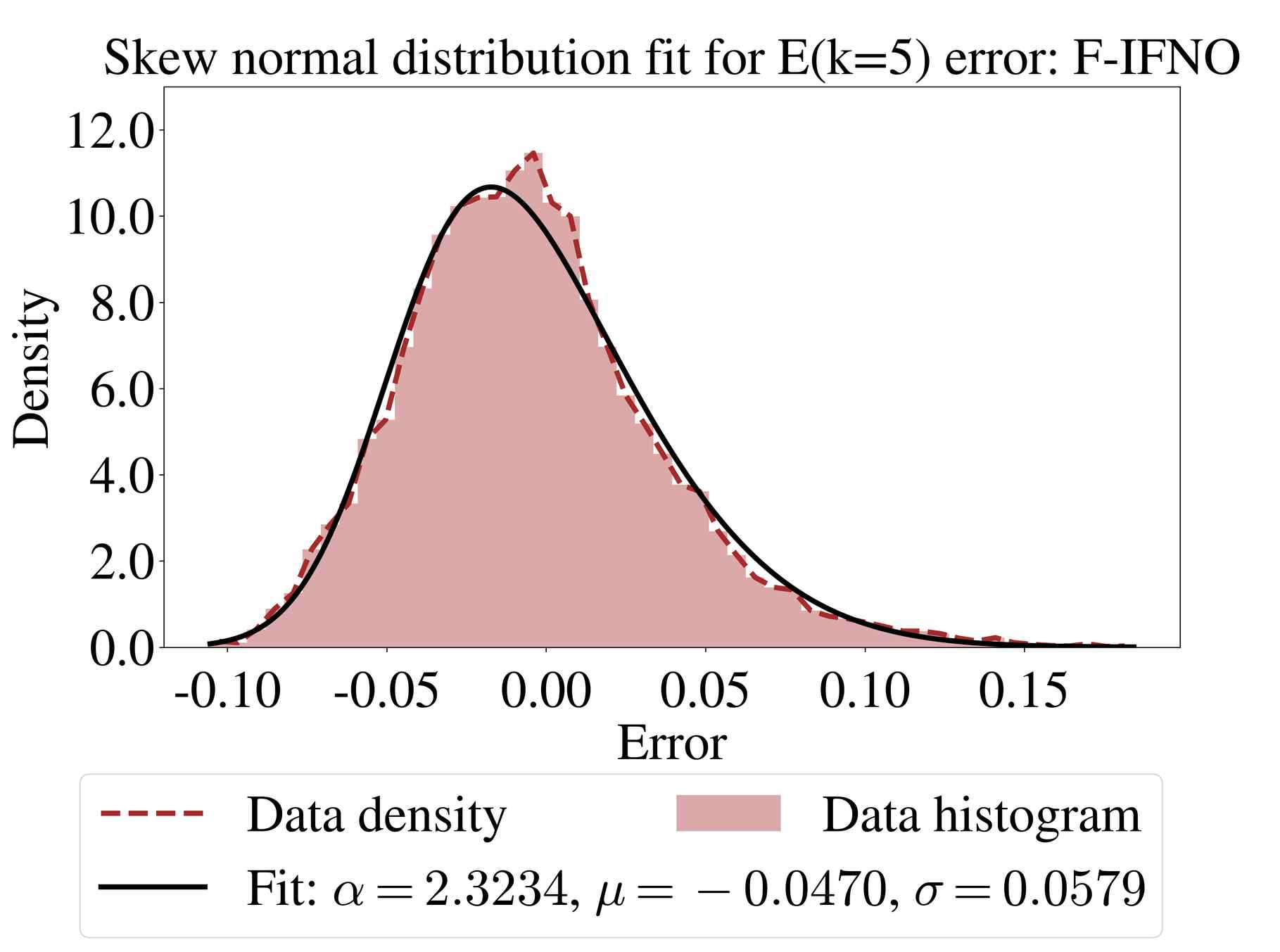}
            \put(-3,60){\small (b)} 
        \end{overpic} 
    \end{subfigure}
    \hfill
    \begin{subfigure}[b]{0.32\textwidth}
        \begin{overpic}[width=1\linewidth]{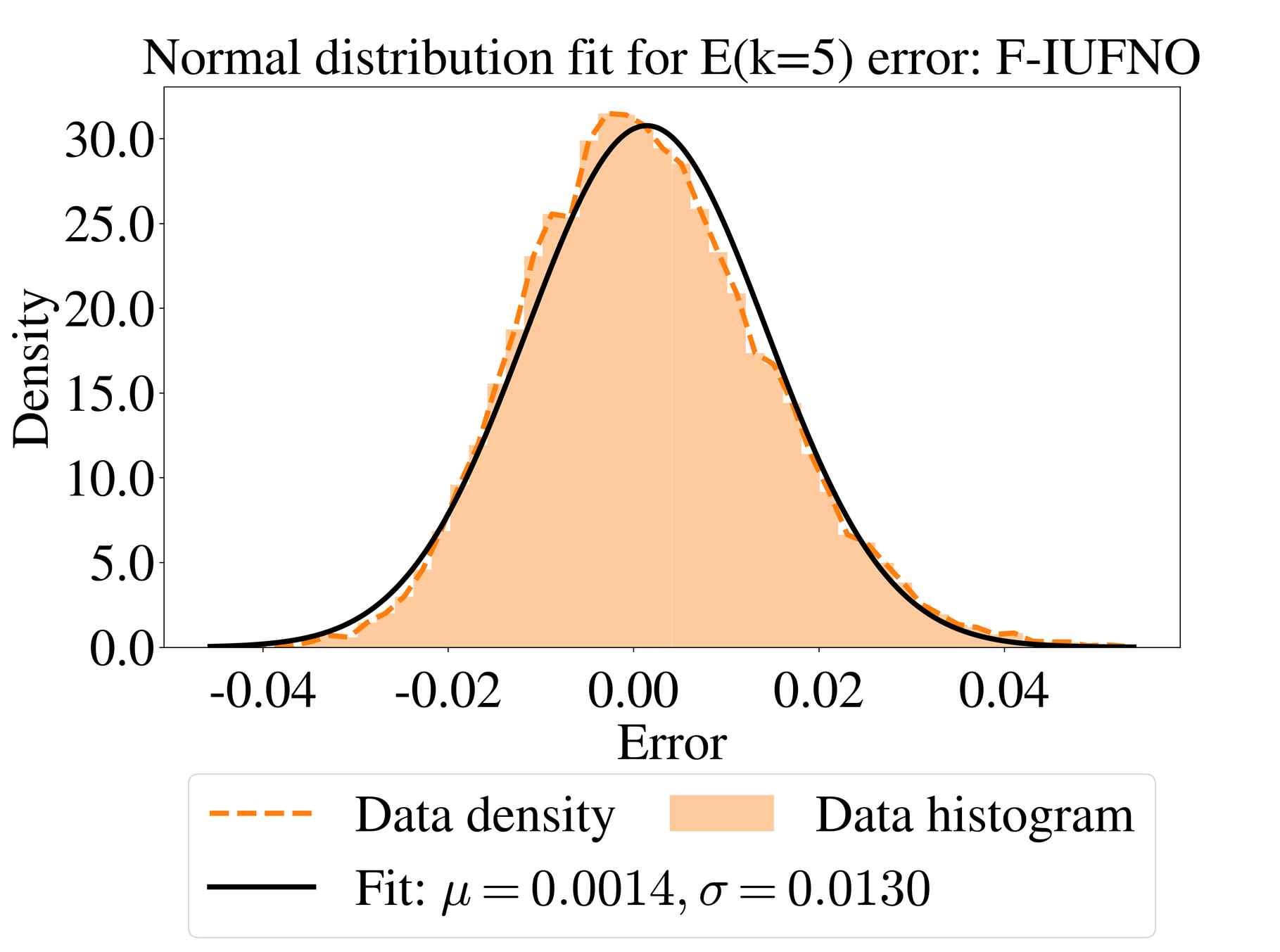}
            \put(-3,60){\small (c)} 
        \end{overpic}
    \end{subfigure}
    \vspace{0.1cm}

    \begin{subfigure}[b]{0.32\textwidth}
        \begin{overpic}[width=1\linewidth]{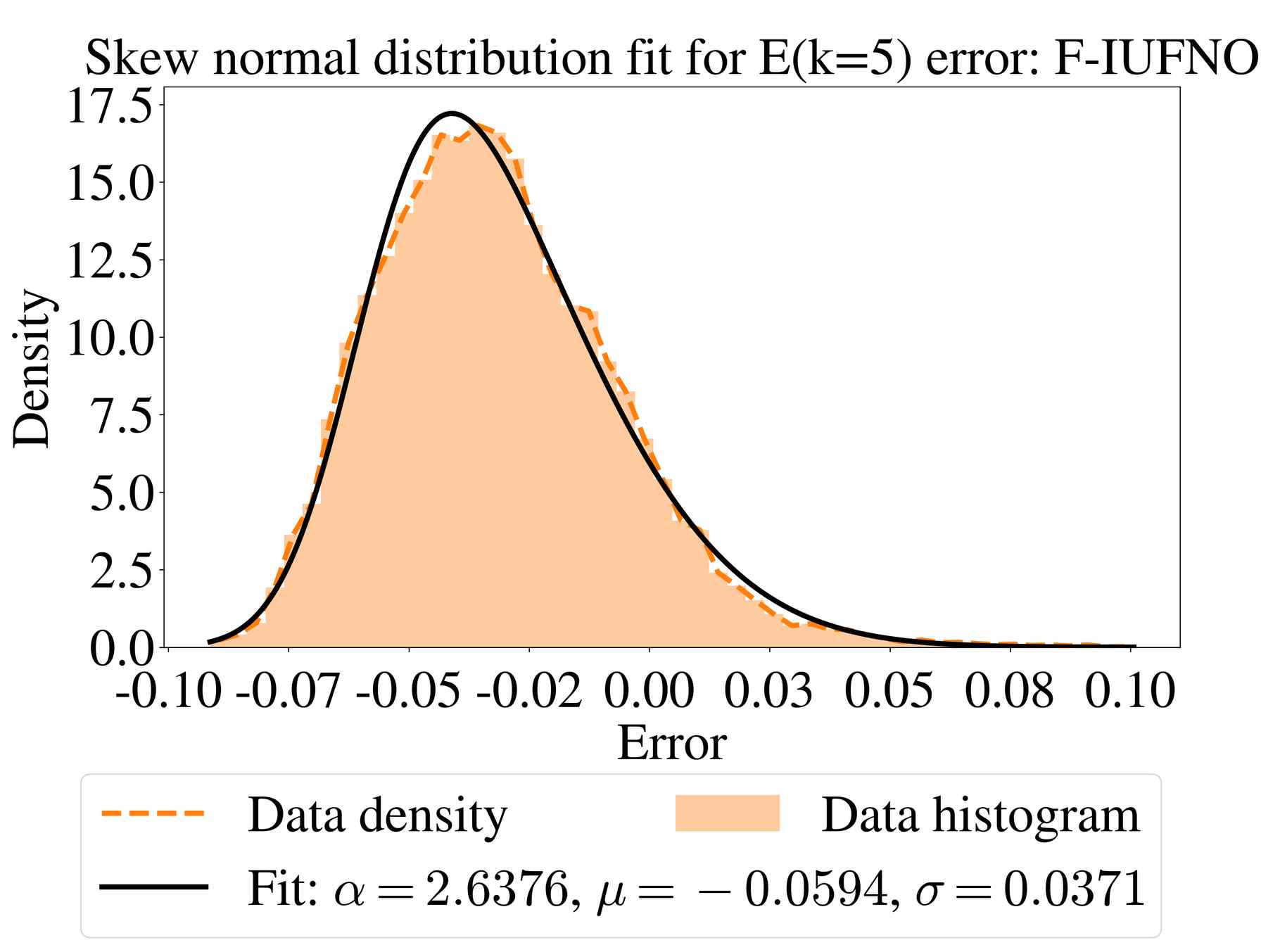}
            \put(-4,60){\small (d)}  
        \end{overpic}
    \end{subfigure}
    \hfill
    \begin{subfigure}[b]{0.32\textwidth}
        \begin{overpic}[width=1\linewidth]{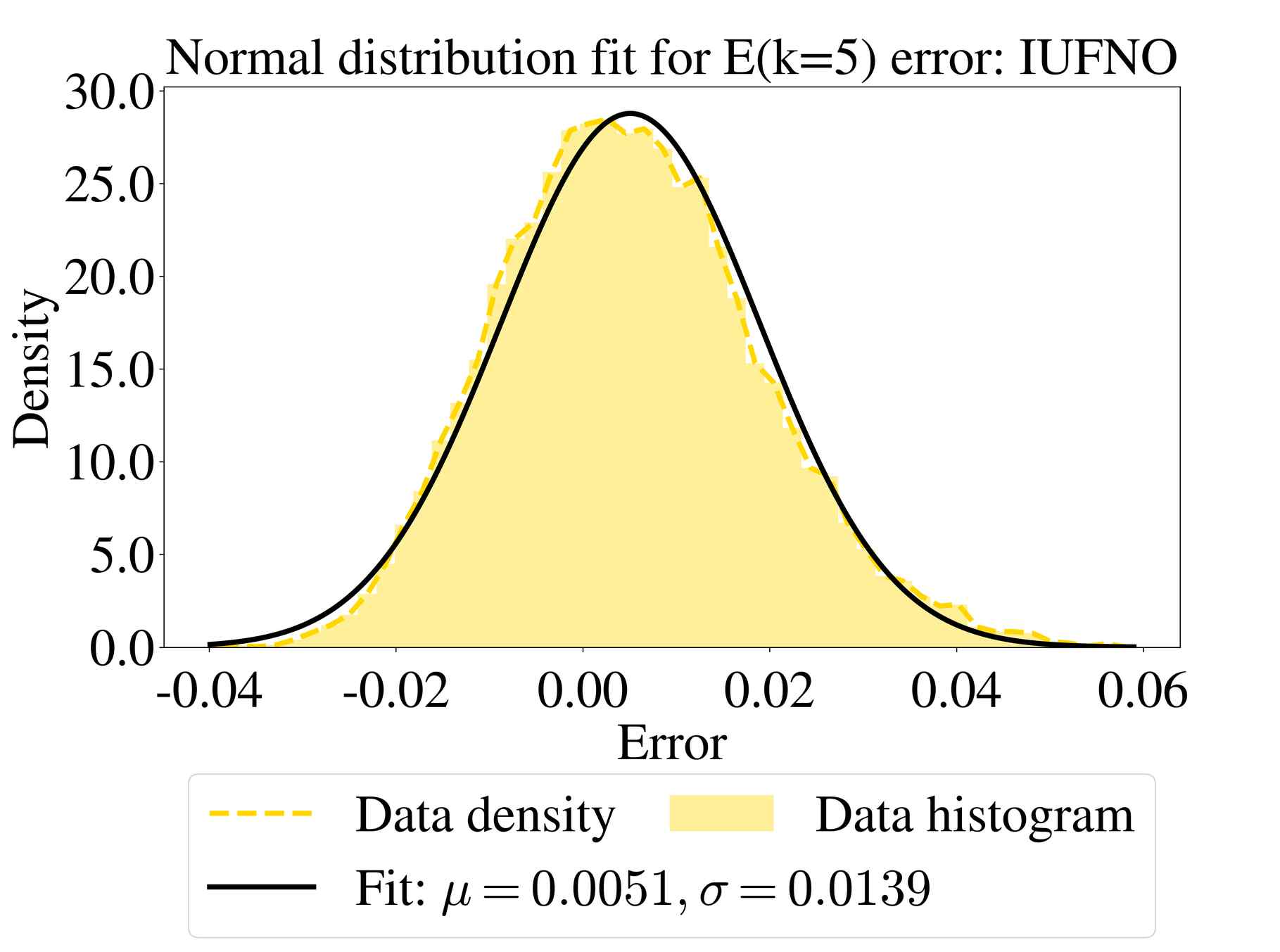}
            \put(-3,60){\small (e)} 
        \end{overpic} 
    \end{subfigure}
    \hfill
    \begin{subfigure}[b]{0.32\textwidth}
        \begin{overpic}[width=1\linewidth]{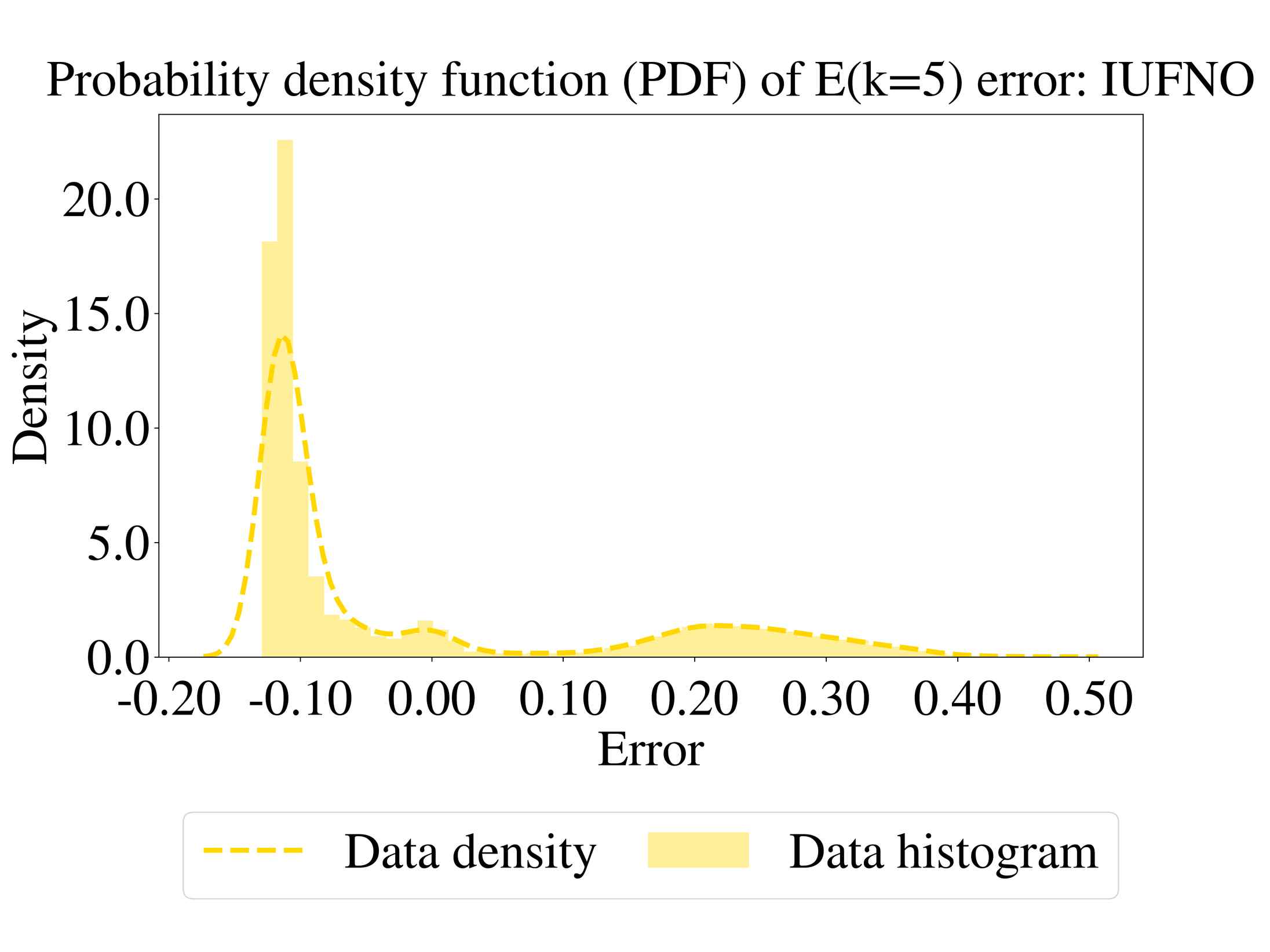}
            \put(-3,60){\small (f)} 
        \end{overpic}
    \end{subfigure}
    \vspace{0.1cm}

    \begin{subfigure}[b]{0.32\textwidth}
        \begin{overpic}[width=1\linewidth]{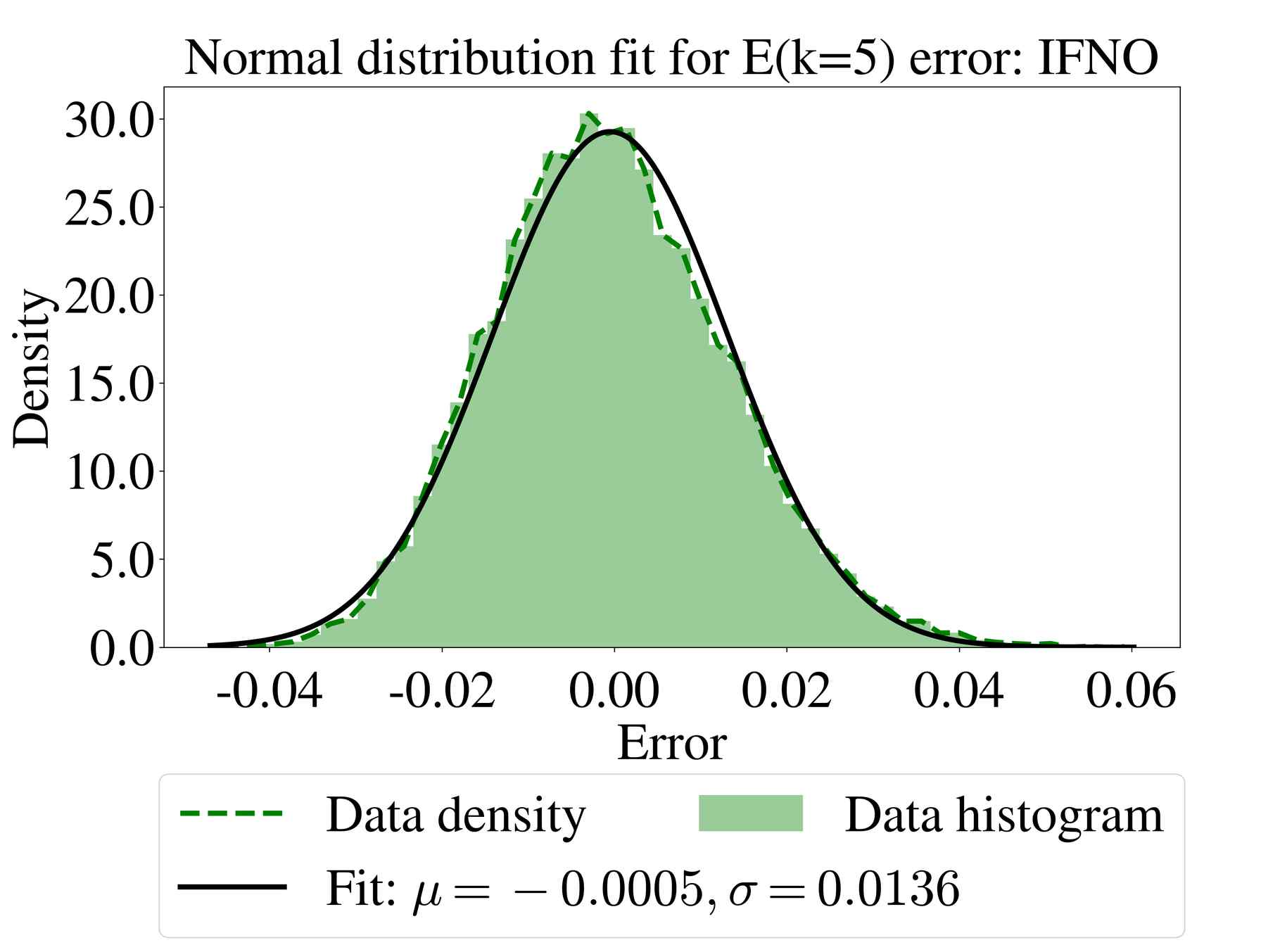}
            \put(-4,60){\small (g)}  
        \end{overpic}
    \end{subfigure}
    \hfill
    \begin{subfigure}[b]{0.32\textwidth}
        \begin{overpic}[width=1\linewidth]{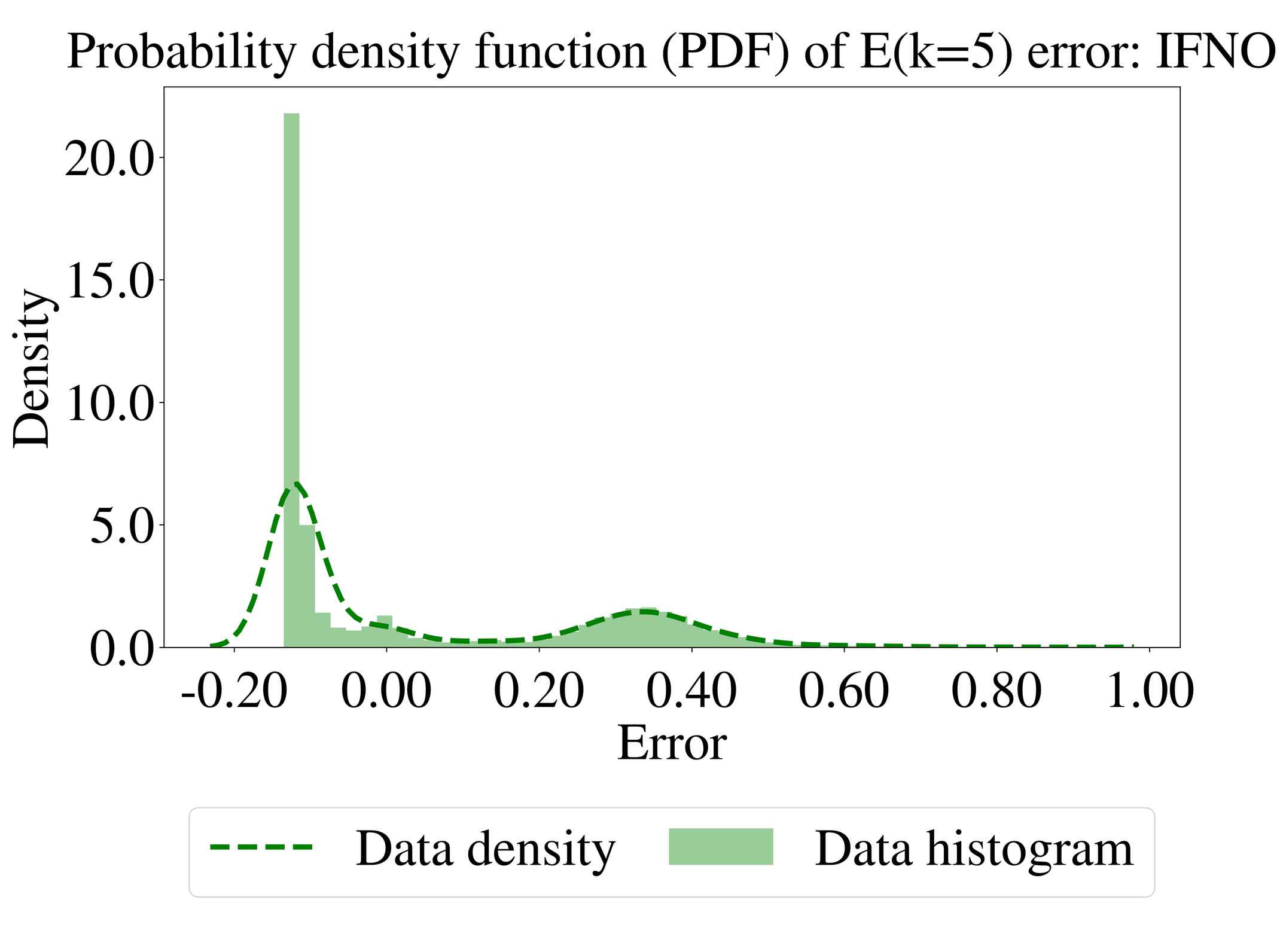}
            \put(-3,60){\small (h)} 
        \end{overpic} 
    \end{subfigure}
    \hfill
    \begin{subfigure}[b]{0.32\textwidth}
        \begin{overpic}[width=1\linewidth]{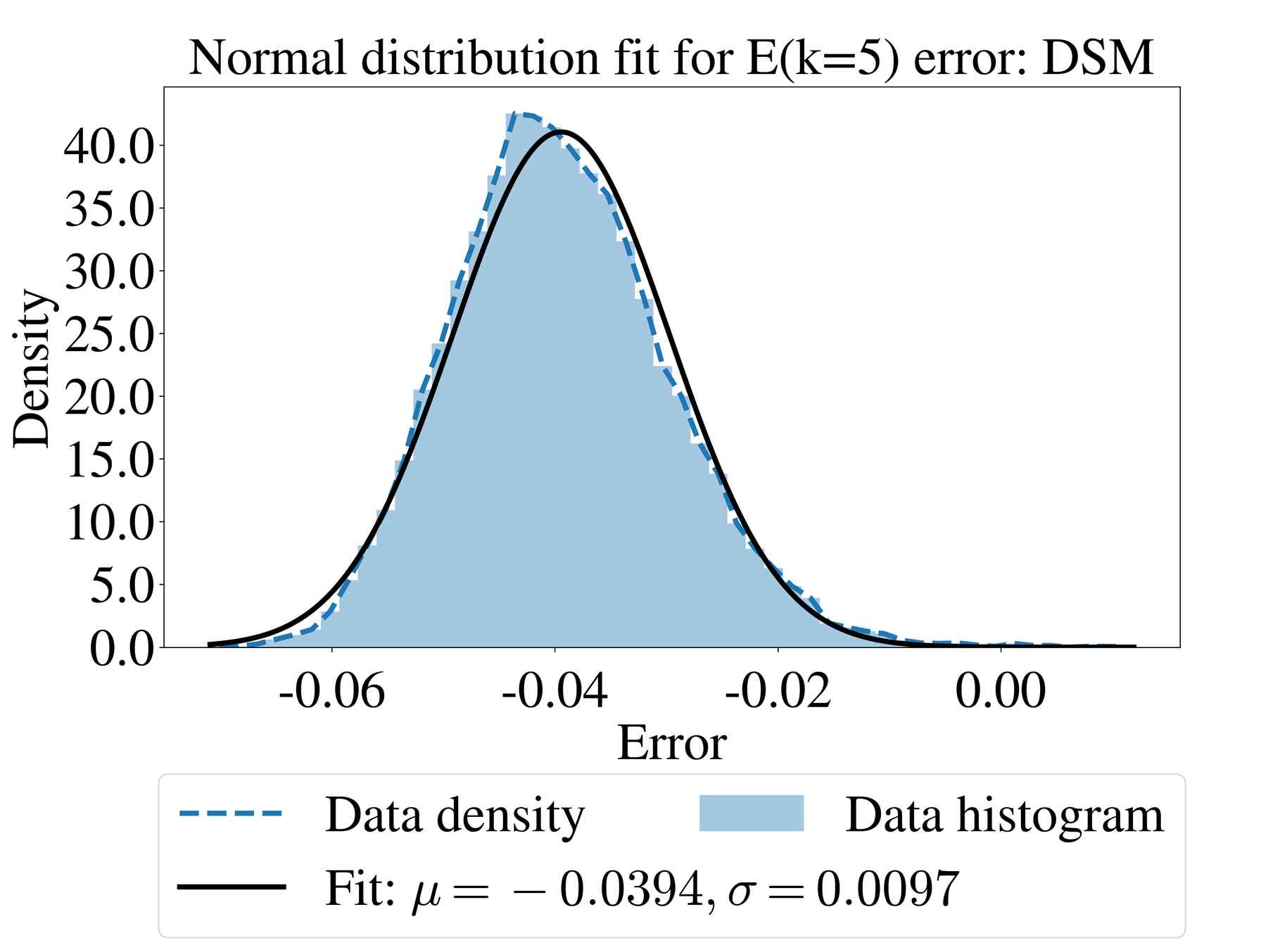}
            \put(-3,60){\small (i)} 
        \end{overpic}
    \end{subfigure}
    \vspace{0.1cm}

    \begin{subfigure}[b]{1\textwidth}
        \centering
        \begin{overpic}[width=0.32\linewidth]{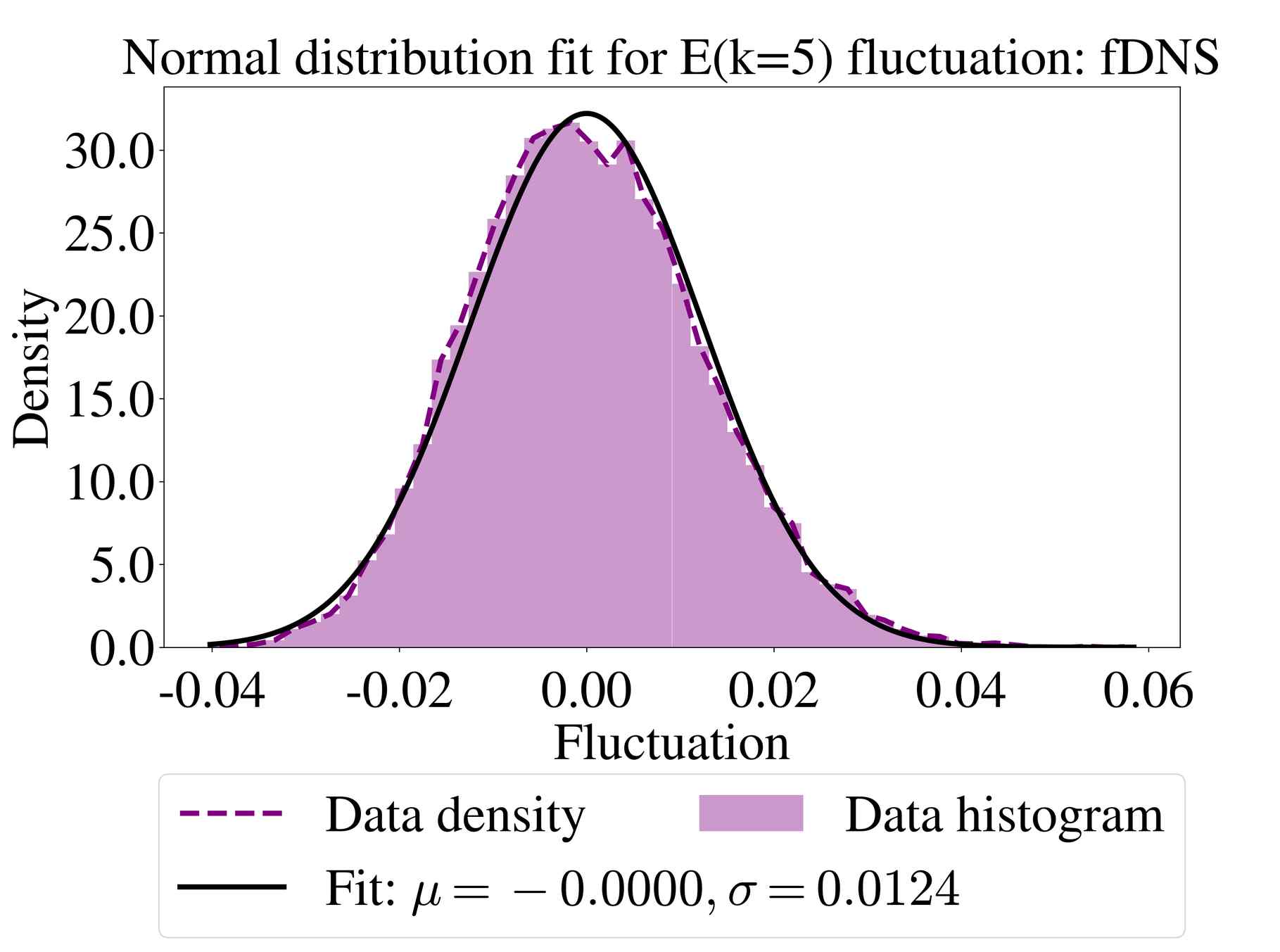}
            \put(-3,60){\small (j)}  
        \end{overpic}
    \end{subfigure}

	\caption{The PDFs of the $E(k=5)$ errors for each method at time interval $\Delta T=0.2\tau$: (a) F-IFNO constrained; (b) F-IFNO unconstrained; (c) F-IUFNO constrained; (d) F-IUFNO unconstrained; (e) IUFNO constrained; (f) IUFNO unconstrained; (g) IFNO constrained; (h) IFNO unconstrained; (i) DSM; (j) fDNS. Note that for fDNS, the values represent natural statistical fluctuations over time, not prediction errors.}\label{fig:19}
\end{figure}

\begin{figure}[ht!]
    \centering
    \begin{subfigure}[b]{0.32\textwidth}
        \begin{overpic}[width=1\linewidth]{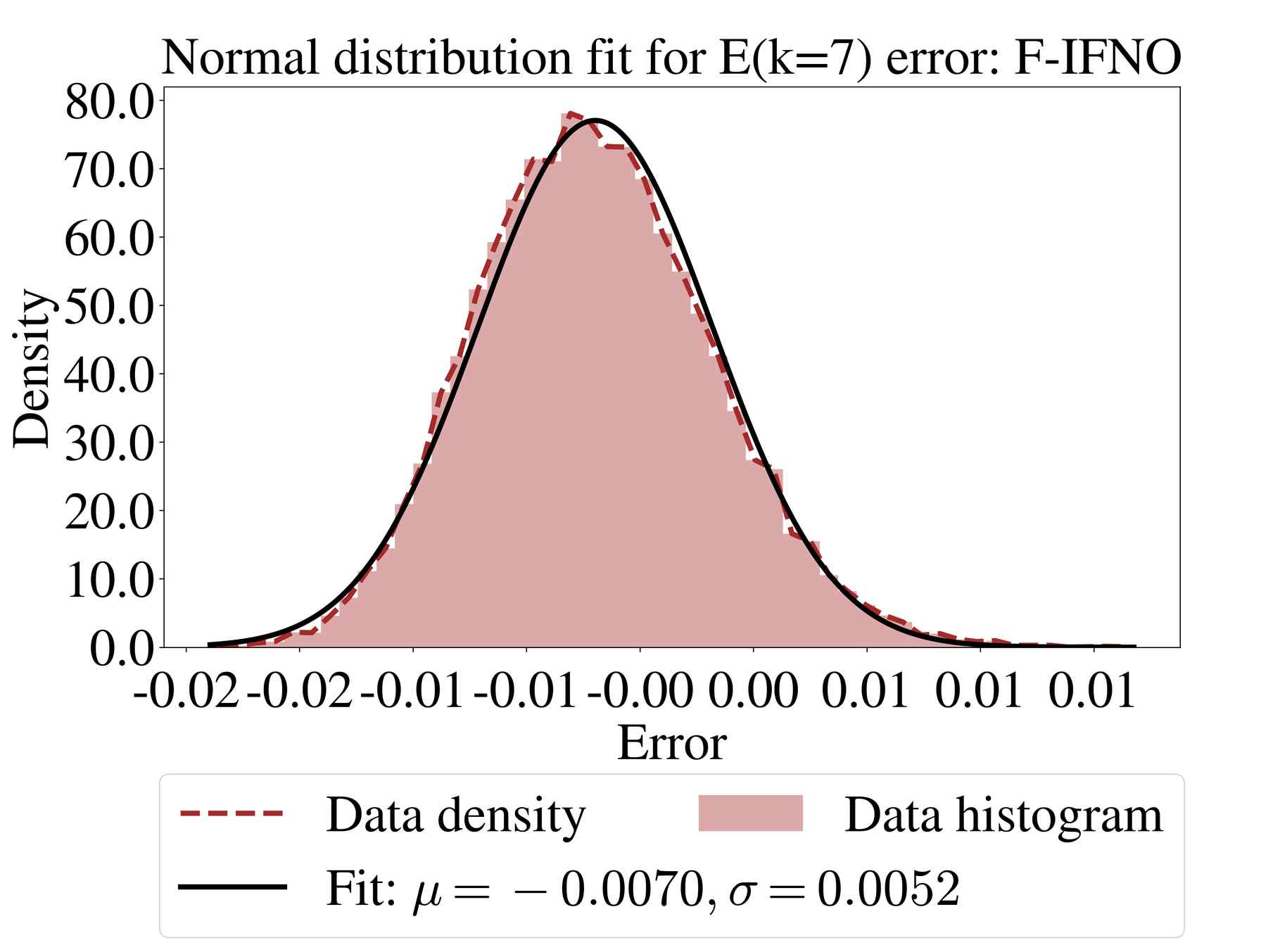}
            \put(-4,60){\small (a)}  
        \end{overpic}
    \end{subfigure}
    \hfill
    \begin{subfigure}[b]{0.32\textwidth}
        \begin{overpic}[width=1\linewidth]{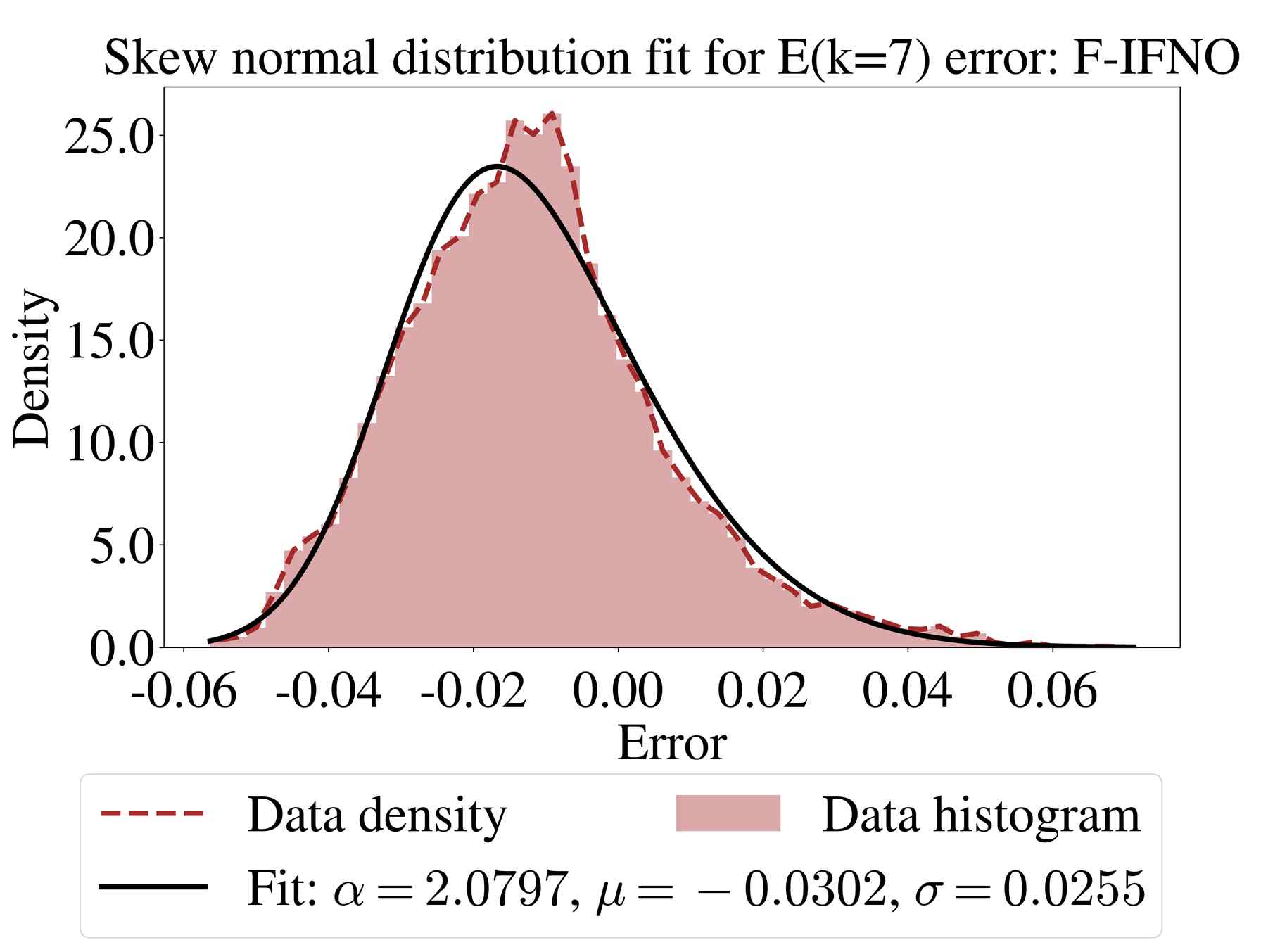}
            \put(-4,60){\small (b)} 
        \end{overpic} 
    \end{subfigure}
    \hfill
    \begin{subfigure}[b]{0.32\textwidth}
        \begin{overpic}[width=1\linewidth]{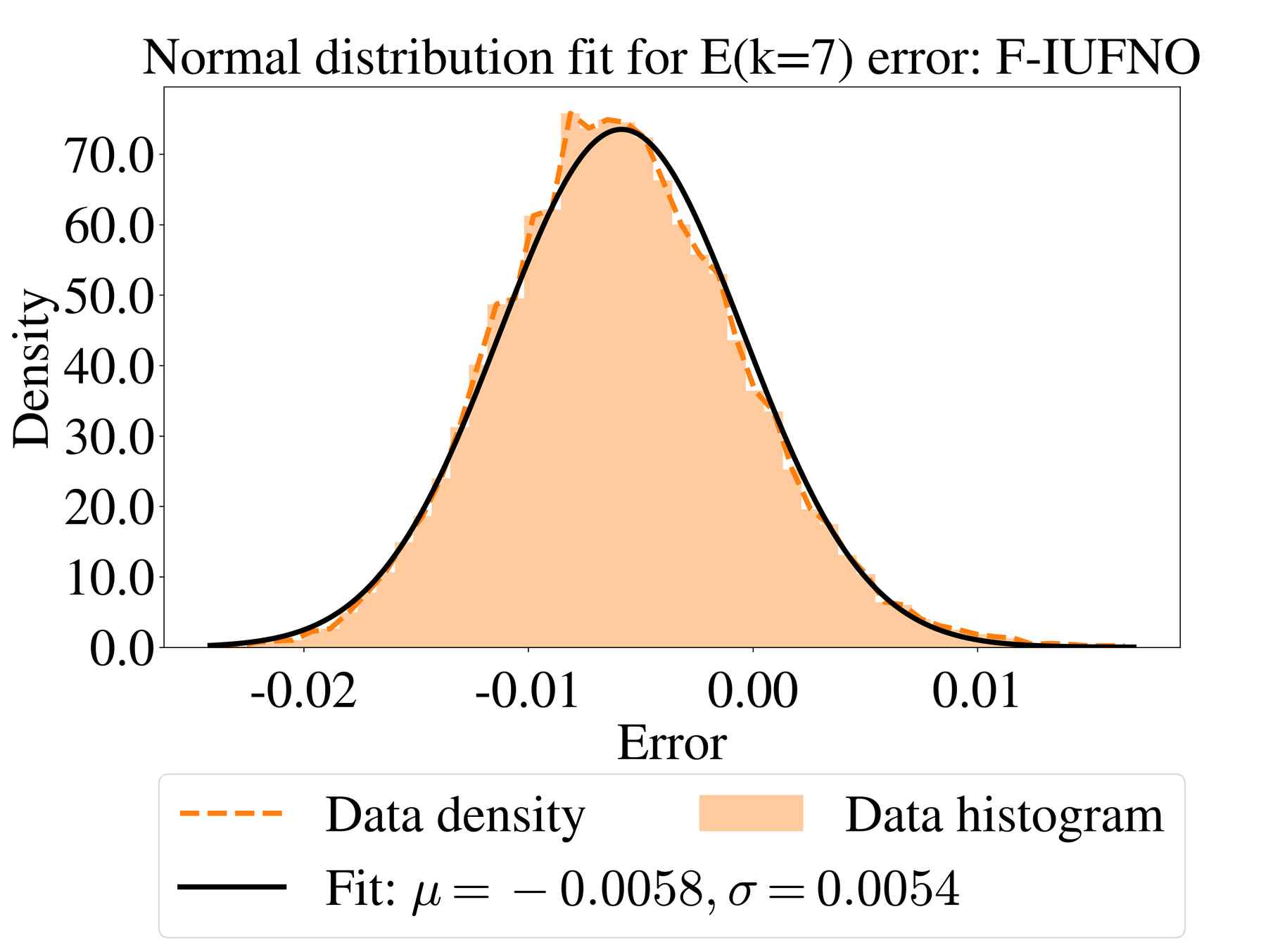}
            \put(-4,60){\small (c)} 
        \end{overpic}
    \end{subfigure}
    \vspace{0.1cm}

    \begin{subfigure}[b]{0.32\textwidth}
        \begin{overpic}[width=1\linewidth]{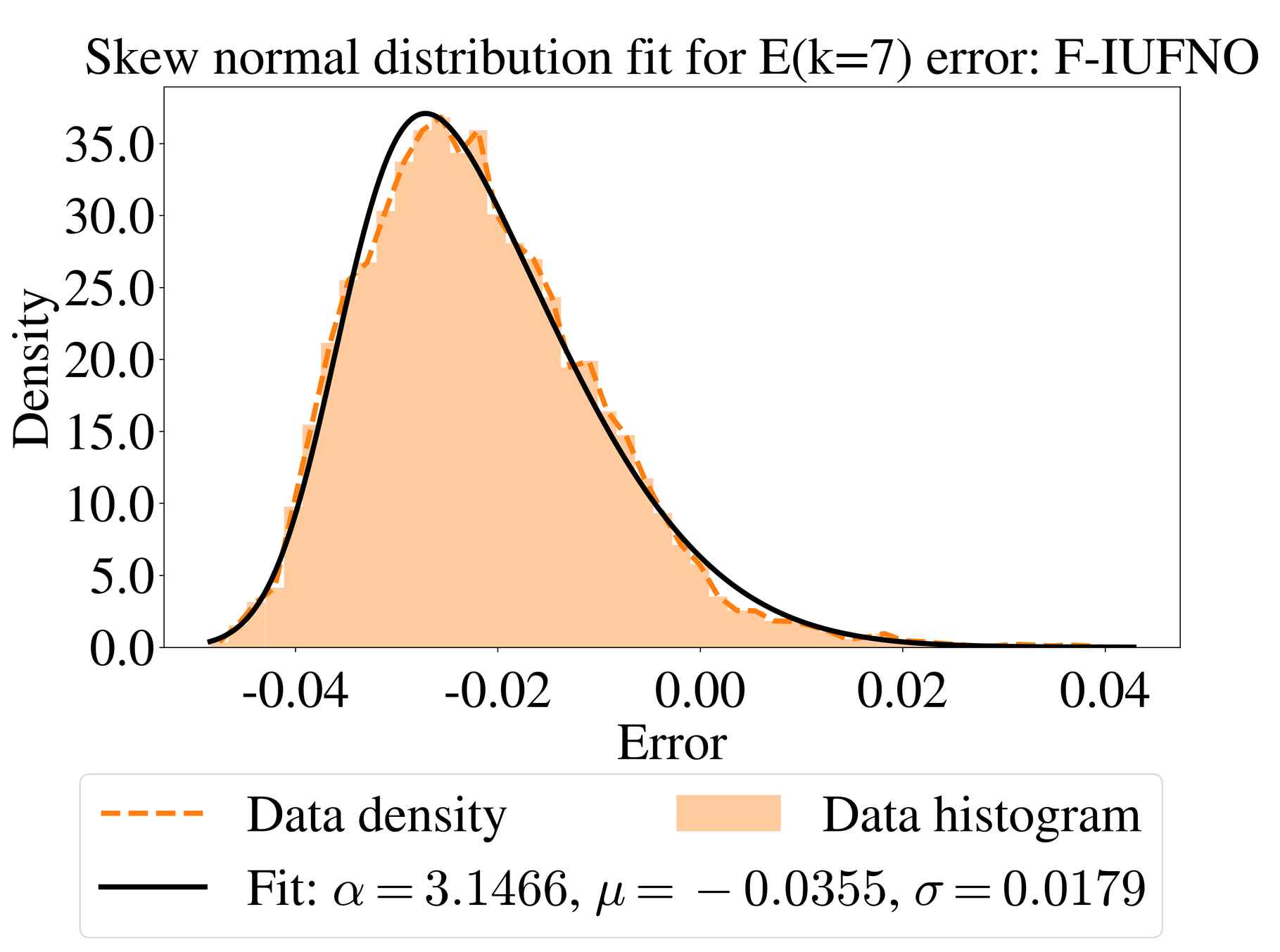}
            \put(-4,60){\small (d)}  
        \end{overpic}
    \end{subfigure}
    \hfill
    \begin{subfigure}[b]{0.32\textwidth}
        \begin{overpic}[width=1\linewidth]{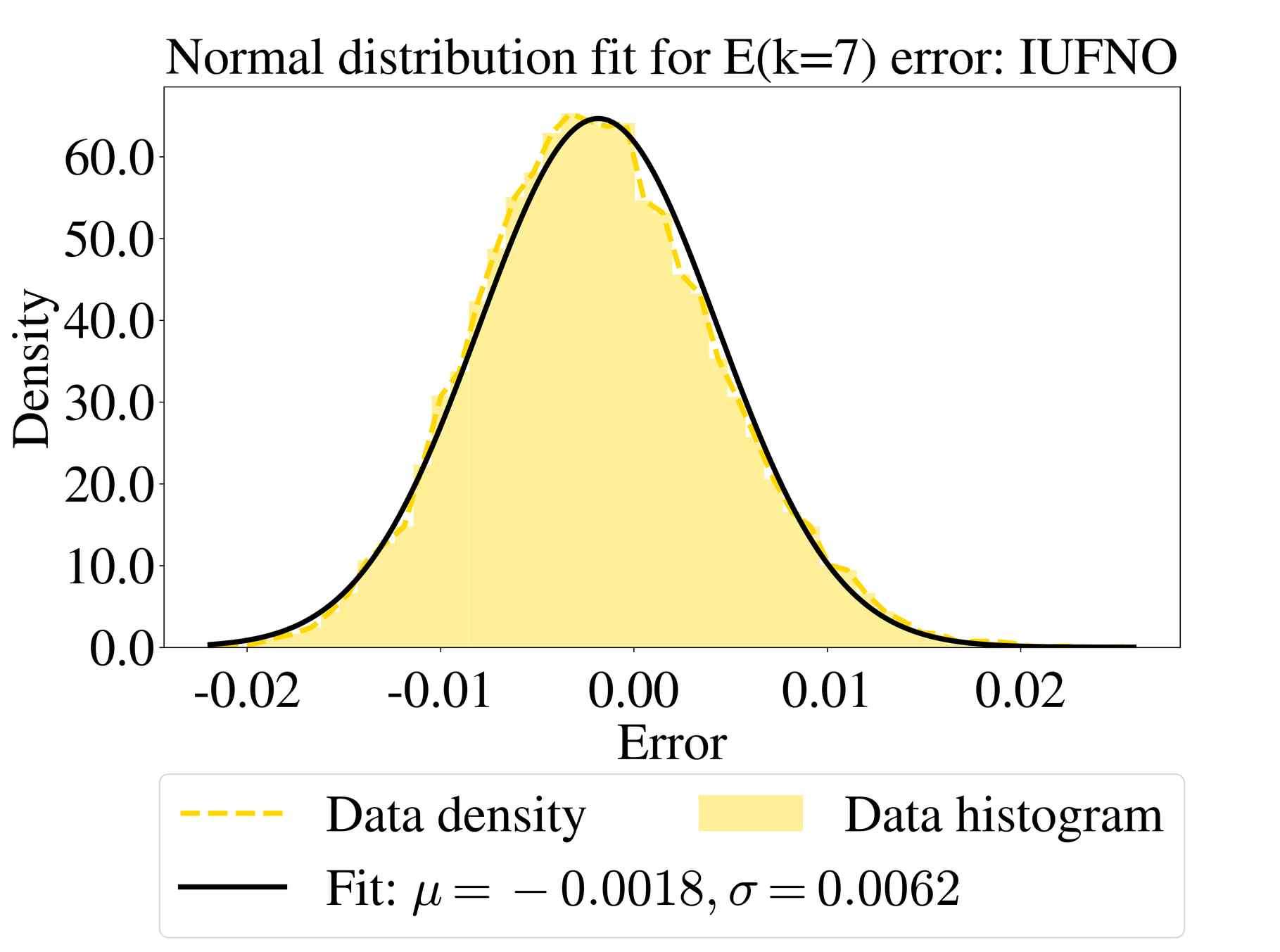}
            \put(-4,60){\small (e)} 
        \end{overpic} 
    \end{subfigure}
    \hfill
    \begin{subfigure}[b]{0.32\textwidth}
        \begin{overpic}[width=1\linewidth]{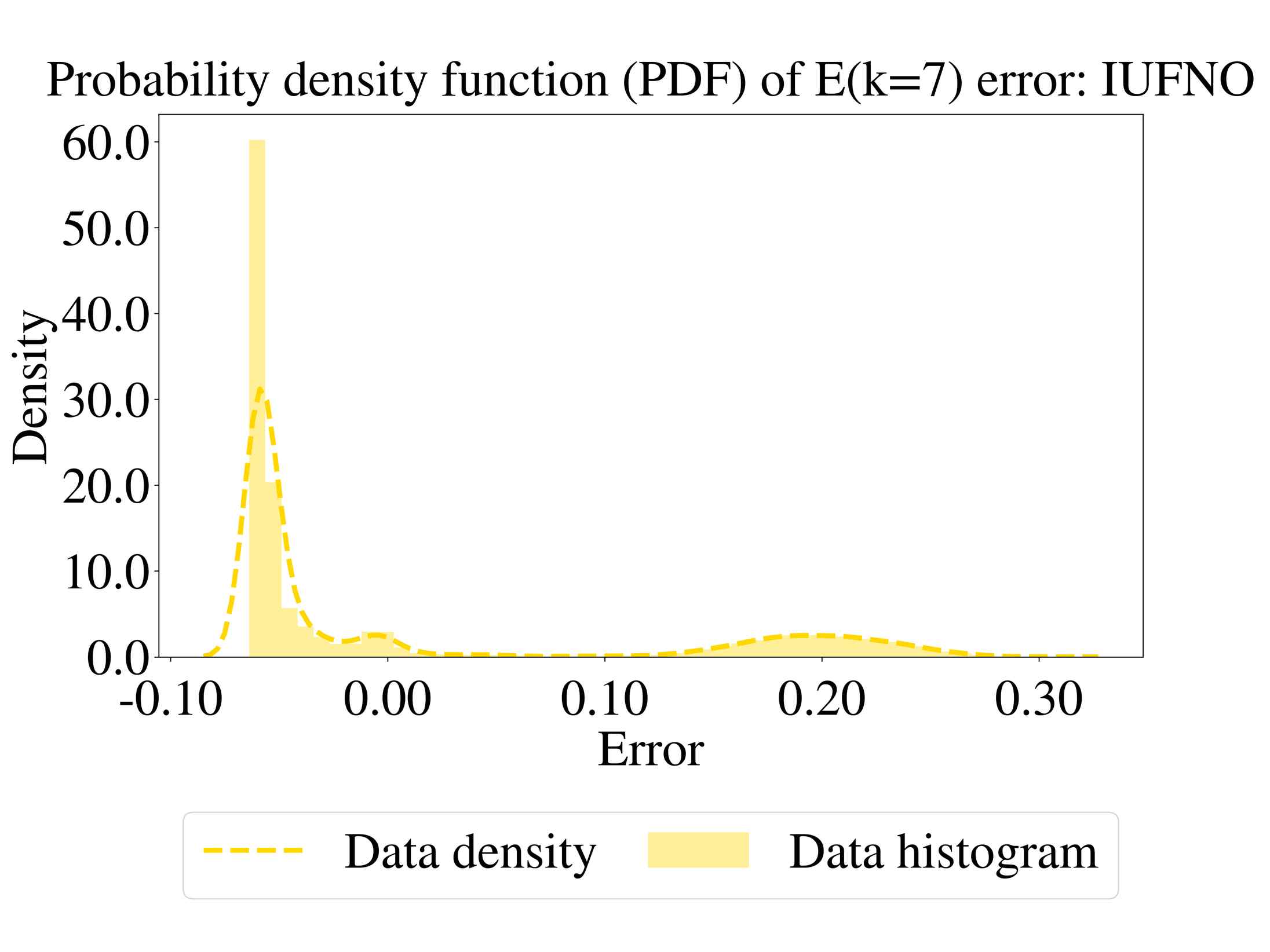}
            \put(-4,60){\small (f)} 
        \end{overpic}
    \end{subfigure}
    \vspace{0.1cm}

    \begin{subfigure}[b]{0.32\textwidth}
        \begin{overpic}[width=1\linewidth]{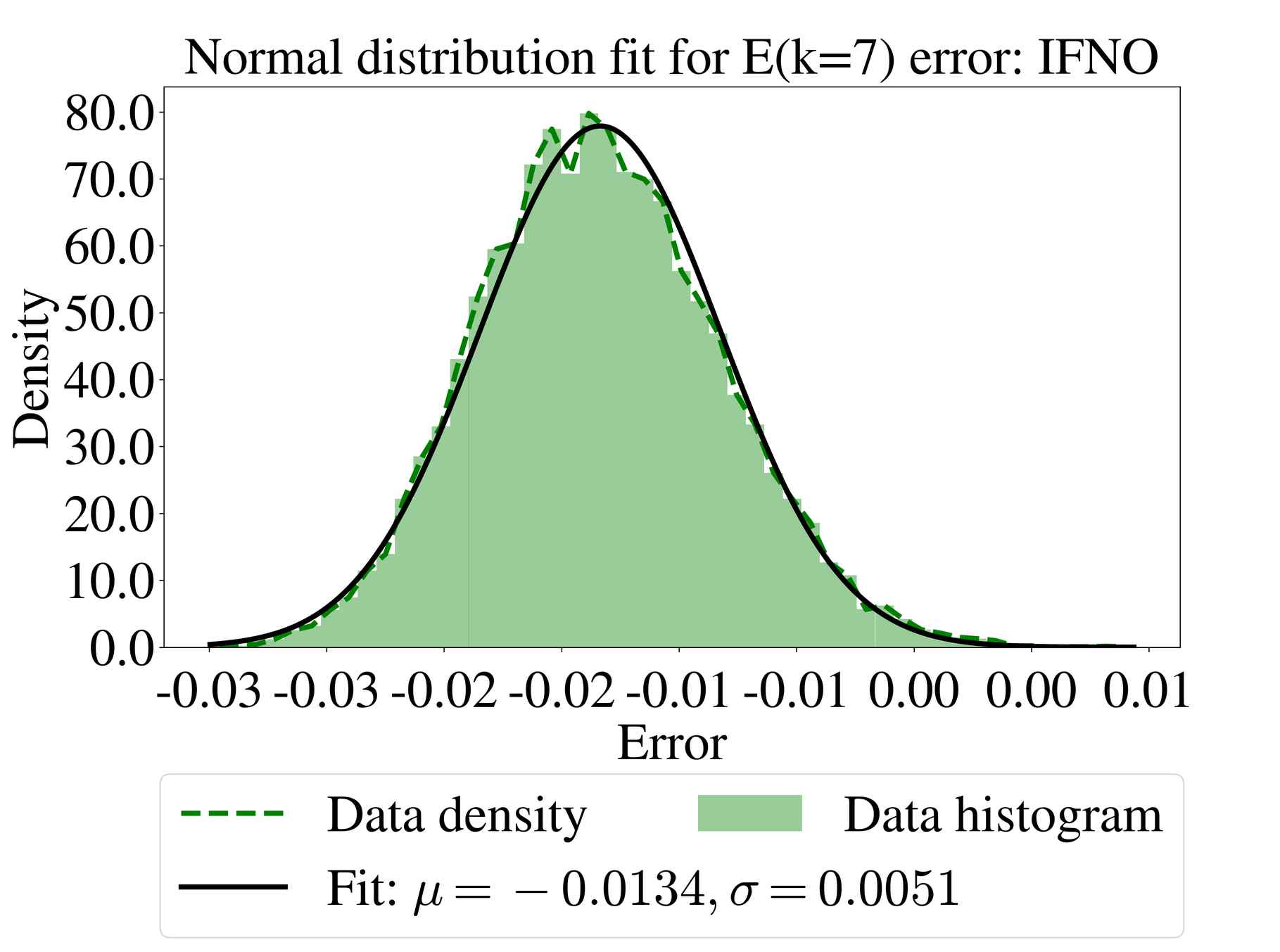}
            \put(-4,60){\small (g)}  
        \end{overpic}
    \end{subfigure}
    \hfill
    \begin{subfigure}[b]{0.32\textwidth}
        \begin{overpic}[width=1\linewidth]{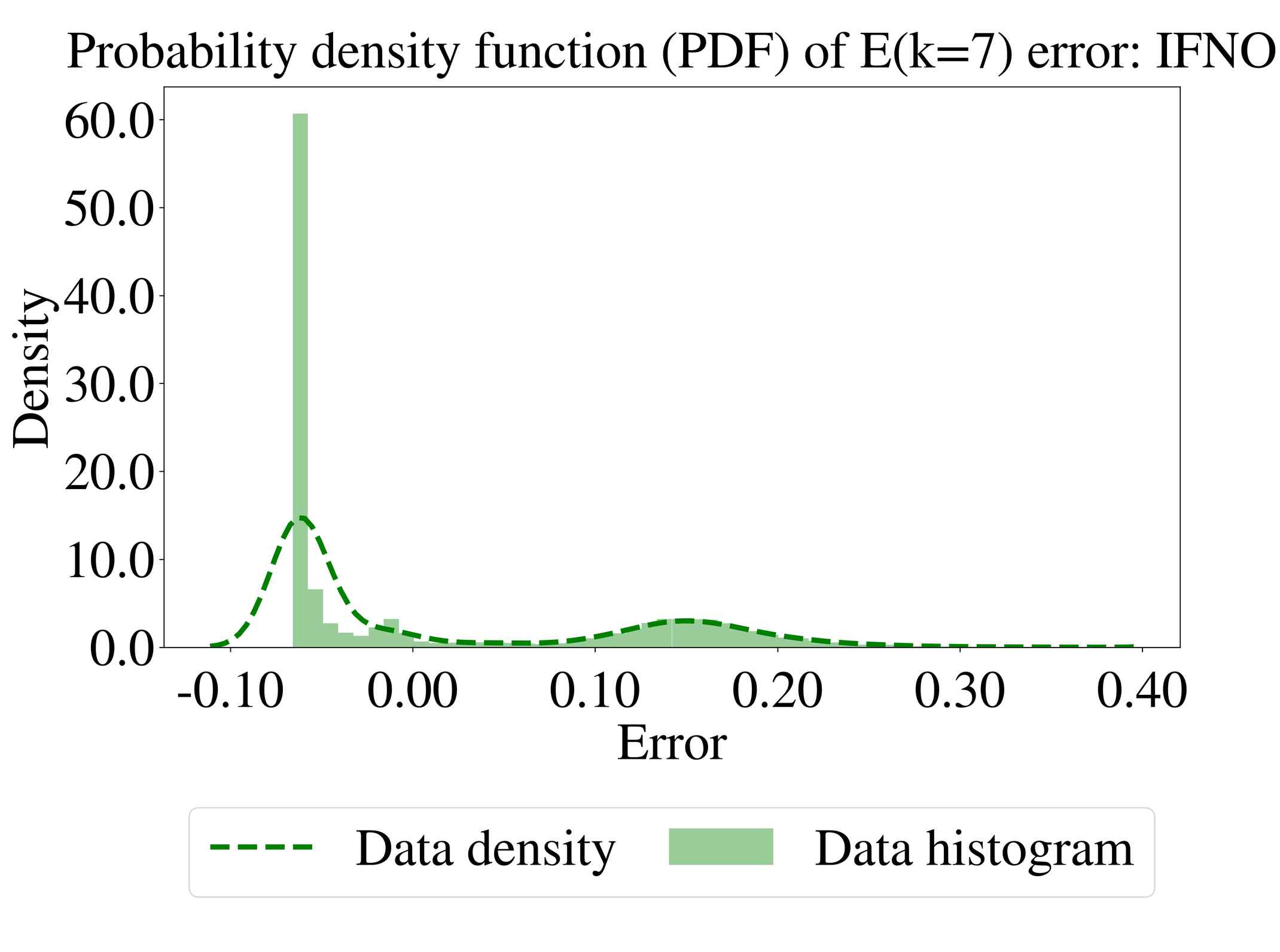}
            \put(-4,60){\small (h)} 
        \end{overpic} 
    \end{subfigure}
    \hfill
    \begin{subfigure}[b]{0.32\textwidth}
        \begin{overpic}[width=1\linewidth]{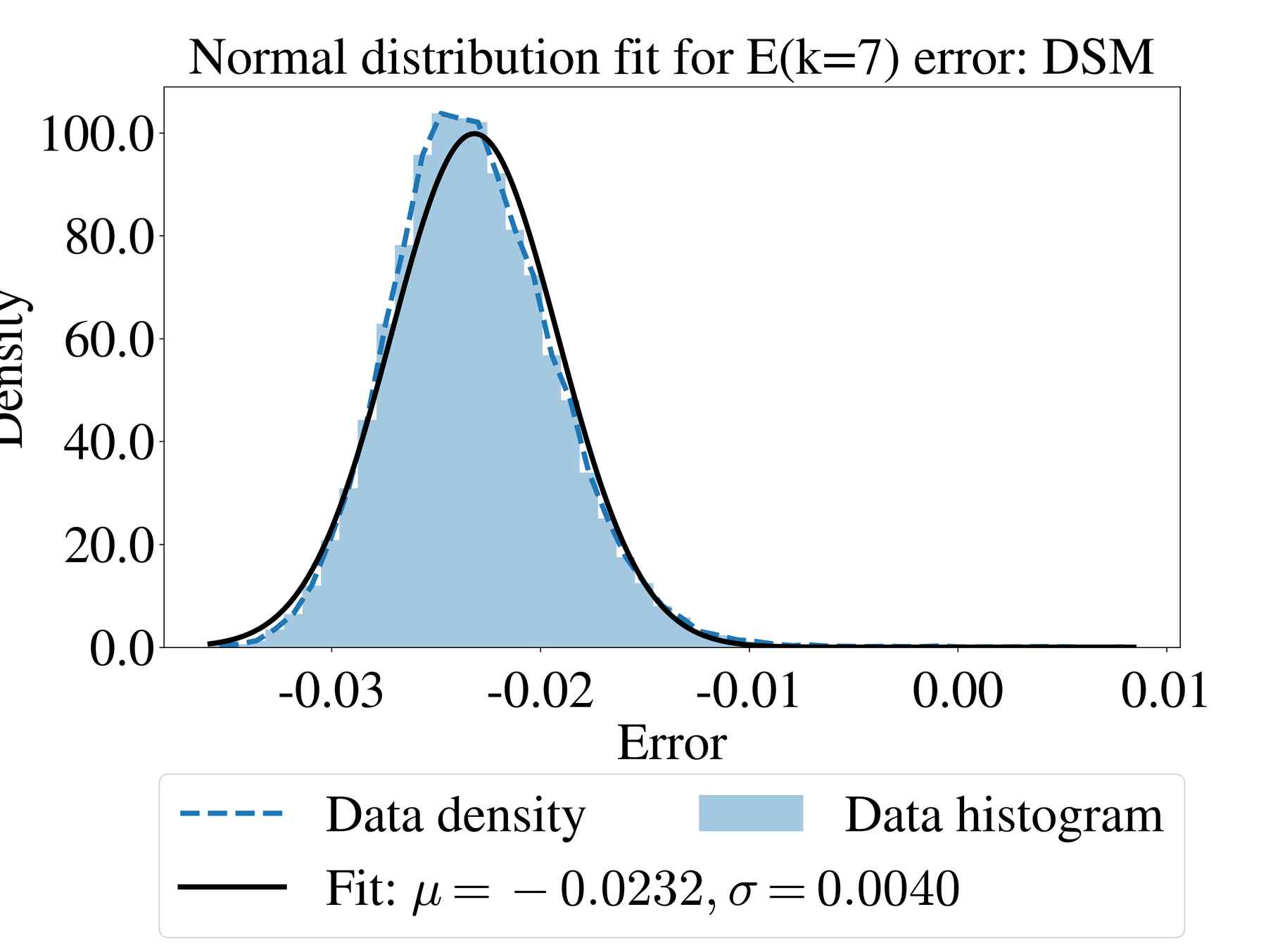}
            \put(-4,60){\small (i)} 
        \end{overpic}
    \end{subfigure}
    \vspace{0.1cm}

    \begin{subfigure}[b]{1\textwidth}
        \centering
        \begin{overpic}[width=0.32\linewidth]{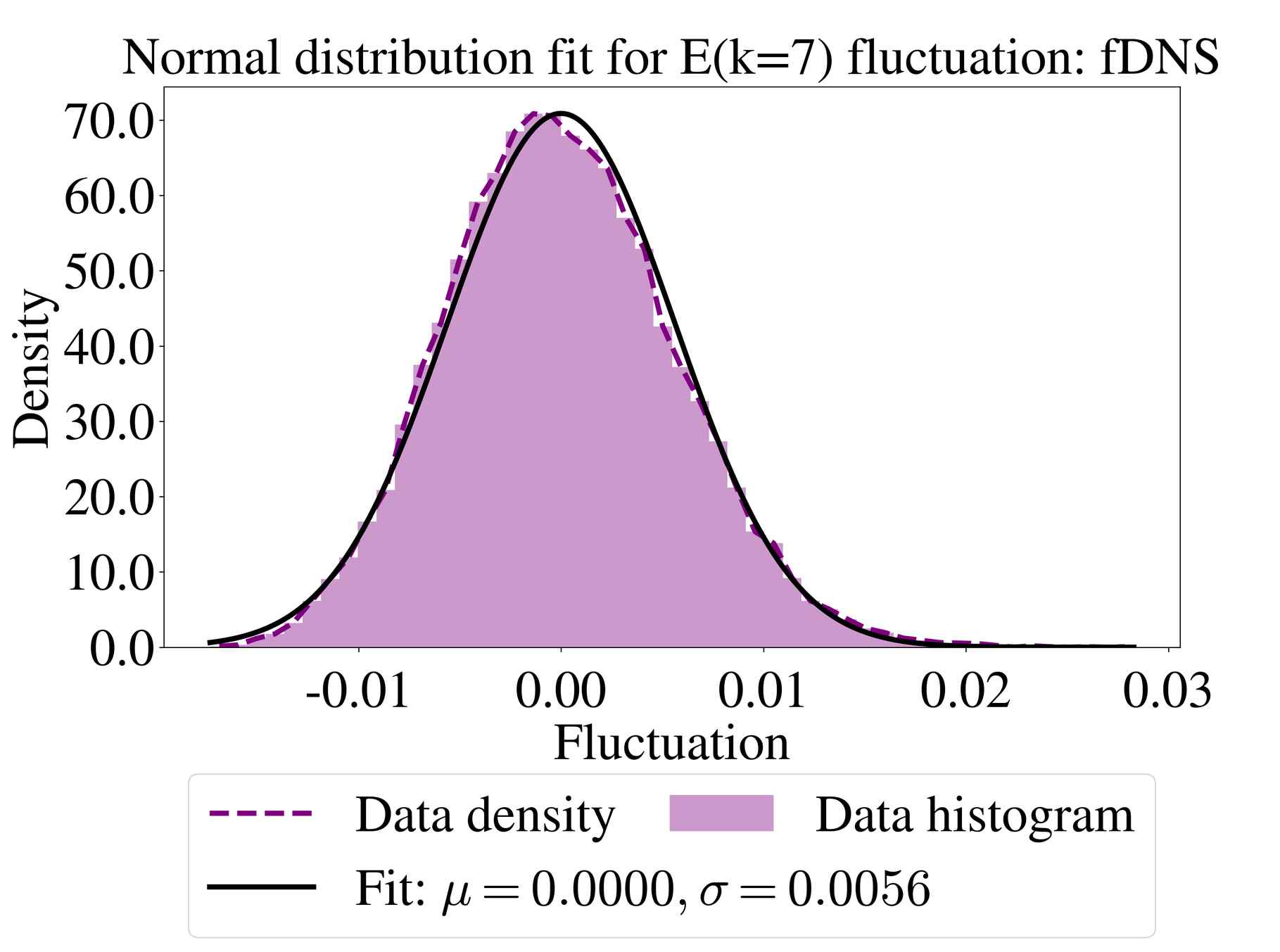}
            \put(-4,60){\small (j)}  
        \end{overpic}
    \end{subfigure}

	\caption{The PDFs of the $E(k=7)$ errors for each method at time interval $\Delta T=0.2\tau$: (a) F-IFNO constrained; (b) F-IFNO unconstrained; (c) F-IUFNO constrained; (d) F-IUFNO unconstrained; (e) IUFNO constrained; (f) IUFNO unconstrained; (g) IFNO constrained; (h) IFNO unconstrained; (i) DSM; (j) fDNS. Note that for fDNS, the values represent natural statistical fluctuations over time, not prediction errors.}\label{fig:21}
\end{figure}

Furthermore, following the methodology used in Fig.~\ref{fig:12}, we also plot the probability density functions (PDFs) of the $E(k=3)$, $E(k=5)$, and $E(k=7)$ errors for each method at the time interval $\Delta T = 0.2\tau$ in Fig.~\ref{fig:17}, Fig.~\ref{fig:19}, and Fig.~\ref{fig:21}, respectively. 
From Figs.~\ref{fig:17}, \ref{fig:19}, and \ref{fig:21}, similar conclusions can be drawn: for the constrained FNO-based models, as well as for fDNS and DSM, the fluctuation or error distributions of $E(k=3)$, $E(k=5)$, and $E(k=7)$ follow a normal distribution. In contrast, the error distributions of unconstrained F-IFNO and F-IUFNO follow a skew normal distribution. Moreover, for unconstrained IFNO and IUFNO, no reasonable distribution (either normal or skew normal) can adequately fit their error distributions.
As a consequence, constrained FNO-based models demonstrate error distribution characteristics and predictive accuracy comparable to fDNS, while unconstrained F-IFNO and F-IUFNO exhibit similar performance to DSM. On the other hand, unconstrained IFNO and IUFNO yield the poorest predictive performance among all the methods.

\begin{figure}[ht!]
    \centering
    \begin{subfigure}[b]{0.32\textwidth}
        \begin{overpic}[width=1\linewidth]{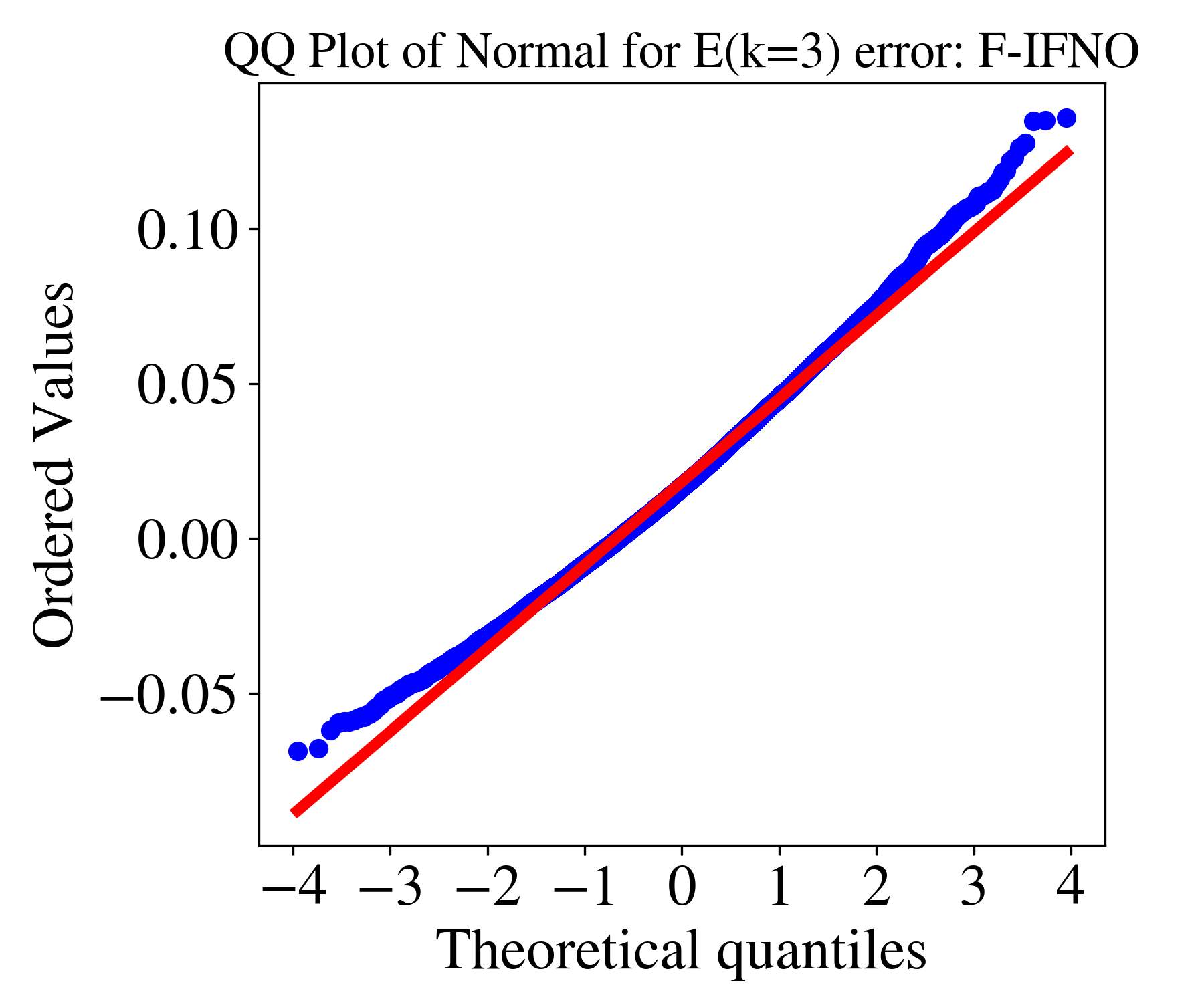}
            \put(-3,65){\small (a)}  
        \end{overpic}
    \end{subfigure}
    \hfill
    \begin{subfigure}[b]{0.32\textwidth}
        \begin{overpic}[width=1\linewidth]{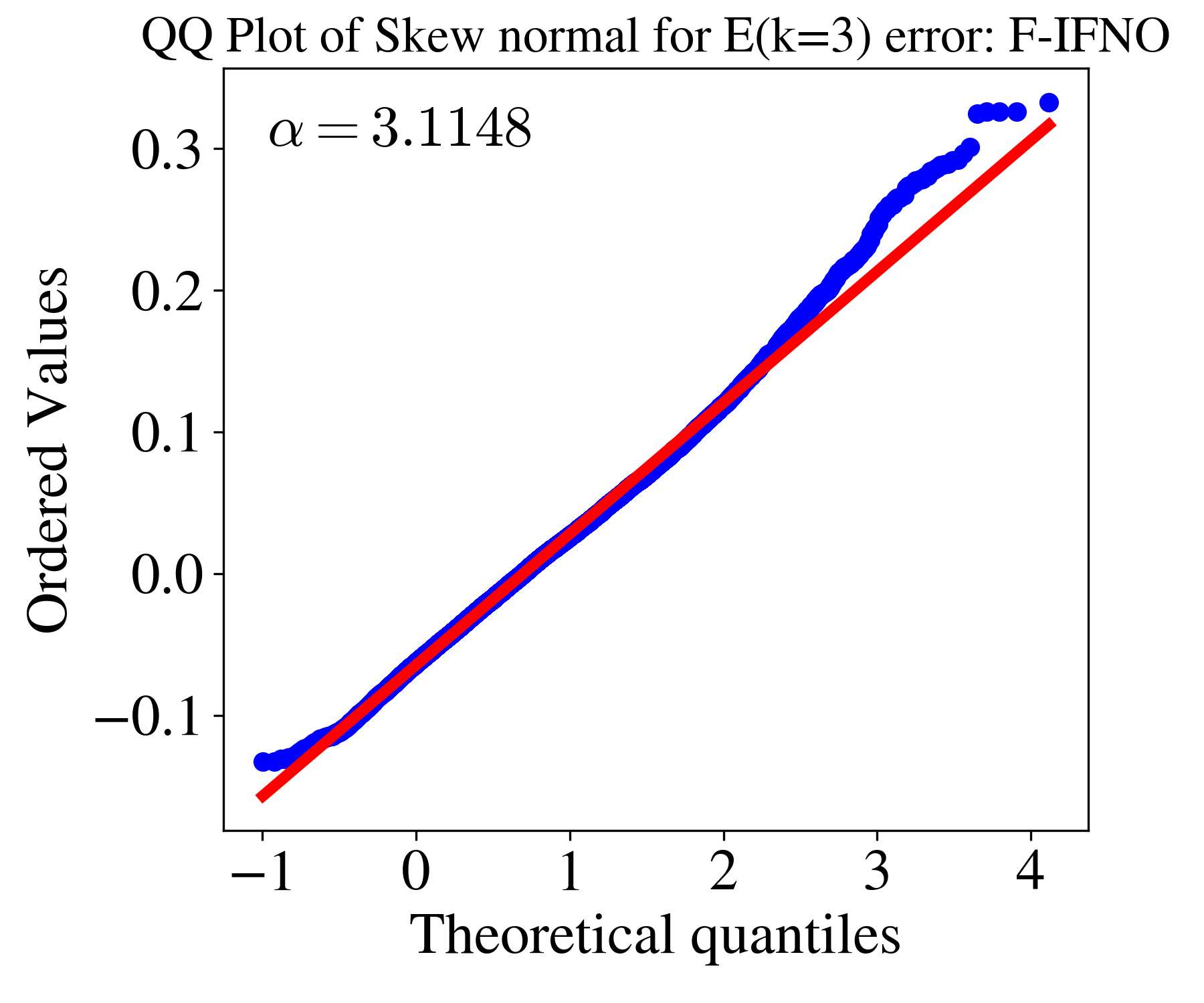}
            \put(-3,65){\small (b)} 
        \end{overpic} 
    \end{subfigure}
    \hfill
    \begin{subfigure}[b]{0.32\textwidth}
        \begin{overpic}[width=1\linewidth]{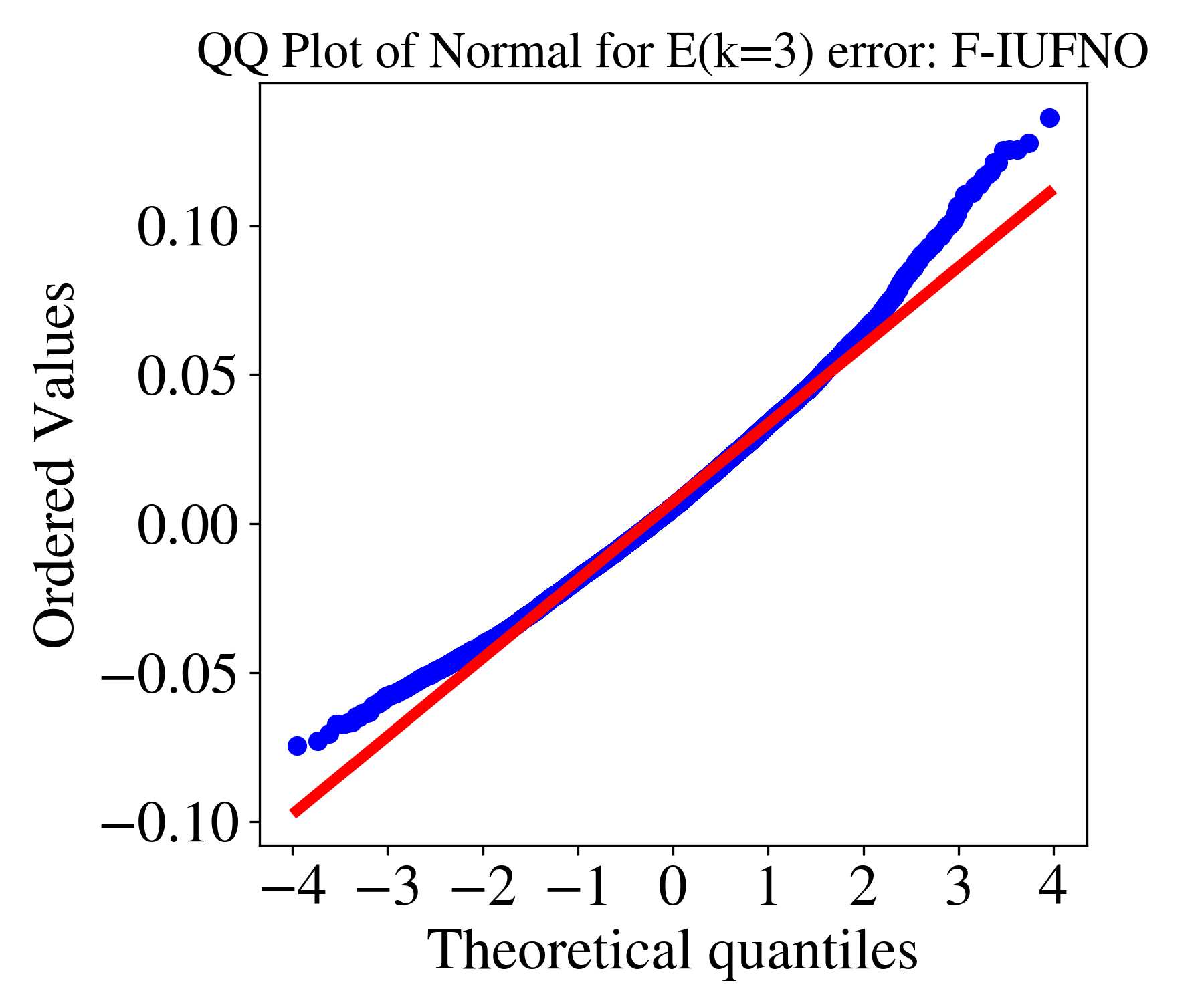}
            \put(-3,65){\small (c)} 
        \end{overpic}
    \end{subfigure}
    \vspace{0.1cm}

    \begin{subfigure}[b]{0.32\textwidth}
        \begin{overpic}[width=1\linewidth]{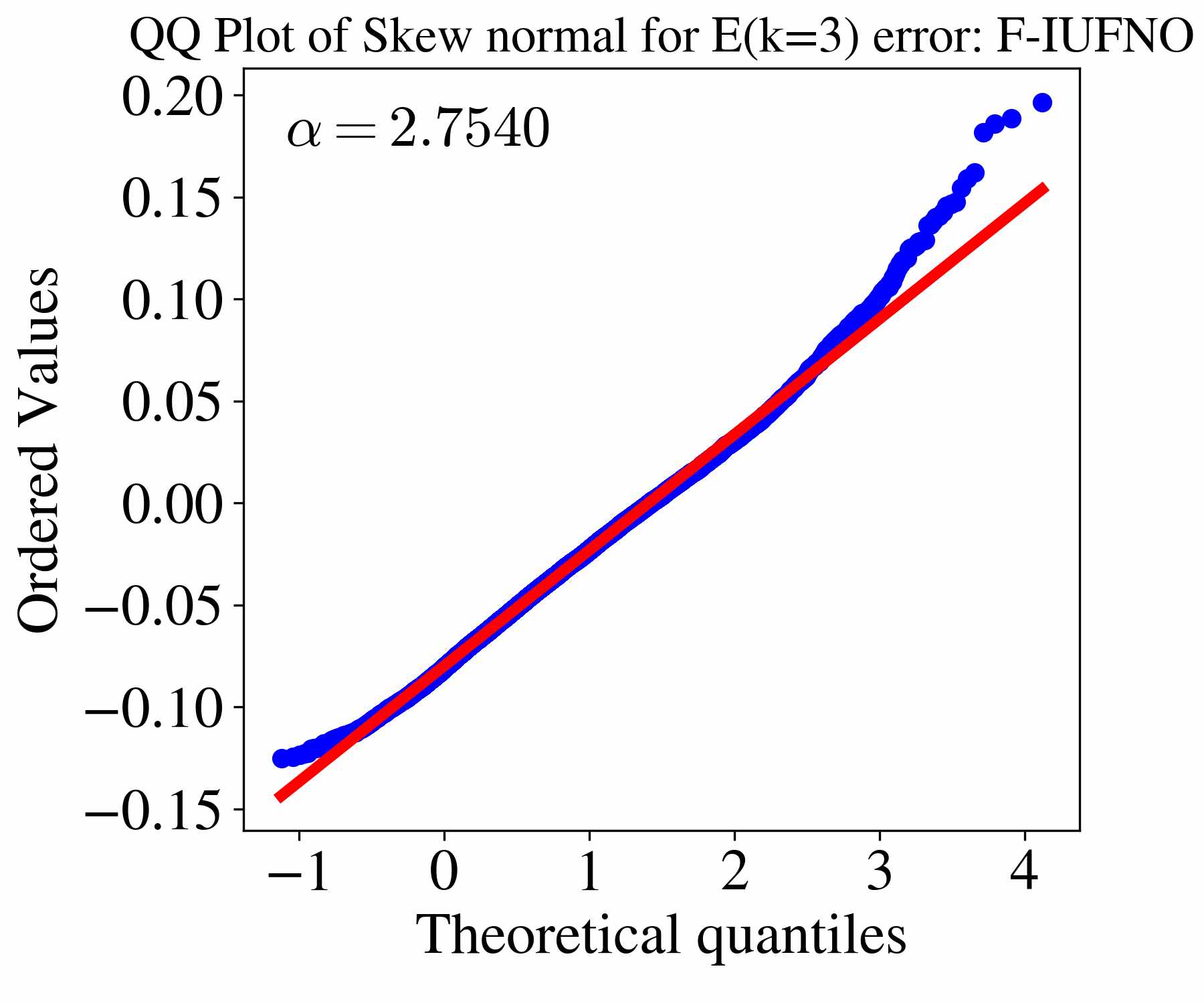}
            \put(-3,65){\small (d)}  
        \end{overpic}
    \end{subfigure}
    \hfill
    \begin{subfigure}[b]{0.32\textwidth}
        \begin{overpic}[width=1\linewidth]{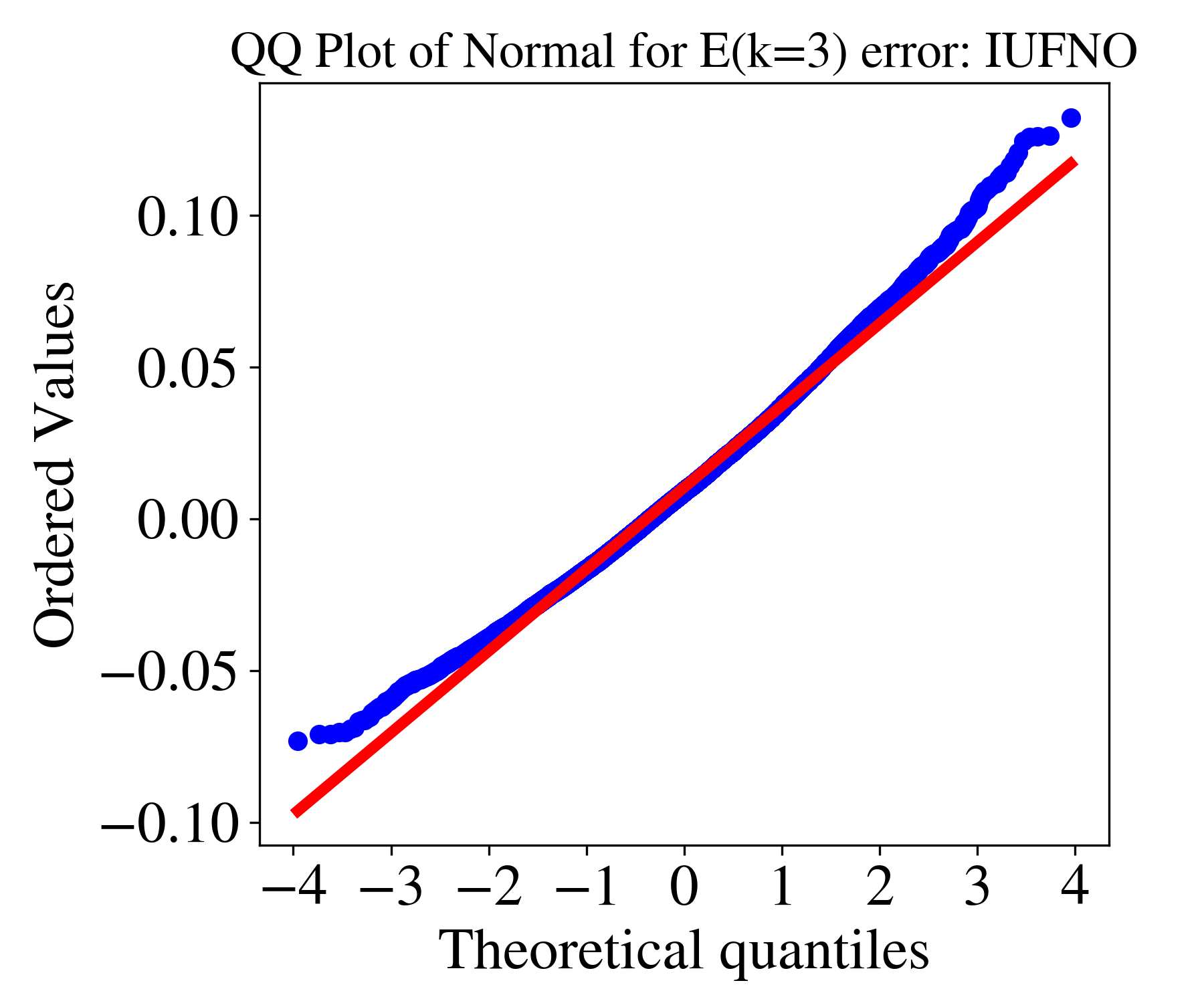}
            \put(-3,65){\small (e)} 
        \end{overpic} 
    \end{subfigure}
    \hfill
    \begin{subfigure}[b]{0.32\textwidth}
        \begin{overpic}[width=1\linewidth]{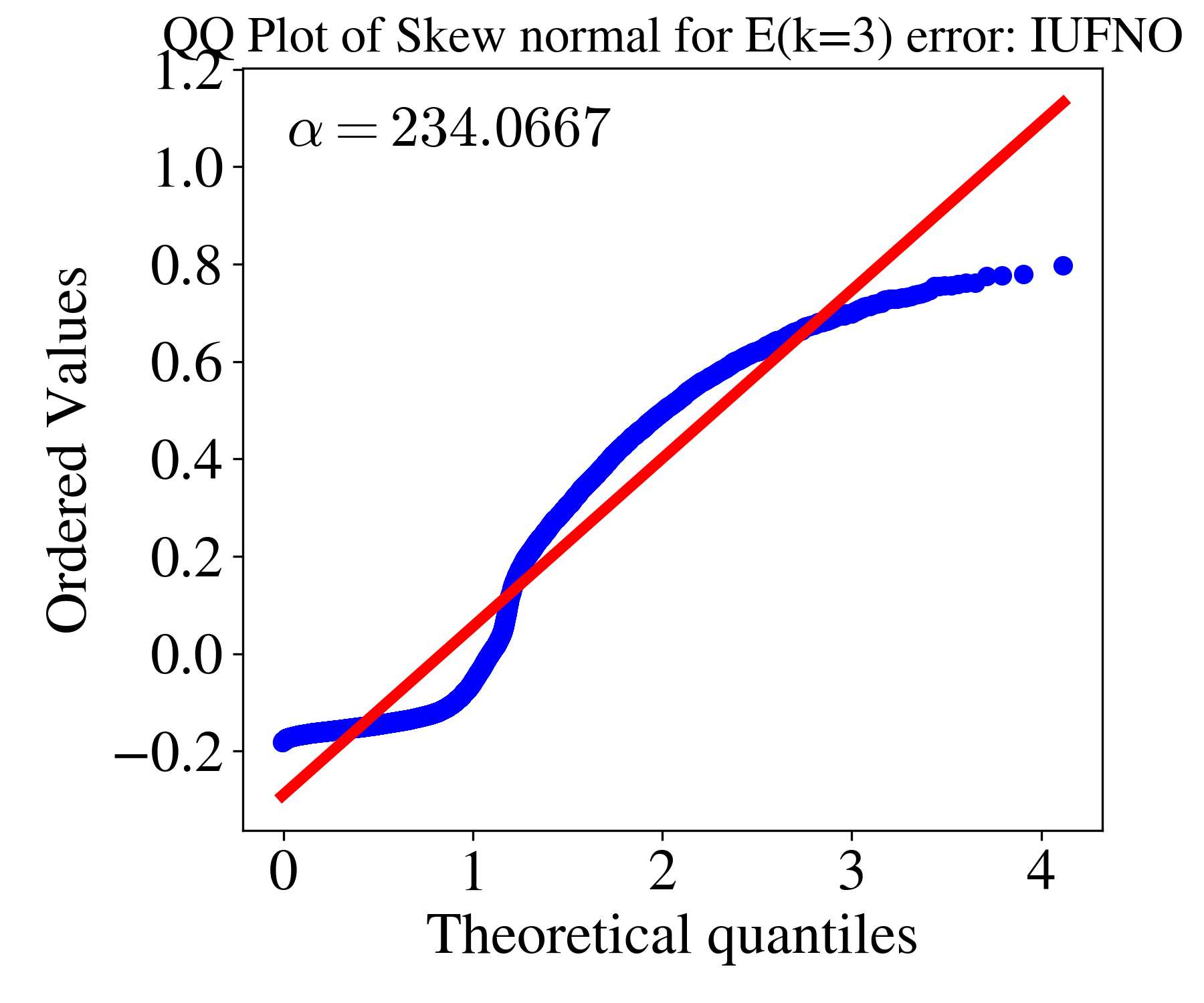}
            \put(-3,65){\small (f)} 
        \end{overpic}
    \end{subfigure}
    \vspace{0.1cm}

    \begin{subfigure}[b]{0.32\textwidth}
        \begin{overpic}[width=1\linewidth]{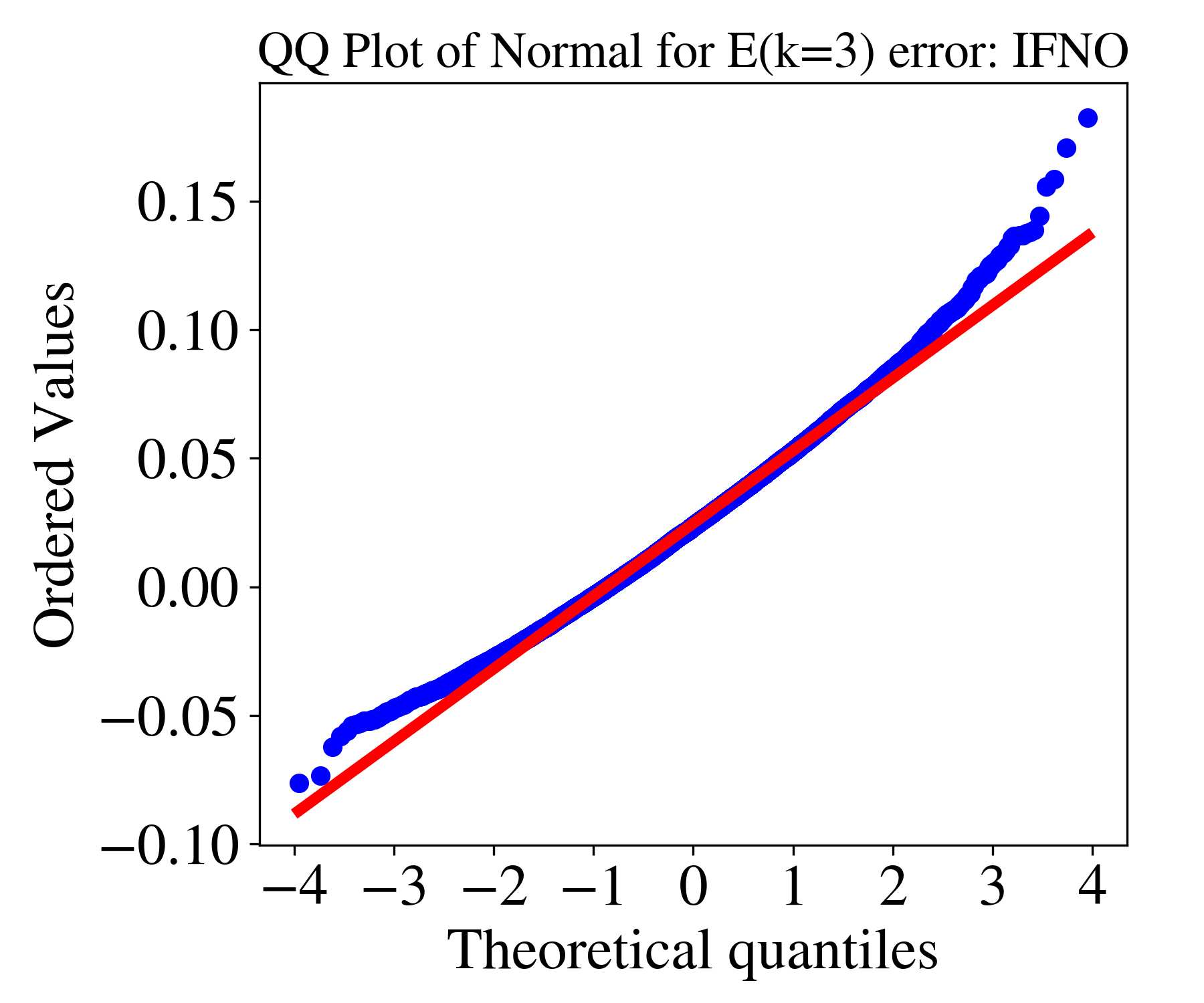}
            \put(-3,65){\small (g)}  
        \end{overpic}
    \end{subfigure}
    \hfill
    \begin{subfigure}[b]{0.32\textwidth}
        \begin{overpic}[width=1\linewidth]{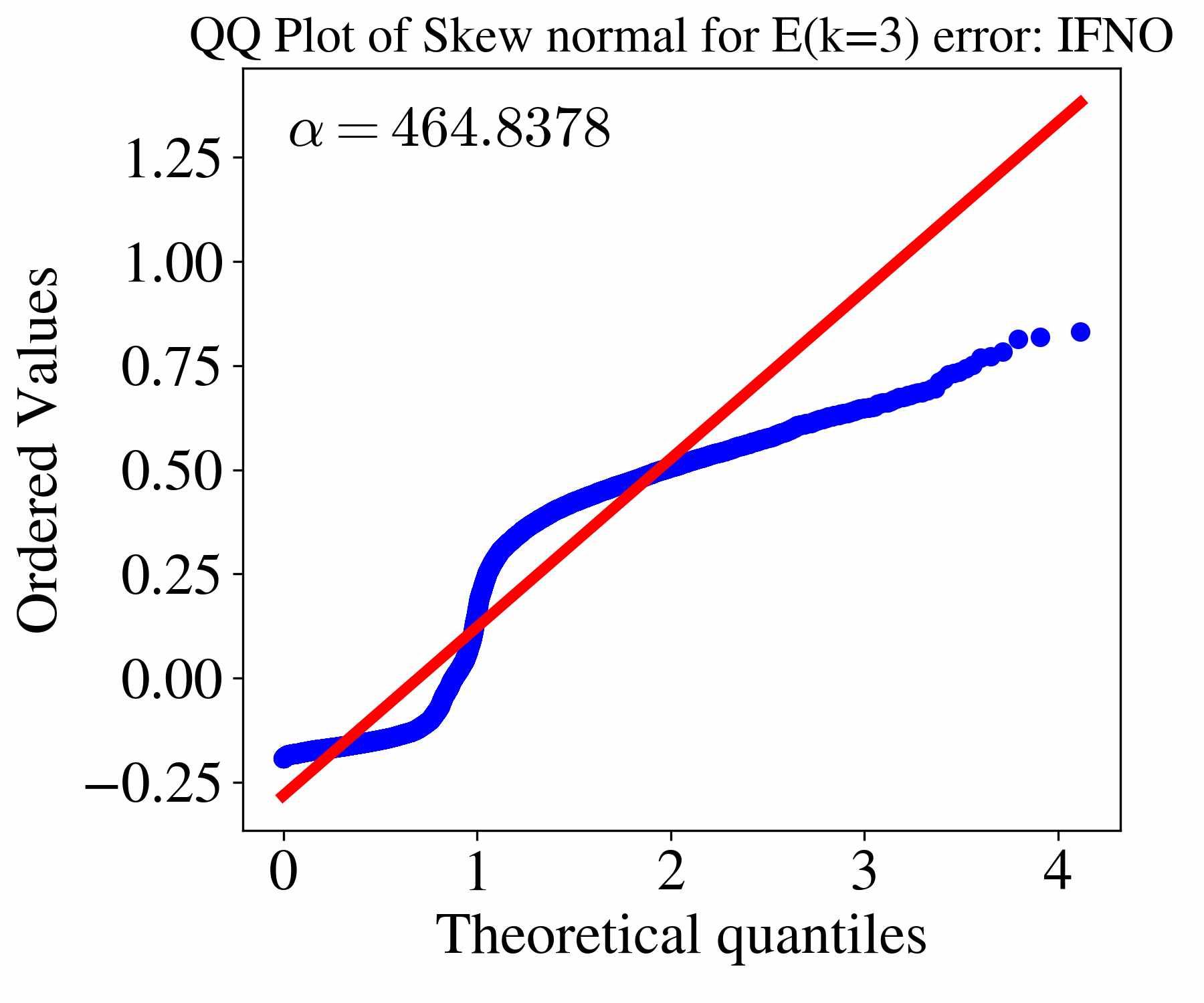}
            \put(-3,65){\small (h)} 
        \end{overpic} 
    \end{subfigure}
    \hfill
    \begin{subfigure}[b]{0.32\textwidth}
        \begin{overpic}[width=1\linewidth]{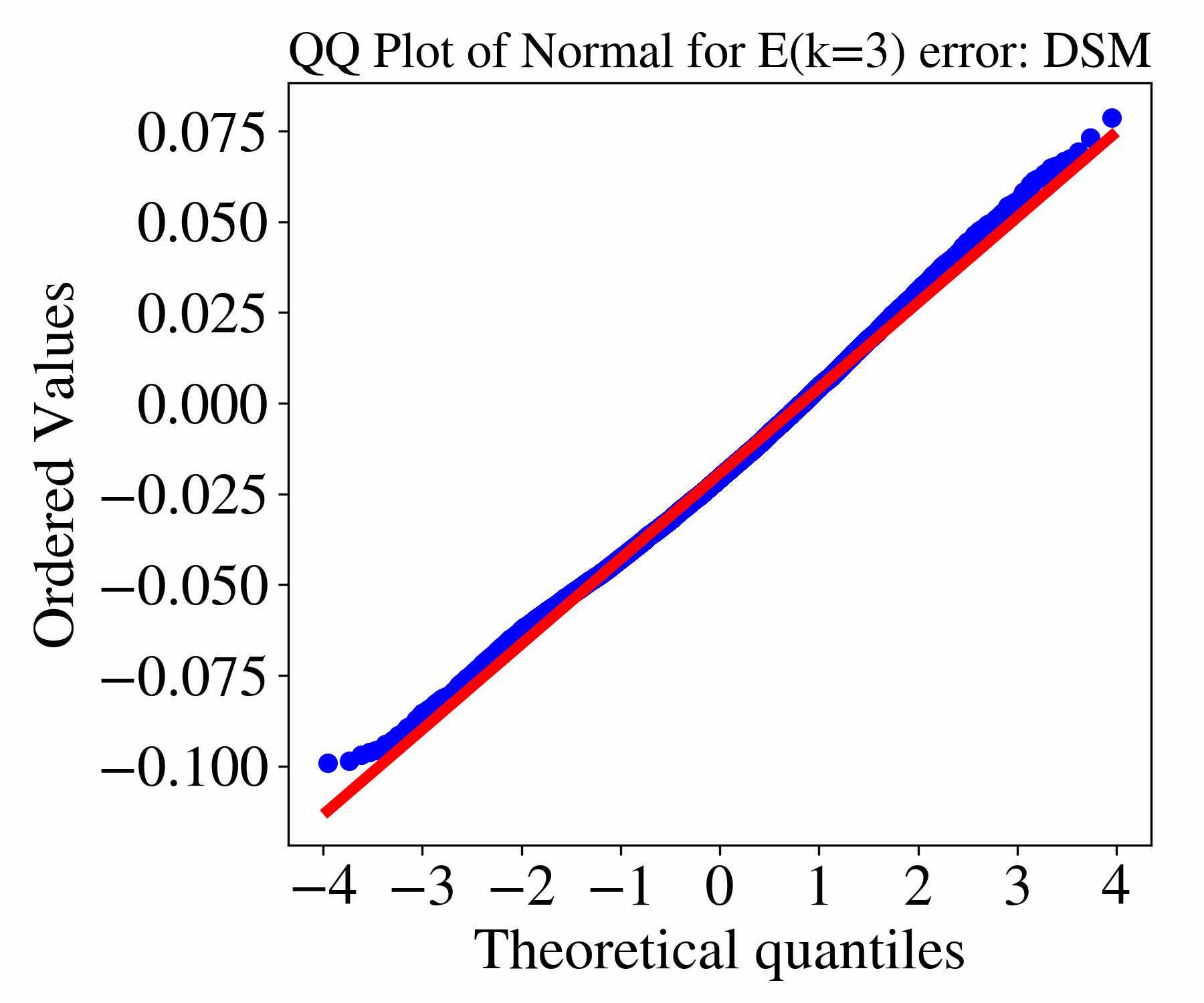}
            \put(-3,65){\small (i)} 
        \end{overpic}
    \end{subfigure}
    \vspace{0.1cm}

    \begin{subfigure}[b]{1\textwidth}
        \centering
        \begin{overpic}[width=0.32\linewidth]{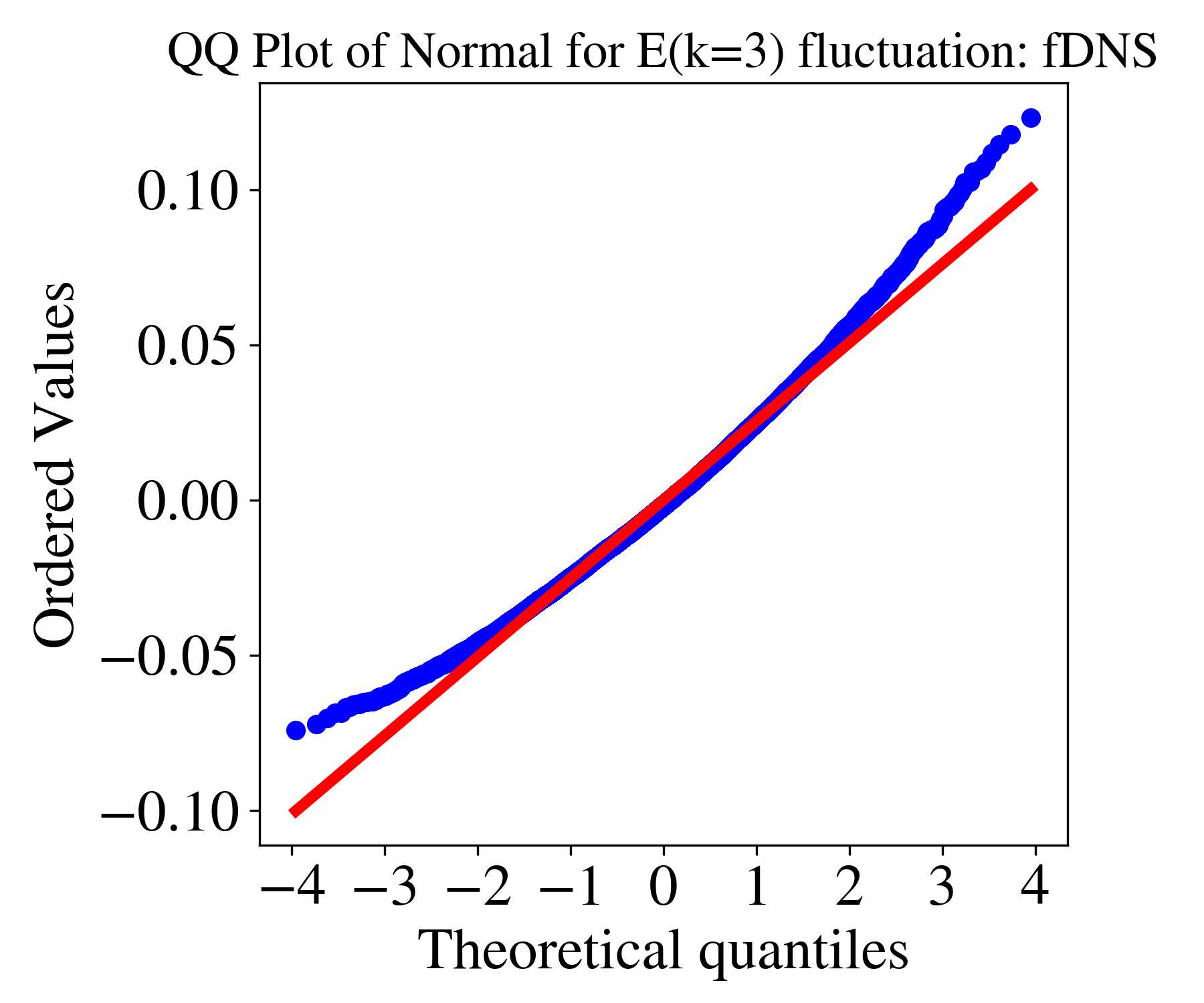}
            \put(-3,65){\small (j)}  
        \end{overpic}
    \end{subfigure}

	\caption{The QQ plots of $E(k=3)$ errors for each method at the representative time interval $\Delta T = 0.2\tau$: (a) F-IFNO constrained; (b) F-IFNO unconstrained; (c) F-IUFNO constrained; (d) F-IUFNO unconstrained; (e) IUFNO constrained; (f) IUFNO unconstrained; (g) IFNO constrained; (h) IFNO unconstrained; (i) DSM; (j) fDNS. Note that for fDNS, the values represent natural statistical fluctuations over time, not prediction errors.}\label{fig:18}
\end{figure}

\begin{figure}[ht!]
    \centering
    \begin{subfigure}[b]{0.32\textwidth}
        \begin{overpic}[width=1\linewidth]{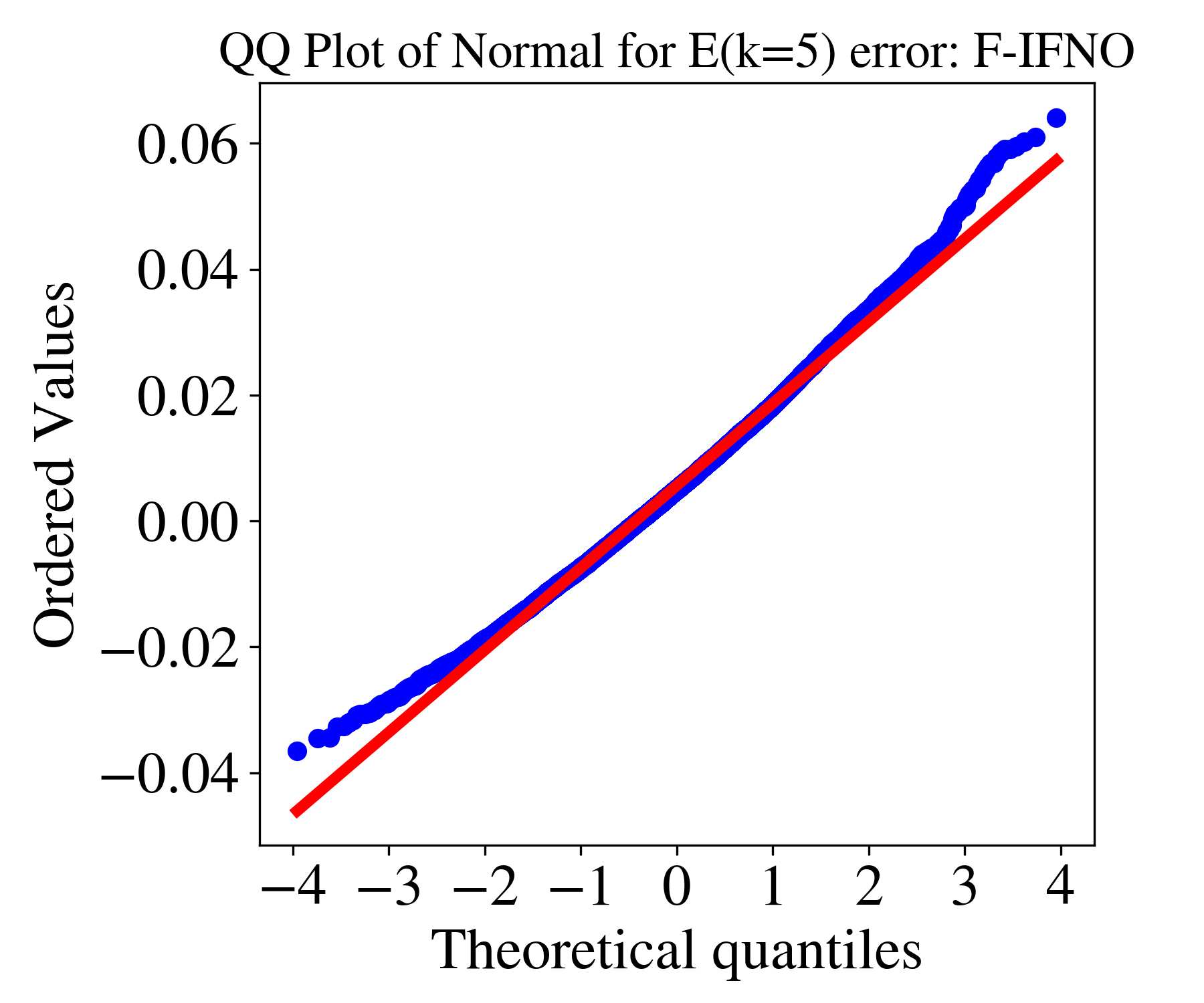}
            \put(-3,65){\small (a)}  
        \end{overpic}
    \end{subfigure}
    \hfill
    \begin{subfigure}[b]{0.32\textwidth}
        \begin{overpic}[width=1\linewidth]{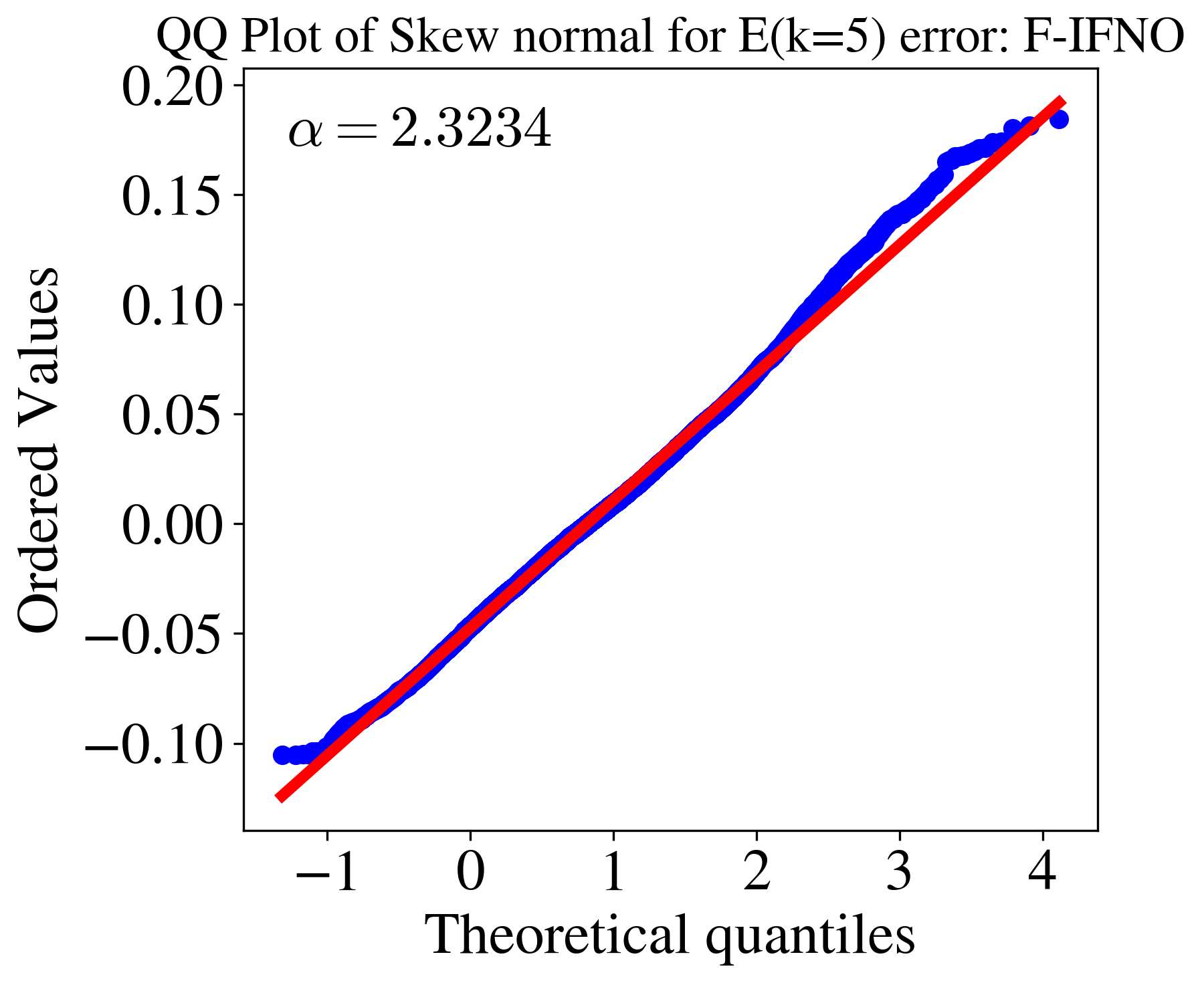}
            \put(-3,65){\small (b)} 
        \end{overpic} 
    \end{subfigure}
    \hfill
    \begin{subfigure}[b]{0.32\textwidth}
        \begin{overpic}[width=1\linewidth]{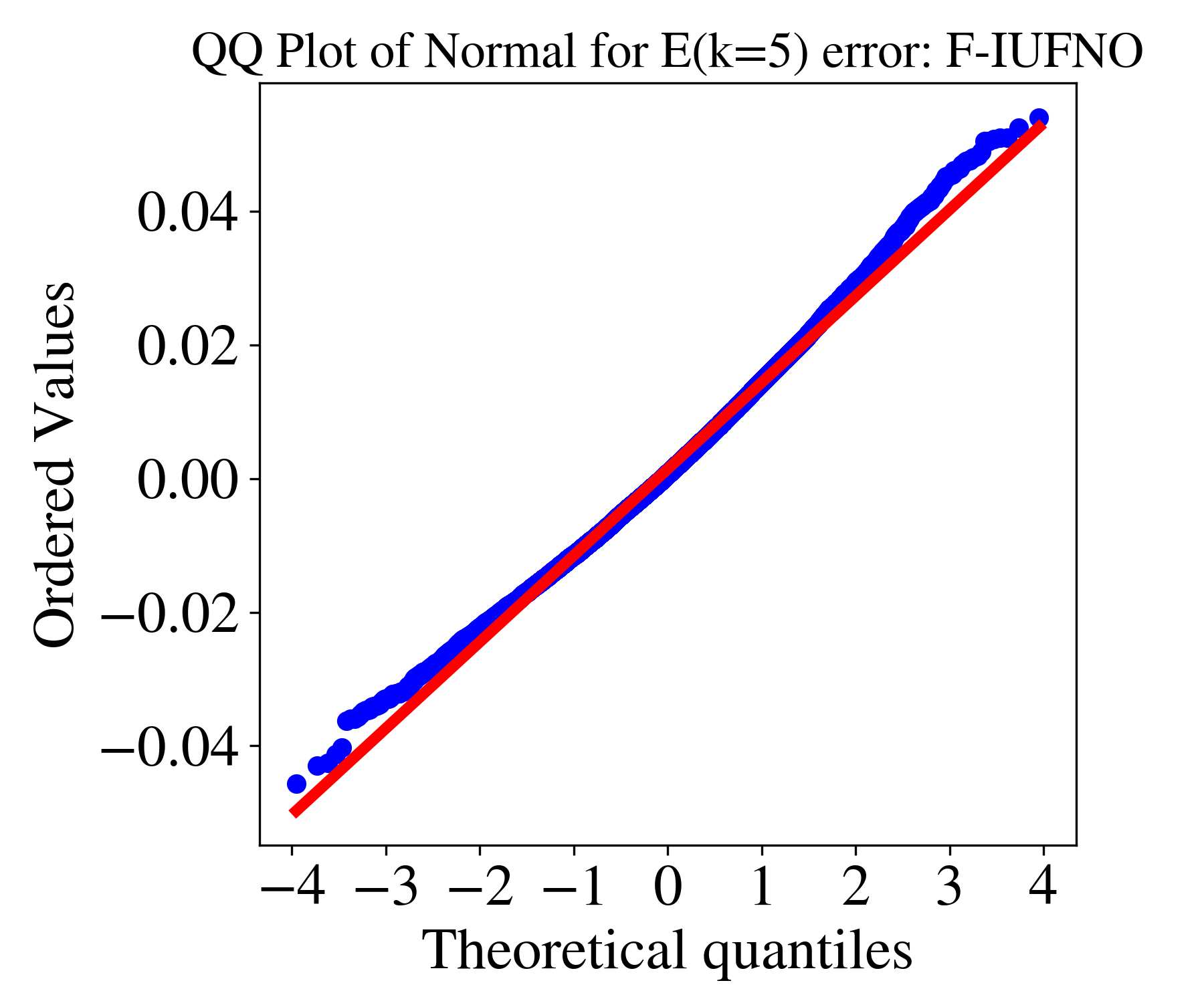}
            \put(-3,65){\small (c)} 
        \end{overpic}
    \end{subfigure}
    \vspace{0.1cm}

    \begin{subfigure}[b]{0.32\textwidth}
        \begin{overpic}[width=1\linewidth]{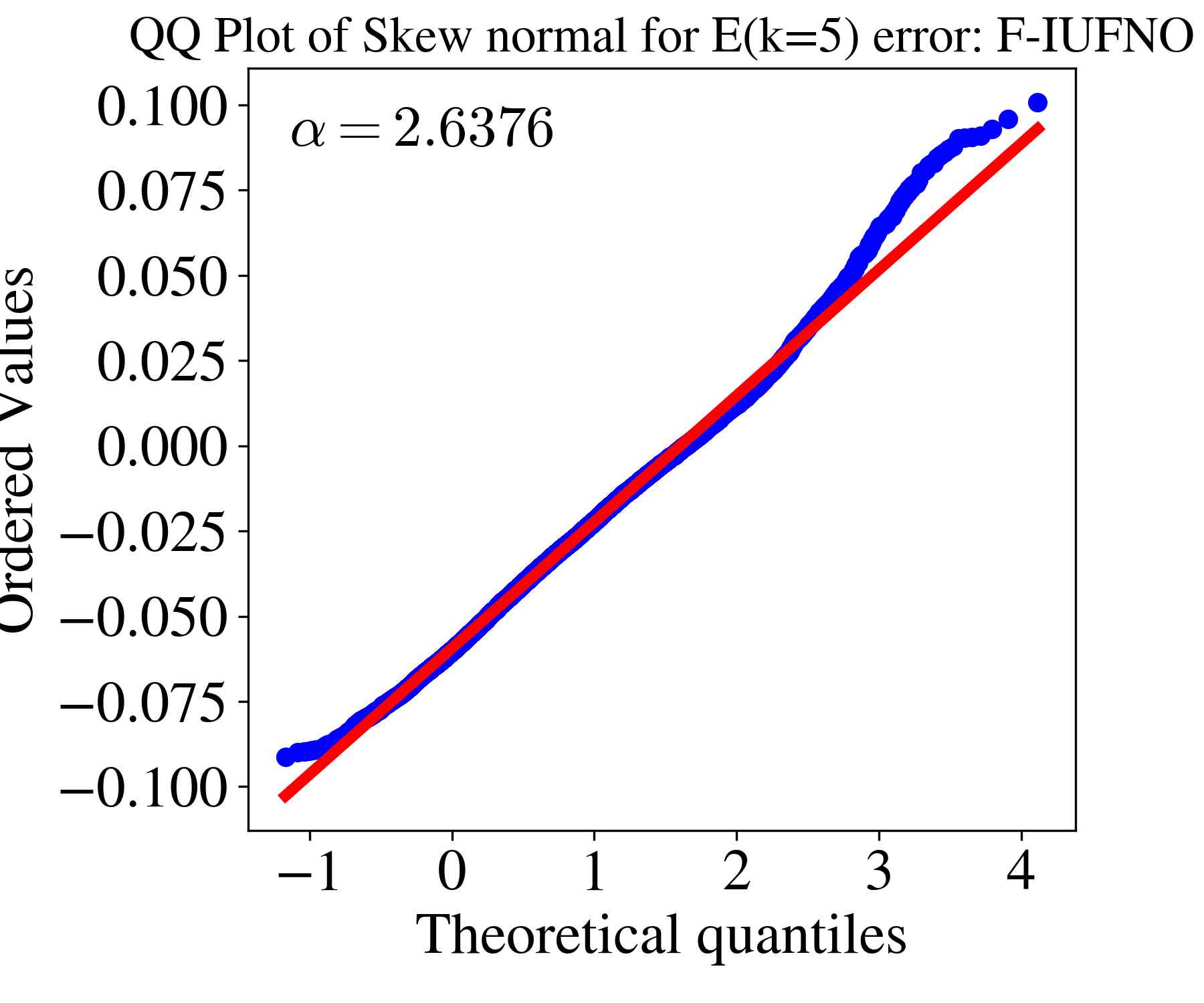}
            \put(-3,65){\small (d)}  
        \end{overpic}
    \end{subfigure}
    \hfill
    \begin{subfigure}[b]{0.32\textwidth}
        \begin{overpic}[width=1\linewidth]{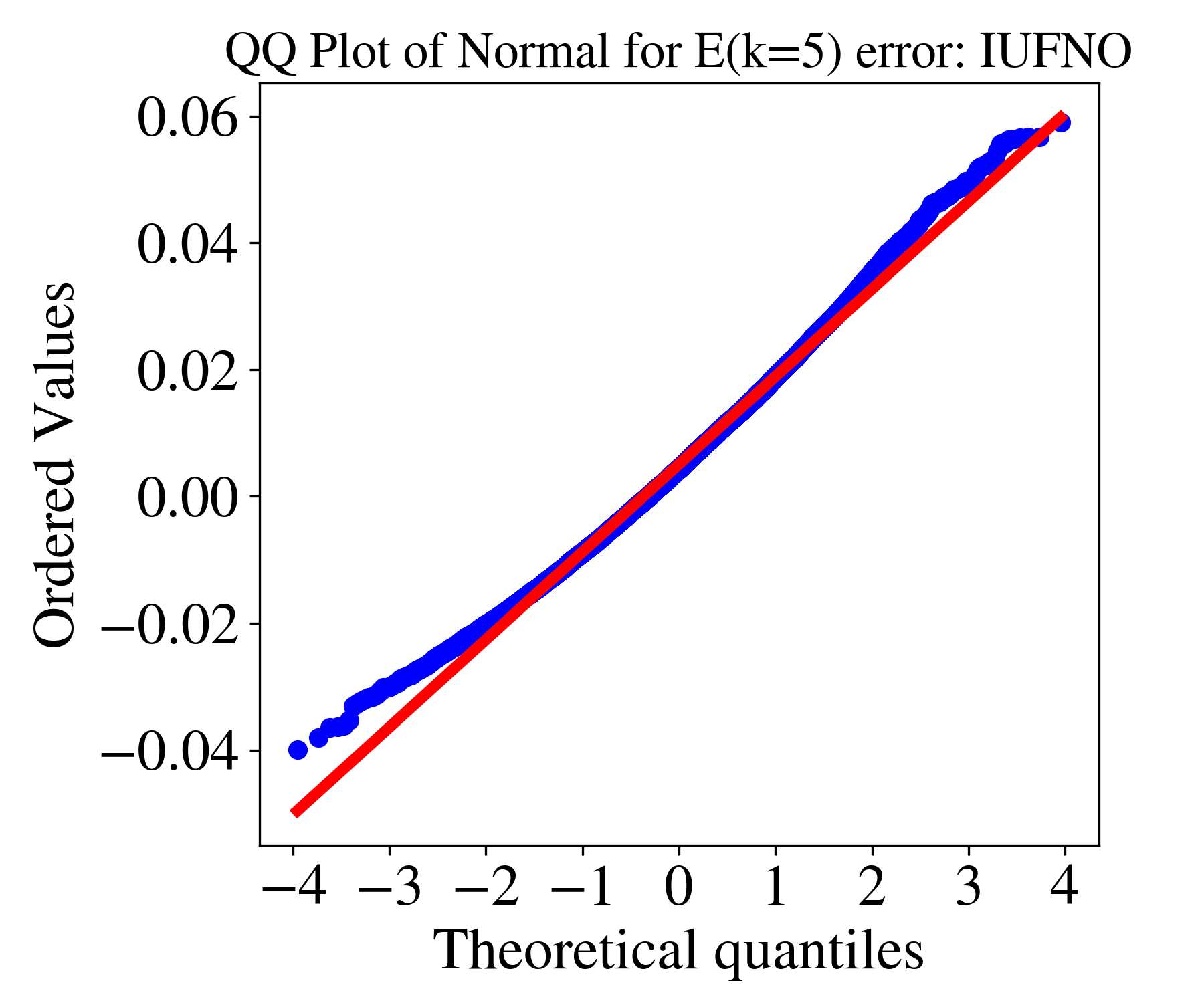}
            \put(-3,65){\small (e)} 
        \end{overpic} 
    \end{subfigure}
    \hfill
    \begin{subfigure}[b]{0.32\textwidth}
        \begin{overpic}[width=1\linewidth]{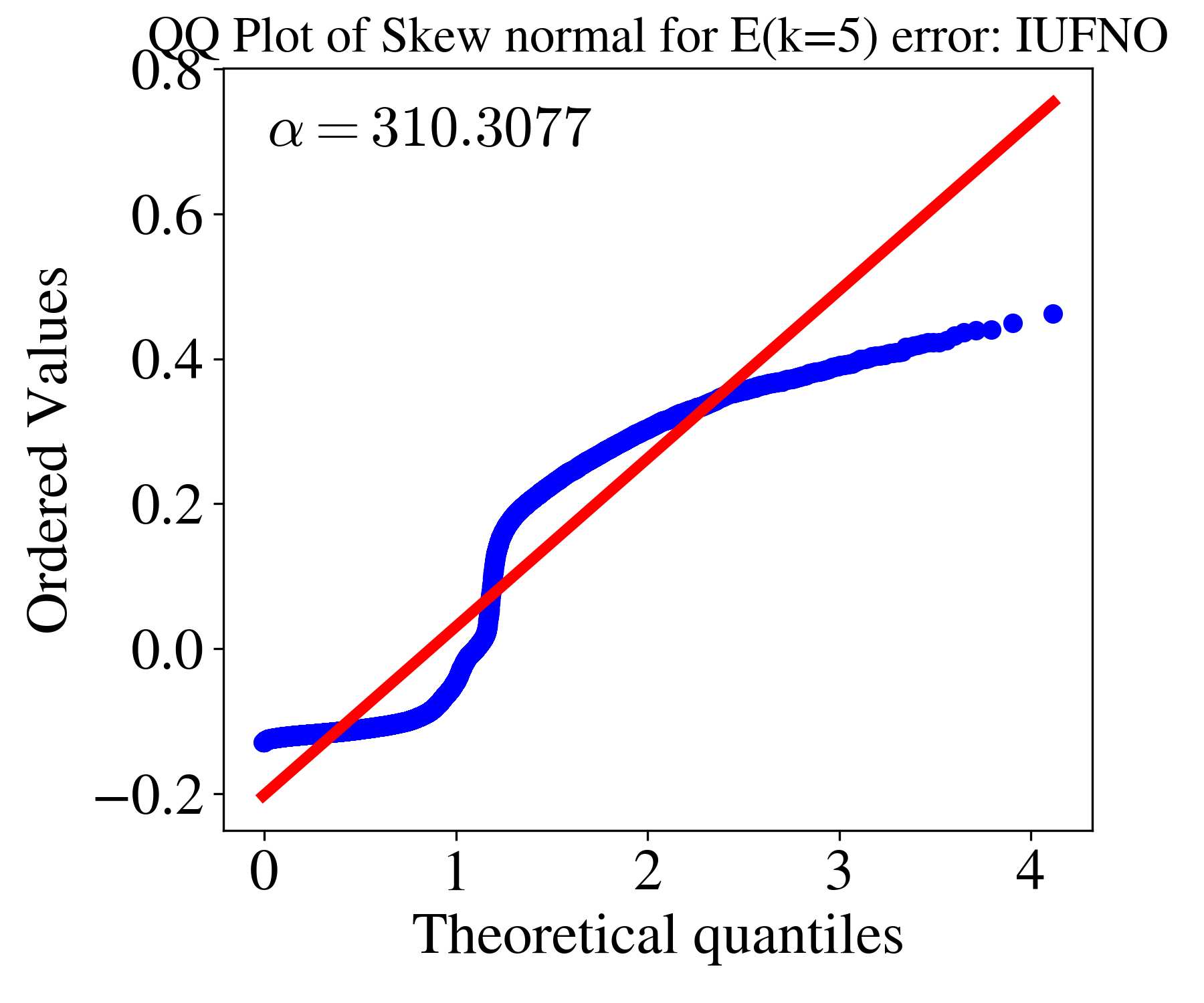}
            \put(-3,65){\small (f)} 
        \end{overpic}
    \end{subfigure}
    \vspace{0.1cm}

    \begin{subfigure}[b]{0.32\textwidth}
        \begin{overpic}[width=1\linewidth]{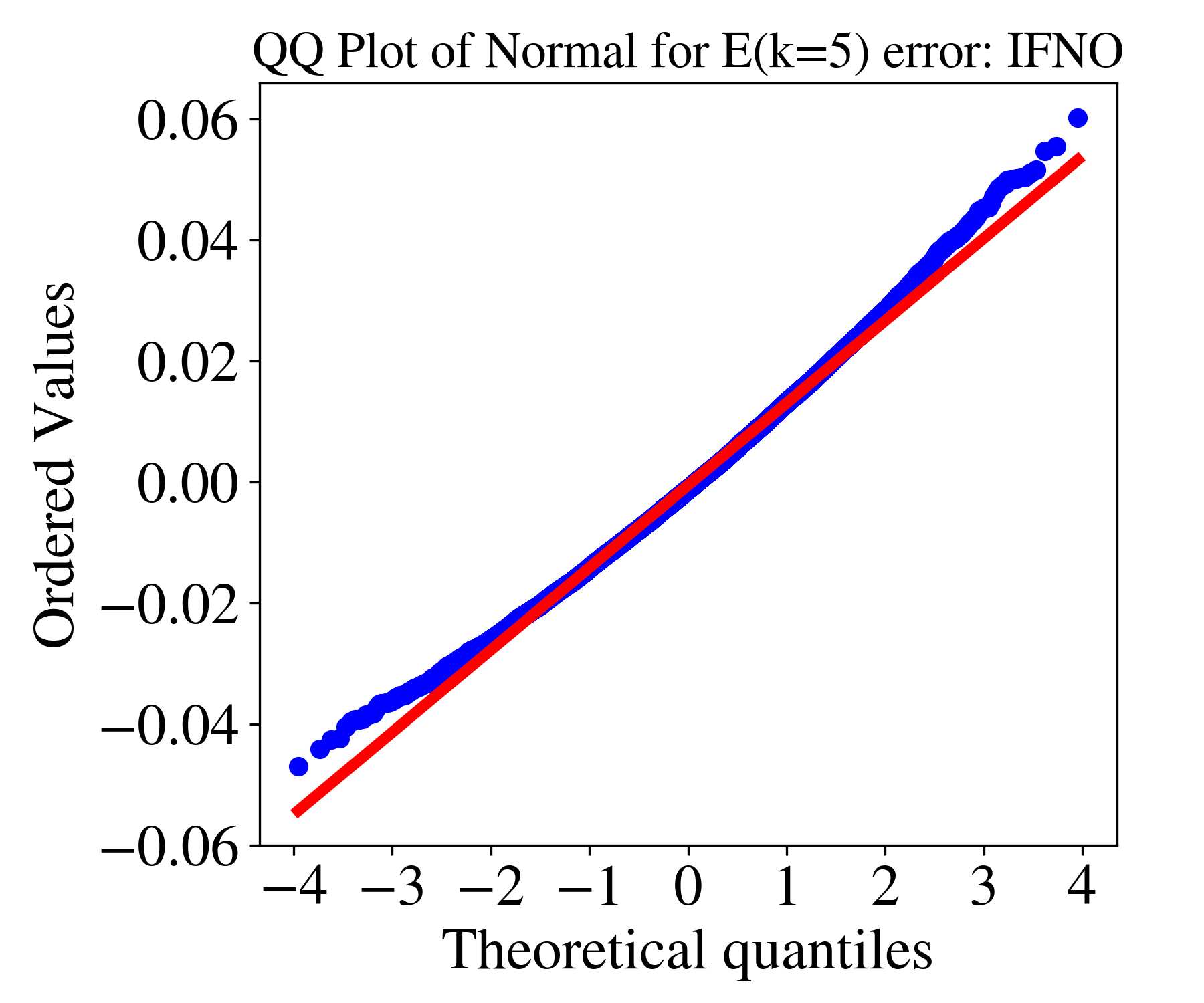}
            \put(-3,65){\small (g)}  
        \end{overpic}
    \end{subfigure}
    \hfill
    \begin{subfigure}[b]{0.32\textwidth}
        \begin{overpic}[width=1\linewidth]{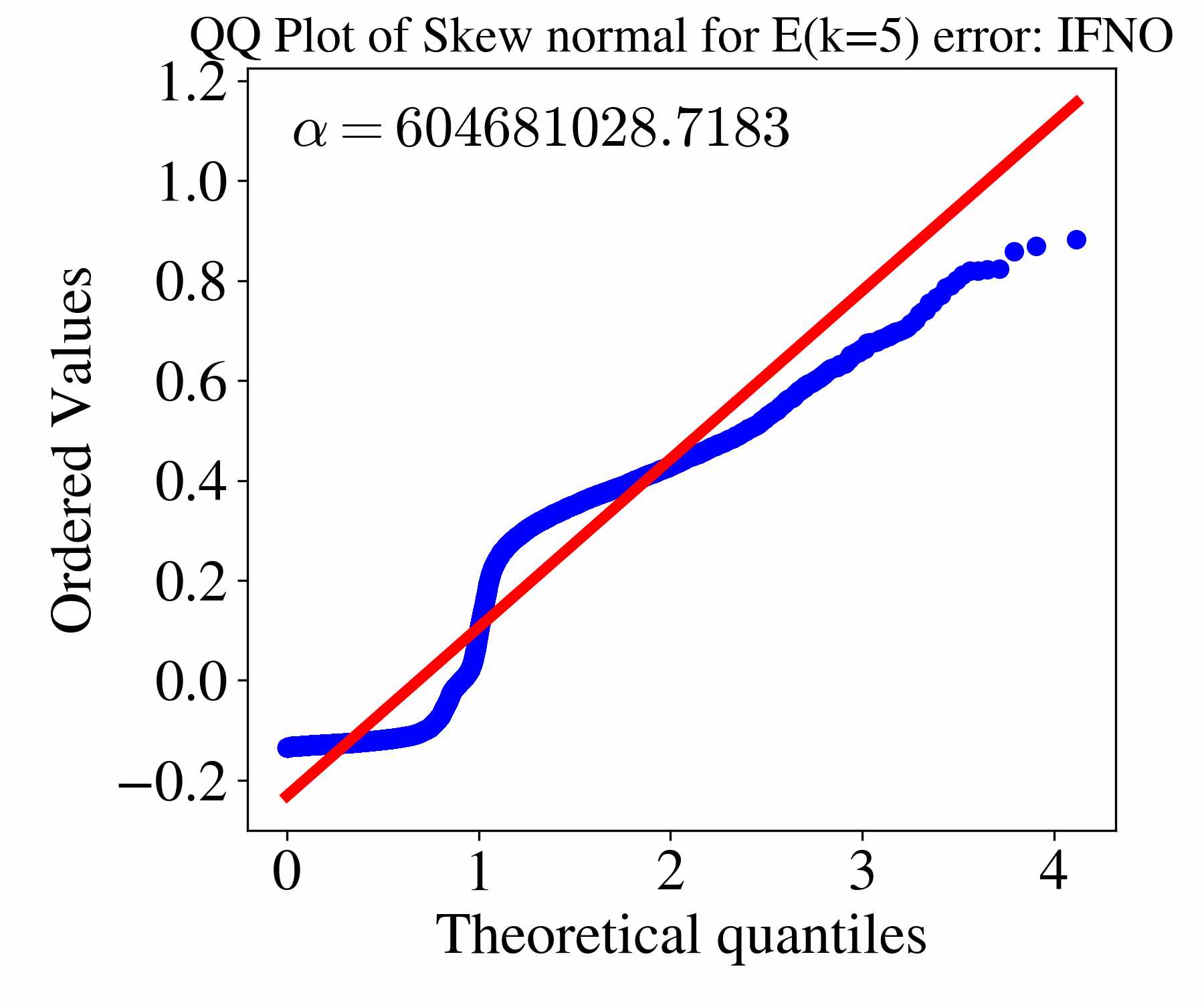}
            \put(-3,65){\small (h)} 
        \end{overpic} 
    \end{subfigure}
    \hfill
    \begin{subfigure}[b]{0.32\textwidth}
        \begin{overpic}[width=1\linewidth]{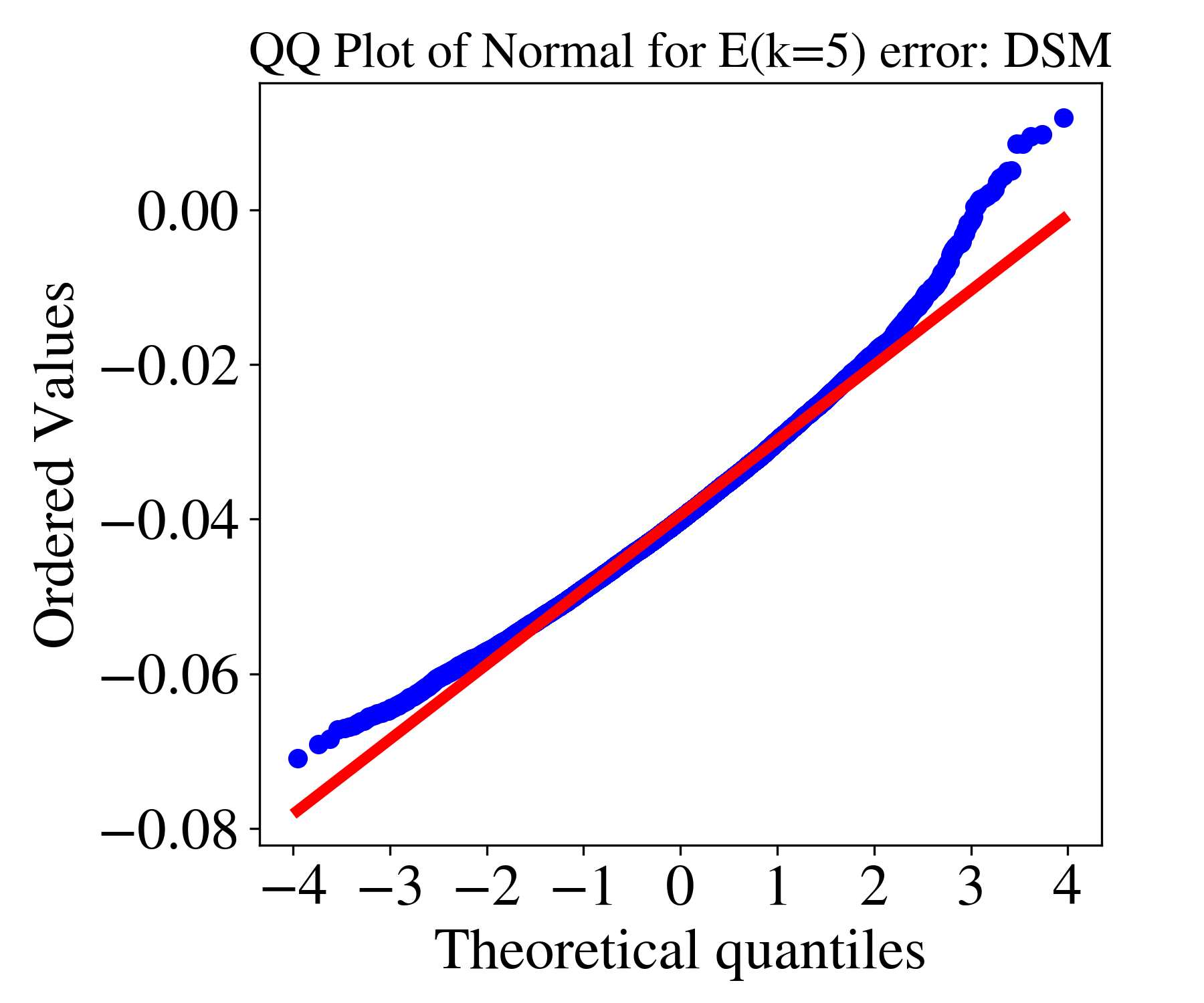}
            \put(-3,65){\small (i)} 
        \end{overpic}
    \end{subfigure}
    \vspace{0.1cm}

    \begin{subfigure}[b]{1\textwidth}
        \centering
        \begin{overpic}[width=0.32\linewidth]{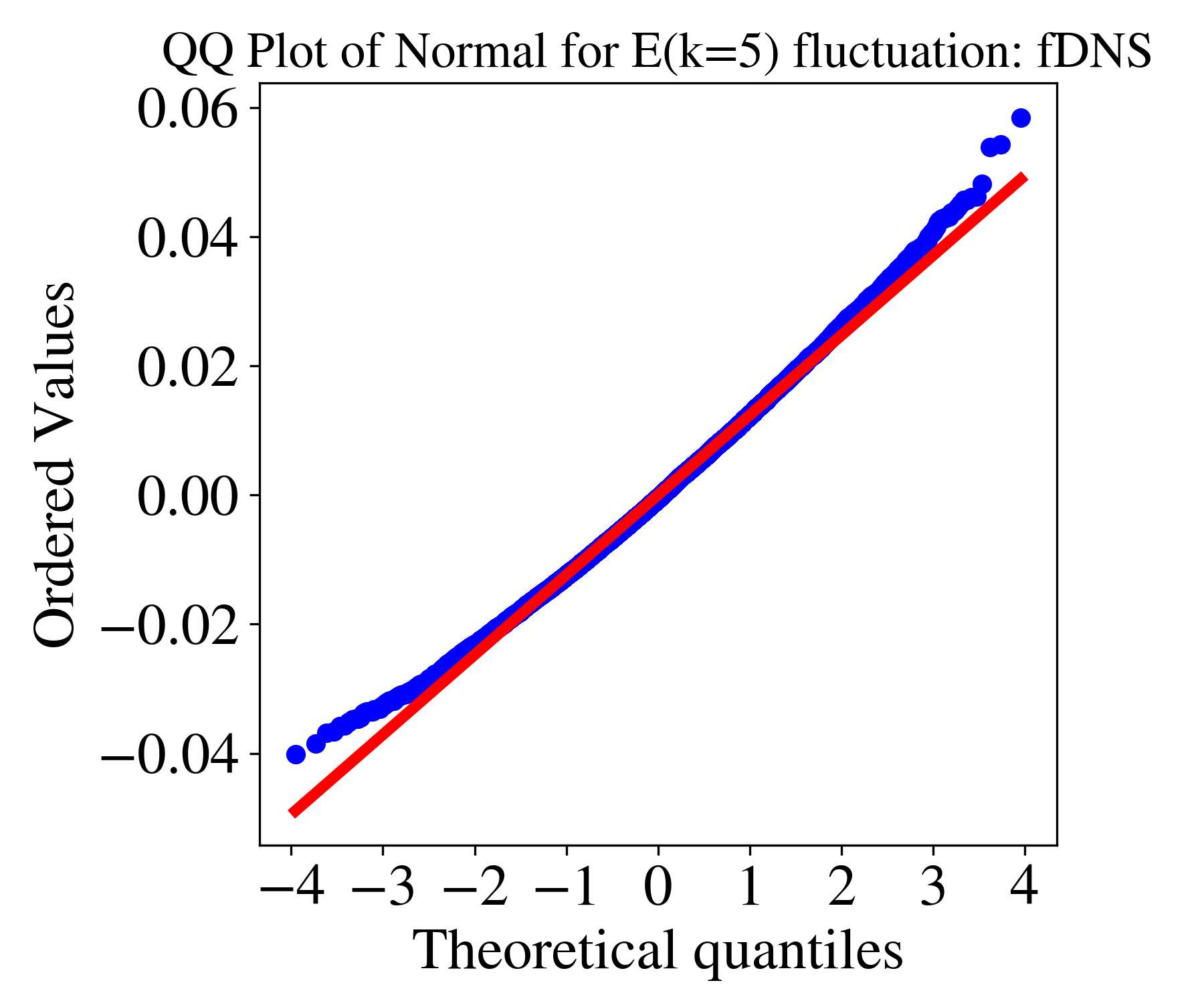}
            \put(-3,65){\small (j)}  
        \end{overpic}
    \end{subfigure}

	\caption{The QQ plots of $E(k=5)$ errors for each method at the representative time interval $\Delta T = 0.2\tau$: (a) F-IFNO constrained; (b) F-IFNO unconstrained; (c) F-IUFNO constrained; (d) F-IUFNO unconstrained; (e) IUFNO constrained; (f) IUFNO unconstrained; (g) IFNO constrained; (h) IFNO unconstrained; (i) DSM; (j) fDNS. Note that for fDNS, the values represent natural statistical fluctuations over time, not prediction errors.}\label{fig:20}
\end{figure}

\begin{figure}[ht!]
    \centering
    \begin{subfigure}[b]{0.32\textwidth}
        \begin{overpic}[width=1\linewidth]{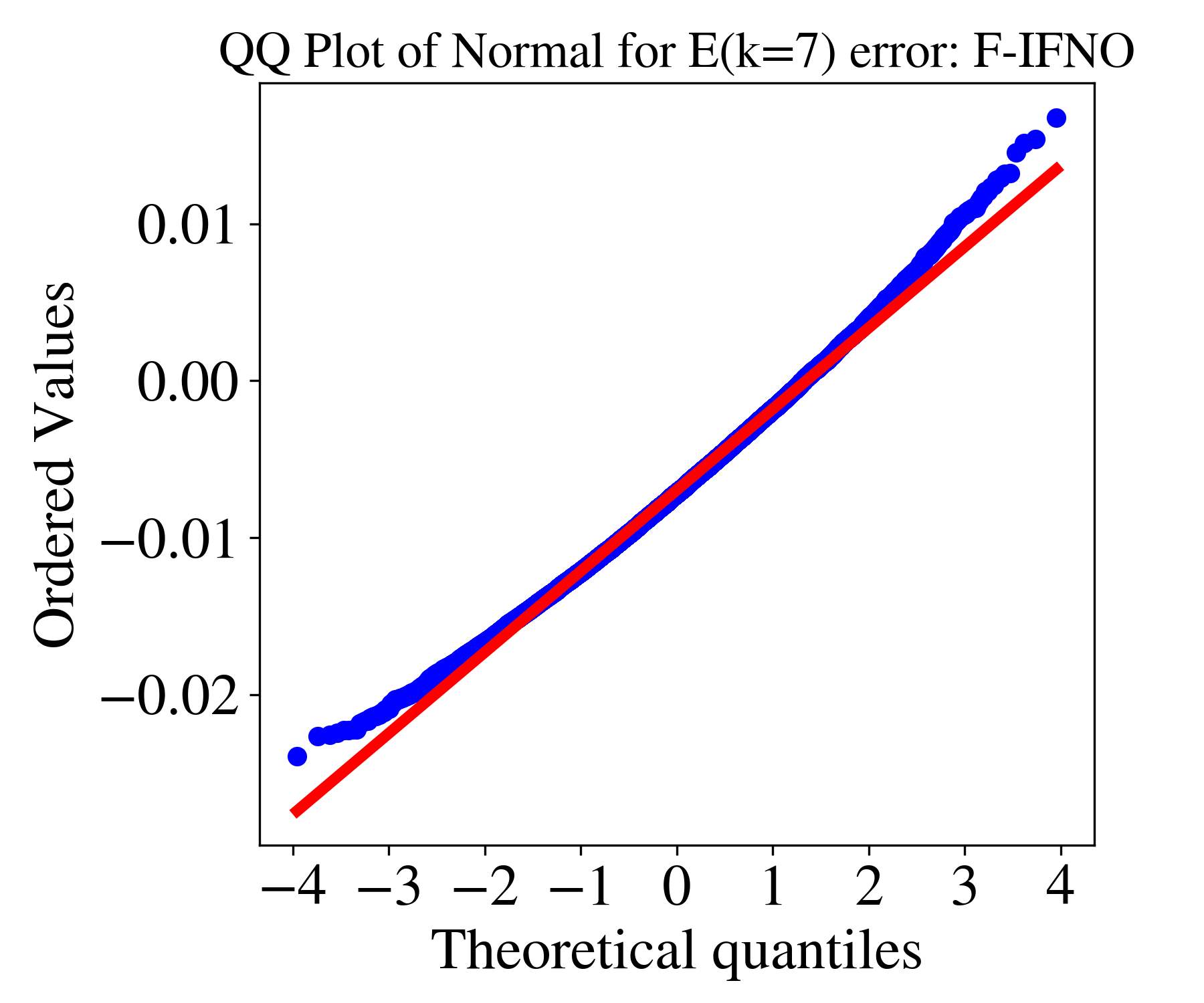}
            \put(-3,65){\small (a)}  
        \end{overpic}
    \end{subfigure}
    \hfill
    \begin{subfigure}[b]{0.32\textwidth}
        \begin{overpic}[width=1\linewidth]{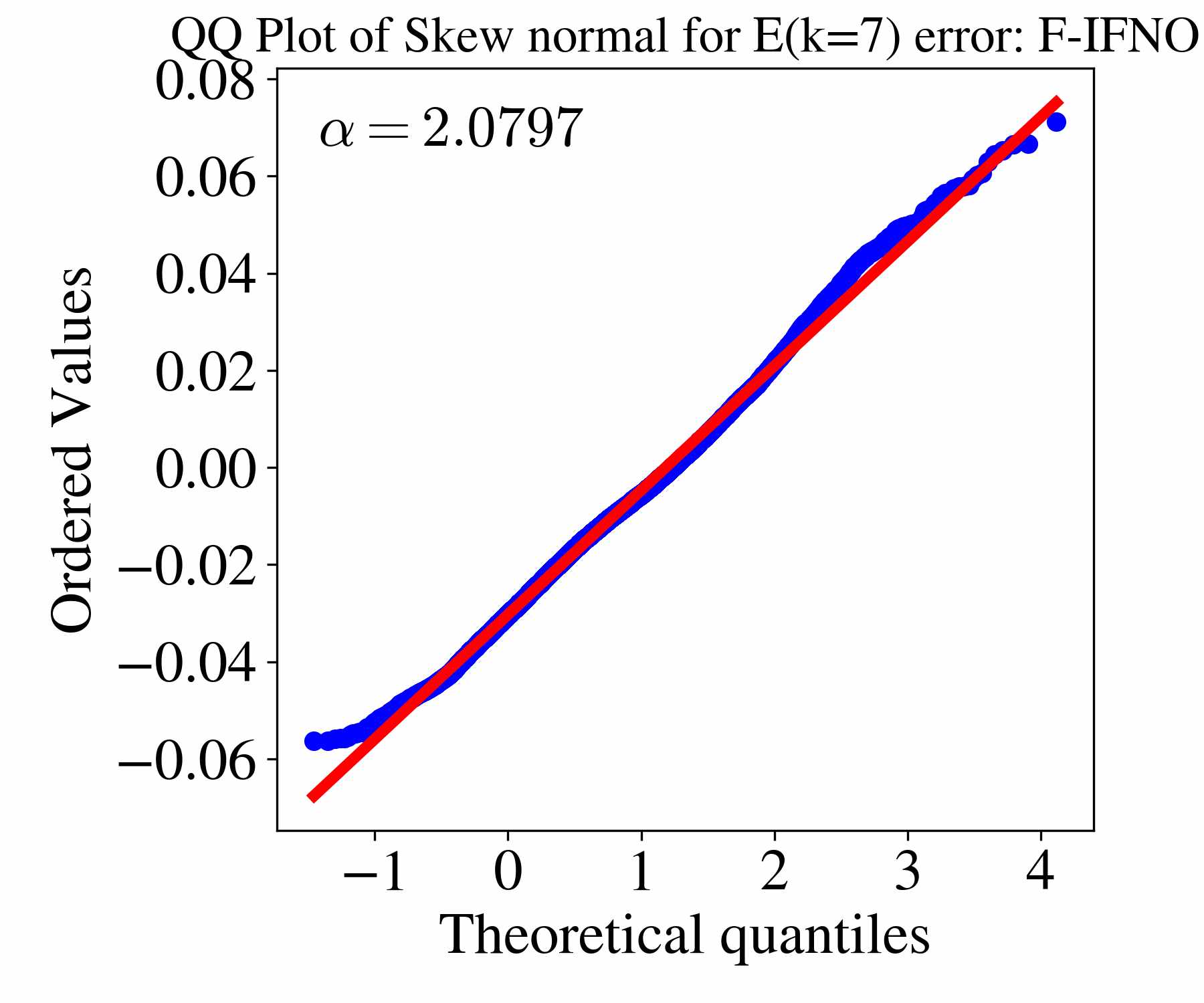}
            \put(-3,65){\small (b)} 
        \end{overpic} 
    \end{subfigure}
    \hfill
    \begin{subfigure}[b]{0.32\textwidth}
        \begin{overpic}[width=1\linewidth]{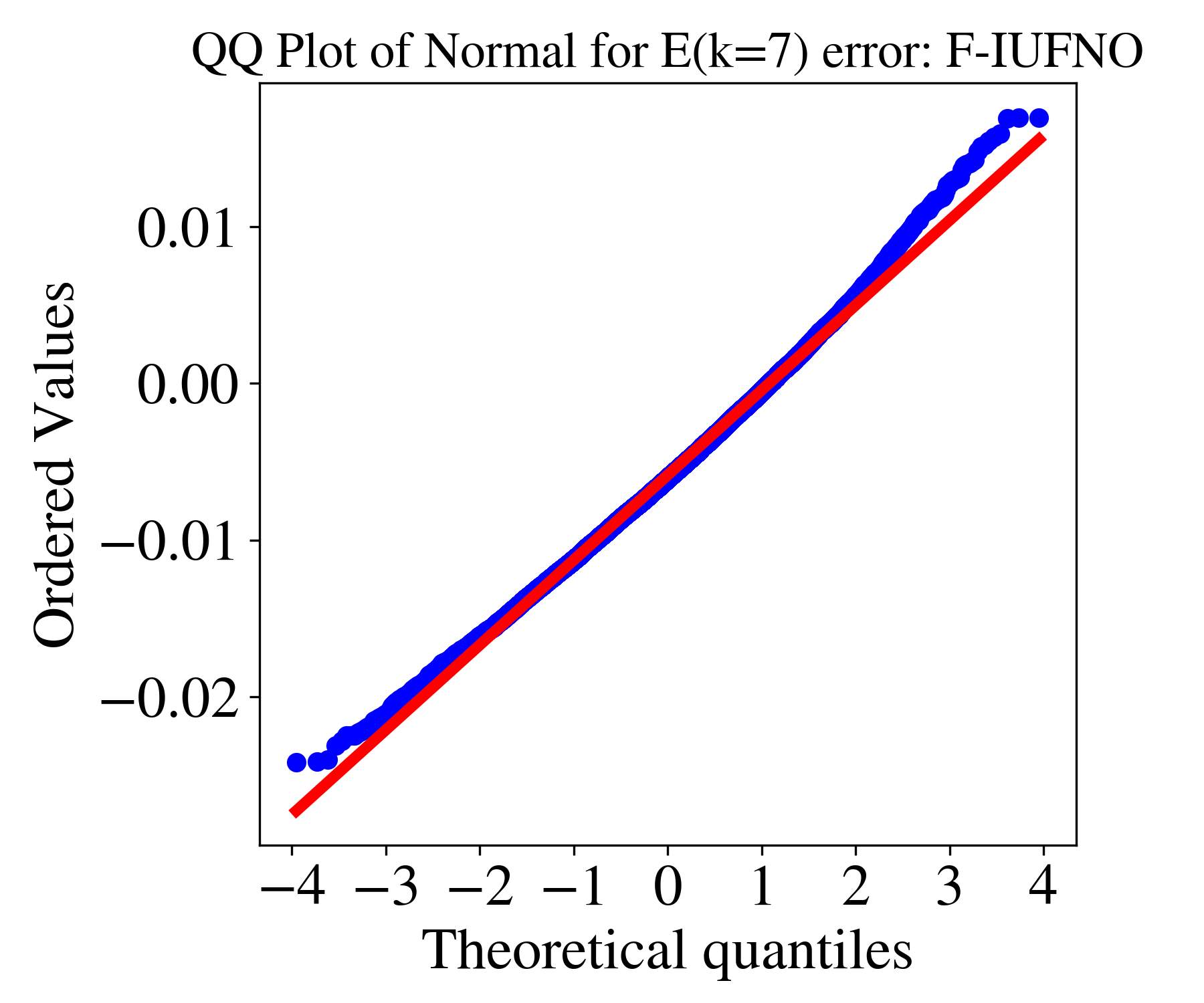}
            \put(-3,65){\small (c)} 
        \end{overpic}
    \end{subfigure}
    \vspace{0.1cm}

    \begin{subfigure}[b]{0.32\textwidth}
        \begin{overpic}[width=1\linewidth]{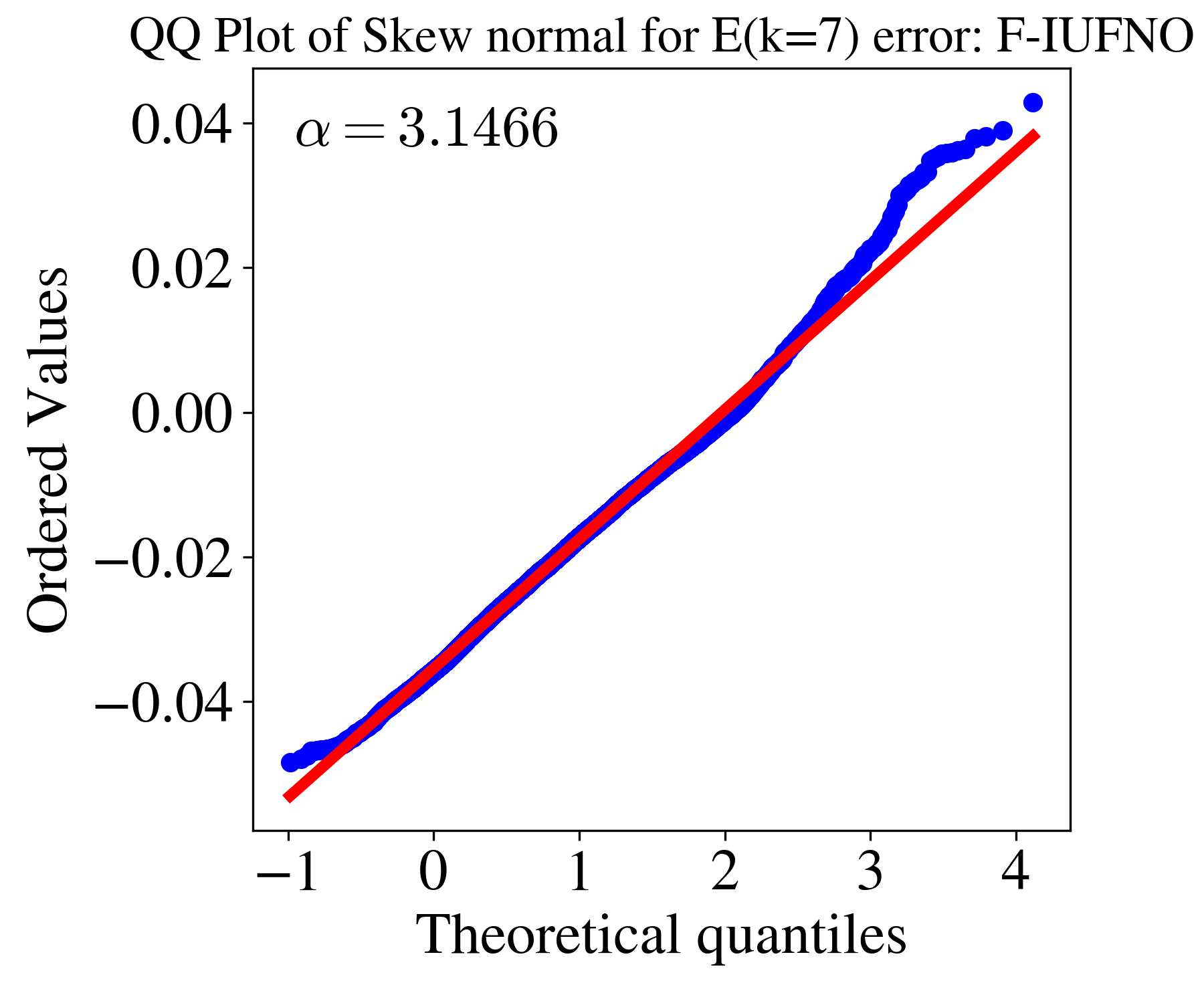}
            \put(-3,65){\small (d)}  
        \end{overpic}
    \end{subfigure}
    \hfill
    \begin{subfigure}[b]{0.32\textwidth}
        \begin{overpic}[width=1\linewidth]{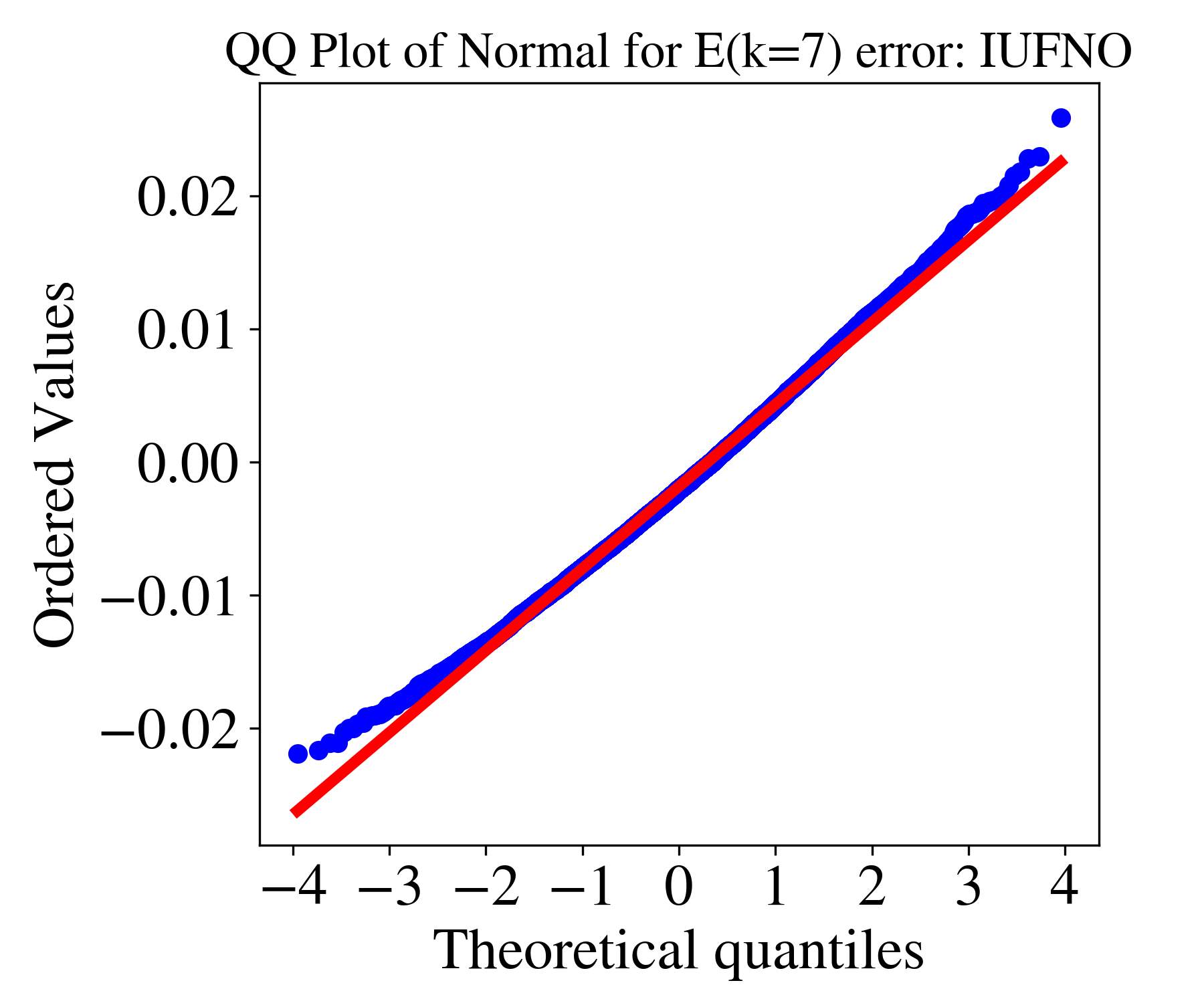}
            \put(-3,65){\small (e)} 
        \end{overpic} 
    \end{subfigure}
    \hfill
    \begin{subfigure}[b]{0.32\textwidth}
        \begin{overpic}[width=1\linewidth]{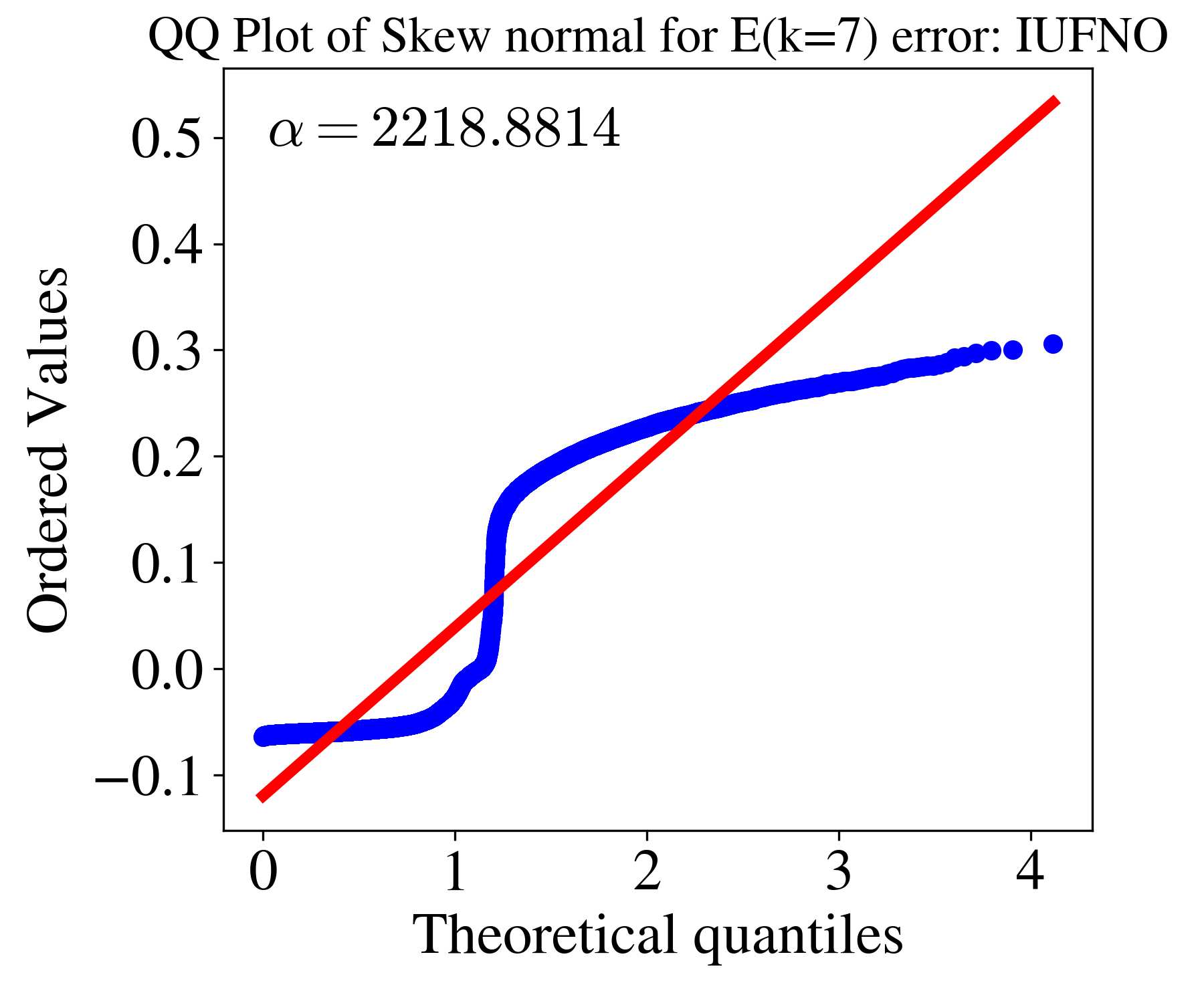}
            \put(-3,65){\small (f)} 
        \end{overpic}
    \end{subfigure}
    \vspace{0.1cm}

    \begin{subfigure}[b]{0.32\textwidth}
        \begin{overpic}[width=1\linewidth]{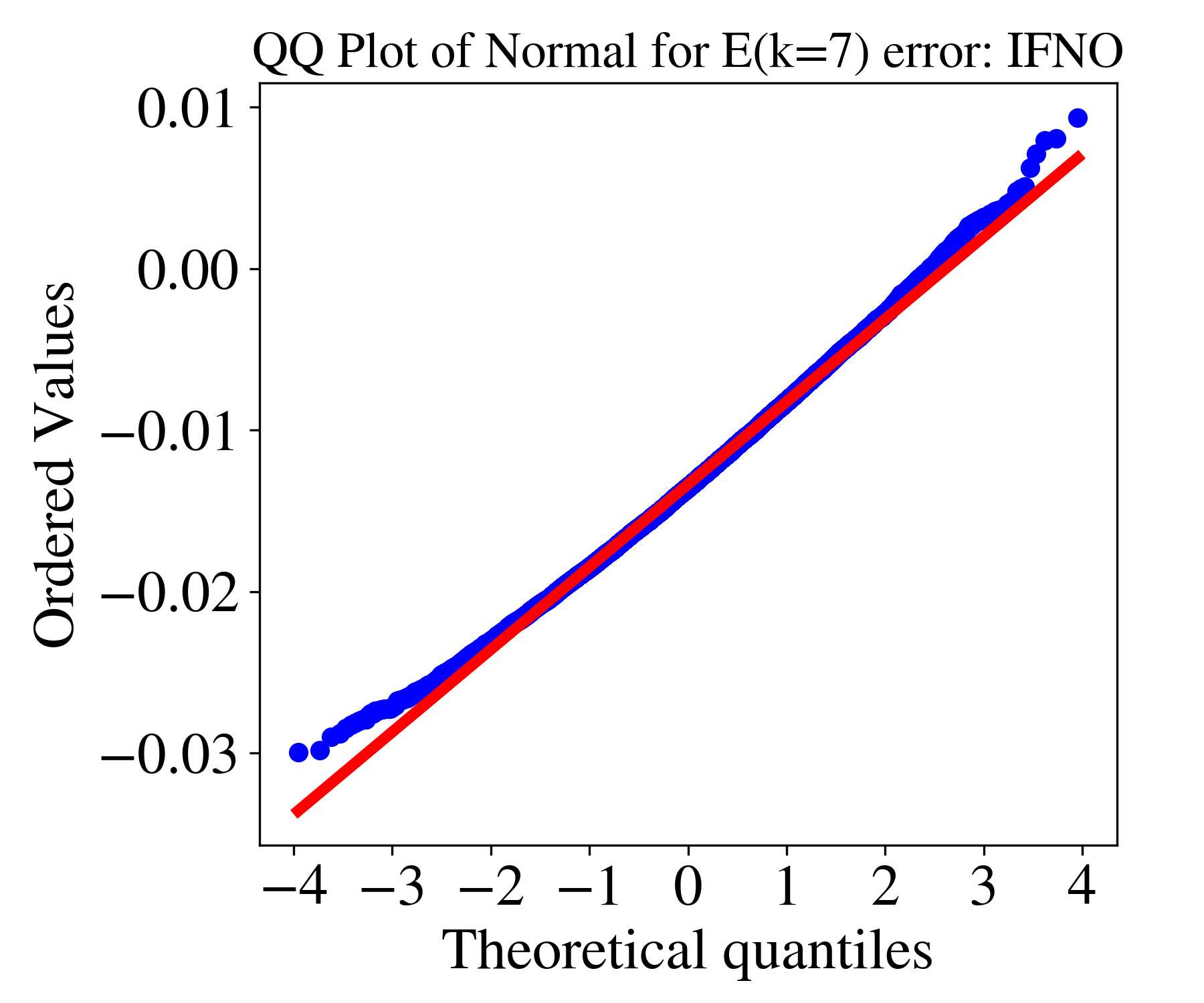}
            \put(-3,65){\small (g)}  
        \end{overpic}
    \end{subfigure}
    \hfill
    \begin{subfigure}[b]{0.32\textwidth}
        \begin{overpic}[width=1\linewidth]{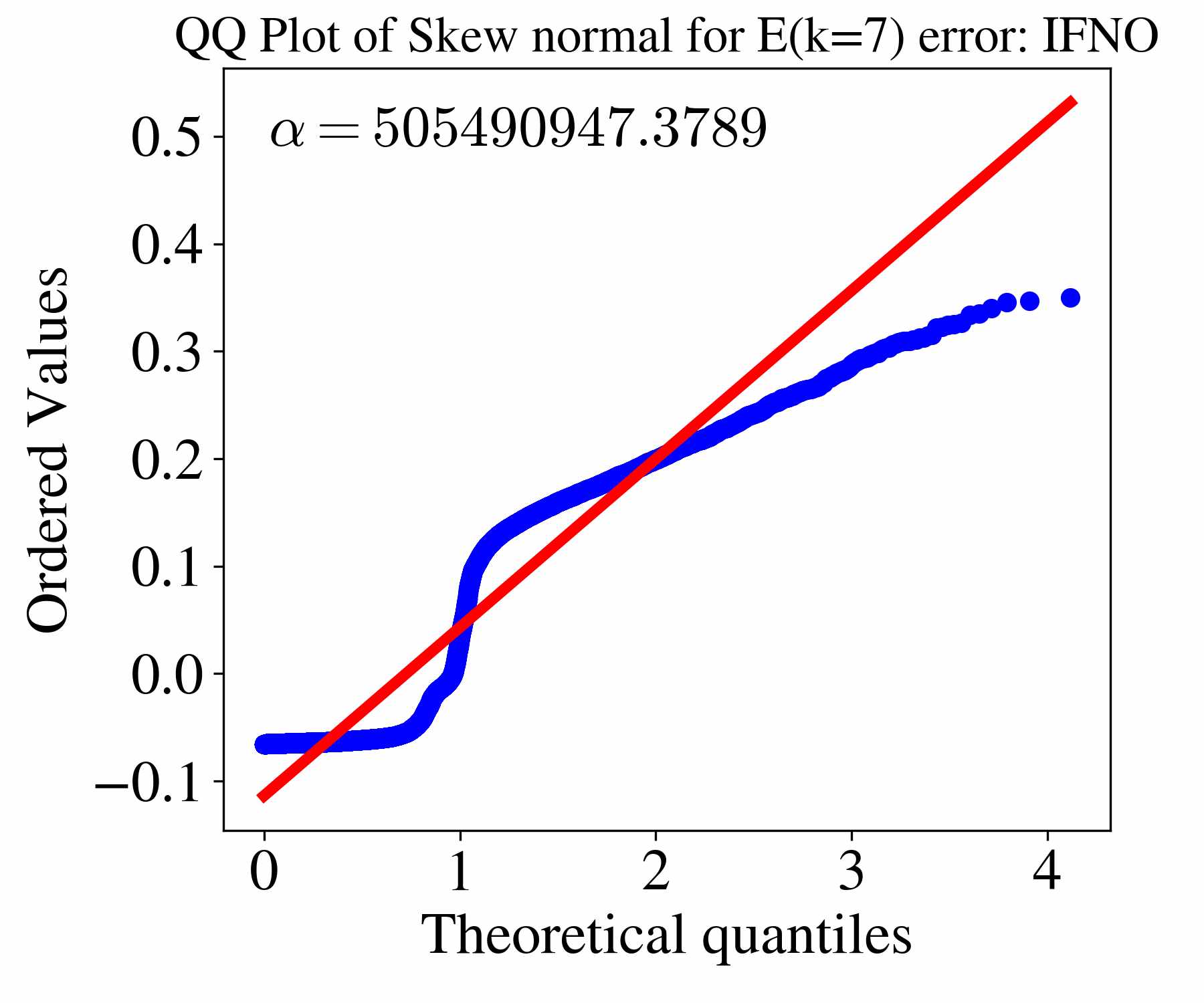}
            \put(-3,65){\small (h)} 
        \end{overpic} 
    \end{subfigure}
    \hfill
    \begin{subfigure}[b]{0.32\textwidth}
        \begin{overpic}[width=1\linewidth]{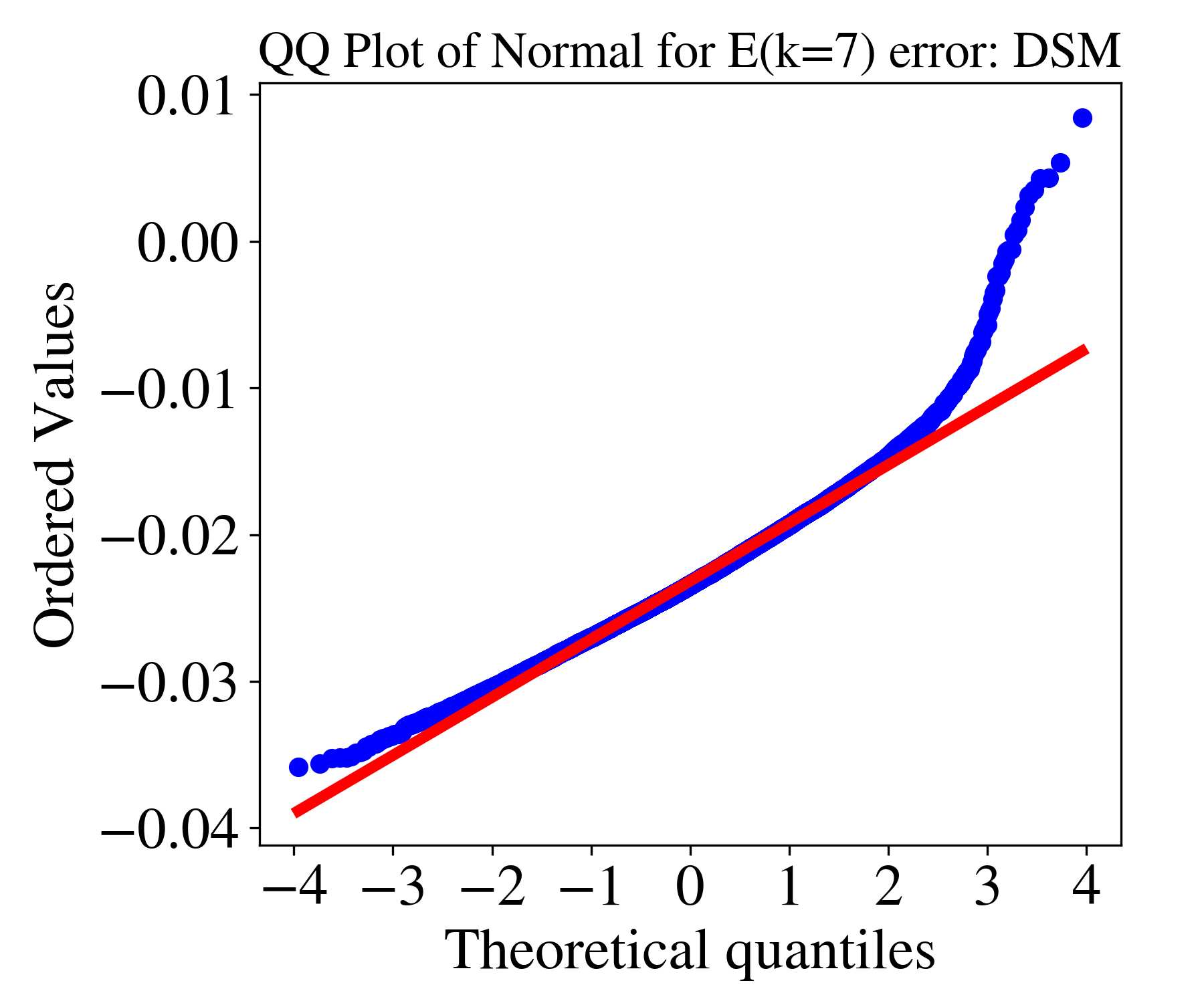}
            \put(-3,65){\small (i)} 
        \end{overpic}
    \end{subfigure}
    \vspace{0.1cm}

    \begin{subfigure}[b]{1\textwidth}
        \centering
        \begin{overpic}[width=0.32\linewidth]{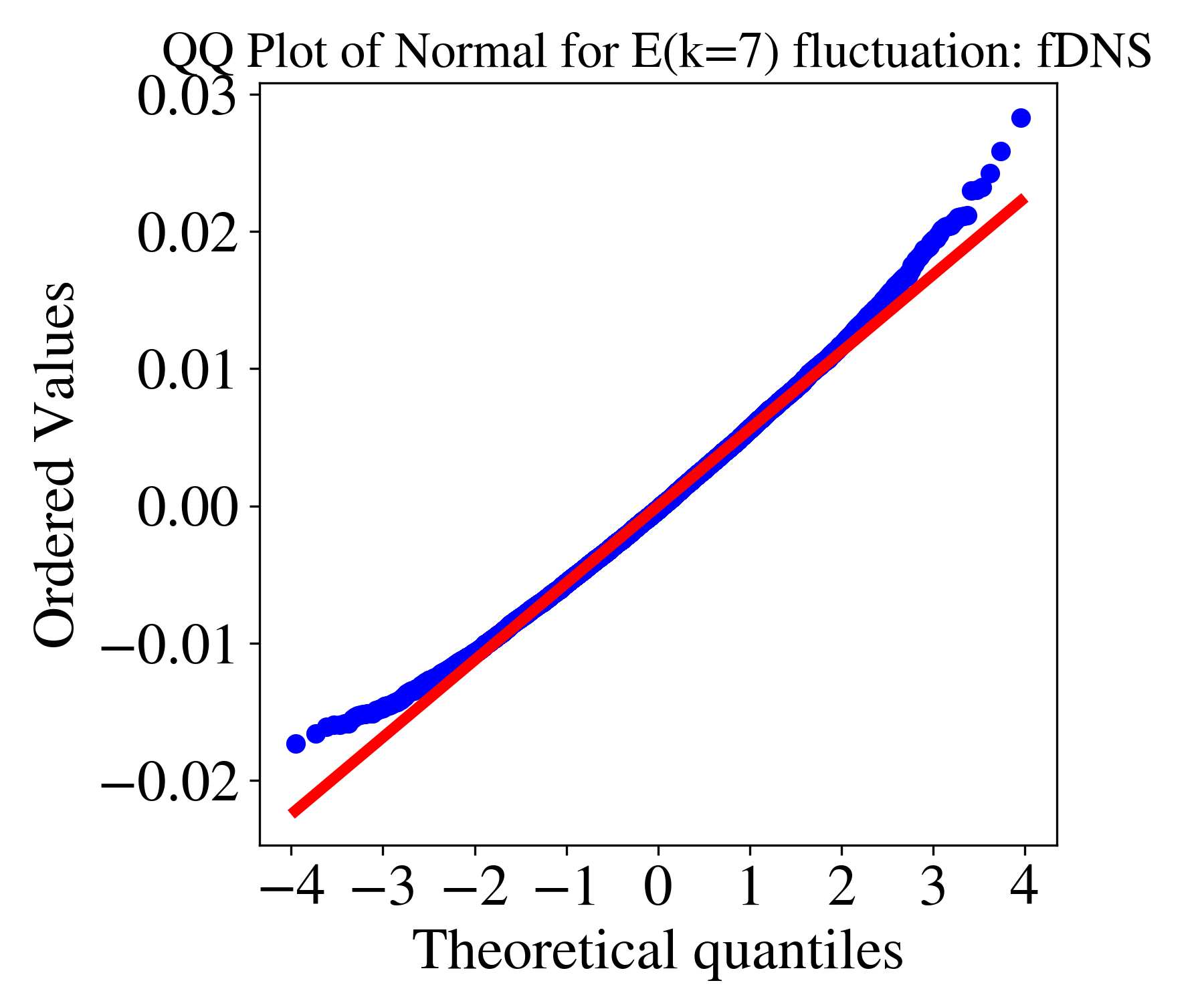}
            \put(-3,65){\small (j)}  
        \end{overpic}
    \end{subfigure}

	\caption{The QQ plots of $E(k=7)$ errors for each method at the representative time interval $\Delta T = 0.2\tau$: (a) F-IFNO constrained; (b) F-IFNO unconstrained; (c) F-IUFNO constrained; (d) F-IUFNO unconstrained; (e) IUFNO constrained; (f) IUFNO unconstrained; (g) IFNO constrained; (h) IFNO unconstrained; (i) DSM; (j) fDNS. Note that for fDNS, the values represent natural statistical fluctuations over time, not prediction errors.}\label{fig:22}
\end{figure}

Following the methodology employed in Fig.~\ref{fig:13}, we further plot the QQ plots of the $E(k=3)$, $E(k=5)$, and $E(k=7)$ errors for each method at the representative time interval $\Delta T = 0.2\tau$ in Fig.~\ref{fig:18}, Fig.~\ref{fig:20}, and Fig.~\ref{fig:22}, respectively, corresponding to the PDFs shown in Fig.~\ref{fig:17}, Fig.~\ref{fig:19}, and Fig.~\ref{fig:21}. 
From Figs.~\ref{fig:18}, \ref{fig:20}, and \ref{fig:22}, it is evident that the error distributions assumed in Figs.~\ref{fig:17}, \ref{fig:19}, and \ref{fig:21} are valid. Specifically, for constrained FNO-based models, DSM, and fDNS, the QQ plots confirm the normality of the distributions. For unconstrained F-IFNO and F-IUFNO, the QQ plots align well with the skew normal distribution. Meanwhile, unconstrained IFNO and IUFNO fail to match either distribution form, consistent with the previous analysis.

In summary, based on the long-term prediction stability and uncertainty quantification (UQ) analysis presented in this subsection, we conclude that both constrained and unconstrained F-IFNO and F-IUFNO, when operated within the optimal time interval, consistently achieve the best overall performance in comparison to DSM and other FNO-based models.

\subsection{Stability analysis of posterior results with perturbations}
\label{subsec4.3}

In this subsection, we apply the perturbation method described in Subsection~\ref{subsec2.3}, where perturbations with different magnitudes ($\tilde{\varepsilon} = 0.1, 0.5, 1, 2, 5, 10$) are added to all Fourier modes for both unconstrained and constrained FNO-based models, as well as DSM, at a fixed time interval $\Delta T = 0.2\tau$. We conduct a statistical stability analysis based on thirty independent datasets, each consisting of 600 time steps, resulting in a total of 18,000 samples per method for each perturbation magnitude. Furthermore, we evaluate the prediction errors of the kinetic energy $E_k$ and the velocity spectra $E(k)$ for each method under different levels of perturbation.

\begin{figure}[ht!]
    \centering
    \begin{subfigure}[b]{0.49\textwidth}
        \begin{overpic}[width=1\linewidth]{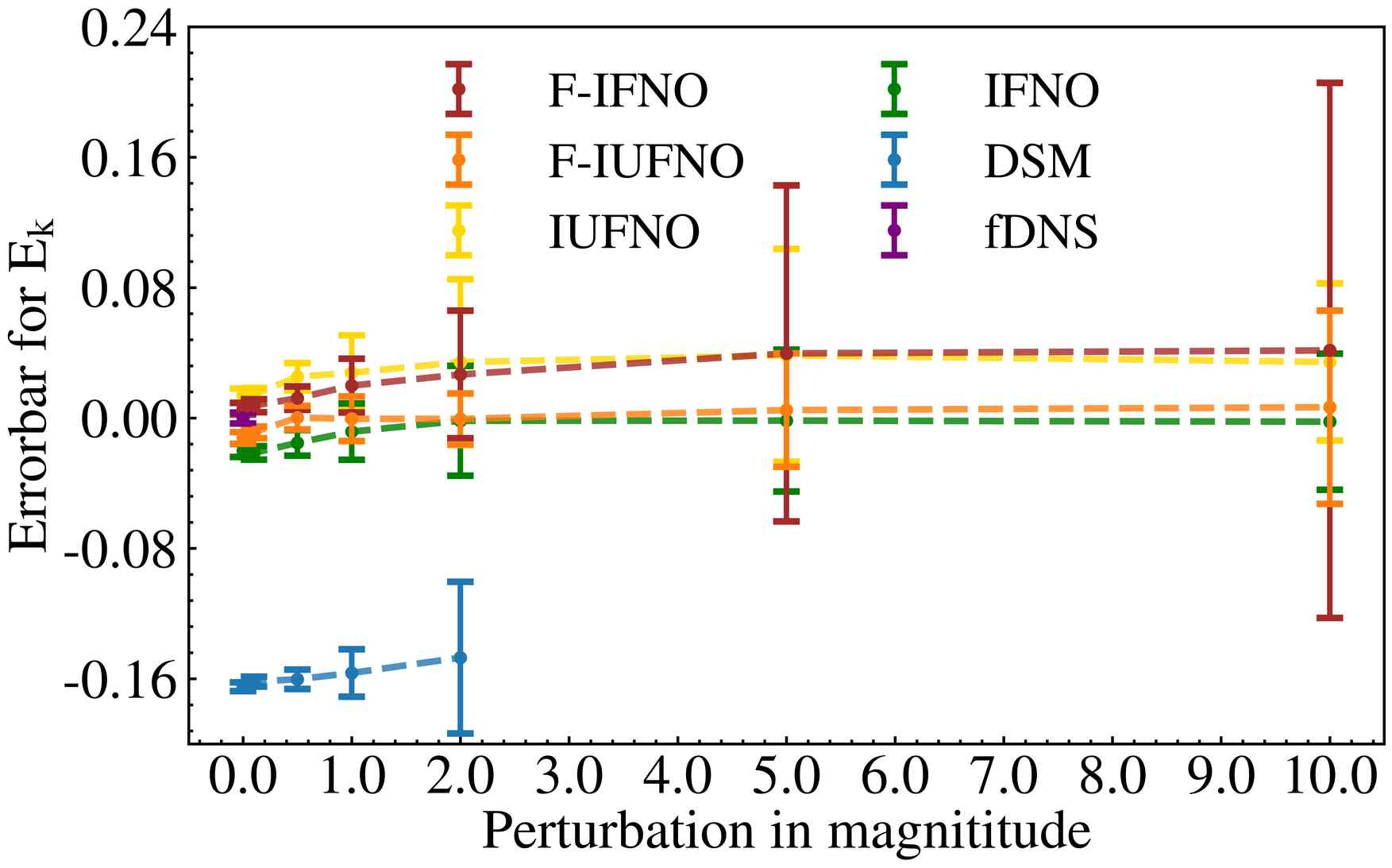}
            \put(0,55){\small (a)}  
        \end{overpic}
    \end{subfigure}
    \hfill
    \begin{subfigure}[b]{0.49\textwidth}
        \begin{overpic}[width=1\linewidth]{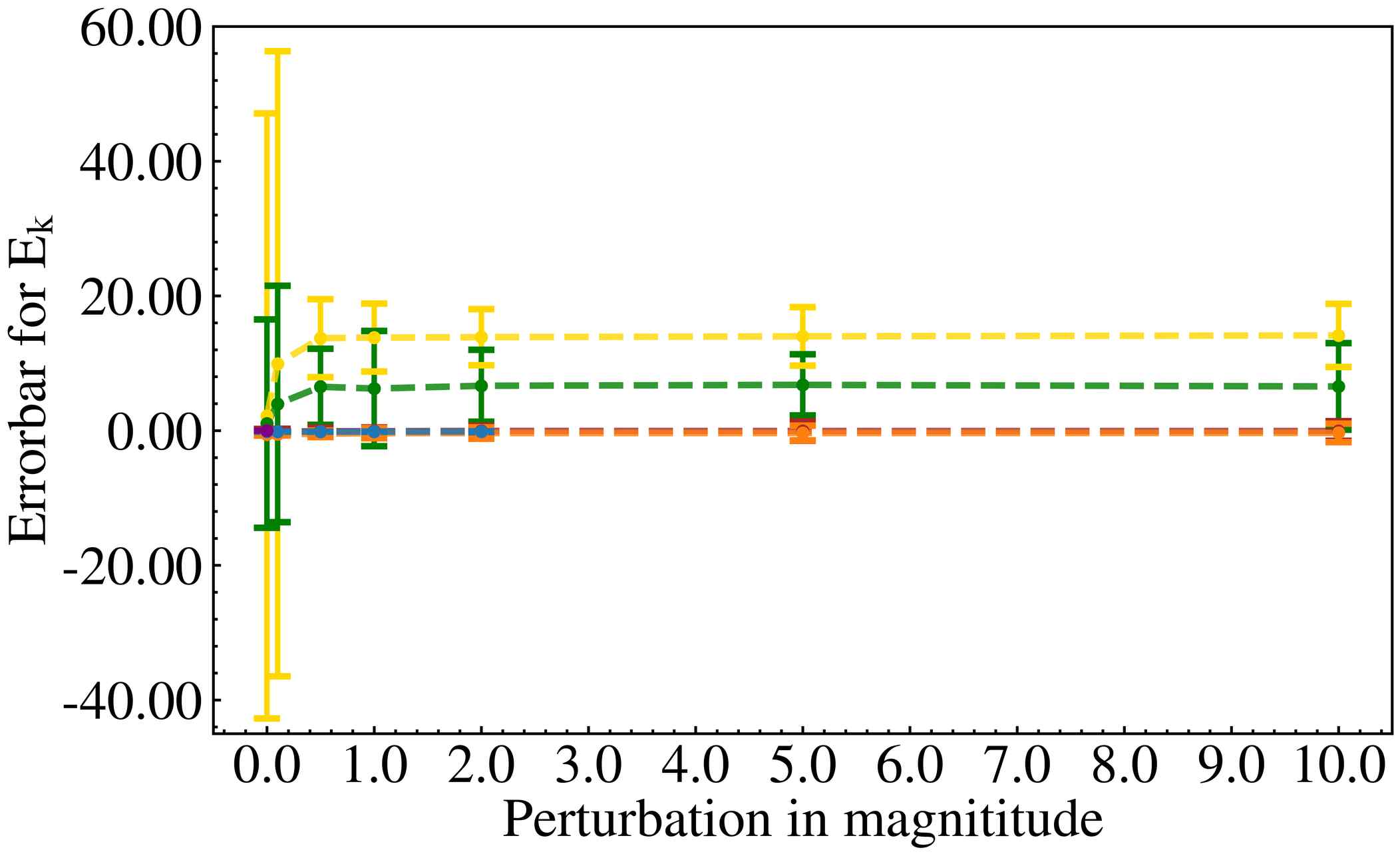}
            \put(0,55){\small (b)} 
        \end{overpic} 
    \end{subfigure}

	\caption{Errorbars of the kinetic energy $E_k$ for various methods as a function of the perturbation magnitude $\tilde{\varepsilon}$: (a) FNO-based models constrained; (b) FNO-based unconstrained. Note that for fDNS, the values represent natural statistical fluctuations over time, not prediction errors.}\label{fig:23}
\end{figure}

As shown in Fig.~\ref{fig:23}, the errorbars of kinetic energy $E_k$ are plotted for various methods under different perturbation magnitudes $\tilde{\varepsilon}$. In Fig.~\ref{fig:23}(a), it is observed that the constrained FNO-based models maintain stable performance even under large perturbations, whereas DSM fails when $\tilde{\varepsilon} = 5$ and $10$. Across all levels of perturbation, the constrained FNO-based models consistently exhibit smaller errorbars with both lower mean and standard deviation compared to DSM. Moreover, for all methods, the errorbars increase with increasing $\tilde{\varepsilon}$, which aligns with physical expectations.
Among all models, F-IUFNO achieves the smallest mean and standard deviation errors across all perturbation magnitudes. In contrast, Fig.~\ref{fig:23}(b) shows that the unconstrained versions, particularly IFNO and IUFNO, suffer from significantly larger errors, indicating poor accuracy and stability. Nevertheless, the unconstrained F-IFNO and F-IUFNO still demonstrate superior accuracy compared to DSM. Overall, F-IFNO and F-IUFNO consistently outperform other methods in terms of stability and robustness under perturbations.

\begin{figure}[ht!]
    \centering
    \begin{subfigure}[b]{0.32\textwidth}
        \begin{overpic}[width=1\linewidth]{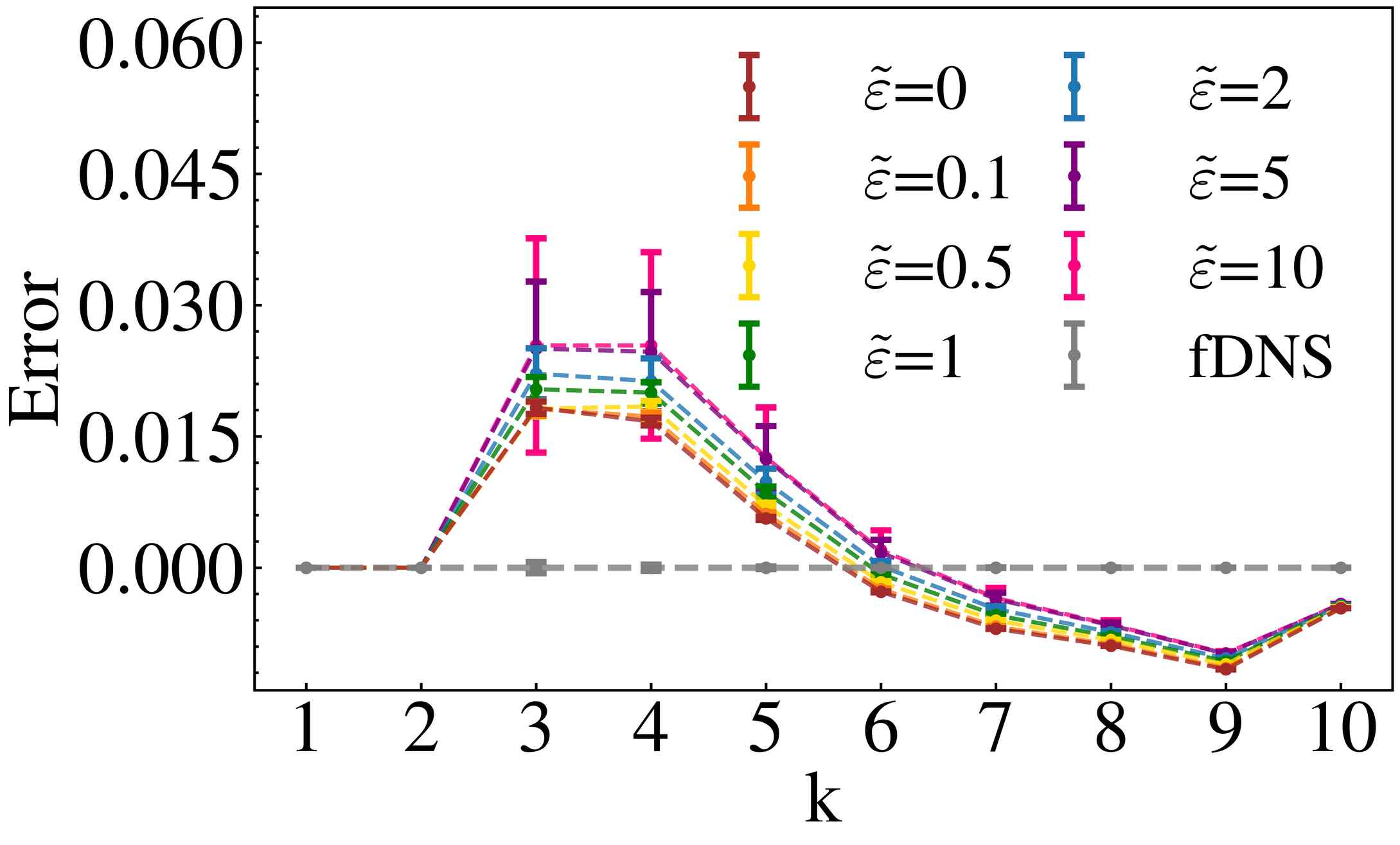}
            \put(-3,55){\small (a)}  
        \end{overpic}
    \end{subfigure}
    \hfill
    \begin{subfigure}[b]{0.32\textwidth}
        \begin{overpic}[width=1\linewidth]{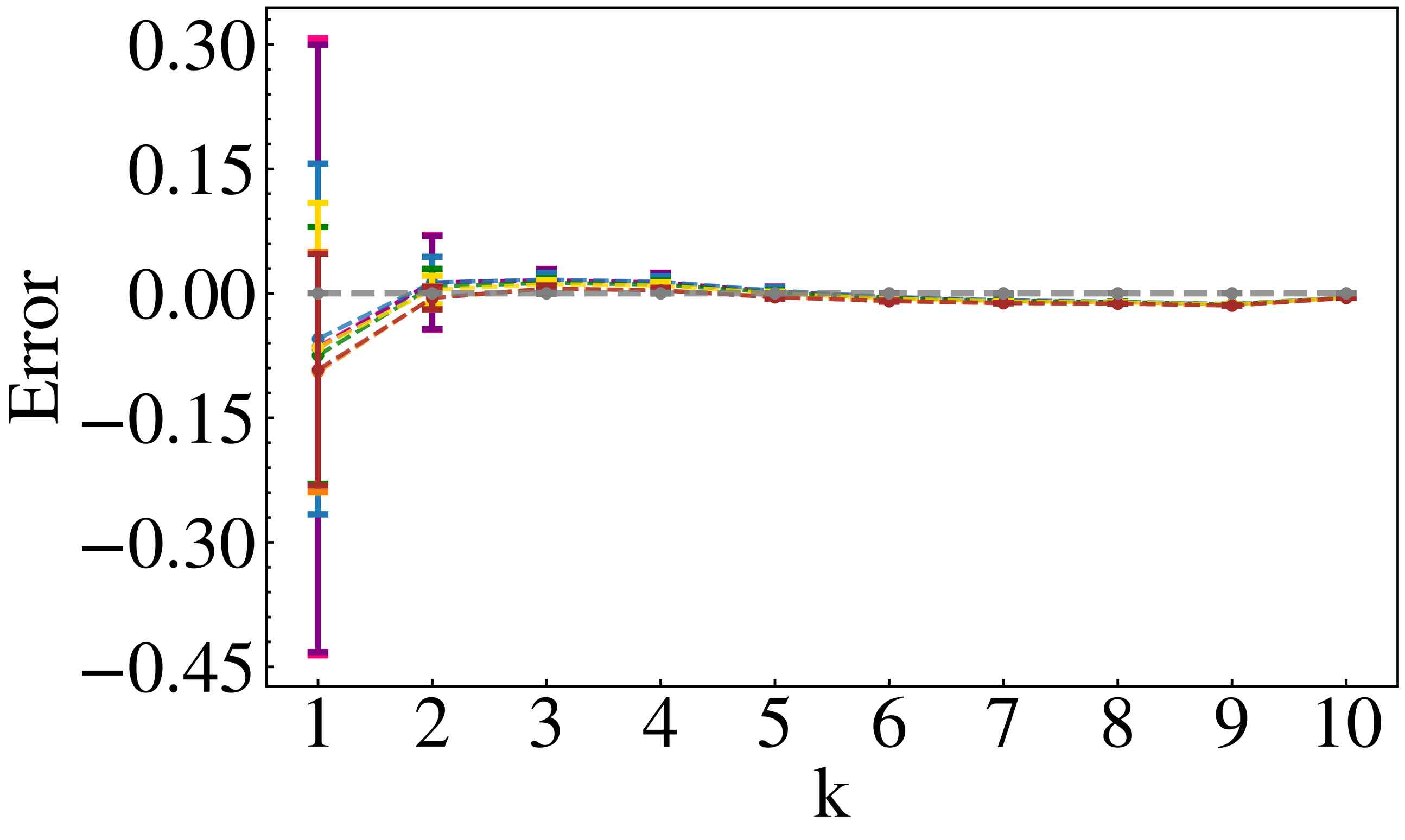}
            \put(-3,55){\small (b)} 
        \end{overpic} 
    \end{subfigure}
    \hfill
    \begin{subfigure}[b]{0.32\textwidth}
        \begin{overpic}[width=1\linewidth]{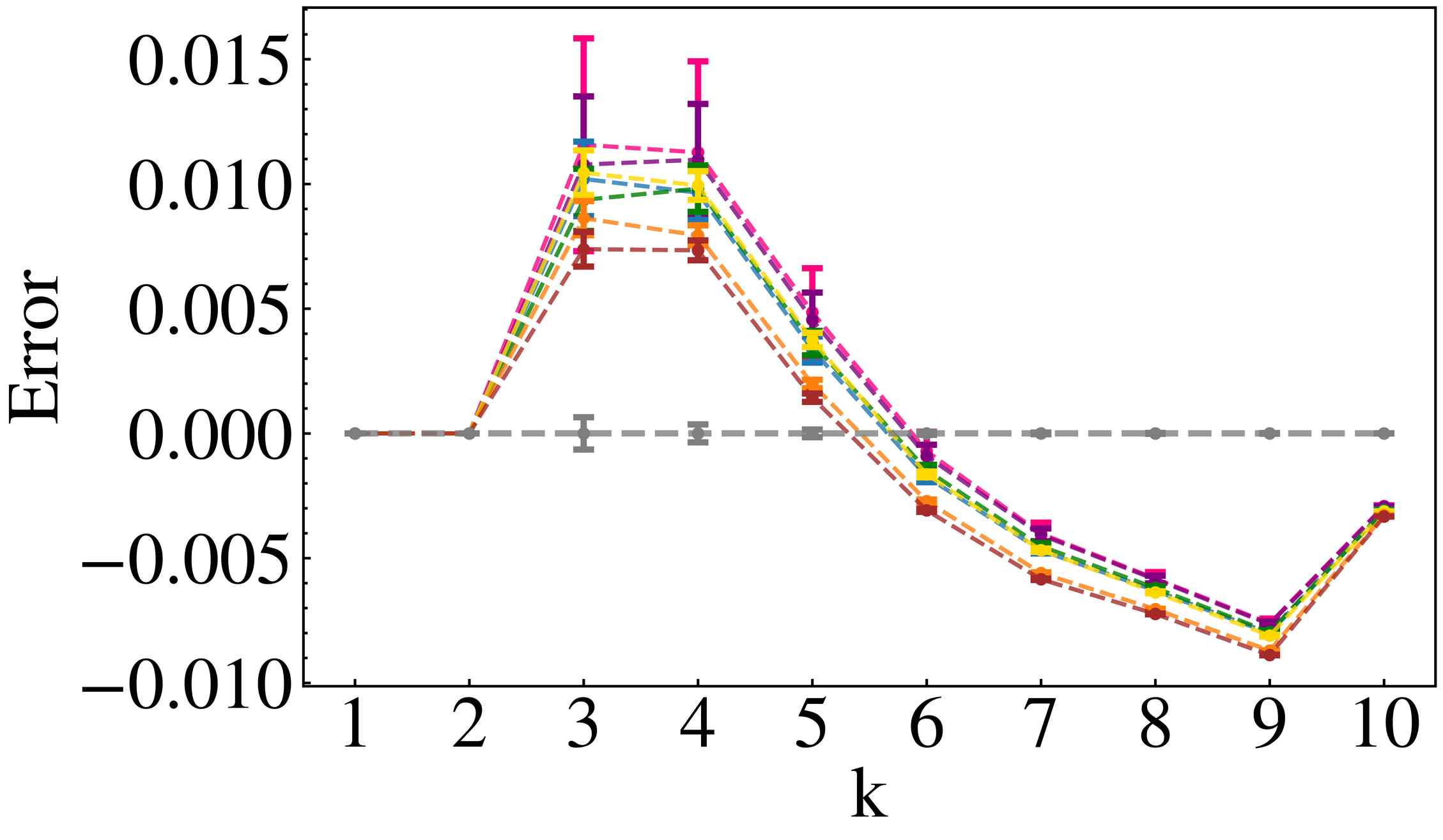}
            \put(-3,55){\small (c)} 
        \end{overpic}
    \end{subfigure}
    \vspace{0.1cm}

    \begin{subfigure}[b]{0.32\textwidth}
        \begin{overpic}[width=1\linewidth]{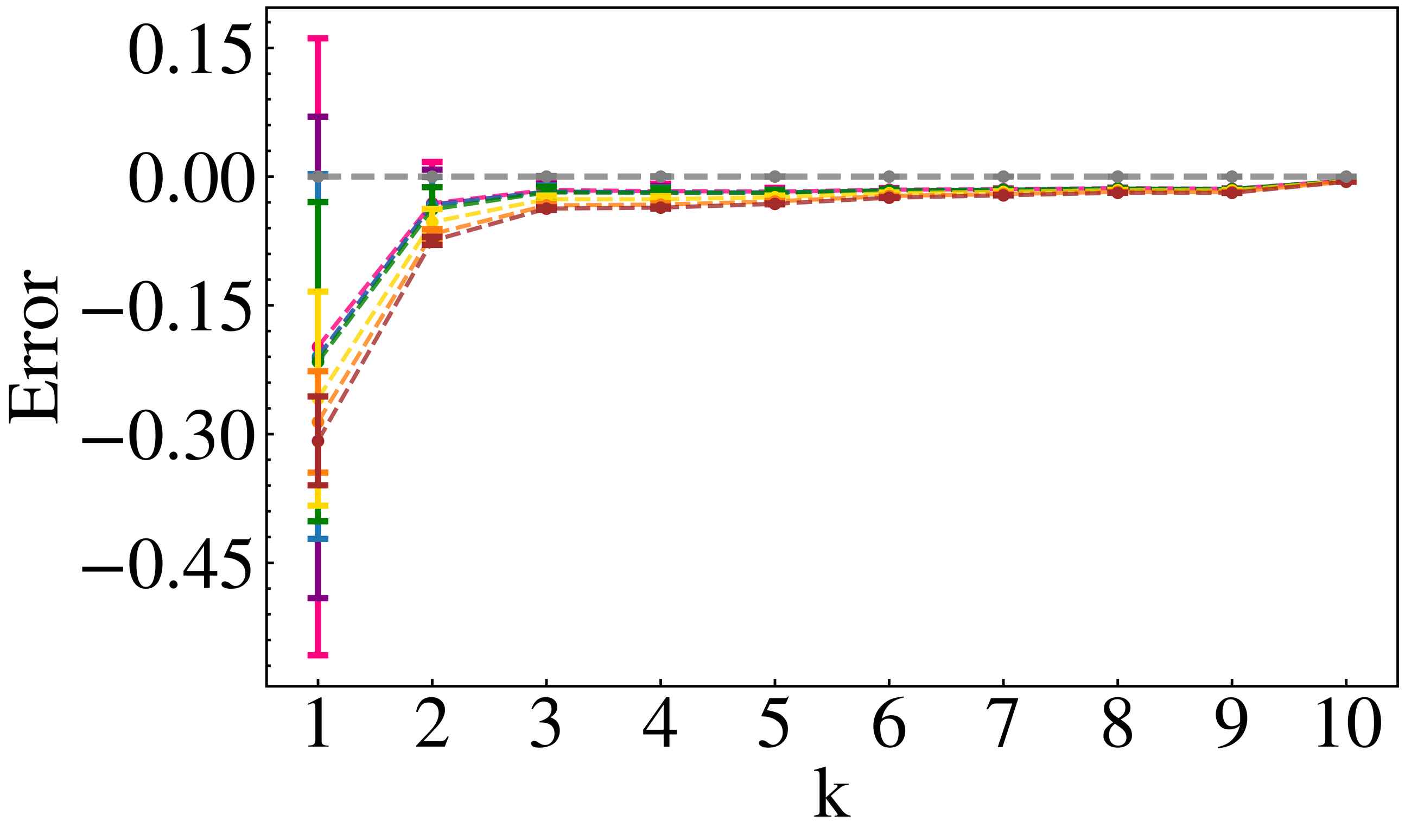}
            \put(-3,55){\small (d)}  
        \end{overpic}
    \end{subfigure}
    \hfill
    \begin{subfigure}[b]{0.32\textwidth}
        \begin{overpic}[width=1\linewidth]{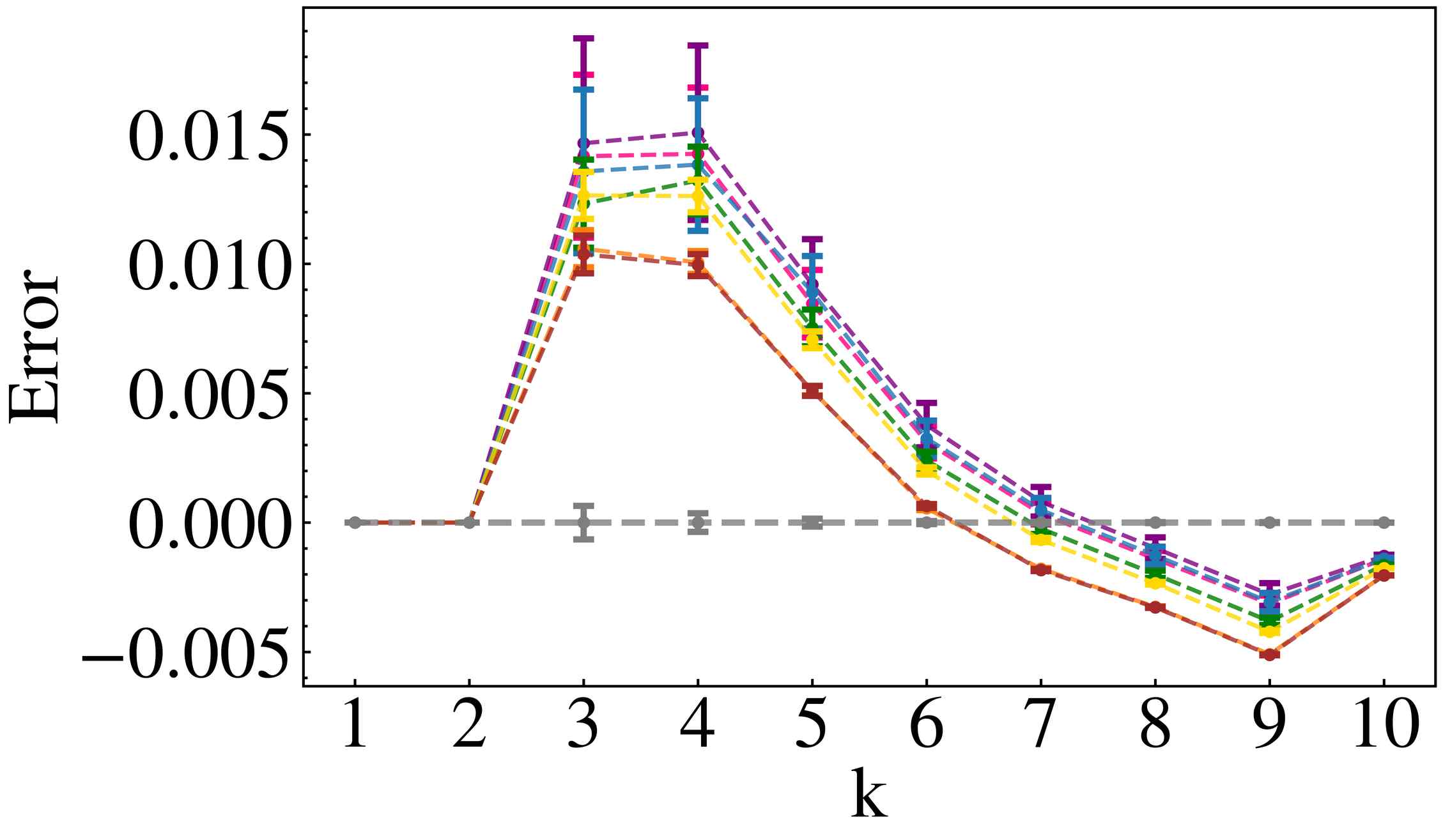}
            \put(-3,55){\small (e)} 
        \end{overpic} 
    \end{subfigure}
    \hfill
    \begin{subfigure}[b]{0.32\textwidth}
        \begin{overpic}[width=1\linewidth]{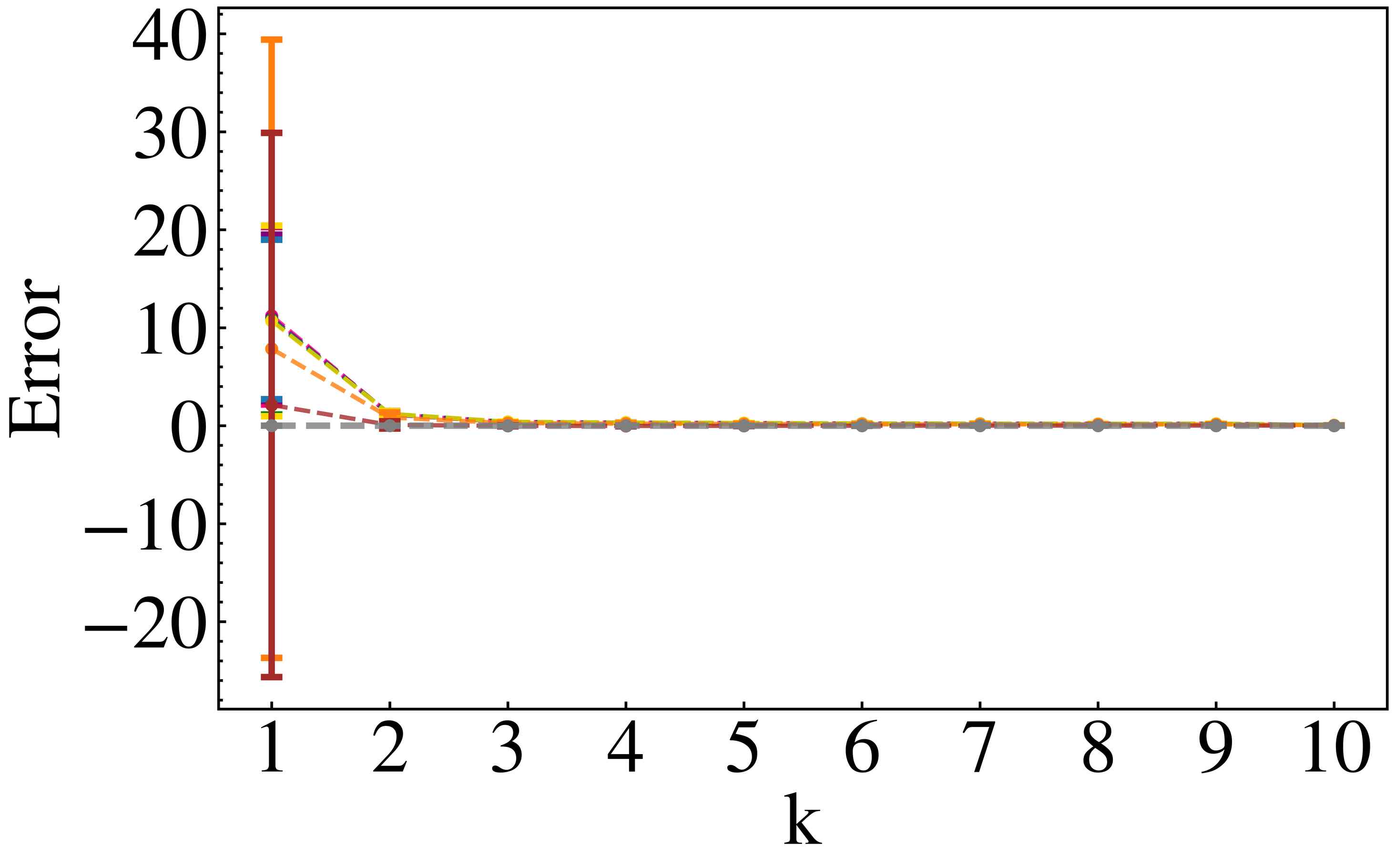}
            \put(-3,55){\small (f)} 
        \end{overpic}
    \end{subfigure}
    \vspace{0.1cm}

    \begin{subfigure}[b]{0.32\textwidth}
        \begin{overpic}[width=1\linewidth]{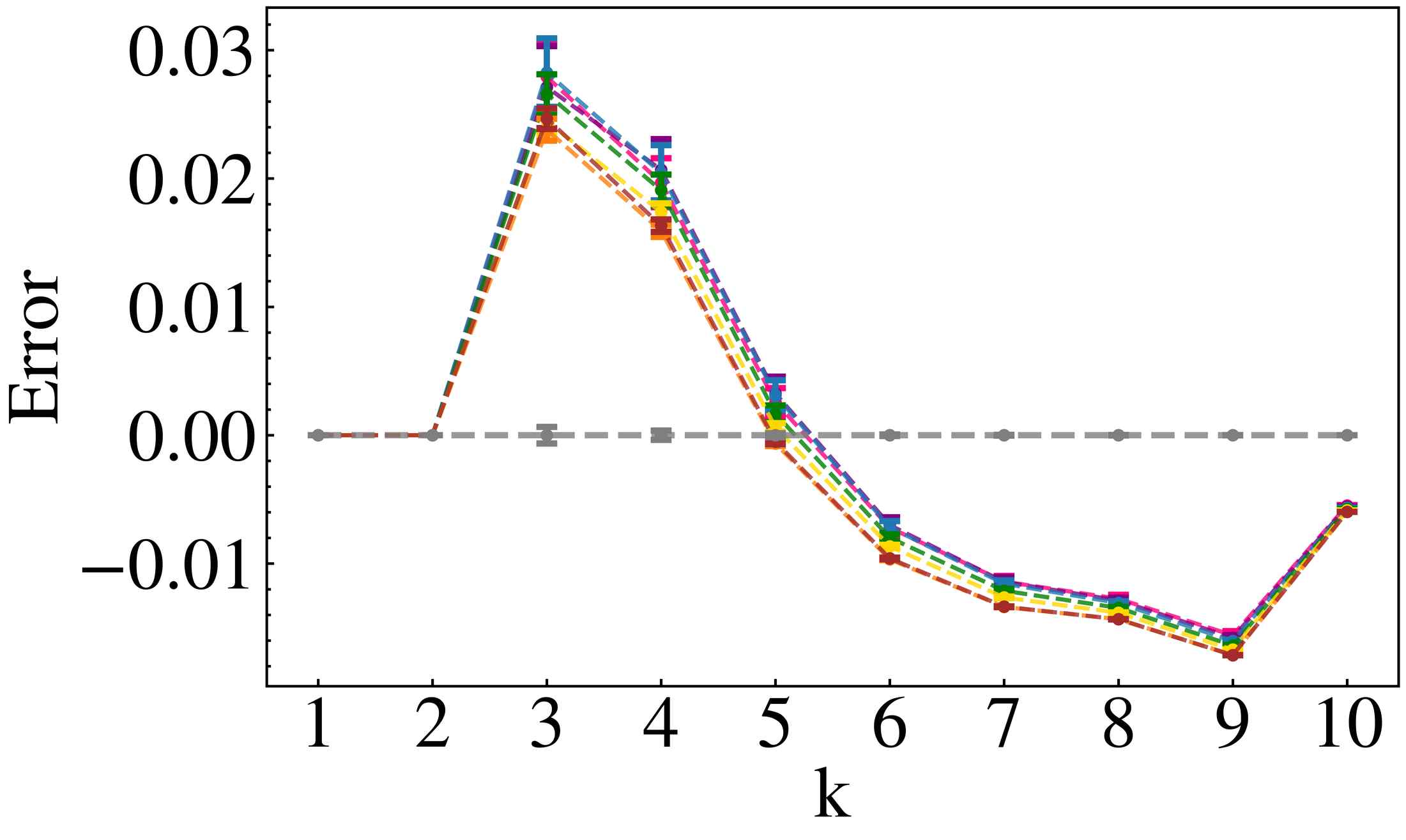}
            \put(-3,55){\small (g)}  
        \end{overpic}
    \end{subfigure}
    \hfill
    \begin{subfigure}[b]{0.32\textwidth}
        \begin{overpic}[width=1\linewidth]{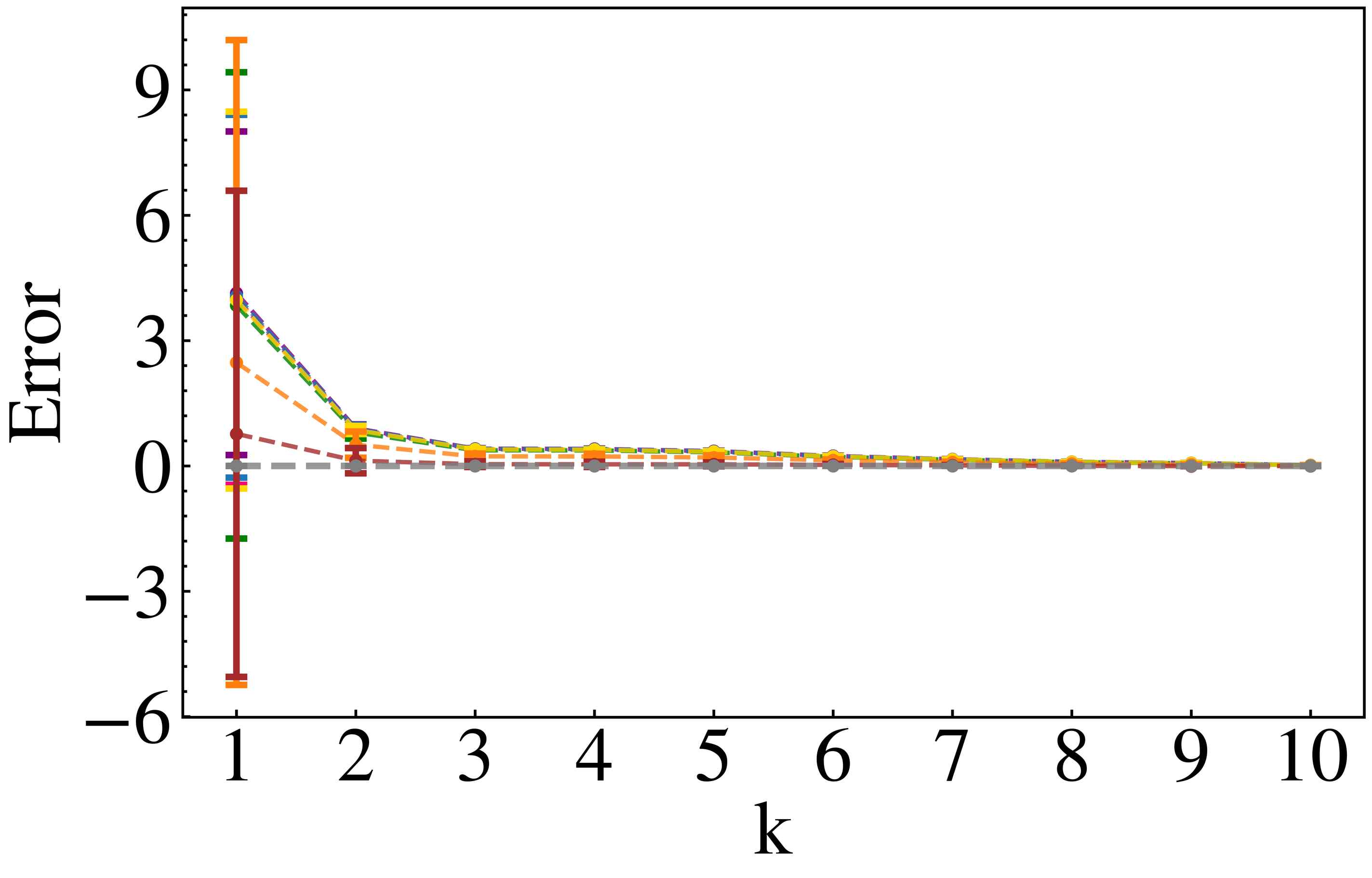}
            \put(-3,55){\small (h)} 
        \end{overpic} 
    \end{subfigure}
    \hfill
    \begin{subfigure}[b]{0.32\textwidth}
        \begin{overpic}[width=1\linewidth]{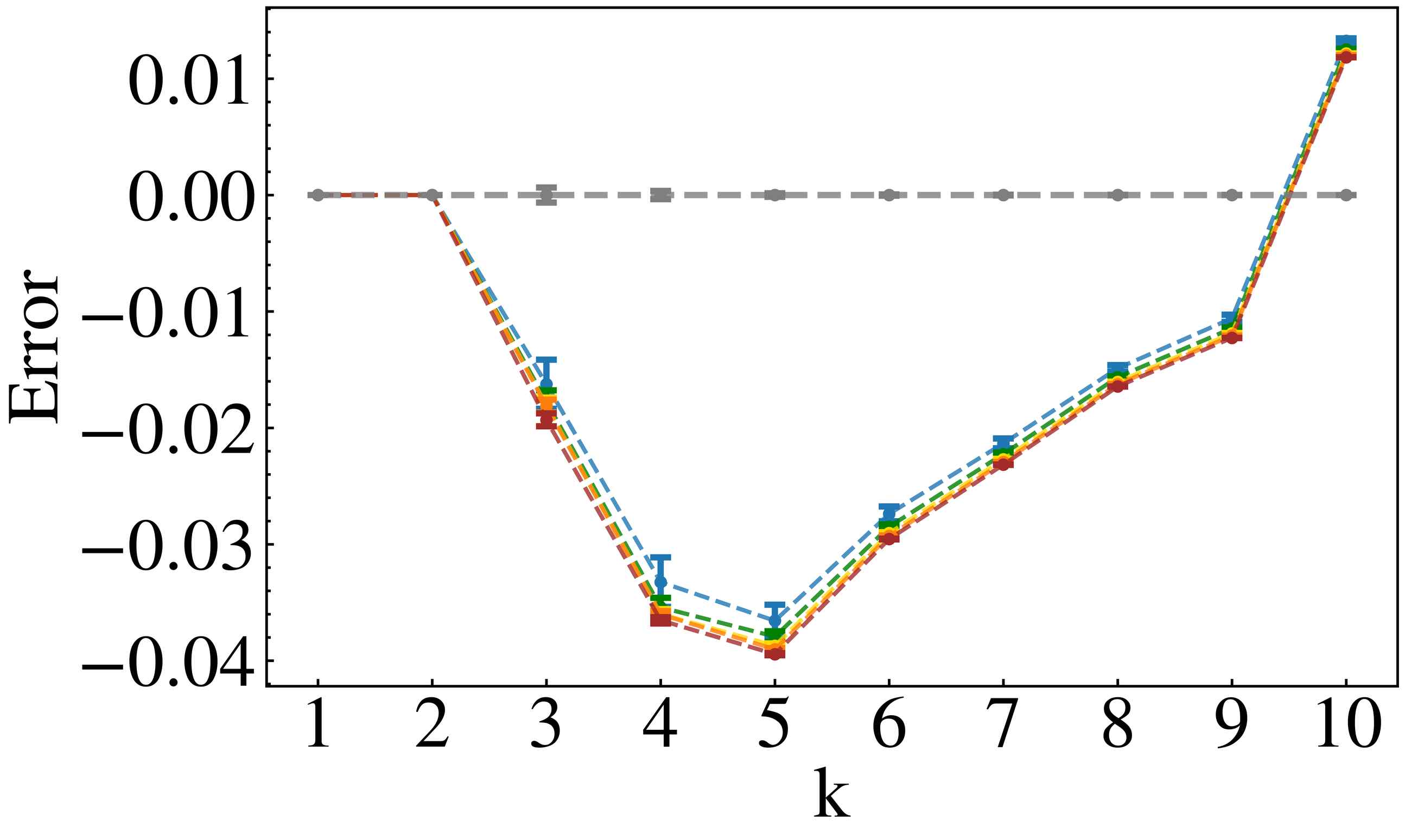}
            \put(-3,55){\small (i)} 
        \end{overpic}
    \end{subfigure}

	\caption{Errorbars of velocity spectra $E(k)$ for various methods with different perturbation magnitude $\tilde{\varepsilon}$: (a) F-IFNO constrained; (b) F-IFNO unconstrained; (c) F-IUFNO constrained; (d) F-IUFNO unconstrained; (e) IUFNO constrained; (f) IUFNO unconstrained; (g) IFNO constrained; (h) IFNO unconstrained; (i) DSM. Note that for fDNS, the values represent natural statistical fluctuations over time, not prediction errors.}\label{fig:24}
\end{figure}

Fig.~\ref{fig:24} presents the errorbars of the velocity spectra $E(k)$ for various methods under different levels of perturbation magnitude $\tilde{\varepsilon}$. Specifically, Fig.~\ref{fig:24}(a) and (b) show results for constrained and unconstrained F-IFNO, (c) and (d) for constrained and unconstrained F-IUFNO, (e) and (f) for constrained and unconstrained IUFNO, (g) and (h) for constrained and unconstrained IFNO, and (i) corresponds to DSM.
For both versions of the FNO-based models and DSM, the errorbars of $E(k)$ generally increase with the perturbation magnitude $\tilde{\varepsilon}$. In the constrained FNO-based models shown in Figs.~\ref{fig:24}(a), (c), (e), and (g), the mean error initially increases from zero at $k=3$, then drops below zero, and eventually rises back toward zero as $k$ increases. At the same time, the standard deviation tends to decrease with increasing $k$.
For DSM (Fig.~\ref{fig:24}(i)), the mean error decreases from zero at $k=3$, reaches its minimum at $k=5$, and then increases to a positive value at $k=10$. Overall, constrained FNO-based models consistently exhibit a much better performance, lower mean and standard deviation of errors than their unconstrained counterparts, demonstrating greater robustness and accuracy under perturbations.

\begin{figure}[ht!]
    \centering
    \begin{subfigure}[b]{0.49\textwidth}
        \begin{overpic}[width=1\linewidth]{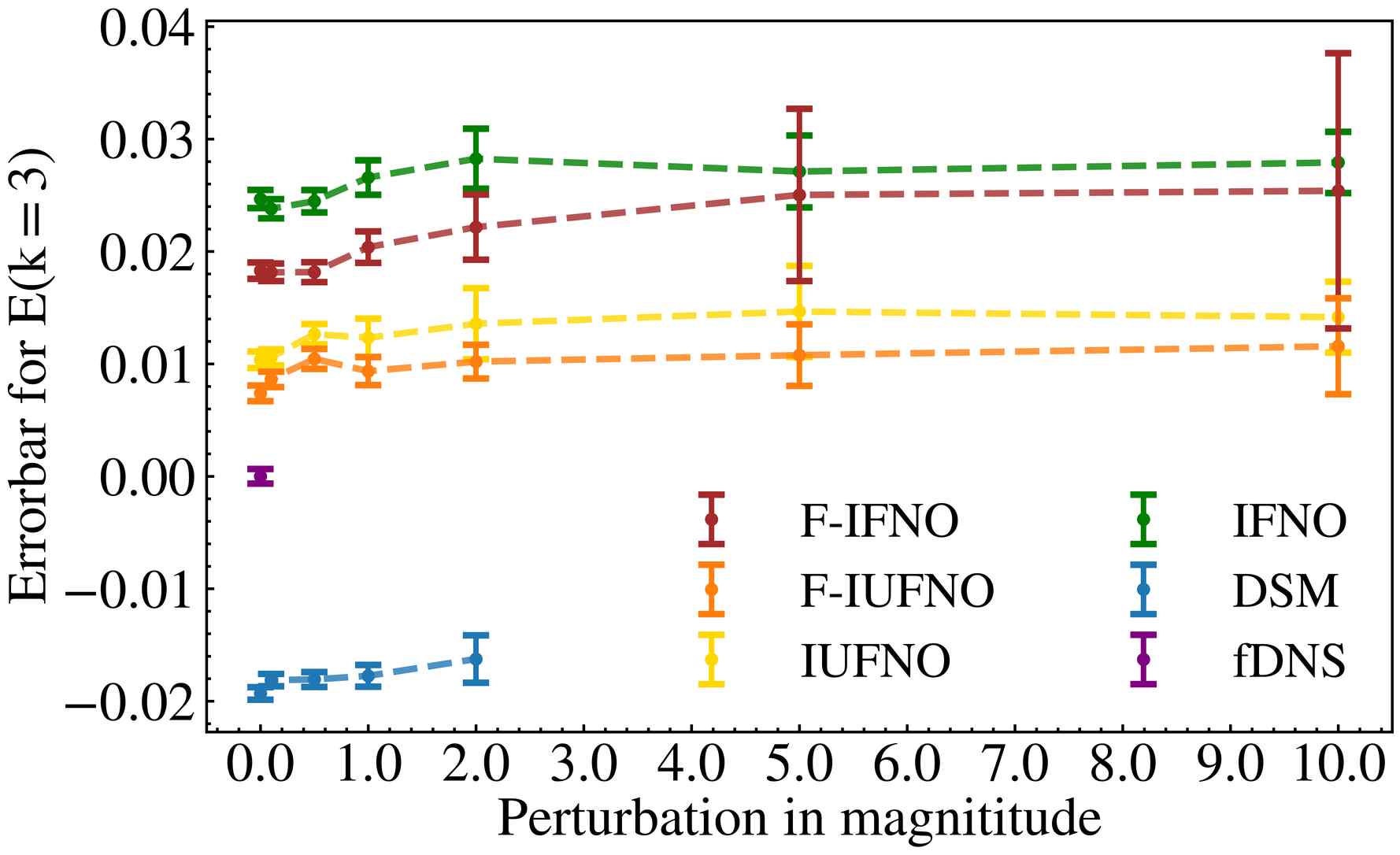}
            \put(-1,55){\small (a)}  
        \end{overpic}
    \end{subfigure}
    \hfill
    \begin{subfigure}[b]{0.49\textwidth}
        \begin{overpic}[width=1\linewidth]{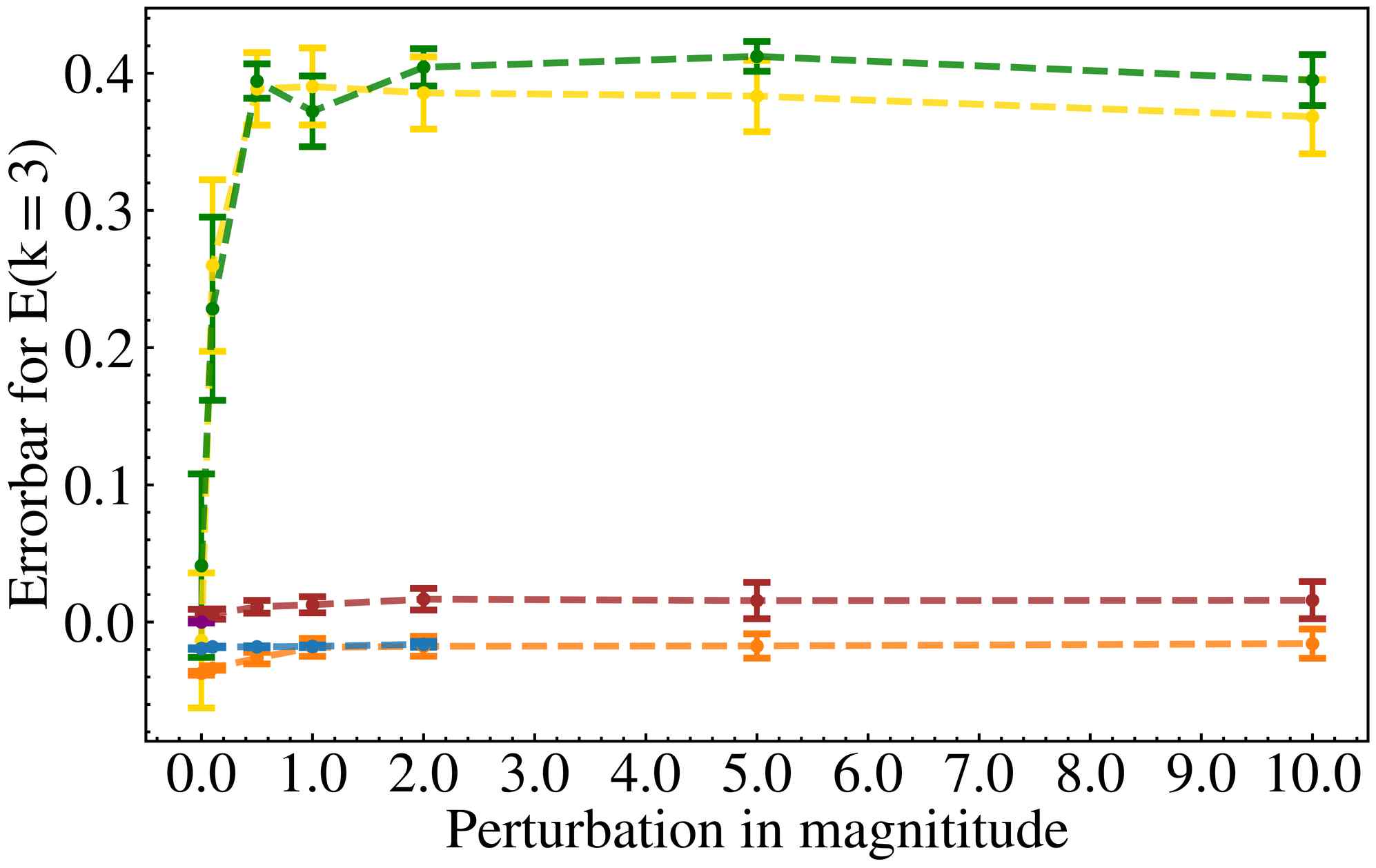}
            \put(-1,55){\small (b)} 
        \end{overpic} 
    \end{subfigure}
    \vspace{0.1cm}

    \begin{subfigure}[b]{0.49\textwidth}
        \begin{overpic}[width=1\linewidth]{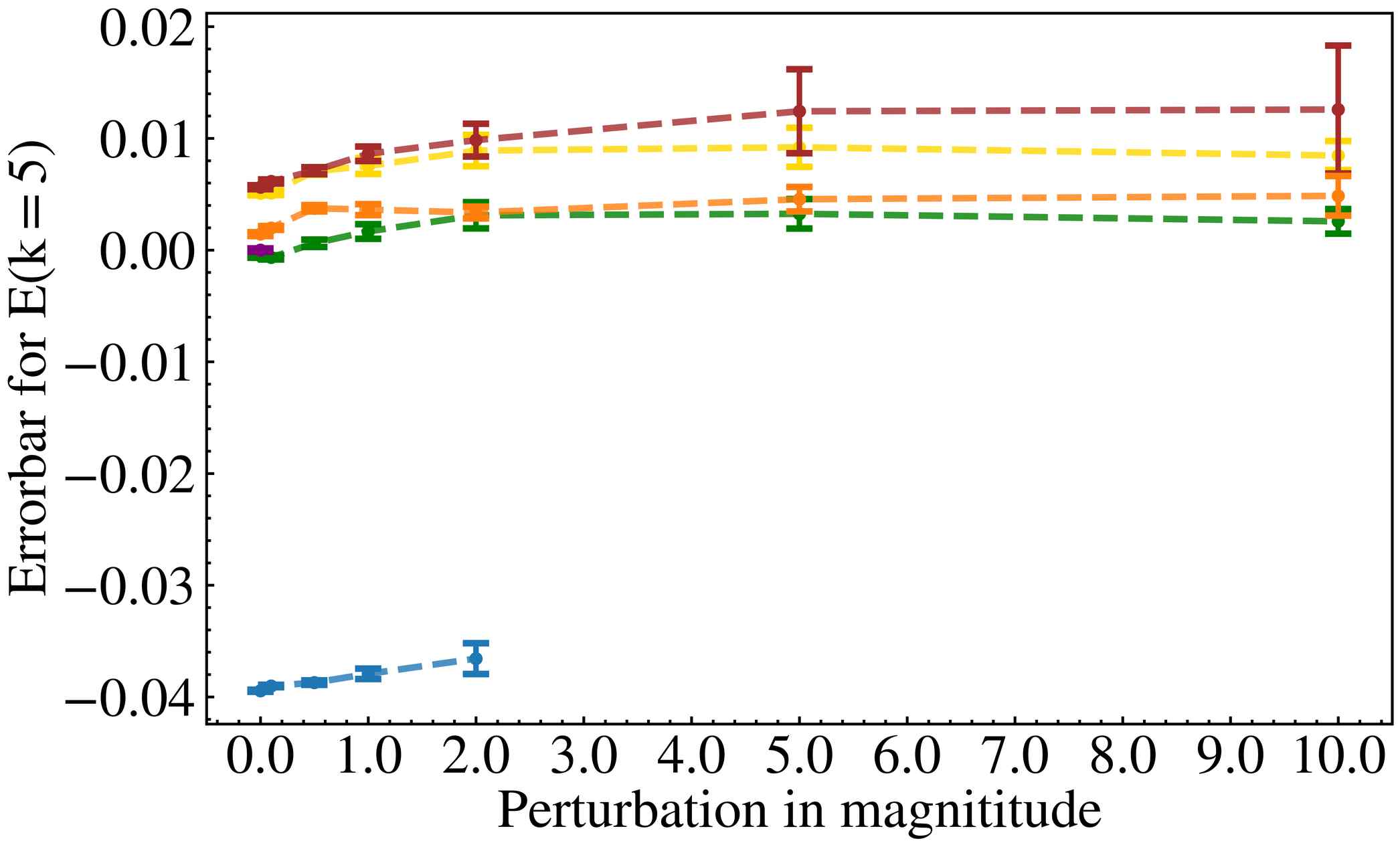}
            \put(-1,55){\small (c)}  
        \end{overpic}
    \end{subfigure}
    \hfill
    \begin{subfigure}[b]{0.49\textwidth}
        \begin{overpic}[width=1\linewidth]{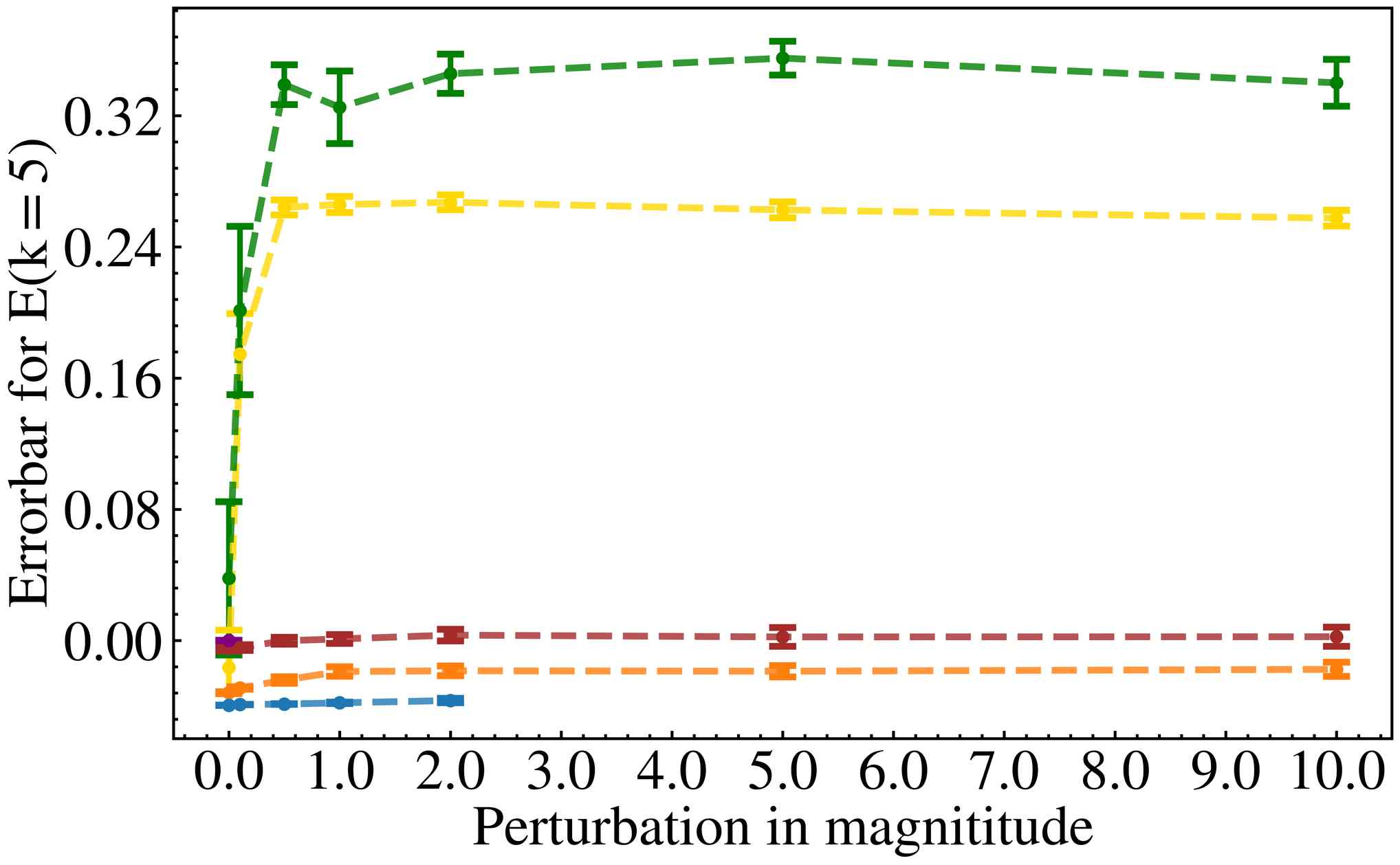}
            \put(-1,55){\small (d)} 
        \end{overpic} 
    \end{subfigure}
    \vspace{0.1cm}

    \begin{subfigure}[b]{0.49\textwidth}
        \begin{overpic}[width=1\linewidth]{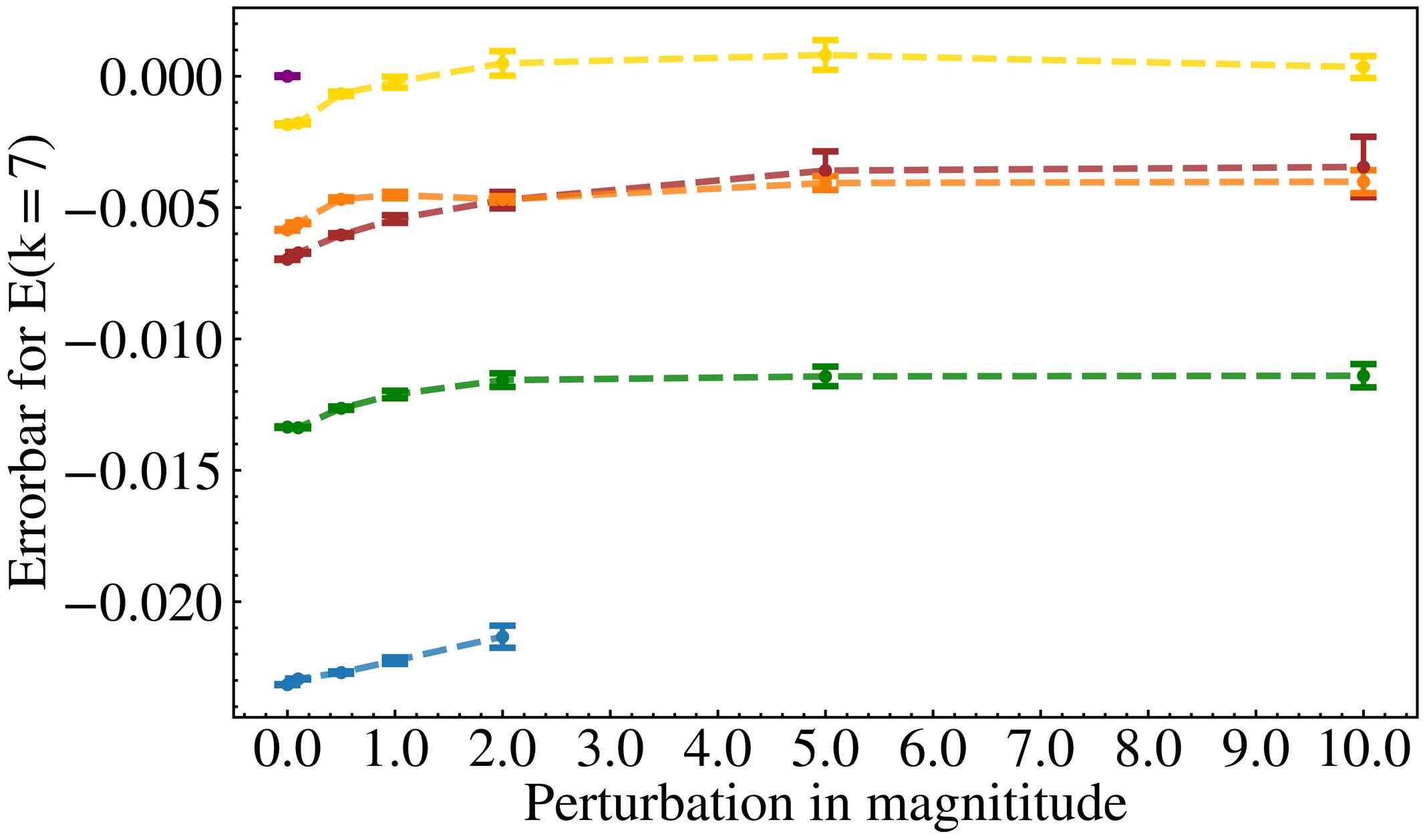}
            \put(-1,55){\small (e)}  
        \end{overpic}
    \end{subfigure}
    \hfill
    \begin{subfigure}[b]{0.49\textwidth}
        \begin{overpic}[width=1\linewidth]{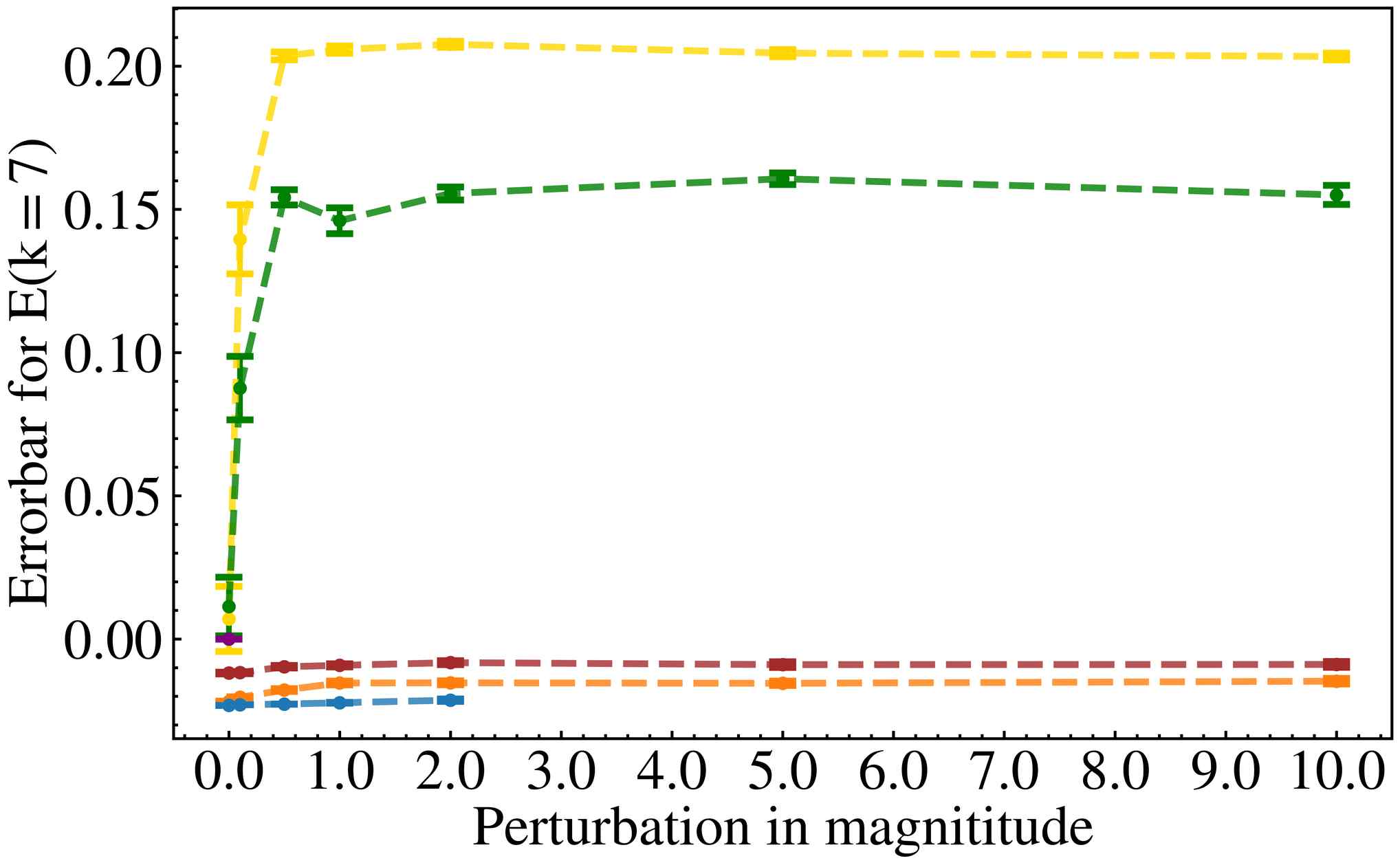}
            \put(-1,55){\small (f)} 
        \end{overpic} 
    \end{subfigure}
    
	\caption{Errorbars of $E(k=3, 5, 7)$ for various methods as a function of the perturbation magnitude $\tilde{\varepsilon}$: (a) $E(k=3)$ for FNO-based models constrained; (b) $E(k=3)$ for FNO-based models unconstrained; (c) $E(k=5)$ for FNO-based models constrained; (d) $E(k=5)$ for FNO-based models unconstrained; (e) $E(k=7)$ for FNO-based models constrained; (f) $E(k=7)$ for FNO-based models unconstrained. Note that for fDNS, the values represent natural statistical fluctuations over time, not prediction errors.}\label{fig:25}
\end{figure}

Furthermore, Fig.~\ref{fig:25} presents the errorbars of $E(k=3, 5, 7)$ for various methods as a function of the perturbation magnitude $\tilde{\varepsilon}$. Specifically, Figs.~\ref{fig:25}(a) and (b) show the errorbars of $E(k=3)$ for constrained and unconstrained FNO-based models, respectively; Figs.~\ref{fig:25}(c) and (d) show those for $E(k=5)$; and Figs.~\ref{fig:25}(e) and (f) correspond to $E(k=7)$.
For the constrained FNO-based models, all variants consistently outperform DSM across all perturbation magnitudes $\tilde{\varepsilon}$, exhibiting smaller mean and standard deviation in $E(k=3, 5, 7)$. Among the unconstrained models, only F-IFNO and F-IUFNO surpass DSM in terms of performance.
Additionally, the constrained FNO-based models show similar performance among themselves for $E(k=3, 5, 7)$, and their results are significantly better than their unconstrained counterparts. Overall, F-IFNO and F-IUFNO demonstrate strong robustness and accuracy under perturbations, outperforming DSM and other FNO-based models.

In summary, the results presented in this subsection demonstrate that both constrained and unconstrained F-IFNO and F-IUFNO exhibit superior robustness and accuracy under various perturbation magnitudes~$\tilde{\varepsilon}$, consistently outperforming DSM and other FNO-based models in terms of prediction stability.

\subsection{Autocorrelation function (ACF) analysis}
\label{subsec4.4}

In this subsection, we analyze the autocorrelation function (ACF) defined in Subsection~\ref{subsec2.4}, which consists of two components: $f_{ac}(\Delta T)$ and $f_{ac}(\Delta T,k)$.

\begin{figure}[ht!]\centering
	\includegraphics[width=0.75\textwidth]{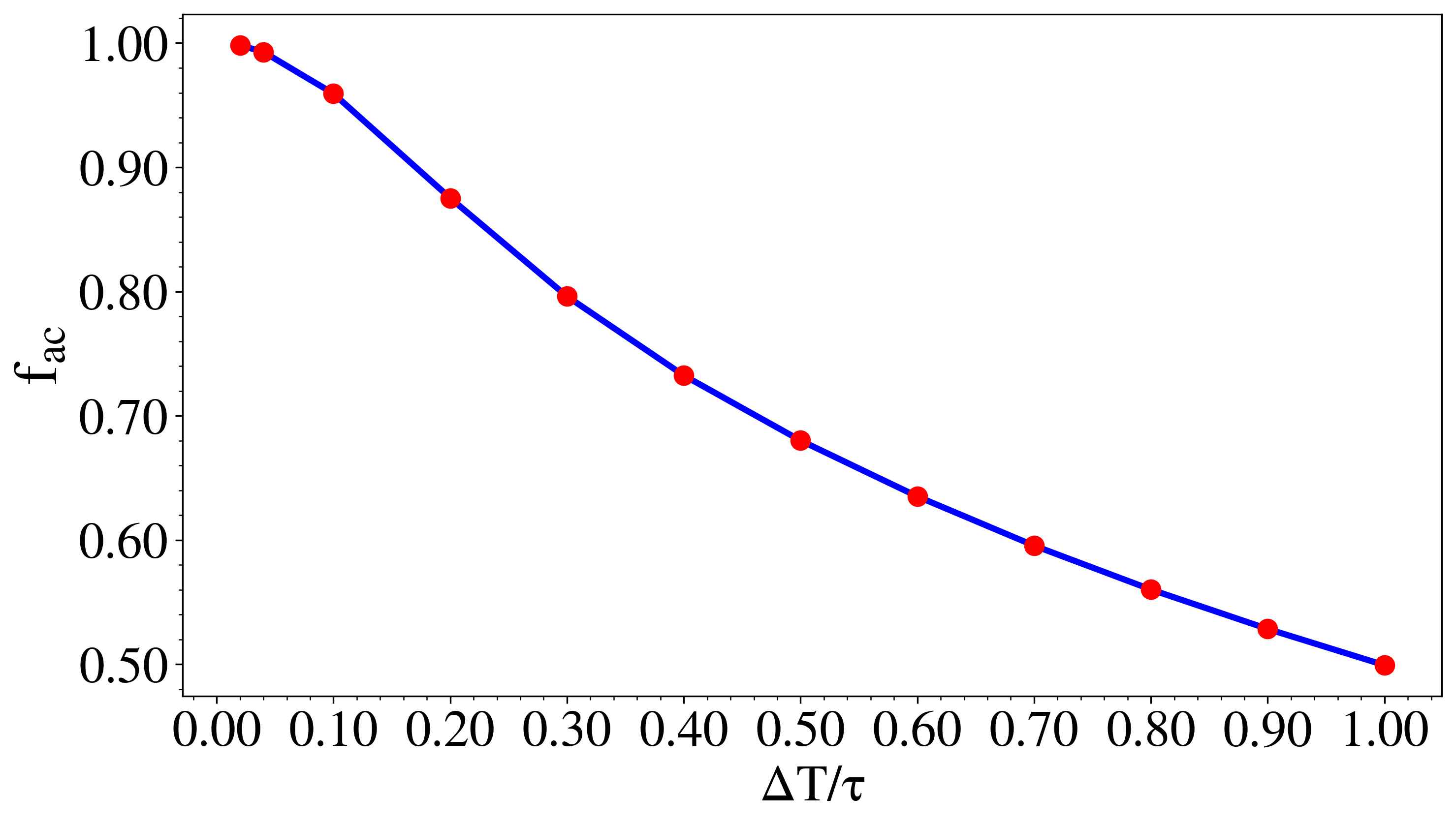}
	\caption{The ACF of the velocity field for forced HIT, over 18,000 fDNS data samples.}\label{fig:26}
\end{figure}

Fig.~\ref{fig:26} presents the ACF of the velocity field, $f_{ac}(\Delta T)$, obtained from fDNS at various time lags $\Delta T = 0.02\tau, 0.04\tau,$ $ 0.1\tau, 0.2\tau, 0.3\tau, 0.4\tau, 0.5\tau, 0.6\tau, 0.7\tau, 0.8\tau, 0.9\tau, 1.0\tau$. The results show a clear monotonic decreasing trend in the ACF as $\Delta T$ increases. When $\Delta T$ is small, the ACF is close to 1, indicating strong temporal correlation. Conversely, for sufficiently large $\Delta T$, the ACF tends toward 0, implying a loss of temporal correlation. Specifically, the ACF values for selected time lags are as shown in Table~\ref{tab:3}.

\begin{table}[ht!]
\centering
\small
\caption{Autocorrelation values $f_{ac}(\Delta T)$ for different time lags $\Delta T$}
\begin{tabular}{cc}
\hline\hline
$\Delta T$ & $f_{ac}(\Delta T)$ \\
\hline
0.02$\tau$   & 0.9982 \\
0.04$\tau$   & 0.9929 \\
0.1$\tau$  & 0.9593 \\
0.2$\tau$  & 0.8752 \\
0.3$\tau$  & 0.7961 \\
0.4$\tau$  & 0.7325 \\
0.5$\tau$  & 0.6801 \\
0.6$\tau$  & 0.6351 \\
0.7$\tau$  & 0.5955 \\
0.8$\tau$  & 0.5602 \\
0.9$\tau$  & 0.5285 \\
$\tau$ & 0.4994 \\
\hline\hline
\end{tabular}
\label{tab:3}
\end{table}

\begin{figure}[ht!]
    \centering
    \begin{subfigure}[b]{0.49\textwidth}
        \begin{overpic}[width=1\linewidth]{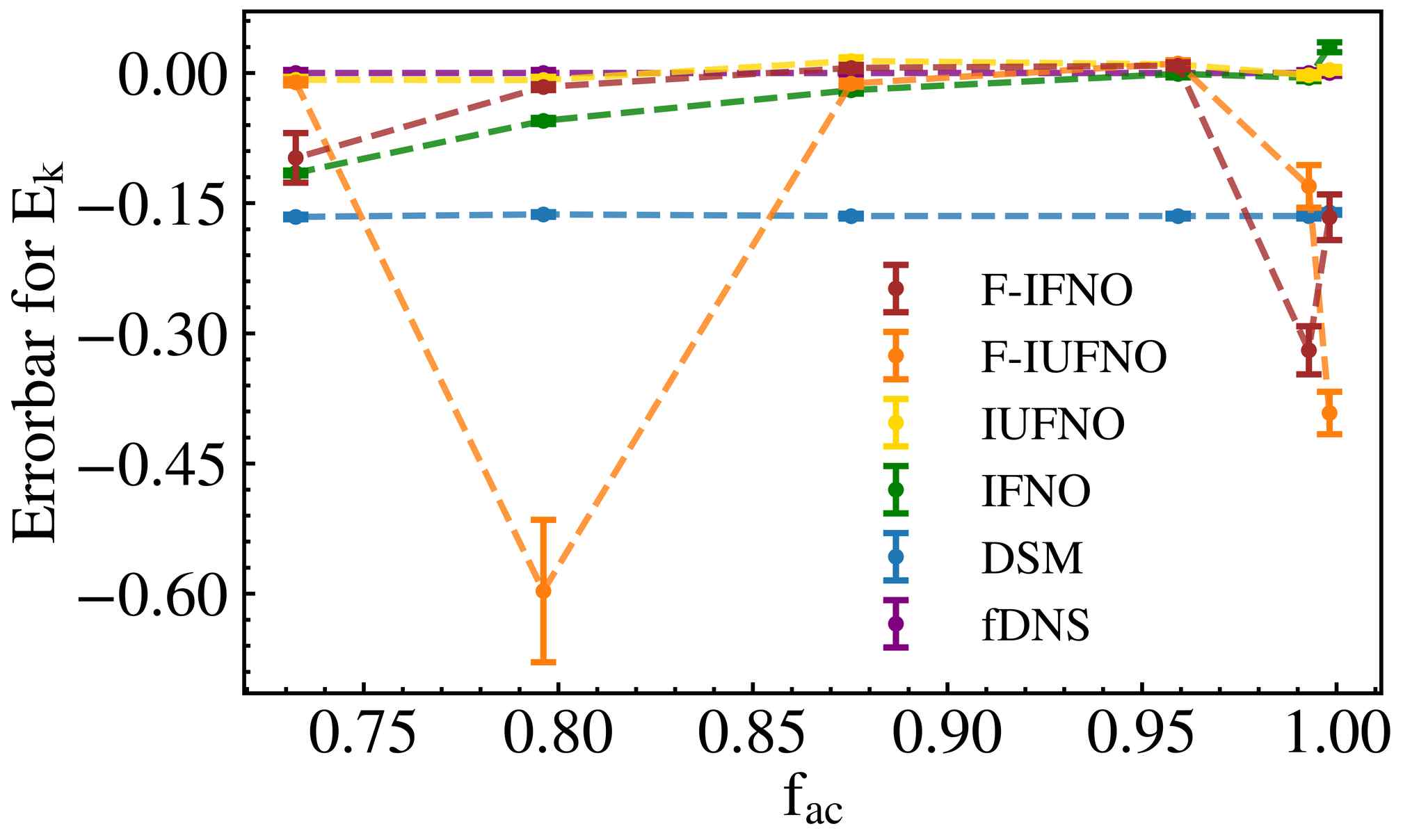}
            \put(-1,55){\small (a)}  
        \end{overpic}
    \end{subfigure}
    \hfill
    \begin{subfigure}[b]{0.49\textwidth}
        \begin{overpic}[width=1\linewidth]{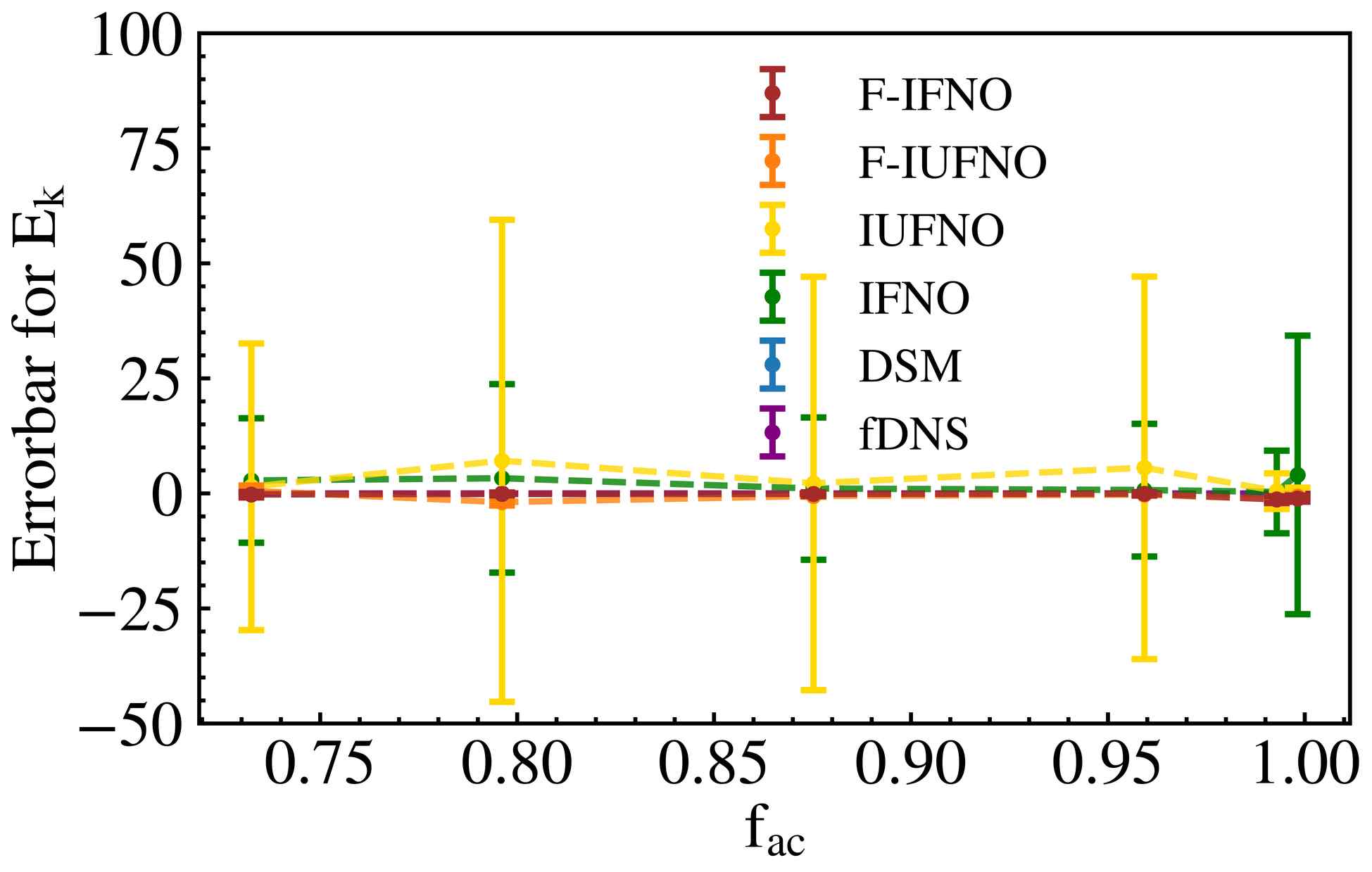}
            \put(-1,55){\small (b)} 
        \end{overpic} 
    \end{subfigure}
    \vspace{0.1cm}

    \begin{subfigure}[b]{0.49\textwidth}
        \begin{overpic}[width=1\linewidth]{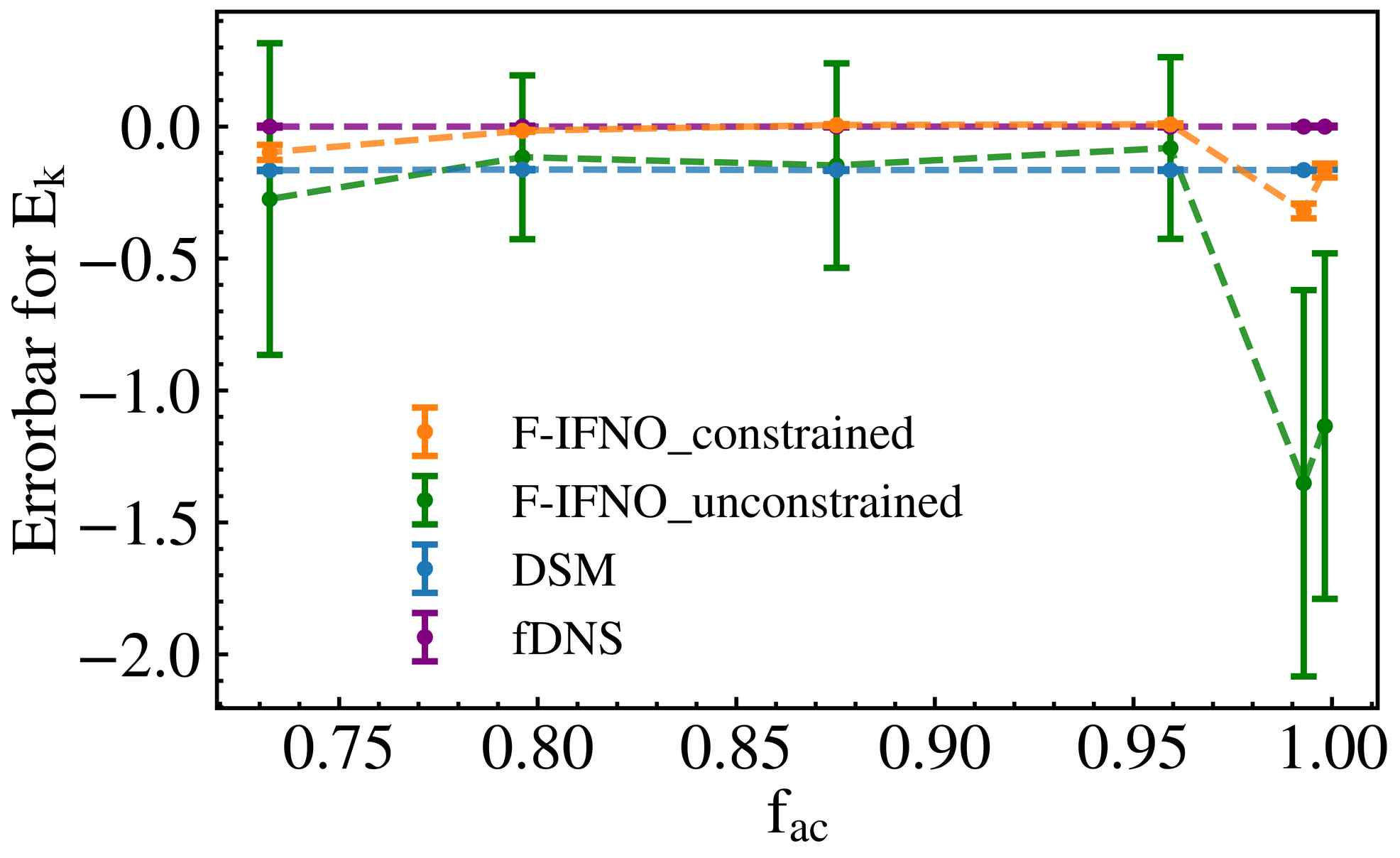}
            \put(-1,55){\small (c)}  
        \end{overpic}
    \end{subfigure}
    \hfill
    \begin{subfigure}[b]{0.49\textwidth}
        \begin{overpic}[width=1\linewidth]{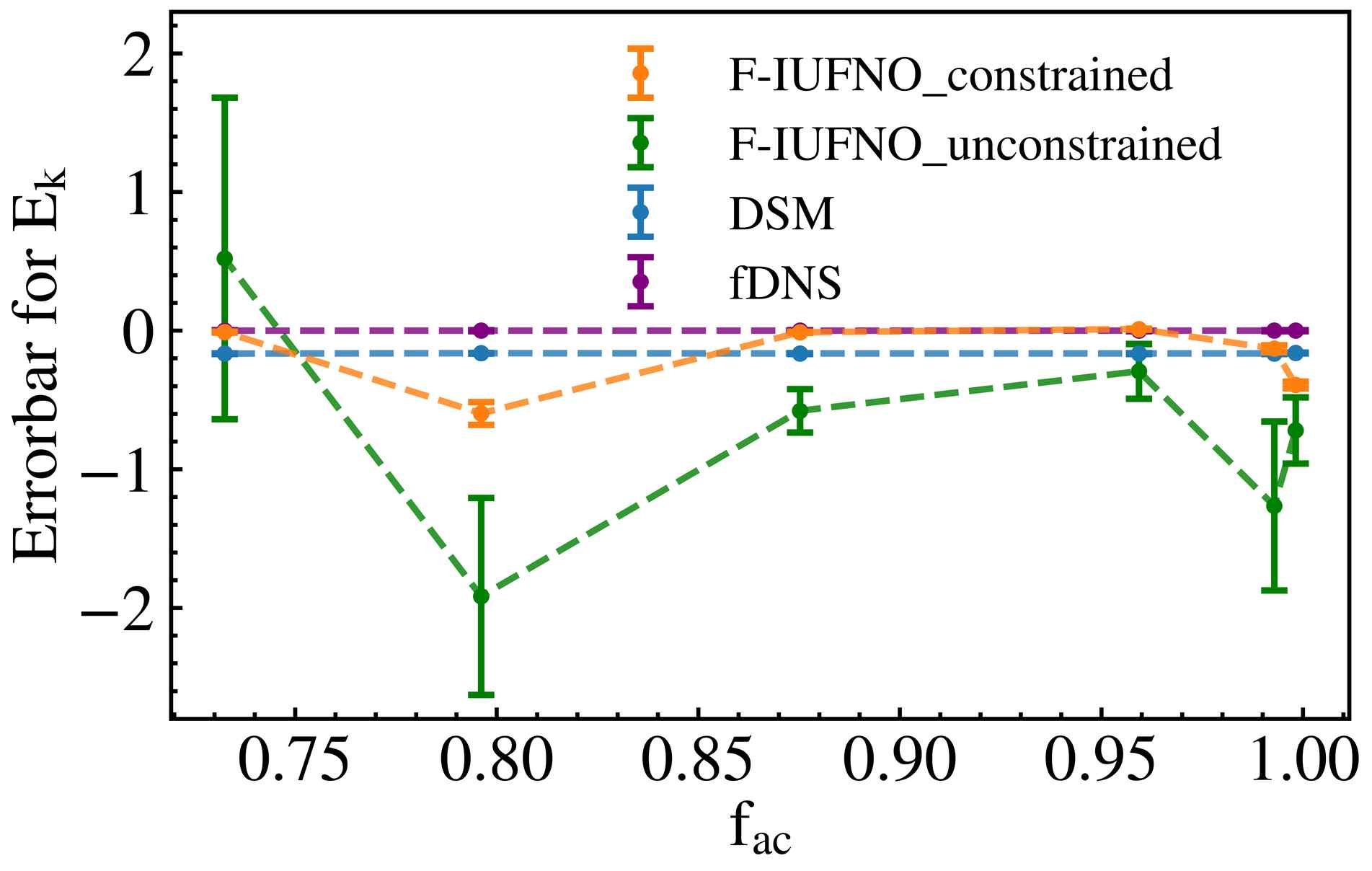}
            \put(-1,55){\small (d)} 
        \end{overpic} 
    \end{subfigure}
    \vspace{0.1cm}

    \begin{subfigure}[b]{0.49\textwidth}
        \begin{overpic}[width=1\linewidth]{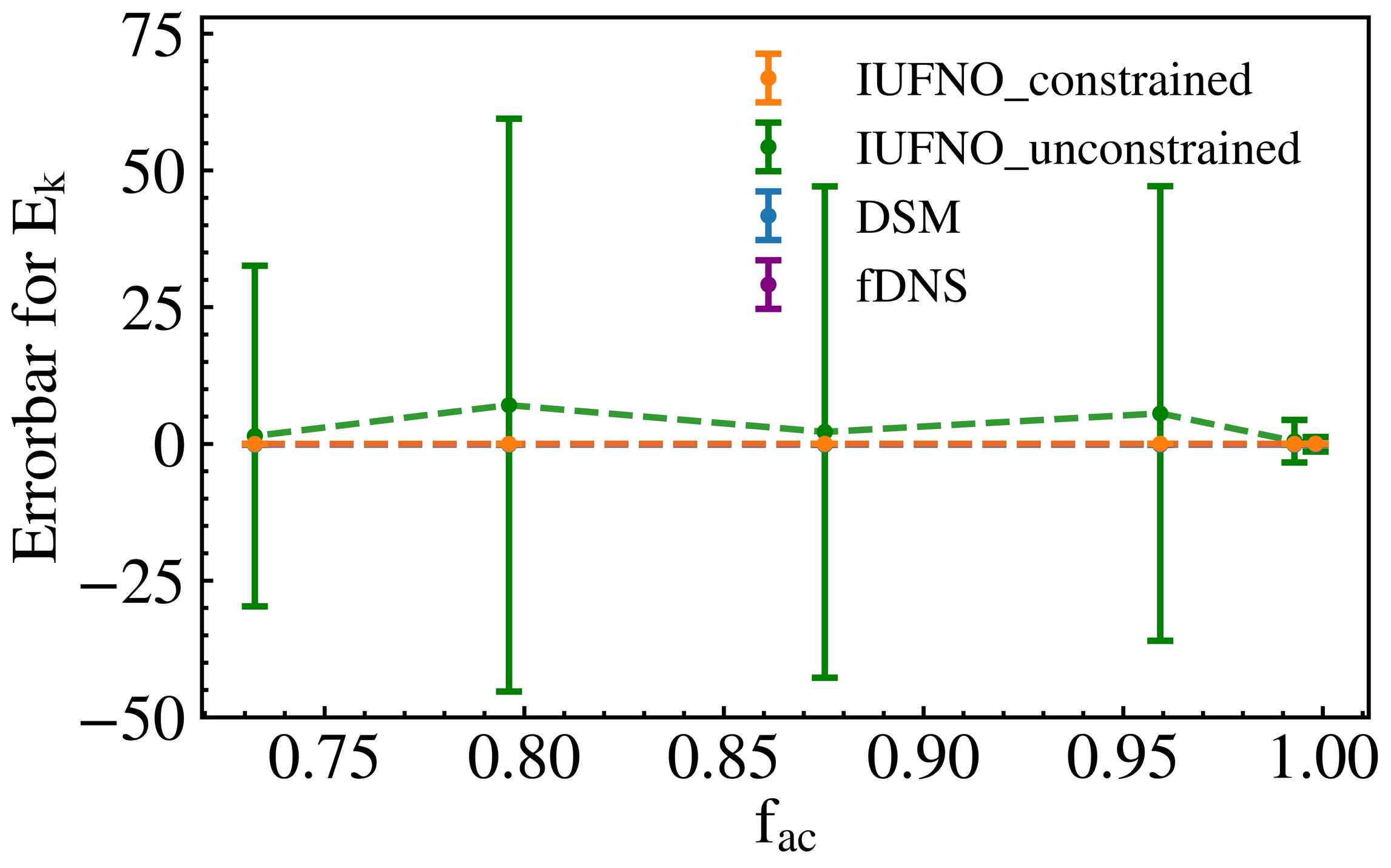}
            \put(-1,55){\small (e)}  
        \end{overpic}
    \end{subfigure}
    \hfill
    \begin{subfigure}[b]{0.49\textwidth}
        \begin{overpic}[width=1\linewidth]{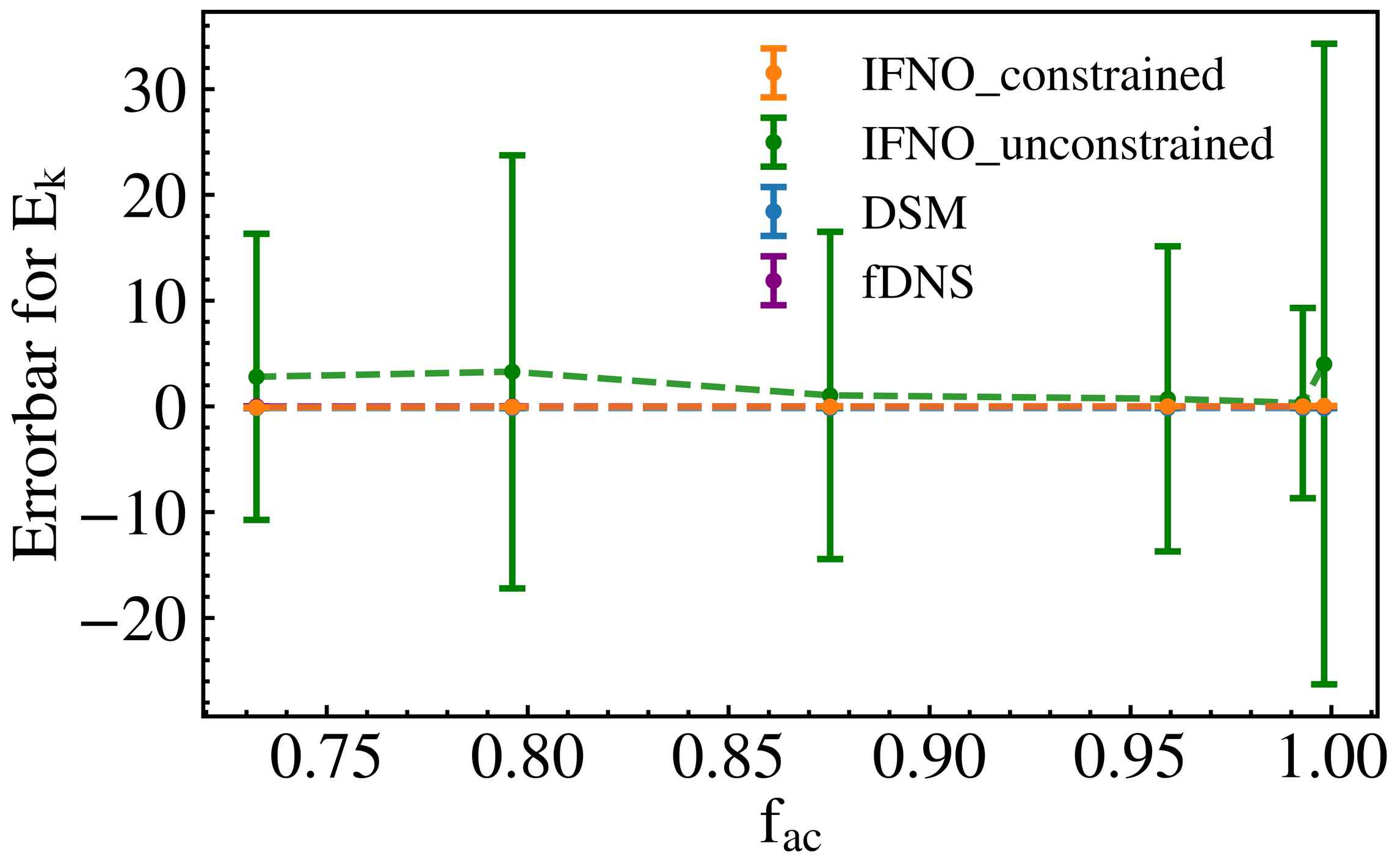}
            \put(-1,55){\small (f)} 
        \end{overpic} 
    \end{subfigure}
    
	\caption{Errorbars of kinetic energy $E_k$ for various methods as a function of $f_{ac}$: (a) FNO-based models constrained; (b) FNO-based unconstrained; (c) constrained and unconstrained F-IFNO; (d) constrained and unconstrained F-IUFNO; (e) constrained and unconstrained IUFNO; (f) constrained and unconstrained IFNO. Note that for fDNS, the values represent natural statistical fluctuations over time, not prediction errors.}\label{fig:27}
\end{figure}

To further investigate the performance of different models in relation to the autocorrelation function $f_{ac}$, Fig.~\ref{fig:27} presents errorbars of the kinetic energy $E_k$ for various methods as a function of $f_{ac}$.
The results suggest that most FNO-based models achieve better accuracy when the temporal autocorrelation falls within a specific range. We define two thresholds, \(C_L = 0.87\) and \(C_U = 0.96\), which represent the lower and upper bounds of effective temporal relevance. Within this interval, both F-IFNO and F-IUFNO demonstrate optimal performance, achieving a relative kinetic energy error below 1.2\% for constrained versions and 33.1\% for unconstrained versions.
For constrained IUFNO, the effective autocorrelation range is broader, with \(f_{ac} \in [C_L = 0.73,\ C_U = 0.998]\), within which the model maintains a relative kinetic energy error below 1.1\%. Constrained IFNO achieves its best performance near \(f_{ac} \approx 0.9929\), corresponding to a minimum relative error of approximately 1.2\%, indicating its reliance on highly coherent temporal information. In contrast, unconstrained IFNO and IUFNO do not exhibit a well-defined optimal range of \(f_{ac}\), and their relative errors remain above 165.8\%, suggesting poor utilization of temporal coherence.
These observations highlight that when \(f_{ac} \in [C_L, C_U]\), most models benefit from a balanced level of temporal coherence, neither overly redundant nor entirely uncorrelated, thereby enhancing their learning capacity and generalization performance.
This observation is consistent with earlier discussions on the choice of training and prediction time interval \(\Delta T\) for FNO-based models. If \(\Delta T\) is too small, the resulting snapshots contain excessive redundancy, which limits the model's ability to capture meaningful temporal dynamics. Conversely, a too-large \(\Delta T\) weakens the temporal correlation, leading to a loss of critical flow features. As a result, each FNO-based model exhibits an optimal \(\Delta T\) that aligns with an ideal level of temporal relevance, enabling reliable predictions with high accuracy and stability.

\begin{figure}[ht!]\centering
	\includegraphics[width=0.7\textwidth]{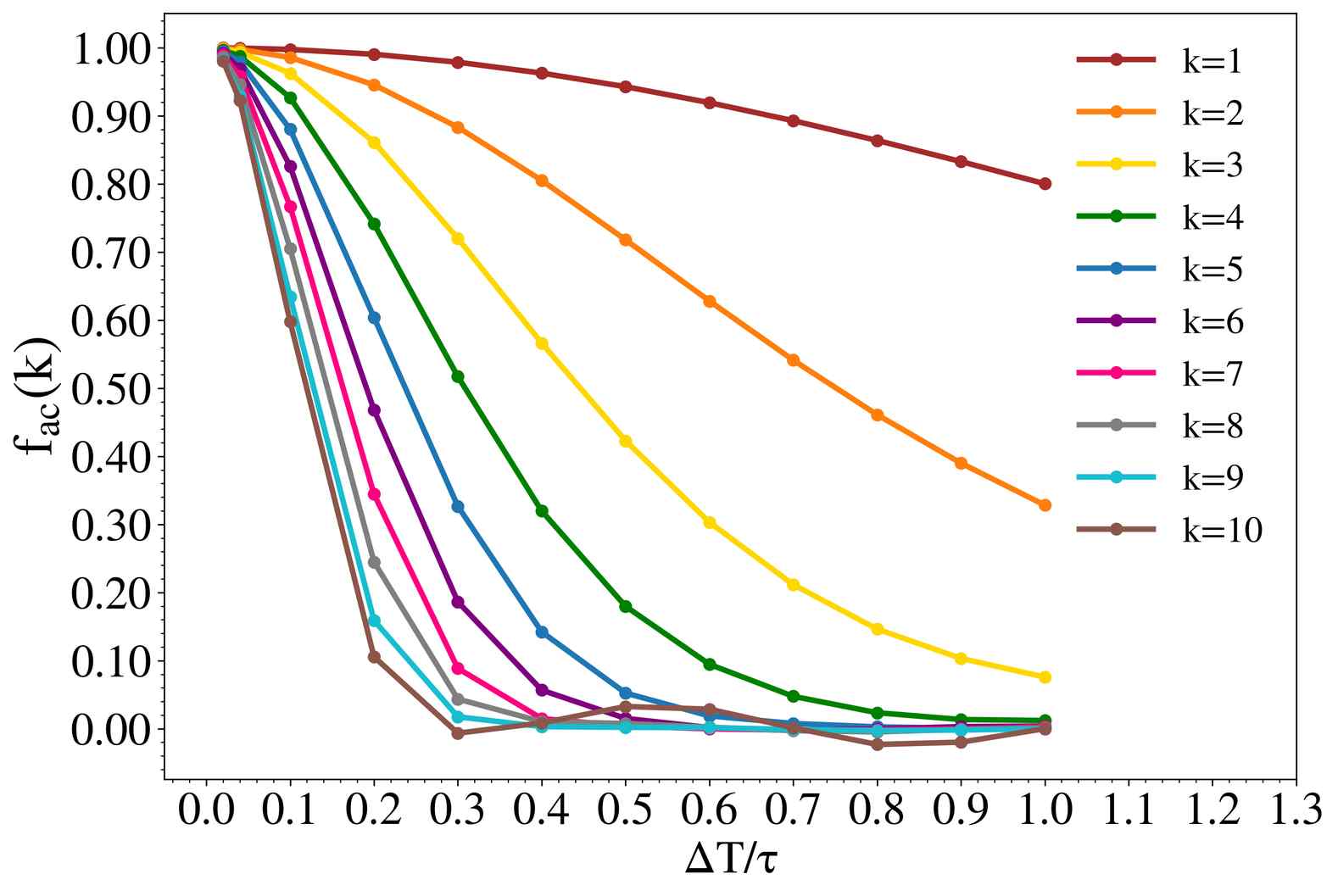}
	\caption{The ACF of the velocity field corresponding to the Fourier mode $k$ for forced HIT, over 18,000 fDNS data samples for each $k$.}\label{fig:28}
\end{figure}

Furthermore, we present the ACF of the velocity field corresponding to different Fourier modes $k$ for forced HIT in Fig.~\ref{fig:28}. Similar to Fig.~\ref{fig:26}, $f_{ac}(\Delta T, k)$ exhibits a decreasing trend as the time lag $\Delta T$ increases for each $k$. Moreover, it is observed that large-scale components (i.e., low-$k$ modes) have higher autocorrelation values compared to small-scale components (high-$k$ modes), indicating that $f_{ac}(\Delta T, k)$ decreases with increasing $k$. This observation is consistent with the characteristics of turbulent flows, where large scales exhibit more temporal coherence. Specifically, Table~\ref{tab:4} provides the detailed ACF values associated with the selected time lags and Fourier modes $k$ illustrated in Fig.~\ref{fig:28}.

\begin{table}[ht!]
\centering
\small
\caption{Autocorrelation values $f_{ac}(\Delta T, k)$ for different time lags $\Delta T$ and Fourier modes $k$}
\begin{tabular}{cccccc}
\hline\hline
$\Delta T$ & $f_{ac}(\Delta T, k=1)$ & $f_{ac}(\Delta T, k=2)$ & $f_{ac}(\Delta T, k=3)$ & $f_{ac}(\Delta T, k=4)$ & $f_{ac}(\Delta T, k=5)$ \\
\hline
0.02$\tau$   & 0.9999 &  0.9994   &  0.9984  & 0.9969 &  0.9948 \\
0.04$\tau$   & 0.9997 &  0.9977   &  0.9938  & 0.9878 &  0.9797  \\
0.1$\tau$  & 0.9977 &  0.9859   &  0.9627  & 0.9269 &   0.8804 \\
0.2$\tau$  & 0.9906 &  0.9457   &  0.8614  & 0.7416 &  0.6040 \\
0.3$\tau$  & 0.9789 &  0.8835   &  0.7201  & 0.5172 &  0.3267  \\
0.4$\tau$  & 0.9629 &  0.8053   &  0.5663  & 0.3198 &  0.1422 \\
0.5$\tau$  & 0.9430 &  0.7181    &  0.4230  & 0.1798  &  0.0523 \\
0.6$\tau$  & 0.9195 &  0.6282   &  0.3033  & 0.0946  &  0.0188  \\
0.7$\tau$  & 0.8930 &  0.5413   &  0.2116  & 0.0475 &  0.0076 \\
0.8$\tau$  & 0.8640 &  0.4612   &  0.1464  & 0.0234 &  0.0027 \\
0.9$\tau$  & 0.8330 &  0.3901   &  0.1032  & 0.0134 &  0.0007  \\
$\tau$ & 0.8004 &  0.3284   &  0.0758  & 0.0120 &  -0.0001 \\
\hline\hline    
$\Delta T$ & $f_{ac}(\Delta T, k=6)$ & $f_{ac}(\Delta T, k=7)$ & $f_{ac}(\Delta T, k=8)$ & $f_{ac}(\Delta T, k=9)$ & $f_{ac}(\Delta T, k=10)$ \\
\hline
0.02$\tau$   & 0.9922 & 0.9895    &  0.9866  & 0.9818 & 0.9806   \\
0.04$\tau$   & 0.9696 &  0.9585  &  0.9463  & 0.9296  &  0.9228   \\
0.1$\tau$  & 0.8260 &  0.7670   &  0.7051  & 0.6344 &   0.5978 \\
0.2$\tau$  & 0.4683 &  0.3445   &  0.2450  &  0.1591 &  0.1056   \\
0.3$\tau$  & 0.1863 &  0.0886   &  0.0435  & 0.0175 &  -0.0066   \\
0.4$\tau$  & 0.0567 &  0.0149   &   0.0104 & 0.0031 &  0.0085   \\
0.5$\tau$  & 0.0151 &  0.0040  &  0.0076  & 0.0020 &   0.0326 \\
0.6$\tau$  & 0.0017 &  -0.0002   &  0.0023 &  0.0016 &  0.0287  \\
0.7$\tau$  & -0.0024 &  -0.0021   &  -0.0029  & -0.0012 &   0.0013 \\
0.8$\tau$  & -0.0004 &  -0.0016 &  -0.0047  & -0.0028 &  -0.0231  \\
0.9$\tau$  & 0.0031 &  -0.0003   &  -0.0010  & -0.0019 &  -0.0198  \\
$\tau$ & 0.0038 &  0.0027   &  0.0012  & -0.0001 &   0.0000 \\
\hline\hline
\end{tabular}
\label{tab:4}
\end{table}

\begin{figure}[ht!]
    \centering
    \begin{subfigure}[b]{0.32\textwidth}
        \begin{overpic}[width=1\linewidth]{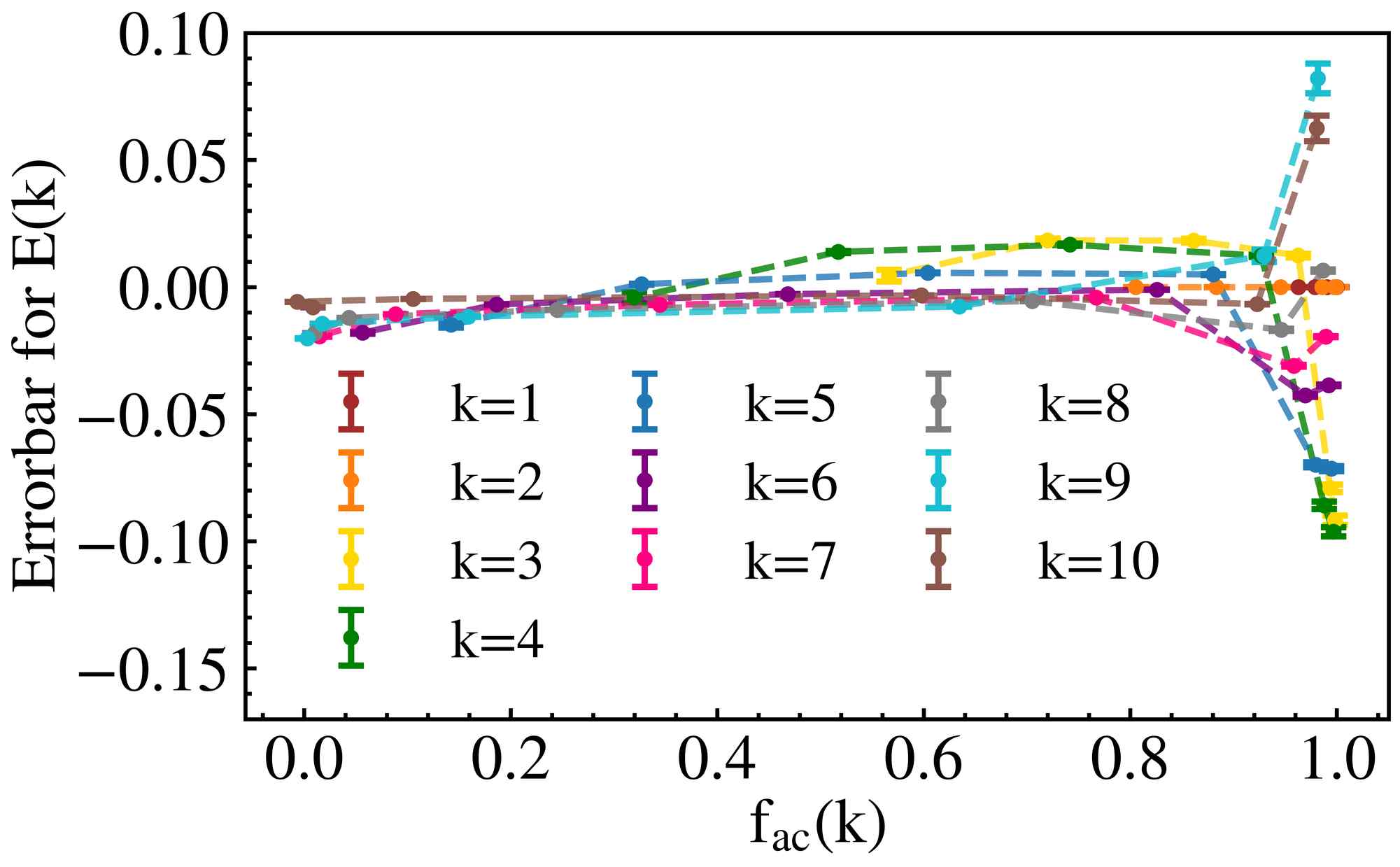}
            \put(-1,58){\small (a)}  
        \end{overpic}
    \end{subfigure}
    \hfill
    \begin{subfigure}[b]{0.32\textwidth}
        \begin{overpic}[width=1\linewidth]{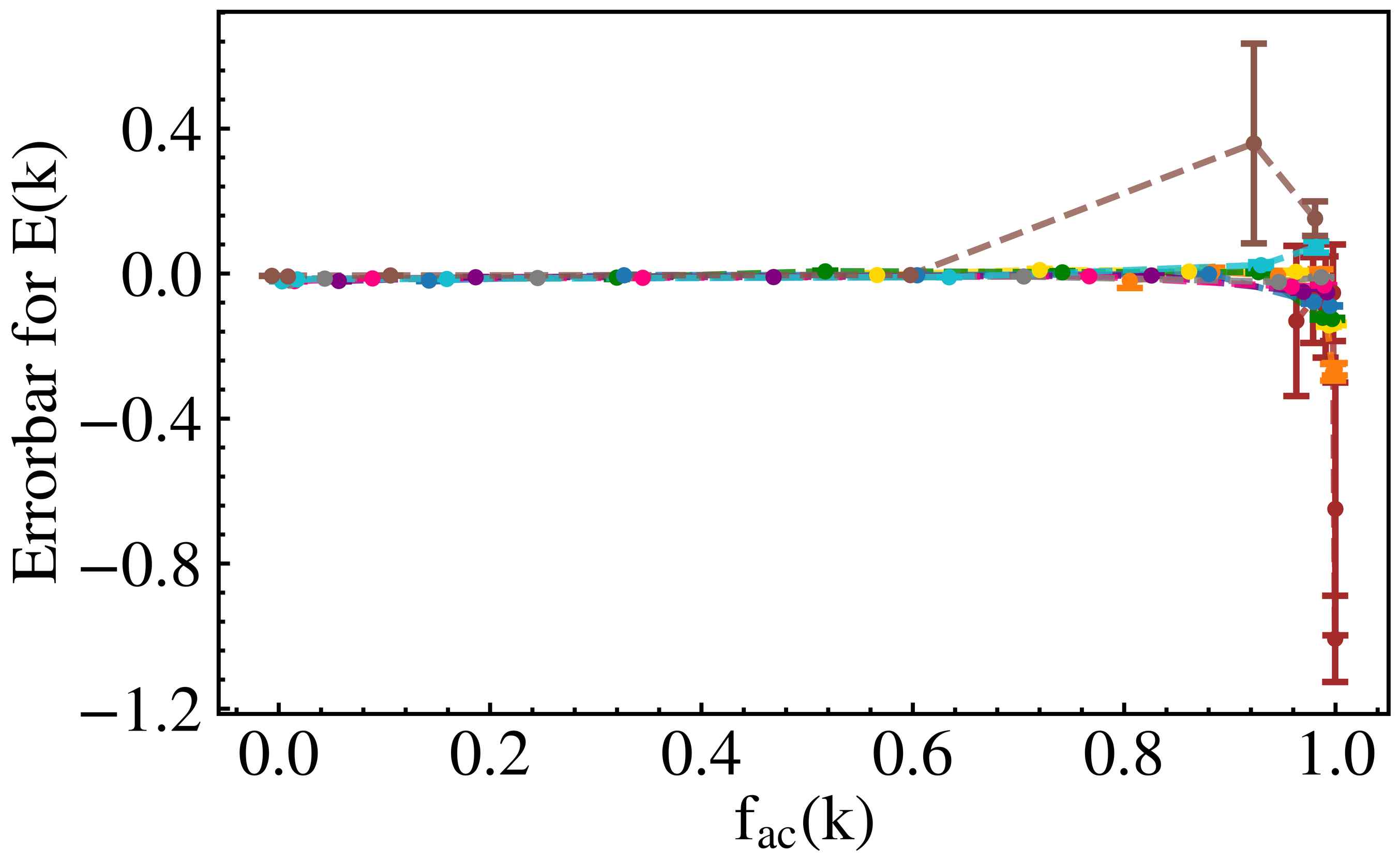}
            \put(-1,58){\small (b)} 
        \end{overpic} 
    \end{subfigure}
    \hfill
    \begin{subfigure}[b]{0.32\textwidth}
        \begin{overpic}[width=1\linewidth]{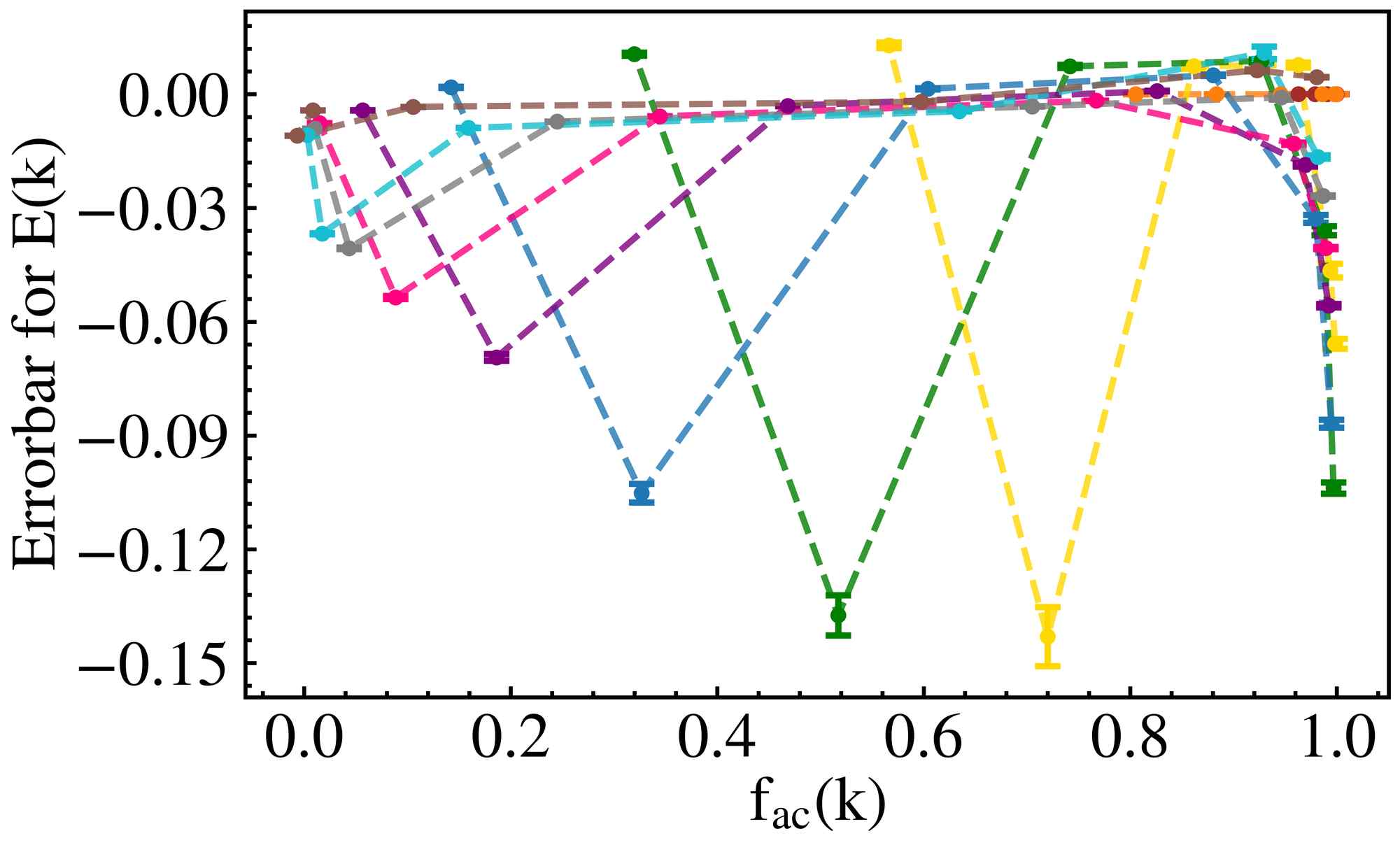}
            \put(-1,58){\small (c)} 
        \end{overpic}
    \end{subfigure}
    \vspace{0.1cm}

    \begin{subfigure}[b]{0.32\textwidth}
        \begin{overpic}[width=1\linewidth]{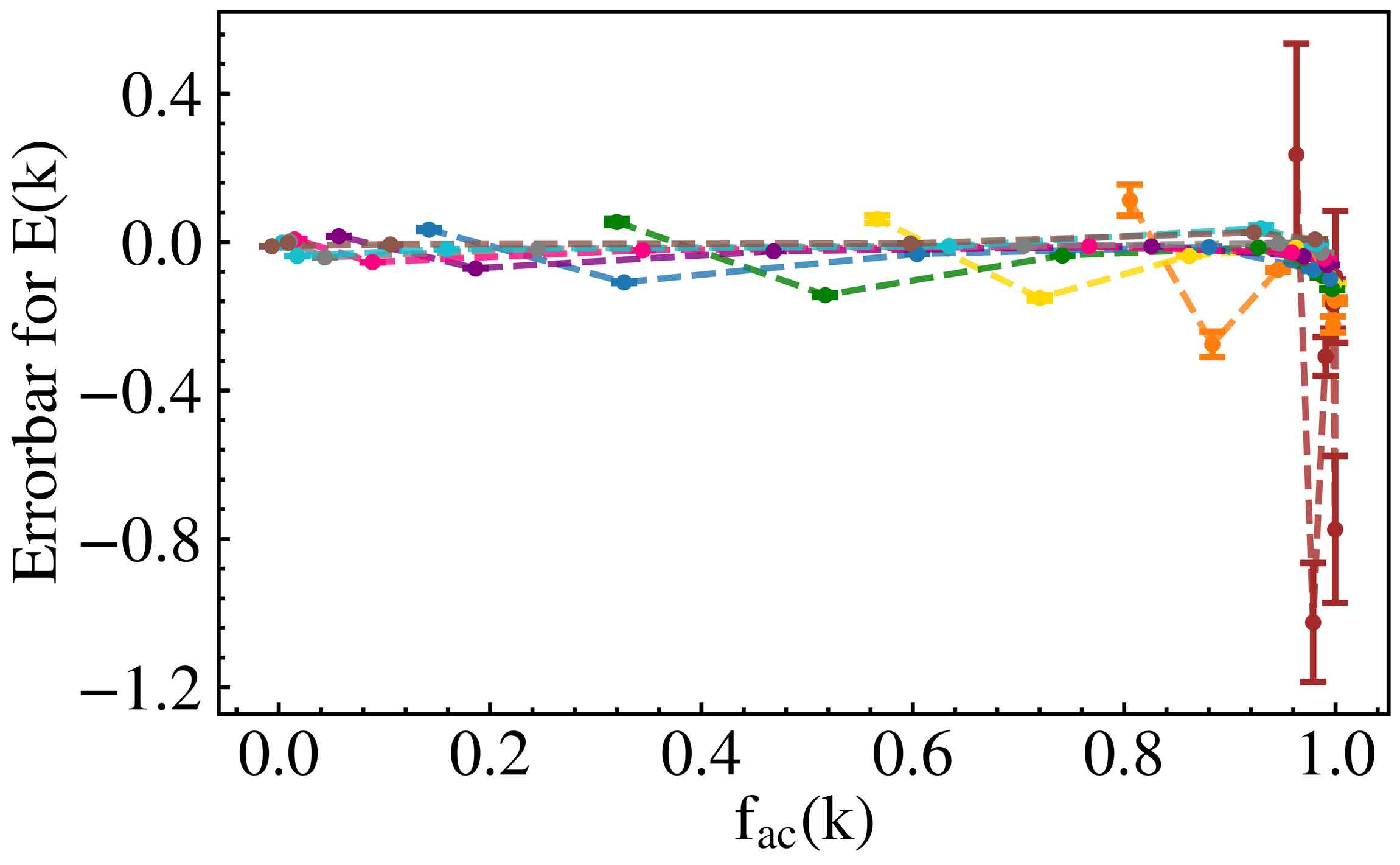}
            \put(-1,58){\small (d)}  
        \end{overpic}
    \end{subfigure}
    \hfill
    \begin{subfigure}[b]{0.32\textwidth}
        \begin{overpic}[width=1\linewidth]{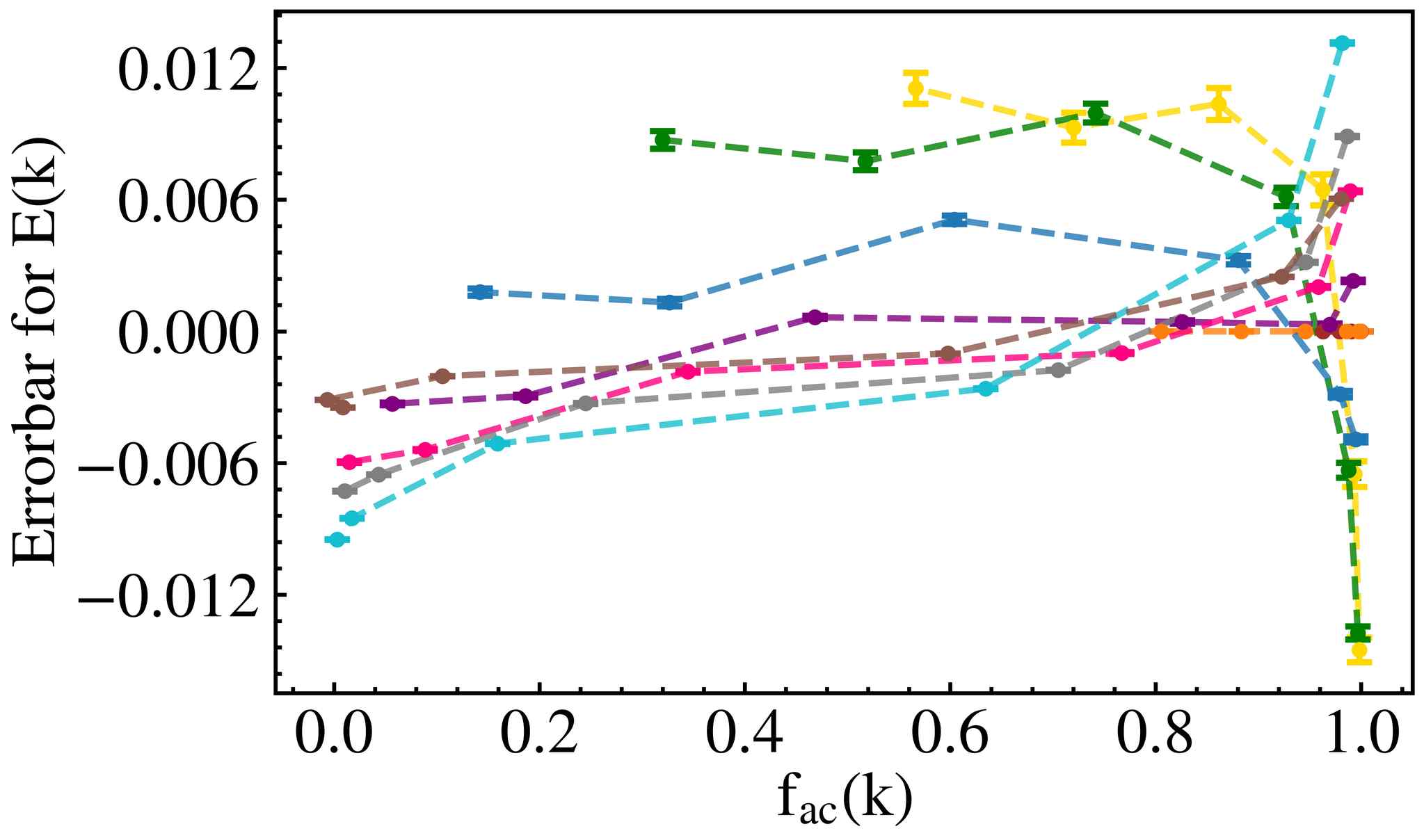}
            \put(-1,58){\small (e)} 
        \end{overpic} 
    \end{subfigure}
    \hfill
    \begin{subfigure}[b]{0.32\textwidth}
        \begin{overpic}[width=1\linewidth]{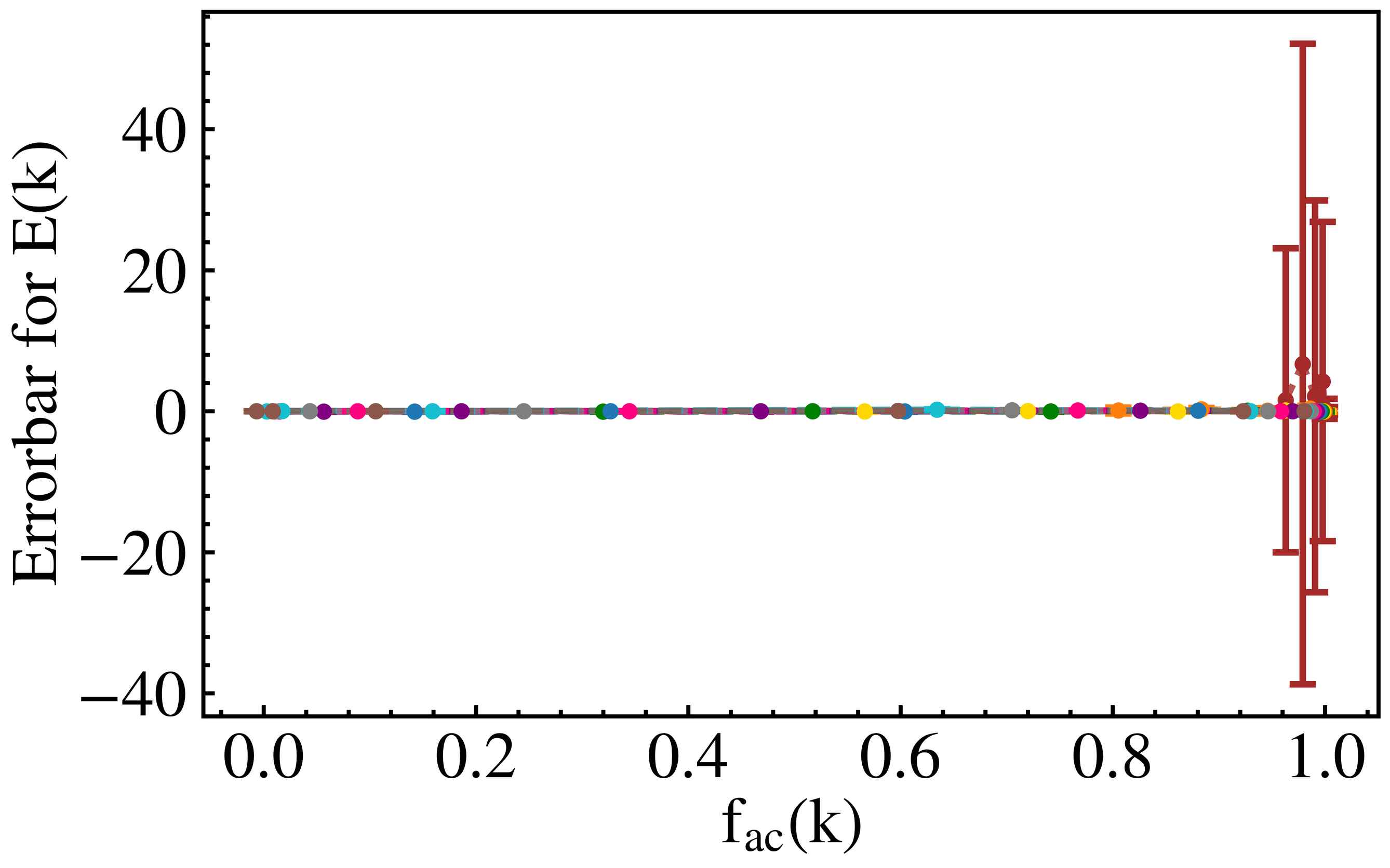}
            \put(-1,58){\small (f)} 
        \end{overpic}
    \end{subfigure}
    \vspace{0.1cm}

    \begin{subfigure}[b]{0.32\textwidth}
        \begin{overpic}[width=1\linewidth]{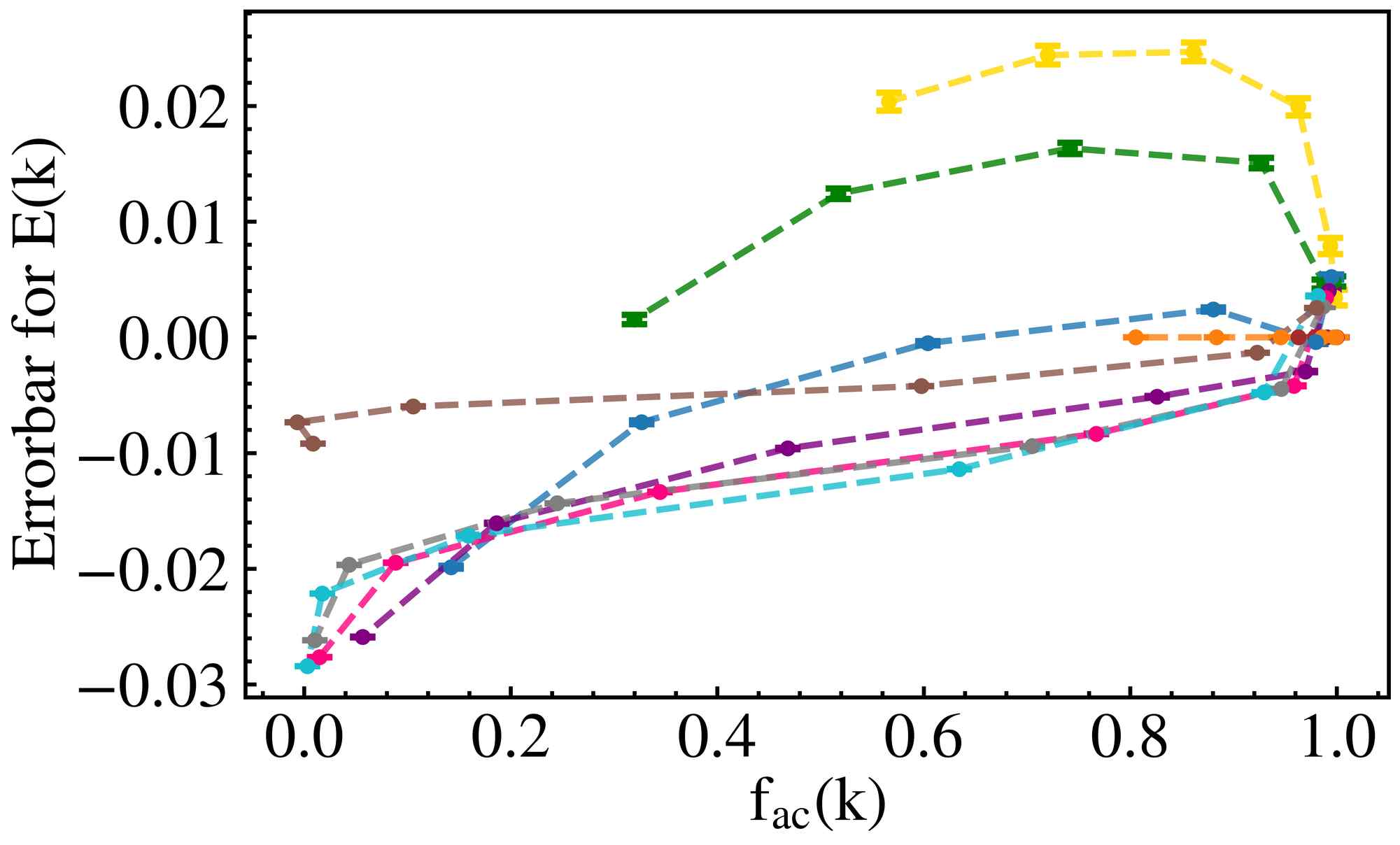}
            \put(-1,58){\small (g)}  
        \end{overpic}
    \end{subfigure}
    \hfill
    \begin{subfigure}[b]{0.32\textwidth}
        \begin{overpic}[width=1\linewidth]{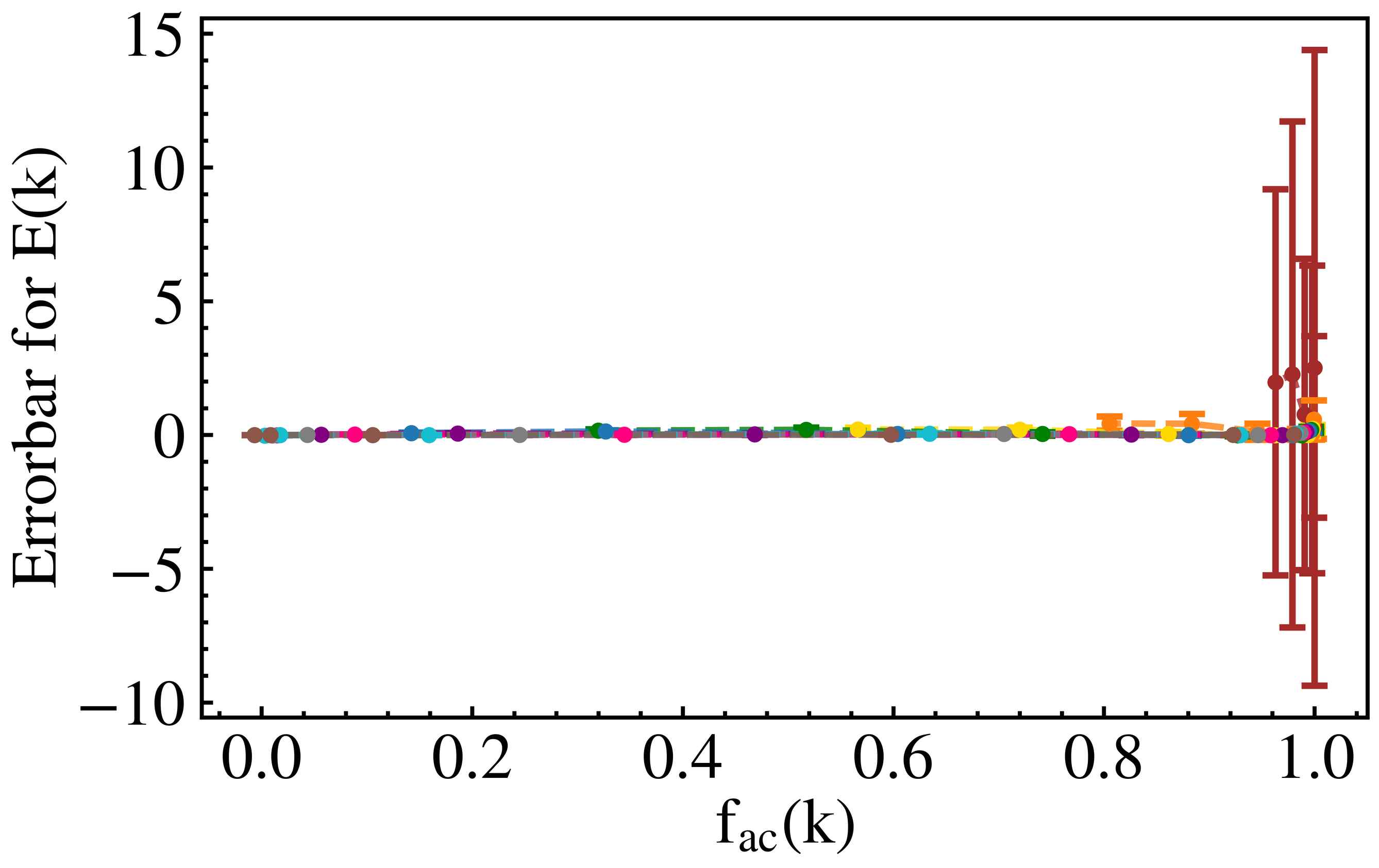}
            \put(-1,58){\small (h)} 
        \end{overpic} 
    \end{subfigure}
    \hfill
    \begin{subfigure}[b]{0.32\textwidth}
        \begin{overpic}[width=1\linewidth]{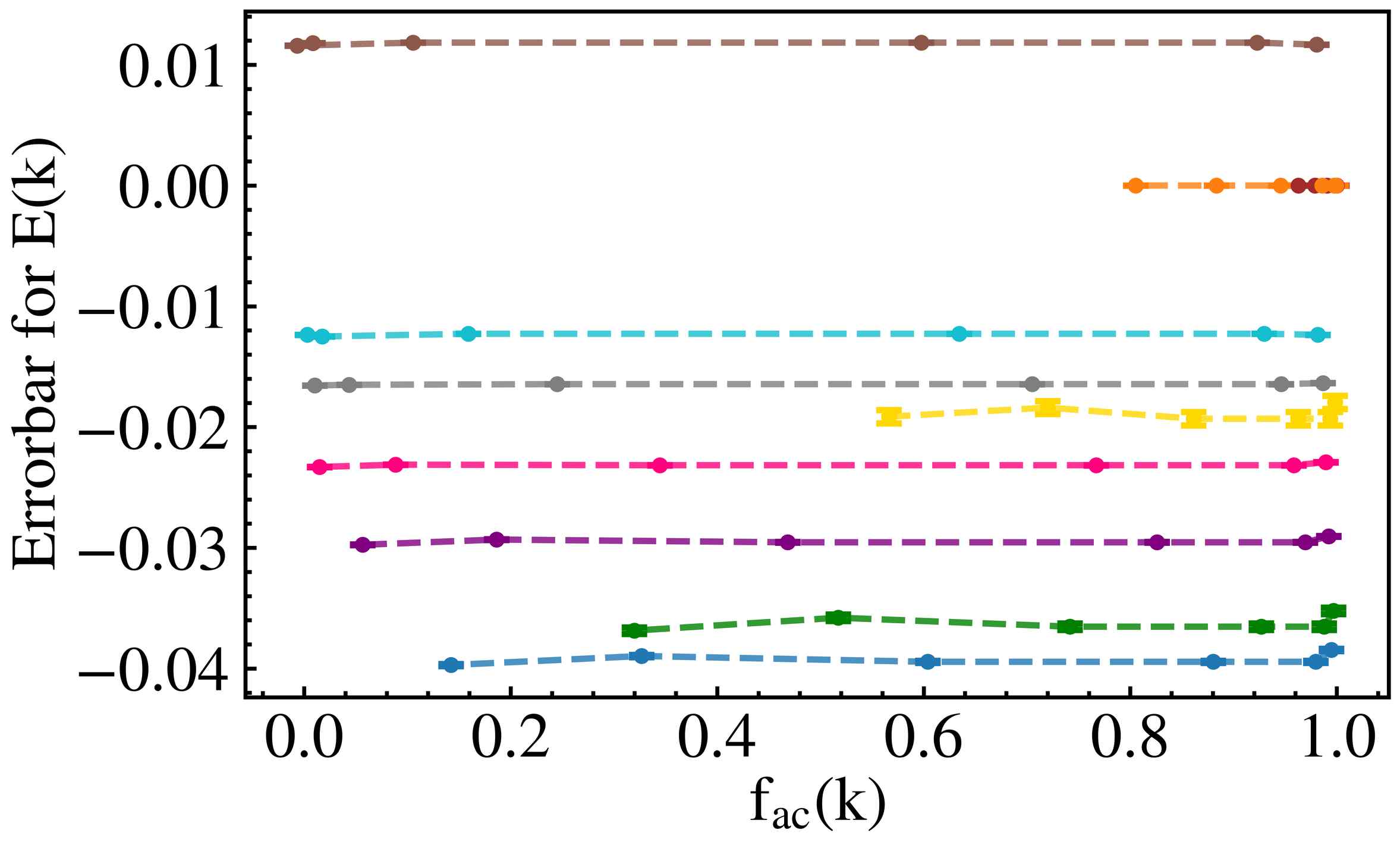}
            \put(-1,58){\small (i)} 
        \end{overpic}
    \end{subfigure}
    \vspace{0.1cm}

    \begin{subfigure}[b]{1\textwidth}
        \centering
        \begin{overpic}[width=0.32\linewidth]{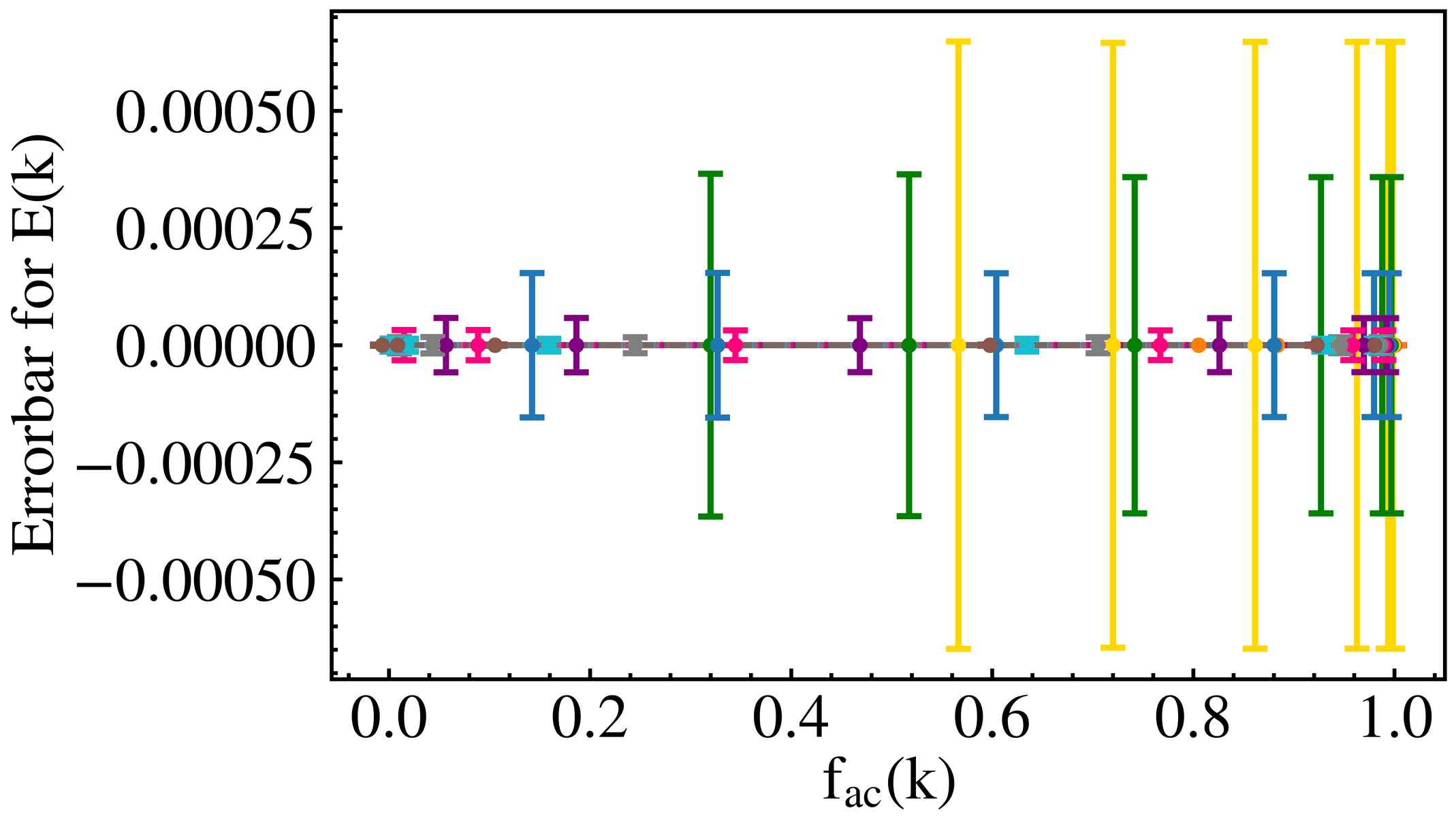}
            \put(-1,58){\small (j)}  
        \end{overpic}
    \end{subfigure}

	\caption{Errorbars of velocity spectra $E(k)$ for various methods as a function of $f_{ac}(k)$: (a) F-IFNO constrained; (b) F-IFNO unconstrained; (c) F-IUFNO constrained; (d) F-IUFNO unconstrained; (e) IUFNO constrained; (f) IUFNO unconstrained; (g) IFNO constrained; (h) IFNO unconstrained; (i) DSM; (j) fDNS. Note that for fDNS, the values represent natural statistical fluctuations over time, not prediction errors.}\label{fig:29}
\end{figure}

In Fig.~\ref{fig:29}, we present the errorbars of the velocity spectra $E(k)$ as a function of the autocorrelation $f_{ac}(k)$ for various models. For unconstrained FNO-based models, the primary errors are observed at large scales, corresponding to small Fourier modes $k$, consistent with previous observations. For constrained F-IFNO, the prediction errors increase across all modes when $f_{ac}(k)$ approaches 1. In contrast, constrained F-IUFNO exhibits elevated errors across all modes when $f_{ac}(k)$ is either too large or too small. A similar trend is observed in constrained IUFNO, where the errors rise significantly when $f_{ac}(k)$ is close to either 0 or 1, indicating that both excessively small and large $\Delta T$ values are detrimental. Constrained IFNO achieves the lowest errors when $f_{ac}(k)$ is near 1 across all modes. In comparison, the errors for DSM and the fluctuations for fDNS remain stable across varying $f_{ac}(k)$ values and depend solely on the mode $k$. These results further confirm that the choice of $\Delta T$, which controls $f_{ac}(k)$, is crucial for FNO-based models: overly small $\Delta T$ (large $f_{ac}(k)$) or both overly large $\Delta T$ (small $f_{ac}(k)$) lead to significant performance degradation.

Conclusionly, the ACF analysis offers a quantitative perspective on the temporal relevance between flow snapshots, which directly impacts the performance and stability of FNO-based models. The observed monotonic decay of $f_{ac}(\Delta T)$ and $f_{ac}(\Delta T, k)$ validates the need for selecting an appropriate time interval $\Delta T$ that balances information redundancy and decorrelation. This complements our earlier findings on the robustness of F-IFNO and F-IUFNO under varying perturbation magnitudes, highlighting that model reliability is closely tied to the degree of temporal correlation in the training data and prediction procedures. Overall, the ACF analysis serves as a fundamental tool to guide the design of temporally consistent data sampling strategies for reliable learning and prediction in data-driven turbulence modeling.

\subsection{Computational efficiency}
\label{subsec4.5}

In this study, neural operators are trained and tested on an NVIDIA V100 GPU, with the CPU being an Intel\textsuperscript{\textregistered} Xeon\textsuperscript{\textregistered} E5-4627 v3 @ 2.60\,GHz. The DSM simulations are conducted on a computing cluster equipped with Intel Xeon Gold 6148 CPUs, each with 20 cores running at 2.40\,GHz. All neural operator simulations are implemented using \texttt{PyTorch}.
Table~\ref{tab:5} presents a comparison of the computational cost for a single prediction step ($\Delta T = 0.2\tau$), including the number of model parameters, GPU memory usage, and both GPU and CPU computation times. For FNO-based models, predictions are performed on the aforementioned GPU. Compared to IFNO, IUFNO requires more parameters and GPU memory due to the additional U-Net component. F-IFNO, which utilizes factorized Fourier transforms, requires the fewest parameters and the least GPU memory, while F-IUFNO also reduces parameter count and memory usage relative to IUFNO.
Among all FNO-based models, IFNO achieves the lowest GPU time, while IUFNO and F-IFNO have comparable GPU times. F-IUFNO incurs the highest GPU time. To compare the efficiency of FNO-based models with the DSM model, we also measured the CPU computation time of the FNO-based models using the same CPU type as the DSM simulations. In this comparison, FNO-based models were run on 2 CPU cores, while the DSM model utilized 32 cores. For fairness, the CPU$\cdot$s metric in Table~\ref{tab:5} represents the total CPU time multiplied by the number of cores used.
The results demonstrate that FNO-based models are significantly more efficient than the DSM model. This efficiency is mainly due to the fact that DSM requires iterative solutions with very small time steps, whereas neural operators can directly predict the flow field over a larger time interval. Moreover, for neural operator models, changes in $\Delta T$ do not affect the number of parameters, GPU memory usage, or computation time, since their predictions are data-driven. In contrast, for the DSM model, the CPU time increases with larger $\Delta T$ due to its reliance on a fixed, small time step. Therefore, we report results using a baseline time interval of $\Delta T = 0.2\tau$ for fair comparison.

Considering both prediction reliability and computational efficiency, the proposed F-IFNO exhibits an outstanding performance. It maintains high accuracy and stability across a reasonable range of time intervals, while significantly enhancing efficiency. Specifically, the F-IFNO model reduces the number of parameters and GPU memory usage by approximately $98.84\%$ and $74.69\%$, respectively, compared to the IFNO model; by approximately $98.92\%$ and $75.75\%$, respectively, compared to the IUFNO model; and by approximately $86.39\%$ and $20.19\%$, respectively, compared to the F-IUFNO model.

\begin{table}[ht!]
	\begin{center}
		\caption{Comparison of computational efficiency among different methods in forced HIT with time interval $\Delta T = 0.2\tau$.}\label{tab:5}
		\begin{tabular*}{1\textwidth}{@{\extracolsep{\fill}} lccccc }
			\hline\hline
			\small    
			Model & Numbers of parameters & GPU memory usage (MB) & GPU$\cdot$s & CPU$\cdot$s \\ \hline
			DSM & N/A    & N/A   & N/A  & 39.72\\            
			IFNO & 55,996,023    & 610.47   & 0.309  & 7.43 \\
			IUFNO & 60,119,283    & 637.06   & 0.561 & 13.48   \\
			F-IFNO & 649,533    & 154.48   & 0.562 & 13.52  \\           
			F-IUFNO & 4,772,793   & 193.51  & 0.762 & 18.30 \\ \hline\hline
		\end{tabular*}%
	\end{center}
\end{table}

\section{Discussion}
\label{sec5}

\subsection{Importance of UQ, long-term stability, and ACF in PDEs}
\label{subsec5.1}

In engineering applications, solving nonlinear partial differential equations (PDEs) remains a significant challenge, especially in three dimensions. While many effective and practical data-driven approaches have been developed for one-dimensional and two-dimensional PDEs, reliable solutions for three-dimensional PDEs are relatively few. Recently, the Fourier neural operator (FNO) has demonstrated its potential as an effective surrogate model for solving PDEs, including three-dimensional cases, due to its remarkable ability to capture the nonlinear and chaotic nature of high-dimensional systems \cite{li2021fourierneuraloperatorparametric,LI2022100389FNO}. Building on the FNO framework, several advanced models have been proposed for various PDE problems \cite{YOU2022115296,WEN2022104180,liLongtermPredictionsTurbulence2023a,tranFactorizedFourierNeural2023d}. However, most existing studies focus primarily on short-term prediction accuracy, while neglecting the uncertainty in model predictions and the stability of long-term forecasts. For complex systems such as high-dimensional PDEs, long-term statistical consistency is often more important than short-term trajectory accuracy \cite{liLearningDissipativeDynamics2022a}.

When applying neural operator-based models to high-dimensional nonlinear PDEs, uncertainty quantification (UQ) plays a critical role in evaluating model performance and trustworthiness. Through UQ, we can assess the distribution of prediction errors across different time intervals $\Delta T$ and Fourier modes $k$. In classical numerical methods including finite difference, finite volume, and spectral methods, numerical stability is a well-defined concept in time-marching schemes. Similarly, for neural operator-based models, which also operate in a time-marching manner, the accumulation of prediction errors leads to stability concerns. Stability, in this context, also encompasses the model's resilience to initial perturbations. The autocorrelation function (ACF) serves as a valuable diagnostic tool to evaluate the temporal relevance of input data and the stability of model predictions. Despite the increasing adoption of FNO-based models in scientific machine learning, there has been little systematic investigation into the interrelations between UQ, stability, and ACF, particularly in the context of three-dimensional turbulence. This study aims to address this gap.

\subsection{Key findings: insights from UQ, stability, and ACF analysis}
\label{subsec5.2}

From the results presented in Section~\ref{sec4}, several key findings can be drawn regarding the predictive behavior of different FNO-based models. 
First, constrained FNO-based models (e.g., F-IFNO and F-IUFNO) exhibit significantly improved long-term stability and reduced uncertainty when the autocorrelation $f_{ac}(\Delta T)$ falls within a moderate range. 
Second, while unconstrained FNO-based models including F-IFNO and F-IUFNO may still perform reasonably well when $f_{ac}(\Delta T)$ lies within an optimal range, the performance of IFNO and IUFNO degrades substantially as $f_{ac}(\Delta T) \to 1$ or $f_{ac}(\Delta T) \to 0$. This degradation is due to their sensitivity to either excessive similarity or excessive dissimilarity between the velocity fields at two consecutive time steps.
Third, the distribution of errors across different Fourier modes $k$ indicates that large-scale components exhibit significantly greater statistical uncertainty and instability compared to small-scale ones. Across all FNO-based models, their constrained versions consistently demonstrate improved stability and lower uncertainty, reinforcing the finding that large-scale dynamics are the dominant sources of error and instability in the statistical sense.
Fourth, when subject to initial perturbations, F-IFNO and F-IUFNO display strong robustness, maintaining stable performance across various perturbation magnitudes $\tilde{\varepsilon}$. In contrast, IFNO, IUFNO, and the traditional DSM perform poorly when the perturbations become too large.

Additionally, in terms of computational efficiency, the proposed F-IFNO model is the most efficient. Compared to IFNO, it reduces the number of parameters and GPU memory usage by approximately $98.84\%$ and $74.69\%$, respectively. Compared to IUFNO, the reductions are approximately $98.92\%$ and $75.75\%$, respectively. When compared to F-IUFNO, the savings are $86.39\%$ in parameters and $20.19\%$ in memory usage. Moreover, in terms of runtime efficiency, F-IFNO requires only 0.562 GPU$\cdot$s per prediction step, compared to 39.72 CPU$\cdot$s for DSM. This significant difference illustrates the computational advantage of machine learning models when leveraging GPU acceleration.
In conclusion, F-IFNO outperforms all other models when considering uncertainty quantification, stability, and computational efficiency.

\subsection{Limitations and future work}
\label{subsec5.3}

From the results above, we provide a systematic analysis of the uncertainty and stability of FNO-based models in three-dimensional forced homogeneous isotropic turbulence (HIT). However, several limitations of the current framework need to be addressed in future work.
First, this study is limited to forced HIT and has not been validated on more complex flow configurations, including turbulent channel flows, flows over periodic hills and more complex wall-bounded turbulent flows. Extending the analysis to these scenarios is essential for assessing the broader applicability of the neural operator models. 
Second, although the autocorrelation function (ACF) serves as a useful metric, it captures only linear temporal correlations. Future work could explore nonlinear and multiscale temporal diagnostics, including mutual information or informative/non-informative decomposition \cite{arranzInformativeNoninformativeDecomposition2024a, lozano-duranInformationtheoreticFormulationDynamical2022a}, to provide a more comprehensive characterization of model dynamics.
Third, the generalizability of the proposed models across different Reynolds numbers and grid resolutions remains untested and should be systematically examined.
Lastly, integrating \textit{Bayesian operator learning} into the framework may offer more systematic UQ for neural operator methods of turbulence in the future \cite{ZOU2025117479,PSAROS2023111902}.

\section{Conclusions}
\label{sec6}

In this paper, we present a systematic analysis of uncertainty quantification (UQ) and prediction stability for FNO-based models, with the aim of rigorously evaluating their reliability and trustworthiness in three-dimensional forced homogeneous isotropic turbulence (HIT). We perform UQ and stability diagnostics on both the kinetic energy $E_k$ and the velocity spectra $E(k)$. The UQ analysis captures the error distribution across various time intervals $\Delta T$ and Fourier modes $k$ for different methods. Meanwhile, the stability analysis assesses the models' ability to resist error accumulation under both time-marching predictions and initial perturbations. To quantify temporal correlation, we introduce the autocorrelation function $f_{ac}(\Delta T)$ and its scale-dependent counterpart $f_{ac}(\Delta T, k)$, enabling an in-depth examination of model behavior under varying temporal coherence conditions. 

Our results show that the proposed F-IFNO and F-IUFNO models outperform both traditional DSM and other FNO-based models, particularly in terms of long-term stability, robustness to initial perturbations, and reduced predictive uncertainty when $f_{ac}(\Delta T)$ lies within a moderate range. Moreover, the F-IFNO model achieves a favorable balance between predictive accuracy and computational efficiency, significantly reducing parameter count and memory usage while maintaining robust performance. 
Our findings suggest that incorporating prediction constraints and optimizing time interval choices are critical to improve the robustness and reliability of FNO-based models. Furthermore, the error analysis across different Fourier modes indicates that large-scale modes are more prone to instability and uncertainty in the statistical sense, while constrained models are better at mitigating these issues.

Overall, this study provides valuable insights into the interplay between UQ, stability, and temporal correlation within the framework of operator learning. It highlights the importance of integrating prediction constraints and designing informed temporal strategies to enhance the robustness and generalization of learning-based surrogates, especially for multi-scale turbulent flows and other high-dimensional nonlinear systems.

\section*{CRediT authorship contribution statement}
\textbf{Xintong Zou}: Writing - review \& editing, Writing - original draft, Methodology, Conceptualization, Investigation, Coding, Data curation, Visualization, Validation. 
\textbf{Zhijie Li}: Methodology, Conceptualization, Investigation, Data curation.
\textbf{Yunpeng Wang}: Conceptualization, Investigation, Data curation.
\textbf{Huiyu Yang}: Methodology, Conceptualization, Investigation.
\textbf{Jianchun Wang}: Writing - review \& editing, Supervision, Project administration, Funding acquisition, Methodology, Conceptualization, Investigation.

\section*{Declaration of competing interest}
The authors declare that they have no known competing financial interests or personal relationships that could have appeared to influence the work reported in this paper.

\section*{Acknowledgements}
This work was supported by the National Natural Science Foundation of China (NSFC Grant Nos. 12172161, and 12302283), by NSFC Excellence Research Group Program for 'Multiscale Problems in Nonlinear Mechanics' (No. 12588201), by the Shenzhen Science and Technology Program (Grant No. KQTD20180411143441009), and by Department of Science and Technology of Guangdong Province (Grant Nos. 2019B21203001, 2020B1212030001, and 2023B1212060001).
Additional support was provided by the Innovation Capability Support Program of Shaanxi (Program No. 2023-CX-TD-30) and the Center for Computational Science and Engineering of Southern University of Science and Technology.

\section*{Code availability}
Code used to train models and to produce analysis computational results along with datasets in this work can be accessed at \href{https://github.com/cc0429/UQ4Turbu}{\textit{https://github.com/cc0429/UQ4Turbu}}.

\section*{Data availability}
Data will be made available on request.


\appendix
\section{Details of neural operator architectures in Subsection~\ref{subsec3.2}}
\label{appendix1}

\subsection{Details of implicit Fourier neural operator}
\label{subappendix1.1}

\begin{figure}[ht!]\centering
	\includegraphics[width=1\textwidth]{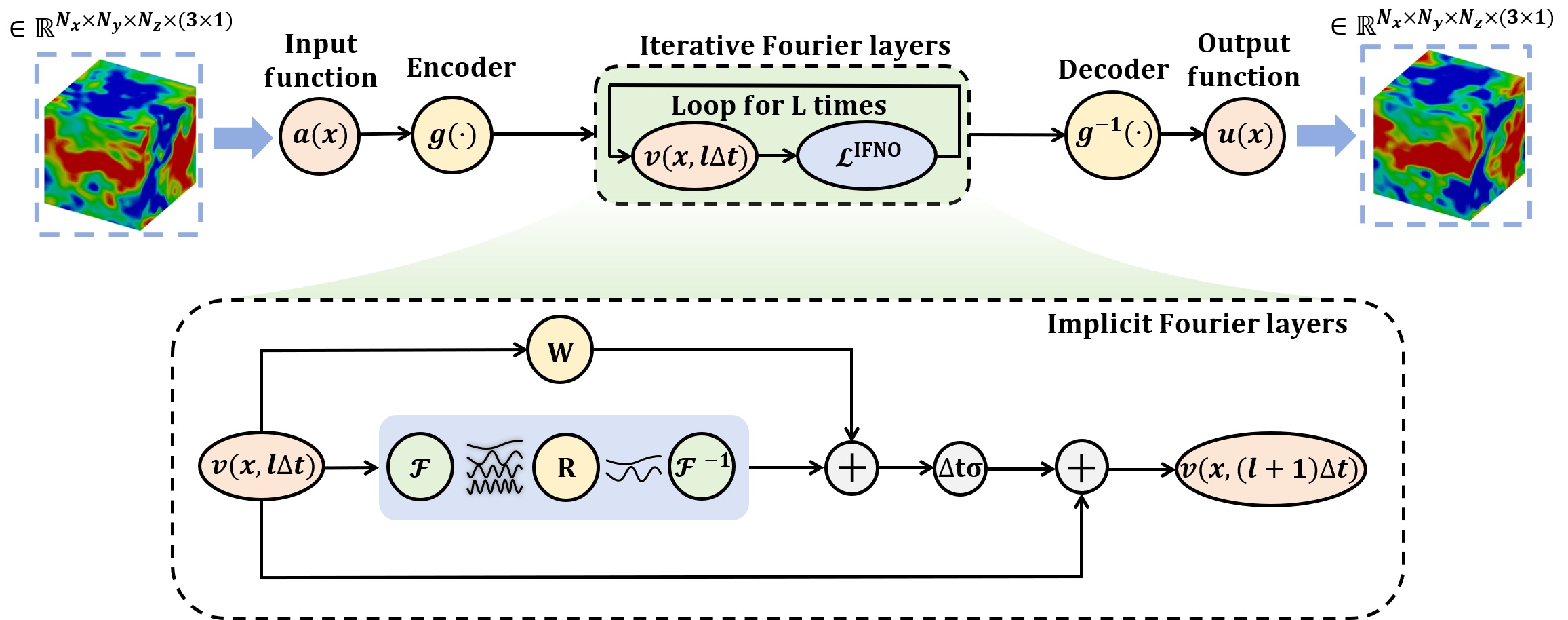}
	\caption{The architecture of the implicit Fourier neural operator (IFNO) model.}\label{fig:A.30}
\end{figure}

Consider a bounded open domain \( D \subset \mathbb{R}^d \), with Banach spaces \( \mathcal{A} = \mathcal{A}(D; \mathbb{R}^{d_a}) \) and \( \mathcal{U} = \mathcal{U}(D; \mathbb{R}^{d_u}) \) denoting the input and output function spaces, respectively \cite{beauzamy2011introduction}. The implicit Fourier neural operator (IFNO) aims to learn a parametric mapping \( \theta \in \Theta \) that approximates the operator \( \mathcal{A} \to \mathcal{U} \), where the optimal parameters \( \theta^{\dagger} \) are determined by error minimization over training data \cite{vapnik1999overview}. As illustrated in Fig.~\ref{fig:A.30}, the IFNO architecture consists of three main computational stages:

1) Encoder: The input function \( a \in \mathcal{A} \) is first lifted into a higher-dimensional latent space via a convolutional neural network (CNN) layer, yielding the encoded representation \( v(x, l\Delta t) = g(a(x)) \) \cite{li2021fourierneuraloperatorparametric}.

2) Implicit iterative kernel integration: The encoded representation is then updated iteratively for \( L \) time steps as follows \cite{YOU2022115296}:
\begin{equation}
v(x, (l+1)\Delta t) = \mathcal{L}^{IFNO}[v(x, l\Delta t)] := v(x, l\Delta t) + \Delta t \, \sigma\left(W v(x, l\Delta t) + \mathcal{K}(a; \phi) v(x, l\Delta t)\right), \quad \forall x \in D,
\label{eq:A1}
\end{equation}
where \( \mathcal{K}: \mathcal{A} \times \Theta_{\mathcal{K}} \to \mathcal{L}(\mathcal{U}(D; \mathbb{R}^{d_v}), \mathcal{U}(D; \mathbb{R}^{d_v})) \) denotes a bounded linear operator parameterized by \( \phi \in \Theta_{\mathcal{K}} \). The matrix \( W: \mathbb{R}^{d_v} \to \mathbb{R}^{d_v} \) represents a learnable linear transformation, and \( \sigma: \mathbb{R} \to \mathbb{R} \) is a nonlinear activation function \cite{YOU2022115296}.

3) Decoder: The final output \( u \in \mathcal{U} \) is obtained by projecting the latent variable from the last iteration back to the output space through a CNN layer: \( u(x) = g^{-1}(v(x, (l+1)\Delta t)) \), where \( g^{-1}: \mathbb{R}^{d_v} \to \mathbb{R}^{d_u} \) denotes the decoding map.

Let \( \mathcal{F} \) and \( \mathcal{F}^{-1} \) represent the Fourier and inverse Fourier transforms of a function \( f: D \to \mathbb{R}^{d_v} \), respectively. By expressing the kernel operator \( \mathcal{K} \) in Eq.~\eqref{eq:A1} as a convolution in Fourier space, we obtain:
\begin{equation}
\mathcal{K}(\phi)v(x, l\Delta t) = \mathcal{F}^{-1}(R_\phi \cdot \mathcal{F}(v(x, l\Delta t))), \quad \forall x \in D,
\label{eq:A2}
\end{equation}
where \( R_\phi \) is the Fourier transform of a periodic kernel \( \mathcal{K}: \bar{D} \to \mathbb{R}^{d_v \times d_v} \), parameterized by \( \phi \in \Theta_{\mathcal{K}} \). The Fourier modes are indexed by \( k \in \mathbb{Z}^d \), and a truncated Fourier series is employed with cutoff frequency index \( k_{\max} \), defined by \( Z_{k_{\max}} = \{k \in \mathbb{Z}^d : |k_j| \le k_{\max, j},\, j=1,\dots,d\} \).

Upon discretizing the domain \( D \) into \( n \in \mathbb{N} \) grid points, the function \( v(x, l\Delta t) \in \mathbb{R}^{n \times d_v} \) is transformed via the Fourier transform into \( \mathcal{F}(v(x, l\Delta t)) \in \mathbb{C}^{n \times d_v} \), and subsequently truncated to \( \mathcal{F}(v(x, l\Delta t)) \in \mathbb{C}^{k_{\max} \times d_v} \). The kernel \( R_\phi \in \mathbb{C}^{k_{\max} \times d_v \times d_v} \) is a complex-valued, learnable tensor representing Fourier coefficients \cite{li2021fourierneuraloperatorparametric}. Its application is computed as:
\begin{equation}
(R_\phi \cdot \mathcal{F}(v(x, l\Delta t)))_{k, l} = \sum_{j=1}^{d_v} R_{\phi, k, l, j} \, \mathcal{F}(v(x, l\Delta t))_{k, j}, \quad k = 1, \dots, k_{\max}, \quad l = 1, \dots, d_v.
\label{eq:A3}
\end{equation}

\subsection{Details of implicit U-Net enhanced Fourier neural operator}
\label{subappendix1.2}

\begin{figure}[ht!]\centering
	\includegraphics[width=1\textwidth]{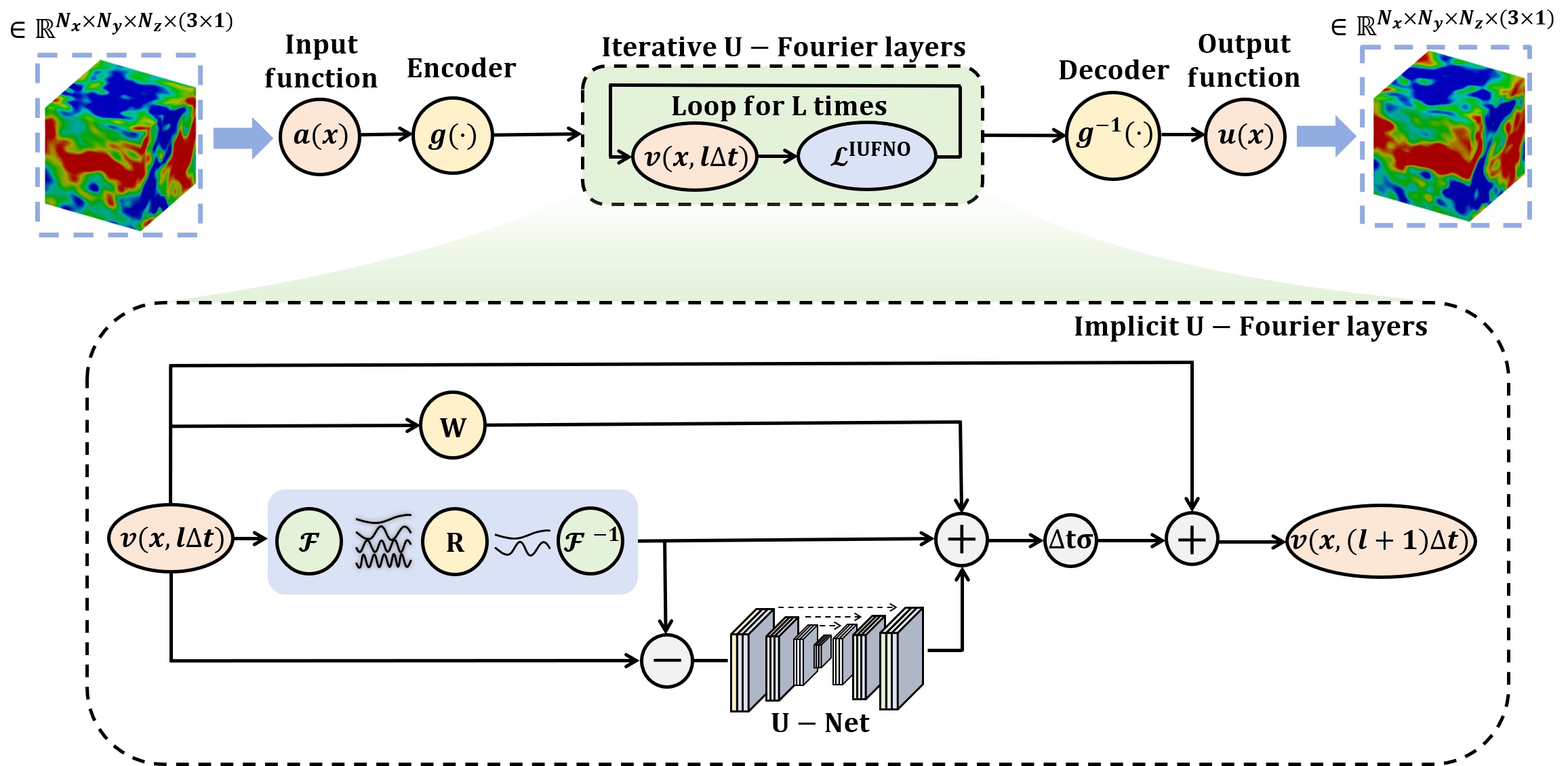}
	\caption{The architecture of the implicit U-Net enhanced Fourier neural operator (IUFNO) model.}\label{fig:A.31}
\end{figure}

The architecture of implicit U-Net enhanced Fourier neural operator (IUFNO) is illustrated in Fig.~\ref{fig:A.31}. Apart from the structure of the implicit Fourier layers, IUFNO retains the same architecture as IFNO in all other aspects. As shown in the figure, the update rule for the implicit U-Fourier layer can be formulated as follows:
\begin{equation}
v(x,(l+1)\Delta t)=\mathcal{L} ^{IUFNO}[v(x,l\Delta t)]: = v(x,l\Delta t)+\Delta t\sigma (c(x,l\Delta t)),\quad \forall x\in D,
\label{eq:A4}
\end{equation}
\begin{equation}
c(x,l\Delta t): = Wv(x,l\Delta t)+\mathcal{F} ^{-1}(R_{\phi}\cdot \mathcal{F}(v(x,l\Delta t)))+\mathcal{U}^{*}s(x,l\Delta t), \quad \forall x\in D,
\label{eq:A5}
\end{equation}
\begin{equation}
s(x,l\Delta t): = v(x,l\Delta t)-\mathcal{F}^{-1}(R_{\phi}\cdot \mathcal{F}(v(x,l\Delta t))), \quad \forall x\in D.
\label{eq:A6}
\end{equation}
Here, $c(x,l\Delta t) \in \mathbb{R}^{d_v}$ represents the global-scale flow features by combining the large-scale components extracted via FFT with the small-scale features $s(x,l\Delta t)$ learned by the U-Net network $\mathcal{U}^{*}$. The small-scale component $s(x,l\Delta t) \in \mathbb{R}^{d_v}$ is obtained by subtracting the large-scale representation from the complete field information $v(x,l\Delta t)$, as described in Eq.~\eqref{eq:A6}. The network $\mathcal{U}^{*}$ is based on the U-Net architecture, a convolutional neural network featuring a symmetric encoder-decoder structure. The encoder progressively reduces the spatial resolution to extract hierarchical features, while the decoder restores the resolution through upsampling operations, enabling the capture of both coarse and fine-scale information \cite{WEN2022104180,ronnebergerUnetConvolutionalNetworks2015a}. This design makes U-Net particularly effective in preserving multiscale spatial features, which is crucial for modeling the intricate structures of turbulent flows \cite{ronnebergerUnetConvolutionalNetworks2015a}. Moreover, by incorporating the implicit design, the IUFNO model significantly reduces the number of trainable parameters while achieving superior performance compared to IFNO in three-dimensional turbulence prediction tasks \cite{liLongtermPredictionsTurbulence2023a}.

\subsection{Details of factorized-implicit Fourier neural operator}
\label{subappendix1.3}

\begin{figure}[ht!]\centering
	\includegraphics[width=1\textwidth]{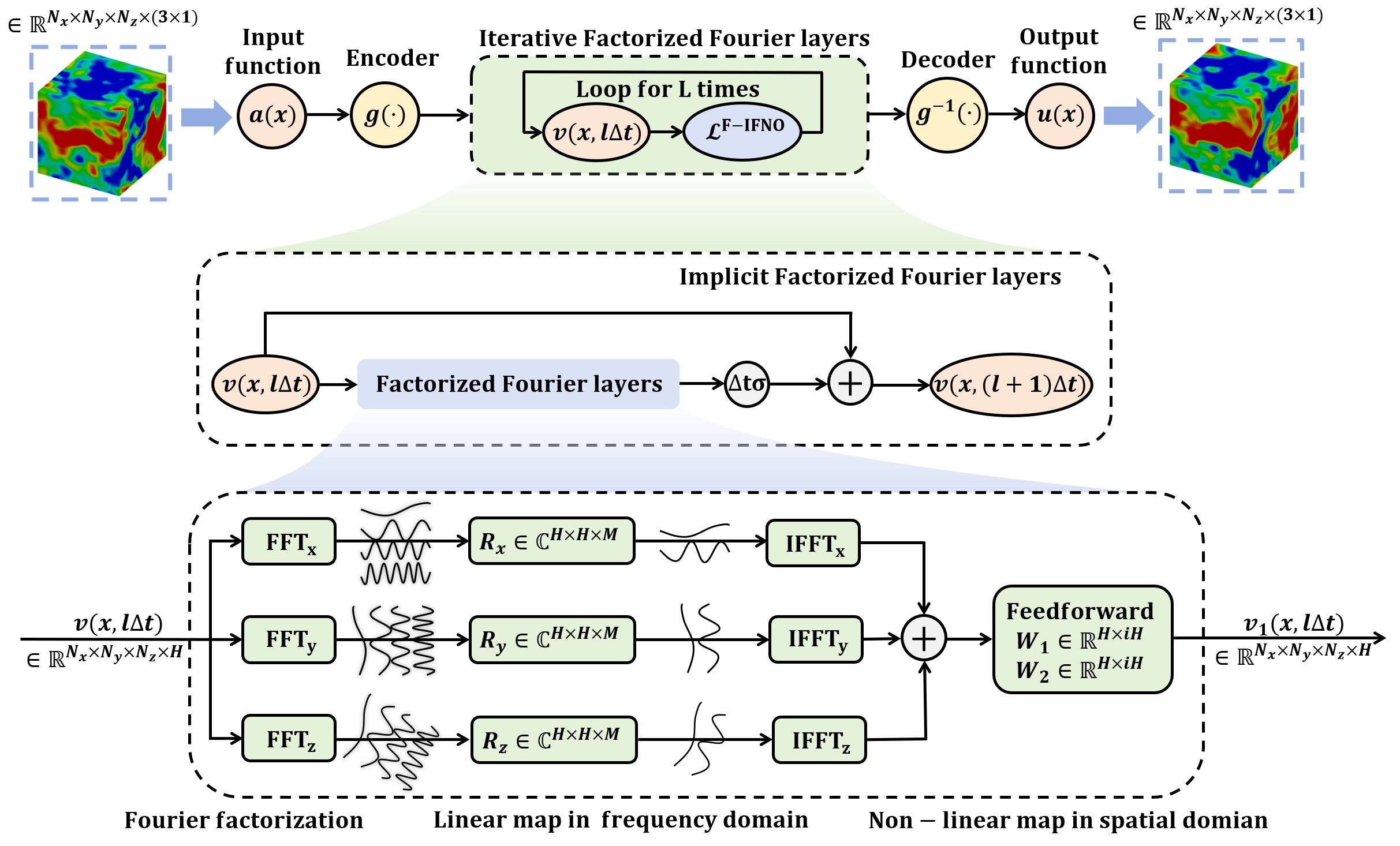}
	\caption{The architecture of the factorized-implicit Fourier neural operator (F-IFNO) model.}\label{fig:A.32}
\end{figure}

As illustrated in Fig.~\ref{fig:A.32}, the architecture of factorized-implicit Fourier neural operator (F-IFNO) closely resembles that of IFNO, with the only distinction being the structure of the implicit Fourier layers. The iterative update rule for F-IFNO is expressed as:
\begin{equation}
v(x, (l+1)\Delta t) = \mathcal{L}^{F-IFNO}[v(x, l\Delta t)] := v(x, l\Delta t) + \Delta t \, \sigma (\sum_{i\in d} \mathcal{F}_{i}^{-1}(R_{\phi i}\cdot \mathcal{F}_{i}(v(x, l\Delta t)))),
\quad \forall x \in D.
\label{eq:A7}
\end{equation}
Here, the replacement of $R_{\phi}$ with $R_{\phi i}$ in Fourier layers reduces parameter numbers from $\mathcal{O}(LH^2 M^d)$ to $\mathcal{O}(LH^2Md)$, where $H$ denotes the hidden dimension, $M$ is the number of retained Fourier modes, and $d$ represents the spatial dimensionality of the problem. The $\Delta t$ in Eq.~\eqref{eq:A7} represents the feature increment captured by the implicit factorized Fourier layer. By combining Fourier transform factorization with implicit iteration, F-IFNO achieves superior prediction accuracy compared to both IFNO and IUFNO, while maintaining a significantly smaller number of parameters, thereby reducing the overall computational cost.

\subsection{Details of factorized-implicit U-Net enhanced Fourier neural operator}
\label{subappendix1.4}

\begin{figure}[ht!]\centering
	\includegraphics[width=1\textwidth]{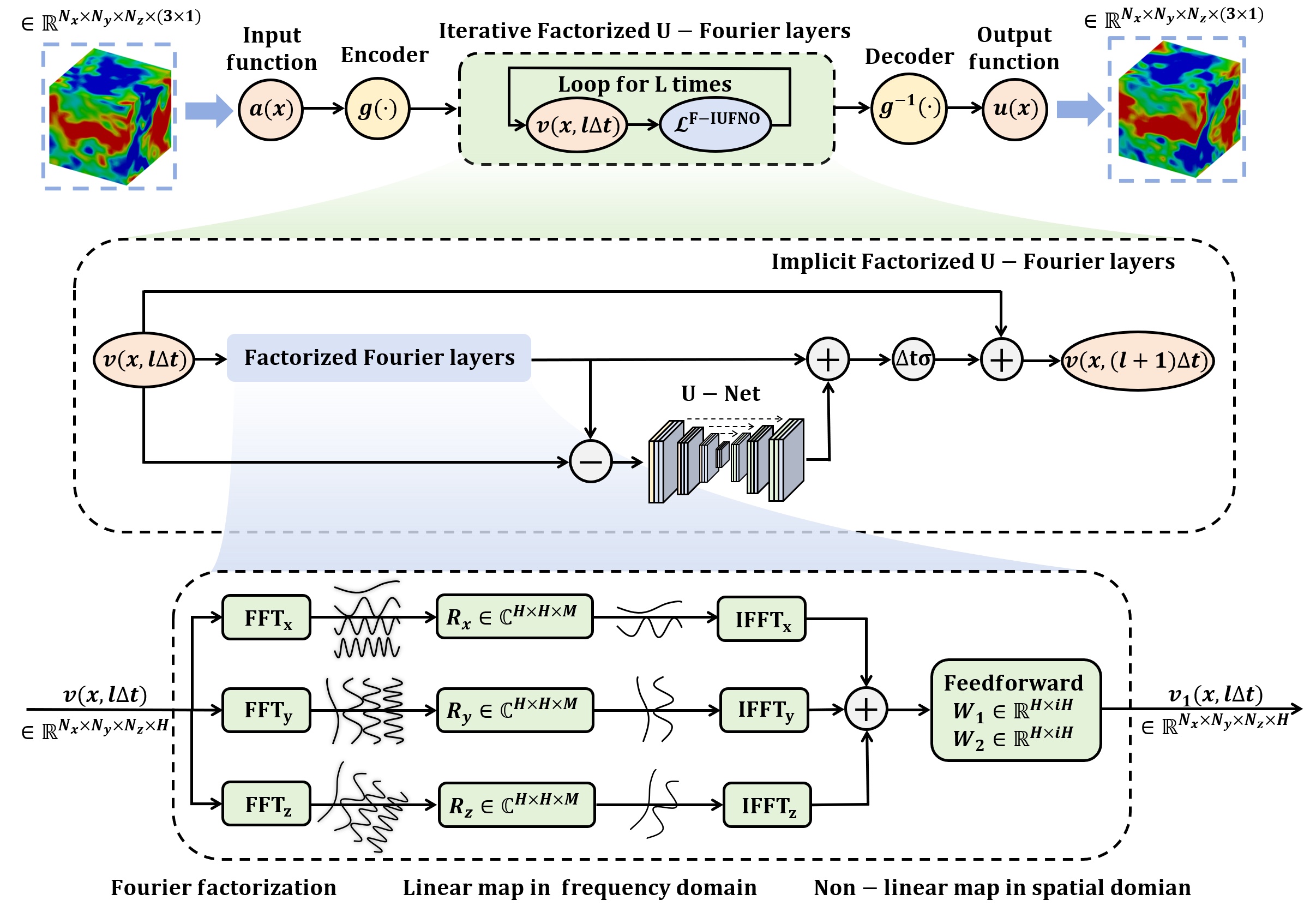}
	\caption{The architecture of the factorized-implicit U-Net enhanced Fourier neural operator (F-IUFNO) model.}\label{fig:A.33}
\end{figure}

The detailed architecture of factorized-implicit U-Net enhanced Fourier neural operator (F-IUFNO) is illustrated in Fig.~\ref{fig:A.33}. Both F-IFNO and F-IUFNO model the feature increment in an integrator-like manner. The only difference between them is that F-IUFNO includes an additional residual component in the implicit Fourier layers, specifically, a U-Net-based residual path designed to capture more small-scale information \cite{liLongtermPredictionsTurbulence2023a}. The iterative network update of F-IUFNO is given as:
\begin{equation}
v(x, (l+1)\Delta t) = \mathcal{L}^{F-IUFNO}[v(x, l\Delta t)] := v(x, l\Delta t) + \Delta t \, \sigma (\sum_{i\in d} \mathcal{F}_{i}^{-1}(R_{\phi i}\cdot \mathcal{F}_{i}(v(x, l\Delta t)))+\mathcal{U}^{*}s(x,l\Delta t)), \quad \forall x \in D,
\label{eq:A8}
\end{equation}
\begin{equation}
s(x,l\Delta t): = v(x,l\Delta t)-\sum_{i\in d} \mathcal{F}_{i}^{-1}(R_{\phi i}\cdot \mathcal{F}_{i}(v(x, l\Delta t))), \quad \forall x\in D.
\label{eq:A9}
\end{equation}
Here, $\mathcal{U}^{*}$ denotes the same U-Net architecture used in the original IUFNO. Similar to IUFNO, the term $s(x, l\Delta t)$ captures small-scale flow components through a residual learning strategy \cite{he2015deepresiduallearningimage}. While F-IUFNO and F-IFNO exhibit comparable performance in terms of prediction accuracy and stability, F-IFNO is significantly more efficient due to its reduced number of trainable parameters.

\bibliographystyle{elsarticle-num} 
\bibliography{reference.bib}



\end{document}